\documentclass[12pt]{caltech_thesis}
\usepackage{graphicx,enumitem}
\usepackage{hyperref}\hypersetup{hidelinks}
\usepackage{refcount}
\newcommand{\chapref}[1]{%
  \hyperref[#1]{\MakeUppercase{\romannumeral\numexpr\getrefnumber{#1}\relax}}%
}

\usepackage[numbers,sort&compress]{natbib}
\usepackage[sectionbib]{bibunits}
\defaultbibliographystyle{utphys-modified} 

\usepackage{bibentry}
\nobibliography*
\newcommand{\fullcite}[1]{\bibentry{#1}}

\usepackage{changepage}

\newenvironment{fullnote}[1][1.0\textwidth]
    {%
        \smallskip%
        \begin{adjustwidth}{0pt}{\dimexpr\textwidth-#1\relax}%
        \small%
    }
    {%
        \end{adjustwidth}%
        \smallskip%
    }
    
\newenvironment{Fullnote}[1][0.75\textwidth]
    {%
      \smallskip
      \begin{adjustwidth}
        {\dimexpr(\textwidth-#1)/2\relax}
        {\dimexpr(\textwidth-#1)/2\relax}%
    }
    {%
      \end{adjustwidth}%
      \smallskip
    }
    
\newenvironment{Fullnotew}[1][0.8\textwidth]
    {%
      \smallskip
      \begin{adjustwidth}
        {\dimexpr(\textwidth-#1)/2\relax}
        {\dimexpr(\textwidth-#1)/2\relax}%
    }
    {%
      \end{adjustwidth}%
      \smallskip
    }

\makeatletter
\newenvironment{statement}
  {%
    \par\noindent
    \setlength{\fboxsep}{8pt}
    \setlength{\fboxrule}{0.8pt}
    \begin{lrbox}{\@tempboxa}%
    \begin{minipage}{\dimexpr\linewidth-2\fboxsep-2\fboxrule\relax}%
    \centering
    \bfseries
\vspace{4pt}
  }
  {%
\vspace{4pt}
    \end{minipage}%
    \end{lrbox}%
    \fbox{\usebox{\@tempboxa}}%
  }
\makeatother

\usepackage{caption}
\setsecheadstyle{\normalfont\normalsize\bfseries}
\setsubsecheadstyle{\normalfont\normalsize\bfseries}

\newenvironment{dedication}
  {%
    \clearpage
    \thispagestyle{headings}%
    \begingroup
    \SingleSpacing
    \itshape\footnotesize
    \setlength{\parindent}{0pt}%
    \setlength{\parskip}{0pt}%
    \vspace*{0.30\textheight}%
    \centering
  }
  {%
    \par
    \vfill
    \endgroup
    \clearpage
  }
  
\usepackage{amsmath,amssymb,amsfonts}
\usepackage{multirow}
\usepackage{array,booktabs}

\usepackage{accents}
\usepackage{mathrsfs}
\usepackage{mathtools}
\usepackage{bm}

\usepackage[dvipsnames]{xcolor}
\definecolor{labelcolor}{RGB}{194, 175, 116}
\definecolor{rmkcolor}{RGB}{15,120,255}
\definecolor{bred}{RGB}{240,8,8}
\definecolor{indigo}{RGB}{63,72,204}
\definecolor{linkcolor}{RGB}{78,90,121}
\definecolor{bgrey}{RGB}{78,90,121}
\definecolor{lbgrey}{RGB}{101,117,159}

\definecolor{figred}{RGB}{212,61,34}
\definecolor{figbg}{RGB}{118,136,176}
\definecolor{figsky}{RGB}{170,188,255}

\definecolor{point}{RGB}{59,90,236}

\definecolor{skyblue}{RGB}{34,139,230}
\definecolor{kgreen}{rgb}{0,0.63,0.05}

\usepackage{hyphenat}
\usepackage[export]{adjustbox}
\setlist[itemize]{
    label=\adjustbox{scale=0.7}{$\bullet$}, itemsep=-3pt,topsep=0px
}

\setlist[enumerate]{
    itemsep=-1.5pt,topsep=4px
}

\usepackage[final]{showlabels} 

\usepackage{tikz}
\usetikzlibrary{calc} 
\usetikzlibrary{shapes.geometric} 
\usetikzlibrary{positioning} 
\usetikzlibrary{fit} 
\usepackage[a]{esvect} 
\tikzset{empty/.style = {inner sep = 0pt, outer sep = 0, minimum size = 0}}
\tikzset{b/.style = {inner sep = 2pt, outer sep = 4pt, minimum size = 12pt}}
\tikzset{c/.style = {inner sep = 2pt, outer sep = 4pt, minimum size = 12pt}}
\tikzset{w/.style = {inner sep = 1pt, outer sep = 2pt, minimum size = 12pt, anchor = west}}
\tikzset{s/.style = {inner sep = 2.5pt, outer sep =2.5pt, minimum size = 1pt, font = \small}}
\tikzset{lin/.style = {draw, line width = 0.5pt}}
\usepackage[export]{adjustbox}

\definecolor{sky}{RGB}{144,187,231}
\definecolor{OxyRed}{RGB}{190,70,62}
\definecolor{NitroBlue}{RGB}{91,122,239}
\definecolor{HydrogenLight}{RGB}{245,250,252}

\tikzset{
	line/.style = {draw, line width = 1.1pt, line cap = round, rounded corners = 0.2pt},
	bine/.style = {draw, line width = 1.1pt, line cap = round, rounded corners = 0.0pt, dotted, color=OxyRed},
	dine/.style = {draw, line width = 1.4pt, line cap = round, rounded corners = 0.0pt, double},
	D/.style = {below = -1.1pt, font = \footnotesize\bfseries\sffamily},
	U/.style = {above = -1.2pt, font = \footnotesize\bfseries\sffamily},
	L/.style = {left,  font=\footnotesize},
	R/.style = {right, font=\footnotesize},
	X/.style = {circle, draw=black, fill=HydrogenLight, inner sep=0pt, outer sep=0pt, minimum size=4.5pt, line width=1.1pt},
	Y/.style = {circle, draw=black, fill=NitroBlue, inner sep=0pt, outer sep=0pt, minimum size=4.5pt, line width=1.1pt}
}

\definecolor{feyntext}{RGB}{20,125,233}
\tikzset{
    t/.style = {
        inner sep = 1.5pt, outer sep =1.5pt, minimum size = 1pt,
        font = \small, text = feyntext
    },
    f/.style = {
        inner sep = 1.5pt, outer sep =1.5pt, minimum size = 1pt,
        font = \small, text = feyntext,
        label={[label distance=-4.5pt]right:{\footnotesize\color{feyntext}#1}}
    },
}

\usetikzlibrary{decorations.pathmorphing}
\tikzset{wiggly/.style = {draw, line width = 1.2pt,
    decorate, decoration={snake, 
        amplitude=1.5pt, segment length=5.0pt, post length=0pt, pre length=0pt
    }
}}
\tikzset{
    prop/.style = {
        draw, line width=1.2pt, 
        decoration = { 
            markings, 
            mark = at position 0.5 + 3.2pt with {
                \arrow{>[length=5pt,width=5pt]}
            }
        },
        postaction = {decorate}
    }
}

\usetikzlibrary{patterns}
\usetikzlibrary{patterns.meta}
\tikzdeclarepattern{
  name=mylines,
  parameters={
      \pgfkeysvalueof{/pgf/pattern keys/size},
      \pgfkeysvalueof{/pgf/pattern keys/angle},
      \pgfkeysvalueof{/pgf/pattern keys/line width},
  },
  bounding box={
    (0,-0.5*\pgfkeysvalueof{/pgf/pattern keys/line width}) and
    (\pgfkeysvalueof{/pgf/pattern keys/size},
0.5*\pgfkeysvalueof{/pgf/pattern keys/line width})},
  tile size={(\pgfkeysvalueof{/pgf/pattern keys/size},
\pgfkeysvalueof{/pgf/pattern keys/size})},
  tile transformation={rotate=\pgfkeysvalueof{/pgf/pattern keys/angle}},
  defaults={
    size/.initial=5pt,
    angle/.initial=45,
    line width/.initial=.4pt,
  },
  code={
      \draw [line width=\pgfkeysvalueof{/pgf/pattern keys/line width}]
        (0,0) -- (\pgfkeysvalueof{/pgf/pattern keys/size},0);
  },
}

\tikzset{
    wdot0/.style={
        circle,
        draw=black,
        fill=white,
        inner sep=0pt,
        outer sep=0pt,
        minimum size=6.2pt,
        line width=1.2pt
    },
    wdot/.style={
        wdot0,
        label={[overlay, label distance=-.8pt]below:{\footnotesize\color{feyntext}#1}}
    }
}
\tikzset{
    bdot0/.style = {
        circle,
        draw=black,
        fill=black,
        inner sep=0pt,
        outer sep=0pt,
        minimum size=6.2pt,
        line width=1.2pt,
        opacity=0.8,
        postaction={
            circle, draw=black, inner sep=0pt, outer sep=0pt, minimum size=6.2pt, line width=1.2pt
            ,pattern={mylines[size=1.5pt, line width=0.8pt, angle=45]},
            pattern color=white,
            opacity=1.0,
        }
    },
    bdot/.style={
        bdot0,
        label={[overlay, label distance=-.8pt]below:{\footnotesize\color{feyntext}#1}}
    }
}

\tikzset{
    ndot0/.style = {
        circle,
        draw=black,
        fill=black,
        inner sep=0pt,
        outer sep=0pt,
        minimum size=2.5pt,
        line width=1.2pt
    },
    ndot/.style={
        ndot0,
        label={[overlay, label distance=-.8pt]below:{\footnotesize\color{feyntext}#1}}
    }
}

\tikzset{
    wdot'0/.style={
        regular polygon, regular polygon sides=4,
        draw=black,
        fill=white,
        inner sep=0pt,
        outer sep=0pt,
        minimum size=7.5pt,
        line width=1.2pt
    },
    wdot'/.style={
        wdot'0,
        label={[overlay, label distance=-.8pt]below:{\footnotesize\color{feyntext}#1}}
    }
}
\tikzset{
    bdot'0/.style = {
        regular polygon, regular polygon sides=4,
        draw=black,
        fill=black,
        inner sep=0pt,
        outer sep=0pt,
        minimum size=7.5pt,
        line width=1.2pt,
        opacity=0.8,
        postaction={
            circle, draw=black, inner sep=0pt, outer sep=0pt, minimum size=6.0pt, line width=1.2pt
            ,pattern={mylines[size=1.5pt, line width=0.8pt, angle=45]},
            pattern color=white,
            opacity=1.0,
        }
    },
    bdot'/.style={
        bdot'0,
        label={[overlay, label distance=-.8pt]below:{\footnotesize\color{feyntext}#1}}
    }
}

\tikzset{
    ndot'0/.style = {
        regular polygon, regular polygon sides=4,
        draw=black,
        fill=black,
        inner sep=0pt,
        outer sep=0pt,
        minimum size=3pt,
        line width=1.2pt
    },
    ndot'/.style={
        ndot'0,
        label={[overlay, label distance=-.8pt]below:{\footnotesize\color{feyntext}#1}}
    }
}

\tikzset{
    swdot0/.style={
        circle,
        draw=black,
        fill=white,
        inner sep=0pt,
        outer sep=0pt,
        minimum size=4.5pt,
        line width=1.2pt
    },
    swdot/.style={
        swdot0,
        label={[overlay, label distance=-.8pt]below:{\footnotesize\color{feyntext}#1}}
    }
}
\tikzset{
    sbdot0/.style = {
        circle,
        draw=black,
        fill=black,
        inner sep=0pt,
        outer sep=0pt,
        minimum size=4.5pt,
        line width=1.2pt,
        opacity=0.8,
        postaction={
            circle, draw=black, inner sep=0pt, outer sep=0pt, minimum size=4.5pt, line width=1.2pt
            ,pattern={mylines[size=1.5pt, line width=0.8pt, angle=45]},
            pattern color=white,
            opacity=1.0,
        }
    },
    sbdot/.style={
        sbdot0,
        label={[overlay, label distance=-.8pt]below:{\footnotesize\color{feyntext}#1}}
    }
}

\tikzset{
    ta/.style={
        empty,
        label={[overlay, label distance=-.65pt]above:{\footnotesize\color{gray}#1}}
    }
}

\usetikzlibrary{arrows, decorations.markings}
\makeatletter
\def\pgf@lib@dec@parsenum#1{%
    \gdef\pgf@lib@dec@computed@width{0 pt}%
    \tsx@pgf@lib@dec@parsenum#1+endmarker+%
    \ifdim\pgf@lib@dec@computed@width<0pt\relax%
        \pgfmathparse{\pgfdecoratedpathlength\pgf@lib@dec@computed@width}
        \edef\pgf@lib@dec@computed@width{\pgfmathresult pt}%
    \fi%
}

\def\tsx@pgf@lib@dec@parsenum@endmarker{endmarker}

\def\tsx@pgf@lib@dec@parsenum#1+{
    \def\temp{#1}%
    \ifx\temp\tsx@pgf@lib@dec@parsenum@endmarker%
    \else%
        \tsx@pgf@lib@dec@parsenum@one{#1}%
        \expandafter\tsx@pgf@lib@dec@parsenum%
    \fi%
}

\def\tsx@pgf@lib@dec@parsenum@one#1{%
  \pgfmathparse{#1}%
  \ifpgfmathunitsdeclared%
    \pgfmathparse{\pgf@lib@dec@computed@width + \pgfmathresult pt}%
  \else%
    \pgfmathparse{\pgf@lib@dec@computed@width + \pgfmathresult*\pgfdecoratedpathlength*1pt}%
  \fi%
  \edef\pgf@lib@dec@computed@width{\pgfmathresult pt}%
}
\makeatother

\usetikzlibrary{arrows.meta}
\tikzset{
    lin/.style = {
        draw, line width=1.2pt
    },
    dlin/.style = {
        draw, line width=1.2pt, dotted
    }
}
\tikzset{
    prop/.style = {
        draw, line width=1.2pt, 
        decoration = { 
            markings, 
            mark = at position 0.5 + 3.2pt with {
                \arrow{>[length=5pt,width=5pt]}
            }
        },
        postaction = {decorate}
    },
    dprop/.style = {
        draw, line width=1.2pt, dotted,
        decoration = { 
            markings, 
            mark = at position 0.5 + 3.2pt with {
                \arrow{>[length=5pt,width=5pt]}
            }
        },
        postaction = {decorate}
    }
}

\tikzset{cdot/.style = {circle, draw=black, fill=black, inner sep=0pt, outer sep=0pt, minimum size=2.0pt, line width=1.2pt}}

\newcommand{\Fig}[1]{Fig.\,\ref{#1}}
\providecommand{\fref}{}\renewcommand{\fref}{\Fig}
\newcommand{\frefs}[2]{Figs.\,\ref{#1} and \ref{#2}}
\newcommand{\oldeqref}[1]{Eq.\,(\ref{#1})}
\renewcommand{\eqref}[1]{Eq.\,(\ref{#1})}
\newcommand{\eqrefs}[2]{Eqs.\,(\ref{#1}) and (\ref{#2})}
\newcommand{\eqrefss}[3]{Eqs.\,(\ref{#1}), (\ref{#2}), and (\ref{#3})}
\newcommand{\eqrefsss}[4]{Eqs.\,(\ref{#1}), (\ref{#2}), (\ref{#3}), and (\ref{#4})}
\newcommand{\eqrefsor}[2]{Eqs.\,(\ref{#1}) or (\ref{#2})}

\newcommand{\eqrefsto}[2]{Eqs.\,(\ref{#1}) to (\ref{#2})}

\newcommand{\Chap}[1]{Chap.\,\chapref{#1}}
\newcommand{\Chaps}[2]{Chaps.\,\chapref{#1} and \chapref{#2}}

\newcommand{\Sec}[1]{Sec.\,\ref{#1}}
\newcommand{\Secs}[2]{Secs.\,\ref{#1} and \ref{#2}}
\newcommand{\Secss}[3]{Secs.\,\ref{#1}, \ref{#2}, and \ref{#3}}
\newcommand{\Secsto}[2]{Secs.\,\ref{#1} to \ref{#2}}
\newcommand{\Secsor}[2]{Secs.\,\ref{#1} or \ref{#2}}

\newcommand{\App}[1]{App.\,\ref{#1}}
\newcommand{\rcite}[1]{Ref.\,\cite{#1}}
\newcommand{\rrcite}[1]{Refs.\,\cite{#1}}

\newcommand{\transition}[1]{\,\quad\adjustbox{scale=0.95}{\text{#1}}\quad\,}

\DeclareMathOperator{\diag}{diag}
\DeclareMathOperator{\Tr}{tr}
\def\Re{{\operatorname{Re}}}

\DeclareMathOperator{\sinc}{sinc}

\DeclareMathOperator{\Exp}{Exp}
\DeclareMathOperator{\Lie}{Lie}

\def\mem{\hspace{0.1em}}
\def\hem{\hspace{0.05em}}
\def\nem{\hspace{-0.1em}}
\def\hnem{\hspace{-0.05em}}
\def\hhem{\hspace{0.025em}}
\def\hhnem{\hspace{-0.025em}}
\def\hhhem{\hspace{0.0125em}}

\def\blank{{\,\,\,\,\,}}

\def\iq{{{\implies}\quad}}
\def\qiq{{\quad\implies\quad}}
\def\qfq{{\quad\iff\quad}}

\def\minie{{\tfrac{1}{2}}}

\def\a{\alpha}
\def\b{\beta}
\def\c{\gamma}
\def\g{\gamma}
\def\d{\delta}
\def\e{\epsilon}
\def\ve{\varepsilon}
\def\m{\mu}
\def\n{\nu}
\def\r{\rho}
\def\s{\sigma}
\def\k{\kappa}
\def\l{\lambda}
\def\t{\tau}
\def\x{\xi}
\def\vf{\varphi}
\def\vth{\vartheta}
\def\L{\Lambda}

\def\bgamma{{\bar{\gamma}}}

\def\be{{\bar{\epsilon}}}

\def\bpsi{{\smash{\bar{\psi}}\kern0.02em\vphantom{\psi}}}

\def\bmu{{\bar{\mu}}}

\def\bZ{{\bar{Z}}}
\def\bW{{\bar{W}}}

\def\bR{\bar{R}}

\def\bz{{\bar{z}}}

\def\mathe{{\scalebox{1.025}[1]{$\mathrm{e}$}}}
\def\pt{{\hnem\nem\mathscr{P}}}

\def\rmA{{\mathrm{A}}}
\def\rmB{{\mathrm{B}}}

\def\mwedge{{\mem\wedge\mem\hhem}}
\def\swedge{{\mem{\wedge}\,}}
\def\moplus{{\mem\oplus\mem}}
\def\motimes{{\mem\otimes\mem}}
\def\modot{{\mem\odot\mem}}
\def\mtensor{\motimes}

\def\mplus{{\mem+\mem}}
\def\mminus{{\mem-\mem}}
\def\mtimes{{\mem\times\mem}}
\def\mdot{{\mem\cdot\mem}}

\def\mt{{\mem\times}}

\def\mcdots{{\mem\cdots\mem}}
\def\msubset{{\,\subset\,}}

\def\da{{\dot{\a}}}
\def\db{{\dot{\b}}}
\def\dc{{\dot{\c}}}
\def\dd{{\dot{\d}}}

\newcommand{\wrap}[1]{{\smash{#1}\vphantom{\b}}}

\def\sigu{\hat{\s}}
\def\sigd{\check{\s}}

\renewcommand{\o}{o}
\def\i{\iota}
\def\bo{{\bar{o}}}

\def\lsq{{
    \kern-0.037em
    \adjustbox{scale=0.99,valign=c}{$
        {\lfloor \llap{\reflectbox{\rotatebox[origin=c]{180}{$\lfloor$}}}}
    $}
    \kern-0.04em
}}
\def\rsq{{
    \kern-0.04em
    \adjustbox{scale=0.99,valign=c}{$
        {\rlap{\reflectbox{\rotatebox[origin=c]{180}{$\rfloor$}}} \rfloor}
    $}
    \kern-0.037em
}}

\newcommand{\lp}[1]{{
    \langle #1 \rangle
}}
\newcommand{\rp}[1]{{
    \lsq #1 \rsq
}}

\newcommand{\lket}[1]{
    |\hem #1 \rangle
}
\newcommand{\rket}[1]{
    |\hem #1 \rsq
}

\def\rambda{\tilde{\lambda}}
\def\tmu{\tilde{\mu}}

\def\id{{\rlap{1} \hskip 1.6pt \adjustbox{scale=1.1}{1}}}

\newcommand{\dbar}{
    d\kern-.20em\makebox[0pt][l]{$\bar{}$}\kern.20em
}
\newcommand{\deltabar}{
    \delta\kern-.20em\makebox[0pt][l]{$\bar{}$}\kern.20em
}

\newcommand{\bigbig}[1]{\big(\mem{#1}\mem\big)}
\newcommand{\BB}[1]{\Big(\,{#1}\,\Big)}
\newcommand{\bb}[1]{\bigg(\,{#1}\,\bigg)}

\newcommand{\lrp}[1]{\left(\,{#1}\,\right)}

\newcommand{\bbsq}[1]{\bigg[\,{#1}\,\bigg]}
\newcommand{\lrsq}[1]{\left[\,{#1}\,\right]}

\newcommand{\act}[1]{[\,{#1}\,]}

\def\kerr{{\smash{\text{$\kern-0.075em\sqrt{\text{Kerr\hem}}$}}}}

\def\Kerr{{\smash{\text{Kerr}}}}

\def\AHH{{\mathrm{AHH}}}

\def\Pleb{{{Pleba\'nski}}}

\def\Sing{{\textsc{\small Singlet}}}
\def\Fund{{\textsc{\small Fund.}}}
\def\Adj{{\textsc{\small Adjoint}}}

\def\Co{{\textsc{\small Co}}}

\def\Ric{{\mathrm{Ric}}}

\usepackage[outline]{contour}

\definecolor{thy}{RGB}{64,69,77}
\contourlength{0.3pt}

\newcommand{\thy}[1]{\text{\contour{thy}{\color{thy}#1}}}

\def\SDYM{{\thy{SDYM}}}
\def\SDGR{{\thy{SDGR}}}

\def\NLSM{{\thy{NLSM}}}
\def\BI{{\thy{BI}}}
\def\SG{{\thy{SG}}}

\def\BAS{{\thy{BAS}}}
\def\YM{{\thy{YM}}}
\def\GR{{\thy{GR}}}
\def\GRf{{\thy{GR${}^*$}}}

\def\CS{{\thy{CS}}}

\DeclareMathOperator{\vol}{vol}
\def\Gauge{{\text{Gauge}}}
\newcommand{\D}[1]{\mathcal{D}\hnem{#1}\,}

\newcommand{\expval}[1]{
    \big\langle\hem{
        #1
    }\hem\big\rangle
}

\let\oldexp\exp
\renewcommand{\exp}{\oldexp\nem\hhnem}

\def\Pexp{\mathrm{P}\kern-0.1em\exp}

\def\bPexp{\bar{\mathrm{P}}\kern-0.1em\exp}

\def\bP{
    \mathrlap{\kern0.05em%
    \adjustbox{scale=0.8,raise=0.12em}{
        $\bar{\vphantom{\mathrm{P}}}$
    }}\mathrm{P}\kern-0.1em%
}

\def\M{{\mathcal{M}}}
\def\N{{\mathcal{N}}}

\def\P{{\mathcal{P}}}
\def\CC{{\mathcal{C}}}
\def\S{{\mathcal{S}}}
\def\V{{\mathcal{V}}}
\def\mflat{\mathbb{M}}
\def\F{{\mathcal{F}}}
\def\PT{\mathbb{P}\mathcal{T}}
\def\pt{\mathbb{PT}}
\def\CP{\mathbb{CP}}

\def\O{\mathcal{O}}
\def\A{\mathcal{A}}

\def\tk{\tilde{\kappa}}

\def\tzeta{\tilde{\zeta}}
\def\tchi{\tilde{\chi}}

\def\tdo{\protect\tilde{o}}
\def\ti{\protect\tilde{\iota}}

\def\tm{{\tilde{m}}}

\def\te{{\tilde{\epsilon}}}

\def\tz{\widetilde{z}}

\def\ta{\widetilde{\a}}
\def\tb{\widetilde{\b}}
\def\tc{\widetilde{\c}}

\def\tF{\widetilde{F}}

\def\tC{\tilde{C}}

\def\vex{\vec{x}}
\def\vea{\vec{a}}

\newcommand{\ad}[1]{\mathrm{ad}_{#1}}
\newcommand{\Ad}[1]{\mathrm{Ad}_{#1}}
\newcommand{\coad}[1]{\mathrm{ad}^*_{#1}}
\newcommand{\coAd}[1]{\mathrm{Ad}^*_{#1}}

\def\g{\mathsf{g}}
\def\G{\mathsf{G}}

\def\U{\mathsf{U}}

\def\su{\mathsf{su}}
\def\SU{\mathsf{SU}}

\def\so{\mathsf{so}}
\def\SO{\mathsf{SO}}

\def\PSO{\mathsf{PSO}}

\def\diff{\mathsf{diff}}
\def\Diff{\mathsf{Diff}}

\def\sdiff{\mathsf{sdiff}}
\def\SDiff{\mathsf{SDiff}}

\def\iso{\mathsf{iso}}

\def\GL{\mathsf{GL}}

\def\sl{\mathsf{sl}}
\def\SL{\mathsf{SL}}

\def\Sp{\mathsf{Sp}}

\def\Spin{\mathsf{Spin}}

\def\npad{\kern0.4em}
\def\mpad{\kern0.525em}
\def\wpad{\kern1.75em}
\def\hpad{\kern2.2em}

\def\Cinfty{C^\infty\hnem}
\def\R{\mathbb{R}}
\def\C{\mathbb{C}}
\def\Z{\mathbb{Z}}

\newcommand{\cont}[2]{\big\langle\hem{#1},\hem{#2}\hem\big\rangle}
\def\Tstar{T^*\nem}

\newcommand{\Cont}[2]{\bigg\langle\hem{#1},\hem{#2}\hem\bigg\rangle}

\def\lb{\{\kern-0.15em\{}
\def\rb{\}\kern-0.15em\}}
\newcommand{\pb}[2]{{\{\hem{#1},{#2}\hem\}}}
\newcommand{\dpb}[2]{\lb\hem{#1},{#2}\hem\rb}

\newcommand{\comm}[2]{[\hem{#1},{#2}\hem]}

\newcommand{\TwoByTwoMatrixWide}[4]{
    \lrp{\begin{array}{cc}
        \hphantom{\wpad}\mathclap{
            #1
        }\hphantom{\wpad}&\hphantom{\wpad}\mathclap{
            #2
        }\hphantom{\wpad}
        \\
        \hphantom{\wpad}\mathclap{
            #3
        }\hphantom{\wpad}&\hphantom{\wpad}\mathclap{
            #4
        }\hphantom{\wpad}
    \end{array}}
}

\newcommand{\TwoByTwoMatrixCustom}[5]{
    \lrp{\begin{array}{cc}
        \hphantom{\kern#1}\mathclap{
            #2
        }\hphantom{\kern#1}&\hphantom{\kern#1}\mathclap{
            #3
        }\hphantom{\kern#1}
        \\
        \hphantom{\kern#1}\mathclap{
            #4
        }\hphantom{\kern#1}&\hphantom{\kern#1}\mathclap{
            #5
        }\hphantom{\kern#1}
    \end{array}}
}

\newcommand{\hla}[1]{{\color[RGB]{220,10,10}{}#1}}
\newcommand{\hlb}[1]{{\color[RGB]{226,108,190}{}#1}}

\newcommand{\hls}[1]{{\color[RGB]{0,137,186}{}#1}}

\newcommand{\hlx}[1]{{\color[RGB]{10,110,244}{}#1}}
\newcommand{\hly}[1]{{\color[RGB]{95,103,234}{}#1}}
\newcommand{\hlz}[1]{{\color[RGB]{161,92,192}{}#1}}

\def\Qm{{Q^\star}}

\def\hx{\hat{\xi}}

\def\inn{{\:\in\:}}
\def\eqq{{\:=\:}}
\def\too{{\:\to\:}}

\def\vph{\vphantom{\big|}}
\def\vp{\vphantom{0}}

\let\oldfrac\frac
\renewcommand{\frac}[2]{\oldfrac{#1}{#2\vph}}

\let\oldsqrt\sqrt
\renewcommand{\sqrt}[1]{\oldsqrt{#1\vphantom{|}}}

\def\tell{\tilde{\ell}}
\def\txi{\tilde{\xi}}

\def\zag{{\includegraphics[scale=1.2,valign=c]{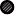}}}
\def\zig{{\includegraphics[scale=1.2,valign=c]{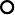}}}

\newif\ifToggleMacros
\ToggleMacrosfalse  
\ifToggleMacros
    \newcommand{\jh}[1]{{\color[RGB]{240,25,25} {JH:#1}}}
\else
   \newcommand{\jh}[1]{}
\fi

\def\sJ{{J^\star}{}}
\def\sT{{T^\star}{}}

\def\ta{{\smash{\tilde{a}}}{}}
\def\tb{{\smash{\tilde{b}}}{}}
\def\tc{{\smash{\tilde{c}}}{}}
\def\td{{\smash{\tilde{d}}}{}}
\def\te{{\smash{\tilde{e}}}{}}
\def\tf{{\smash{\tilde{f}}}{}}

\def\tg{\tilde{\mathsf{g}}}

\def\gk{{\mathsf{g}_{\adjustbox{raise=-0.185em,scale=0.75}{%
    \kern-0.08em\textbf{\itshape ?}}}%
    \kern-0.08em%
}}

\DeclareMathOperator{\link}{link}
\DeclareMathOperator{\intersect}{int}

\def\cen{{\alpha}}
\def\rep{{\hnem\rho}}
\def\fund{{\text{fund}}}

\def\Uinv{U_\cen\kern-0.3em\mathrlap{\adjustbox{raise=0.11em}{$^{-1}$}}\kern0.95em}

\def\GN{{G_\mathrm{N}}}

\def\got{{\mathfrak{g}}}
\def\igot{(\mathfrak{g}^{-1})}

\def\vn{\lambda}

\def\Cech{{\v{C}ech}}
\def\cC{\check{C}}
\def\cZ{\check{Z}}

\def\cH{\check{H}}

\def\CU{U}

\usepackage{dsfont}

\def\E{\mathds{E}}
\def\J{\hem\mathds{J}\hem}
\def\K{\mathds{K}}
\def\KA{\mathds{A}}

\def\Cart{\mathds{C}}

\def\vep{\vec{p}\mem}
\def\vex{\vec{x}}

\def\vp{\vec{\partial}}

\def\rhat{\hem\hat{r}}

\def\sprime{{\mathrlap{\smash{{}^\prime}}{\hspace{0.05em}}}}

\def\robl{{r}}
\def\rprol{{\tilde{r}}}

\def\R{\mathbb{R}}
\def\C{\mathbb{C}}

\def\mflat{\mathbb{M}}

\def\txi{{\protect\tilde{\xi}}}
\def\tzeta{{\protect\tilde{\zeta}}}

\def\tz{{\tilde{z}}}

\def\tm{{\tilde{m}}}

\def\tchi{\tilde{\chi}}
\def\te{\tilde{\epsilon}}

\def\tell{\smash{\tilde{\ell}}}
\def\trho{\smash{\tilde{\rho}}}
\def\ttheta{\smash{\tilde{\theta}}}
\def\teta{\smash{\tilde{\eta}}}

\newcommand{\co}[1]{c^{\hhem#1}}
\newcommand{\cop}[1]{ \mathrlap{c'}\phantom{c}^{\kern0.24em#1} }

\def\acX{\acute{X}}
\def\acY{\acute{Y}}
\def\acQ{\acute{Q}}

\def\i{{\iota}}

\def\Del{\mathit{\Delta}}
\def\tU{\widetilde{U}}

\def\heav{{\adjustbox{raise=-1.5pt,scale=0.7}{\nem$\mathscr{H}$}}}

\def\Et{\mathfrak{E}}

\def\dv{\mathrm{div}}

\def\hx{\hat{x}}
\def\hy{\hat{y}}
\def\hp{\hat{p}}

\def\hdelta{\hat{\delta}}

\def\ga{{\text{G2A}}}

\def\trambda{\smash{\tilde{\lambda}}}

\def\MT{\mathcal{MT}}

\let\del\undefined
\newcommand{\del}[2]{
    \delta^{(4)}\hnem\big(\hem\hhhem{
        {#1}
        {\mem-\mem}
        {#2}
    }\hem\big)
}
\newcommand{\dell}[2]{
    \delta^{(4)}\hnem\big(\hem\hhhem{
        {#1}
        - 
        {#2}
    }\hem\big)
}

\newcommand{\inv}[1]{\mathrlap{\smash{\adjustbox{raise=0.11em}{$^{\mem-1}$}}}_{\mathrlap{#1}\hphantom{\mem-1}}\hnem}

\def\blambda{{\bar{\lambda}}}

\def\mhat{{\smash{\widehat{\mathbb{M}}}}}
\def\mtflat{{\mathbb{MT}}}

\def\tflat{{\mathbb{T}}}
\def\kflat{{\mathbb{K}}}

\def\Mhat{\smash{\widehat{\M}}}

\newcommand{\Ket}[1]{{\hem\big|\hem{#1}\big\rangle}}
\newcommand{\Bra}[1]{{\big\langle{#1}\hem\big|\hem}}
\newcommand{\BraKet}[2]{{\big\langle{#1}\hem\big|\hem{#2}\big\rangle}}

\def\ups{\mathbf{P}\kern-0.0025em}
\def\ps{\mathds{P}}

\def\stheta{\smash{
    \accentset{\adjustbox{scale=0.45}{$\sqrt{}$}}{\theta}\hhnem{}
}}

\def\pffp{{p\hem\vf_3\vf_4p}}
\def\yffp{{y\hem\vf_3\vf_4p}}
\def\ryffp{{y\hem\vf_4\vf_3p}}
\def\yffy{{y\hem\vf_3\vf_4y}}

\def\pffpPP{{p\hem\vf_3^+\nem\hnem\vf_4^+\nem\hnem p}}
\def\pffpPM{{p\hem\vf_3^+\nem\hnem\vf_4^-\nem\hnem p}}
\def\pffpMM{{p\hem\vf_3^-\nem\hnem\vf_4^-\nem\hnem p}}

\def\yffpPP{{p\hem\vf_3^+\nem\hnem\vf_4^+\nem\hnem p}}
\def\ryffpPP{{p\hem\vf_4^+\nem\hnem\vf_3^+\nem\hnem p}}

\def\yffpPM{{y\hem\vf_3^+\nem\hnem\vf_4^-\nem\hnem p}}

\def\pffpPPMM{{p\hem\vf_3^\pm\nem\hnem\vf_4^\pm\nem\hnem p}}

\newcommand{\delvec}[1]{{\d{#1}\mem \frac{\partial}{\partial{#1}}}}

\def\doz{{\dot{0}}}
\def\diz{{\dot{1}}}

\def\LL{\mathscr{L}}

\def\z{\zeta}
\def\dz{\delta\zeta}

\def\Q{\mathcal{Q}}
\def\Hilb{\mathcal{H}}
\DeclareMathOperator{\Ops}{Ops}

\def\tV{\tilde{V}}

\def\ein{e}

\def\Zslash{{Z\mathllap{\adjustbox{scale=1.0,valign=c,raise=0.11em}{\scalebox{0.5}[1.0]{$-$}}\kern0.163em}}}
\def\bZslash{{\bar{Z}\mathllap{\adjustbox{scale=1.0,valign=c,raise=0.11em}{\scalebox{0.5}[1.0]{$-$}}\kern0.163em}}}

\begin{document}\title{\bfseries
    Reimagining Gravity:\\[0.3\baselineskip]
    Generalized Symmetries,
    Double Copy,\\[0.165\baselineskip]
    and Spinning Black Holes
}
\author{\vphantom{.}\\[-0.88\baselineskip]
    Joon-Hwi Kim
}

\degreeaward{Doctor of Philosophy}
\university{California Institute of Technology}
\address{Pasadena, California}
\unilogo{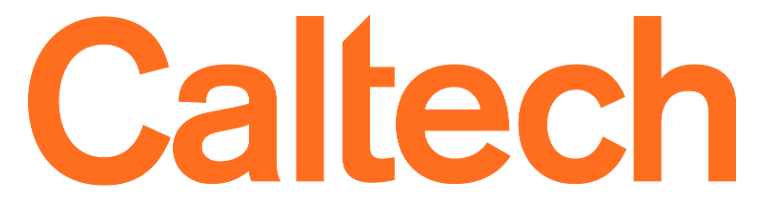}
\copyyear{2027} 
\defenddate{June 29, 2026} 

\orcid{0000-0002-5474-123X}

\rightsstatement{All rights reserved}

\maketitle[logo]

\begin{dedication}
    To my mother, who gave me a canvas of imagination.
\end{dedication}

\begin{acknowledgements}

I would like to thank my advisor,
Clifford Cheung,
for his invaluable guidance,
steadfast support,
and genuine faith in my potential.
I am deeply grateful to Cliff for
his unwavering generosity with his time;
Cliff always welcomed me into his office with an open door,
dedicating countless days to boundless discussions that profoundly enriched my understanding of physics.
I will cherish the moments when 
our creativity resonated;
together, we let our imagination run wild,
skied
through mountains covered with
snowy trees and fiery loops,
and traveled through space portals by storming ahead with field redefinitions.
Doing physics with Cliff was
not just a collaboration, but
an adventure.
I thank Cliff for the most interesting physics discussions I have ever had,
and the contagious $1/\epsilon$ pole in his enthusiasm.
Cliff also showed me how to build anything from scratch,
based on logic and minimal assumptions,
and how to think and express ideas clearly.
More\:than\:an exceptional mentor,
I am truly fortunate to know Cliff as a remarkable human being.
Choosing Cliff as my advisor is one of the best decisions of my life.

\medskip

I would like to thank
my collaborators on the projects
included in this thesis.

I am grateful to
Cliff, Ira Rothstein, Nabha Shah, Maria Derda, and Vinicius Nevoa,
from whom I learned many physical insights
as well as the 
intrinsic
joy of collaboration.
I will remember our---%
in Cliff's words---%
``improvisational jam sessions''
where we melted
twisted gauge transformations,
pancaking,
Burgers vector,
and a delectable variety of strings
into a summer rhapsody.

I am indebted to
Sangmin Lee,
who has been a reliable mentor and anchor since my undergraduate years.
Sangmin has watched my growth with warmth,
constantly nudging me with stimulating problems that invariably widened my academic scope.
His steadfast guidance
greatly supported my development as a physicist.
His comprehensive lectures on relativity remain a defining memory of my freshman year,
providing an enduring inspiration.

\medskip

I would like to thank my committee members,
Cliff,
Saul Teukolsky, David Simmons-Duffin, and Julio Parra-Martinez,
for their guidance and insightful feedback throughout the defense process.
In particular,
I wish to express my heartfelt thanks to Saul.
As someone drawn to gravitation and spinning black holes, I am deeply grateful to have received Saul’s encouragement.
Given his foundational contributions to Kerr perturbation theory, 
his support has been especially meaningful to me.

\medskip

I would like to 
thank
mentors who have guided and inspired me throughout my graduate years.

I am 
deeply grateful 
to Lionel Mason 
for his early encouragement of my work on massive twistors.
Lionel gave me invaluable opportunities to
present my research at a seminar and
participate in various twistor programs, 
which were formative experiences for my growth.
My sincere thanks also go to Tim Adamo for his kindness and intellectual generosity.
Whenever we crossed paths at schools or conferences,
Tim was always open to diving into discussions on 
twistors or self-dual black holes.
His insights and welcoming nature meant a great deal to me as a student.
Overall, I am indebted to both Lionel and Tim
for welcoming me into the twistor community, 
and for generously sharing glimpses into the stories 
of
Ezra {T.\nem} Newman and Roger Penrose.

My deepest and special thanks go to Julio.
The exhilarating discussions and conversations I shared with Julio during the first couple of years of my {Ph.D.\nem} remain among the most beautiful and exciting memories,
providing a source of inspiration that carried through my doctoral journey.
Moreover, Julio’s belief in my independent work 
and instincts
gave me the immense courage and encouragement needed to forge my own research paths.
It has been a true delight to resonate with
Julio's positive energy,
creativity,
and enthusiasm.

I am also thankful to
Jung-Wook Kim
for various collaborations
that widened my horizons,
and for showing me paths toward a postdoctoral career.

\medskip

The vibrant energy and infectious joy within Cliff's group have made my time incredibly special,
and it has been a privilege to be part of such a lively environment.
My sincere thanks go to
Nabha, Maria, Tara, Francesco, and Lihang
for their brilliant presence.
I am also thankful to Allic for
discussions and for his support on various occasions.
I am thankful to
Carol for her wonderful administrative assistance
over the years.
I wish to thank everyone
who supported me throughout my studies
and made my time at Caltech special.

\medskip

I acknowledge the support I received from 
the Walter Burke Institute for Theoretical Physics,
the Department of Energy (Grant No.\:DE-SC0011632),
and Ilju Academy and Culture Foundation.

\medskip

Finally,
I would like to thank
my family for
supporting
my dreams and ambitions.
  
\end{acknowledgements}

\begin{abstract}
    General relativity is a century-old subject.
    Yet modern explorations through generalized symmetries, scattering amplitudes, and effective field theory have raised open problems,
    motivating a reassessment of conventional views on gravitation, spacetime, and spin.
    First, do generalized symmetries exist in dynamical gravity as a low-energy effective field theory?
    Second, is there a field-theoretic explanation for
    the tree-level double copy relationship between
    general relativity and Yang-Mills theory?
    Third, what are the exact equations of motion or Lagrangians of four-dimensional spinning black holes in external fields, in their point-particle effective theory?
    
    In this dissertation,
    three different perspectives on gravity
    are developed to shed light on
    each of these puzzles.
    
    First, we view gravity as a gauge theory of the Lorentz group.
    This refers to the vielbein formulation of gravity.
    It implies that the effective field theory of gravity,
    at the fully nonlinear level,
    exhibits a one-form symmetry valued in the center of the local Lorentz group.
    The charged objects are the Wilson loops of the spin connection, i.e., spin holonomies in various representations.
    The symmetry operator
    is a topological operator
    associated with a certain area measured in Planck units.
    Their linking admits an elegant physical interpretation
    in classical gravitation:
    the symmetry operator materializes a tetradic cousin of cosmic string
    that induces 
    quantized spin precession angles
    as a gravitational Aharonov-Bohm effect.
    Notably, our construction
    identifies a new symmetry of the standard model
    at scales below the lightest neutrino mass.
    The absence of global symmetries in quantum gravity 
    suggests that
    this gravitational one-form symmetry
    is either gauged or explicitly broken,
    the latter of which
    mandates the existence of fermions.

    Second, we investigate how far gravity can be viewed as a gauge theory of diffeomorphisms.
    Based on an established result on Born-Infeld theory,
    we point out that
    color-kinematics duality
    at the equations of motion level
    mandates treating the diffeomorphism algebra
    formally like a gauge algebra 
    in the internal, fiberwise sense of Yang-Mills theory.
    We show that this seemingly radical prescription
    admits a concrete construction within
    a mathematically consistent framework:
    field theory of a dynamical frame field.
    We demonstrate that
    color-kinematics duality
    systematically defines
	diffeomorphism gauge connection, diffeomorphism covariant derivative, and even diffeomorphism Wilson line.
    In this context,
    we revisit the teleparallel formulation of gravity
    and evaluate its relevance 
    to double copy
    in three or four dimensions.
    The Misner string
    is reinterpreted as a topological classical solution in any diffeomorphism gauge theory,
    endable on Newman-Unti-Tamburino charges.

    Third, we view four-dimensional gravity
    as 
    a nonlinear interaction
    between self-dual and anti-self-dual
    parts.
    We begin by reviewing
    the simplicity of self-dual gravity
    in the context of 
    double copy,
    Pleba\'nski's second heavenly equation,
    and
    Mason-Newman or Lax pair formulations.
    This motivates the program of understanding full gravity
    by perturbing away from the simpler self-dual sector,
    an insight often advocated by twistor theorists.
    Chiral formulations of gravity
    are discussed in this context,
    via both Lorentz and diffeomorphism gauge theory perspectives.
    In the former, we review Pleba\'nski gravity.
    In the latter,
    we derive a Chalmers-Siegel version of teleparallel gravity
    that perturbs away from self-dual gravity in the Mason-Newman characterization.

    We then show how this view on four-dimensional gravity
    can be applied
    to the derivation of four-dimensional spinning black hole solutions 
    as well as 
    their dynamics in point-particle effective theory. 

    We begin by
    a pedagogical demonstration of
    the simplicity of self-dual backgrounds.
    The motion of a relativistic charged particle
    in the background of a self-dual dyon
    is shown to be maximally superintegrable
    by being isomorphic to the hydrogen atom.
    Similarly,
    the motion of a relativistic particle
    in a self-dual black hole background
    is maximally superintegrable
    by being isomorphic to the Kepler problem.
    
    A precise relation between these two backgrounds
    is established by the Kerr-Schild double copy.
    We derive a 
    Kerr-Schild metric for the self-dual Taub-Newman-Unti-Tamburino solution
    by double copying a gauge potential of the self-dual dyon.
    A simple complexified coordinate transformation is constructed
    to explicitly show that this new metric
    is diffeomorphic to the well-known Gibbons-Hawking instanton metric.
    
    By taking this result as a crucial lemma,
    it is then shown that
    the Kerr metric
    represents the nonlinear superposition of
    self-dual and anti-self-dual Taub-Newman-Unti-Tamburino solutions.
    This elevates the Newman-Janis algorithm
    to a rigorous derivation of the Kerr metric
    and explicates its origin.
    More generally,
    the five-parameter family of solutions
    including Kerr-Newman and Kerr-Taub-Newman-Unti-Tamburino black holes
    represents
    systems of Taub-Newman-Unti-Tamburino instantons and chiral dyons.
    
    Based on this realization,
    we propose a probe counterpart of the Newman-Janis algorithm,
    which uniquely constrains the effective equations of motion of Kerr and Kerr-Newman black holes
    in external gravitational and electromagnetic fields
    within the self-dual sector.
    It is shown that 
    these effective equations of motion
    are maximally superintegrable
    in the backgrounds of self-dual black holes,
    via dynamical Newman-Janis shifts of conserved charges.
    This identifies
    an exactly solvable subsector of the general-relativistic spinning black hole binary problem.
    It also identifies
    a hidden symmetry 
    as an infrared principle
    that can pinpoint black holes
    among all massive spinning objects
    in the point-particle effective theory,
    within the self-dual sector.
    
    In the Lagrangian formulation,
    we construct a unique class of
    effective worldline actions for
    Kerr and Kerr-Newman black holes
    in generic, non-self-dual external fields.
    We provide their explicit all-orders formulae
    in terms of expansions in curvature and spin length.
    This is facilitated by
    developing an ``in-in'' formalism for all-orders geodesic deviation
    as an alternative to the Synge framework.

    A universal framework arises for
    spinning-particle mechanics
    in which
    spin is added to spacetime as the imaginary part
    while chirality and holomorphy are inherently linked.
    We demonstrate how
    exact spinning black hole equations of motion are derived
    beyond the self-dual sector.

    Finally, we derive the classical Compton scattering amplitudes of spinning black holes,
    directly from the explicit effective point-particle actions
    by implementing a chiral worldline perturbation theory
    in which the plus-helicity spin exponentiation is Lagrangian-level manifest to all orders.
    This provides an invariant characterization of the effective dynamics of spinning black holes.
    Our amplitudes are free of spurious poles and exhibit correct residues on physical factorization channels,
    for both same-helicity and mixed-helicity configurations.
    Physically, our computation
    views the Kerr black hole as
    a dynamical Taub-Newman-Unti-Tamburino pair
    and systematically understands it by expanding around the self-dual Taub-Newman-Unti-Tamburino worldline.
    
    These explorations spark
    a new approach to
    spinning black hole dynamics
    that makes maximal use of
    the simplicity and
    the manifest Newman-Janis property
    of the self-dual sector,
    finding
    applications to post-Minkowskian gravity.
\end{abstract}

\extrachapter{Published Content and Contributions}

Chapters 
\chapref{K1},
\chapref{K3:HYDROGEN},
\chapref{K3:SDTN},
\chapref{K3:NJA},
\chapref{K3:GDE},
\chapref{K3:PROBENJ},
\chapref{K3:OSD1},
\chapref{K3:OSD11},
\chapref{K3:SST},
and
\chapref{K3:BHEOM}
of this thesis are based on the
published articles and publicly available preprints
[1–9],
for all of which I served as the primary author.
Chapter \chapref{J} is a review written specifically for this thesis as an original synthesis of known formulations of gravity;
Sections \ref{Jcx>FRAME} and \ref{Jcx>NLIN} 
go beyond review and contain
original constructions and calculations 
developed by the author.
Chapters \chapref{K2} and \chapref{K3:COMPTON}
are based on 
original research conducted by the author,
unpublished at the time of submission for this thesis.
Appendices A and B are expository reviews
that draw on known results 
as well as research conducted by the author,
independently or in collaboration.

\begin{publishedcontent}[iknowwhattodo]
\nocite{GenSymGrav,note-sdtn,nja,gde,probe-nj,njmagic.1,njmagic.11,sst-asym,univ}
\putbib[references]
\end{publishedcontent}

\newpage
\tableofcontents
\listoffigures
\makenomenclature
\printnomenclature

\mainmatter\setsecnumdepth{subsubsection}
\settocdepth{subsection}
\setlength{\parindent}{1em}
\setlength{\parskip}{2.5pt}

\setbeforesecskip{-\onelineskip}
\setbeforesubsecskip{-\onelineskip}
\setbeforesubsubsecskip{-\onelineskip}

\setaftersecskip{6pt}
\setaftersubsecskip{3pt}
\setaftersubsubsecskip{1pt}

\chapter{Introduction}
\section{ Gravitation from the Infrared to the Ultraviolet}
\label{Ia}

General relativity (GR)\nomenclature{GR}{General Relativity}
is a century-old subject.
It stands as one of the great triumphs of modern physics,
not only for its theoretical breakthroughs
that reformed our conception of time, space, and gravitation
but also for its remarkable experimental success.
Its earliest victories came from
explaining the anomalous perihelion precession of Mercury
and the bending of starlight by the Sun,
dramatically confirmed during the 1919 solar eclipse
\cite{Einstein:1915bz,Dyson:1920cwa,pais2005subtle}.
Later experiments further consolidated this success,
confirming
the gravitational redshift and time delay \cite{pound1959gravitational,Shapiro:1964uw}.
The ability to correctly predict such ``relativistic clock effects''
is necessary for the precision of the global positioning system,
on which many of our everyday activities heavily depend \cite{Ashby:2003vja}.
In stronger gravitational fields,
striking indirect evidence for gravitational radiation
was provided by
detecting a decrease in the orbital period
of a binary pulsar \cite{Hulse:1974eb,Taylor:1994zz},
leading to the 1993 Nobel Prize in physics.
Eventually,
the direct detection of gravitational waves by 
LIGO and Virgo
opened an entirely new observational window onto dynamical spacetime \cite{LIGOScientific:2016aoc,LIGOScientific:2017vwq},
as is recognized by the 2017 Nobel Prize.

\nomenclature{LIGO}{Laser Interferometer Gravitational-Wave Observatory}

Black holes have provided another arena of triumph.
Penrose’s singularity theorem established black hole formation as a robust prediction of GR \cite{Penrose:1964wq}.
A supermassive compact object,
named Sagittarius $\mathrm{A}^*$,
was discovered at the center of the Milky Way
by monitoring stellar orbits over nearly three decades \cite{Ghez:2008ms,Gillessen:2008qv}.
These achievements were 
honored with the 2020 Nobel Prize.
More recently,
horizon-scale black-hole imaging has provided 
further consistency checks on the idea that
the astrophysical black holes we observe are, 
to excellent approximation, 
described by the Kerr metric \cite{EventHorizonTelescope:2019dse}.

Together, these observations throughout a century
have made GR the most successful classical theory of gravity, 
a theory whose geometric vision continues to withstand increasingly stringent tests \cite{Will:2014kxa}.

\medskip
Now turning to the history of the theoretical side,
it should be noted that
GR has been derived and illuminated from two complementary perspectives.
To one tradition,
GR is Einstein's profound \textit{geometrization} of gravity
\cite{Einstein:1916vd}.
To another,
GR is
the unique long-range interacting theory of a massless spin-2 field
\cite{Gupta:1954zz,Deser:1969wk,Wald:1986bj}.
The former refers to Einstein's original derivation,
which describes a highly nonlinear pathway
shaped by paradoxes, thought experiments, and geometric inspirations.
The latter refers to the field theorists' rederivation of Einstein's artwork \textit{from scratch},
which claims to provide a much more streamlined and principled approach.

\begin{figure}[t]
    \centering
    \includegraphics[width=0.2\linewidth]{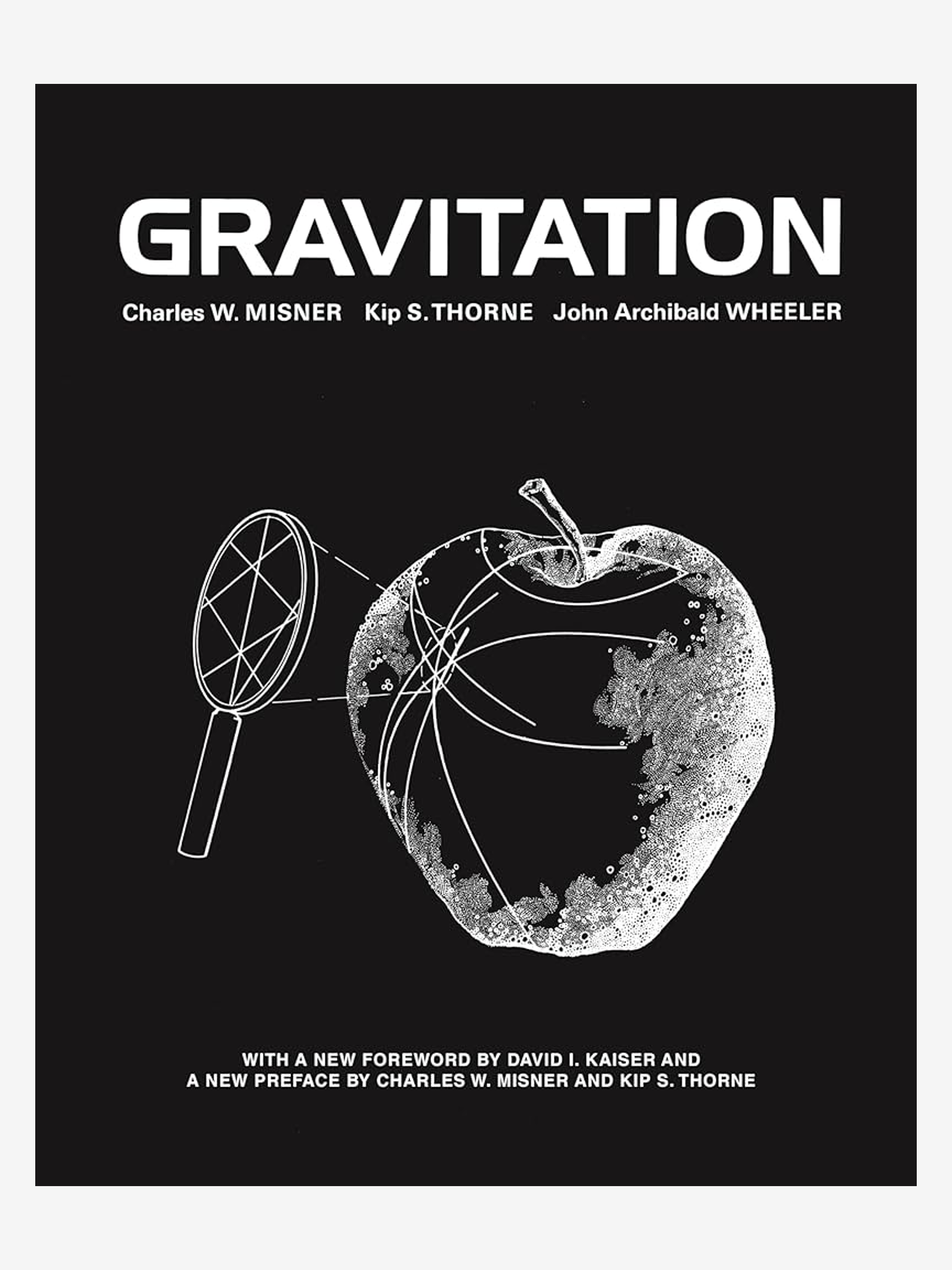}
    \qquad
    \includegraphics[width=0.2\linewidth]{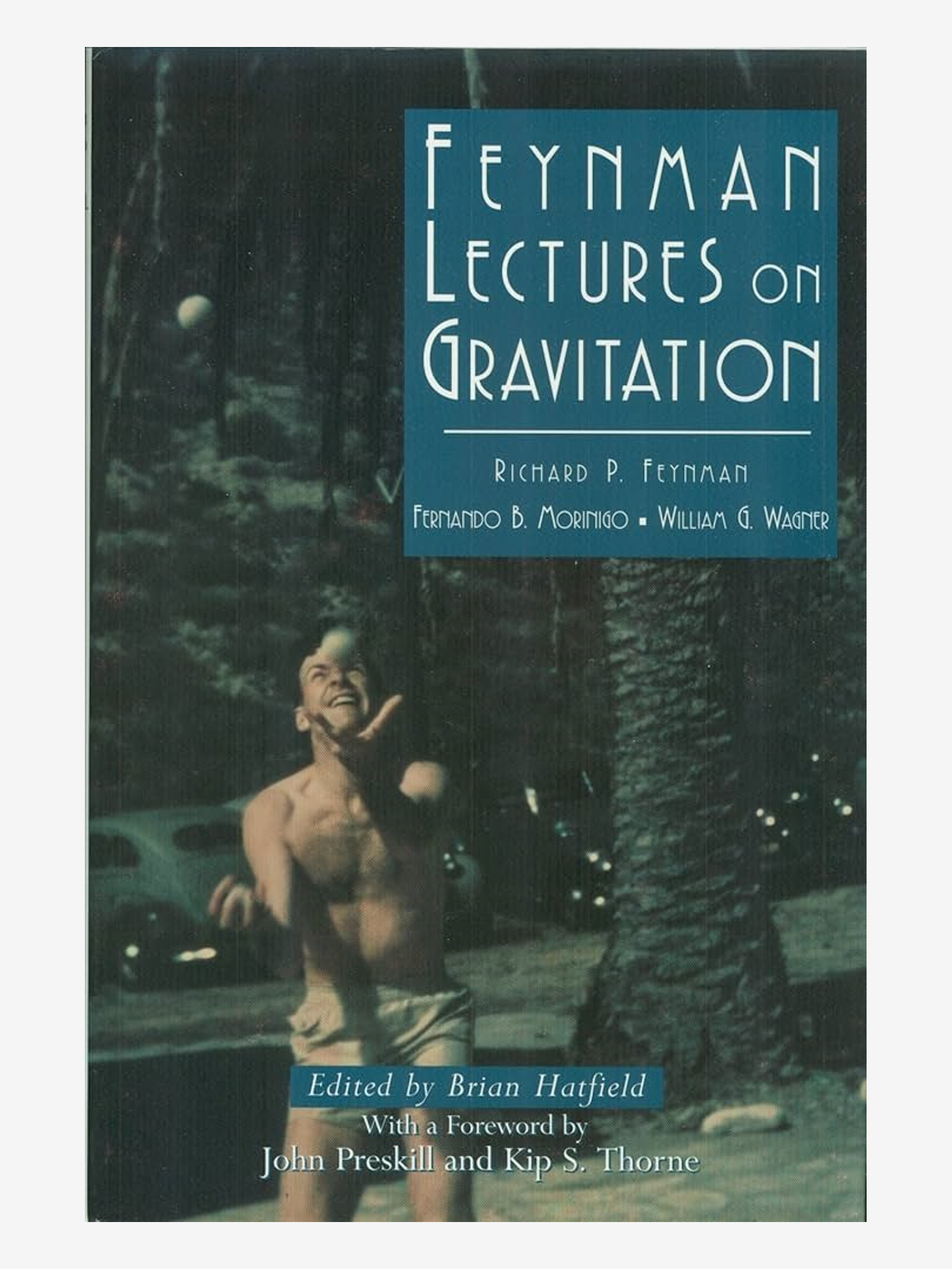}
    \caption{%
        Textbooks on GR
        representing the two schools:
        the relativists
        and
        the field theorists.
    }
    \label{fig:books}
\end{figure}

The former view,
which we may refer to as the \textit{relativist} tradition,
could be well-represented by
the  textbook \cite{mtw}
by Misner, Wheeler, and Thorne.
It is written in its characteristic visual and geometric style,
providing a guided tour
from the concepts of
metric and connection to Riemann curvature and Einstein tensor.
Gravitation is the dynamical geometry of spacetime itself.
Matter tells spacetime how to curve, 
and spacetime tells matter how to move.

The latter view,
which we may refer to as the \textit{field theorist} tradition,
is demonstrated in 
the textbooks written by
Feynman \cite{Feynman:1996kb}
and
Weinberg \cite{weinberg1998gravitation}.
Starting from basic assumptions,
one constructs a relativistic quantum field theory (QFT)
\nomenclature{QFT}{Quantum Field Theory}
of a spin-2 field
and discovers that its unique classical nonlinear completion happens to be Einstein’s theory
\cite{Gupta:1954zz,Deser:1969wk,Wald:1986bj}.
The geometrical visions of Einstein
are deconstructed into
logical necessities
demanded for a consistent QFT.
For example,
instead of imagining space elevators and falling apples,
the weak equivalence principle
is derived by
stipulating
the existence of a Lorentz-invariant and gauge-invariant S-matrix
containing a massless spin-2 particle
and the standard soft-pole factorization of scattering amplitudes
\cite{Weinberg:1964ew}.

The boundary between
the relativist school and the field theorist school
is perhaps not strict these days;
the best modern work often moves fluently between both,
and this bilingual ability 
is beneficial.
Yet the contrast could be sometimes useful to draw,
as it highlights two very different instincts about what GR is.

\newpage
The relativist approach to GR
could be illuminated in the broader context of
formulating and comprehending
physical laws and phenomena
in the geometrical language.\footnote{
    For instance,
    Misner, Thorne, and Wheeler \cite{mtw}, pp.\,370-371,
    draw
    an analogy 
    between gauge theory and gravity
    from the perspective of exterior calculus,
    through diagrams titled
    ``Structure of Electrodynamics/Geometrodynamics in Outline Form.''
}
For gauge theories,
such a geometric paradigm
had been put forward by
Wu, Yang, and Mills
\cite{Yang:1954ek,Wu:1975es,Yang:1977cq,Yang:1978fb,Trautman:1980fb}
by the late twentieth century,
which had sparked fruitful developments
in both physics
\cite{Eguchi:1980jx,Nakahara,tong2018gauge,SkinnerAQFTNotes}
and mathematics
\cite{Atiyah:1983ah,Uhlenbeck:1982,Donaldson:1983wm,Donaldson:1983,Freed:1984ifm,Donaldson:1990gfm}.
Notably, it provided a unified geometrization of
the three fundamental forces \textit{except gravity}:
electromagnetism, the weak, and the strong.

Regarding this separation,
it is worth clarifying that
the geometry of Yang-Mills (YM) theory
\nomenclature{YM}{Yang-Mills}
is quite
different than
that of GR.
YM theory governs
a connection on a principal fiber bundle
over a fixed spacetime,
whereas
GR governs
a dynamical pseudo-Riemannian manifold
as the spacetime itself.
The 
redundancies involved in YM theory
describe \textit{internal} local transformations
and refer to a particular, narrower geometric semantics
that demands structures such as fiber bundles  
\cite{Yang:1954ek,Wu:1975es,Yang:1977cq,Yang:1978fb,Trautman:1979gf,Trautman:1980fb,Eguchi:1980jx,Nakahara,tong2018gauge,SkinnerAQFTNotes}.
The 
redundancies involved in GR
are diffeomorphisms,
which are \textit{external} transformations
that shift the very points of the base manifold.

It is sometimes said that
GR is a gauge theory of 
local translations
or diffeomorphisms
\cite{DeWitt:1964mxt,Feynman:1996kb}.
Strictly speaking,
this is correct only
in the colloquial or broader sense that 
it involves redundancies:
the freedom to choose coordinates.

Despite this crucial discrepancy between
gravity and the other three forces,
historically,
there had been 
aspirations toward reformulating GR 
as a gauge theory
in the narrower sense
\cite{Utiyama:1956sy,Kibble:1961ba,Sciama:1962,Hehl:1976kj,Hayashi:1967se,hayashi1977gauge,hayashi1981addendum,cho1976einstein,cho1976gauge,cho1992gravdiff,%
Grignani:1991nj,Blagojevic:2003cg,%
Percacci:1984bq,%
Blagojevic:2002grg,Blagojevic:2013xpa,Obukhov:2006pgt,Hehl:1994ue,MacDowell:1977jt,Wise:2010sm,Catren:2014mna,Ashtekar:1986yd},
envisioning a unified geometrization of
all fundamental forces.

\medskip
The field theorist assessment of GR
connects to the effective field theory (EFT) of gravity.
\nomenclature{EFT}{Effective Field Theory}
An EFT is a QFT
whose validity is restricted
to a certain range of energy scales
\cite{Wilson:1973jj}.
A paradigmatic example of an EFT is Fermi’s theory of weak interactions,
valid at energies below the W-boson mass.
Although nonrenormalizable,
it gives a systematically improvable low-energy expansion whose breakdown merely signals the appearance of new short-distance degrees of freedom, namely the W and Z bosons.
In the EFT, the effect of these mediator particles is encapsulated and hidden
in a four-fermion contact interaction vertex.
Another paradigmatic example
is chiral perturbation theory,
an EFT of pions below the quark confinement scale.

In the same spirit,
perturbative quantum GR is understood as
a low-energy EFT of the massless spin-2 field,
i.e., the \textit{graviton}
\cite{Donoghue:1994dn,Donoghue:1995cz,Donoghue:2012zc,Burgess:2003jk}.
Its perturbative nonrenormalizability, 
detected via the need of higher-derivative counterterms in loop calculations \cite{Goroff:1985th,vandeVen:1991gw},
signals its own breakdown at the Planck scale,
$M_\text{Pl} \sim 10^{19} \text{GeV}
 \sim \hbar/(10^{-35}\text{m})
$.
Crucially, however,
this does not obstruct its use as a predictive low-energy QFT.
Quantum corrections 
are systematically organized by a derivative expansion,
equivalently 
by the powers of
the low-energy scale over the Planck scale, $E/M_\text{Pl}$.
The EFT framework separates
universal infrared predictions from the unknown ultraviolet physics,
ensuring finite predictability at each fixed order in
the low-energy expansion.

\medskip
When thinking along these lines,
we are inevitably led to the ultimate question
of quantum gravity.
If GR breaks down at the Planck scale,
what can be the ultraviolet (UV) theory of quantum gravity?
\nomenclature{UV}{Ultraviolet}
More generally, what is the full space of 
low-energy EFTs that admits consistent UV completions into quantum gravity?
Motivated by explorations through string theory,
the swampland program \cite{Vafa:2005ui}
has identified and proposed
a collection of properties
that any consistent theory of quantum gravity
should universally satisfy.
They are known as swampland conjectures,
stating, e.g., that
no global symmetries are allowed
\cite{Misner:1957mt,Polchinski:2003bq,Banks:1988yz,Banks:2010zn,Harlow:2018tng},
gravity is the weakest among all forces
\cite{Arkani-Hamed:2006emk},
and
any charge that is not forbidden is mandatory
\cite{Polchinski:2003bq,Banks:2010zn}.
The swampland conjectures 
aim to
tell whether a low-energy EFT
admits a consistent UV completion into quantum gravity
(lies in the ``landscape'')
or not
(lies in the ``swampland'').

\begin{figure}[t]
    \centering
    \includegraphics[width=0.99\linewidth]{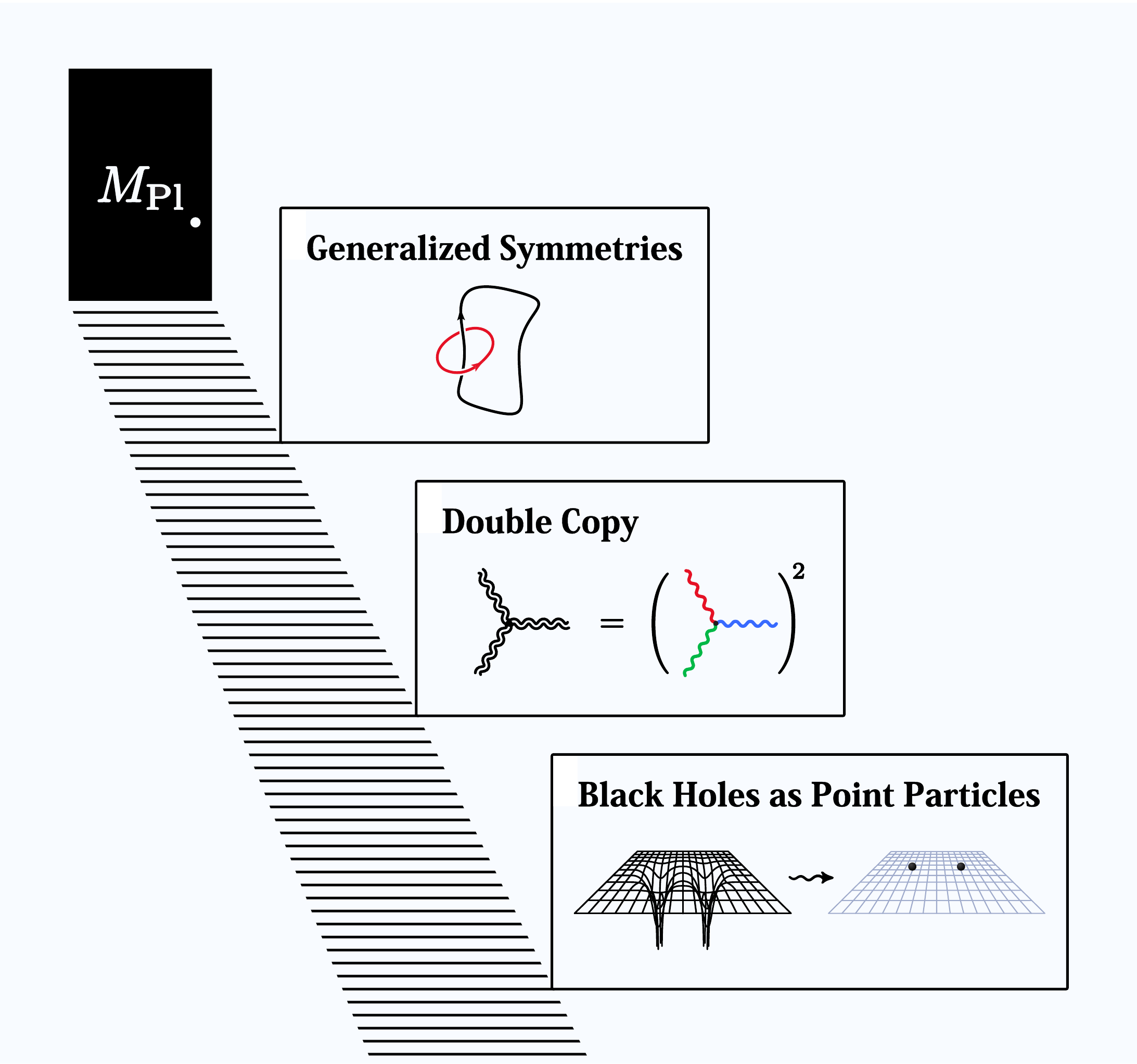}
    \smallskip
    \caption{%
        Topics in GR 
        from the IR to the UV.
    }
    \label{fig:roadmap}
\end{figure}

The no global symmetry conjecture should concern not only
ordinary symmetries acting on local operators
but also
\textit{generalized symmetries}
\cite{Gaiotto:2014kfa,Cordova:2022ruw,Shao:2023gho,Schafer-Nameki:2023jdn,Gomes:2023ahz,%
Brennan:2023mmt,Bhardwaj:2023kri,McGreevy:2022oyu,Simmons-Duffin:2016gjk},
including
higher-form,
noninvertible,
and
higher-group symmetries.
In\:particular,
$p$-form symmetries are
symmetries that act on $p$-dimensional extended operators.
Maxwell theory, for example, exhibits a global one-form symmetry
acting on Wilson loops.
The no global symmetry conjecture
states that\:this one-form symmetry 
in the infrared (IR)
\nomenclature{IR}{Infrared}
will be lost
in the regime where gravity becomes fully dynamical and quantum.
As a result, it must be either gauged or explicitly broken.
The latter implies the existence of charged particles,
as Wilson lines are endable on electric charges.
\rcite{Heidenreich:2021xpr} gives a related discussion.

The existence of a $p$-form symmetry amounts to having
a topological operator of codimension $(p{\,+\mem}1)$.
This is
the modern definition of symmetry due to
\rcite{Gaiotto:2014kfa}.
This topological operator
establishes a Ward identity
by linking with the $p$-dimensional charged operator.
For the above example of Wilson loops in Maxwell theory,
the topological operator
arises by integrating the Hodge dual of the field strength
over closed codimension two surfaces.

\medskip
Returning to the low-energy regime
below the Planck scale,
one learns that two interesting topics
have emerged from
modern investigations of gravity
through the lenses of
scattering amplitudes and EFT:
see \fref{fig:roadmap}.
The first is \textit{double copy},
relevant to pure GR
as a perturbative QFT
\cite{KLT,BCJ1,BCJ2,BCJReview,CCK}.
The second is \textit{point-particle effective theory of black holes}
\cite{
    Cheung:2018wkq,%
    ShR065,Porto:2016pyg,Levi:2018nxp,Kalin:2020mvi,%
    Porto:2005ac,Levi:2015msa,%
    aho2020%
},
which zooms out to a further IR scale.

We may want to
describe the latter first:
point-particle effective theory of black holes.
This topic has found fruitful practical applications
in describing the classical gravitational dynamics of astrophysical bodies,
relevant to the gravitational-wave physics
taking place at
LIGO and Virgo.

\newpage

The general-relativistic two-body problem
stands notoriously unsolved.
Dec\-ades of research have
devised
various approximation schemes
such as
post-Newtonian (PN)
\cite{ShR043,ShR044,ShR045,ShR046,ShR047,ShR048,ShR049,ShR050,ShR051,ShR052,ShR053,ShR054,ShR055,ShR056,ShR057},
post-Minkowskian (PM)
\cite{Bern:2019nnu,ShR030,ShR031,ShR032,ShR033,ShR034,ShR035,ShR036,ShR037,ShR038,Damour:2016gwp,Damour:2017zjx},
and
self-force (SF)
\cite{ShR058,ShR059}
expansions,
\nomenclature{PN}{Post-Newtonian}
\nomenclature{PM}{Post-Minkowskian}
\nomenclature{SF}{Self-Force}
while utilizing
numerical relativity \cite{ShR062,ShR063,ShR064},
effective one-body formalism \cite{ShR060,ShR061},
and
EFT methods
\cite{ShR065,Porto:2016pyg,Levi:2018nxp,%
    Cheung:2018wkq,%
    Porto:2005ac,Levi:2015msa,%
ShR066,ShR067,ShR068,ShR069,ShR070,ShR071,ShR072,ShR073,ShR074,%
    Neill:2013wsa,%
    Bern:2019nnu,%
    Bern:2019crd,%
    Cristofoli:2019neg,%
    Kalin:2020mvi,Kalin:2020fhe,%
    Bjerrum-Bohr:2021din,%
    Bern:2021dqo,%
    Bjerrum-Bohr:2022blt%
}.

How could an effective field theorist
approach the general-relativistic two-body problem?
Take a black hole binary, for instance.
In the inspiral phase,
the distance between the black holes,
say $b$,
is much larger than their typical Schwarzschild radius,
say $Gm$.
This separation of scales,
$b \gg Gm$,
facilitates an effective theory
that replaces 
the curved, dynamical geometry describing the black hole binary
with two point particles put on a flat Minkowski space
with fields,
as illustrated in \fref{fig:roadmap}.
This is the point-particle effective theory
\cite{
    ShR065,Porto:2016pyg,Levi:2018nxp,Kalin:2020mvi,%
    Cheung:2018wkq,%
    Porto:2005ac,Levi:2015msa,%
    aho2020%
},
organized in the PM expansion 
in $Gm/b$.
This effective theory integrates out
scales smaller than the Schwarzschild radii
while taking the exact general relativistic description of the black hole binary geometry
as the UV theory.

As is well-reviewed in \rcite{Bern:2019crd},
the EFT approach to PM gravity
establishes an efficient computational framework
by importing state-of-the-art 
techniques
in scattering amplitudes
such as recursion 
\cite{Britto:2004ap,Britto:2005fq},
generalized unitarity
\cite{Bern:1994zx,Bern:1994cg,Bern:2011qt},
and double copy
\cite{KLT,BCJ1,BCJ2,BCJReview}.
Especially, the double copy
provides radical shortcuts to gravitational amplitudes.

\medskip

Our attention now turns to the double copy.
The double copy is a remarkable property of perturbative quantum GR,
observed and established at the level of scattering amplitudes
so far
\cite{KLT,BCJ1,BCJ2,BCJReview}.
Roughly speaking, it states that
gravity is somehow the ``square'' of the strong force:
``$\text{GR} = (\text{YM})^2$.''
It suggests that
the entire nonlinearity of gravitational waves---%
usually represented in terms of
an infinite collection of vertices
of all valences
in the graviton Feynman rules
via the Einstein-Hilbert action---%
is in fact encoded in a
sole \textit{cubic} vertex.

\nomenclature{KLT}{Kawai-Lewellen-Tye}
\nomenclature{BCJ}{Bern-Carrasco-Johansson}
The double copy has 
an important string-theoretic precursor,
known as Kawai-Lewellen-Tye (KLT) relations \cite{KLT}.
Later, Bern, Carrasco, and Johansson (BCJ) \cite{BCJ1, BCJ2}
showed that a purely field-theoretic formulation is viable.
Their construction is based on
the idea of
color-kinematics (CK) duality:
\nomenclature{CK}{Color-Kinematics}
YM amplitudes can be represented 
as sums over cubic graphs
in which
the kinematic numerators obey the same algebraic relations 
(Jacobi identity)
as the color factors.
Replacing the color factors by a second copy of kinematic numerators then yields the graviton amplitudes.
See \rcite{BCJReview} for a detailed review.

Cheung and Mangan \cite{CCK} demonstrated that
tree-level CK duality can be made manifest to all multiplicities 
if one identifies the relevant kinematic algebra and current 
at the level of classical equations of motion (EoM),
\nomenclature{EoM}{Equations of Motion}
in the framework of
Berends-Giele (BG) recursion \cite{BerendsGiele}
(perturbiner expansion \cite{Rosly:1996vr,Selivanov:1997an,Mizera:2018jbh}).
\nomenclature{BG}{Berends-Giele}
When applied to GR,
one encounters a surprisingly bold conjecture:
one may be able to recast
Einstein's equations
in the form of
YM equations.

The physical significance is twofold.
For the geometers,
it proposes that
there indeed can exist a unified 
geometric formulation
that puts
all fundamental forces
on the same footing,
not as a mere analogy or a formal mathematical similarity
but a precise, quantitative correspondence
leading to concrete computational outputs.
In this sense,
the double copy fixes a unique goalpost
for the geometrization program
and also provides explicit criteria for testing it.
For field theory practitioners,
the double copy
is a powerful computational tool for gravitational amplitudes
that circumvents the pain involved
in using the Einstein-Hilbert action.

The nonperturbative aspects of double copy
are far less explored
\cite{Cheung:2022mix,Armstrong-Williams:2022apo}.
Yet correspondences between exact classical solutions
in Maxwell theory and GR
have been discussed
under the name of ``classical double copy.''
In particular,
a version known as 
Kerr-Schild (KS) double copy
\cite{monteiro2014black}
\nomenclature{KS}{Kerr-Schild}
ignited a modern discourse on KS metrics
\cite{Berman:2018hwd,bahjat2020monopoles,note-sdtn,nja,%
Carrillo-Gonzalez:2017iyj,CarrilloGonzalez:2019gof,Bahjat-Abbas:2017htu,Banerjee:2019saj,elor2020newman,bah2020kerr,Monteiro:2020plf,Alfonsi:2020lub,Monteiro:2021ztt,gabriel1,%
AAS}
and established a systematic use
\cite{vines2018scattering,nja}
of its effective linearization 
\cite{xanthopoulos1978exact,Harte:2016vwo}
property.
For example,
the KS double copy identifies
the Schwarzschild solution
as the gravitational counterpart of
the electric monopole
(Coulomb solution)
\cite{monteiro2014black}.
Similarly,
the electromagnetic field of the Kerr-Newman solution
\cite{Newman:1965my-kerrmetric,Lynden-Bell:2002dvr},
dubbed ``{\kerr}\mem'' solution \cite{aho2020},
double copies to
the Kerr solution
\cite{monteiro2014black}.
See also \rrcite{Luna:2015paa,Luna:2018dpt,Chacon:2021wbr,white2021twistorial,Luna:2022dxo}
for alternative constructions using double KS metrics, Weyl tensor, or twistor space.
\medskip
To sum up,
we have provided an overview of GR,
outlining
its experimental successes
as well as
its systematic theoretical organization
from the IR to the UV.
While introducing
both the geometric and field-theoretic
perspectives,
we have encountered various topics
as illustrated in \fref{fig:roadmap}.
To question the UV,
one utilizes
the swampland conjectures
with generalized symmetries.
To predict gravitational phenomena in the IR
for LIGO and Virgo,
one utilizes
the point-particle effective theory of black holes
as well as 
the efficient shortcuts
provided by amplitudes methods,
such as the double copy.
Having traveled a\\ century,
we now turn to remaining puzzles,
which are the subject of this thesis.\newpage
\section{ Open Puzzles in Gravitation}

Modern explorations of gravity through
generalized symmetries,
scattering amplitudes,
and EFT
have raised open problems.
They motivate a reassessment of 
conventional views on gravitation,\:spacetime,\:and\:spin.
Below, we state three.
\begin{enumerate}[leftmargin=1.5em]
    \item 
        Do generalized symmetries exist in dynamical gravity as a low-energy EFT?
    \item
        Is there a field-theoretic explanation for
        the tree-level double copy relation between
        GR and YM theory?
    \item
        What is the exact point-particle effective theory of four-dimensional spinning black holes
        in external fields?
\end{enumerate}

The objective of this thesis
is to answer these questions partially,
or at least provide
new approaches
that can be useful or insightful.
Our conclusions are summarized below.
The progress on Problem 2 may be limited,
but nevertheless,
it served as an important element
for tackling Problems 1 and 3
by providing unique perspectives and inspirations.
\begin{enumerate}[leftmargin=1.5em]
    \item 
        We found at least one generalized symmetry 
        in dynamical gravity as a low-energy EFT
        \cite{GenSymGrav}.
        This identifies a new symmetry of nature
        at scales below the lightest neutrino mass.
    \item
        We did not succeed in finding such an explanation.
        However, we were able to make
        a useful
        observation
        that gives a geometrical picture
        to the double copy
        for Born-Infeld (BI) theory,
        in connection with teleparallel gravity.
    \item
        We identified 
        a hidden symmetry
        in the IR 
        (point-particle effective theory)
        that uniquely characterizes spinning black holes
        among all massive spinning objects,
        yet within the self-dual (SD) sector.
        
        Beyond the SD sector,
        we constructed minimal proposals for
        the
        exact---meaning all orders in spin and coupling---%
        effective actions of
        Kerr and (nonabelian) Kerr-Newman black holes
        in backgrounds with gauge and gravitational fields,
        in the conservative sector.
        We obtained their explicit formulae
        by developing a new formalism for all-orders geodesic deviation.
        We proposed a chiral approach
        to spinning black hole effective dynamics
        that maximizes the utility of the simplicity in the SD sector.
\end{enumerate}
\nomenclature{BI}{Born-Infeld}
\nomenclature{SD}{Self-Dual}
\nomenclature{ASD}{Anti-Self-Dual}

Below, we define the above three problems in detail
and clarify the necessary or sufficient conditions
for their resolution.

\subsection{Generalized Symmetries in Dynamical Gravity}\label{Ib1}
\vspace{8pt}
\begin{statement}
    Do generalized symmetries exist in dynamical gravity\\
    as a low-energy EFT?
\end{statement}

\paragraph{Goal}
Find generalized symmetries in pure GR
(with higher-curvature corrections)
by treating it as a low-energy EFT of gravity.
There may or may not be more than one.
This problem presumes working in the regime
where the dimension and topology of spacetime are strictly fixed.
The gravitational path integral will 
not implement the sum over spacetime topologies.

\paragraph{Problem solved if}
\begin{enumerate}[label=(\alph*)]
    \item 
        The charged operators and their transformation behavior are identified,
    \item
        The topological symmetry operator
        is identified,
    \item
        The Ward identity between the two
        is concretely established
        in either the covariant (path integral) or the canonical (Hamiltonian) formalisms.
\end{enumerate}
We may focus on
higher-form symmetries
for simplicity,
although the possibilities of finding
more general symmetries
should not be precluded.
A degree of concreteness
may be desired or demanded,
in which case one has to 
construct the charged operators and the symmetry operator
explicitly in terms of the local degrees of freedom
of the gravitational EFT.

\paragraph{Significance}

\begin{itemize}[leftmargin=1.25em]
    \item 
        The no global symmetry conjecture implies that
        any global symmetry observed in a low-energy EFT must either be gauged or explicitly broken.
        In the latter scenario,
        one could argue that
        specific types of matter fields
        are required
        to break the gravitational higher-form symmetry.
    \item
        If found in $d {\:=\:} 4$ spacetime dimensions,
        one can be reporting a new symmetry in our universe
        that is unbroken below some mass scale.
    \item
        It may shed a new light on
        the definition of
        gravitational charge in general relativity,
        which has been a classic subject to relativists
        \cite{Arnowitt:1959ah,Arnowitt:1962hi,Komar:1959wp,Bondi:1962px,Sachs:1962wk,Sachs:1962zza,Regge:1974zd,Iyer:1994ys,Wald:1999wa,Penrose:1982wp}
        (see, e.g., \rrcite{Kmec:2025ftx,Freidel:2026iia}
        for recent attempts).
\end{itemize}

\newpage

\subsection{Double Copy}\label{Ib2}
\vspace{8pt}
\begin{statement}
    Is there a field-theoretic explanation for the 
    tree-level\\ double copy relationship between
    YM theory and GR?
\end{statement}

\paragraph{Problem solved if}
\begin{enumerate}[label=(\alph*)]
    \item 
        Vacuum Einstein's equations
        with no cosmological constant
        can be perturbatively reformulated as
        YM equations on flat background
        with the same spacetime dimension $d {\:>\:} 2$
        whose gauge algebra is replaced with a Lie algebra $\gk$
        that may or may not be infinite-dimensional.
\end{enumerate}
This appears as a bold claim,
but it is indeed a valid conjecture 
for explaining the BCJ duality \cite{BCJ1,BCJ2,BCJReview}
according to \rcite{CCK}'s formulation.
The mystery Lie algebra $\gk$ is referred to as the \textit{kinematic algebra} of YM theory.
It is plausible that
$\gk$ can be infinite-dimensional,
as it eventually has to take
momenta and polarizations
as the adjoint indices
to construct the kinematic numerators.
The reformulation should be invertible on
external (asymptotically free) states.

\paragraph{Problem solved also if}
\begin{enumerate}[label=(\alph*),resume]
    \item
        YM equations 
        can be perturbatively reformulated as
        bi-adjoint scalar (BAS) equations
        by replacing one of its Lie algebras
        with $\gk$.
        \nomenclature{BAS}{Bi-Adjoint Scalar}
\end{enumerate}
Here, flat spacetime is assumed
with dimension $d{\:>\:}2$.
A BAS field is a scalar field $\Phi^{a\ta}$
carrying
two adjoint indices $a \eqq {1,{\cdots\mem},\dim\g}$ and $\ta \eqq {1,{\cdots\mem},\dim\tg}$
for two Lie algebras $\g$ and $\tg$.
Their structure constants
$f^a{}_{bc}$ and $\tf^\ta{}_{\tb\tc}$
must satisfy the Jacobi identity
by definition.
The BAS equations are
\begin{align}
    \label{BAS}
    \BAS:\quad
    \Box\mem \Phi^{a\ta}
    \,=\,
    \tfrac{1}{2}\,
        f^a{}_{bc}\mem 
        \tf^\ta{}_{\tb\tc}\,
            \Phi^{b\tb}\mem \Phi^{c\tc}
    \,.
\end{align}

\vspace{1pt}
\paragraph{Problem solved iff}\kern-1em
one is able to generate tree-level BCJ numerators \cite{BCJ1,BCJ2}
for YM theory and GR
to any number of external legs.
\vspace{-5pt}

\paragraph{Significance}

\begin{itemize}[leftmargin=1.25em]
    \item 
        Will bring immense practical benefit
        by radically simplifying 
        gluon/graviton perturbation theory,
        summarizing all interactions into a single universal cubic vertex
        that merely states/squares the structure constant of $\gk$.
    \item
        Might bring an unforeseen
        geometric interpretation of GR
        that understands gravity
        in the same fashion as the strong force,
        not merely as an analogy.
\end{itemize}

\newpage

\subsection{Effective Dynamics of Spinning Black Holes}\label{Ib3}
\vspace{8pt}
\begin{statement}
    What is the exact point-particle effective theory of\\
    four-dimensional spinning black holes
    in external fields?
\end{statement}

\paragraph{Background}
The Kerr solution provides an excellent approximation to
real-world astrophysical black holes
\cite{EventHorizonTelescope:2019dse}.
While the point-particle effective theory of non-spinning black holes
stands rather well-established
\cite{
    ShR065,Porto:2016pyg,Levi:2018nxp,Kalin:2020mvi,%
    Cheung:2018wkq,%
    Porto:2016zng%
},
the point-particle effective theory of spinning black holes
\cite{
    Porto:2005ac,Levi:2015msa,%
    aho2020%
}
is far from complete.

\paragraph{Problem solved iff}
\begin{enumerate}[label=(\alph*)]
    \item
        The point-particle effective action of
        the Kerr black hole
        in gravitational backgrounds
        is determined to
        all orders in spin length and coupling.
    \item
        The point-particle effective EoM of
        the Kerr black hole
        in gravitational backgrounds
        are determined to
        all orders in spin length and coupling.
\end{enumerate}
This problem
represents a simple enough yet fundamental curiosity:
what is the law of motion governing black holes?
Concretely,
the EoM of Kerr can be given in the following form: 
\begin{align}
    \label{KerrEoM}
    \thy{Kerr}:\quad
    \dot{x}^\m
    \,=\,\,
        ?
    \,,\quad
    \frac{Dp_\m}{d\t}
    \,=\,\,
        ?
    \,,\quad
    \frac{Dy^\m}{d\t}
    \,=\,\,
        ?
\end{align}
In the effective theory,
the Kerr black hole is represented by a massive spinning particle.
In \eqref{KerrEoM},
$x^\m$ and $p_\m$ are the position and momentum of this particle
while 
$y^\m \eqq {*}S^{\m\n} p_\n/(-p^2)$
is the particle's Pauli-Lubanski spin pseudovector normalized in the units of length,
known as \textit{spin length pseudovector}.
Namely,
$y^\m$ recasts
the ring radius $a = S/M$
as a four-pseudovector.
Specifying the time evolution of $x^\m$, $p_\m$, and $y^\m$
as in \eqref{KerrEoM}
uniquely determines
the classical dynamics of the Kerr black hole
in the effective theory.

\paragraph{Previous Results}

At the linear order in the gravitational coupling
(linearized gravity),
the most general point-particle effective action
for a relativistic massive spinning body
was established by Levi and Steinhoff \cite{Levi:2015msa}.
In our presentation,
their interaction action is
\begin{align}
    \label{LSA}
    \int\,
        \sum_{\ell=2}^\infty\mem
        \frac{C_\ell}{\ell!}\,\mem
            p_\m\,
                {*^\ell}\hnem R^\m{}_{\r_1\r_2\s;\r_3;\cdots;\r_\ell}\hnem(x)\,
            y^{\r_1} y^{\r_2} y^{\r_3}\nem\cdots y^{\r_\ell}\,
                \dot{x}^\s
                \, d\t
    \,,
\end{align}
where 
${*}^\ell R^\m{}_{\n\r\s}$ denotes the Riemann tensor
acted by the Hodge star $\ell$ times.
\newpage

The dynamical variables
in \eqref{LSA}
are $x^\m$, $p_\m$, and $y^\m$.
The constants $C_\ell$ are Wilson coefficients
capturing spin-induced mass $2^\ell$-pole moments
$C_\ell\mem |p|\hhnem |y|^\ell$
(electric for even $\ell$ and magnetic for odd $\ell$).

For the Kerr black hole,
classic works
\cite{%
    Hansen:1974zz,Geroch:1970cd,%
    Newman:1965tw-janis,janis1965structure,%
    Hernandez:1967zza,%
    Thorne:1980ru%
}
had established that
the $2^\ell$-pole moment of the Kerr black hole
is $Ma^\ell$,
which implies $C_\ell \eqq 1$.
This ``unity'' of multipole coefficients
has also been derived via
modern amplitudes methods
\cite{ahh2017,Guevara:2018wpp,Guevara:2019fsj,chkl2019,aho2020}.

At the quadratic order in the coupling,
however,
Arkani-Hamed, Huang, and Huang (AHH)
\nomenclature{AHH}{Arkani-Hamed, Huang, and Huang}
\cite{ahh2017}
reported a difficulty in fixing the effective theory
from amplitudes methods,
formulated in terms of a mixed-helicity Compton amplitude.
Since then, a substantial body of works
\cite{ahh2017,%
Guevara:2018wpp,Guevara:2019fsj,chkl2019,aho2020,%
Johansson:2019dnu,Aoude:2020onz,Lazopoulos:2021mna,%
Bern:2022kto,Aoude:2022trd,%
Falkowski:2020aso,Chiodaroli:2021eug,%
Ochirov:2022nqz,Cangemi:2022bew,Cangemi:2023bpe,%
fabian1,fabian2,zihan23,Saketh:2023bul,%
Cangemi:2022abk,Alessio:2025nzd%
} has been devoted to the problem.
In works \cite{aho2020,gmoov},
the Newman-Janis (NJ) property \cite{Newman:1965tw-janis}
of the Kerr solution
has been identified as an important clue.

\paragraph{Criteria}
\begin{enumerate}[label=(\alph*)]
    \item
        The Kerr effective action
        should reduce to
        the Levi-Steinhoff \cite{Levi:2015msa} action
        at the leading order in the gravitational coupling.
    \item
        The Kerr effective EoM should
        reduce to an all-orders-in-spin
        extension of
        Mathisson-Papapetrou-Dixon (MPD) \cite{Mathisson:1937zz,Papapetrou:1951pa,Dixon:1970zza} equations 
        (cf. Refs. \cite{Steinhoff:2009tk,Harte:2011ku,Vines:2016unv,Compere:2023alp,Ramond:2026fpi,%
        khriplovich1989particle,Yee:1993ya,Khriplovich:1997ni,Thorne:1984mz})
        for $C_\ell=1$
        at the leading order in the gravitational coupling.
        \nomenclature{MPD}{Mathisson-Papapetrou-Dixon}
    \item
        The Kerr effective action or EoM
        may reflect the NJ
        property,
        which
        will assign certain meanings
        to the complex combinations $x^\m {\,\pm\:} iy^\m$.
\end{enumerate}

\paragraph{Problem solved if}
\begin{enumerate}[label=(\alph*)]
    \item 
        An IR principle
        (e.g., symmetry)
        is shown to constrain the effective theory entirely.
    \item
        All Wilson coefficients
        are determined by matching with the UV theory,
        which is GR.
        This entails
        solving
        black hole perturbation theory
        \cite{Teukolsky:1972my,Teukolsky:1973ha,%
        fabian1,fabian2,zihan23,Saketh:2023bul}.
    \item
        One is able to predict the gravitational Compton amplitudes of the Kerr black hole
        to any graviton multiplicities and helicities.
\end{enumerate}

\paragraph{Significance}

\begin{itemize}[leftmargin=1.25em]
    \item 
        Quenches a fundamental curiosity in classical gravitation
        and relativistic spinning-particle mechanics.
    \item 
        Is directly relevant to the relativistic two-body problem for spinning black holes.
    \item
        Is directly relevant to
        precision PM calculations
        if one is interested in the
        large spin or spin-resummed regime.
        The gravitational Compton amplitude,
        for instance,
        can be directly fed into
        the 2PM two-body eikonal.
\end{itemize}

\paragraph{Related Problems}

\begin{enumerate}[label=(\alph*)]
    \item
        Investigate spinning black holes
        with electromagnetic or YM hairs.
        The abelian {\kerr} solution 
        is the Maxwell analog of
        the Kerr black hole
        by classical double copy,
        whose trivial embedding into YM theory is
        the nonabelian {\kerr} solution
        \cite{monteiro2014black}.
        These are the $\GN\too0$ limit of the Kerr-Newman solution \cite{Newman:1965my-kerrmetric,aho2020}
        and its nonabelian generalization  \cite{Perry:1977wk}.
        Determine the point-particle effective action or EoM
        of abelian/nonabelian {\kerr} or Kerr-Newman black holes
        in external fields.
    \item
        The Newman-Janis algorithm (NJA)
        \cite{Newman:1965tw-janis}
        \nomenclature{NJA}{Newman-Janis Algorithm}
        is a solution-generating technique
        that wishes to add angular momentum to four-dimensional black holes.
        In a sense, it ``derives'' the Kerr metric
        by applying an imaginary displacement ``$+ia$'' to the Schwarzschild metric.
        This procedure was originally introduced as an ad hoc trick
        without a clear physical basis
        \cite{Newman:1965tw-janis}.
        However, it holds a historical significance as
        the very crucial method
        that enabled the derivation of the very Kerr-Newman solution \cite{Newman:1965my-kerrmetric}.
        Figure out
        whether the NJA is
        a mere mathematical trick
        or an indication of a deeper physical structure
        (cf. \rrcite{Talbot:1969bpa,Drake:1998gf,schiffer1973kerr,Finkelstein:1974nr,Gurses:1975vu,flaherty1976hermitian,Rajan:2016zmq,giampieri1990introducing,Erbin:2014aya,Erbin:2016lzq},
        \rrcite{newman1974curiosity,newman1974collection,Newman:1973afx,Newman:2004ba,Newman:1973yu,Newman:2002mk,Newman:1976gc,ko1981theory},
        \rrcite{ahh2017,Guevara:2018wpp,Guevara:2019fsj,chkl2019,aho2020},
        and \rrcite{Crawley:2021auj,Adamo:2023fbj}).
        If there is a physical content,
        investigate if it can help determine the point-particle effective theory of spinning black holes.
    \item
        The relativistic two-body problem
        is notoriously hard to solve.
        In particular,
        one does not expect to solve the relativistic two-body problem for a spinning binary formed by random stellar objects.
        Black holes, however, are believed to be special
        as 
        ``the most perfect macroscopic objects'' in the universe
        (in the words of Chandrasekhar \cite{chandrasekhar1983mathematical}).
        Do spinning black holes
        exhibit a certain ideal property
        that allows room for integrability
        in their binary problem,
        under particular assumptions
        or in specific regimes?
        If so, how is that property
        stated in the point-particle effective theory?
\end{enumerate}\newpage
\section{ Plan of This Thesis}

This thesis is organized as follows.
\begin{enumerate}[
    start=0,%
    leftmargin=5em,%
    label={Part \arabic*.\,}%
]
    \item 
        Reimagining Gravity
    \item
        Generalized Symmetries
        \hfill
        {\small
            (\textsc{Problem 1} in \Sec{Ib1})
        }
    \item
        Double Copy
        \hfill
        {\small
            (\textsc{Problem 2} in \Sec{Ib2})
        }
    \item
        Spinning Black Holes
        \hfill
        {\small
            (\textsc{Problem 3} in \Sec{Ib3})
        }
\end{enumerate}
The contents of each part are summarized below.

\begin{enumerate}[
    start=0,%
    leftmargin=5em,%
    label={Part \arabic*.\,}%
]
    \item
        \Chap{J}
        reviews, derives, and envisions
        various formulations of GR
        in general and four dimensions:
        metric-based formulations (\Sec{Ja}),
        vielbein formulation (\Sec{J2}),
        teleparallel formulation (\Sec{Jb}),
        SD gravity and its simplicity (\Sec{Jc}),
        chiral and related formulations (\Sec{Jcx}).
        This reimagines gravity
        as a Lorentz gauge theory,
        a diffeomorphism gauge theory,
        or
        a tapestry of SD and anti-self-dual (ASD) parts.
    \item
        \Chap{K1}
        constructs a one-form symmetry in dynamical gravity.
        This identifies a new symmetry of our nature
        at scales below the lightest neutrino mass.
        It views gravity
        as a Lorentz gauge theory.
        It intro\-duces a tetradic cousin of cosmic string
        as a local Lorentz vortex.
    \item
        \Chap{K2}
        investigates the idea of diffeomorphism gauge theory as an implication of EoM level CK duality.
        It wishes to view gravity
        as a diffeomorphism gauge theory,
        but results are only exact in three dimensions or in the SD sector in four dimensions.
        It illuminates the Misner string
        as a diffeomorphism vortex.
    \item
        \Chap{K3:HYDROGEN}
        offers a pedagogical demonstration of
        the simplicity of SD black holes.
        The motion of a relativistic test particle 
        in the backgrounds of a SD black hole
        is maximally superintegrable
        by hidden symmetries,
        reincarnating the Kepler problem (hydrogen atom).

        \Chap{K3:SDTN}
        derives a previously unknown KS metric
        for the SD Taub-Newman-Unti-Tamburino (NUT) black hole
        by classical double copy.
        \nomenclature{NUT}{Newman-Unti-Tamburino}
        An explicit diffeomorphism proves its equivalence with a previously known metric due to Gibbons and Hawking.

        \Chap{K3:NJA}
        shows that the Kerr metric
        represents a pair of SD and ASD Taub-NUT black holes.
        This elevates the NJA to a rigorous derivation of the Kerr metric with a definite physical origin.
        Real black holes are derived by
        nonlinear superpositions of
        SD and ASD building blocks
        (Taub-NUT instantons and chiral dyons).
        
        The systematic program of investigating
        four-dimensional black holes---%
        as \textit{background geometries}---%
        in terms of SD and ASD building blocks
        is established.
        
        The systematic program of investigating
        four-dimensional black holes---%
        as \textit{dynamical objects} in point-particle effective theory---%
        in terms of SD and ASD building blocks
        is proposed.

        To this end,
        \Chap{K3:GDE}
        carries out an essential development
        on all-orders geodesic deviation
        as a preliminary.
        A new formalism based on 
        tangent bundle
        is established
        as an alternative to Synge calculus.
        Explicit expressions
        for the geodesic deviation equation (GDE)
        \nomenclature{GDE}{Geodesic Deviation Equation}
        and its Lagrangian formulation
        are obtained up to tenth order.

        \Chap{K3:PROBENJ}
        introduces the probe counterpart of NJA,
        which Wick rotates the all-orders GDE
        into a part of exact spinning-particle EoM.
        The point-particle effective action of the Kerr black hole
        is completely constrained in the SD sector
        for hidden symmetries,
        which identifies a physical IR principle.
        
        \Chap{K3:OSD1}
        applies the probe counterpart of NJA
        to derive a candidate exact point-particle effective action
        of the Kerr black hole
        in gravitational backgrounds.
        
        \Chap{K3:OSD11}
        applies the probe counterpart of NJA
        to derive a candidate point-particle effective action
        of the Kerr-Newman black hole
        in Einstein-Maxwell backgrounds.
        
        \Chap{K3:SST}
        establishes a framework for spinning-particle mechanics
        that unifies spacetime and spin into a complex four-manifold,
        resurrecting an idea that had been
        envisioned by Newman \cite{newman1974curiosity,newman1974collection,Newman:1973afx,Newman:2004ba,Newman:1973yu,Newman:2002mk,Newman:1976gc,ko1981theory}.
        
        \Chap{K3:BHEOM}
        demonstrates and outlines
        the derivation of
        exact
        spinning black hole EoM
        beyond the SD sector
        in the spinspacetime framework.
\phantom{
    \cite{Adamo:2023fbj,Adamo:2024xpc,Adamo:2025fqt,Crawley:2021auj,Crawley:2023brz,Guevara:2023wlr,Guevara:2024edh,Guevara:2025psg,note-sdtn,nja,AAS,Skvortsov:2025ohi,Doran:2026bng}
    \cite{Atiyah:1978wi,Gibbons:1987sp,Nozawa:2015qea}
}

        An integrable subsector is identified for
        the general-relativistic binary problem
        of Kerr or Kerr-Newman black holes,
        again exhibiting the hidden symmetries of
        the Kepler problem.

        \Chap{K3:COMPTON}
        gives a Lagrangian derivation of
        the Compton scattering amplitudes of spinning black holes
        for all helicity configurations.
        A chiral perturbation theory is developed,
        achieving Kerr by expanding around SD Taub-NUT.
\end{enumerate}

\chapter{Reimagining Gravity}\label{J}

According to its original roots,
the term relativity
refers to
the investigation into 
how physics appear in different frames of reference.
Newtonian mechanics
was built upon
the Galilean notion of relativity \cite{galilei-dialogue},
postulating absolute time but relative space.
Special relativity was born
from the conceptual clash between Galilean relativity and Maxwell's electrodynamics
\cite{einstein1905electrodynamics},
leading to a unification of space and time.
GR then 
fully embraced
the principle of general covariance,
investigating
how physical laws appear 
in arbitrary 
coordinate systems
on curved spacetimes.

On the one hand,
the freedom to change one’s description of physics is a blessing.
By using
different coordinate systems or observers,
one can manifest or highlight different aspects of the same physical system.
On the other hand,
the same freedom can turn into a trap
leading into a maze
of fictitious or redundant structures.

Consider the Lagrangian
$\mathcal{L} = 
    - \tfrac{1}{2}\mem (\partial\phi)^2
    - g\phi\mem (\partial\phi)^2
    - \tfrac{1}{2}\mem g^2\phi^2\mem (\partial\phi)^2
$,
where $g$ is a small coupling constant.
Apparently, it defines a perturbative QFT
where particles interact via
cubic and quartic vertices.
This, however, turns out to be a mirage.
Computation shows that scattering amplitudes all vanish,
despite the fact that
the number of Feynman diagrams
explodes 
with the number of external legs.
This demonstrates the falseness 
inherent in
the Feynman diagram representation of scattering amplitudes.

The resolution is simple.
Take 
$\mathcal{L} \eqq
    {- \tfrac{1}{2}\mem (\partial\Phi)^2}
$,
which evidently defines a free theory.
Our earlier quartic Lagrangian is reproduced by 
putting
$\Phi = \phi + g\mem \phi^2 \nem/2$,
which is a \textit{field redefinition}
that preserves the external scattering states.
The \textit{S-matrix equivalence theorem}
\cite{Chisholm:1961tha,Kamefuchi:1961sb,Arzt:1993gz,Cohen:2023ekv,Haag:1958vt} states that
scattering amplitudes are
physical observables that are 
invariant under such field redefinitions.

The above demonstration
is a typical example
well-known in the field theory community \cite{Cheung:2017pzi}.
To the eyes of a geometer,
it is interesting to note that
field redefinitions
could be thought of as
coordinate transformations
in the ``field space''
\cite{Cheung:2021yog,Cohen:2024bml,DeWitt:1984sjp,Vilkovisky:1984st,Coleman:1969sm,Callan:1969sn}.
In particular,
$\Phi = \phi + g\mem \phi^2 \nem/2$
is a local change of coord\-inate
on a one-dimensional line.
The free Lagrangian
$\mathcal{L} =
    {- \tfrac{1}{2}\mem (\partial\Phi)^2}
$
is how physics appears to an ``inertial observer''
(nice field basis),
while
$\mathcal{L} = 
    - \tfrac{1}{2}\mem (\partial\phi)^2
$ $
    - g\phi\mem (\partial\phi)^2
    - \tfrac{1}{2}\mem g^2\phi^2\mem (\partial\phi)^2
$ is how a ``noninertial observer''
(bad field basis)
sees the physics.
The latter is plagued by
virtual
interacting
particles as
fictitious, ``noninertial'' artifacts
creating the false impression of an interacting QFT.

In short,
a field basis declares one's choice of coordinates on the field space.
Scattering amplitudes are physical observables
independent of such choice.
In this view,
the subjectivity involved in Feynman rules/diagrams
can be seen as
a higher-level 
reincarnation of
GR's general covariance.

\medskip

At this moment,
we may want to
describe how CK duality,
introduced in \Sec{Ia},
could be revisited in this perspective.
Again, the angle is
the treachery of Feynman diagrams.
Scattering amplitudes are gauge-invariant physical observables.
The physical meaning of
Lehmann–Symanzik–Zimmermann\:(LSZ) reduction
\cite{Lehmann:1955zz}
\nomenclature{LSZ}{Lehmann–Symanzik–Zimmermann}
is bulk-to-boundary propagation
(cf. \rcite{Cheung:2022pdk}):
particles escaping to the boundary are on-shell and physical,
while
particles remaining in the bulk are off-shell and virtual.
The bulk physics is field basis dependent.

Take the usual YM Lagrangian, which describes a quartic perturbation theory.
The computation of a tree-level gluon scattering amplitude $\A$
via Feynman diagrams
implements a sum over
tree graphs,
where each vertex carries degree three or four.
Crucially, this diagrammatic representation of $\A$
is field basis dependent.
By a field redefinition
that preserves the external states,
the Feynman rules can change arbitrarily
while leaving the scattering amplitudes invariant.
As a result, one can obtain the same $\A$
by imposing a totally different rule for
how virtual gluons interact in the bulk,
while keeping the external on-shell gluons the same.
This demonstrates the falseness of virtual gluons
in the Feynman diagram representation of scattering amplitudes.

CK duality claims that
there may exist a particularly nice 
coordinate system (``observer'') on the field space of YM theory,
such that
the virtual gluons only interact via a universal cubic vertex
composed of two Lie algebra factors.
The one is the structure constant of the very color Lie algebra,
such as $\su(N)$.
The other is the structure constant of the mystery Lie algebra $\gk$,
which takes the kinematic data (momenta and polarizations) as ``indices.''\footnote{
    To be precise,
    CK duality states that
    for each triplet of cubic graphs
    $(\Gamma_i,\Gamma_j,\Gamma_k)$
    whose color factors 
    are related by the Jacobi identity
    $c_i + c_j + c_k = 0$,
    the same relation holds for the numerator factors as
    $n_i + n_j + n_k = 0$
    \cite{BCJ1,BCJ2,BCJReview}.
    The simplest explanation for this fact
    is the existence of a Lie algebra $\gk$
    that constructs the BCJ numerators $(n_i,n_j,n_k)$.
}
In this sense it is said that
YM theory exhibits
a duality between color and kinematics.

Once such a cubic representation of a gluon amplitude is obtained,
one performs the diagram-wise operation of
discarding the color factor
and doubling the kinematic factor.
Remarkably,
this produces a graviton amplitude.
This is the very double copy correspondence
between YM theory and GR.

The implication is that
there may exist a particularly nice choice of 
a ``coordinate system
on the field space of gravity,''
such that the virtual gravitons
exhibit only a cubic interaction
through a universal vertex
that simply squares the structure constant of $\gk$.

Note how scattering amplitudes
serve as concrete, invariant anchors
when navigating the vast sea of field spaces.

\medskip

In this chapter,
we would like to explore the field spaces for gravity.
This is an investigation into how GR appears in various
frames of descriptions,
each manifesting or highlighting
different aspects of the same physics: gravitation.

\section{ Gravity as Metric Geometry}
\label{Ja}

According to the standard textbook view, 
GR is a geometrical theory where the metric
is the ``foundation of all'' \cite{mtw}.
At the risk of boring the reader,
let us quickly review how the metric determines all geometrical structures in GR:
Levi-Civita connection,
Riemann tensor,
Ricci tensor,
and Ricci scalar.

A metric $g$ 
on a manifold $\M$
is a symmetric rank-two tensor field $g_{\m\n} = g_{\n\m}$.
It defines the Levi-Civita connection $\nabla$
as the unique torsion-free affine connection
on the tangent bundle $T\M$
compatible with it:
for arbitrary vector fields $U$, $V$, and $W$,
\begin{subequations}
\label{LCconditions}
\begin{align}
    \label{indexfree.torsionfree}
    \nabla_V W - \nabla_W V \,&=\, \comm{V}{W}
    \,,\\
    \label{indexfree.metricity}
    \nabla_U\hnem \bigbig{
        g(V,W)
    }
    \,&=\,
        g\bigbig{
            \nabla_U V,\mem W
        }
        + g\bigbig{
            V,\mem \nabla_U W
        }
    \,.
\end{align}
\end{subequations}
Here, $\comm{V}{W}$ denotes the Lie bracket between vector fields,
\begin{align}
    \comm{V}{W}
    \,=\,
        \BB{
            V^\r \partial_\r W^\m
            - W^\r \partial_\r V^\m
        }\mem
        \partial_\m
    \,.
\end{align}
The Christoffel symbols are
the connection coefficients 
$\Gamma^\m{}_{\n\r} = \cont{dx^\m}{\nabla_{\partial_\r} \partial_\n}$
of the Levi-Civita connection
in the coordinate frame.
By plugging in the coordinate basis vector fields,
\eqref{LCconditions} translates to
$\Gamma^\m{}_{\r\s} = \Gamma^\m{}_{\s\r}$
and $\partial_\r g_{\m\n} = g_{\m\k}\mem \Gamma^k{}_{\n\r} + g_{\k\n}\mem \Gamma^\k{}_{\m\r}$.
The solution is unique:
\begin{align}
    \label{christoffel}
    \Gamma^\m{}_{\r\s}
    \,&=\,
        \frac{1}{2}\, (g^{-1})^{\m\n}\mem
        \BB{
            - \partial_\n g_{\r\s}
            + \partial_\s g_{\n\r}
            + \partial_\r g_{\n\s}
        }
    \,.
\end{align}
Although nonstandard,
we denote the inverse metric as $(g^{-1})^{\m\n}$
for clarity.

\newpage

The curvature tensor of an affine connection $\nabla$ is defined as
\begin{align}
    \label{indexfree.riemann}
    R(V,W)\mem U
    \,=\,
        \nabla_V \nabla_W U
        - \nabla_W \nabla_V U
        - \nabla_{[V,W]} U
    \,.
\end{align}
By plugging in coordinate basis vector fields, this translates to
\begin{align}
    \label{riemann}
    R^\m{}_{\n\r\s}
    \,=\,
        \partial_\r \Gamma^\m{}_{\n\s}
        - \partial_\s \Gamma^\m{}_{\n\r}
        + \Gamma^\m{}_{\k\r} \Gamma^\k{}_{\n\s}
        - \Gamma^\m{}_{\k\s} \Gamma^\k{}_{\n\r}
    \,.
\end{align}
The Ricci tensor, Ricci scalar, and Einstein tensor
are then defined as
\begin{align}
    \label{ricci}
    \Ric_{\n\s}
    \,=\,
        R^\m{}_{\n\m\s}
    \,,\quad
    \Ric
    \,=\,
        \Ric_{\m\n} (g^{-1})^{\m\n}
    \,,\quad
    G_{\m\n}
    \,=\,   
        \Ric_{\m\n}
        - \tfrac{1}{2}\, \Ric\mem g_{\m\n}
    \,.
\end{align}

If $\nabla$ is Levi-Civita,
\eqref{riemann} defines the Riemann tensor $R^\m{}_{\n\r\s}$.
In this case,
it follows that the Riemann and Ricci tensors exhibit the index symmetries
\begin{align}
    \label{Rindexsym}
    R_{\m\n\r\s} 
    \,=\,
        R_{[\m\n][\r\s]}
    \,=\,
        R_{[\r\s][\m\n]}
    \,,\quad
    \Ric_{\m\n}
    \,=\,
        \Ric_{\n\m}
    \,,
\end{align}
where indices are raised and lowered by $(g^{-1})^{\m\n}$ and $g_{\m\n}$.
The Riemann tensor satisfies
the algebraic and differential Bianchi identities:
\begin{align}
    \label{Rbianchis}
    R^\m{}_{[\n\r\s]} \,=\, 0
    \,,\quad
    \nabla_{[\k|} R^\m{}_{\n|\r\s]} \,=\, 0
    \,.
\end{align}

\subsection{Einstein-Hilbert Action}

The Einstein-Hilbert action reads
\begin{align}
    \label{EHaction}
    S[g]
    \,=\,
        \frac{1}{2\k^2}
        \int d^dx\,
            \sqrt{-\nem\det g}\:
            \Ric[g]
    \,,
\end{align}
which governs a nondegenerate Lorentzian-signature metric
$g_{\m\n}$
as a sole dynamical symmetric tensor field
given to a $d$-dimensional spacetime manifold.
We have assumed vanishing cosmological constant
for simplicity
and identified 
\begin{align}
    \label{kappa}
    \k \,=\, \sqrt{8\pi\GN}
\end{align}
as the gravitational coupling.
Variation of \eqref{EHaction} derives
the vacuum Einstein's equations,
\begin{align}
    \label{EinsteinEq}
    \Ric_{\m\n}[g]
    - \frac{1}{2}\, \Ric[g]\mem g_{\m\n}
    \,=\,
        0
    \,.
\end{align}
Here, $\Ric_{\m\n}[g]$ and $\Ric[g]$ denote the Ricci tensor and the Ricci scalar 
of the metric $g$,
respectively.

\subsection{Landau-Lifshitz Equations}
\label{Ja>LL}

In the perturbation theory point of view,
the metric formulation of GR
is practically a nightmare.
Expanding the metric $g_{\m\n}$ around a flat background
produces an infinite number of vertices with all valences,
each of which describes a complicated expression.

Regarding this matter,
it is a well-known wisdom
that utilizing the densitized inverse metric
(``gothic metric'')
tends to be helpful:
\begin{align}
    \label{got-def}
    \got^{\m\n}
    \,=\,
        \sqrt{-\nem\det g}\, (g^{-1})^{\m\n}
    \,.
\end{align}
The gothic metric is utilized in
the Landau-Lifshitz (LL) formulation of Einstein's equations
\cite{LL:1975vol2},
\nomenclature{LL}{Landau-Lifshitz}
nicely reviewed in Section 6.1 of Poisson and Will \cite{PoissonWill:2014}.
Note that
\eqref{got-def} is a local, algebraic field redefinition:
a simple change of coordinates
on the field space.

Consider the tensor density
\begin{align}
    \label{LL:Htensor}
    H^{\m\n\a\b}
    \,=\,
        \got^{\m\a} \got^{\n\b}
        - \got^{\m\b} \got^{\n\a}
    \,,
\end{align}
which exhibits the same $\boxplus$-tableaux index symmetry
as the Riemann tensor.
Computation shows that the Einstein tensor,
when densitized and index-raised,
can be identically rewritten as
\begin{align}
    \label{LLid}
    (-2\det g)\, G^{\m\a}
    \,=\,
        \partial_\n \partial_\b H^{\m\n\a\b}
        - \t_\text{LL}^{\m\a}
    \,,
\end{align}
where $\t_\text{LL}^{\m\a}$ describes a complicated expression
involving $\got^{\m\n}$, $\got^{\m\n}{}_{,\r}$, and $\igot_{\m\n}$.
Explicitly,
$\tfrac{1}{2}\mem \t_\text{LL}^{\m\a}$ is given by
\begin{align}
\label{LL.tau}
\nonumber
    &
    \BB{
        \tfrac{1}{4}\,
            \got^{\m\a}\mem
                \igot_{\c\d}\mem
                \got^{\r\c}{}_{,\s}\mem
                \got^{\s\d}{}_{,\r}\mem
        + \tfrac{1}{2}\,
            \got^{\m\k}{}_{,\r}\mem
            \igot_{\k\l}\mem
            \got^{\l\a}{}_{,\s}\mem
            \got^{\r\s}
        -
            \got^{\r(\m}\mem \got^{\a)\k}{}_{,\n}\mem
            \igot_{\k\l}\mem \got^{\l\n}{}_{,\r}
    }
\nonumber\\
    &
        + \tfrac{1}{8}\,
        \BB{
                \igot_{\k(\r} 
                \igot_{\s)\l}
            - \tfrac{1}{d-2}\mem
                \igot_{\k\l}\mem
                \igot_{\r\s}
        }
        \BB{
            2\mem \got^{\m(\c} \got^{\d)\a}
            - \got^{\m\a} \got^{\c\d}
        }\mem
        \got^{\k\l}{}_{,\c}\mem
        \got^{\r\s}{}_{,\d}
\nonumber\\
    &
        + \tfrac{1}{2}\,
        \BB{
            \got^{\m\a}{}_{,\n}\mem
            \got^{\n\b}{}_{,\b}
            -
            \got^{\m\n}{}_{,\n}\mem
            \got^{\a\b}{}_{,\b}
        }
    \,.
\end{align}

The identity in
\eqref{LLid} implies that
\eqref{EinsteinEq} is equivalent to
\begin{align}
    \label{LL.eom}
    \partial_\n \partial_\b H^{\m\n\a\b}
    \,=\,
        \t_\text{LL}^{\m\a}
    \,.
\end{align}
\eqref{LL.eom} is the LL formulation of
vacuum Einstein's equations,
which we might simply refer to as LL equations.

It should also be clear that 
a stress-energy pseudotensor
of the gravitational field,
conserved in terms of the partial derivative,
can be constructed with the aid of $\t_\text{LL}^{\m\a}$
\cite{LL:1975vol2,PoissonWill:2014}.

The LL formulation gets along well with the harmonic (de Donder) gauge,
\begin{align}
    \label{deDonder}
    \partial_\m \got^{\m\a}
    \,=\,
        0
    \qfq
    \Gamma^\m{}_{\r\s}\mem (g^{-1})^{\r\s}
    \,=\,
        0
    \,.
\end{align}
This kills the last line in \eqref{LL.tau}, for instance.
Still, the perturbation theory due to 
the LL equations in \eqref{LL.eom}
describes an infinite number of vertices 
of all valences,
as can be seen by plugging in the expansion 
$\got^{\m\a} = \eta^{\m\a} + h^{\m\a}$.

\subsection{Cheung-Remmen Action}

Notably,
Cheung and Remmen \cite{CheungRemmen:2017}
provided a cubic, first-order Lagrangian formulation of GR
that employs the gothic metric
$\got^{\m\n} = \got^{\n\m}$
together with an auxiliary field
$\mathfrak{A}^\m{}_{\r\s} = \mathfrak{A}^\m{}_{\s\r}$:
\begin{align}
    \label{CheungRemmen}
    S[\got,\mathfrak{A}]
    \,=\,
        -\frac{1}{2\k^2}
        \int d^dx\,
            \lrp{
            \begin{aligned}[c]
                &
                    \mathfrak{A}^\m{}_{\r\s}\, \partial_\m \got^{\r\s}
                \\[-0.075\baselineskip]
                &
                    +
                    \BB{
                        \mathfrak{A}^\m{}_{\n\r}\mem \mathfrak{A}^\n{}_{\m\s}
                        - \tfrac{1}{d-1}\,
                            \mathfrak{A}^\m{}_{\m\r}\mem
                            \mathfrak{A}^\n{}_{\n\s}
                    }\mem \got^{\r\s}
            \end{aligned}
            }
    \,.
\end{align}
Integrating out $\mathfrak{A}$
shows that \eqref{CheungRemmen} is exactly equivalent to
\begin{align}
    \label{CheungRemmen2}
    S[\got]
    \,=\,
        \frac{1}{2\k^2}
        \int d^dx\,
            \lrp{
            \begin{aligned}[c]
            &
                \tfrac{1}{2}\,
                    \igot_{\m\n}\mem
                    \partial_\r \got^{\m\s}\mem
                    \partial_\s \got^{\n\r}
            \\[-0.15\baselineskip]&
                - \tfrac{1}{4}\,
                    \igot_{\m\n}\mem
                    \igot_{\r\s}\mem
                    \got^{\k\l}\mem
                    \partial_\k \got^{\m\r}\mem
                    \partial_\l \got^{\n\s}
            \\[-0.15\baselineskip]&
                + \tfrac{1}{4}\mem \tfrac{1}{d-2}\mem
                    \igot^{\k\l}\mem
                    (\partial_\k\log\det\got)
                    (\partial_\l\log\det\got)
            \end{aligned}
            }
    \,.
\end{align}
Brute-force calculation verifies that
\eqref{CheungRemmen2}
is equivalent to the Einstein-Hilbert action in \eqref{EHaction}
upon plugging in \eqref{got-def}
as an algebraic field redefinition.
It is also insightful to note that
a further algebraic field redefinition,
$\mathfrak{A}^\m{}_{\r\s} = \mathfrak{B}^\m{}_{\r\s} - \delta^\m{}_{(\r}\mem \mathfrak{B}{}^\k{}_{\s)\k}$,
brings \eqref{CheungRemmen}
to a form reminiscent of
the metric-affine formulation \cite{Dadhich:2012htv} of GR,
as is pointed out in \rcite{CheungRemmen:2017}.

\section{ Gravity as Lorentz Gauge Theory}
\label{J2}

The metric formulation of GR describes the dynamics of a pseudo-Riemannian manifold $(\M,g)$
in terms of the Einstein-Hilbert action.
The vielbein formulation of GR
offers a different view
in which the metric is 
a derived object.

\paragraph{Coframe}

A coframe $e$ is a local trivialization of the cotangent bundle $\Tstar\M$.
Explicitly,
one thinks of a collection of differential one-forms
$e^A = e^A{}_\m\mem dx^\m$ on $\M$
for $A = 0,1,{\cdots},(d\mminus1)$.
Suppose $|e| := \det e$
is nonzero at any point in $\M$.

In a pseudo-Riemannian manifold $(\M,g)$,
an orthonormal coframe
refers to a coframe that represents
the metric $g$ as
\begin{align}
    \label{g=ee}
    g_{\m\n}
    \,=\,
        \eta_{AB}\,
            e^A{}_\m\mem e^B{}_\n
    \,,
\end{align}
where $\eta_{AB} = \diag(-1,+1,{\cdots},+1)$ describes a constant matrix.
\newpage

This representation, however, is redundant.
Two orthonormal coframes represent the same metric by the equivalence
\begin{align}
    \label{LLT.e}
    e^A{}_\m
    \,\,\,\sim\,\,\,
    \Omega^A{}_B\mem e^B{}_\m
    \,,
\end{align}
where $\Omega^A{}_B$ is
an orthogonal matrix
assigned to each point in $\M$.
This describes a redundancy
involved in the description of pseudo-Riemannian manifolds
by the orthonormal coframe.
\eqref{LLT.e} is referred to as a local Lorentz transformation.

The vielbein formulation
aims to view GR as
governing a dynamical coframe $e$
subject to the equivalence relation in \eqref{LLT.e}.
This takes $e$ as fundamental, while
the pseudo-Riemannian metric $g$
is a derived structure 
via \eqref{g=ee}.
The vielbein formulation may also be called
the coframe formulation of GR.

Since \eqref{LLT.e} describes a fiberwise transformation,
the coframe formulation of GR
conforms to the grammar of gauge theory
in the precise YM sense:
GR is a gauge theory of the Lorentz group,
although
the diffeomorphism redundancy
has to be taken into account eventually.
The frame indices $A, B, \cdots$ 
are the \textit{internal} indices,
raised and lowered by $\eta_{AB}$ and its inverse $\eta^{AB}$.

Note that there exist various choices for
the local Lorentz group.
For instance, it could be
the orthogonal group $\mathsf{O}(1,d\mminus1)$,
the identity component of $\SO(1,d\mminus1)$,
or
the spin group $\Spin(1,d\mminus1)$.
For concreteness and the perturbative context,
we may presume $\SO(1,d\mminus1)$ below,
in which case
the internal epsilon tensor
$\e_{A_1\cdots A_d}$ becomes additionally available as an invariant structure.

\paragraph{Frame}

A frame $E$ is a local trivialization of the tangent bundle $T\M$.
Explicitly, it is a collection of vector fields $E_A = E^\m{}_A\mem \partial_\m$ on $\M$
for $A = 0,1,{\cdots},(d-1)$.
Suppose $|E| := \det E$ does not vanish at any point in $\M$.

Suppose a frame subject to the equivalence relation
\begin{align}
    \label{LLT.E}
    E^\m{}_A
    \,\,\,\sim\,\,\,
        E^\m{}_B\mem (\Omega^{-1})^B{}_A
    \,,
\end{align}
for $\Omega^A{}_B$ valued in a Lorentz group.
In a pseudo-Riemannian manifold $(\M,g)$,
an orthonormal frame refers to such a frame
such that
\begin{align}
    g(E_A,E_B)
    \,=\,
        g_{\m\n}\hem E^\m{}_A\hem E^\n{}_B
    \,=\,
        \eta_{AB}
    \,.
\end{align}
We can take it as the dual of
the orthonormal coframe introduced in \eqref{g=ee}:
\begin{align}
    e^A{}_\m\mem E^\m{}_B
    \,=\,
        \delta^A{}_B
    \,,\quad
    E^\m{}_A\mem e^A{}_\n
    \,=\,
        \delta^\m{}_\n
    \,,
\end{align}
so the two local Lorentz transformations in
\eqrefs{LLT.e}{LLT.E}
are linked.

\newpage

\paragraph{Spin Connection}

Next, let $\nabla$ be the Levi-Civita connection of
the pseudo-Riemannian metric $g$.
The spin connection coefficients
are the connection coefficients of $\nabla$
in the orthonormal frame:
\begin{align}
    \label{gamma-def}
    \gamma^A{}_{BC}
    \,=\,
        \cont{e^A}{
            \nabla_{E_C} E_B
        }
    \,=\,
        \BB{
            - \partial_\r e^A{}_\n 
            + e^A{}_\m\mem \Gamma^\m{}_{\n\r}
        }\mem E^\n{}_B\mem E^\r{}_C
    \,.
\end{align}
As a one-form, the spin connection 
in the current context
refers to
\begin{align}
    \gamma^A{}_B
    \,=\,
        \gamma^A{}_{BC}\mem e^C
    \,.
\end{align}
Under \eqref{LLT.e},
the spin connection transforms as
\begin{align}
    \label{LLT.gamma}
    \gamma^A{}_B
    \,\,\,\sim\,\,\,
        \Omega^A{}_C\mem \gamma^C{}_D\mem (\Omega^{-1})^D{}_B
        + \Omega^A{}_C\mem d(\Omega^{-1})^C{}_B
    \,.
\end{align}
Certainly, \eqref{LLT.gamma} is the very behavior
of a connection one-form, Lie algebra valued.
The gauge-covariant exterior derivative of 
a vector-valued gauge-covariant differential form $\a^A$
is defined as
\begin{align}
    D\a^A
    \,=\,
        d\a^A + \gamma^A{}_B \wedge \a^B
    \,,
\end{align}
while
it is easy to extend this definition
to differential forms valued in any representation of the local Lorentz group.

\begin{subequations}
The torsion-free property of $\nabla$ translates to the equation
\begin{align}
    \label{De=0}
    0
    \,=\,  
        De^A
    \,=\,
        de^A + \gamma^A{}_B \wedge e^B
    \,,
\end{align}
which can also be directly shown from \eqref{gamma-def}.
The metric-preserving property of $\nabla$ translates to the equation
\begin{align}
    \label{Deta=0}
    0 
    \,=\,
        D\eta_{AB}
    \,=\,
        \gamma_{AB}
        + \gamma_{BA}
    \,,
\end{align}
meaning that $\gamma^A{}_B$ is valued in the adjoint of Lorentz.
\end{subequations}

\paragraph{Curvature}

The curvature two-form 
associated with the spin connection is
\begin{align}
    \label{Rform}
    R^A{}_B
    \,=\,
        d\gamma^A{}_B + \gamma^A{}_C \wedge \gamma^C{}_B
    \,=\,
        \frac{1}{2}\,
            R^A{}_{B\r\s}\mem
                dx^\r \swedge dx^\s
    \,.
\end{align}
It follows that
$R^A{}_{B\r\s}$
is the Riemann tensor in \eqref{riemann}
with two of its indices converted to the local Lorentz frame.
Under the local Lorentz transformation in \eqref{LLT.e},
this transforms as
\begin{align}
    \label{LLT.R}
    R^A{}_B
    \,\,\,\sim\,\,\,
        \Omega^A{}_C\mem R^C{}_D\mem (\Omega^{-1})^D{}_B
    \,,
\end{align}
assuming that $\Omega^A{}_B$ is single-valued.

\newpage

For any vector-valued, gauge-covariant differential form $\a^A$,
it holds that
\begin{align}
    \label{DDa}
    D^2 \a^A
    \,=\,
        R^A{}_B \wedge \a^B
    \,.
\end{align}
With this understanding,
\begin{subequations}
\eqref{De=0} implies
\begin{align}
    \label{DDe=0}
    0 
    \,=\, D^2e^A 
    \,=\,
        R^A{}_B \wedge e^B
    \,=\,
        \frac{1}{3!}\,
            \BB{
                3\mem R^A{}_{[\n\r\s]}
            }\mem
                dx^\n \swedge dx^\r \swedge dx^\s        
    \,,
\end{align}
while
\eqref{Deta=0} implies
\begin{align}
    \label{DDeta=0}
    0
    \,=\, D^2\eta_{AB}
    \,=\,
        R_{AB} + R_{BA}
    \,.
\end{align}
\end{subequations}
\eqref{DDe=0} derives the algebraic Bianchi identity in \eqref{Rbianchis}.
\eqref{DDeta=0} states that the curvature two-form is Lorentz-valued.

To derive the differential Bianchi identity,
consider computing 
$D^3\a^A$ as $D^2(D\a^A)$ and $D(D^2\a^A)$.
By equating the two, one obtains
\begin{align}
    \label{DR}
    0
    \,=\,
        DR^A{}_B
    \,=\,
        \frac{1}{3!}\,
            \BB{
                3\mem D_{[\n|} R^A{}_{B|\r\s]}
            }\mem
                dx^\n \swedge dx^\r \swedge dx^\s    
    \,.
\end{align}

\paragraph{Einstein Tensor}

The internal Hodge dual, $\star$,
is defined
via contracting the internal indices with
the internal epsilon tensor.
A useful identity follows that
\begin{align}
\begin{split}
    \label{Gid}
    \mathrm{Eins\hnem}_A
    \,&:=\,
        G^\m{}_A\,
        \bb{
            \frac{1}{(d\mminus1)!}\,
            \ve_{\m\r_1\cdots\r_{d-1}}\,
            dx^{\r_1} \swedge \mcdots \swedge dx^{\r_{d-1}}
        }
    \\
    &\,=\,
        {-\frac{1}{(d\mminus3)!}}\:
        {\star R}_{A B_1\cdots B_{d-3}} 
            \swedge e^{B_1} \swedge \mcdots \swedge e^{B_{d-3}}
    \,.
\end{split}
\end{align}
Here, we have denoted
the dynamical volume form as
\begin{align}
    \ve_{\m_1\cdots\m_d} 
    \,=\, |e|\mem \e_{\m_1\cdots\m_d}
    \,,
\end{align}
where
$\e_{\m_1\cdots\m_d}$ is the permutation symbol
taking values in $\{+1,-1,0\}$
such that
$\e_{01\cdots (d-1)} = +1$.
To reiterate, $\e_{\m_1\cdots\m_d}$ is the ``bare'' Levi-Civita symbol whose values are merely constants,
whereas $\ve_{\m_1\cdots\m_d}$ is dynamical
and gives the diffeomorphism-invariant integration measure.
Namely,
\begin{align}
    \ve
    \,=\,
        \frac{1}{d!}\,
            \e_{A_1\cdots A_d}\,
            e^{A_1} \swedge {\cdots} \swedge e^{A_d}
    \,=\,
        \sqrt{-\nem\det g}\,
            d^dx
    \,.
\end{align}

Note that we have named the differential form in \eqref{Gid} as $\mathrm{Eins\hnem}_A$,
which may be called the Einstein $(d\mminus1)$-form.
Another notable identity in this framework is
the covariant $d$-form equation,
\begin{align}
    \label{DE}
        0 \,=\, D \hem\mathrm{Eins\hnem}_A
        \,=\,
            \BB{
                (\hhem\nabla_\m G^\m{}_\n\hnem)\mem E^\n{}_A
            }\, |e|\mem d^dx
    \,,
\end{align}
which encodes the covariant conservation of the Einstein tensor.
This can be seen by
combining \eqref{DR}
with the metric- and $\star$-preserving nature of $D$.

\newpage

\subsection{Electric and Magnetic Equations, I}
\label{Ja>EM1}

\eqref{Gid} shows that
the vacuum Einstein's equations
can be stated
in the language of differential forms as
$
    {\star R}_{A B_1\cdots B_{d-3}}
        \swedge e^{B_1} \swedge \mcdots \swedge e^{B_{d-3}}
    = 0
$.
From physical grounds,
it is helpful to juxtapose this equation with
the algebraic Bianchi identity in \eqref{DDe=0}:
\begin{align}
    \label{GEM0}
    {\star R}_{A B_1\cdots B_{d-3}} 
        \swedge e^{B_1} \swedge \mcdots \swedge e^{B_{d-3}}
    \,=\, 0
    \,,\quad
    R^A{}_B \wedge e^B
    \,=\, 0
    \,.
\end{align}
These are $(d\mminus1)$-form and $3$-form equations,
respectively.
Note how these two equations
involve the same number of derivatives
(on the coframe)
but differ by involving a Hodge star.
The first is a dynamical equation stating the absence of stress-energy.
The second
is an identity
if $R$ is defined as a derived object from $e$.
By analogy,
we may identify these as the
electric and magnetic vacuum equations of GR,
respectively.
The reasoning is shown below.

In Maxwell theory,
the abelian gauge connection $A$
serves as the electromagnetic potential.
The curvature $F = dA$
is the electromagnetic field strength.
This is the Wu-Yang \cite{Wu:1975es} dictionary.
The source-free Maxwell's equations are
\begin{align}
    \label{EM0}
    d\,{*}F \,=\, 0
    \,,\quad
    dF \,=\, 0
    \,,
\end{align}
which are $(d\mminus1)$-form and $3$-form equations,
respectively.
These two equations involve the same number of derivatives
but differ via the Hodge star $*$ acting on spacetime indices.
The first is a dynamical equation stating the absence of electric charges.
The second equation 
encodes the absence of magnetic charges,
which is an identity
if $F$ is defined as a derived object from $A$.

It could be helpful to elaborate 
more on the physical semantics.
Let us do so via the magnetic equations.
Consider the following sequences of differential forms.
For electromagnetism, we have
\begin{align}
\label{sequence.EM}
    A
    \,\,\,\xrightarrow[\qquad]{
        d
    }\,\,\,
    F
    \,\,\,\xrightarrow[\qquad]{
        d
    }\,\,\,
    dF = 0
    \,.
\end{align}
For gravity, we have
\begin{align}
\label{sequence.GEM}
    e^A
    \,\,\,\xrightarrow[\qquad]{
        d
    }\,\,\,
    - \gamma^A{}_B \wedge e^B
    \,\,\,\xrightarrow[\qquad]{
        d
    }\,\,\,
    - R^A{}_B \wedge e^B = 0
    \,.
\end{align}
The Bianchi identity $dF = 0$ is automatic
when the field strength $F$ is derived from the potential $A$.
Earlier, we have sketched
a similar intuition around \eqref{Rbianchis}.
The metric $g$ can be thought of as describing the gravitational potential.
The Levi-Civita connection $\Gamma$
describes
gravitational acceleration
(${\sim\,}9.8 \text{m}/\text{s}^2$)
and frame-dragging forces
measured in a laboratory.
The Riemann tensor $R$
realizes the idea of gravitational field
in a more invariant fashion,
encoding tidal effects.
When the potential $g$ defines everything,
the algebraic Bianchi identity is automatic.
Interestingly, the sequence in \eqref{sequence.GEM}
makes this intuition mathematically precise
in terms of the identity $d^2 = 0$ of exterior calculus
while identifying the coframe $e$ as the gravitational potential.

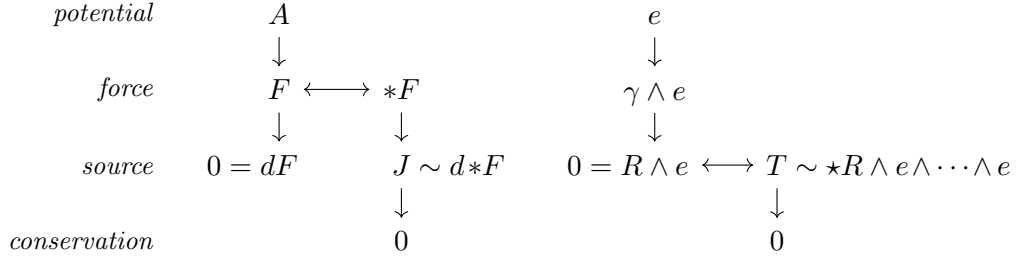
\begin{figure}[t]
    \centering
\begin{align*}
\adjustbox{valign=c}{
    \hspace{-7.0em}
    \adjustbox{valign=t}{\begin{tikzpicture}
        \node[empty] (O) at (0,0) {};
        \node[empty] (nX) at (-2.618, 0) {};
        \node[empty] (X) at (1.618, 0) {};
        \node[empty] (Y) at (0, -1.0) {};
        \node[s] (00) at ($(O)$) {$A$};
        \node[s] (10) at ($(O)+1*(Y)+0*(X)$) {$F$};
        \node[s] (20) at ($(O)+2*(Y)+0*(X)$) {$\phantom{\smash{
            dF
        }\vphantom{d*AFJ}}\mathllap{
        \smash{
            0= {dF}
        }\vphantom{d*AFJ}}$};
        \node[s] (30) at ($(O)+3*(Y)+0*(X)$) {$\phantom{dF}$};
        \node[s] (11) at ($(O)+1*(Y)+1*(X)$) {${\ast}F$};
        \node[s] (21) at ($(O)+2*(Y)+1*(X)$) {$
            \mathrlap{
                \phantom{J} \sim d\mem{*}F
            }J
        $};
        \node[s] (31) at ($(O)+3*(Y)+1*(X)$) {$0$};
        \draw[->] (00)--(10) node[midway,left] {};
        \draw[->] (10)--(20) node[midway,left] {};
        \draw[<->] (10)--(11) node[midway,above] {};
        \draw[->] (11)--(21) node[midway,left] {};
        \draw[->] (21)--(31) node[midway,left] {};
        \node[empty] (t0) at ($(00)+(nX)$) {\footnotesize\phantom{\textit{conservation}}\llap{\textit{potential}}};
        \node[empty] (t1) at ($(10)+(nX)$) {\footnotesize\phantom{\textit{conservation}}\llap{\textit{force}}};
        \node[empty] (t2) at ($(20)+(nX)$) {\footnotesize\phantom{\textit{conservation}}\llap{\textit{source}}};
        \node[empty] (t3) at ($(30)+(nX)$) {\footnotesize\phantom{\textit{conservation}}\llap{\textit{conservation}}};
    \end{tikzpicture}}
    \qquad\qquad\quad\,\,
    \adjustbox{valign=t}{\begin{tikzpicture}
        \node[empty] (O) at (0,0) {};
        \node[empty] (X) at (1.618, 0) {};
        \node[empty] (Y) at (0, -1.0) {};
        \node[s] (00) at ($(O)$) {$\smash{
            e
        }\vphantom{d*AFJ}$};
        \node[s] (10) at ($(O)+1*(Y)+0*(X)$) {$\smash{
            \gamma \wedge e
        }\vphantom{d*AFJ}$};
        \node[s] (20) at ($(O)+2*(Y)+0*(X)$) {$\phantom{\smash{
            R \wedge e
        }\vphantom{d*AFJ}}\mathllap{
        \smash{
            0= {R \wedge e}
        }\vphantom{d*AFJ}}$};
        \node[s] (21) at ($(O)+2*(Y)+1*(X)$) {$
            \mathrlap{
                \phantom{T} \sim 
                    {\star}R \wedge e \swedge {\cdots} \swedge e
            }T
        $};
        \node[s] (31) at ($(O)+3*(Y)+1*(X)$) {$0$};
        \draw[->] (00)--(10) node[midway,left] {};
        \draw[->] (10)--(20) node[midway,left] {};
        \draw[<->] (21)--(20) node[midway,above] {};
        \draw[->] (21)--(31) node[midway,left] {};
    \end{tikzpicture}}
}
\end{align*}
    \caption{%
        Structure of electromagnetism and geometrodynamics in outline form.
        The vertical direction corresponds to differentiation.
        The horizontal direction describes Hodge duality.
    }
    \label{mtwsequence}
\end{figure}

In fact,
this seems to be exactly what
Misner, Thorne, and Wheeler \cite{mtw}
aim to envision
in their diagrams
mentioned in \Sec{Ia}
(Chapter 15, ``Bianchi Identities and the Boundary of a Boundary'').
In \fref{mtwsequence},
we have provided our version of their diagrams
while being totally schematic about
indices, signs, proportionality factors, etc.
For simplicity,
we have assumed the case
where electric sources are present
but not magnetic sources.

\medskip
In the presence of 
both electric and magnetic sources,
the Maxwell's equations
(on a $d$-dimensional background spacetime)
are 
\begin{subequations}
\label{EM}
\begin{align}
    \label{EM.e}
    d\,{*}F \,&=\,
        J^\m\,
        \bb{
            \frac{1}{(d\mminus1)!}\,
            \ve_{\m\r_1\cdots\r_{d-1}}\,
            dx^{\r_1} \swedge \mcdots \swedge dx^{\r_{d-1}}
        }
    \,,\\
    \label{EM.m}
    dF \,&=\,
        \sJ^{\m_1\cdots\m_{d-3}}\,
        \bb{
            \frac{1}{(d\mminus3)!\hem3!}\,
            \ve_{\m_1\cdots\m_{d-3}\r_1\r_2\r_3}\,
            dx^{\r_1} \swedge 
            dx^{\r_2} \swedge 
            dx^{\r_3}
        }
    \,.
\end{align}
\end{subequations}
The electric current is a $(d\mminus1)$-form or equivalently a vector.
The magnetic current is a 3-form or equivalently a $(d\mminus3)$-vector.

In the presence of ordinary stress-energy $T_{\m\n}$,
the Einstein's equations are
\begin{subequations}
\label{GEM}
\begin{align}
\begin{split}
    \label{GEM.e}
    &
    {-\frac{1}{\k^2}}\,
    {\frac{1}{(d\mminus3)!}}\:
    {\star R}_{A B_1\cdots B_{d-3}} 
        \swedge e^{B_1} \swedge \mcdots \swedge e^{B_{d-3}}
    \\
    &\,=\,
        T^\m{}_A\,
        \bb{
            \frac{1}{(d\mminus1)!}\,
            \ve_{\m\r_1\cdots\r_{d-1}}\,
            dx^{\r_1} \swedge \mcdots \swedge dx^{\r_{d-1}}
        }
    \,,
\end{split}
\end{align}
as can be seen from \eqref{Gid}.
Here, we have converted one spacetime index of $T_{\m\n}$
into an internal index via the coframe/frame.
It becomes clear that the natural definition of \textit{magnetic stress-energy} will be
(see, e.g., \cite{cho1991magnetic})
\begin{align}
\begin{split}
    \label{GEM.m}
    &
    {- \frac{1}{\k^2}}\, R_{AB} \wedge e^B
    \\[-0.3\baselineskip]
    &\,=\,
        \sT^{\m_1\cdots\m_{d-3}}{}_A\,
        \bb{
            \frac{1}{(d\mminus3)!\hem3!}\,
            \ve_{\m_1\cdots\m_{d-3}\r_1\r_2\r_3}\,
            dx^{\r_1} \swedge 
            dx^{\r_2} \swedge 
            dx^{\r_3}
        }
    \,.
\end{split}
\end{align}
\end{subequations}
Note how \eqref{GEM} provides the inhomogeneous version of \eqref{GEM0}.
The electric stress-energy is
recast as
a $(d\mminus1)$-form
carrying one internal index,
while
the magnetic stress-energy is 
formulated as
a 3-form
also
carrying one internal index.

\subsection{Coframe Formulation of General Relativity (Second-Order)}

We are now ready to describe Lagrangian formulations.
Consider the action
\begin{align}
    \label{CoframeAction}
    S[e]
    \,=\,
        \frac{1}{\k^2}
    \int_\M\,
        \frac{
            1
        }{2!\hem (d\mminus2)!}\,
        \e_{A_1\cdots A_d}
            R^{A_1A_2}[e]
                \wedge
            e^{A_3} \swedge \mcdots \swedge e^{A_d}
    \,,
\end{align}
which takes 
an orthonormal coframe $e^A$
as the sole dynamical field.
Here, $R^{A_1A_2}[e]$ is the curvature two-form
associated with the coframe $e^A$
via \eqrefs{gamma-def}{Rform}.
\eqref{CoframeAction}
is manifestly invariant under
diffeomorphisms on $\M$
as well as
the local Lorentz transformations in \eqref{LLT.e}.

\eqref{CoframeAction} is the action
for the coframe formulation of GR.
By plugging in \eqref{Rform},
one equates \eqref{CoframeAction}
with the Einstein-Hilbert action in \eqref{EHaction}.
Variation of \eqref{CoframeAction} with respect to $e$
derives the vacuum Einstein's equations 
in the form stated in \eqref{GEM0}.

Note that the cosmological constant $\L$ can be readily implemented
by adding the following term in the integrand:
a multiple of the dynamical volume form,
\begin{align}
    \label{cc}
    - \frac{1}{\k^2}\,
    \L\,
    \bb{
    \frac{1}{d!}\,
        \e_{A_1\cdots A_d}\,
        e^{A_1} \swedge \mcdots \swedge e^{A_d}
    }
    \,=\,
        - \frac{1}{\k^2}\,
            \L\mem |e|\mem d^dx
    \,.
\end{align}

\subsection{Coframe Formulation of General Relativity (First-Order)}

Now consider the action
\begin{align}
    \label{Palatini}
    S[e,\gamma]
    \,=\,
        \frac{1}{\k^2}
    \int_\M\,
        \frac{
            1
        }{2!\hem (d\mminus2)!}\,
        \e_{A_1\cdots A_d}
            R^{A_1A_2}[\gamma]
                \wedge
            e^{A_3} \swedge \mcdots \swedge e^{A_d}
    \,,
\end{align}
which takes $e^A$ and $\gamma^A{}_B$ as \textit{independent} dynamical fields.
Here, $e^A$ is an orthonormal coframe
while $\gamma^A{}_B$ is a Lorentz-valued connection.
Here, $R^{A_1A_2}[\gamma]$
is viewed as a sole functional of $\gamma$
defined by the formula in \eqref{Rform}.

\eqref{Palatini} is quadratic in $\gamma$.
Hence $\gamma$ can be exactly integrated out.
By using the Palatini variation identity
\cite{palatini1919deduzione},
one finds that 
the variation of \eqref{Palatini} with respect to $\gamma$ imposes
\begin{align}
    \label{Deee=0}
    D\bigbig{
        e^{A_3} \swedge \mcdots \swedge e^{A_d}
    }
    \,=\,
        0
    \,.
\end{align}
Given $e$ is nondegenerate,
it can be seen that \eqref{Deee=0} is algebraically equivalent to the vanishing of torsion, $De^A = 0$, in \eqref{De=0}.
As a result, $\gamma$ is dynamically equated to the spin connection of $e$.
Therefore, \eqref{Palatini} is equivalent to \eqref{CoframeAction}.

\eqref{Palatini} is a first-order action of GR,
known as the vielbein-Palatini action.
Note that the variation of \eqref{Palatini} with respect to $e$
immediately derives
$
{\star R}_{A B_1\cdots B_{d-3}}[\gamma]
        \swedge e^{B_1} \swedge \mcdots \swedge e^{B_{d-3}}
    = 0
$.

Since the top form in \eqref{cc} describes a sole functional of the coframe, $e$,
the cosmological constant is straightforwardly incorporated
in the Palatini action in \eqref{Palatini}.

\section{ Gravity as Frame Geometry}
\label{Jb}

The coframe formulation of GR
involves two sorts of redundancies:
diffeomorphisms and local Lorentz transformations.
Can there be a formulation
that focuses on the diffeomorphism side,
rather than local Lorentz?
A formulation known as 
teleparallel gravity
\cite{weitzenbock1923invarianten,einstein1925einheitliche,einstein1928riemann,einstein1929einheitliche,einstein1956appendix,cartan1979lettres,Goenner:2004se,sauer2014einstein_unified_field_theory,Maluf:2013gaa,Aldrovandi:2013wha,Ferraro:2016wht,TrinityHeisenberg:2019}
will achieve this idea,
which views GR as a theory of a dynamical frame field $E$.

\paragraph{Magnetic Equation Revisited}

It could be instructive to begin our exposition
from the magnetic equation of gravity
in \eqref{GEM0},
i.e.,
the algebraic Bianchi identity.
The sequence of differential forms in \eqref{sequence.GEM}
describes
\begin{align}
    \label{des}
    - de^A
    \,=\,
        \gamma^A{}_B \wedge e^B
    \,,\quad
    - dde^A
    \,=\,
        R^A{}_B \wedge e^B
    \,.
\end{align}
Let us examine the content of these equations
in the internal frame.

First of all,
a well-known gymnastics \cite{Eguchi:1980jx} unfolds as
\begin{align}
    \label{gym.1}
    de^C(E_A,E_B)
    \,=\,
        \cont{
            \i_{E_A} de^C
        }{E_B}
    \,=\,
        \cont{
            \pounds_{E_A} e^C
        }{E_B}
    \,=\,
        - \cont{
            e^C
        }{
            \comm{E_A}{E_B}
        }
    \,,
\end{align}
where we have used the Cartan magic formula :
on any differential form $\a$,
the Lie derivative $\pounds_V$
with respect to a vector field $V$
is given by
\begin{align}
    \label{magic}
    \pounds_V \a
    \,=\,
        d\mem \i_V \a
        + \i_V\hem d\a
    \,,
\end{align}
where $\i_V$ denotes the interior product with respect to $V$.

\newpage

Given a frame $E$,
the \textit{anholonomy coefficients}
are defined as
\begin{align}
    \label{anholonomy-def}
    \Omega^C{}_{AB}
    \,=\,
        \cont{
            e^C
        }{
            \comm{E_A}{E_B}
        }
    \,,
\end{align}
measuring the noncoordinate nature of $E$.
\eqref{gym.1} implies that
the anholonomy coefficients are related to the spin connection coefficients in \eqref{gamma-def} as
\begin{align}
    \frac{1}{2}\:
        \Omega^C{}_{AB}\,
        e^A \swedge e^B
    \,=\,
        -de^C
    \,=\,
        \gamma^C{}_B \wedge e^B
    \,=\,
        \gamma^C{}_{BA}\mem e^A \wedge e^B
    \,.
\end{align}
That is,
\begin{align}
    \label{Om=sc-sc}
    \Omega^C{}_{AB}
    \,=\,
        \gamma^C{}_{BA} - \gamma^C{}_{AB}
    \,=\,
        -2\hem \gamma^C{}_{[AB]}
    \,.
\end{align}

Second of all,
the three-form equation in \eqref{des}
can be now written as
\begin{align}
    \label{dO=-Re}
    d\mem\bb{
        \frac{1}{2}\:
        \Omega^D{}_{AB}\,
        e^A \swedge e^B
    }
    \,=\,
        \frac{1}{3!}\: \BB{
            3\mem R^D{}_{[CAB]}
        }\,
            e^A \wedge e^B \wedge e^C
    \,.
\end{align}
For any two-form $\a$ and vector fields $U,V,W$,
it holds that \cite{spivak1970comprehensive}
\begin{align}
    \label{3id}
    d\a(U,V,W)
    \,=\,
        \bb{
            U\act{
                \a(V,W)
            }
            + \a\bigbig{
                U ,
                \comm{V}{W}
            }
        }
        + \text{cyclic}
    \,,
\end{align}
which sums over the three cyclic permutations of $(U,V,W)$.
Here, $U\act{f} = \cont{df}{U} = U^\m \partial_\m f$ denotes the action of the vector field $U$ as a differential operator on the scalar field $f$.
Applying \eqref{3id} to \eqref{dO=-Re}, we find
\begin{align}
\begin{split}
    R^D{}_{[CAB]}
    \,&=\,
        E_{[C}\act{
            \Omega^D{}_{AB]}
        }
        + \Omega^D{}_{[C|E}\mem \Omega^E{}_{|AB]}
    \,=\,
        \cont{e^D}{
            \comm{
                E_{[C}
            }{
                \comm{
                    E_A
                }{
                    E_{B]}
                }
            }
        }
    \,.
\end{split}
\end{align}
This shows that the algebraic Bianchi identity
has secretly encoded the Jacobi identity of the Lie bracket between vector fields:
\begin{align}
    \label{EEE=0}
    \comm{
        E_A
    }{
        \comm{
            E_B
        }{
            E_C
        }
    }^\m
    +
    \comm{
        E_B
    }{
        \comm{
            E_C
        }{
            E_A
        }
    }^\m
    +
    \comm{
        E_C
    }{
        \comm{
            E_A
        }{
            E_B
        }
    }^\m
    \,=\,
        0
    \,.
\end{align}
Note that this result can also be directly derived 
from \eqrefs{indexfree.torsionfree}{indexfree.riemann}:
\begin{align}
\begin{split}
    \label{gymnasticsREEE}
    &
    R(E_A,E_B)\mem E_C
    + \text{cyclic}
    \\
    \,&=\,
        \nabla_{E_A} \nabla_{E_B} E_C
        - \nabla_{E_B} \nabla_{E_A} E_C
        - \nabla_{\comm{E_A}{E_B}} E_C
    + \text{cyclic}
    \,,\\
    \,&=\,
        \nabla_{E_A} \bigbig{
            \nabla_{E_B} E_C - \nabla_{E_C} E_B
        }
        - \nabla_{\comm{E_A}{E_B}} E_C
    + \text{cyclic}
    \,,\\
    \,&=\,
        \nabla_{E_A} \bigbig{
            \comm{E_B}{E_C}
        }
        - \nabla_{\comm{E_B}{E_C}} E_A
    + \text{cyclic}
    \,,\\
    \,&=\,
        \comm{E_A}{
            \comm{E_B}{E_C}
        }
    + \text{cyclic}
    \,.
\end{split}
\end{align}

To sum up,
we have shown that the magnetic equation of gravity
can be equivalently formulated as \eqref{EEE=0},
based on 
the Lie bracket between 
the frame fields.
Note that another name for
the Lie algebra of vector fields on $\M$
is the diffeomorphism algebra $\diff(\M)$.
In \Sec{K2:tele},
we will use this mathematical fact
to interpret \eqref{EEE=0}
as the Bianchi identity for
a ``gauge theory of the diffeomorphism group.''

\newpage

\paragraph{Diffeomorphism Yang-Mills Equations}

How about the electric equation of GR
in \eqref{GEM0},
$
    {\star R}_{A B_1\cdots B_{d-3}}
        \wedge e^{B_1} \swedge \mcdots \swedge e^{B_{d-3}}
    = 0
$\mem ?
One could wonder if it could be equivalent to
\begin{align}
    \label{MNE}
    \comm{E^B}{\comm{E_B}{E_A}}^\m
    \,=\,
        0
    \,,
\end{align}
which traces back to
Mason and Newman \cite{MasonNewman:1989}.
In the modern context of double copy,
this is a fantastic proposal
suggesting that
the double copy problem of GR is solved
by $\gk = \diff(\M)$;
see \Chap{K2}.

Unfortunately,
the electric equation of GR
fails to be \eqref{MNE}.
To address this explicitly,
we may now want to switch to the Lagrangian formalism.

\subsection{Topological Limit}
\label{J:topoGR}

Our point of departure is the Palatini action in \eqref{Palatini},
which splits as
\begin{align}
    S[e,\gamma]
    \,=\,
        \frac{1}{\k^2}
        \int_\M
            L_1[e,\gamma]
        \,\mem+\,
        \frac{1}{\k^2}
        \int_\M
            L_2[e,\gamma]
    \,.
\end{align}
Here, $L_1[e,\gamma]$ and $L_2[e,\gamma]$ are top forms
given as
\begin{subequations}
\begin{align}
    \label{telder.L1}
    L_1[e,\gamma]
    \,&=\,
        \frac{
            1
        }{2!\hem (d\mminus2)!}\,
        \e_{A_1\cdots A_d}\,
            d\gamma^{A_1A_2}
                \wedge
            e^{A_3} \swedge \mcdots \swedge e^{A_d}
    \,,\\
    \label{telder.L2}
    L_2[e,\gamma]
    \,&=\,
        \frac{
            1
        }{2!\hem (d\mminus2)!}\,
        \e_{A_1\cdots A_d}\,
            \gamma^{A_1}{}_C \wedge \gamma^{CA_2}
                \wedge
            e^{A_3} \swedge \mcdots \swedge e^{A_d}
    \,.
\end{align}
\end{subequations}
Our goal is to reformulate the Palatini action
such that the frame field $E$ is taken as 
one of the fundamental degrees of freedom.
We find it instructive to first focus on the first part, \eqref{telder.L1}.
With hindsight,
we refer to this case as the topological limit.\footnote{
    The decoupling between \eqrefs{telder.L1}{telder.L2}
    naturally arises
    by implementing a rescaling of the Lorentz connection,
    $\gamma^A{}_B \mapsto \k^2\mem \gamma^A{}_B$.
    Amusingly, this mimics a typical move
    \cite{GravityMHVTwistors,Abou-Zeid:2005zfo}
    that one encounters in a different context:
    see \Sec{Jcx>COFRAME}.
}

\medskip

In the topological limit,
the Lorentz connection is merely a Lagrange multiplier.
By discarding a total derivative,
$L_1[e,\gamma]$ in \eqref{telder.L1}
is brought to
\begin{align}
    \label{telder.L1>a}
        \bb{
            \frac{1}{(d\mminus3)!}\,
                {\star\gamma}_{A_3\cdots A_d}
                \wedge e^{A_4} \swedge \mcdots \swedge e^{A_d}
        }
        \wedge de^{A_3}
    \,.
\end{align}
The identity explored in \eqref{gym.1},
$E^\m{}_C\, de^C = - \tfrac{1}{2}\, \comm{E_A}{E_B}^\m\, e^A \swedge e^B$,
then brings \eqref{telder.L1>a} to
\begin{align}
    \label{telder.L1>b}
        \frac{1}{2}\,
        B^{AB}{}_\m\,
        \comm{E_A}{E_B}^\m
        \, d^dx
    \,,
\end{align}
where we have employed an algebraic field redefinition,
\begin{align}
    \label{telBdef.form}
    B^{AB}{}_\m\, d^dx
    \,=\,
        \bb{
            -\frac{1}{(d\mminus3)!}\,
                {\star\gamma}_{CC_4\cdots C_d}
                \wedge e^{C_4} \swedge \mcdots \swedge e^{C_d}
            \wedge e^A \swedge e^B
        }\, e^C{}_\m
    \,.
\end{align}
According to
\Sec{K2>CaseBF},
\eqref{telder.L1>b} describes a diffeomorphism $BF$ theory.
Its saddles describe
\textit{coordinate} frame fields,
free of anholonomy:
$\comm{E_A}{E_B} = 0$.

\subsection{Frame Formulation of General Relativity (First-Order)}

Now let us incorporate the second part of the Palatini action in \eqref{telder.L2}
to derive the full action.
First of all, \eqref{telBdef.form} boils down to
\begin{subequations}
\begin{align}
\begin{split}
    \label{telBdef}
    B^{AB}{}_\m
    \,&=\,
        \BB{
            -3\mem 
            \gamma^{[AB}{}_K\mem \delta^{K]}{}_C\mem 
        }\, 
        |e|\mem e^C{}_\m
    \,.
\end{split}
\end{align}
The inverse transformation is
\begin{align}
\begin{split}
    \label{telBdef.inv}
    \gamma^{AB}{}_C
    \,&=\,
        -|E|\,
        \bb{
            B^{AB}{}_\m\mem E^\m{}_C
            + \frac{2}{d\mminus2}\,
                \delta^{[A}{}_C\,
                B^{B]D}{}_\r\mem E^\r{}_D
        }
    \,.
\end{split}
\end{align}
\end{subequations}
Direct computation shows that
\begin{align}
\begin{split}
    L_2[e,\gamma]
    \,&=\,
        \tfrac{1}{2}\mem\BB{
            \gamma_{ACB}\mem \gamma^{ABC}
            - \gamma^{CA}{}_C\mem \gamma^D{}_{AD}
        }\,
        |e|\mem d^dx
    \,,\\
    \,&=\,
        - \tfrac{1}{2}\,
            \mathcal{G}^{\m\n}_{\nem ABCD}[E]\,
            B^{AB}{}_\m\mem B^{CD}{}_\n
        \,d^dx
    \,,
\end{split}
\end{align}
where we have defined
\begin{align}
    \label{telGdef}
    \mathcal{G}^{\m\n}_{\nem A_1A_2B_1B_2}[E]
    \,=\,
        4\hem 
        |E|\mem \bb{
            E^\m{}_{[B_1} \eta_{B_2][A_1} E^\n{}_{A_2]}
            - \frac{1}{d\mminus2}\,
                E^\m{}_{[A_1} \eta_{A_2][B_1} E^\n{}_{B_2]}
        }
    \,.
\end{align}

Collecting all the results so far,
we conclude
that the Palatini action in \eqref{Palatini},
up to discarding a total derivative,
is equivalent to
\begin{align}
    \label{FrameAction1}
    \kern-0.3em
    S[B,E]
    \mem=\mem
        \frac{1}{\k^2}
        \int d^dx\,\,
        \bb{
            \frac{1}{2}\,
                B^{AB}{}_\m\,
                \comm{E_A}{E_B}^\m
            - \frac{1}{8}\,
                \mathcal{G}^{\m\n}_{\nem ABCD}[E]\,
                B^{AB}{}_\m\mem B^{CD}{}_\n
        \nem}
    \,,
    \kern-0.6em
\end{align}
where $\mathcal{G}^{\m\n}_{ABCD}[E]$
is defined in \eqref{telGdef}.
To clarify,
\eqref{FrameAction1}
is a first-order action of GR
that governs two dynamical fields.
The first is a frame field, $E^\m{}_A$.
The second is an auxiliary field, $B^{AB}{}_\m$.

We should clarify on the
gauge redundancies as well.
A careful analysis will tell us that there are two.
The first kind describes the transformation
\begin{align}
\begin{split}
    \label{tele.diff}
    \d_\xi E^\m{}_A
    \,&=\,
        \xi^\r\mem \partial_\r E^\m{}_A
        - E^\r{}_A\mem \partial_\r \xi^\m
    \,,\\
    \d_\xi B^{AB}{}_\m
    \,&=\,
        \xi^\r\mem \partial_\r B^{AB}{}_\m
        + B^{AB}{}_\r\mem \partial_\m \xi^\r
        + B^{AB}{}_\m\mem \partial_\r \xi^\r
    \,.
\end{split}
\end{align}
Clearly, \eqref{tele.diff} 
will be interpreted as
the action of an infinitesimal diffeomorphism.
It states that
$E^\m{}_A$ is a vector
while
$B^{AB}{}_\m$ is a covector density of weight $+1$.
It is important that $B^{AB}{}_\m$ is a density,
since otherwise 
terms such as $B^{AB}{}_\m\, \comm{E_A}{E_B}^\m\mem d^dx$
cannot be diffeomorphism invariant.

\newpage

Note that \eqref{tele.diff} induces
\begin{align}
    \d_\xi |E|
    \,=\,
        -|E|^2\, \partial_\m\bigbig{
            |E|^{-1}\mem \xi^\m
        }
    \,.
\end{align}
The action in \eqref{FrameAction1} changes
by the boundary term,
\begin{align}
    \frac{1}{\k^2}
    \int d^dx\,\,
    \partial_\r\bbsq{
        \xi^\r\mem
        \bb{
            \frac{1}{2}\,
                B^{AB}{}_\m\,
                \comm{E_A}{E_B}^\m
            - \frac{1}{8}\,
                \mathcal{G}^{\m\n}_{\nem ABCD}[E]\,
                B^{AB}{}_\m\mem B^{CD}{}_\n
        \nem}
    }
    \,,
\end{align}
which vanishes as the local parameter $\xi^\r$ dies off at the infinities.

The second kind is subtler
and is more nontrivial to see.
It describes the transformation
\begin{align}
\begin{split}
    \label{tele.LLT}
    \d_\vth E^\m{}_A
    \,&=\,
        - E^\m{}_B\mem \vartheta^B{}_A
    \,,\\
    \d_\vth B^{AB}{}_\m
    \,&=\,
        2\mem \vth^{[A|}{}_C\mem B^{C|B]}{}_\m
        + \frac{1}{|E|}\,
            \BB{
                \partial_\m \vth^{AB}
                + 2\mem E^\r{}_C\mem
                    \partial_\r \vth^{C[A}
                    \, e^{B]}{}_\m
            }
    \,,
\end{split}
\end{align}
where $\vth^{AB} = -\vth^{BA}$.
Under \eqref{tele.LLT},
$
    \tfrac{1}{2}\,
        B^{AB}{}_\m\,
        \comm{E_A}{E_B}^\m
$ changes by
\begin{align}
    \partial_\m\BB{
        E^\m{}_A\mem E^\r{}_B\mem
            \partial_\r\vth^{AB}
    }
    - B^{AB}{}_\r\mem E^\r{}_C\,
        E^\m{}_A\mem \partial_\m \vth^C{}_B
    \,,
\end{align}
whereas
$
    - \tfrac{1}{8}\,
        \mathcal{G}^{\m\n}_{\nem ABCD}[E]\,
        B^{AB}{}_\m\mem B^{CD}{}_\n
$ changes by
$
    + B^{AB}{}_\r\mem E^\r{}_C\,
        E^\m{}_A\mem \partial_\m \vth^C{}_B
$.
Hence
the full action in \eqref{FrameAction1}
changes by the boundary term,
\begin{align}
    \label{tele.varS.LLT}
    \frac{1}{\k^2}
    \int d^dx\,\,
        \partial_\m\BB{
            E^\m{}_A\mem E^\r{}_B\mem
                \partial_\r\vth^{AB}
        }
    \,,
\end{align}
which vanishes 
provided the derivative $\partial_\r\vth^{AB}$
dies off at the infinities.

Crucially, 
this gauge redundancy is not enjoyed by the topological limit
but emerges in the full theory defined by \eqref{FrameAction1}.
It can be seen that \eqref{tele.LLT}
precisely arises from
the local Lorentz transformation
given in \eqrefs{LLT.e}{LLT.gamma}
(cf. \rcite{Ferraro:2016wht}).
Note how it is significantly non-manifest
in \eqref{FrameAction1}.

\subsection{Frame Formulation of General Relativity (Second-Order)}

The action in \eqref{FrameAction1} is quadratic in the auxiliary field $B$.
Hence it can be exactly integrated out as
\begin{align}
\begin{split}
    \label{teleB=?}
    B^{AB}{}_\r
    \,=\,
        \frac{1}{2|E|}\,
            \BB{
                \Omega^{ABC} + \Omega^{BCA} - \Omega^{CAB}
            \hnem}\, e_{C\r}
        - \frac{1}{|E|}\,
            2\mem \Omega^C{}_C{}\vphantom{\Omega}^{[A}\mem e^{B]}{}_\r
    \,.
\end{split}
\end{align}
Plugging in \eqref{teleB=?} to \eqref{FrameAction1} gives
\begin{align}
    \label{FrameAction2}
    \kern-0.3em
    S[E]
    \mem=\mem
        -\frac{1}{2\k^2}
        \int 
        \frac{d^dx}{|E|}\mem
        \bb{
            \frac{1}{4}\mem \Omega^{ABC}\mem \Omega_{ABC}
            + \frac{1}{2}\mem \Omega^{BAC}\mem \Omega_{ABC}
            - \Omega^A{}_{AC}\mem \Omega^B{}_B\vphantom{\Omega}^C
        \nem}
    \,,
    \kern-0.6em
\end{align}
where $\Omega^C{}_{AB}$
is viewed as the functional of $E$ 
defined by \eqref{anholonomy-def}.

\eqref{FrameAction2}
establishes a formulation of GR
that takes the frame field $E^\m{}_A$
as the sole dynamical field.
We will refer to this
as the \textit{frame formulation of GR}.

\newpage

\paragraph{Equations of Motion}

As remarked earlier,
we see that Einstein's equations
cannot be formulated as \eqref{MNE}
as explored by Mason and Newman \cite{MasonNewman:1989}.
The frame formulation of GR
does enjoy the local Lorentz symmetry
as shown in \eqref{tele.varS.LLT}.
In contrast,
\eqref{MNE} is not invariant under local Lorentz transformations.
Explicitly, one can simply vary the action in 
\eqref{FrameAction2}
to directly find that the EoM disagree with \eqref{MNE}.

In fact,
it is not clear whether
one can integrate
equations like \eqref{MNE}
into an action.
Naively, one could envision a YM-like action
that squares $\comm{E_A}{E_B}^\m$,
in which case the spacetime index $\m$ needs to be contracted.
This crucially requires
an invariant inner product
(a Killing form)
on the diffeomorphism algebra.
Note also how $\mathcal{G}^{\m\n}_{ABCD}[E]$ in \eqref{FrameAction1} is dynamical.

\paragraph{Derivation by Bootstrap}

To reiterate,
our analysis shows that
the frame formulation of GR
enjoys local Lorentz symmetry,
although non-manifest.
While the frame field $E^\m{}_A$
provides $d^2$ components
in $d$ dimensions,
local Lorentz symmetry subtracts
$d(d-1)/2$ degrees of freedom,
retrieving the usual number $d(d+\nem1)/2$
in the metric formulation;
see also \rcite{Ferraro:2016wht}.

In fact, it seems that
the action in \eqref{FrameAction2}
can be derived
from the bottom-up
as the \textit{unique} action
of a dynamical frame field $E^\m{}_A$
that is invariant under both diffeomorphisms
in \eqref{tele.diff}
and local Lorentz transformations
in \eqref{tele.LLT}
while being quadratic in the anholonomy coefficients
$\Omega^A{}_{BC}$.
First, there are three quadratic contractions of the anholonomy coefficients:
$\Omega^{ABC}\mem \Omega_{ABC}$,
$\Omega^{BAC}\mem \Omega_{ABC}$,
and
$\Omega^A{}_{AC}\mem \Omega^B{}_B$.
Second, diffeomorphism invariance
demands placing an overall $|E|^{-1}$
in the Lagrangian.
Note that
\begin{align}
    \d_\xi \Omega_{ABC}
    \,=\,
        \xi^\m\mem \partial_\m \Omega_{ABC}
    \,,\quad
    \d_\xi\bigbig{
        |E|^{-1}
    }
    \,=\,
        \partial_\m\bigbig{
            |E|^{-1}\xi^\m
        }
    \,.
\end{align}
Third,
one examines local Lorentz invariance
with fuzzy factors $a_1,a_2,a_3$:
\begin{align}
\begin{split}
    \label{fuzzya123}
    &
    \d_\vth\mem\bbsq{
        \frac{1}{2\hem|E|}\mem
        \bb{
            a_1\, \Omega^{ABC}\mem \Omega_{ABC}
            + a_2\, \Omega^{BAC}\mem \Omega_{ABC}
            + a_3\, \Omega^A{}_{AC}\mem \Omega^B{}_B\vphantom{\Omega}^C
        \nem}
    }
    \\
    &
    \,\sim\,
        \vth_{AB}\mem
        \bbsq{
            \bigbig{
                a_2 \mminus 2a_1
            }\mem
                \partial_\m\bb{
                    \frac{
                        E^\m{}_C\mem \Omega^{ABC}
                    }{|E|}
                }
            +
            \bigbig{
                a_3 \mplus 2a_2
            }\mem
            \bb{\frac{
                E_A\act{
                    \Omega^C{}_{CB}
                }
            }{|E|}}
        }
    \,,
\end{split}
\end{align}
where $\sim$ means we discarded a total derivative
in a way that removes derivatives on the local parameter $\vth^{AB}$.
Here, the following identity can be helpful:
\begin{align}
    \label{divO-id}
    \Omega^C{}_{CA}
    \,&=\,
    \partial_\m\bigbig{
        |E|^{-1} E^\m{}_A
    }
    \,=\,
        \dv_\ve\hem E_A
    \,,\quad
    \pounds_{E_A} \ve
    \,=\,  
        \Omega^C{}_{CA}\mem \ve
    \,.
\end{align}
Apparently,
\eqref{fuzzya123}
pinpoints
$-4a_1 = -2a_2 = a_3$
as the unique solution
to this bootstrap.
Taking $(a_1,a_2,a_3) = (-1/4,-1/2,1)$
reproduces \eqref{FrameAction2}.

\paragraph{Teleparallel Gravity}

In the literature,
the action in \eqref{FrameAction2} 
is known
to provide the
``teleparallel equivalent of GR''
\cite{Maluf:2013gaa,Aldrovandi:2013wha,Ferraro:2016wht,TrinityHeisenberg:2019}.
A \textit{teleparallel theory}
refers to a field theory of a dynamical frame $E^\m{}_A$.
A frame amounts to a notion of absolute parallelism,
meaning a flat connection
known as Weitzenb\"ock connection
\cite{weitzenbock1923invarianten}.
The Weitzenb\"ock connection
declares that
a vector $E^\m{}_A(x_1)\mem v^A$ at point $x_1$
is parallel to
$E^\m{}_A(x_2)\mem v^A$ at point $x_2$:
the internal index $A$ serves as
a ``global'' index,
so the parallel propagator from point $x_1$ to $x_2$
is defined
as $E^\m{}_A(x_2)\mem e^A{}_\n(x_1)$.
``Teleparallelism'' is another name for ``absolute parallelism.''

Teleparallel gravity
was one of the approaches put forward by Einstein himself
during his late-stage research on
unified field theories.
See \rrcite{cartan1979lettres,einstein1925einheitliche,einstein1928riemann,einstein1929einheitliche,einstein1956appendix,Goenner:2004se,sauer2014einstein_unified_field_theory,Maluf:2013gaa,Aldrovandi:2013wha,Ferraro:2016wht,TrinityHeisenberg:2019}
and also a remark made in
Mason and Newman \cite{MasonNewman:1989}, Sec.\,5.

Given our clarifications above,
we may prefer the name ``frame formulation of GR''
over ``teleparallel formulation of GR,''
in fact:
local Lorentz redundancy is secretly present.

\paragraph{Flat Connection Formulation}

Provided a frame $E^\m{}_A$ and its inverse $e^A{}_\m$,
the Weitzenb\"ock connection
is defined by the following connection coefficients:
\begin{align}
    \label{weit}
    \widetilde{\Gamma}^\m{}_{\n\r}
    \,=\,
        E^\m{}_A\mem \partial_\r e^A{}_\n
    \,.
\end{align}
It is easy to show that
the Weitzenb\"ock connection is flat, metric-preserving, and torsionful:
\begin{subequations}
\begin{align}
    \label{weit.R}
    0\,&=\,
        \partial_\r \widetilde{\Gamma}^\m{}_{\n\s}
        - 
        \partial_\s \widetilde{\Gamma}^\m{}_{\n\r}
        + 
        \widetilde{\Gamma}^\m{}_{\k\r}\mem
        \widetilde{\Gamma}^\k{}_{\n\s}
        -
        \widetilde{\Gamma}^\m{}_{\k\s}\mem
        \widetilde{\Gamma}^\k{}_{\n\r}
    \,,\\
    \label{weit.g}
    0\,&=\,
        \partial_\r g_{\m\n}
        - g_{\m\k}\hem \widetilde{\Gamma}^\k{}_{\n\r}
        - g_{\k\n}\hem \widetilde{\Gamma}^\k{}_{\m\r}
    \,,\\
    \label{weit.T}
    \widetilde{T}^\m{}_{\r\s}
    \,&=\,
        \widetilde{\Gamma}^\m{}_{\s\r}
        -
        \widetilde{\Gamma}^\m{}_{\r\s}
    \,=\,
        - E^\m{}_A\, \Omega^A{}_{BC}\mem\hhem e^B{}_\r\mem e^C{}_\s
    \,.
\end{align}
\end{subequations}
By using these facts,
\eqref{FrameAction2}
can be rewritten as
\begin{align}
    \label{FrameAction2.coords}
        -\frac{1}{2\k^2}
        \int 
        \sqrt{-\det g}\, d^dx\,
        \bb{
            \frac{1}{4}\mem \widetilde{T}^{\m\n\r}\mem \widetilde{T}_{\m\n\r}
            + \frac{1}{2}\mem \widetilde{T}^{\n\m\r}\mem \widetilde{T}_{\m\n\r}
            - \widetilde{T}^\m{}_{\m\r}\mem \widetilde{T}^\n{}_\n\vphantom{\widetilde{T}}^\r
        \nem}
    \,,
\end{align}
where internal indices are completely ``hidden.''
\eqref{FrameAction2.coords}
is the commonly known formula for the teleparallel equivalent of GR.
Yet, to truly regard it as a functional of
$g_{\m\n}$ and $\widetilde{\Gamma}^\m{}_{\r\s}$
as fundamental field variables,
we might be
imposing
the flatness and metric-preserving conditions in
\eqrefs{weit.R}{weit.g}
via Lagrange multipliers.

\newpage

\newpage

\section{ Simplicity of Self-Dual Gravity}
\label{Jc}

In the remaining parts of this chapter,
we specialize in
the phenomenologically relevant case of
$d=4$ spacetime dimensions.
It is known that
four-dimensional gravitation
admits a remarkably simpler subsector:
SD gravity
\cite{MasonNewman:1989,penrose1976nonlinear,Penrose:1976js,plebanski1975some,Plebanski:1977zz,GravityMHVTwistors,%
chakravarty1991canonical,chacon2019self,gindikin1986construction,dunajski2000hyper,%
Adamo:2021bej,boyer1983geometry,hull1991geometry,%
Plebanski:1994qi,Park:1990fp,park19922d,Husain:1993dp}.
In particular,
Penrose's nonlinear graviton theorem
\cite{penrose1976nonlinear,Penrose:1976js}
establishes that
SD gravity is classically integrable.
This realization has served as one of the motivations behind
chiral approaches
to gravity
that aim to understand the full dynamics of four-dimensional GR
by perturbing away from 
SD gravity.

\nomenclature{SDYM}{Self-Dual Yang-Mills}
\nomenclature{SDGR}{Self-Dual Gravity}

This section aims to
provide a brief sketch of SD gravity,
from the angle of
double copy \cite{monteiro2011kinematic}.
We review two senses in which
SD gravity can be ``derived'' from SDYM theory:
{\Pleb} formulation \cite{plebanski1975some}
and
Mason-Newman formulation \cite{MasonNewman:1989}.
\subsection{Self-Dual Yang-Mills and Double Copy}
\label{SDYM}

First of all, we define SDYM theory and SD gravity.

YM theory governs the dynamics of a connection
on a principal fiber bundle over spacetime.
In YM theory,
the curvature two-form $F^a
$ satisfies
\begin{align}
    \label{YM.em}
    D\hem {*} F^a
    \,=\,
        0
    \,,\quad
    D F^a
    \,=\,
        0
    \,.
\end{align}
These are the \textit{electric and magnetic equations} of YM theory.
The former is a dynamical equation.
The latter is the Bianchi identity
and is identically satisfied
when the gauge connection is the foundation of all.

Via complexification,
SDYM theory studies configurations such that
\begin{align}
    \label{SDYM.def}
    {*} F^a
    \,=\,
        +i\hem F^a
    \,,
\end{align}
where ${*}$ denotes the Hodge dual
acting on spacetime indices.
Notably, such SD configurations
are automatically on-shell:
$D\hem{*}F^a = i\mem DF^a = 0$.
Also, \eqref{SDYM.def} describes a first-order differential equation on the gauge potential $A^a$
while \eqref{YM.em} describes second-order differential equations.
In these senses,
SDYM theory
identifies a simpler subsector inside
complexified
YM theory.

Next, recall the electric and magnetic equations of GR,
identified in \eqref{GEM0}.
In four dimensions, we have
\begin{align}
    \label{GR.em}
    {\star R}^A{}_B \wedge e^B
    \,=\, 0
    \,,\quad
    R^A{}_B \wedge e^B
    \,=\, 0
    \,.
\end{align}
Again, 
the former is a dynamical equation
while
the latter 
is
identically satisfied
when the metric/coframe is the foundation of all.

Via complexification,
SD gravity studies configurations such that
\begin{align}
    \label{SDGR.def}
    {\star R}^A{}_B
    \,=\,
        +i\hem R^A{}_B
    \,,
\end{align}
where ${\star}$ is the internal Hodge dual.\footnote{
    Note that SD gravity is readily formulated
    without a need for the coframe/frame.
    In the pure metric formulation of GR,
    the SD gravity equations are
    ${*} R_{\m\n\r\s} = +i\hem R_{\m\n\r\s}$,
    where the spacetime Hodge dual ${*}$
    acts either on the first set $[\m\n]$ of antisymmetric indices
    or the second set $[\r\s]$.
}
Notably, such SD configurations
are automatically on-shell:
${\star}R^A{}_B \wedge e^B = i\mem R^A{}_B \wedge e^B = 0$.
Moreover, \eqref{SDGR.def} describes a first-order differential equation on the coframe $e^A$
while \eqref{GR.em} describes second-order differential equations.
In these senses,
SD gravity already
identifies a simpler subsector inside GR.

Remarkably, it turns out that
at the heart of SDYM theory
on flat spacetime
there lies an infinite-dimensional Lie algebra,
and by ``doubling'' it
one derives SD gravity.
In this sense 
there exists a concrete connection between
the SD sectors of YM theory and GR:
\begin{align}
    \label{dc.SD}
    \bigbig{
        \SDYM
    }^{\nem2}
    \,=\,
        \SDGR
    \,.
\end{align}

The astute reader will point out a slight mismatch
in the parallel between
SDYM theory and SD gravity
implied above:
while 
\eqref{YM.em} is differential on the curvature $F$
that is subject to \eqref{SDYM.def},
\eqref{GR.em} is algebraic on $R$
that is subject to \eqref{SDGR.def}.
This discomfort indeed detects an important point
in correctly grasping the double copy relation in \eqref{dc.SD},
as will be revisited 
in \eqref{DYMV}.

We shall also clarify on the meaning of complexification
in the current context.
The discourse on complex spacetimes
traces back to
Penrose \cite{penrose1976nonlinear,Penrose:1976js}
Newman \cite{shaviv1975general,Newman:1976gc},
and
{\Pleb} \cite{plebanski1975some,Plebanski:1977zz};\footnote{
    Newman \cite{shaviv1975general,Newman:1976gc}
    and
    {\Pleb} \cite{plebanski1975some,Plebanski:1977zz}
    coined the term ``heaven''
    for referring to SD spacetimes.
}
see Newman \cite{newman1988remarkable}
and Flaherty \cite{grg207flaherty}
for nice reviews.
The celebrated works \cite{Plebanski:1975xfb,PD}
of {\Pleb} and Demia\'nski
also demonstrate how one
naturally arrives at the idea of
complexified coordinate transformations and spacetimes.

Penrose \cite{penrose1976nonlinear},
for instance,
argued that the notion of a quantum spacetime is inherently complex-geometrical.
The spacetime of a single graviton in a helicity eigenstate
is half-flat,
meaning either the 
SD or ASD
curvature is vanishing.
This necessitates complexification
and leads to a holomorphic metric (not K\"ahler \cite{penrose1976nonlinear}) geometry. 

To clarify, the ``barred''
coordinates or indices never enter in this discussion
\cite{newman1988remarkable,grg207flaherty,penrose1976nonlinear}.
Mathematically speaking,
this means to work in the holomorphic, complexified category of manifolds
so that the coordinate transformations (transition functions) are holomorphic maps
\cite{Plebanski:1975xfb,PD,Adamo:2023fbj}.
One then studies
holomorphic metrics or gauge fields
solving complexified field equations.

\subsection{{\Pleb} Formulation: Second Heavenly Equation}
\label{Jc>SDG>HEAV}

SDYM theory has been well-studied in the lightcone gauge
\cite{parkes1992cubic,bardeen1996self,cangemi1997self}.

Flat spacetime $\mflat = (\C^4,\eta)$
is the complexified Minkowski space,
i.e., the space $\C^4$ equipped with a holomorphic metric $\eta$ that exhibits Lorentzian signature
on the real section.
Let $S^+\mflat$ and $S^-\mflat$ be the SD and ASD spinor bundles on $\mflat$, respectively.
Their fibers
are equipped with
the epsilon tensors
$\e^{\a\b}$ and $\te^{\da\db}$.
The tangent bundle $T\mflat$ is isomorphic to
the direct product bundle $S^+\mflat \otimes S^-\mflat$.
This isomorphism describes a tensor
$(\sigu_\m)^{\da\a}$,
while its inverse describes a tensor
$(\sigd^\m)_{\a\da}$.
These invariant tensors are used to convert
vector indices to spinor indices 
and vice versa.
Various conventions exist in the literature,
the detailed discussion on which
we might want to omit.

Let $x^\m$ be Cartesian coordinates for the flat spacetime $\mflat$.
Lightcone coordinates refer to
\begin{align}
    \label{V2Sa}
    x^{\da\a}
    \,=\,
        (\sigu_\m)^{\da\a}\mem x^\m
    \,\propto\,
        \TwoByTwoMatrixWide{
            x^0 \mplus x^3
        }{
            x^1 \mminus ix^2
        }{
            x^1 \mplus ix^2
        }{
            x^0 \mminus x^3
        }
    \,.
\end{align}
We also employ
constant dyads:
$\o^\a \doteq \delta_0{}^\a$,
$\i^\a \doteq \delta_1{}^\a$,
$\tdo^\da \doteq \delta^\da{}_0$,
$\ti^\da \doteq \delta^\da{}_1$.

Now consider YM theory on $\mflat$
for a 
finite-dimensional
gauge group $\G$.
The gauge potential describes $A^a{}_{\a\da}$,
where $a,b,c,d,\cdots$ are adjoint indices
of the gauge algebra $\g = \Lie(\G)$.
Typically, one has
$\G = \SU(N)$ and $\g = \su(N)$,
for instance.
The lightcone gauge sets
$\o^\a A^a{}_{\a\da} = A^a{}_{0\da} = 0$
for a fixed reference spinor $\o^\a$.
Let us denote the remaining components of the gauge potential as
\begin{align}
    j^a{}_\da
    \,:=\,
        \i^\a A^a{}_{\a\da}
    \,=\,
        A^a{}_{1\da}
    \,.
\end{align}

The SDYM equations, \eqref{SDYM.def},
impose vanishing of
the ASD field strength $F^a{}_\wrap{\a\b}$.
This describes
\begin{align}
    \label{SDYM.spinorfull}
    \te^{\da\db}\hem
    \BB{
        \partial_\wrap{\a\da} A^a{}_\wrap{\b\db}
        -
        \partial_\wrap{\b\db} A^a{}_\wrap{\a\da}
        + f^a{}_{bc}\mem
        A^b{}_\wrap{\a\da}\mem A^c{}_\wrap{\b\db}
    }
    \,=\,
        0
    \,.
\end{align}
In the lightcone gauge,
the nontrivial contents of
\eqref{SDYM.spinorfull} are
\begin{align}
    \label{SDYM.j}
    \partial_\wrap{0\da} j^{a\da}
    \,=\,
        0
    \,,\quad
    \partial_\wrap{1\da} j^a{}_\db
    - \partial_\wrap{1\db} j^a{}_\da
    + f^a{}_{bc}\mem j^b{}_\da\mem j^c{}_\db
    \,=\,
        0
    \,.
\end{align}
\newpage

The first condition in \eqref{SDYM.j}
states that
$j^{a\da}$ is ``divergence-free,''
implying the existence of a $\g$-valued scalar field $\Phi^a$ such that
\begin{align}
    \label{j-phi}
    j^{a\da}
    \,=\,
        \te^{\da\db}\mem \partial_\wrap{0\db} \Phi^a
    \,.
\end{align}
The second condition in \eqref{SDYM.j}
states that
$j^a{}_\da$ is a ``flat connection,''
which is solved by
$j^a{}_\da = 
    (U^{-1}\hem
    \partial_{1\da} U)^a
$
for a $\G$-valued field $U$.

By plugging in \eqref{j-phi}
to \eqref{SDYM.j},
one finds
\begin{align}
    \label{heavenly1}
    \Box\mem \Phi^a
    \,=\,
        f^a{}_{bc}\mem
            \te^{\da\db}\mem 
            \partial_\wrap{0\da} \Phi^b
            \mem
            \partial_\wrap{0\db} \Phi^c
    \,,
\end{align}
where $\Box$ is the four-dimensional d'Alembertian operator.
\eqref{heavenly1} is a \textit{cubic} formulation of SDYM theory
in terms of a colored \textit{scalar} field $\Phi^a$;
see, e.g., \rcite{parkes1992cubic}.
This reveals the simplicity of SDYM theory.

Notably,
we see that \eqref{heavenly1}
identifies a (degenerate) Poisson structure
on the spacetime $\mflat$:
\begin{align}
    \label{Pi-heav}
    \Pi_\heav
    \,:=\,
        \tfrac{1}{2}\,
        \te^{\da\db}\hem
        \partial_\wrap{0\da}
        \swedge
        \partial_\wrap{0\db}
    \,.
\end{align}
The Poisson bracket,\footnote{
    This gives a fiberwise Poisson bracket
    via a $2+2$ splicing of spacetime:
    see \App{COTANGENT}.
}
\begin{align}
    \label{pb-lightcone}
    \pb{F}{G}_\heav
    \,:=\,
        \te^{\da\db}\mem 
        \partial_\wrap{0\da} F
        \mem
        \partial_\wrap{0\db} G
    \,,
\end{align}
satisfies the standard axioms:
skew symmetry,
Leibniz rule,
and Jacobi identity.
With this understanding,
\eqref{heavenly1}
can be written as
\begin{align}
    \label{heavenly1.pb}
    \SDYM(\g):\quad
    \Box\mem \Phi^a
    \,=\,
        f^a{}_{bc}\mem
            \pb{\Phi^b}{\Phi^c}_\heav
    \,.
\end{align}
In this way, \eqref{heavenly1}
describes two Lie algebras:
the color Lie algebra $\su(N)$ and the Poisson Lie algebra $\diff(\C^4,\Pi_\heav)$.

Now consider replacing the color Lie algebra
with the Poisson Lie algebra.
This maps \eqref{heavenly1.pb} to
\begin{align}
    \label{heavenly2.dpb}
    \SDGR:\quad
    \Box\mem \Phi
    \,=\,
        \dpb{\Phi}{\Phi}_\heav
    \,,
\end{align}
where the ``double Poisson bracket''\footnote{
    This is the $B_2$ bracket
    in \rcite{kontsevich}
    for Moyal deformation quantization.
} is defined as
\begin{align}
    \label{dpb-lightcone}
    \dpb{F}{G}_\heav
    \,:=\,
        \te^{\da\db}\mem 
        \te^{\dc\dd}\mem 
        \partial_\wrap{0\da} \partial_\wrap{0\dc} F
        \mem
        \partial_\wrap{0\db} \partial_\wrap{0\dd} G
    \,,
\end{align}
literally doubling the Poisson bracket in \eqref{pb-lightcone}.
Remarkably,
\eqref{heavenly2.dpb}
is a formulation of SD gravity
due to {\Pleb} \cite{plebanski1975some},
known as 
the \textit{second heavenly equation}.
Interested readers may refer to
\App{APP:PLEBHE}
for its derivation.
This reveals the simplicity of SD gravity:
it admits a \textit{cubic} formulation
in terms of a single \textit{scalar} field $\Phi$.

\newpage

The above findings can be summarized in terms of the BAS equation,
\begin{align}
    \label{heavenly0}
    \BAS(\g,\tg):\quad
    \Box\mem \Phi^{a\ta}
    \,=\,
        f^a{}_{bc}\mem
        \tf^\ta{}_{\tb\tc}\mem
            \Phi^{b\tb}\mem
            \Phi^{c\tc}
    \,,
\end{align}
which describe two Lie algebras $\g$ and $\tg$.
The formulation of SDYM theory in \eqref{heavenly1.pb}
implies that
SDYM theory 
describes an instance of BAS theory
for $\tg = \diff(\C^4,\Pi_\heav)$:
\begin{align}
    \label{CKD:sdym}
    \SDYM(\g)
    \,\,\cong\,\,
        \BAS\bigbig{
            \g,\diff(\C^4,\Pi_\heav)
        \hnem}
    \,.
\end{align}
The {\Pleb} formulation of SD gravity in \eqref{heavenly2.dpb}
implies that
SD gravity describes
an instance of SDYM theory
that takes the Poisson algebra
$\diff(\C^4,\Pi_\heav)$
as the color algebra:
\begin{align}
\begin{split}
    \label{CKD:sdgr}
    \SDGR
    &\,\,\cong\,\,
        \BAS\bigbig{
            \diff(\C^4,\Pi_\heav)
            \hem,\mem
            \diff(\C^4,\Pi_\heav)
        \hnem}
    \,,\\
    &\,\,\cong\,\,
        \SDYM\bigbig{
            \diff(\C^4,\Pi_\heav)
        \hnem}
    \,.
\end{split}
\end{align}

The isomorphism in \eqref{CKD:sdym}
is a manifestation of 
the very CK duality of YM theory
in the SD sector,
identifying the kinematic algebra $\gk$
as the Poisson algebra:
\begin{align}
    \bigbig{\,\text{$\gk$ in four dimensions, in the SD sector}\mem}
    \,\,\cong\,\,
        \diff(\C^4,\Pi_\heav)
    \,.
\end{align}
The correspondences in \eqref{CKD:sdgr} then
explain the double copy relationship between
YM theory and gravity
in the SD sector.
These facts were
established by
the seminal work \cite{monteiro2011kinematic}
by Monteiro and O'Connell;
see also \rcite{bjerrum2012algebras}.

\subsection{Mason-Newman Formulation}
\label{Jc.MN}

In component terms,
the SDYM equations in \eqref{SDYM.def} describe
\begin{align}
    \label{MNE1}
    \frac{1}{2}\,
        \e_{ABCD}\mem 
        \comm{D^C}{D^D}
    \,=\,
        i\mem \comm{D_A}{D_B}
    \,.
\end{align}
Here, the field strength
is understood as an operator $F_{AB} = \comm{D_A}{D_B}$
arising from
the commutator between covariant derivatives.
Also,
for this moment,
we have temporarily used $A,B,C,D,\cdots$
for denoting the \textit{global} Lorentz indices of flat spacetime $\mflat$,
raised and lowered via the flat metric $\eta_{AB}$.

\begin{subequations}
Remarkably,
Mason and Newman \cite{MasonNewman:1989}
established that
SD gravity
on a four-manifold $\M$
can be formulated
by the following equation:
\begin{align}
    \label{MNE2}
    \frac{1}{2}\,
        \e_{ABCD}\mem 
        \comm{V^C}{V^D}
    \,=\,
        i\mem \comm{V_A}{V_B}
    \,.
\end{align}
Here, $V_A = V^\m{}_A\mem \partial_\m$
are four linearly independent vector fields 
on $\M$
that preserve a volume form $\nu$:
\begin{align}
    \label{Vvolpres}
    \pounds_{V_A} \nu
    \,=\,
        0
    \,.
\end{align}
$A,B,C,D,\cdots$ are frame indices,
raised and lowered via the flat metric $\eta_{AB}$.
\end{subequations}

Specifically,
given a volume-preserving frame $V^\m{}_A$
that satisfies \eqref{MNE2},
one constructs
another frame $E^\m{}_A$ via
\begin{align}
    \label{MNrecipe}
    E_A
    \,=\,
        f^{-1}\mem V_A
    \transition{where}
    f^2
    \,=\,
        \nu(V_0,V_1,V_2,V_3)
    \,.
\end{align}
The statement is that this $E^\m{}_A$ 
defines
a metric $g_{\m\n} = \eta_{AB}\mem e^A{}_\m\mem e^B{}_\n$
whose curvature is SD,
where $e^A{}_\m E^\m{}_B = \delta^A{}_B$.

Conversely, Mason and Newman established that
an SD spacetime $(\M,g)$
always admits an orthonormal frame $E_A$
and a nonvanishing function $f$
such that
$V_A = f\mem E_A$
satisfy \eqrefs{MNE2}{Vvolpres}
for some volume form $\nu$.

Mason and Newman's formulation of SD gravity
generalizes
an earlier result by
Ashtekar, Jacobson, and Smolin
\cite{Ashtekar:1987qx}.\footnote{
    To this end,
    Mason and Newman \cite{MasonNewman:1989}
    assume that the spacetime $\M$ is topologically a product of a three-manifold and a line.
    Here, we may work locally or simply take $\M \cong \C^4$.
}
Below, we wish to review the core part of their derivation
of the converse statement.

In the complexified setup,
spacetime is
an oriented complex four-manifold $\M$
equipped with a holomorphic metric $g$.
Let $\ve$ be the metric volume form,
and let $\M \cong \C^4$ for simplicity.
$(\M,g)$ is said to be a SD spacetime
if the metric $g$
describes SD Riemannian curvature
as
${*} R_{\m\n\r\s} = +i\hem R_{\m\n\r\s}$.

Suppose an orthonormal frame $E_A$
in a SD spacetime.
Firstly, we recall the definition of the anholonomy coefficients in \eqref{anholonomy-def}:
$\comm{E_A}{E_B} = \Omega^C{}_{AB}\mem E_C$.
Upon a rescaling $V_A = f\mem E_A$,
we find
\begin{align}
    \label{VVcomm}
    \comm{V_A}{V_B}
    \,=\,
        \BB{
            \Omega^C{}_{AB}
            + E_A\act{\log f}\mem \delta^C{}_B
            - E_B\act{\log f}\mem \delta^C{}_A
        }\mem f^2 E_C
    \,.
\end{align}

\begin{subequations}
Secondly,
by utilizing the freedom of local Lorentz transformations,
the ASD part of the spin connection can be made to vanish:
\begin{align}
    \label{sd-on-sc}
    \star\gamma_{ABC}
    \,=\,
        +i\mem \gamma_{ABC}
    \,.
\end{align}
\eqref{sd-on-sc} implies
$
    \gamma_{ABC}\mem \e^{ABCD}\nem/2
    = -{\star \gamma}^{CD}{}_C
    = -i\mem \gamma^{CD}{}_C
$,
or equivalently
\begin{align}
    \label{sd-on-sc3}
    3\hem \gamma_\wrap{[ABC]}
    \,=\,
        -i\hem \gamma^{KD}{}_K\mem
        \e_{ABCD}
    \,.
\end{align}
For the anholonomy coefficients,
this self-duality condition translates to
\begin{align}
    \label{sd-on-Om}
    \Omega_{CAB}
    \,=\,
        \gamma_{ABC}
        - 3\hem \gamma_\wrap{[ABC]}
    \,=\,
        \gamma_{ABC}
        + i\hem \gamma^{KD}{}_K\mem
        \e_{ABCD}
    \,,
\end{align}
where we have used \eqref{Om=sc-sc}.
Since $\gamma^K{}_{AK} = \Omega^K{}_{KA}$,
\eqref{sd-on-Om}
implies that
the following combination is SD on the antisymmetric pair of indices $[AB]$:
\begin{align}
    \label{sdcomb-Om}
    \Omega^C{}_{AB}
    \mem+\mem
    \Omega^K{}_{KA}\mem \delta^C{}_B
    -
    \Omega^K{}_{KB}\mem \delta^C{}_A
    \,.
\end{align}
\end{subequations}

Combining \eqrefs{VVcomm}{sdcomb-Om},
one establishes \eqref{MNE2}
if there exists a scalar $f$ such that
\begin{align}
    \label{O-Elogf}
    \Omega^K{}_{KA}
    \,=\,
        E_A\act{
            \log f
        }
    \,.
\end{align}
If \eqref{O-Elogf} holds,
the identity stated earlier in \eqref{divO-id}
implies that
\begin{align}
    \label{dVcheck}
    \pounds_{V_A}\bigbig{
        f^{-2}\hem \ve
    }
    \,=\,
        \pounds_{E_A}\bigbig{
            f^{-1}\hem \ve
        }
    \,=\,
        f^{-1}\mem
        \BB{
            \Omega^C{}_{CA}
            - E_A\act{\log f}
        }\mem \ve
    \,=\,
        0
    \,,
\end{align}
where the first identity can be seen by the Cartan magic formula.
Hence one also establishes \eqref{Vvolpres}
by taking 
\begin{align}
    \nu \,=\, f^{-2}\hem \ve
    \qiq
    \nu(V_0,V_1,V_2,V_3)
    \,=\,
        f^2\mem \ve(E_0,E_1,E_2,E_3)
    \,=\,
        f^2
    \,,
\end{align}
which explains the recipe in \eqref{MNrecipe}.

The condition in \eqref{O-Elogf} 
is not generally true,
but Mason and Newman
show that
a concrete construction exists in a SD spacetime
by utilizing a harmonic function
and the {\Pleb} ASD two-forms
\cite{Plebanski:1977zz,Capovilla:1991qb}.\footnote{
    Note that one will be able to fix
    $\nu = d^4x$,
    i.e., $\nu_{\m\n\r\s} = \e_{\m\n\r\s}$,
    by a choice of coordinates.
    In this case, one can take $f = |E|^{-1/2}$,
    so $V_A = |E|^{-1/2} E_A = \text{``}\mem\Et_A\text{''}$
    precisely becomes the densitized frame
    that serves as the ``square root'' of the 
    LL gothic metric
    in \Sec{Ja>LL}:
    $\got^{\m\n} = \Et^\m{}_A\mem \Et^\n{}_B\mem \eta^{AB}$.
}

\medskip

Several remarks can be given here.
Firstly,
note that \eqref{MNE2}
follows from \eqref{MNE1}
by replacing the color Lie bracket
with the Lie bracket
between vector fields,
i.e.,
the diffeomorphism Lie bracket.
As will be made further explicit in \Chap{K2},
in this sense
the Mason-Newman formulation of SD gravity 
describes that
\begin{align}
\begin{split}
    \label{CKD:mn}
    \SDGR
    &\,\,\cong\,\,
        \SDYM\bigbig{
            \diff(\C^4,\nu)
        \hnem}
    \,,
\end{split}
\end{align}
where $\diff(\C^4,\nu)$
denotes the Lie algebra of diffeomorphisms of $\C^4$
that preserve the volume form $\nu$,
i.e., the volume-preserving diffeomorphism algebra
(also denoted as $\sdiff(\C^4)$).

Secondly,
note an interesting ``role reversal''
between the spacetime and internal indices
in establishing the relationship in \eqref{CKD:mn}.
The frame indices of SD gravity,
$A,B,C,D,\cdots$,
are made to correspond to
the flat spacetime indices 
of SDYM theory on $\mflat$.
It seems that this point could be generally relevant
since YM equations crucially employ
a flat metric $\eta^{AB}$ to contract indices:
$\eta^{AB} D_A F^a{}_{BC} = 0$.

\begin{subequations}\label{DYMV}
Thirdly,
it is easily seen that
\eqref{MNE2}
implies
\begin{align}
    \label{DYMV.e}
    \comm{V^A}{\comm{V_A}{V_B}}
    \,=\,
        0
    \,,
\end{align}
when combined with the Jacobi identity
\begin{align}
    \label{DYMV.m}
    \comm{V_A}{\comm{V_B}{V_C}}
    + \comm{V_B}{\comm{V_C}{V_A}}
    + \comm{V_C}{\comm{V_A}{V_B}}
    \,=\,
        0
    \,.
\end{align}
\end{subequations}
As will be elaborated in \Chap{K2},
we can identify \eqrefs{DYMV.e}{DYMV.m}
as the electric and magnetic equations
of volume-preserving diffeomorphism YM theory.
Unfortunately,
Mason and Newman \cite{MasonNewman:1989} report
that \eqref{DYMV.e}
does not describe GR
beyond the SD sector;
recall our earlier exploration in \Sec{Jb}.

Lastly,
with the aid of an auxiliary ASD spinor $\l_\a$,
the SDYM and SD gravity equations in
\eqrefs{MNE1}{MNE2}
can be stated as
\begin{align}
    \label{LaxLL}
    \comm{L_\da}{L_\db}
    \,=\,
        0
    \,,
\end{align}
via $L_\da = \l^\a\hhem D_{\a\da}$
or $L_\da = \l^\a\hhhem\mem V_{\a\da}$.
Geometrically, \eqref{LaxLL} implies
integrability of (gauge-covariant) translations
along SD null planes,
known as
$\a$-surfaces.
This gives the Lax pair 
(and eventually twistor)
formulations
of SDYM and SD gravity
\cite{Penrose:1976js,GravityMHVTwistors,ward1991twistor,huggett1994introduction}.

\section{ Gravity as a Tapestry of Self-Dual and Anti-Self-Dual Parts}
\label{Jcx}

In \Sec{Jc},
we have briefly reviewed
the simplicity of SD gravity.
{\Pleb}'s heavenly equation
\cite{plebanski1975some}
shows that
a single scalar field $\Phi$
dictates the full dynamics of SD gravity.
This also reveals that
a Poisson algebra
$\diff(\C^4,\Pi_\heav)$
lies at the heart of perturbative gravity
in the SD sector,
from which one sees that
SD gravity is a double copy of SDYM theory.
In Mason and Newman \cite{MasonNewman:1989}'s
version,
SD gravity is understood as
an instance of SDYM theory
for a volume-preserving diffeomorphism algebra.

As we have remarked in a footnote,
Newman \cite{shaviv1975general,Newman:1976gc}
and
{\Pleb} \cite{plebanski1975some,Plebanski:1977zz}
coined the technical term ``heaven''
for referring to SD spacetimes.
This terminology 
is to be taken as acknowledging
the simplicity and ideal nature of SD gravity.
The nonlinear dynamics of full gravity,
however,
is much more complicated.
Newman \cite{shaviv1975general,Newman:1976gc}
and
{\Pleb} \cite{plebanski1975some,Plebanski:1977zz}
refer to spacetimes with both SD and ASD curvatures
as ``earth.''

\begin{figure}[t]
    \centering
    \includegraphics[
        valign=c,
        width=0.9\linewidth
    ]{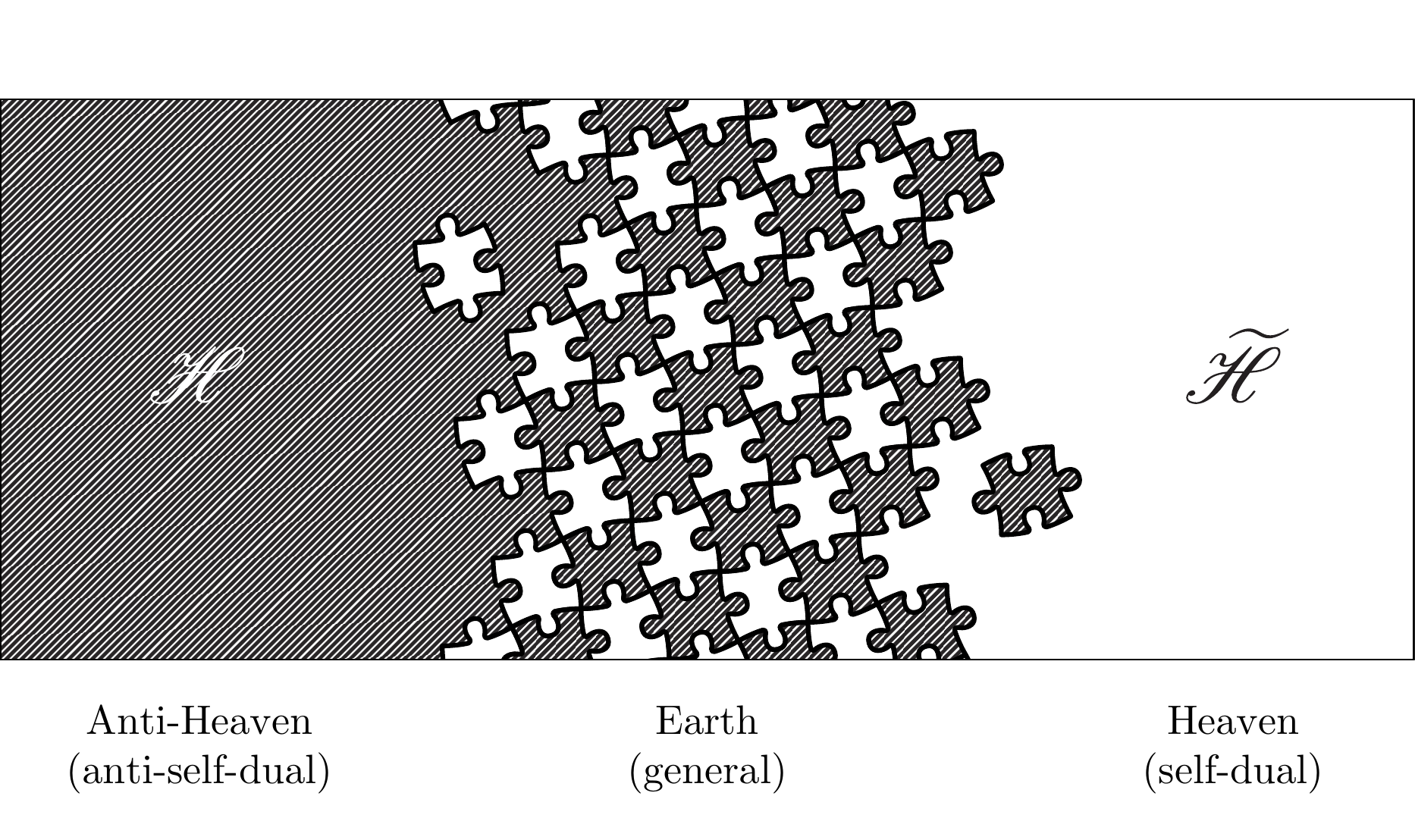}
    \caption[%
        \textit{Heaven} refers to 
            SD spacetimes.
        \textit{Earth} refers to
            spacetimes with both SD and ASD curvatures nonvanishing.
        Right-handed gravitons inhabit heaven;
        left-handed gravitons inhabit anti-heaven;
        both coexist on earth.
    ]{%
        \textit{Heaven} refers to 
            SD spacetimes.
        \textit{Earth} refers to
            spacetimes with both SD and ASD curvatures nonvanishing.
        Right-handed gravitons inhabit heaven;
        left-handed gravitons inhabit anti-heaven;
        both coexist on earth
        \cite{bialynicki1981note,ashtekar1986note}.
    }
    \label{fig:heaven-earth}
\end{figure}

In {\Pleb}'s world view,
a real spacetime
is a tapestry of 
SD and ASD building blocks:
an interacting mixture of heaven and anti-heaven,
so to speak.
The idea of ``patching'' SD and ASD spacetimes together
was the dream of the {\Pleb} school
for an era
\cite{plebanski1998linear,robinson-TN44-03,robinson1987some}.\footnote{
    The author thanks Maciej Dunajski for this anecdote.
}
An artist's impression 
(in the style of M.\,C. Escher)
is given
in \fref{fig:heaven-earth}.

Accordingly,
the program of
\textit{understanding earth
by maximally utilizing the simplicity of heaven}
has existed as
a reasonable slogan for four-dimensional gravitation.
This approach has been systematized and largely advocated by
twistor theorists.
In this context,
two perspectives have been pursued.
\begin{itemize}[leftmargin=1.5em]
    \item
        The \textit{googly} agenda
        \cite{Penrose:2015lla:Palatial,penr04-googly,Witten:2003nn,Adamo:2017qyl}.---%
        In essence,
        solving the googly problem means to
        describe ASD modes
        in the same variables in which SD modes are formulated.
        Though considered incomplete,
        a viable approach to this problem
        is to
        understand the full theory
        by perturbing away from the SD sector.
        
    \item
        The \textit{ambitwistor} approach
        \cite{Witten:1978xx,Isenberg:1978kk,Yasskin:1982dv,lebrun1985ambi,Mason:2005kn}.---%
        This line of thinking seeks to formulate
        SD and ASD modes on an equal footing,
        by employing both ``untilded'' and ``tilded'' variables.
        It is also open to
        describing theories in
        general dimensions
        and employing 
        non-spinorial variables.
\end{itemize}

With this broad remark made,
in this final section for
reimagining gravity
we would like to describe
two threads
on grasping the full dynamics of GR
as a nonlinear interaction of
SD and ASD modes.
From \Secsto{Jcx>CS}{Jcx>FRAME},
we implement the chiral, ``googly'' philosophy
that perturbatively adds ASD modes
into SD gravity.
From \Secsto{Jcx>CCK}{Jcx>NLIN},
we follow the ``ambitwistor'' spirit
to treat SD and ASD modes symmetrically.
We clarify, however, that our explorations are strictly field-theoretical
in spacetime,
not in twistor spaces.

\subsection{Chalmers-Siegel Action}
\label{Jcx>CS}

We begin with the YM analog.
YM theory
in four dimensions
describes the Lagrangian four-form
\begin{align}
    \label{YML.A}
    L[A]
    \,=\,
        -\frac{1}{2g^2}\,
            \delta_{ab}\mem F^a[A] \wedge {*}F^b[A]
    \,,
\end{align}
where $\delta_{ab}$ is a Killing form,
and $g$ is the coupling.
A well-known identity 
on the Chern-Simons (CS) three-form
\nomenclature{CS}{Chern-Simons}
reads
\begin{align}
\label{CS3form}
    d\mem\BB{
        \delta_{ab}\mem A^a \wedge dA^b
        + \tfrac{1}{3}\mem
            f_{abc}\mem A^a \wedge A^b \wedge A^c
    }
    \,=\,
        \delta_{ab}\mem F^a[A] \wedge F^b[A]
    \,,
\end{align}
so the Lagrangian four-form in \eqref{YML.A}
can be modified to
\begin{align}
    \label{YML.A+}
    L[A]
    \,=\,
        \frac{i}{g^2}\, \delta_{ab}\mem 
            F^a[A] \wedge F^-{}^b[A]
    \,,
\end{align}
without changing
the Euler-Lagrange equations.
\eqref{YML.A+} is perturbatively equivalent to \eqref{YML.A}.
Here, $F^\pm{}^a[A] = \tfrac{1}{2}\mem (1 \mp i\hem{*})\hem F^a[A]$.
Switching to the first-order formulation gives
\begin{align}
    \label{CS1}
    L[A,B]
    \,=\,
        -i\hem B_a \wedge F^a\hnem[A]
        \,-\, \frac{ig^2}{4}\,
            \delta^{ab}\mem
            B_a \wedge B_b
    \,,
\end{align}
where $B_a$ is a coadjoint-valued ASD two-form:
\begin{align}
    \label{Basd1}
    {*}B_a \,=\, -i\hem B_a
    \,.
\end{align}
\eqref{CS1} is known to provide
the Chalmers-Siegel \cite{Chalmers:1996rq}
formulation of YM theory
in four dimensions.

In the limit $g \to 0$,
the Chalmers-Siegel Lagrangian four-form in \eqref{CS1}
reduces to $B_a \swedge F^a[A]$,
which describes SDYM theory:
the $B$-field is a Lagrange multiplier
stipulating that the ASD part of $F^a[A]$ vanishes.
In this sense the Chalmers-Siegel formulation understands full YM theory
as a perturbation away from SDYM theory.
This approach exactly resonates with
the perturbative solution to the googly problem
in spirit
(see, e.g., \rcite{Adamo:2017qyl}).


The EoM of \eqref{CS1} are
\begin{align}
    \label{CS1.eom}
    F^-{}^a[A]
    \,=\,
        -\frac{g^2}{2}\mem B^a
    \,,\quad
    D\hnem B_a
    \,=\,
        0
    \,.
\end{align}
In the $g\to0$ limit,
$F^-{}^a[A] = 0$,
so
one can interpret the $A$-field as describing SD (positive-helicity) gluons,
on the background of which the ``cotangent'' $B$-field
propagates as
a linearized ASD (negative-helicity) perturbation.
For $g \neq 0$,
this ASD perturbation
``backreacts'' to the $A$-field.

\subsection{Chiral Coframe Action ({\Pleb} Action)}
\label{Jcx>COFRAME}

For gravity,
the {\Pleb} formulation of GR
\cite{Plebanski:1977zz}
has been widely identified
as 
the gravitational analog
of the Chalmers-Siegel formulation;
see, e.g., \rcite{GravityMHVTwistors}.
Note that this is different from
the {\Pleb} formulation of SD gravity
(heavenly equation)
\cite{plebanski1975some}
described in \Sec{Jc>SDG>HEAV}.

\paragraph{Nonchiral Version}

Let us first review the nonchiral formulation
before 
reproducing the chiral version
of {\Pleb} \cite{Plebanski:1977zz}.
The point of departure
is the coframe formulation of GR.
In four dimensions,
\eqref{Palatini}
describes the Lagrangian four-form
\begin{align}
    \label{GRL.e}
    L[e,\gamma]
    \,=\,
        \frac{1}{2\k^2}\mem
        \bb{
            \frac{1}{2}\,
            \e_{ABCD}\mem 
            e^C \swedge e^D
        \nem}
        \wedge
            R^{AB}[\gamma]
    \,.
\end{align}
The key observation is that
the two-form in parenthesis in \eqref{GRL.e}
emerges as a solution to the following ``simplicity constraint'':
\begin{align}
\begin{split}
    \label{simplicity.nonchiral}
    B_\wrap{AB} \wedge B_\wrap{CD}
    \,=\,
        B_\wrap{[AB} \swedge B_\wrap{CD]}
    \,.
\end{split}
\end{align}
With this understanding,
the proposed reformulation for \eqref{GRL.e} is
\begin{align}
    \label{PL.nonchiral0}
    L[B,\gamma,\Psi]
    \,=\,
        \frac{1}{2\k^2}\mem 
        B_{AB} \wedge R^{AB}[\gamma]
        - \frac{1}{8\k^2}\mem
            \Psi^{ABCD}\hem B_{AB} \swedge B_{CD}
    \,,
\end{align}
where $\Psi^{ABCD} = \Psi^{[AB][CD]} = \Psi^{[CD][AB]}$
is a Lagrange multiplier subject to the condition
(assuming vanishing cosmological constant)
\begin{align}
    \e_{ABCD}\mem
    \Psi^{ABCD}
    \,=\,
        0
    \,,
\end{align}
hence imposing
$6(6{\,+\,}1)/2-1 = 20$ constraints.
As a result,
the $36$ components of $B_{AB\m\n}$
are reduced down
to $16$ degrees of freedom,
which precisely match with
the number of components in the tetrad $e^A{}_\m$.

To be precise,
the general solution to \eqref{simplicity.nonchiral} is
\cite{DePietri:1998hnx,Buffenoir:2004vx,Celada:2016bf}
\begin{align}
\begin{split}
    \label{eq:simplicity-impl}
    B_{AB}
    \,\,=\,\,
        \pm\mem\frac{1}{2}\, \e_{ABCD}\mem e^C \swedge e^D
    \quad\text{or}\quad
        \pm e_A \swedge e_B
    \quad\text{or}\quad
        0
    \,.
\end{split}
\end{align}
These three branches of solutions 
may be referred to as
the gravitational, topological, and degenerate sectors,
respectively
\cite{Buffenoir:2004vx}.
One may
assume that the
vacuum expectation value of the metric
(see \eqref{Urbantke})
chooses the positive gravitational branch,
$B_{AB}
= + \tfrac{1}{2}\mem \e_{ABCD}\mem e^C \swedge e^D$,
as a superselection sector.
With this assumption,
\eqref{PL.nonchiral0}
can be brought back to
the tetradic Palatini formulation in
\eqref{GRL.e}
by integrating out the $\Psi$-field.

In the meantime,
from the variations of \eqref{PL.nonchiral0}
with respect to the $B$- and $\gamma$-fields,
one finds
\begin{align}
    R^{AB}[\gamma]
    \,=\,
        \frac{1}{2}\, \Psi^{ABCD}\hem B_{CD}
    \,,\quad
    D\hnem B_{AB}
    \,=\,
        0
    \,.
\end{align}
The first equation tells us that
$\Psi^{ABCD}$ encode the components of the Weyl tensor
in the orthonormal frame.
The second equation encodes the torsion-free condition
for the spin connection $\gamma$.

We see that
an interesting perspective toward gravity
has emerged.
Firstly,
the area element $B_{AB}$ is taken as
a fundamental variable for describing Riemannian geometry,
instead of lengths or orthonormal frames.
Intriguingly,
the line element is retrieved via a remarkable formula due to Urbantke
\cite{Urbantke:1984eb,Capovilla:1991qb}:
\begin{align}
\begin{split}
    \label{Urbantke}
    g_{\m\n}
    \,&=\,
        -\frac{1}{12|e|}\mem
        B^A{}_{B\m\k}\mem
        B^B{}_{C\n\l}\mem
        B^C{}_{A\r\s}\mem
        \e^{\k\l\r\s}
    \,,\\
    |e|
    \,&=\, \frac{1}{96}\mem
    \e^{ABCD}\mem 
    B_{AB\m\n}\mem
    B_{CD\r\s}\mem
    \e^{\m\n\r\s}
    \,,
\end{split}
\end{align}
where we have assumed the positive gravitational branch.
For the topological and degenerate branches,
\eqref{Urbantke} gives $g_{\m\n} = 0$.

Secondly,
the graviton is ``carved out'' from a topological $BF$ theory
by constraints.
This illustrates how gravity as a diffeomorphism-invariant theory
is close to topological field theories.
In a sense, gravitons are being viewed as ``impurities''
in a topological theory.
The $B$-field could have been
a Lagrange multiplier
stipulating flatness $R^{AB}[\gamma] = 0$
on the spin connection,
but it is made ``defective''
due to the $\Psi$-field.
In consequence, $\gamma$ acquires the freedom to be not perfectly flat.
In fact, performing a rescaling $B_{AB} \mapsto \k^2\hem B_{AB}$
brings \eqref{PL.nonchiral0} to
\begin{align}
    \label{PL.nonchiral}
    L[B,\gamma,\Psi]
    \,=\,
        B_{AB} \wedge R^{AB}[\gamma]
        - \frac{\k^2}{8}\mem
            \Psi^{ABCD}\hem B_{AB} \swedge B_{CD}
    \,,
\end{align}
in which case the $\k \to 0$ limit
describes the $BF$ theory.

\paragraph{Chiral Version}

We are now ready to describe the chiral version.
In the tetradic Palatini formalism,
the Nieh-Yan identity
\cite{Nieh:1981ww}
reads
\begin{align}
    \label{NiehYan}
    d\mem\bigbig{
        e_A \wedge T^A
    }
    \,=\,
        T_A \wedge T^A
        - e_A \wedge e_B \wedge R^{AB}
    \,,
\end{align}
where $T^A = De^A = de^A + \gamma^A{}_B \wedge e^B$ is the torsion two-form.
Based on this fact,
a simplified alternative for
\eqref{GRL.e}
has been proposed as
\begin{align}
    \label{GRL.e+}
    L[e,\gamma]
    \,=\,
        -\frac{i}{\k^2}\,
            B_{AB}[e]
        \wedge
            R^{AB}[\gamma]
    \,,
\end{align}
which is attributed to Holst \cite{Holst:1995pc}.
Here, $B_{AB}[e]$ denotes the two-form
\begin{align}
    \label{Be-PL}
    B_{AB}[e]
    \,=\,
        \frac{1}{2}\mem\bb{
            e_A \wedge e_B
            + \frac{i}{2}\mem \e_{ABCD}\mem e^C \swedge e^D
        }
    \,,
\end{align}
which is ASD on the pair of indices $[AB]$:
\begin{align}
    \label{Basd2.e}
    {\star}B_{AB} \,=\, -i\hem B_{AB}
    \,.
\end{align}
Note that \eqref{Basd2.e} could be compared to \eqref{Basd1}.

The Holst action is
classically equivalent to 
the tetradic Palatini action
in the sense that it gives the same bulk EoM,
provided that the coframe $e$ is nondegenerate.\footnote{
    Variation of $\gamma$ imposes vanishing torsion
    $T^A[e,\gamma] = 0$
    provided $e$ is nondegenerate.
    As a result,
    dropping the torsion-squared term in \eqref{GRL.e+}
    does not alter the classical EoM.
}
However, the two actions are not equivalent off shell,
and
further subtleties may arise
in the presence of fermions or spin sources.

The ASD two-form in \eqref{Be-PL}
is called the {\Pleb} two-form
\cite{Plebanski:1977zz,Capovilla:1991qb}.
In the spinor notation, it boils down to
\begin{align}
    \label{B[e]asd}
    B^{\a\b}[e]
    \,=\,
        \te_\wrap{\da\db}\mem (e^{\da\a} \swedge e^{\db\b})
    \,.
\end{align}
With this understanding,
{\Pleb} \cite{Plebanski:1977zz}
reformulates \eqref{GRL.e+} as
\begin{align}
    \label{CS2.basic}
    L[B,\gamma,\Psi]
    \,=\,
        -\frac{2i}{\k^2}\mem
        \bb{
            B_{\a\b}
        \wedge
            R^{\a\b}[\gamma]
        - \frac{1}{2}\,
            \Psi^{\a\b\c\d}
            B_{\a\b} \wedge B_{\c\d}
        }
    \,,
\end{align}
where $B_{\a\b} = B_{(\a\b)}$ is now taken as a fundamental field.
The Lagrange multiplier $\Psi^{\a\b\c\d} = \Psi^{(\a\b\c\d)}$
imposes $5$ constraints,
reducing the $18$ components of $B_{(\a\b)[\m\n]}$
down to $13 = 16-3$.
This precisely corresponds to the degrees of freedom in $e^{\da\a}{}_\m$
modulo SD local Lorentz transformations
to which \eqref{B[e]asd} is ignorant.


It should also be clarified that
$\gamma$ in \eqref{CS2.basic}
refers to an ASD spin connection,
and the SD part completely drops out:
\begin{align}
    R_\a{}^\b[\gamma]
    \,=\,
        d\gamma_\a{}^\b 
        + \gamma_\a{}^\c \swedge \gamma_\c{}^\b
    \,.
\end{align}

Finally,
by a rescaling 
$\gamma \mapsto \k^2\hem \gamma$
\cite{GravityMHVTwistors,Abou-Zeid:2005zfo},
together with
$B \mapsto iB/2$
and $\Psi \mapsto -2\hem i\k^2\mem \Psi$,
\eqref{CS2.basic} is brought to
\begin{align}
    \label{CS2}
    L[B,\gamma,\Psi]
    \,=\,
        B_{\a\b}
        \wedge
            \BB{
                d\gamma^{\a\b}
                + \k^2\mem 
                    \gamma^{\a\c} \swedge \gamma_\c{}^\b
            }
        - \frac{1}{2}\,
            \Psi^{\a\b\c\d}
            B_{\a\b} \wedge B_{\c\d}
    \,.
\end{align}

The interpretation of \eqref{CS2}
is detailed in \rrcite{GravityMHVTwistors,Abou-Zeid:2005zfo},
for instance.
In the limit $\k\too0$,
the saddle of \eqref{CS2}
describes
$dB_{\a\b} \eqq 0$
with $B_{\a\b}$ subject to the simplicity constraint
$B_\wrap{(\a\b} \swedge B_\wrap{\c\d)} = 0$.
It is not difficult to see that
this implies a coframe $e^{\da\a}$
whose ASD spin connection coefficients vanish.
Hence
the $B$-field defines a SD metric
via the (spinorial version of) Urbantke formula.

The {\Pleb} formulation of GR 
in \eqref{CS2}
will be compared with
the Chalmers-Siegel formulation of YM theory 
in \eqref{CS1}.
In both cases,
the full theory
is formulated in a way that perturbs away from the SD sector.

\subsection{Chiral Frame Action}
\label{Jcx>FRAME}

The naive perturbation theory of
the Lagrangian in \eqref{CS2}
might be somewhat confusing
when taken literally,
however.
The $B$-field is subject to the simplicity constraint.
It gains a vacuum expectation value,
so the $\Psi BB$ term produces a tadpole.
Note also that,
while
the variation of \eqref{CS1} directly yields
the SDYM equations in \eqref{MNE1},
$F^a{}_{\a\b}[A] = 0$,
the variation of \eqref{CS2}
does not directly yield
the SD gravity equations
in the form of \eqref{MNE2}.
Hence the map between 
YM and gravity
in the SD sector,
stated in \eqref{CKD:mn},
is not direct.

In this light,
we may want to envision an alternative approach,
based on the frame formulation
instead of the coframe formulation.

The point of departure is the Holst formulation
in \eqref{GRL.e+}.
This time, we present it as
\begin{align}
    \label{GRL'.e+}
    L[e,\gamma^-]
    \,=\,
        -i\,
            e_A \wedge e_B
        \wedge
            d\gamma^-{}^{AB}
        -i\k^2\,
            e_A \wedge e_B
        \wedge
            \gamma^-{}^A{}_C \wedge \gamma^-{}^{CB}
    \,,
\end{align}
where it is emphasized that $\gamma^-$
is an ASD connection by fiat:
\begin{align}
    \star \gamma^-{}^{AB}
    \,=\,
        -i\hem \gamma^-{}^{AB}
    \,.
\end{align}
We have also implemented the rescaling
$\gamma^- \mapsto \k^2\hem \gamma^-$.


While {\Pleb} has oriented us to view $e \wedge e$ as the $B$-field,
here we wish to
view $\gamma^-$ as the $B$-field instead.
Hence, via integration by parts,
we bring \eqref{GRL'.e+} to
the sum of the following two terms:
\begin{subequations}
\begin{align}
    \label{xtelder.L1}
    L_1[e,\gamma^-]
    \,&=\,
        -i\hem
            \gamma^-_{AB}
        \wedge
            d\bigbig{
                e^A \wedge e^B
            }
    \,,\\
    \label{xtelder.L2}
    L_2[e,\gamma^-]
    \,&=\,
        -i\k^2\,
            e_A \wedge e_B
        \wedge
            \gamma^-{}^A{}_C \wedge \gamma^-{}^{CB}
    \,.
\end{align}
\end{subequations}
With 
$
    \ve =
    \tfrac{1}{4!}\,
    \e_{ABCD}\mem e^A \swedge e^B \swedge e^C \swedge e^D
$,
it holds that
\begin{subequations}
\begin{align}
    \label{dee.id1}
    d\bigbig{
        e^A \swedge e^B
    }
    \,=\,
        \frac{1}{2}\,
        \e^{ABCD}\mem 
            d\hem \i_{E_C} \i_{E_D} \ve
    \,,
\end{align}
while Cartan calculus shows that
\begin{align}
\begin{split}
    \label{dee.id2}
    d\hem \i_{E_C} \i_{E_D} \ve
    \,&=\,
        \pounds_{E_C} \i_{E_D} \ve
        - \i_{E_C} d\hem \i_{E_D} \ve
    \,,\\
    \,&=\,
        \i_{\comm{E_C}{E_D}} \ve
        + \i_{E_D} \pounds_{E_C} \ve
        - \i_{E_C} \pounds_{E_D} \ve
    \,,\\
    \,&=\,
    \BB{
        \Omega^I{}_{CD}
        + \Omega^K{}_{KC}\mem \delta^I{}_D
        - \Omega^K{}_{KD}\mem \delta^I{}_C
    }\mem \i_{E_I} \ve
    \,.
\end{split}
\end{align}
\end{subequations}
In the last equality, we have recalled \eqref{divO-id}.
Via \eqrefs{dee.id1}{dee.id2},
\eqref{xtelder.L1} is brought to
\begin{subequations}
\begin{align}
    \label{xtelder.L1'}
    L_1[E,\gamma^-]
    \,&=\,
        \BB{
            \Omega^C{}_{AB}
            + \Omega^K{}_{KA}\mem \delta^C{}_B
            - \Omega^K{}_{KB}\mem \delta^C{}_A
        }\,
        \gamma^-{}^{AB}{}_C\,
        |E|^{-1}\hem d^4x
    \,.
\end{align}
We observe the same combination
that appeared previously in \eqref{sdcomb-Om}.
Meanwhile, 
\eqref{xtelder.L2}
boils down to
\begin{align}
    \label{xtelder.L2'}
    L_2[E,\gamma^-]
    \,&=\,
        \k^2\,
        \BB{
            \gamma^-_{ABC}\mem \gamma^-{}^{ACB}
            - \gamma^-{}^B{}_{AB}\mem \gamma^-{}^{CA}{}_C
        }
        \, |E|^{-1}\hem d^4x
    \,.
\end{align}
\end{subequations}

The sum of \eqrefs{xtelder.L1'}{xtelder.L2'} enjoys
three kinds of gauge redundancies.
The first is diffeomorphisms.
The second is SD local Lorentz transformations,
which alter $E$
but fix $\gamma^-$.
The third is ASD local Lorentz transformations,
which alter both $E$ and $\gamma^-$.
Only the first and the second are robust
in the decoupling limit $\k\too0$.

With this understanding, let us introduce a new scalar degree of freedom $f$,
by plugging in $E_A = f^{-1}\hem V_A$
while declaring the redundancy
\begin{align}
    \label{stuck.scaling}
    f \,\,\sim\,\,
        \mathe^{\varsigma}\mem f
    \,,\quad
    V_A \,\,\sim\,\,
        \mathe^{\varsigma}\mem V_A
    \,,
\end{align}
for a local scalar parameter $\varsigma$
(cf. Stueckelberg trick \cite{stuckelberg1938wechselwirkungskrafte}).
Now our field basis is $(V,\gamma^-,f)$.
This brings \eqrefs{xtelder.L1'}{xtelder.L2'} to
\begin{subequations}
\begin{align}
    \label{xtelder.L1''}
    L_1[V,\gamma^-,f]
    \,&=\,
        \BB{
            v^C{}_\m\mem \comm{V_A}{V_B}^\m
            + 2\hem \Delta_\wrap{[A}\mem \delta^C{}_\wrap{B]}
        }\,
        \gamma^-{}^{AB}{}_C\,
        |V|^{-1}\hem d^4x
    \,,\\
    \label{xtelder.L2''}
    L_2[V,\gamma^-,f]
    \,&=\,
        \k^2\,
        \BB{
            \gamma^-_{ABC}\mem \gamma^-{}^{ACB}
            - \gamma^-{}^B{}_{AB}\mem \gamma^-{}^{CA}{}_C
        }
        \, f\mem |V|^{-1}\hem d^4x
    \,,
\end{align}
\end{subequations}
where
$v^A{}_\m\mem V^\m{}_B = \delta^A{}_B$,
and
\begin{align}
\begin{split}
    \label{DelfV}
    \Delta_A[V,f]
    \,&=\,
        f\mem 
        \BB{
            \Omega^K{}_{KA} 
            - E_A\act{
                \log f
            }
        }
    \,,\\
    \,&=\,
        f^2\mem v^K{}_\m\mem
        \comm{f^{-1}\hhem V_K}{f^{-1}\hhem V_A}^\m
        - V_A\act{
            \log f
        }
    \,.
\end{split}
\end{align}
Note that the computation performed in \eqref{dVcheck} implies
\begin{align}
    \pounds_{V_A}\hnem\bigbig{
        f^{-2}\hem \ve
    }
    \,=\,
        \Delta_A[V,f]\mem
        \bigbig{
            f^{-2}\hem \ve
        }
    \,.
\end{align}

Now
the sum of \eqrefs{xtelder.L1''}{xtelder.L2''} exhibits
four kinds of gauge redundancies.
The first kind is diffeomorphisms.
The second kind is SD local Lorentz transformations.
The third kind is ASD local Lorentz transformations.
The fourth kind is the scaling transformation in \eqref{stuck.scaling}.
The first, the second, and the fourth are robust
in the decoupling limit $\k\too0$.

Direct computation verifies that
the set of four functionals in \eqref{DelfV},
$\Delta_A[V,f]$,
changes nontrivially under
the scaling and local Lorentz transformations.
Hence,
at the classical level,
we might envision an explicitly gauge-fixed action
by introducing the term
\begin{align}
    \label{xtelder.Lgf}
    L_\text{gf}[V,\gamma^-,f,\rho]
    \,=\,
        - \rho^A\mem \Delta_A[V,f]
        \, f^{-2}\hem \ve
    \,=\,
        -\rho^A\mem 
        \pounds_{V_A}\hnem\bigbig{
            f^{-2}\hem \ve
        }
    \,,
\end{align}
where $\rho^A$ is a Lagrange multiplier,
and
we have used
the dynamical volume form
$\ve 
=
    \tfrac{1}{4!}\,
    f^4\,
    \e_{ABCD}\mem v^A \swedge v^B \swedge v^C \swedge v^D
$
to prevent interference with
the diffeomorphism redundancies.
Four combinations of the scaling and local Lorentz redundancies will be
eliminated by \eqref{xtelder.Lgf}.
See \rcite{Grant:1994vp} for 
a remark on
the existence of this $f$ 
for off-shell geometries.

The sum of \eqrefss{xtelder.L1''}{xtelder.L2''}{xtelder.Lgf}
is equivalent to
\begin{align}
\begin{split}
    \label{xFrameAction1}
    &
    L[V,B,f,\trho]
    \\
    &
    =\,
        \frac{1}{2}\mem B^{AB}{}_\m\mem 
            \comm{V_A}{V_B}^\m
        + \frac{\k^2}{8}\,
            f\mem \mathcal{G}_{ABCD}^{\m\n}[V]\mem
            B^{AB}{}_\m\mem B^{CD}{}_\n
        - \trho^A\mem \Delta_A[V,f]
    \,,
\end{split}
\end{align}
where we have performed 
algebraic field redefinitions,
\begin{align}
    \gamma^-_{ABC}
    \,\mapsto\,
        \frac{1}{2}\mem
        B_{AB\m}\mem V^\m{}_C\mem
        |V|
    \,,\quad
    \rho^A
    \,\mapsto\,
        f\hem |V|\, \trho^A
    \,,
\end{align}
and also defined
\begin{align}
    \label{xtelGdef}
    \mathcal{G}^{\m\n}_{\nem A_1A_2B_1B_2}[V]
    \,=\,
        4\hem |V|\,
        V^{[\m}{}_{[A_1}\hem
        \eta_{A_2][B_1}
        V^{\n]}{}_{B_2]}\mem
    \,.
\end{align}
\eqrefs{xFrameAction1}{xtelGdef}
provide a chiral frame formulation of GR,
which is to be compared with
the nonchiral frame formulation given in
\eqrefs{FrameAction1}{telGdef}.
The action due to \eqref{xFrameAction1}
will enjoy
the diffeomorphism redundancy
as well as three more redundancies.

In the decoupling limit $\k \to 0$,
one will obtain a Lagrangian formulation
of the Mason-Newman \cite{MasonNewman:1989} formulation of SD gravity:
the ASD Lagrange multiplier $\star B^{AB}{}_\m = -i\hem B^{AB}{}_\m$
imposes \eqref{MNE2},
while the other Lagrange multiplier $\trho^A$
imposes the volume-preserving condition
in \eqref{Vvolpres}
with $\nu = f^{-2}\hem \ve$.
In this manner, \eqref{xFrameAction1}
describes full gravity
by perturbing away from SD gravity as 
a SDYM theory for volume-preserving diffeomorphisms.
It is a Chalmers-Siegel version of teleparallel gravity, so to speak.

Speaking of which, the Chalmers-Siegel action in
\eqref{CS1} unpacks to
\begin{align}
    \label{CS1.unpacked}
    L[A,B]
    \,=\,
        \frac{1}{2}\,
            B_a{}^{AB}\mem F^a{}_{AB}[A]
        \,+\, \frac{g^2}{8}\,
            \BB{
                \delta^{ab}\mem
                \eta_{A[C} \eta_{D]B}
            }\mem
            B_a{}^{AB}\mem B_{b}{}^{CD}
    \,,
\end{align}
where we have used $A,B,\cdots$ for the global Lorentz indices;
suppose we install a trivial frame $\delta^\m{}_A$ on the flat spacetime
to convert $\m,\n,\cdots$ to $A,B,\cdots$.
\eqref{CS1.unpacked} can be compared with \eqref{xFrameAction1}.
The correspondence is only exact
in the decoupling limit,
since $f\mem \mathcal{G}^{\m\n}_{ABCD}[V]$
in \eqref{xFrameAction1}
is dynamical.\footnote{
    It will be interesting to
    integrate out the $B$-field
    in \eqref{xFrameAction1}
    as an auxiliary
    and understand 
    the discarded term
    (around \eqref{NiehYan})
    in terms of frame variables.
}

\subsection{Covariant Color-Kinematics Duality}
\label{Jcx>CCK}

We now wish to treat SD and ASD modes symmetrically.
There can be many approaches,
but our goal is to demonstrate one concrete version.

To begin with,
consider YM theory on a $d$-dimensional flat spacetime.
The electric and magnetic equations of YM theory,
shown in \eqref{YM.em},
are
\begin{align}
    \label{YM.cem}
    D^\m F^a{}_{\m\n}
    \,=\,
        0
    \,,\quad
    D_\wrap{[\r} F^a{}_\wrap{\m\n]}
    \,=\,
        0
    \,.
\end{align}
For the moment, 
let us assume generic $d$.
We will later specialize in $d = 4$.

\newpage

In the usual treatment,
the gauge potential $A^a{}_\m$
is taken as the fundamental field.
By plugging in the definition of the field strength,
\begin{align}
    \label{Jd:fsdef}
    F^a{}_{\m\n}
    \,=\,
        \partial_\m A^a{}_\n - \partial_\n A^a{}_\m
        + f^a{}_{bc}\mem A^b{}_\m\mem A^c{}_\n
    \,,
\end{align}
to the electric equation in \eqref{YM.cem},
one derives the second-order EoM,
\begin{align}
    \label{YM.waveA}
    \Box\hem A^{a\m}
    + \partial^\m (\partial\mdot\nem\hnem A^a)
    \,=\,
    {}
\begin{aligned}[t]
    &
        f^a{}_{bc}\mem \BB{
            A^{b\r} \partial_\r A^{c\m}
            - A^{c\r} \partial_\r A^{b\m}
        }
        - f^a{}_{bc}\mem A^{b\r}\mem \partial^\m\hnem A^c{}_\r
    \\
    &
        - f^a{}_{cb}\mem A^{c\m}\mem (\partial\mdot\nem\hnem A^b)
        + f^a{}_{bc}\mem f^c{}_{de}\mem 
            A^b \mdot A^d\mem
            A^{e\m}
    \,,
\end{aligned}
\end{align}
where $\Box := -\eta^{\m\n} \partial_\m \partial_\n$.

As a classical field theory,
pure YM theory
describes
the dynamics of ``color waves.''
The second-order equation in \eqref{YM.waveA}
describes a \textit{nonlinear wave equation}.
At the linearized order,
the solutions to \eqref{YM.waveA} 
take the form
\begin{align}
    \label{YM.waveA:lin}
    A^{a\m}
    \,=\,
        g\,
        \sum_{i}\:
            \e_i\,
            c_i^a\hem e_i^\m
            \,\mathe^{ik_ix}
        \,+\, \O(g^2)
    \,,
\end{align}
where $k_i{}_\m$, $c_i^a$, and $e_i^\m$ are
the 
momentum,
color polarization,
and 
kinematic polarization
of the $i$\textsuperscript{th} color plane wave,
which are subject to the massless on-shell condition
$k_i^2 = 0$ and $k_i \mdot e_i = 0$.
Also, we have chosen to employ
formal parameters $\e_i$ such that ${\e_i}^2 = 0$.
With $k_{ij} := k_i + k_j$,
the generic perturbative solution to \eqref{YM.waveA}
is then given by
\begin{align}
\begin{split}
    \label{YM.waveA:nlin}
    A^{a\m}
    \,=\,
    {}
        &
        g\,
            \sum_{i}\:
                \e_i\,
                c_i^a\hem e_i^\m
                \mathe^{ik_ix}
        \:+\:
        g^2\,
            \sum_{i,j}
                \e_i\hhem \e_j\,
                \frac{
                    c_{ij}^a\hem e_{ij}^\m
                }{
                    {k_{ij}}^2
                }\,
                \mathe^{ik_{ij}x}
        \:+\:
        \O(g^3)
    \,,
\end{split}
\end{align}
which takes the linearized solution in \eqref{YM.waveA:lin}
as a ``seed''
and constructs an on-shell ``nonlinear completion''
to a desired order in the coupling $g$.
\eqref{YM.waveA:nlin}
represents the ``nonlinear superposition'' of color plane waves.

This is the idea of
BG recursion
\cite{BerendsGiele}
(perturbiner method \cite{Rosly:1996vr,Selivanov:1997an,Mizera:2018jbh}):
solve nonlinear wave equations perturbatively.
Diagrammatically,
the perturbative solution at each order in $g$
is given as a sum over
tree graphs
where each vertex exhibits degree $3$ or $4$.
In this way,
one sees that
\eqref{YM.waveA}
describes cubic and quartic interactions
between color plane waves (gluons).

\medskip

Meanwhile,
suppose one wants to see
how SD and ASD color waves interact
in $d = 4$ dimensions.
The field basis $A^{a\m}$
is not a good choice,
since it does not split cleanly into SD and ASD parts.

To this end,
we \textit{reimagine YM theory}
such that
the field strength $F^a{}_{\m\n}$ serves as a fundamental field variable.
Since $F^a{}_{\m\n}$ is a two-form,
it can be cleanly split into SD and ASD parts in $d = 4$ dimensions.

First of all,
a second-order EoM
should be derived
for the field strength.
The electric and magnetic equations in \eqref{YM.cem}
can be combined
as follows,
yielding a \textit{covariant nonlinear wave equation} for YM theory:
\begin{align}
\begin{split}
	\label{CCK.YM.deriv}
		D^2 F^a{}_{\m\n}
	\,&=\,
		- D^\r D_\m F^a{}_{\n\r}
		+ D^\r D_\n F^a{}_{\m\r}
	\,,\\
	\,&=\,
		- f^a{}_{bc}\mem F^b{}^\r{}_\m F^c{}_{\n\r}
		+ f^a{}_{bc}\mem F^b{}^\r{}_\n F^c{}_{\m\r}
	\,,\\
	\,&=\,
		-2\mem f^a{}_{bc}\mem F^b{}_{\m\r} F^{c\r}{}_\n
	\,=\,
		- f^a{}_{bc}\mem \BB{
			F^b{}_{\m\r} F^{c\r}{}_\n
			- F^c{}_{\m\r} F^{b\r}{}_\n
		}
	\,.
\end{split}
\end{align}
The first line uses the magnetic equation.
The second line uses the electric equation
while commuting covariant derivatives.
This gymnastics is due to 
Cheung and Mangan \cite{CCK}.

Second of all,
the field strength should be promoted to a fundamental variable.
To this end,
one can employ the axial gauge.
By introducing a reference vector $\eta^\m$,
one imposes
\begin{align}
	\label{axialgauge-A}
	A^a{}_\m\mem \eta^\m 
	\,=\,0
	\qiq
	A^a{}_\m
	\,=\,
		\frac{1}{\eta\mdot\partial}\,
            \eta^\r F^a{}_{\r\m}
	\,.
\end{align}
After this gauge fixing,
the $A$-field
is viewed
as a nonlocal functional of the field strength:
$A^a{}_\m[F]$.

Cheung and Mangan \cite{CCK}
make an intriguing observation:
the YM field strength $F^{a\m\n}$
can be viewed as a BAS field $F^{a\ta}$
that takes the Lorentz algebra
as its tilded Lie algebra,
i.e.,
\begin{align}
    \tg
    \,=\,
        \so(1,d-1)
    \,.
\end{align}
That is, the antisymmetric pair of indices $[\m\n]$
is identified as an adjoint index $\ta$
via the generators of the Lorentz algebra,
say $(\Sigma_\ta)^{\m\n}$:
\begin{align}
    (\Sigma_\ta)^\m{}_\r\mem (\Sigma_\tb)^\r{}_\n
    -
    (\Sigma_\tb)^\m{}_\r\mem (\Sigma_\ta)^\r{}_\n
    \,=\,
        \tf^\tc{}_{\ta\tb}\mem
        (\Sigma_\tc)^\m{}_\n
    \,.
\end{align}
With this realization,
the nonlinear wave equation in \eqref{CCK.YM.deriv}
appears as
\begin{align}
    \label{CCK1}
    -D^2 F^{a\ta}
    \,=\,
        f^a{}_{bc}\mem \tf^\ta{}_{\tb\tc}\mem 
            F^{b\tb}\mem F^{c\tc}
    \,,
\end{align}
which very much resembles
the BAS equation in \eqref{BAS}.
In \eqref{CCK1},
the covariant derivative gauges the untilded Lie algebra only:
\begin{align}
    \label{gbas-D}
    D_\r F^{a\ta}
    \,=\,
        \partial_\r F^{a\ta}
        + f^a{}_{cb}\mem A^c{}_\r\hem F^{b\ta}
    \,.
\end{align}
Expanding out the covariant derivative in \eqref{CCK1}
by \eqref{gbas-D}
gives
\begin{align}
\begin{split}
    \label{CCK1.expanded}
    \Box\hem F^{a\ta}
    \,=\,
    {}&{}
        f^a{}_{bc}\mem \tf^\ta{}_{\tb\tc}\mem 
        F^{b\tb}\mem F^{c\tc}
        + \BB{
            2f^a{}_{bc}\mem A^b \mdot \partial F^{c\ta}
            + f^a{}_{bc}\mem 
                \partial\mdot\nem A^b\mem
                F^{c\ta}
        }
    \\
    {}&{}
        + f^a{}_{be}\mem f^e{}_{cd}\mem
        (A^b\mdot A^c)\mem F^{d\ta}
    \,,
\end{split}
\end{align}
which explicitly spells out how exactly
the nonlinear wave equation for $F^{a\ta}$
is mismatched from the BAS equation in \eqref{BAS}.
In this sense, 
one finds
an \textit{approximate} notion of CK duality:
\begin{align}
    \label{CCKD.YM}
    \YM(\g)
    \,\,\cong\,\,
    \BAS(\g,\so(1,d\mminus\nem1))
    \text{\,\,with corrections}
    \,.
\end{align}

In Cheung and Mangan \cite{CCK}'s version,
the $A$-field and the $F$-field
are treated as a priori independent objects.
Then \eqref{CCK1} can be interpreted as describing
a ``color-Lorentz'' scalar wave $F^{\a\ta}$
propagating in the background of the $A$-field
while exhibiting cubic self-interaction.
Namely, it is pointed out that 
\eqref{CCK1}
is isomorphic to a partially gauged BAS theory.
This approximate version of CK duality was dubbed
covariant CK duality.

For our purposes,
it could be simpler to
view $F^{a\ta}$ as the sole dynamical field
by taking the $A$-field as a dependent field via 
\eqref{axialgauge-A}.
In $d=4$ dimensions,
the axial gauge condition in \eqref{axialgauge-A}
allows us to split the $A$-field
\textit{linearly} into SD and ASD parts as
\begin{align}
	A^\pm{}^{a\m}
	\,=\,
		\frac{1}{\eta\mdot\partial}\,
            F^{a\ta}\mem 
            (\eta\Sigma^\pm_\ta)^\m
    \qiq
    A^{a\m}
    \,=\,
        A^+{}^{a\m}
        +
        A^-{}^{a\m}
    \,,
\end{align}
where $(\Sigma^\pm_\ta)$ are the SD and ASD Lorentz generators,
and $(\eta\Sigma^\pm_\ta)^\m = \eta_\r (\Sigma^\pm_\ta)^{\r\m}$.
More details will be explored in
\Sec{Jcx>NLIN}.

\subsection{Electric and Magnetic Equations, II}
\label{Jcx>EM2}

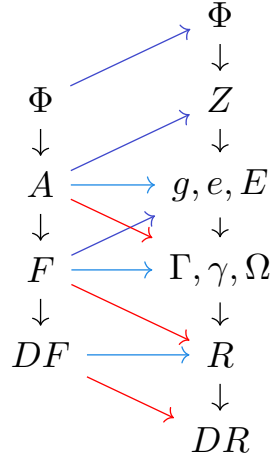
\begin{figure}[t]
    \centering
\begin{align*}
\adjustbox{valign=c,scale=1.25}{
    \adjustbox{valign=t}{\begin{tikzpicture}
        \node[empty] (O) at (0,0) {};
        \node[empty] (X) at (1.9, 0) {};
        \node[empty] (Y) at (0, -.9) {};
        \node[s] (Z0) at ($(O)-1*(Y)+0*(X)$) {$\Phi$};
        \node[s] (00) at ($(O)+0*(Y)+0*(X)$) {$A$};
        \node[s] (10) at ($(O)+1*(Y)+0*(X)$) {$F$};
        \node[s] (20) at ($(O)+2*(Y)+0*(X)$) {$DF$};
        \node[s] (W1) at ($(O)-2*(Y)+1*(X)$) {$\Phi$};
        \node[s] (Z1) at ($(O)-1*(Y)+1*(X)$) {$Z$};
        \node[s] (01) at ($(O)+0*(Y)+1*(X)$) {$g,e,E$};
        \node[s] (11) at ($(O)+1*(Y)+1*(X)$) {$\Gamma,\gamma,\Omega$};
        \node[s] (21) at ($(O)+2*(Y)+1*(X)$) {$R$};
        \node[s] (31) at ($(O)+3*(Y)+1*(X)$) {$DR$};
        \draw[->] (Z0)--(00) {};
        \draw[->] (00)--(10) {};
        \draw[->] (10)--(20) {};
        \draw[->] (W1)--(Z1) {};
        \draw[->] (Z1)--(01) {};
        \draw[->] (01)--(11) {};
        \draw[->] (11)--(21) {};
        \draw[->] (21)--(31) {};
        \draw[indigo,->] (Z0)--(W1) {};
        \draw[indigo,->] (00)--(Z1) {};
        \draw[indigo,->] (10)--(01) {};
        \draw[skyblue,->] (00)--(01) {};
        \draw[skyblue,->] (10)--(11) {};
        \draw[skyblue,->] (20)--(21) {};
        \draw[red,->] (00)--(11) {};
        \draw[red,->] (10)--(21) {};
        \draw[red,->] (20)--(31) {};
    \end{tikzpicture}}
}
\end{align*}
    \caption[%
        Parallels between
        gauge theory and gravity
        in the context of double copy,
        organized
        in terms of the derivative count.
        (a) The SD double copy
        via the second heavenly equation
        (\Sec{Jc>SDG>HEAV})
        climbs up one level.
        (b) The Mason-Newman formulation of SD gravity
        (\Sec{Jc.MN})
        preserves the level.
        (c) The covariant CK duality or Penrose wave equation
        (\Sec{Jcx>EM2})
        step down by one level.
        For classical double copies,
        the SD double copy in terms of 
        Tod's construction
        describes case (a),
        the KS double copy describes case (b),
        and
        the Weyl double copy describes case (c).
        Recall also \fref{mtwsequence},
        the structure of electromagnetism and geometrodynamics in outline form.
    ]{%
        Parallels between
        gauge theory and gravity
        in the context of double copy,
        organized
        in terms of the derivative count.
        (a) The SD double copy
        via the second heavenly equation
        (\Sec{Jc>SDG>HEAV})
        climbs up one level.
        (b) The Mason-Newman formulation of SD gravity
        (\Sec{Jc.MN})
        preserves the level.
        (c) The covariant CK duality or Penrose wave equation
        (\Sec{Jcx>EM2})
        step down by one level.
        For classical double copies,
        the SD double copy in terms of 
        Tod's theorem \cite{Tod:1982mmp,AAS}
        describes case (a),
        the KS double copy \cite{monteiro2014black} describes case (b),
        and
        the Weyl double copy \cite{Luna:2018dpt,Chacon:2021wbr} describes case (c).
        Recall also \fref{mtwsequence},
        the structure of electromagnetism and geometrodynamics in outline form.
    }
    \label{dersequence}
\end{figure}

In \Sec{Ja>EM1},
we identified the electric and magnetic equations
of pure GR
as Ricci flatness and the algebraic Bianchi identity.
In terms of the number of derivatives on the potential field,
for instance,
this identification
aligned with
the electric and magnetic equations of gauge theory,
\eqref{YM.cem}.

In \Sec{SDYM},
this parallel was reviewed 
and was refashioned into
alternative proposals in 
\Secs{Jc.MN}{Jc>SDG>HEAV},
whose validity is yet restricted within the SD sector.
As illustrated in \fref{dersequence},
these proposals reduce or preserve the derivative count
(if the gauge potential $A$ and the metric $g$ are declared as carrying derivative count zero).

However, 
an indisputable parallel between gauge theory and gravity
that shall be also seen as classic
is the obvious parallel between
the connection $A$ and the Levi-Civita connection $\Gamma$,
or
the curvature $F$ and the curvature $R$.
In particular,
this parallel is the very insight 
encoded in
Ashtekar variables \cite{Ashtekar:1986yd}
whose covariant incarnation is
{\Pleb} gravity reviewed in \Sec{Jcx>COFRAME}.

In this light,
the electric and magnetic equations of gauge theory
in \eqref{YM.cem}
shall be paralleled with
covariant first-order differential equations on the Riemannian curvature.
It can be shown that
the Riemannian curvature satisfies
\begin{align}
    \label{GR.cem}
    \nabla^\r R^{\m\n}{}_{\r\s}
    \,=\,
        0
    \,,\quad
    \nabla_\wrap{[\k} R^{\m\n}{}_\wrap{\r\s]}
    \,=\,
        0
    \,,
\end{align}
in generic $d$ spacetime dimensions.
\eqref{GR.cem} provides our second version
of the \textit{electric and magnetic equations of GR}.

The magnetic equation in \eqref{GR.cem}
is the differential Bianchi identity in \eqref{Rbianchis}.
The electric equation in \eqref{GR.cem} is implied by
the index symmetry in \eqref{Rindexsym},
the differential Bianchi identity,
and 
Ricci flatness:
\begin{align}
	\nabla^\r R^{\m\n}{}_{\r\s}
	\,&=\,
		- \nabla^\m R^{\n\r}{}_{\r\s}
		- \nabla^\n R^{\r\m}{}_{\r\s}
	\,=\,
		\nabla^\m \Ric^\n{}_\s
		- \nabla^\n \Ric^\m{}_\s
	\,=\,
		0
	\,.
\end{align}

Again,
a second-order equation on
the curvature tensor
can be derived by combining 
the electric and magnetic equations in \eqref{GR.cem}.
We implement the following gymnastics:
\begin{align}
	\label{CCK.GR.deriv}
    &
	g^{\k\l}\mem \nabla_\k \nabla_{\nem\l}
		R^\m{}_\n{}^\r{}_\s
    \\
	\,&=\,
		\nabla^\l
		\BB{
			 \nabla_{\nem\s} R^\m{}_\n{}^\r{}_\l
			 - (\r \leftrightarrow \s)
		}
    \nonumber
	\,,\\
	\,&=\,
		\BB{
		 	  R^\m{}_\k{}^\l{}_\s R^\k{}_\n{}^\r{}_\l
  		 	- R^\m{}_\k{}^\r{}_\l R^\k{}_\n{}^\l{}_\s 
		 	+ R^\r{}_\k{}^\l{}_\s R^\m{}_\n{}^\k{}_\l
			- R^\m{}_\n{}^\r{}_\k R^\k{}_\l{}^\l{}_\s
		}
		 - (\r \leftrightarrow \s)	
	\nonumber
	\,,\\
	\,&=\,
		2\mem \BB{
 	  		R^\m{}_\k{}^\l{}_\s R^\k{}_\n{}^\r{}_\l
		 	- R^\m{}_\k{}^\r{}_\l R^\k{}_\n{}^\l{}_\s 
		}
		 	+ R^\l{}_\k{}^\r{}_\s R^\m{}_\n{}^\k{}_\l
	\nonumber
	\,,\\
	\,&=\,
		- 2\mem \BB{
 	  		R^\m{}_\k{}^\r{}_\l R^\k{}_\n{}^\l{}_\s 
		 	- R^\m{}_\k{}^\l{}_\s R^\k{}_\n{}^\r{}_\l
		}
		 	+ R^\m{}_\n{}^\k{}_\l R^\l{}_\k{}^\r{}_\s
	\,.
	\nonumber
\end{align}
The first equality uses the magnetic equation in \eqref{GR.cem}.
The second equality uses the electric equation in \eqref{GR.cem}.
The third equality uses the algebraic Bianchi identity and Ricci flatness.

\eqref{CCK.GR.deriv}
is known as
the Penrose wave equation
\cite{Penrose:1960eq,Ryan:1974nt}.
By installing an orthonormal frame $E^\m{}_A$,
\eqref{CCK.GR.deriv} boils down to
\begin{align}
\begin{split}
    \label{CCK.GR}
	D^2 R^A{}_B{}^C{}_D
	\,=\,
        {}&{}
		{-2}\mem \BB{
 	  		R^A{}_E{}^C{}_F R^E{}_B{}^F{}_D
		 	- R^A{}_E{}^F{}_D R^E{}_B{}^C{}_F
		}
    \\
        {}&{}
        + R^A{}_B{}^E{}_F R^F{}_E{}^C{}_D
	\,.
\end{split}
\end{align}
Again,
by identifying the antisymmetrized pair of indices $[AB]$ with a Lorentz Lie algebra index $\ta$,
we view the curvature tensor $R^{ABCD} = R^{[AB][CD]} = R^{[CD][AB]}$
as a symmetric BAS field $R^{\ta\tb} = R^{\tb\ta}$
that takes 
the Lorentz Lie algebra
as both of its Lie algebras:
\begin{align}
    \g 
    \,=\,
        \tg
    \,=\,
        \so(1,d-1)
    \,.
\end{align}
With this understanding,
\eqref{CCK.GR} boils down to
\begin{align}
	\label{GGBAS}
	- D^2
	R^{\ta_1\ta_2}
	\,=\,
		\tf^{\ta_1}{}_{\tb_1\tc_1}\mem \tf^{\ta_2}{}_{\tb_2\tc_2}\,
			R^{\tb_1\tb_2}\mem R^{\tc_1\tc_2}
		\mem+\mem
        2\mem R^{\ta_1\td}\mem R_\td{}^{\ta_2}
	\,,
\end{align}
where
$\eta^{AB} D_A D_B$
utilizes the \textit{flat} metric,
and
the covariant derivative describes
\begin{align}
    \label{ggbas-D}
	D_C \phi^{\ta}
	\,=\,
		E^\r{}_C\mem
			\partial_\r \phi^{\ta}
        + f^\ta{}_{\tc\tb}\,
            \gamma^\tc{}_C\, \phi^{\tb}
    \,,
\end{align}
where $\gamma^\ta{}_C$ is the spin connection coefficient,
$\gamma^{AB}{}_C$.

Some crucial clarifications shall be made.
First,
        in \eqref{GGBAS},
        all indices are gauged.
        The tilded color indices
        are local Lorentz indices
        coupled to the spin connection as a Lorentz-valued connection $A^\ta{}_\r$.
Second,
        the covariant derivative in \eqref{ggbas-D}
        differs from \eqref{gbas-D}
        because it is dressed by the vielbein $E^\r{}_C$.\footnote{
            When expanded around a flat background,
            this gives a diffeomorphism connection component;
            see \Chap{K2}.
        }
        That is,
        the covariant derivative in \eqref{ggbas-D}
        does not map to the covariant derivative in \eqref{gbas-D}
        by the mere replacement of Lorentz with color.
With these understandings,
\eqref{GGBAS}
may be seen as a doubly gauged instance of a symmetric BAS theory.\footnote{
    Note that Cheung and Mangan \cite{CCK}'s covariant CK duality
    instead concerned Einstein-YM theory,
    not pure GR.
    Nevertheless, for our purposes here
    we study the Penrose wave equation in \eqref{GGBAS}
    as the nonlinear wave equation for GR.
}

Again,
\eqref{GGBAS}
provides an approximate idea for double copy:
\begin{align}
    \label{CCKD.GR}
    \GR
    \,\,\cong\,\,
    \BAS(\so(1,d\mminus\nem1),\so(1,d\mminus\nem1))
    \text{\,\,with corrections}
    \,.
\end{align}
The mismatch, however, is more serious than the YM case in \eqref{CCKD.YM}.
Besides the subtle difference in the covariant derivative
and the doubly gauged nature,
the term 
$2\mem R^{\ta_1\td}\mem R_\td{}^{\ta_2}$
in \eqref{GGBAS}
explicitly ruins index factorization.
Namely, even with the covariantization,
the right-hand side of \eqref{GGBAS}
does not take the form of
the right-hand side of the BAS equation in \eqref{BAS}.

Despite this mismatch,
the Penrose wave equation
in the form of \eqref{GGBAS}
offers an interesting perspective on GR.
It reimagines gravitational waves as
``Lorentz-Lorentz'' waves,
exhibiting cubic and higher-order
self-interactions
in terms of \eqref{GGBAS}.
It facilitates a suggestive parallel with YM theory
as a theory of ``color-Lorentz'' waves.

\subsection{Nonlinear Superposition of Self-Dual and Anti-Self-Dual Modes}
\label{Jcx>NLIN}

In \Secs{Jcx>CCK}{Jcx>EM2},
we have reviewed the
covariant wave equations for the curvature field
in YM theory and GR
\cite{CCK,Penrose:1960eq,Ryan:1974nt}.
We learned that they describe
nonlinear waves exhibiting
color and Lorentz polarizations:
\begin{subequations}
\label{FRwaves}
\begin{align}
    \label{Fwave}
    \YM(\g):\quad
    \qquad\qquad
    \mathllap{-D^2 F^{a\ta}}
    \,&=\,
        f^a{}_{bc}\mem \tf^\ta{}_{\tb\tc}\mem 
            F^{b\tb}\mem F^{c\tc}
    \,,\\
    \label{Rwave}
    \GR:\quad
    \qquad\qquad
	\mathllap{- D^2 R^{\ta\tb}}
	\,&=\,
		\tf^{\ta}{}_{\tc\tilde{e}}\mem 
        \tf^{\tb}{}_{\td\tf}\mem
			R^{\tc\td}\mem R^{\tilde{e}\tf}
		\mem+\mem
        2\hem R^{\ta\tc}\mem R_\tc{}^{\tb}
	\,.
\end{align}
\end{subequations}
The purpose of this last subsection
is to concretely demonstrate that
the above equations
uniquely determine
the nonlinear superposition between
plane waves in YM theory and GR
via BG recursion.
Crucially, we take the curvature fields as fundamental
by employing the axial gauge.
We investigate the nonlinear superposition of two plane waves
at the leading order in coupling.

\paragraph{Nonlinear Superposition of Yang-Mills Waves}

We investigate the nonlinear superposition of plane waves in YM theory
as color-Lorentz waves.

The perturbative solution to 
\eqref{Fwave} takes the form
\begin{align}
\begin{split}
    \label{Fwave:nlin}
    F^{a\ta}
    \,=\,
    {}
        &
        \frac{ig}{\sqrt{2}}\,
            \sum_{i}\:
                \e_i\,
                c_i^a\hem \vf_i^\ta
            \,
                \mathe^{ik_ix}
        \:-\:
        g^2\,
            \sum_{i,j}
                \e_i\hhem \e_j\,
                \frac{
                    c_{ij}^a\hem \vf_{ij}^\ta
                }{
                    {{k_{ij}}^2}
                }
            \,
                \mathe^{ik_{ij}x}
        \:+\:
        \O(g^3)
    \,.
\end{split}
\end{align}
At the linearized level,
$k_i{}_\m$, $c_i^a$, and $\vf_i^\ta$,
are the momentum, color polarization, and Lorentz polarization of the $i$\textsuperscript{th} color-Lorentz plane wave.
The on-shell condition imposes
\begin{align}
    {k_i}^2 \,=\, 0
    \,,\quad
    k_i^\m\mem \varphi_i{}_{\m\n} \,=\, 0
    \,,\quad
    k_i{}_\wrap{[\r}\hem \varphi_i{}_\wrap{\m\n]}
    \,=\,
        0
    \,,\quad
    \varphi_i{}_\m{}^\r\mem \varphi_i{}_{\r\n}
    \,=\,
        0
    \,.
\end{align}
The Lorentz polarization $\varphi_i^{\m\n}$
recasts
the usual vector polarization $e_i^\m$
in a gauge-invariant way.
It
will also be called the bivector polarization.
Here, we work in the convention 
\begin{align}
	\varphi_i^\ta\, (\Sigma_\ta)_{\m\n}
	\,=\,
    	\varphi_i{}_{\m\n}
	\,=\,
		\sqrt{2}\mem (k_i \swedge e_i)_{\m\n}
	\,.
\end{align}

By imposing the axial gauge condition
in \eqref{axialgauge-A},
we pursue a ``flipside view''
where the color potential $A^{a\m}$ is a nonlocal functional of the curvature field $F^{\a\ta}$.
For simplicity, consider two incident plane waves
labeled with $3$ and $4$.
By plugging in \eqref{Fwave:nlin}
to \eqref{Fwave}
and examining terms at the order $\O(g^2)$,
the perturbative solution
is determined up to $\O(g^2)$ as
\begin{align}
	\label{F34}
	F^{a\ta}
	\,=\,
		{}&{}
		\frac{ig}{\sqrt{2}}\,
			c_3^a\, \varphi_3^{\ta}\,
				\mathe^{ik_3x}
		+
		\frac{ig}{\sqrt{2}}\,
			c_4^a\, \varphi_4^{\ta}\,
				\mathe^{ik_4x}
    \\
		{}&{}
		-
		g^2\,
			\comm{c_3}{c_4}^a\mem
			\bb{
				\comm{\varphi_3}{\varphi_4}^{\ta}
				- \varphi_3^{\ta}\,
					\frac{
						\eta\hhem\varphi_4\hnem k_3
					}{
						k_4\eta
					}
				+ \varphi_4^{\ta}\,
					\frac{
						\eta\hhem\varphi_3\hnem k_4
					}{
						k_3\eta
					}
			}\mem
				\frac{1}{{k_{34}}^2}\,
				\mathe^{ik_{34}x}
		+ \O(g^3)
        \,,
    \nonumber
\end{align}
where $k_{34} := k_3 + k_4$,
$\comm{c_3}{c_4}^a := f^a{}_{bc}\mem c_3^b\hem c_4^c$,
and
$\comm{\varphi_3}{\varphi_4}^{\ta} := \tf^\ta{}_{\tb\tc}\mem \varphi_3^\tb\hem \varphi_4^\tc$.
Accordingly,
the color potential 
$A^a{}^\m$
is determined as
\begin{align}
	\label{A34}
		&{}
		\frac{g}{\sqrt{2}}\,
			c_3^a\, 
			\frac{(\eta\varphi_3)^\m}{k_3\eta}
				\,\mathe^{ik_3x}
		+
		\frac{g}{\sqrt{2}}\,
			c_4^a\, 
			\frac{(\eta\varphi_4)^\m}{k_4\eta}
				\,\mathe^{ik_4x}
		\\
		&{}
		+i
		g^2\,
			\comm{c_3}{c_4}^a\mem
			\bb{
				\comm{\varphi_3}{\varphi_4}^{\ta}
				- \varphi_3^{\ta}\,
					\frac{
						\eta\hhem\varphi_4\hnem k_3
					}{
						k_4\eta
					}
				+ \varphi_4^{\ta}\,
					\frac{
						\eta\hhem\varphi_3\hnem k_4
					}{
						k_3\eta
					}
			}\mem
				\frac{(\eta\Sigma_\ta)^\m}{k_{34}\eta}
				\mem
				\frac{1}{{k_{34}}^2}
				\,\mathe^{ik_{34}x}
		+ \O(g^3)
		\,,
\nonumber
\end{align}
where 
$(\eta\varphi_i)^\m = \eta_\r \varphi_i{}_{\r\m}$.
When listing these solutions,
we have replaced the formal parameters $\e_i$ with $1$
for simplicity.

\eqref{F34} is an explicit description of
how two plane waves in YM theory 
interact nonlinearly.
It describes that two incident plane waves
create a new plane wave at the next order in the coupling,
whose color and Lorentz polarizations can be
respectively
$c_{34}^a = \comm{c_3}{c_4}^a$
and
\begin{align}
    \label{vf34.YM}
    \varphi_{34}^\ta
    \,=\,
        \comm{\varphi_3}{\varphi_4}^{\ta}
        - \varphi_3^{\ta}\,
            \frac{
                \eta\hhem\varphi_4\hnem k_3
            }{
                k_4\eta
            }
        + \varphi_4^{\ta}\,
            \frac{
                \eta\hhem\varphi_3\hnem k_4
            }{
                k_3\eta
            }
    \,,
\end{align}
up to stripping off the pole $1/{k_{34}}^2$.
The Lie bracket term in \eqref{vf34.YM}
arises from
the double Lie bracket structure
in the right-hand side of \eqref{Fwave},
while the reference-dependent terms
trace back to
the covariantization involved in
the left-hand side.

\paragraph{Color-Lorentz Waves in Helicity Basis}

In $d=4$ dimensions,
we can specialize \eqref{vf34.YM}
into two subcases:
same-helicity configuration,
$(3^+,4^+)$,
and
opposite-helicity configuration,
$(3^+,4^-)$.
Crucially, the Lorentz polarizations
cleanly split into SD and ASD parts,
facilitating a systematic examination of the nonlinear superposition in the helicity (more precisely chirality) basis.

For the same-helicity configuration
$(3^+,4^+)$,
we see that all contributions in \eqref{vf34.YM} survive,
and the resulting $\vf_{34}^\ta$
also exhibits positive helicity:
\begin{align}
    \label{collision34.YM++}
    \varphi^+_{34}
    \,=\,
        \comm{\varphi^+_3}{\varphi^+_4}
        - \varphi^+_3\,
            \frac{
                \eta\hhem\varphi_4\hnem k_3
            }{
                k_4\eta
            }
        + \varphi^+_4\,
            \frac{
                \eta\hhem\varphi_3\hnem k_4
            }{
                k_3\eta
            }
    \,,\quad
    \varphi^-_{34}
    \,=\,
        0
    \,.
\end{align}
This is because the Lorentz algebra
$\tg = \so(1,3)$
decomposes into the SD and ASD sectors,
each of which is closed by itself.

For the opposite-helicity configuration,
we see that
the Lorentz interaction term in \eqref{vf34.YM} 
vanishes:
$\comm{\varphi_3^+}{\varphi_4^-} = 0$.
Again, this is because the SD and ASD sectors of the Lorentz algebra
do not talk to each other.
As a result,
we find that
SD and ASD color-Lorentz waves
are \textit{invisible} to each other
in the sense of
the right-hand side of \eqref{Fwave}.
Yet, 
the left-hand side effect
(contributions from covariantization)
survives
to give
\begin{align}
    \label{collision34.YM+-}
    \varphi_{34}^+
    \,=\,
        - \varphi_3^+\,
            \frac{
                \eta\hhem\varphi^-_4\hnem k_3
            }{
                k_4\eta
            }
    \,,\quad
    \varphi_{34}^-
    \,=\,
        \varphi_4^-\,
            \frac{
                \eta\hhem\varphi^+_3\hnem k_4
            }{
                k_3\eta
            }
    \,.
\end{align}
Hence the collision of SD and ASD waves
produces both SD and ASD modes.

Overall,
we find two types of cubic interaction vertices
in the helicity basis
at this order:
$(+,+,-)$ and $(-,-,+)$,
in the all-incoming convention.
Surely, this is consistent with a well-known fact about YM theory.
Explicitly, one should be able to find the three-point gluon amplitudes
by implementing a proper LSZ reduction recipe
on the one-point function.

To reiterate,
note how the perturbation theory
in which the curvature is the foundation of all
systematically facilitates viewing YM theory as
a nonlinearly interacting system of SD and ASD gluons.
In {\Pleb} and Newman's language in \fref{fig:heaven-earth},
we just have witnessed
how
heaven and anti-heaven
meet and 
create
earth.

\paragraph{Nonlinear Superposition of Gravitational Waves}

We investigate the nonlinear superposition of gravitational waves in GR
as Lorentz-Lorentz waves.

The perturbative solution to 
\eqref{Rwave} takes the form
\begin{align}
\begin{split}
    \label{Rwave:nlin}
    R^{\ta\tb}
    \,=\,
    {}
        &
        \frac{\k}{2}\,
            \sum_{i}\:
                \e_i\,
                \vf_i^\ta\vf_i^\tb
            \,
                \mathe^{ik_ix}
        \:+\:
        \frac{\k^2}{2}\,
            \sum_{i,j}
                \e_i\hhem \e_j\,
                \frac{
                    \vf_{ij}^{\ta\tb}
                }{
                    {k_{ij}}^2
                }
            \,
                \mathe^{ik_{ij}x}
        \:+\:
        \O(\k^3)
    \,,
\end{split}
\end{align}
the linearized part of which describes a superposition of Lorentz-Lorentz plane waves
characterized by
momentum $k_i{}_\m$ and Lorentz polarization $\vf_i^\ta$.
The on-shell condition imposes
${k_i}^2 \eqq 0$,
$k_i^A\mem \varphi_i{}_{AB} \eqq 0$,
$k_i{}_\wrap{[C}\hem \varphi_i{}_\wrap{AB]} \eqq 0$,
and
$\varphi_i{}_A{}^C \varphi_i{}_{CB} = 0$,
where $A,B,\cdots$ are flat indices.
At the quadratic order (and beyond),
our perturbation theory would not
manifestly factorize
$\vf_{ij}^{\ta\tb}$
into a tensor product of two Lorentz polarizations,
in general.

\eqref{Rwave:nlin}
represents the nonlinear superposition of
gravitational waves
in terms of the curvature tensor.
In textbook approaches,
gravitational waves and their nonlinear interaction
are described in terms of 
the metric perturbation
or the gothic metric perturbation
$h^{\m\n} = \got^{\m\n}\nem\hnem - \eta^{\m\n}$;
see \cite{PoissonWill:2014}, for instance.
However,
we would like to see if
a perturbation theory is viable
in which the curvature field
is fundamental and is
directly governed by
the Penrose wave equation in \eqref{Rwave}.
Again, a charm of this pathway
is that
a systematic decomposition into SD and ASD modes
will be possible in $d=4$ dimensions.
The symmetric field $h^{\m\n}$
does not split nicely into SD and ASD parts.

\medskip

To this end,
we devise an \textit{axial gauge prescription
for tetradic Palatini gravity}.
We have found that the following recipe
can be reasonable.
First,
we must stipulate that
the coframe $e^A{}_\m$ exhibits 
a nondegenerate vacuum expectation value $\delta^A{}_\m$,
the components of which 
can simply describe
the Kronecker delta.
Second,
we
employ a reference \textit{vector}
whose components $\eta^\m$ in the spacetime frame
are constants.
For clarity,
its index will not be lowered
by any means.
On 
the spin connection one-form $\gamma^\ta{}_\m$
and
the \textit{coframe} $e^A{}_\m$,
we impose
\begin{align}
    \label{axialgravity}
	\gamma^\ta{}_\m\mem \eta^\m \,=\, 0
	\,,\quad
	e^A{}_\m\mem \eta^\m
	\,=\,
		\delta^A{}_\m\hem \eta^\m
	\,,
\end{align}
where the vacuum expectation value 
$\delta^A{}_\m$
is exploited as a reference coframe.
These stipulations together
determine $\gamma^\ta{}_\m$ and the frame $E^\m{}_A$ as
\begin{subequations}
\label{axialgravitysol}
\begin{align}
\label{axialgravitysol.A}
	\gamma^\ta{}_\n
	\,&=\,
		\frac{1}{\eta\mdot\partial}\,
		\BB{
			R^{\ta\tb}\,
			(\Sigma_\tb)_{AB}\mem
				(\delta^A{}_\m\hem \eta^\m)
				\mem
				e^B{}_\n
		}
	\,,\\
\label{axialgravitysol.E}
	E^\m{}_B
	\,&=\,
		\delta^\m{}_B
		+ 
		\frac{1}{\eta\mdot\partial}\,
		\BB{
			E^\m{}_A\mem
			\gamma^A{}_C{}_\n\mem
            (\delta^C{}_\m\hem \eta^\m)\mem
			E^\n{}_B
		}
	\,,
\end{align}
\end{subequations}
which follow from the relations
\begin{align}
\begin{split}
	R^{\ta\tb}\mem
	(\Sigma_\tb)_{CD}\mem
	e^C{}_\r\mem e^D{}_\s
	\,&=\,
		\partial_\r \gamma^\ta{}_\s
		- \partial_\s \gamma^\ta{}_\r
		+ \tf^\ta{}_{\tb\tc}\mem \gamma^\tb{}_\r\mem \gamma^\tc{}_\s
	\,,\\
	E^\m{}_\wrap{B,[\r}\mem e^B{}_\wrap{\s]}
	\,&=\,
		E^\m{}_A\mem
		\gamma^A{}_B{}_\wrap{[\r}\,
		e^B{}_\wrap{\s]}
	\,,
\end{split}
\end{align}
which are \eqrefs{Rform}{gamma-def}.
In \eqref{axialgravitysol.E},
$\delta^\m{}_B$
is the inverse of $\delta^B{}_\m$.

\eqref{axialgravitysol}
determines the spin connection,
the frame, and the coframe
as nonlocal functionals of the orthonormal-frame curvature $R^{\ta\tb}$,
around the flat background
$\delta^A{}_\m$.
In this way,
a flipside view is provided on
tetradic Palatini gravity:
curvature as the foundation of all.

By plugging in \eqref{axialgravitysol}
to
\eqref{Rwave},
imposing the ansatz in \eqref{Rwave:nlin},
and solving the $\O(\k^2)$ part,
it can be found that
\begin{align}
\begin{split}
    \label{vf34.GR}
    \vf_{34}^{\ta\tb}
    \,\,=\,\mem
    {}&{}
        \bb{\displaystyle
            \comm{\vf_3}{\vf_4}
            - \varphi_3\, \frac{\eta\hhem\varphi_4\hnem k_3}{k_4\eta}
        }^{\nem\nem\hnem\ta}
    \mem
        \bb{\displaystyle
            \comm{\vf_3}{\vf_4}
            - \varphi_3\, \frac{\eta\hhem\varphi_4\hnem k_3}{k_4\eta}
        }^{\nem\nem\hnem\tb}
    \\
    {}&{}
        {}+
        \bb{\displaystyle
            \comm{\vf_3}{\vf_4}
            - \varphi_4\, \frac{\eta\hhem\varphi_3\hnem k_4}{k_3\eta}
        }^{\nem\nem\hnem\ta}
    \mem
        \bb{\displaystyle
            \comm{\vf_3}{\vf_4}
            - \varphi_4\, \frac{\eta\hhem\varphi_3\hnem k_4}{k_3\eta}
        }^{\nem\nem\hnem\tb}
    \\
    {}&{}
        {}
        -
            \comm{\vf_3}{\vf_4}^\ta
            \mem
            \comm{\vf_3}{\vf_4}^\tb
        \,+\,
        \varphi_3\mdot\varphi_4\,
            \BB{
                \varphi_3^\ta\mem \varphi_4^\tb
                +
                \varphi_4^\ta\mem \varphi_3^\tb
            }
    \,,
\end{split}
\end{align}
where $\varphi_3\mdot\varphi_4 = \varphi_3{}_\ta\mem \varphi_4^\ta = \tfrac{1}{2}\, \varphi_3^{AB}\mem \varphi_4{}_{AB}$.
Again, we assume two incident gravitational waves
for simplicity,
labeled with $3$ and $4$.

Again, 
the squaring of the factor
$\comm{\vf_3}{\vf_4}$
in \eqref{vf34.GR}
arises by the
double Lie bracket structure
in the right-hand side of \eqref{Rwave}.
We recall the two mismatch factors
that made \eqref{Rwave} deviate from the BAS grammar:
first,
the covariantization of the d'Alembertian (the left-hand side effect),
and second,
the loss of index factorization by an additional term in the right-hand side.
The reference-dependent terms
in \eqref{vf34.GR}
will be attributed to the former
while the term
$\varphi_3\mdot\varphi_4\mem (\vf_3^\ta \vf_4^\tb + \vf_4^\ta \vf_3^\tb)$
will be traced back to the latter.

\paragraph{Lorentz-Lorentz Waves in Helicity Basis}

In $d=4$ dimensions,
we can specialize \eqref{vf34.GR}
into two subcases:
same-helicity configuration,
$(3^+,4^+)$,
and
opposite-helicity configuration,
$(3^+,4^-)$.
Again, this analysis explicates the content of 
the nonlinear superposition of gravity
in the chirality basis.

For the same-helicity configuration
$(3^+,4^+)$,
all contributions in \eqref{vf34.GR} survive,
and the resulting $\vf_{34}^{\ta\tb}$
is also SD on each of its Lorentz indices.

For the opposite-helicity configuration,
the decomposition of the Lorentz algebra implies
$\comm{\vf_3}{\vf_4} = 0$
and $\vf_3 \mdot \vf_4 = 0$.
Hence
\eqref{vf34.GR} boils down to
\begin{align}
\begin{split}
    \label{vf34.GR.oppo}
    \vf_{34}
    \,\,=\,\mem
    {}&{}
        \vf_3^+ \motimes \vf_3^+
        \,
        \bb{\nem
            \frac{\eta\hhem\varphi_4\hnem k_3}{k_4\eta}
        \nem}^{\nem\nem\hnem2}
    \mem+\,
        \vf_4^- \motimes \vf_4^-
        \,
        \bb{\nem
            \frac{\eta\hhem\varphi_3\hnem k_4}{k_3\eta}
        \nem}^{\nem\nem\hnem2}
    \,,
\end{split}
\end{align}
which represents
a combination of
SD and ASD spin-$2$ modes.
Physically,
\eqref{vf34.GR.oppo}
describes that
the collision of SD and ASD gravitational waves
produces
both SD and ASD modes.

Again, we find that
$(+,+,-)$ and $(-,-,+)$ interaction vertices
are encoded in this computation,
in the all-incoming convention.
Surely, this is consistent with 
a well-known fact about GR.
Explicitly, one should be able to find the three-point graviton amplitudes
via a proper LSZ reduction,
which will describe a square of gluon amplitudes.

As a consistency check,
it should also be remarked that
the curvature field $R^{\ta\tb}$
has been subject to algebraic constraints:
\begin{align}
    R^{\ta\tb} \,=\, R^{\tb\ta}
    \,,\quad
    \delta_{\ta\tb}\mem R^{\ta\tb}
    \,=\,
        0
    \,,\quad
    \star_{\ta\tb}\mem R^{\ta\tb}
    \,=\,
        0
    \,.
\end{align}
Here, $\star^\ta{}_\tb$ is the internal Hodge star
on the space of antisymmetric rank-two tensors.
The second condition is by vanishing Ricci scalar,
while the third condition is by algebraic Bianchi.
It can be checked that these conditions are dynamically preserved
and are also enjoyed by 
$\vf_{34}^{\ta\tb}$
in \eqref{vf34.GR}.

The above computation
should
provide a concrete and explicit demonstration
of 
the nonlinearity of GR in the chirality basis.
This gives a quantitative exploration of 
the view
presented in \fref{fig:heaven-earth},
of {\Pleb} and Newman:
four-dimensional
GR as a nonlinearly interacting system of SD and ASD modes.

\paragraph{Curvature as the Foundation of All}

We end with a broad remark.
Consider how electromagnetism is taught in our classrooms.
The very first concepts
are charges, currents, and electromagnetic forces.
This leads to the definition of
electric and magnetic fields
as the physical fields that we directly experience and measure.
The scalar and vector potentials
are then introduced later,
through involving some degree of mathematical sophistication,
and are portrayed as
secondary ``human'' inventions
subject to unphysical redundancies.
In this way,
the field strength $F$
is viewed as more fundamental
than the manmade construct $A$,
the gauge potential:\footnote{
    Of course,
    the celebrated debate
    \cite{aharonov1959significance}
    arises to
    demonstrate a tension between
    manifest locality and manifest gauge invariance.
}
\begin{align}
    \label{pedagogy.EM}
    F
    \quad\rightsquigarrow\quad
    A
    \,.
\end{align}

In comparison,
consider how GR is taught in our classrooms.
The very first concept
is the \textit{metric}.
The metric is the foundation of all,
defining
Christoffel symbols
(the inertial/gravitational forces---acceleration and frame dragging---%
experienced in a local laboratory),
the Riemann curvature
(the tidal forces),
and finally the Ricci and Einstein tensors
(the sources of gravitation):
\begin{align}
    \label{pedagogy.GR0}
    g
    \quad\rightsquigarrow\quad
    \Gamma
    \quad\rightsquigarrow\quad
    R
    \,.
\end{align}
Amusingly, this describes
the exact \textit{opposite} approach
of the pedagogy pursued for electromagnetism.
Geometrically, the metric seems to be a fundamental measurable quantity, i.e., length.
Physically, however,
it is also the potential field for gravitation.

As an experiment,
it is interesting to envision
an alternative approach to teaching GR
that parallels the pedagogy of electromagnetism in \eqref{pedagogy.EM}:
\begin{align}
    \label{pedagogy.GR}
    R
    \quad\rightsquigarrow\quad
    \Gamma
    \quad\rightsquigarrow\quad
    g
    \,.
\end{align}
In this flipside view,
one starts with
stress-energies
and the measurement of
tidal forces.
While taking the curvature as fundamental,
the Christoffel symbols
are introduced as a secondary construct,
from which
the metric is eventually reconstructed.\footnote{
    If GR was taught in this way,
    the gravitational Aharonov-Bohm (AB) effect
    \nomenclature{AB}{Aharonov-Bohm}
    where
    the proper time
    gives the quantum phase of a particle
    \cite{Linet:1976vw,Stodolsky:1978ks,Hohensee:2011yt,Colella:1975dq,Overstreet:2021hea}
    (see also \rrcite{Alfonsi:2020lub,Alawadhi:2021uie})
    would be perceived as more shocking.
    The AB effect \cite{aharonov1959significance}
    gives surprise because
    it seems to necessitate the gauge potential $A$, 
    seen as a human invention.
    Length and time, in contrast,
    are usually not viewed as human inventions.
}

Of course, this creative attempt
can turn out to be a terrible idea
in reality,
given the mathematical subtleties and complexities.
However,
suppose
length was somehow not an everyday concept
straightforwardly available to human intuition.
If length was not 
an easy, visual, and tangible concept
to process,
then the only way to discover the metric
could be
starting from the tidal forces
and then inventing secondary constructs in order.

Notably,
the axial gauge formalism established 
in \eqrefs{axialgravity}{axialgravitysol}
has concretely conducted this experiment.
There may be more elegant 
and geometrically more interesting
approaches,
representing the frame and spin connection
with different nonlocal functionals.
Some relevant ideas could be
Wilson lines,
gravitational dressing,
``dynamical'' lightcone gauge,
etc.,
to record.

\section{ Ending Remarks}

In this chapter,
we reviewed, derived, or envisioned
various formulations of GR
in general dimensions
and in four dimensions,
at the EoM level or at the action level.
Certainly, this is not an exhaustive exploration on the subject;
see, e.g., \rcite{Krasnov:2020lku}.
The fascinating mystery of double copy
motivated and guided us
throughout this journey
as a concrete 
anchor.
We have encountered a vast pool of options
for the field space for gravity:
a metric or its densitization,
a coframe (Lorentz gauge theory),
a frame (teleparallel gravity),
a single scalar field
({\Pleb} heavenly equation),
area elements
({\Pleb} gravity),
and the
curvature as a gauged BAS field
(Penrose wave equation).

Gravitation was once a manifestation of
a dynamical metric,
but now we are presented with a decentralized view.
None of the formulations,
however,
seem to explain the double copy.
It is unclear when
the sought-after manifest double copy formulation of GR,
if it exists,
will make its entrance onto the scene
and provide a new invariant center.

One might have to perform
field redefinitions
that are
more aggressively nonlocal.\footnote{
    A related idea is loop space \cite{Polyakov:1980ca}.
}
Surely,
twistor space
\cite{Penrose:1974di,penrose1975aims,Penrose:1980hi,penrose1987origins}
has been an aspiration.
Given that the double copy exists 
beyond the SD sector
and
in general $d$ dimensions,
the ambitwistor approaches 
\cite{Witten:1978xx,Isenberg:1978kk,Yasskin:1982dv,lebrun1985ambi,Mason:2005kn}
can be more relevant.
Ambitwistor space is the space of complexified null geodesics
in curved spacetime
\cite{Mason:2013sva}.
The ambitwistor string models of \rrcite{adamo-ym,adamo-gr}
have provided an explanation for
the Cachazo-He-Yuan formulae \cite{Cachazo:2013gna,Cachazo:2013hca},
which
represent scattering amplitudes of massless particles
as moduli space integrals of 
a factorized integrand.
This factorization yields concrete expressions for 
the KLT \cite{KLT} version of double copy
\cite{Geyer:2022cey}.
For the BCJ \cite{BCJ1,BCJ2} version of double copy,
however, it has been emphasized that
a purely field-theoretic explanation
would be desirable.
It could be interesting to see if
the worldsheet constructions of 
\rrcite{adamo-ym,adamo-gr}
have worldline counterparts
\cite{Shi:2021qsb,Bonezzi:2025bgv,Bonezzi:2024emt,Bonezzi:2020jjq}.
In this case the kinematic algebra $\gk$
may be realized as the Poisson algebra
of ambitwistor space
\cite{dpb}.

The frame formulations of GR
apparently do not confirm that $\gk$
is the diffeomorphism algebra of spacetime.
Still, 
explorations of the frame formulations
could turn out to be meaningful,
as diffeomorphism algebras
seem to be the best (and perhaps the only)
candidates for kinematic algebras
regarding the double copy.
On a related note,
our demonstration of the 
nonlinear interaction between SD and ASD modes
in \Sec{Jcx>NLIN}
was based on the Lorentz gauge theory picture.
In the teleparallel picture,
Grant \cite{Grant:1994vp}
pursued
an ambitwistor construction for four-dimensional gravity
that generalizes the Mason-Newman construction in
\eqref{DYMV.e},
finding an equation of the schematic form
\begin{align}
    \comm{E^A}{\comm{E_A}{E_B}}
    \,\,\sim\,\,
        \Omega^+\mem \Omega^-
    \,,
\end{align}
where $\Omega^\pm$
are the SD and ASD components of
the anholonomy $\Omega$
(by applying the internal Hodge star on its naturally antisymmetric pair of flat indices).
Certainly, $\Omega^\pm$ encode the SD and ASD components of the ``diffeomorphism field strength.''
More explorations will be given in \Chap{K2}.

\chapter{Generalized Symmetries in Dynamical Gravity}
\label{K1}

We explore generalized symmetry in the context of nonlinear dynamical gravity. Our basic strategy is to transcribe known results from YM theory directly to gravity via the tetrad formalism, which recasts general relativity as a gauge theory of the local Lorentz group.
By analogy, we deduce that gravity exhibits a one-form symmetry implemented by an operator $U_\cen$ labeled by a center element $\cen$ of the Lorentz group and associated with a certain area measured in Planck units. 
The corresponding charged line operator $W_\rep$ is the holonomy in a spin representation $\r$, which is the gravitational analog of a Wilson loop. 
The topological linking of $U_\cen$ and $W_\rep$ has an elegant physical interpretation in classical gravitation: the former materializes a tetradic version of a cosmic string whose quantized spin precession angle is measured by the latter.
Notably, our conclusions imply that the standard model exhibits a new symmetry of nature at scales below the lightest neutrino mass. More generally, the absence of global symmetries in quantum gravity suggests that the gravitational one-form symmetry is either gauged or explicitly broken; 
the latter mandates the existence of fermions. Finally, we comment on various generalizations,
such as magnetic higher-form or
higher-group gravitational symmetries.

\begin{fullnote}
    Contents of this chapter are adapted from 
    \rcite{GenSymGrav}, \fullcite{GenSymGrav}.
\end{fullnote}

\section{ Introduction}

Symmetry has long been a vital tool for investigating complex physical systems, particularly at strong coupling.
Historically, most efforts in this expansive subject have focused on conventional symmetries, which act on \textit{local} operators.  The standard model of physics exhibits numerous exact and approximate symmetries of this type, for example relating to charge in electromagnetism and chiral symmetry in the strong interactions.

In the past decade, however, the fundamental concept of symmetry has broadened considerably \cite{Alford:1991vr,Alford:1990fc,Alford:1992yx,Bucher:1991bc,Pantev:2005rh,Pantev:2005wj,Pantev:2005zs,Hellerman:2006zs,Nussinov:2009zz,Aharony:2013hda}. As described in the seminal work \cite{Gaiotto:2014kfa}, it is now understood that the traditional formulation of symmetry is actually the tip of a colossal iceberg.  Rather, there exists a rich patchwork of so-called higher-form symmetries whose distinguishing feature is that they act intrinsically on extended objects described by \textit{nonlocal} operators supported on lines, surfaces, and membranes.   Since higher-form symmetries act trivially on local operators, their physical implications are sometimes quite subtle to diagnose.  From this point of view, the standard symmetries found in most QFT textbooks are brusquely relegated to the special case of zero-form symmetry.

The growing body of work on generalized symmetries has revealed new perspectives on a broad spectrum of assorted phenomena in QFT, including phase transitions \cite{Iqbal:2021rkn,Wen:2018zux,Levin:2004mi,Levin:2004js,Hastings:2005xm,Shimizu:2017asf}, anomalies \cite{Choi:2022jqy,Cordova:2022ieu,Gaiotto:2017yup,Wan:2018zql,Delacretaz:2019brr,Hsin:2018vcg,Tanizaki:2017mtm,Cordova:2019bsd,Wan:2018djl,Cordova:2019jnf,Cordova:2019uob,Delmastro:2022pfo}, and symmetry breaking \cite{Kovner:1992pu,Hofman:2018lfz,Lake:2018dqm,Sogabe:2019gif,GarciaEtxebarria:2022jky}. Recent work has even explored new opportunities for physics beyond the standard model, for example in the context of flavor physics \cite{Cordova:2022qtz}, neutrinos \cite{Cordova:2022fhg}, and axions \cite{Hidaka:2020iaz,Hidaka:2020izy,Brennan:2020ehu,Choi:2022fgx,Yokokura:2022alv,Brennan:2023kpw,Choi:2023pdp,Cordova:2023her,Reece:2023iqn,Agrawal:2023sbp}. Such efforts are a welcome development, as they attempt to draw an explicit connection between highly formal developments in mathematical physics and high-energy physics of actual experimental relevance.  That said, the constraints imposed by generalized symmetry on particle physics models tend to be explicable via more conventional means.  This is perhaps not so surprising---these models are easily embedded within renormalizable theories in which all is calculable and there are no surprises to be had or which require explanation.

Gravity, on the other hand, is another story.  Far less is understood about its putative UV completion.
Consequently, the only truly theory-agnostic approach is to retreat to safely low energies, where gravitational dynamics are described universally by an EFT of gravitons on a fixed background, augmented by possible higher-derivative corrections.  For example, see \rrcite{Donoghue:2022eay,Burgess:2003jk} for a review of this perspective.  
The EFT of gravity is clearly a natural target for understanding generalized symmetry in a refreshingly different context.  There has, however,  been relatively little effort in this vein.\footnote{The bulk of work that makes reference to both gravity and generalized symmetries has focused on the implications of swampland conjectures. 
In this picture, one posits a QFT that exhibits certain generalized global symmetries.  The conjectured absence of global symmetries in a theory of quantum gravity %
then imposes constraints on the theory in order to explicitly break or gauge these symmetries.}
Some notable exceptions include interesting recent work studying the higher-form symmetries associated with  parity \cite{McNamara:2022lrw} and topology change \cite{McNamara:2019rup}, as well as generalizations of continuous higher-form symmetries to \textit{linearized gravity} \cite{Hinterbichler:2022agn,Benedetti:2021lxj,Benedetti:2023ipt}.

In this work, we extend the now well-established insights of higher-form symmetry in gauge theory to the effective theory of \textit{nonlinear gravity}  in four-dimensional spacetime.  Our key ingredient is the well-known fact that gravity can itself be recast in gauge theoretic language.   As history would have it, this perspective carries dual meanings.
On the one hand, gravity is a theory of diffeomorphisms, nonlinearly realized by a self-interacting, massless spin-two field.   Since diffeomorphisms are a redundancy, they are on occasion referred to as a gauge symmetry, though colloquially and not in the strict technical sense.
On the other hand, it is well-known that gravity can also be described by a bona fide gauge theory of local Lorentz transformations, which is the so-called Palatini formalism for the tetrad and spin connection. 
Formally, these descriptions are equivalent since gauge symmetry is, after all, pure redundancy and no redundancy is more valid than any other.\footnote{
    At low energies, metric gravity and tetradic Palatini gravity are perturbatively equivalent classically and quantum mechanically since they both reproduce a local EFT of a massless spin-two particle. 
    As is well-known, the dynamics of such a theory is uniquely fixed, up to unknown Wilson coefficients.
}
As we will see, tetradic Palatini gravity is perfectly suited to our purposes because we can work in lockstep analogy with the familiar approach taken in gauge theory. 
When discussing gravity, we work in the phenomenologically relevant case of four spacetime dimensions.
Our conclusions are as follows. 

First and foremost, our central claim is that tetradic Palatini gravity exhibits an electric one-form symmetry described by the center subgroup $Z(\G)$ of the Lorentz group $\G$.
This one-form symmetry depends crucially on the signature and global structure of $\G$.
For example, in 
Euclidean signature the center is nontrivial when we consider $Z(\SO(4)\hnem)=\mathbb{Z}_2$ or $Z(\Spin(4)\hnem)=\mathbb{Z}_2 {\mem\times\mem} \mathbb{Z}_2$, while in Lorentzian signature the center is nontrivial for $Z(\SL(2,\mathbb{C})\hnem)=\mathbb{Z}_2$.  In all cases, these center subgroups have a zero-form symmetry action as various parities on Lorentz vector and spinor indices.

Second, we show how the one-form symmetry of gravity is implemented by a topological symmetry operator $U_\cen$. 
This object is constructed explicitly in terms of the local degrees of freedom
as the exponential of a certain area operator for a closed surface measured in Planck units and labeled by an element of the center $\cen$.  
The symmetry operator $U_\cen$ acts on a line operator $W_\rep$ known as the spin holonomy, which is simply the Wilson loop for the spin connection computed in a spin representation $\rep$ along a chosen contour.
The symmetry transformation 
precisely takes the form of a local Lorentz transformation,
but \textit{multivalued} such that
a nontrivial winding is exhibited,
precluding it from being a genuine gauge transformation.
By using this multivalued local Lorentz transformation,
we show that $W_\rep$ transforms by a center-valued phase that depends only on the topological linking of the surface and curve which define $U_\cen$ and $W_\rep$, respectively.
In turn,
we prove the Ward identity for $U_\cen$ and $W_\rep$ using both covariant and canonical approaches.  
This proof is valid to all orders in perturbation theory, at least within the context of the EFT description of gravity\footnote{As is well-known, quantum corrections are perfectly well-defined even within a low-energy EFT, provided one enforces systematic power counting.  In the EFT of gravity, most quantum corrections are ultraviolet sensitive and thus absorbed into incalculable counterterms.  However, there also exist calculable long-distance quantum corrections \cite{Donoghue:2022eay}.} where the topology and dimension of spacetime are preserved.

Thirdly, we show that the interplay of $U_\cen$ and $W_\rep$ has a remarkably simple interpretation in terms of \textit{classical gravitation}.  The symmetry operator $U_\cen$ creates a vortex geometry in spacetime that materializes
a tetradic version of a cosmic string, 
serving as a certain gravitational analog of the Dirac string.
The strength of this vortex is quantized so as to induce a $\pi$ spin precession angle which is directly measured by the spin holonomy $W_\rep$ as the center-valued linking number.
This linking number can be computed in various backgrounds
with or without matter.
The topological nature of $U_\cen$ implies that its linking with  $W_\rep$ arises purely from contributions at leading order in the SF expansion, where $U_\cen$ is treated as a nondynamical background.  Furthermore, this implies that higher-order SF corrections vanish, so evidently the classical spin precession angle is not quantum corrected at any perturbative order.

Last but not least, we discuss the breaking of the gravitational one-form symmetry.
 As expected, explicit breaking requires a local operator in the representation $\rep$ that renders the spin holonomy $W_\rho$ ``endable,'' thus unspooling its  linking with $U_\cen$.  Physically, this corresponds to the screening of the spin holonomy by spinning particles.
Interestingly, the spin holonomy in the vector representation is automatically screened in pure gravity by orbital angular momentum.  This mirrors the phenomenon in gauge theory where adjoint Wilson lines are screened by the gluon field itself.  On the other hand, holonomies in the spinor representation are endable only by local fermionic operators.  If no such operators exist, then the one-form gravitational symmetry is exact.  Remarkably, this implies the emergence of a hidden symmetry of the real world: below the lightest neutrino mass, there is a gravitational one-form symmetry under which spinor holonomies are charged. 
More generally, in a theory in which the gravitational one-form symmetry is not gauged, the conjectured absence of exact global symmetries in quantum gravity directly implies the existence of fermions.

\section{ The Symmetry Triangle}
\label{K1a}

The modern view states that
having a $p$-form symmetry amounts to
having a topological operator supported on closed oriented codimension $p\mem\mplus1$ submanifolds
\cite{Gaiotto:2014kfa};
see \rrcite{Cordova:2022ruw,Shao:2023gho,Schafer-Nameki:2023jdn,Gomes:2023ahz,%
Brennan:2023mmt,McGreevy:2022oyu,Bhardwaj:2023kri,Simmons-Duffin:2016gjk} 
for reviews.
Concretely,
establishing a $p$-form symmetry
means to identify the following:
\bigskip

\hrule
\medskip
\begin{itemize}[itemsep=1pt]
    \item 
        Charged operators,
            $W(\CC_p)$,
    \item
        Symmetry operator,
            $U(\S_{d-p-1})$,
    \item
        Ward identity,
        $
            \expval{
                W(\CC_p)\,
                U(\S_{d-p-1})
            }
            \,=\,
                (\text{charge})^{\link(\CC_p,\S_{d-p-1})}\,
                \expval{
                    W(\CC_p)
                }
        $\hem.
\end{itemize}
\medskip\vspace{1pt}
\hrule\vspace{-3pt}
\bigskip\noindent
Here, $\link(\CC_p,\S_{d-p-1})$ denotes the linking number between
two closed oriented submanifolds
$\CC_p$ and $\S_{d-p-1}$,
while $d$ is the spacetime dimension.

\begin{figure}[t]
    \centering
    \includegraphics[width=0.7\linewidth]{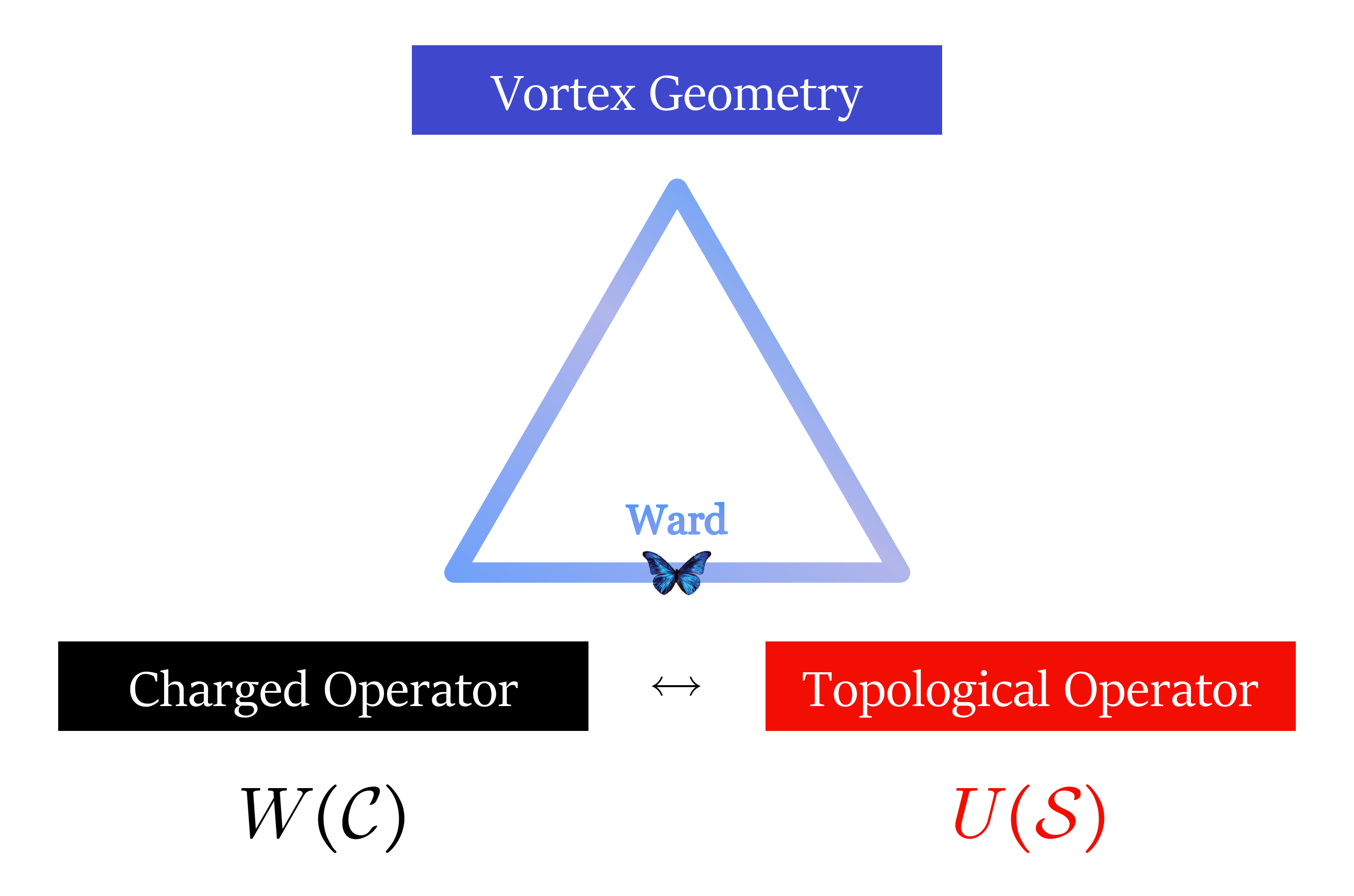}
    \caption{%
        The ``Symmetry Triangle.''
        The linking between
        the charged operator $W(\CC)$ 
        and the topological symmetry operator $U(\S)$
        establishes a Ward identity.
        Physically,
        $U(\S)$ is
        the creation operator for 
        a vortex configuration
        whose topological nontriviality
        is measured by $W(\CC)$.
    }
    \label{SymTriangle}
\end{figure}

A useful physical insight is that
the symmetry operator $U(\S_{d-p-1})$
serves as the creation operator for a vortex 
\cite{Engelhardt:1999xw,Reinhardt:2002mb,Mandelstam:1974pi,tHooft:1979rtg,tHooft:1977nqb} configuration,
on the background of which
the charged operators $W(\CC_p)$
measure the degree of ``twist'' or ``swirl''.
Notably,
this geometrical identification establishes
the core part of the Ward identity
via the path integral formalism.
(In this sense,
it may be said that
the problem of
establishing higher-form symmetries
practically
boils down to the task of finding vortex configurations in the theory.)
This relationship between
vortex geometry, charged operators, and symmetry operator
is schematically visualized in \fref{SymTriangle}:
a ``Symmetry Triangle.''
Our goal is to demonstrate
this template as a recurring structure
and apply it to dynamical gravity.

\subsection{An Abelian Primer}
\label{K1ap}

As a prototype demonstration,
let us consider
abelian $p$-form gauge theories
defined on a Lorentzian $d$-dimensional background spacetime $\M$.
In the first-order formulation,
the Lagrangian $d$-form can be taken as
\begin{align}
    \label{pform.L}
    L[A^{(p)},B^{(d-p-1)}]
    \,=\,
        B^{(d-p-1)}
        \wedge dA^{(p)}
        + f[B^{(d-p-1)}]
    \,,
\end{align}
where $f[B^{(d-p-1)}]$ is 
a $d$-form functional
of the auxiliary $(d\mminus p \mminus 1)$-form field $B^{(d-p-1)}$.
The action is the integration of the Lagrangian $d$-form
over $\M$.

The gauge transformations are
\begin{align}
    \label{pform.gt}
    A^{(p)}
    \,\,\mapsto\,\,
    A^{(p)}
    + d\xi^{(p-1)}
    \,,\quad
    B^{(d-p-1)}
    \,\,\mapsto\,\,
    B^{(d-p-1)}
    \,,
\end{align}
for a single-valued $(p\mminus1)$-form parameter $\xi^{(p-1)} \in \Omega^{(p-1)}(\M)$.
\eqref{pform.gt}
leaves the Lagrangian $d$-form in \eqref{pform.L}
invariant.

Now consider the transformation
\begin{align}
    \label{pform.tgt}
    A^{(p)}
    \,\,\mapsto\,\,
    A^{(p)}
    + 2\pi n\mem \a^{(p)}
    \,,\quad
    B^{(d-p-1)}
    \,\,\mapsto\,\,
    B^{(d-p-1)}
    \,,
\end{align}
where $n \inn \R$
takes any real value.
This assumes
a closed oriented codimension $p\mem\mplus1$ submanifold $\S_{d-p-1} \msubset \M$,
so
the $p$-form
$\a^{(p)}$ represents
the integral de Rham cohomology element
$[\a^{(p)}] \in H^{p}(\M{\setminus}\S_{d-p-1},\mathbb{Z})$
such that
\begin{align}
    \label{a-linking}
    \oint_{\CC_p}
        \a^{(p)}
    \,=\,
        \link(\CC_p,\S_{d-p-1})
    \,,
\end{align}
for any $p$-dimensional submanifold
$\CC_p \msubset \M$
that 
is closed, oriented,
and
does not intersect or overlap with
$\S_{d-p-1}$.

Note that there are two ways to think about the representative $\a^{(p)}$.
Mathematicians would like to view it as a closed but not exact $p$-form
defined on the restricted domain
$\M{\setminus}\S_{d-p-1}$.
Physicists, on the other hand,
will prefer to \textit{formally} view it as a $p$-form on the entire space $\M$
whose exterior derivative produces
a ``Dirac delta form'' supported on $\S_{d-p-1}$
\cite{Gomes:2023ahz,Brennan:2023mmt,Bhardwaj:2023kri,Lechner:2001xk,Moore:2011gagt},
\begin{align}
    \label{da=delta}
    d\a^{(p)}
    \,=\,
        \delta(\S_{d-p-1})
    \,.
\end{align}
Namely, $\a^{(p)}$ is a solenoidal field generated by a distributional current.
For any oriented $p$-dimensional submanifold $\N_p \msubset \M$,
the Dirac delta form is defined as
the $(d \mminus p)$-form such that,
for any test $p$-form field $\varphi^{(p)}$,
\begin{align}
    \label{deltadef}
    \int_{\N_p} 
        \varphi^{(p)}
    \,=\,
        \int_\M 
            \varphi^{(p)} \nem\wedge \delta(\N_p)
    \,.
\end{align}
These two views are equivalent;
see, e.g., Nakahara \cite{Nakahara},
Secs.\:1.9 and 10.5.

It can be seen that
\eqref{pform.tgt} 
is a transformation that creates
a codimension $p+1$ vortex configuration.
This vortex describes
a thin flux tube of strength $n$
such that the Wilson loops,
\begin{align}
    \label{pform.W}
    W_m(\CC_p)
    \,=\,
        \exp\bb{
            im \oint_{\CC_p}\nem\hnem A^{(p)}
        \nem}
    \,,
\end{align}
gain AB
\cite{aharonov1959significance}
phases as
\begin{align}
\begin{split}
    \label{pform.Wtransf}
    W_m(\CC_p)
    \,\,\,\mapsto\,\,\,
    W'_m(\CC_p)
    \,=\,
        W_m(\CC_p)
        \,
        \exp\hnem\bigbig{
            2\pi i mn
        }^{\nem\link(\CC_p,\S_{d-p-1})}
    \,,
\end{split}
\end{align}
where $m\inn\Z$
is a $\U(1)$ representation
(electric charge).
As is well known,
the AB phase is invisible
when the strength $n$ of the flux tube is quantized as $n\inn\Z$.

The above observation
boils down to the following path integral identity:
\begin{align}
    \label{pform.wardderiv}
    \nonumber
    &
    \int \D{A}\D{B}\,\,
        \exp\bb{
            i \int_\M  
                B^{(d-p-1)} \wedge d\bigbig{
                    A^{(p)}
                    - 2\pi n\mem \a^{(p)}
                }
                + f[B^{(d-p-1)}]
        }
        \,
        W_m(\CC_p)
    \\
    &\,=\,
        \int \D{A}\D{B}\,\,
            \exp\bb{
                i \int_\M  
                    B^{(d-p-1)} \wedge dA^{(p)}
                    + f[B^{(d-p-1)}]
            }
            \,
            W'_m(\CC_p)
    \,.
\end{align}
This is established by using \eqref{pform.tgt}
as a change of path integration variables.

By definition,
the right-hand side of
\eqref{pform.wardderiv}
is the expectation value
of the transformed Wilson loop,
$W'_m(\CC_p)$.
The left-hand side, on the other hand,
describes
the expectation value of
$W_m(\CC_p)$
put together with the operator
\begin{align}
    \label{pform.Uderiv}
    U_n(\S_{d-p-1})
    \,=\,
        \exp\bb{
            - 2\pi i n \int_\M  
                B^{(d-p-1)} \wedge d\a^{(p)}
        }
    \,.
\end{align}
This operator is supported on the submanifold $\S_{d-p-1}$.
In view of \eqrefs{da=delta}{deltadef},
\eqref{pform.Uderiv} boils down to
\begin{align}
    \label{pform.U}
    U_n(\S_{d-p-1})
    \,=\,
        \exp\bb{
            - 2\pi i n \int_{\S_{d-p-1}}\nem\nem
                B^{(d-p-1)}
        }
    \,.
\end{align}
The Lagrangian $d$-form in \eqref{pform.L}
describes that
the saddles of the theory satisfy
$dB^{(d-p-1)} = 0$.
Hence \eqref{pform.U} is a topological operator.

In conclusion,
the path integral identity in \eqref{pform.wardderiv}
establishes
the $p$-form symmetry
of $p$-form abelian gauge theories
in terms of the Ward identity
\begin{align}
    \label{pform.ward}
    \expval{
        W_m(\CC_p)\mem U_n(\S_{d-p-1})
    }
    \,=\,
        \exp\bigbig{
            2\pi i mn
        }^{\nem\link(\CC_p,\S_{d-p-1})}
    \,
        \expval{W_m(\CC_p)}
    \,.
\end{align}
The charged operators are the Wilson loops
$W_m(\CC_p)$
in \eqref{pform.W},
which transform as \eqref{pform.Wtransf}.
The symmetry operator
$U_n(\S_{d-p-1})$
is explicitly constructed in \eqref{pform.U},
which is topological
under smooth deformations of its support.

We see that \eqref{pform.tgt},
the transformation for creating a codimension $p+1$ vortex,
is the very $p$-form symmetry transformation
of abelian $p$-form gauge theories.
Note how
the symmetry operator
$U_n(\S_{d-p-1})$
is systematically deduced as
the operator that
induces \eqref{pform.tgt}
when inserted inside the path integral.
That is,
it is the operator
that precisely gets
absorbed into the exponentiated action
when
using \eqref{pform.tgt}
as a change of path integration variables.

\smallskip
A few remarks are in order.
First,
the Ward identity in \eqref{pform.ward}
admits another physical interpretation as
measuring the charge of $W_m(\CC_p)$
by a Gaussian pillbox $\S_{d-p-1}$;
\eqref{pform.U}
exponentiates the charge
as the Gaussian flux integral.
In nonabelian gauge theories,
however,
the AB phase interpretation
turns out to be more robust than
the Gaussian pillbox intuition.

Second, note how the ``$B\mem dA$'' structure
of \eqref{pform.L}
provides a covariant reincarnation of the canonical formalism 
(in which ``$p\mem dx$'' gives rise to $\comm{x}{p} = i$)
such that the charged and symmetry operators
develop a conjugate-type relationship
as functionals of $A$ and $B$,
serving as the very basis of 
\eqref{pform.ward}.

Third,
an explicit formula for
the Dirac delta form
in \eqref{deltadef}
reads
\begin{align}
    \label{eq:delta-param}
    \delta(\N_p)_{\m_1\cdots\m_{d-p}}
    \,=\,
    \int d^p\s\,\,
        \delta^{(d)}\hnem\bigbig{
            x{\,-\,}X(\s)
        }\,
        \frac{\partial X^{\l_1}}{\partial \s^1}
        \cdots
        \frac{\partial X^{\l_p}}{\partial \s^p}
    \,
        \e_{\l_1\cdots\l_p\m_1\cdots\m_{d-p}}
    \,,
\end{align}
where
$\delta^{(d)}\hnem(x{\,-\,}X(\s))$ is the ordinary Dirac delta function,
$X^\m(\s)$ parametrizes $\N_k$,
and
$\e_{\m_1\cdots\m_d}$ is the permutation symbol taking values from $\{+1,-1,0\}$.
The ordinary Dirac delta
is reproduced for $p \eqq 0$
as an invariant top form.
The Dirac delta form of
a top-dimensional submanifold
is its characteristic function.

Mathematically,
$\delta(\N_p)$
is the distributional Poincar\'e dual of the embedded oriented submanifold $\N_p$:
a singular current representative of the Poincaré dual class
\cite{deRham:1984,bott1982differential,Thom:1954,Lechner:2001xk,Moore:2011gagt,Bhardwaj:2023kri}.
This facilitates a consistent definition of
the linking and intersection numbers
between various oriented submanifolds
and a derivation of their relations,
with one's favorite sign conventions.
A notable identity is
$d\hem \delta(\N_p) = (-1)^p\mem\hem \delta(\partial\N_p)$,
implied by Stokes' theorem.

Last but not least,
note that the symmetry transformation in \eqref{pform.tgt}
can be interpreted as
a \textit{multivalued gauge transformation}.
Locally, it looks like the small gauge transformation in \eqref{pform.gt}
since a single-valued $(p\mminus1)$-form $\xi^{(p-1)}$
such that $\a^{(p)} = d\xi^{(p-1)}$
can be constructed within each small patch
in view of Poincar\`e lemma.
However, the global existence of such a single-valued $(p\mminus1)$-form gauge parameter
is obstructed,
which is precisely the content of
the de Rham cohomology
$H^{p}(\M{\setminus}\S_{d-p-1},\mathbb{Z})$.
If one insists on taking $\a^{(p)} = d\xi^{(p-1)}$
for some $\xi^{(p-1)}$,
one inevitably has to take $\xi^{(p-1)}$ as multivalued.

\subsection{Multivalued Gauge Transformation}

In gauge theories,
one can argue that
any higher-form symmetry transformation
must locally look like
a gauge transformation.
This is because
it has to preserve all local, i.e., pointlike gauge-invariant operators.
Meanwhile,
to describe a genuine global symmetry
(not redundancy),
the higher-form symmetry transformation
must also be globally nontrivial
at the same time.

A multivalued gauge transformation
is a field redefinition that
takes the form of a gauge transformation
when restricted to each small patch in spacetime.
When viewed as a whole,
however,
it glues
such local gauge transformations
together
into a nontrivial configuration.
\begin{center}
    \bfseries
    Higher-form symmetry transformations are\\
    multivalued gauge transformations.
\end{center}

A comparison will be drawn between
multivalued gauge transformations and
large gauge transformations
(in the sense of \rcite{tong2018gauge}, not \rcite{Strominger:2017zoo}).
Crucially,
the latter demands that the gauge parameter is single-valued.
The former is a symmetry transformation,
while the latter is a gauge transformation.

Take Maxwell theory, for instance.
The Dirac string is a magnetic flux tube
or equivalently a thin solenoid,
which describes a codimension two surface
in the spacetime view
\cite{Dirac:1931kp,Dirac:1948um,Shnir:2011zz}.
Its flux is quantized
when generated by
a single-valued $\U(1)$ gauge transformation
exhibiting a nontrivial winding
\cite{Nakahara}.

In contrast,
a multivalued $\U(1)$ gauge transformation
generates a Dirac string
whose flux is improperly quantized.
This serves as
the electric one-form symmetry transformation
of Maxwell theory.
The symmetry operator
creates an improperly quantized Dirac string
as the vortex configuration.
The charged operators are the Wilson loops,
measuring the AB phases
around this thin solenoid.
An explicit demonstration of these points
will appear in \Sec{K1mv}.

\section{ Universal Template for One-Form Symmetry}
\label{K1b}

Gathering the insights from
\Sec{K1a},
this section shows how
any nonabelian gauge theory
of a finite-dimensional (matrix) Lie group $\G$,
whose Lagrangian top form
admits a first-order formulation in the form
\begin{align}
    \label{1form.L}
    L[A,B]
    \,=\,
        \frac{1}{g^2}\,\BB{
            B_a
            \wedge F^a[A]
            \,+\, f[B]
        }
    \,,
\end{align}
exhibits a one-form symmetry
valued in the center $Z(\G)$
of the gauge group.
In \eqref{1form.L}, $F^a[A]$ denotes the field strength of the nonabelian gauge connection,
\begin{align}
    \label{eq:fsdef}
    F^a[A]
    \,=\,
        dA^a + \frac{1}{2}\, f^a{}_{bc}\, A^b \swedge A^c
    \,,
\end{align}
while $f[B]$ is
a gauge-invariant algebraic $d$-form functional
of the auxiliary, coadjoint-valued $(d\mminus 2)$-form field $B_a$.
We assume that the theory is defined on a $d$-dimensional spacetime manifold $\M$.
We use $a,b,c,d,\cdots$
for adjoint indices and
$i,j,k,l,\cdots$
for fundamental indices.

The gauge transformations are\footnote{
    Of course, 
    $\Omega$ should properly approach the identity at the infinities
    if $\M$ is noncompact.
}
\begin{align}
    \label{1form.gt}
    A
    \,\,\mapsto\,\,
        \Ad{\Omega^{-1}} A
        + \Omega^{-1} d\Omega
    \,,\quad
    B
    \,\,\mapsto\,\,
        \coAd{\Omega^{-1}} B
    \,,
\end{align}
where 
$\Omega$ is a single-valued zero-form parameter
valued in the gauge group $\G$.
\eqref{1form.gt}
leaves the Lagrangian $d$-form in \eqref{1form.L}
invariant.

\subsection{Multivalued Gauge Transformation}

Recalling 
our earlier discussion via
\eqrefs{pform.gt}{pform.tgt},
we expect that
the one-form symmetry transformation
will be obtained by promoting $g$ in \eqref{1form.gt}
to a multivalued parameter $\Omega_\S$
that exhibits a nontrivial winding around 
a closed oriented codimension two submanifold $\S \mem\msubset \M$:
\begin{align}
    \label{1form.tgt}
    A
    \,\,\mapsto\,\,
        \Ad{\Omega_\S^{-1}} A
        + \Omega_\S^{-1} d\Omega_\S
    \,,\quad
    B
    \,\,\mapsto\,\,
        \coAd{\Omega_\S^{-1}} B
    \,.
\end{align}

\begin{figure}[t]
    \centering
    \includegraphics[scale=1.15]{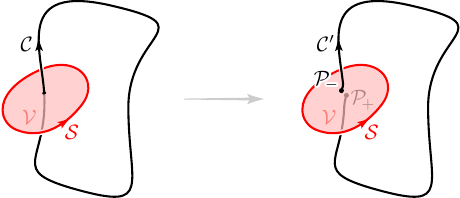}
    \medskip
    \caption{%
        Infinitesimal opening of the closed contour $\CC$
        at the intersection point $\CC \cap \V$.
    }
    \label{fig:W'}
\end{figure}

A detailed elaboration on
the geometry and definition of $\Omega_\S$
is postponed to \Sec{K1mv}.
For the current moment,
it suffices to note that
the multivaluedness of $\Omega_\S$
can be characterized by
the discontinuity exhibited across a branch sheet
$\V$ such that $\S = \partial\V$.
This means to stipulate the condition
\begin{align}
    \label{eq:lambda-disc1}
    \lim_{
        \P_+ \to\, \P_-
    }\hem
        \Omega_\S(\P_{\hnem+})
        \mem 
        \Omega_\S^{-1}\hnem(\P_{\hnem-})
    \,=\,
    \cen
    \,,
\end{align}
where $\P_{\hnem+}$ and $\P_{\hnem-}$ are points
infinitesimally displaced away from the same point on the branch sheet $\V$,
but in opposite directions:
see \fref{fig:W'}.

\subsection{Charged Operator}
\label{K1b.W}

Next, we identify the charged operator
as the Wilson loop $W_\rep(\CC)$
for a representation $\r$ of the gauge group $\G$,
defined by the path-ordered integral formula\footnote{
    For technical reasons,
    within this chapter
    we let
    Wilson lines solve the
    right-multiplication transport equation,
    unlike in \Chap{K2}
    where the left-multiplication convention is taken.
    Hence
    the path ordering in \eqref{1form.W}
    places
    earlier points on the left
    and later points to the right
    as indicated by an overbar.
}
\begin{align}
    \label{1form.W}
    W_\rho(\CC)
    \,=\,
        \Tr_\rho
        \bP
        \exp\bb{
            \oint_\CC A
        }
    \,.
\end{align}
Here, $\CC$ is
a closed oriented one-dimensional submanifold in $\M$.

To show that $W_\rep(\CC)$
is charged under the proposed symmetry transformation in \eqref{1form.tgt},
it suffices to first consider the case
when $\CC$ links with $\S$ once.
With this assumption, we consider
an auxiliary open contour $\CC'$
that arises by cutting the loop $\CC$
by a branch sheet $\V$,
as is illustrated in \fref{fig:W'}.
Let $\partial\CC' = \P_+ - \P_-$,
where $\P_+$ and $\P_-$ are infinitesimally separated
away from the exact intersection point
$\P = \CC \cap \V$.

\begin{figure}[t]
    \centering
    \includegraphics[width=0.95\linewidth]{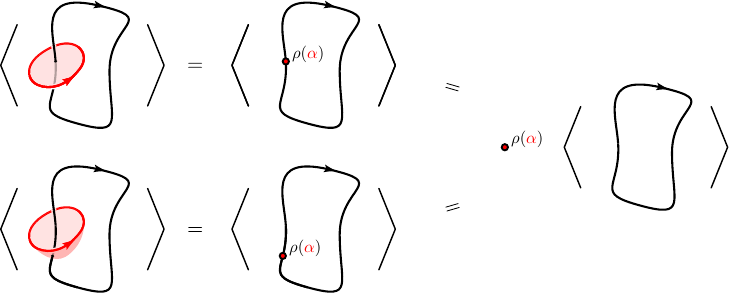}
    \medskip
    \caption{%
        The one-form symmetry operator inserts 
        the group element $\cen$ inside the trace of the Wilson loop.
        The location of this insertion can be arbitrary
        if $\cen$ belongs to the center of the gauge group: $\cen \inn Z(\G)$.
    }
    \label{fig:cuts}
\end{figure}

Consider the Wilson \textit{line}
about the open contour $\CC'$,
which outputs a group element:
$W(\CC') \eqq \bPexp\bigbig{\int_{\CC'}\nem A}$.
Since
Wilson lines exhibit
the bilocal transformation behavior,
\eqref{1form.tgt}
transforms $W(\CC')$ as
\begin{align}
    W(\CC')
    \,\,\,\mapsto\,\,\,
    \Omega_\S^{-1}(\P_-)\mem W(\CC')\,\Omega_\S(\P_+)
    \,.
\end{align}
It follows that
the Wilson loop $W_\rep(\CC)$
transforms under \eqref{1form.tgt} as
\begin{align}
\begin{split}
    \label{eq:W'}
    W_\rep(\CC)
    \,\,\,\mapsto\,\,\,
    {}&{}
    \Tr_\rep
    \lim_{\P_\pm \to \P} \BB{
        \Omega_\S^{-1}(\P_-)\mem W(\CC')\, \Omega_\S(\P_+)
    }
    \,,\\
    \,=\,
    {}&{}
        \Tr_\rep
        \lim_{\P_\pm \to \P} \BB{
            \Omega_\S(\P_+)\mem \Omega_\S^{-1}(\P_-)\, W(\CC')
        }
    \,,\\
    \,=\,
    {}&{}
        \Tr_\rep\nem
            \BB{
                \lim_{\P_\pm \to \P}
                    \a\mem
                    W(\CC')
            }
    \,,
\end{split}
\end{align}
which shows that the effect of the multivalued gauge transformation
is
the insertion of the group element $\a$
into the Wilson loop
at the intersection point $\P = \CC \cap \V$.
See \fref{fig:cuts}.

At this point,
we remind ourselves that
the branch sheet $\V$
has been an utterly arbitrary construct.
The transformation behavior of the Wilson loop
should not depend on one's arbitrary choice of a branch sheet.

\begin{figure}[t]
    \centering
    \includegraphics[scale=1.15]{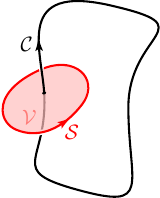}
    \qquad\qquad
    \includegraphics[scale=1.15]{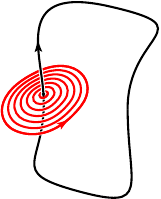}
    \medskip
    \caption{%
        \textit{Left:}
        The linking number between 
        a one-dimensional loop $\CC$ and an exact codimension two surface $\S = \partial\V$ 
        encodes
        the number of
        intersections 
        between 
        $\CC$ and
        a branch sheet $\V$.
        \textit{Right:}
        The branch sheet $\V$
        defines
        a homotopy for shrinking the surface $\S$ to a point on $\CC$.
        Physically, this precisely represents
        the operation of measuring the charge.
    }
    \label{fig:CS}
\end{figure}

Physically, the surface $\S$
will be the support of the symmetry operator.
Crucially, the symmetry operator needs to be \textit{topological}.
As illustrated in \fref{fig:CS},
the branch sheet choice
precisely corresponds to
one's prescription
of collapsing the symmetry operator
down to a point on the Wilson loop.
If this prescription
affects the outcome of charge measurement,
the very premise that the symmetry operator is topological
gets violated.

It is also well-known that
any higher-form symmetry algebra must be abelian,
since there does not exist a canonical ordering
between submanifolds whose codimension is greater than one.
It is not difficult to see that
the above ``branch sheet independence''
is a consistency condition
needed for this abelianity of higher-form symmetry algebra
as well.

In conclusion,
we must stipulate that
the transformed Wilson loop
is ignorant of
the choice of a branch sheet $\V$.
As illustrated in \fref{fig:cuts},
this means that
the group element $\a$
should commute with the rest of the Wilson line
regardless of the location of its insertion.
This is viable if $\a$ belongs to the center $Z(\G)$ of the gauge group:
\begin{align}
    \label{eq:lambda-disc1.cen}
    \lim_{
        \P_+ \to\, \P_-
    }\hem
        \Omega_\S(\P_{\hnem+})
        \mem 
        \Omega_\S^{-1}\hnem(\P_{\hnem-})
    \,=\,
    \cen
    \,\in\,
        Z(\G)
    \,.
\end{align}
The implication is that
the multivaluedness of $\Omega_\S$
must be valued in $Z(\G)$.

Finally,
provided this center stipulation,
\eqref{eq:W'} boils down to
$W_\rep(\CC) \mapsto \rep(\cen)\, W_\rep(\CC)$,
where $\rep(\cen)$ denotes 
the representation of the center element $\cen$
as a complex phase.
This establishes that Wilson loops 
do transform under
the multivalued gauge transformation in \eqref{1form.tgt}
by the factor $\rep(\cen)$,
for any branch sheet choice.
Recalling the fact that
we have assumed unit linking between $\CC$ and $\S$
for simplicity,
we conclude that,
for generic $\CC$ and $\S$
that do not intersect or overlap,
\begin{align}
    \label{1form.Wtransf}
    W_\rep(\CC)
        \,\,\,\mapsto\,\,\,
    \rep(\cen)^{\link(\CC,\S)}\mem
    W_\rep(\CC)
    \,.
\end{align}

\subsection{Ward Identity}

To recapitulate,
we have established that
Wilson loops are charged under
the multivalued gauge transformation in \eqref{1form.tgt}
as in \eqref{1form.Wtransf},
provided the center condition in \eqref{eq:lambda-disc1.cen}.
This result
boils down to the following path integral identity:
\begin{align}
    \label{1form.wardderiv}
    \nonumber
    &
    \int \D{A}\D{B}\,\,
        \exp\bb{
            \frac{i}{g^2} \int_\M  
                B_a \wedge \BB{
                    F^a[A] - (\Omega^{-1}_\S dd\Omega_\S)^a
                }
                + f[B]
        }
        \,
        W_\rep(\CC)
    \\
    &\,=\,
        \int \D{A}\D{B}\,\,
            \exp\bb{
                \frac{i}{g^2} \int_\M  
                    B_a \wedge F^a[A]
                    + f[B]
            }
            \,\mem
            \rep(\cen)^{\link(\CC,\S)}\mem
            W_\rep(\CC)
    \,.
\end{align}
Recall \eqref{pform.wardderiv}.
Again, \eqref{1form.wardderiv}
is established by using \eqref{1form.tgt}
as a change of path integration variables.
Note that \eqrefs{eq:fsdef}{1form.tgt} imply
\begin{align}
    \label{eq:YM.Ftransf}
    F[
        \Ad{\Omega_\S^{-1}}{A}
        + \Omega_\S^{-1} d\Omega_\S
    ]
    \,=\,
    \Ad{\Omega_\S^{-1}}{F[A]}
    \mem+\mem
        \Omega_\S^{-1} dd\Omega_\S
    \,,
\end{align}
where $dd\Omega_\S$
should not be equated to zero
due to the multivaluedness of $\Omega_\S$;
it will eventually produce the Dirac delta form $\delta(\S)$.

By definition,
the right-hand side of
\eqref{1form.wardderiv}
is the expectation value
of the transformed Wilson loop.
The left-hand side, on the other hand,
describes
the expectation value of
$W_\rep(\CC)$
put together with the operator
\begin{align}
    \label{1form.Uderiv}
    U_\a(\S)
    \,=\,
        \exp\bb{
            - \frac{i}{g^2}  \int_\M  
                B_a \wedge (\Omega_\S^{-1} dd\Omega_\S)^a
        }
    \,.
\end{align}
The invariant geometrical data characterizing this operator are $\S$, the Dirac delta support, 
and $\a$, the multivaluedness of $\Omega_\S$.
Hence it is named $U_\a(\S)$.

To conclude,
the path integral identity in \eqref{1form.wardderiv}
establishes
the one-form symmetry
of nonabelian gauge theories
in terms of the Ward identity
\begin{align}
    \label{1form.ward}
    \expval{
        W_\rep(\CC)\mem U_\cen(\S)
    }
    \,=\,
        \rho(\cen)^{\link(\CC,\S)}
    \,
        \expval{W_\rep(\CC)}
    \,.
\end{align}
The charged operators are the Wilson loops
$W_\rep(\CC)$
defined in \eqref{1form.W},
which transform as \eqref{1form.Wtransf}.
The symmetry operator 
is stipulated to be the operator $U_\cen(\S)$ defined in \eqref{1form.Uderiv}.
This operator
is topological
because its action
on $W_\rep(\CC)$
does not change
under arbitrary smooth deformations of its support $\S$,
provided that the symmetry group is 
an abelian subgroup
of the gauge group.

Again,
note how
the symmetry operator $U_\cen(\S)$
is systematically deduced as
the operator that
induces the multivalued gauge transformation \eqref{1form.tgt}
when inserted inside the path integral.
It also follows that $U_\cen(\S)$
is the creation operator for a codimension two vortex configuration
that induces 
AB phases for colored particles
(as Wilson loops)
around $\S$.
It should also be clear that
\begin{align}
\begin{split}
    \label{eq:arbWard}
    \expval{
        U_\cen(\S)\mem \O
    }
    \,=\,
    \expval{
        \O_{\Omega_\S}
    }  
\end{split}
\end{align}
for any operator $\O$ in the theory,
if $\O_{\Omega_\S}$ denotes the image of $\O$
under the multivalued gauge transformation in \eqref{1form.tgt}.
For gauge-invariant local operators,
$\O_{\Omega_\S} = \O$.

\subsection{Symmetry Operator}
\label{K1>1form>U}

Note that 
\eqref{1form.Uderiv}
is a concrete formula
that expresses the one-form symmetry operator 
$U_\cen(\S)$
in terms of the local degrees of freedom of the theory.
A further explicit formula
can be obtained by choosing a specific representative for the multivalued gauge parameter $\Omega_\S$.
As will be discussed in detail in \Sec{K1mv},
there are many choices one can make.
One nice choice is
\begin{align}
    \label{Omega=exp(nphi)}
    \Omega_\S
    \,=\,
        \exp\bigbig{
            \vn\mem \phi_\S
        }
    \transition{where}
    \mathe^{2\pi \vn}
    \,=\,
        \a
    \,\in\,
        Z(\G)
    \,,
\end{align}
where $\phi_\S$ is a multivalued function
that exhibits a unit winding from $0$ to $2\pi$
around $\S$,
and $\vn^a$ is an element of the gauge algebra
$\g \eqq \Lie(\G)$
that exponentiates to the center element $\a$.
We take $\vn^a$ as a constant.
With this choice, it follows that
\begin{align}
    dd\phi_\S
    \,=\,
        2\pi\hem \delta(\S)
    \qiq
    (\Omega_\S^{-1} dd\Omega_\S)^a
    \,=\,
        2\pi \vn^a\,
            \delta(\S)
    \,,
\end{align}
so \eqref{1form.Uderiv} boils down to
\begin{align}
    \label{1form.U}
    U_\a(\S)
    \,=\,
        \exp\bb{
            - \frac{2\pi i}{g^2}  \int_\S  
                B_a\mem \vn^a
        }
    \,,
\end{align}
manifesting the surface support on $\S$.

\eqref{1form.U} is one possible specific realization of the symmetry operator.
This formula, \eqref{1form.U},
has appeared in the literature in several contexts.
In
Eq.\,(46) of \rcite{Engelhardt:1999xw}
and
Eq.\,(30) of \rcite{Reinhardt:2002mb},
it was introduced
as a vortex creation operator.
As a one-form symmetry operator,
Ref.\,\cite{Gomes:2023ahz}
provided its realization
in a particular setup
where
temporal winding is utilized.
See \rcite{Brennan:2023mmt}
for a more sophisticated realization;
\eqref{1form.U} is also a special case of the family of symmetry operators constructed in \rcite{Cordova:2022rer}.
For a lattice counterpart,
see \rcite{Cherman:2022eml}, for instance.

One may worry that \eqref{1form.U}
utilizes a color reference $\vn^a$
on its right-hand side,
which is not unique for a given $\a$ on the left-hand side.
Even worse, $\vn^a$ seems to specify an arbitrary direction in color space that naively violates gauge invariance.
However, this dependence on $\vn^a$ is spurious
because the properties of $U_\cen(\S)$
are dictated entirely by its action on Wilson loops
via the Ward identity in \eqref{1form.ward},
which depends only on $\cen$.
Hence any distinct choices of $\vn^a$ with the same twist $\cen$ are regarded as physically equivalent.\footnote{
    This is similar to what occurs in the case of instantons in gauge theory \cite{Nair:2005iw}. 
    These pure gauge configurations necessarily specify some explicit path in color space, and hence appear naively color breaking.  
    However, only the topological winding number is the invariant label on these configurations.
    If desired,
    one could implement a sum over group orbit for $\vn^a$, for instance.
}

It remains to elaborate on the interpretation of the vortex geometry
which $U_\a(\S)$ creates:
a colored Dirac string.
We wish to do so while revealing more details on the multivalued gauge transformation and its mathematical definition.
Yet, readers may skip the following and directly jump to \Sec{K1.grav}.

\section{ More on Multivalued Gauge Transformations}
\label{K1mv}

Let us show how
the multivalued gauge transformation
in abelian and nonabelian gauge theories,
or equivalently the geometry of the Dirac strings,
is described and defined by
the physicists
and
the mathematicians.
See \rcite{Nakahara},
Secs.\:1.9 and 10.5,
and also 
\rrcite{Shnir:2011zz,Kleinert:2008zzb},
for instance.
For simplicity and concreteness,
we take $\M$ as the four-dimensional space of coordinates $(t,x,y,z)$
and take the support $\S$ of the Dirac strings as the $t$-$z$ plane,
for a moment.

\subsection{Physicists' Approach}
\label{K1mv.phys}

To a physicist,
a Dirac string
in an abelian gauge theory
means the following gauge potential:
\begin{align}
    \label{adirac.A}
    A
    \,=\,
        n\, d\phi_\S
    \,=\,
        n\,
            \frac{x\mem dy - y\mem dx}{
                x^2 + y^2
            }
    \,.
\end{align}
Here, $\phi_\S(x,y)$ is the azimuthal angle:
a \textit{multivalued function} of $x$ and $y$
with periodicity $2\pi$.
By formally treating \eqref{adirac.A} as 
a one-form field on the entire space $\M$,
the physicist computes its curl as
\begin{align}
    \label{adirac.F}
    F
    \,=\,
        dA
    \,=\,
        n\, dd\phi_\S
    \,=\,
        2\pi n\,
            \delta(x)\hem \delta(y)
        \, dx \wedge dy
    \,.
\end{align}
This describes 
the magnetic field of 
an infinitely thin solenoid,
placed along the $z$-axis.
The total magnetic flux flowing inside the solenoid is $2\pi n$.
This can be either quantized or not quantized:
$n \inn \R$.
The AB phase of a particle with electric charge $m \inn \Z$
is $\mathe^{2\pi i mn}$,
which can be obtained by computing
either $\oint A$ or $\int F$
on account of the
Stokes theorem.

Clearly, \eqref{adirac.A} is generated by applying a gauge transformation
with the \textit{multivalued parameter} $n\phi_\S$
to the trivial configuration $A = 0$.
This is why the field strength $F$ is almost zero everywhere,
except on a measure-zero support.

\medskip
The nonabelian generalization is straightforward.
A \textit{colored Dirac string}
is the following field configuration
in a nonabelian gauge theory:
\begin{align}
    \label{nadirac.A}
    A^a
    \,=\,
        \vn^a\mem d\phi_\S
    \,=\,
        \vn^a\,
            \frac{x\mem dy - y\mem dx}{x^2+y^2}
    \,.
\end{align}
Here, $\vn^a$ is a constant color factor:
a fixed element in the gauge algebra $\g = \Lie(\G)$.
The field strength easily computes to
\begin{align}
    \label{nadirac.F}
    F^a
    \,=\,
        \vn^a\, dd\phi_\S
    \,=\,
        2\pi \vn^a\,
            \delta(x)\hem \delta(y)
        \, dx \wedge dy
    \,,
\end{align}
which is simply the colored version of \eqref{adirac.F}.
Clearly, this describes 
the nonabelian magnetic field of 
an infinitely thin solenoid,
placed along the $z$-axis.
The total color flux flowing inside the solenoid is $2\pi \vn^a$.
It describes a color vortex
of codimension two.

The analog of the AB phase is the Wilson loop
for a representation $\rho$ of the gauge group $\G$,
defined in \eqref{1form.W}.
For instance,
let $\CC$ be
the circular contour
of constant radius $R$,
parametrized as
$(x,y,z) = (R\cos\s,R\sin\s,0)$
where $\s$ runs from $0$ to $2\pi$.
By plugging in \eqref{nadirac.A} to \eqref{1form.W},
one obtains
$W_\rep(\CC) = \Tr_\rep(\mathe^{2\pi \vn})$.

As per the conclusion in \Sec{K1b.W},
we consider the physically relevant case of
a \textit{center} vortex
\cite{Engelhardt:1999xw,Reinhardt:2002mb,Mandelstam:1974pi,tHooft:1979rtg,tHooft:1977nqb}.
To this end, we stipulate the condition $\mathe^{2\pi \vn} = \cen \in Z(\G)$ in \eqref{Omega=exp(nphi)}.
Concretely,
suppose $\G = \SU(N)$, so $Z(\G) = \Z_N$.
In this case,
we can take \cite{Gomes:2023ahz}
\begin{align}
    \label{gomesnt}
    \vn^a
    \,=\,
        \frac{k}{N}\, \t^a
    \,,\quad
    \t^i{}_j
    \,=\,
        i\mem \diag(1,1,{\cdots},1,-N{\mem+\mem}1)
    \,,
\end{align}
where $k \inn \Z$.
With \eqref{gomesnt},
the Wilson loop
for the fundamental representation,
for instance,
evaluates to
\begin{align}
    W_\fund(\CC)
    \,=\,
        \mathe^{2\pi ik/N}\mem N
    \,.
\end{align}

Note that shifting $k$ by 
an integer multiple of $N$
does not change this result.
Physically,
this means that
two solenoids of different ``strengths''
can be indistinguishable
in terms of the AB phase.
In fact, it will become self-evident that
such a quantized shift of color flux
corresponds to a \textit{single-valued} gauge transformation
(a genuine gauge transformation describing physical equivalence)
that is not continuously connected to the identity,
i.e., a large gauge transformation in the sense of \rcite{tong2018gauge}.

\medskip
Clearly,
\eqrefs{nadirac.A}{nadirac.F}
can be written as
\begin{align}
    \label{1form.AFO}
    A
    \,=\,
        \Omega_\S^{-1} d\Omega_\S
    \,,\quad
    F
    \,=\,
        \Omega_\S^{-1} dd\Omega_\S
    \,,\quad
    \Omega_\S
    \,=\,
        \exp\bigbig{
            \vn\hem \phi_\S
        }
    \,.
\end{align}
This shows that the colored Dirac string is
created by 
acting on a multivalued gauge transformation
to the trivial configuration $A^a \eqq 0$,
by the parameter $\Omega_\S$.

\begin{figure}[t]
    \centering
    \includegraphics[scale=0.85]{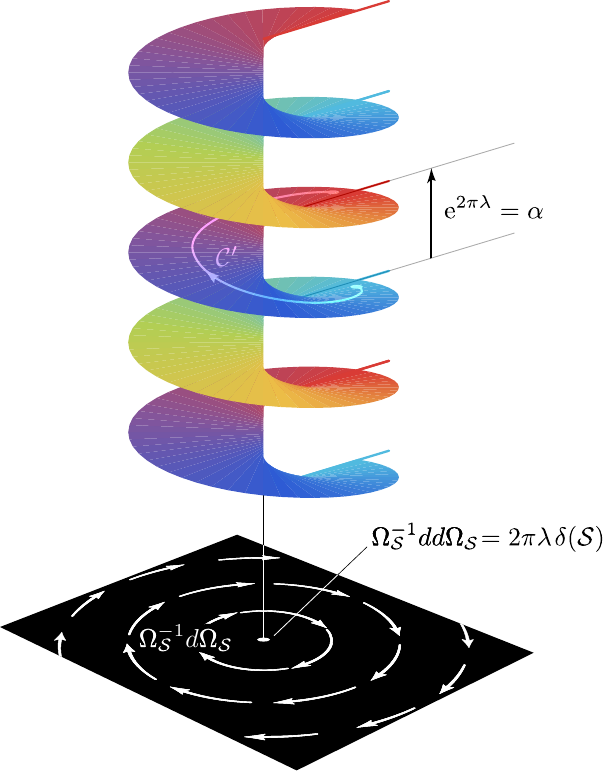}
    \medskip
    \caption{%
        The twist
        of a multivalued gauge parameter $\Omega_\S$
        is characterized by
        the discontinuity $\cen$ 
        exhibited across a branch cut.
        Physicists approach this twist
        as the localization of
        the curl $\Omega_\S^{-1}dd\Omega_\S$ of the flat connection $\Omega_\S^{-1}d\Omega_\S$
        along $\S$.
    }
    \label{fig:swirls}
\end{figure}

The computation of Wilson loops,
demonstrated above,
can be revisited from this angle.
Consider a Wilson \textit{line}
$W(\CC')$
about a contour $\CC'$,
obtained as an infinitesimal opening of $\CC$
by a branch cut:
$(x,y,z) = (R\cos\s,R\sin\s,0)$
for $\s \inn [\e,2\pi\mminus\e]$;
this repeats the approach in \eqref{eq:W'}.
In the trivial configuration $A^a = 0$,
$W(\CC')$ simply computes to the identity element,
$\id$.
The bilocal transformation behavior of Wilson lines
implies that
the multivalued gauge transformation by $\Omega_\S \eqq \exp\bigbig{\vn\hem \phi_\S}$
transforms it to
\begin{align}
    W(\CC')
    \,=\,
        \Omega^{-1}_\S\hnem(\P_-)\mem \Omega_\S(\P_+)
    \,=\,
        \mathe^{2(\pi-\e) \vn}
    \,,
\end{align}
if $\P_\pm$ are the endpoints of $\CC'$
such that $\partial\CC' = \P_+ - \P_-$.
In the limit $\e \to 0$,
the contour $\CC'$ approaches the loop $\CC$,
so
\begin{align}
    W(\CC)
    \,=\,
        \lim_{\CC'\to\CC}
            \Tr_\rep W(\CC')
    \,=\,
        \Tr_\rep(\mathe^{2\pi \vn})
    \,=\,
        \rep(\cen)
        \mem\dim\r
    \,,
\end{align}
reproducing the earlier result.
This reinterprets
the AB phase in terms of
the discontinuity $\cen$
of the multivalued gauge parameter $\Omega_\S$
exhibited at a branch cut.
That is,
the multivaluedness of $\Omega_\S$
is measured by
the jump entailed in ``gluing'' its values across a branch cut.

\subsection{Mathematician's Approach}
\label{K1mv.math}

To a mathematician,
the Dirac string gauge potential
in \eqref{adirac.A}
is perhaps best formulated
by not mentioning the multivalued function $\phi_\S$.
It is rather directly understood as
\begin{align}
    A
    \,=\,
        2\pi n\mem \a_\S
    \,,
\end{align}
with the aid of 
a representative for
an integral de Rham cohomology element:
\begin{align}
    \label{1form.ainH}
    \a_\S
    \,=\,
        \frac{x\mem dy - y\mem dx}{
            x^2 + y^2
        }
    \,,\quad
    [\a_\S]
    =
        1
    \,\in\,
    H^1(\M{\setminus}\S,\Z)
    \cong
        \Z
    \,.
\end{align}
It should be clear that
this exactly is the formalization
given around \eqref{a-linking}.
In particular,
for any closed loop $\CC$ in $\M{\setminus}\S$,
\eqref{1form.ainH} states that
\begin{align}
    \label{1form.a-linking}
    \oint_{\CC}
        \a_\S
    \,=\,
        \link(\CC,\S)
    \,.
\end{align}

\medskip
However, suppose one insists that
a satisfactory formulation
should also be given to
the multivalued function $\phi_\S$.
There can be many ways to achieve so,
but our favorite version might be the following.

In essence,
$\phi_\S$ is
a circle-valued function
defined over the punctured domain:
$\phi_\S \inn \Cinfty(\M{\setminus}\S,2\pi\R/\Z)$.
To characterize its winding,
mathematicians study how $\phi_\S$
arises by gluing local data.

To this end,
suppose an open cover $\{\CU_i\}$ of $\M{\setminus}\S$.
For each patch $\CU_i$,
let $\phi_i \inn \Cinfty(\CU_i,\R)$ be the restriction of 
the azimuthal angle
as a single-valued function
via a branch choice.
It follows that
each overlapping region $\CU_i \cap \CU_j$
is equipped with an integer $\n_{ij}$ via
$\phi_i \mminus \phi_j \eqq 2\pi \n_{ij}$.
It also follows that
$\n_{ij} + \n_{jk} + \n_{ki} = 0$
on each triple overlap $\CU_i \cap \CU_j \cap \CU_k$.
Note that $\n_{ij} = -\n_{ji}$.

A collection of data from patch overlaps
such as $\{\n_{ij}\}$
is referred to as a {\Cech} 1-cochain:
$\{\n_{ij}\} \in \cC^1(\M{\setminus}\S,\Z)$.
The condition 
$(\delta\n)_{ijk} = \n_{ij} + \n_{jk} + \n_{ki} = 0$
is known as {\Cech} cocycle condition,
where $\delta$ is the {\Cech} coboundary operator.
Hence 
it is said that
$\{\n_{ij}\}$ defines
not only a {\Cech} 1-cochain but
a {\Cech} 1-cocycle:
$\{\n_{ij}\} \in \cZ^1(\M{\setminus}\S,\Z)$.
See, e.g., \rrcite{hatcher2002algebraic,bott1982differential}.

\begin{figure}[t]
    \centering
    \includegraphics[width=0.96\linewidth]{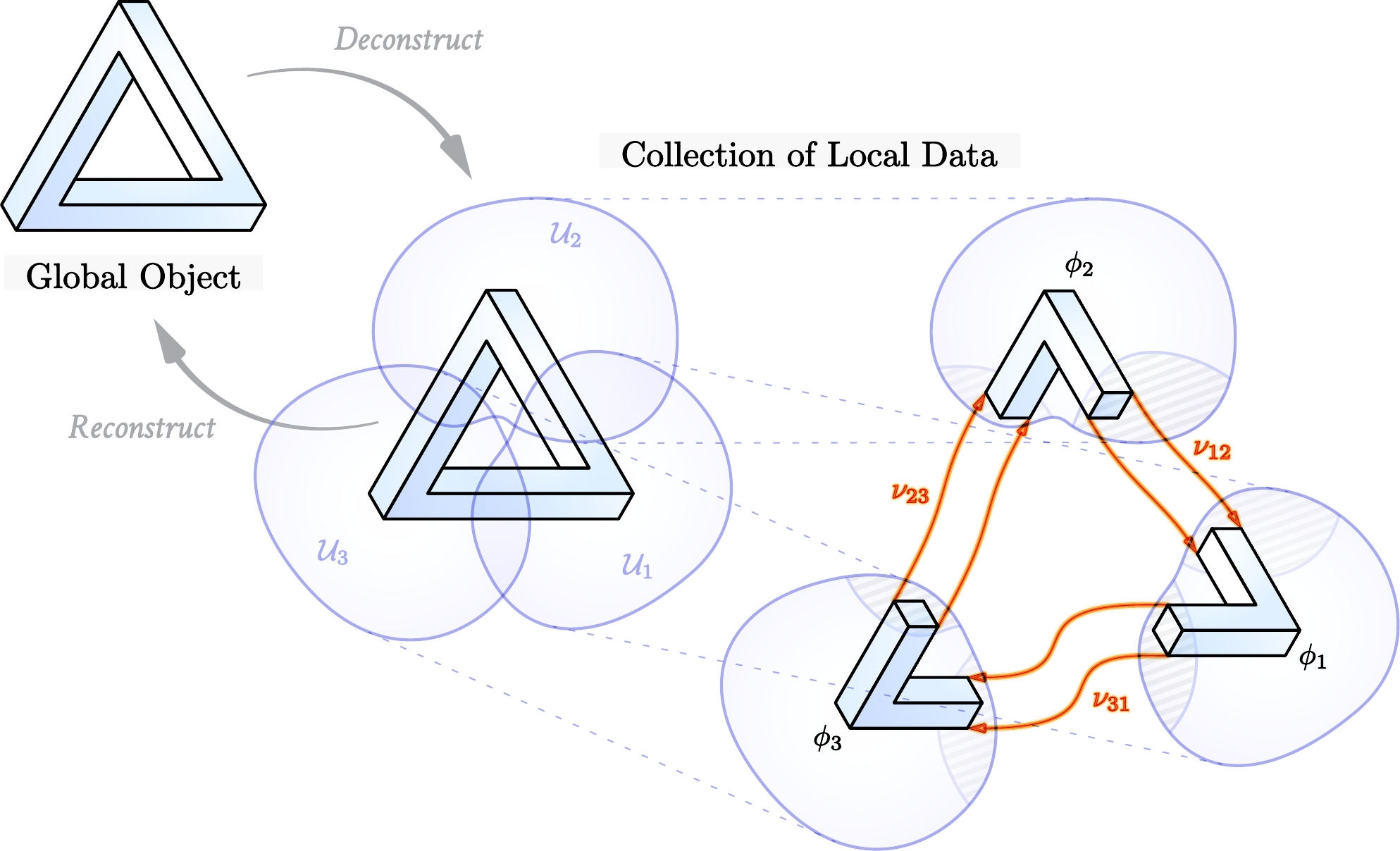}
    \vspace{.18\baselineskip}
    \caption{%
        {\Cech} cohomology studies
        whether a collection of local data
        can be consistently glued into a global object.
    }
    \label{fig:Cech}
\end{figure}

\rcite{penr04}, Sec.\:33.9
provides some helpful visual intuitions for {\Cech} cohomology.
See also \fref{fig:Cech},
which recalls \rcite{penrose1992cohomology}.

Surely, there is an arbitrariness in picking $\phi_i$ for each $\CU_i$,
which is inherited to $\n_{ij}$:
$\{ \n_{ij} \} \sim \{ \n'_{ij} \} $
if there exists a collection of integers $\{\eta_i\} \in \cC^0(\M{\setminus}\S,\Z)$ 
assigned to each patch
such that
$\n'_{ij} - \n_{ij} = (\delta\eta)_{ij} = \eta_i - \eta_j$.

Invariantly, however,
it turns out that
$\{\n_{ij}\}$
is a {\Cech} 1-cocycle
that is not a {\Cech} coboundary of any 0-cocycle.
It represents a nontrivial element
of a {\Cech} cohomology,
\begin{align}
    [\{ \n_{ij} \}]
    = 1
    \,\in\,
        \cH^1(\M{\setminus}\S,\Z)
    \cong \Z
    \,.
\end{align}

We should talk in terms of explicit examples.
Suppose $\M{\setminus}\S$ is covered by
four patches,
\begin{align}
\begin{split}
    \label{fourpatches}
    U_1 
    \mem&=\mem
        \big\{\,
            (r\cos\s_1,r\sin\s_1,z) \inn \R^3
        \,\big|\,
            r > 0
            \mem,\,
            \s_1 \inn ( \mminus\e, \pi/2 \mplus\e)
            \mem,\,
            z \inn \R
        \,\big\}
    \,,\\
    U_2
    \mem&=\mem
        \big\{\,
            (r\cos\s_2,r\sin\s_2,z) \inn \R^3
        \,\big|\,
            r > 0
            \mem,\,
            \s_2 \inn ( \pi/2 \mminus\e, \pi \mplus\e)
            \mem,\,
            z \inn \R
        \,\big\}
    \,,\\
    U_3
    \mem&=\mem
        \big\{\,
            (r\cos\s_3,r\sin\s_3,z) \inn \R^3
        \,\big|\,
            r > 0
            \mem,\,
            \s_3 \inn ( \pi \mminus\e, 3\pi/2 \mplus\e)
            \mem,\,
            z \inn \R
        \,\big\}
    \,,\\
    U_4
    \mem&=\mem
        \big\{\,
            (r\cos\s_4,r\sin\s_4,z) \inn \R^3
        \,\big|\,
            r > 0
            \mem,\,
            \s_4 \inn ( 3\pi/2 \mminus\e, 2\pi \mplus\e)
            \mem,\,
            z \inn \R
        \,\big\}
    \,,
\end{split}
\end{align}
where $\e > 0$ is infinitesimal.
In this case,
the collection of integers
on patch overlaps,
$\{\n_{ij}\} \in \cZ^1(\M{\setminus}\S,\Z)$,
can be given as
\begin{align}
    \label{cech.nlist}
    (\n_{12},\n_{23},\n_{34},\n_{41})
    \,=\,
        (0,0,0,1)
    \mem,\,\,
        (0,-1,0,2)
    \mem,\,\,
        (-3,7,-2,-1)
    \mem,\,\,
    \cdots
    \,.
\end{align}
The first case can arise by choosing
$\phi_i = \s_i$.
The second case can arise by choosing
$\phi_1 = \s_1 - 2\pi$,
$\phi_2 = \s_2 - 2\pi$,
$\phi_3 = \s_3$,
and
$\phi_4 = \s_4$.
The third case can arise by choosing
$\phi_1 = \s_1 + 4\pi$,
$\phi_2 = \s_2 + 10\pi$,
$\phi_3 = \s_3 - 4\pi$,
and
$\phi_4 = \s_4$.
Invariantly, we find $\n_{12} + \n_{23} + \n_{34} + \n_{41} = 1$.

The equivalence between $\{\n_{ij}\}$
in \eqref{cech.nlist}
can also be demonstrated explicitly.
For example,
take the difference between the second and third realizations:
$(3,-8,2,3)$.
It can be seen that this describes the {\Cech} coboundary of
$(\eta_1,\eta_2,\eta_3,\eta_4) = (5,2,10,8)$,
for instance.
This explicitly constructs a ``gauge transformation'':
$(\delta\eta_{12},\delta\eta_{23},\delta\eta_{34},\delta\eta_{41}) = (3,-8,2,3)$.

\medskip

It remains to provide the formalization for the nonabelian case.
Earlier,
we have introduced
$\Omega_\S$
as a multivalued gauge parameter
that exhibits a branch cut discontinuity
$\cen \inn Z(\G)$.
In essence,
this is a $\G/Z(\G)$-valued map,
$\Omega_\S \in \Cinfty(\M{\setminus}\S,\G/Z(\G))$,
locally lifted to $\G$
with $Z(\G)$-valued transition jumps
such that a net monodromy $\cen$ is exhibited around $\S$.

In the {\Cech} picture,
this global object $\Omega_\S$ is 
deconstructed into local data $\{\Omega_i\} \inn \cC^0(\M{\setminus}\S,\G)$
by again supposing an open cover $\{\CU_i\}$ of $\M{\setminus}\S$.
For each patch $\CU_i$,
we choose a group-valued function
$\Omega_i \inn \Cinfty(\CU_i,\G)$.
On each overlapping region $\CU_i \cap \CU_j$,
we stipulate that
\begin{align}
    \label{OO=z}
    \Omega_i\hem \Omega_j^{-1}
    \,=\,
        z_{ij}
    \,\in\,
        Z(\G)
    \,,
\end{align}
where $\{ z_{ij} \} \in \cZ^1(\M{\setminus}\S,Z(\G))$ is a center-valued {\Cech} 1-cocycle.
This means that,
on each triple overlap $\CU_i \cap \CU_j \cap \CU_k$,
we impose
\begin{align}
    z_{ij}\mem z_{jk}\mem z_{ki}
    \,=\,
        \id
    \,.
\end{align}
The topological nontriviality of $\{\Omega_i\}$
arises
by stipulating that
$\{ z_{ij} \}$ represents
a nontrivial {\Cech} cohomology element,
\begin{align}
    \label{cechalpha}
    [\{z_{ij}\}]
    \,=\,
        \cen
    \,\in\,
        \cH^1(\M{\setminus}\S,Z(\G))
    \cong
        Z(\G)
    \,.
\end{align}
This characterizes the multivaluedness of $\Omega_\S$
as an obstruction to a single-valued $\G$-lift
by $\cen \inn Z(\G)$.

A few remarks are in order.
Firstly,
note that the adjoint action of $\Omega_\S$
is globally defined as a single-valued operation.
This can be seen from \eqref{OO=z}:
$\Ad{\Omega_i}\nem = \Ad{\Omega_j}$.

Secondly,
it should be clear that
this construction 
describes a generic multivalued gauge transformation parameter $\Omega_\S$,
so the earlier
$
    \Omega_\S
    =
        \exp\mem(\vn\hem \phi_\S)
$
with $\mathe^{2\pi \vn} = \cen$
from \eqref{1form.AFO}
has provided just one simple realization
of $\Omega_\S$
among many possibilities;
note how the color reference $\vn^a$ appears \textit{nowhere} in the above definition of $\Omega_\S$.
Explicitly, recall the covering of $\M{\setminus}\S$ in \eqref{fourpatches}.
Assuming $\G = \SU(3)$ for concreteness,
the center-valued {\Cech} $1$-cocycle
can be chosen as
$(z_{12},z_{23},z_{34},z_{41})
= (\id,\mathe^{-2\pi i/3}\mem \id,\id,\mathe^{4\pi i/3}\mem \id)
$, so
\begin{align}
    [\{z_{ij}\}]
    \,=\,
        \mathe^{2\pi i/3}\mem \id
    \,\in\,
    Z(\G) = \Z_3
    \,.
\end{align}
A possible local deconstruction of $\Omega_\S$
can be
\begin{align}
    \Omega_1
    \mem=\mem
        g(\s_1)
    \,,\quad
    \Omega_2
    \mem=\mem
        g(\s_2)
    \,,\quad
    \Omega_3
    \mem=\mem
        \mathe^{2\pi i/3}\mem
        g(\s_3)
    \,,\quad
    \Omega_4
    \mem=\mem
        \mathe^{2\pi i/3}\mem
        g(\s_4)
    \,,
\end{align}
where $g(\s)$ is given by
\begin{align}
    \kern-0.3em
        \lrp{\,\begin{matrix}
            \mathe^{i\s/3} \cos^2\s + \mathe^{-2i\s/3} \sin^2\s
            &
            0
            &
            \bigbig{
                \mathe^{i\s/3} - \mathe^{-2i\s/3}
            }\mem \sin\s \cos\s 
            \\
            0 & \mathe^{i\s/3} & 0
            \\
            \bigbig{
                \mathe^{i\s/3} - \mathe^{-2i\s/3}
            }\mem \sin\s \cos\s 
            &
            0
            &
            \mathe^{-2i\s/3} \cos^2\s + \mathe^{i\s/3} \sin^2\s
        \end{matrix}\,\,}
    \,,
    \kern-0.3em
\end{align}
so $g(2\pi) = \mathe^{2\pi i/3}$.
Note how this example
describes a complicated instance of $\Omega_\S$
that encodes a continuously (up to the center jumps) varying direction in the color space,
instead of using a fixed color direction $\vn^a$
as in \eqref{1form.AFO}.

Last but not least,
the {\Cech} picture ensures that
the composition
$\Omega_\S\hem \tilde{\Omega}_\S$
of two multivalued gauge parameters
is well-defined.
Let $\{\Omega_i\} \inn \cC^0(\M{\setminus}\S,\G)$
and $\{\tilde{\Omega}_i\} \in \cC^0(\M{\setminus}\S,\G)$
be local deconstructions of
$\Omega_\S$ and $\tilde{\Omega}_\S$,
respectively.
It follows that
$\{\Omega_i\hem \tilde{\Omega}_i\} \inn \cC^0(\M{\setminus}\S,\G)$
also qualifies as a local deconstruction
of a multivalued gauge parameter.
\eqref{OO=z} implies that
$
    (\Omega_i \tilde{\Omega}_i)(\Omega_j \tilde{\Omega}_j)^{-1}
    = \Omega_i\mem \tilde{z}_{ij}\mem \Omega_{i}^{-1}
    \eqq z_{ij} \tilde{z}_{ij}
$, so
$\{ z_{ij} \tilde{z}_{ij} \}$
can be shown to
define a center-valued $1$-cocycle
that represents
$[\{ z_{ij} \tilde{z}_{ij} \}] = \omega\mem \tilde{\omega} \in \cH^1(\M{\setminus}\S,Z(\G))$.
The composition 
$\Omega_\S\hem \tilde{\Omega}_\S$
is defined as
the multivalued function
whose local deconstruction is
$\{\Omega_i\hem \tilde{\Omega}_i\}$.

For instance,
suppose
$\Omega_\S = \exp\bigbig{\vn\hem \phi_\S}$
and
$\tilde{\Omega}_\S = \exp\bigbig{\tilde{\vn}\hem \phi_\S}$.
The composition of these two gauge parameters
is 
$
    \Omega_\S\hem \tilde{\Omega}_\S
    = \exp\bigbig{\vn\hem \phi_\S} \exp\bigbig{\tilde{\vn}\hem \phi_\S}
$,
whose multivaluedness does lie within $Z(\G)$.

On a related note,
the group multiplication rule for the symmetry operator,
\begin{align}
    U_{\cen}(\S)\, U_{\tilde\cen}(\S) 
    \,=\,
        U_{\cen\tilde\cen}(\S)
    \,,
\end{align}
is required axiomatically.
Since our symmetry operator implements a multivalued gauge transformation,
the composition of two such transformations automatically yields
a third, 
so we know a priori that
the group composition law is valid.\footnote{
    Establishing this 
    at the level of the representatives
    could be yet subtle.
    Suppose $U_{\cen}(\S)$ and $U_{\tilde\cen}(\S)$
    realized with color references $\vn$ and $\tilde \vn$
    in terms of \eqref{1form.U}.
    In the path integral formalism,
    they naively merge into
    $\exp\bigbig{ \int_\S B_a\mem (\vn \mplus \tilde \vn)^a }$.
    Confusingly,
    the multivaluedness of
    $\exp\bigbig{\hnem(\vn\mplus\tilde{\vn})\hem \phi_\S}$
    is not guaranteed to
    lie within $Z(\G)$:
    irrational period.
    To this end, one can realize
    $\cen$ and $\tilde\cen$
    as
    $\exp\bigbig{
        2\pi \Ad{\tilde{\Omega}^{-1}} \vn
    }$
    and
    $\exp\bigbig{2\pi \vn}$.
}

\subsection{Explicit Examples}

\paragraph{Wiggly String}

We end by providing more explicit examples
of the multivalued gauge transformation.
Our last example demonstrated that the direction of the color factor can vary.
The following example instead fixes the color direction
but deforms the shape of the Dirac string
to a ``wiggly'' shape:
\begin{align}
    \label{Omega.wiggly}
    \Omega_\S
    \,=\,
    \exp\bigg(\mem{
        \frac{k}{N}\mem
        \t\mem
            \phi_\S
    }\mem\bigg)
    \transition{where}
    \tan\phi_\S = \frac{y{\,-\,}Y(z)}{x{\mem-\hem}X(z)}
    \,.
\end{align}
Here, the string is realized along
$\{\mem (X(z),Y(z),z) \mem|\mem z\in\R\, \}$.
Indeed, a simple calculation verifies that 
the field strength
$F = \Omega^{-1}dd\Omega$,
arising by applying the multivalued gauge transformation due to \eqref{Omega.wiggly}
to the flat background,
is proportional to
\begin{align}
        \delta\bigbig{
            x{\mem-}X(z)
        }\mem \delta\bigbig{
            y{\mem-}Y(z)
        }
        \,
        d\bigbig{
            x{\mem-}X(z)
        } \swedge\mem d\bigbig{
            y{\mem-}Y(z)
        }
    \,,
\end{align}
precisely detecting the wiggly support.
It is obvious that the AB phase
for an electric charge looping around the string
is not modified by these wiggles,
demonstrating the topological nature
of the Dirac string.

\paragraph{Gauge Theory at Finite Temperature}%
Another interesting example is gauge theory at finite temperature, described by a compact product manifold with compactified Euclidean time:
\begin{align}
    \M \,=\, \M_3 \times \mathrm{S}^1
    \,,\quad
    t
        \,\sim\,
    t + \beta
    \,.
\end{align}
Here, $\M_3$ is a three-manifold.
In addition to the trivial vacuum, there is an infinite set of gauge equivalence classes for the background gauge field.
For example, consider
\begin{align}
    \label{eq:YM.exB.gf}
        A^a 
        \,=\,
            p\,
            \frac{2\pi \t^a}{\beta}\mem dt
    \transition{where}
        p \in \mathbb{Z}
    \,.
\end{align}
In this background,
Wilson loops winding about the thermal circle
are trivial.
Meanwhile, consider the following multivalued gauge transformation:
\begin{align}
    \label{eq:YM.exB}
    \Omega
    \,=\,
    \exp\bb{
        \frac{2\pi k \t}{N}
        \frac{t}{\beta}
    }
    \,.
\end{align}
Via \eqref{1form.tgt},
\eqref{eq:YM.exB} maps the background in \eqref{eq:YM.exB.gf} to
\begin{align}
   \label{eq:YM.exB.gf'}
        A^a &= \bb{
            p {\,+\,} \frac{k}{N}
        }\mem \frac{2\pi \t^a}{\beta}\mem dt
    \,,
\end{align}
so the Wilson loops gain a nontrivial phase factor of
$\mathe^{2\pi ik/N}$
per each thermal circle.
Hence, the multivalued gauge transformation has induced a monodromy in the time direction.

Note that \eqrefs{eq:YM.exB.gf}{eq:YM.exB.gf'}
can be obtained by
identifying the ends of
a flat gauge field configuration in $\M_3$ times an interval
with a twisted boundary condition.
An implementation of this construction in Lorentzian signature can be found in \rcite{Gomes:2023ahz}.

This example exhibits several interesting features,
originating from the topological nontriviality of
the spacetime $\M$.
A multivalued gauge transformation 
is constructed without addressing
the singular support $\S$.
The Wilson loops do not generally admit a coboundary,
so they can be closed but not exact.

\paragraph{Compact Support}%
Lastly, we wish to provide an example that explicitly demonstrates that
the support $\S$ of the Dirac string can be compact.
This is meaningful because
the multivalued gauge transformation for the one-form symmetry
takes $\S$ as a closed surface.
We take $\M \cong \R^4$,
coordinatized by $(t,x,y,z)$.

We employ what is known as toroidal coordinates
$(\eta,\sigma,\phi)$,
which foliate the three-dimensional space of $(x,y,z)$ according to
a ``reference ring'' on the $x$-$y$ plane.
For our purposes, we take its radius to be a function of time, $a(t)$:
\begin{align}
\begin{split}
    (x,y) \,=\, \frac{
        a(t) \sinh\eta
    }{\cosh\eta - \cos\sigma}
        (\cos\phi,\sin\phi)
    \,,\quad
    z \,=\, \frac{
        a(t) \sin\sigma
    }{\cosh\eta - \cos\sigma}
    \,.
\end{split}
\end{align}
While $\phi$ is the azimuthal angle,
surfaces of fixed $\eta \geq 0$ label concentric two-tori which enclose the reference ring, while surfaces of fixed $-\pi {\,\leq\,} \sigma {\,<\,} \pi$ label two-spheres which intersect the reference ring.
Also, spatial infinity corresponds to $\sigma=0$.
Let us choose the branch cut for $\sigma$
such that the discontinuity at $\sigma=\pm \pi$
develops on the disc enclosed by the reference ring.

Crucially,
$\sigma$ describes an angular coordinate that winds like a solenoid about the reference ring.
With this understanding, consider
\begin{align}
    \label{eq:Omega-amp}
    \Omega_\S
    = 
    \exp\bb{
        \frac{k}{N}\mem
        \t\mem
        \sigma
    }
    \,.
\end{align}
It can be seen that \eqref{eq:Omega-amp}
is the multivalued gauge parameter
that creates a Dirac string along
$\S = \{\, (t,x,y,z) \in \R^4 \,|\, x^2+y^2 = a^2(t) \mem,\, z \eqq 0 \,\}$.
Clearly, it induces a rephasing for any Wilson loop that wraps the reference ring.
In the spacetime view,
the branch cut for \eqref{eq:Omega-amp} resides on a volume $\V$ corresponding to the region swept out by the disc enclosed by the reference ring.
Furthermore, $\Omega_\S$ nicely approaches the identity at spatial infinity.

By choreographing the function $a(t)$ to smoothly
increase from and decrease to zero
within a time interval $t \inn [t_1,t_2]$,
the surface $\S$ is made to take finite support in both time and space,
parameterized by
$t \inn [t_1,t_2]$ and $\phi\inn[0,2\pi[$
as
$(t, a(t) \cos\phi, a(t) \sin\phi, 0)$.

\paragraph{Existence Argument}

In general, one wants to construct
the multivalued gauge parameter $\Omega_\S$
for an arbitrarily shaped codimension two surface $\S$
in an arbitrary manifold $\M$,
with or without boundaries.
How does one know that such an $\Omega_\S$ always exists, given some choice of $\S$ and $\M$?

It can be argued that
an $\Omega_\S$ can always be found
for a given $\S$ and $\M$,
on account of an analogous question in \textit{classical magnetostatics}.
Deducing $\Omega_\S$ from $\S$ and $\M$ is mathematically identical to deducing the static magnetic field and potential
of an electric current loop.
Imagine we are experimentalists who construct a loop of electrical line current $\mathbf{J}$, built to specification according to some arbitrary contour.
On account of  Amp\`ere's law, ${\nabla}{\hem\times\mem} \mathbf{H} = \mathbf{J}$, we can deduce the magnetic field $\mathbf{H}$ by simply ``measuring'' it. 
In regions away from the current, we can then reconstruct a magnetic scalar potential via $\mathbf{H} = {-\mathbf{\nabla}\Psi}$.
If $\M$ is not closed, we make its boundary $\partial\M$ superconducting to enforce the boundary condition $\mathbf{H}_\perp {\hem=\,} 0$, in which case $\Psi$ will be set to a constant over $\partial\M$ that we fix to zero.
Finally, $\Omega_\S$ arises by exponentiating $\Psi$
while $\mathbf{H}$ and $\mathbf{J}$
correspond to
$\Omega^{-1}_\S d\Omega_\S$ and
$\Omega^{-1}_\S dd\Omega_\S$,
respectively.

\section{ Application to Gravity}
\label{K1.grav}

Do gravitational higher-form symmetries exist?
Suppose gravity admits a higher-form symmetry.
On account of the symmetry triangle in \fref{SymTriangle},
the symmetry operator will generate a vortex configuration.
One can argue that
this vortex
must \textit{gravitate topologically},
since otherwise
the symmetry operator cannot be topological.
For instance,
the vortex
must not produce a Newtonian $1/r$ gravitational field.
One should not be able to deduce
the presence of the vortex
by measurements bound to each local laboratory.
The vortex
will be detected
only through
global interferometer-type experiments.

This provides an intriguing top-down angle
unique to gravity.
To relativists,
it has been known that
string defects in gravity fall into two categories:
\begin{itemize}[itemsep=3pt,topsep=5pt]
    \item 
        \textit{Cosmic string}
        \cite{Kibble:1976sj,Hindmarsh:1994re}.
        A concentration of stress-energy,
        characterized by the tension $\mu$
        (energy per unit length).
        Detected by
        measuring the conical deficit angle
        $\Del\theta = 8\pi G \mu$
        in terms of scalar particle geodesics.
        Can be obtained via removing a ``wedge'' from spacetime and performing a gluing.
        Entails a distributional Ricci curvature
        and is sourced by the Nambu-Goto worldsheet.
        See \rrcite{Vilenkin:1981zs,Gott:1984ef,Hiscock:1985uc,Linet:1985rw}
        and also
        \rrcite{Garfinkle:1985ax,Geroch:1987qn,Vickers:1998se,Vilenkin:2000jqa}.
    \item
        \textit{Misner string} \cite{Misner:1963flatter,Bonnor:1969ala,sackfield1971physical,Mazur:1986gb,GP_2009_Ch341}.
        A thin flux tube of NUT charge,
        characterized by the total gravitomagnetic flux $N$.
        Serves precisely as the gravitational counterpart of the Dirac string
        when the NUT charge is seen as the gravitomagnetic charge
        \cite{demianski1966combined,Plebanski:1975xfb}.
        Detected by measuring the
        time monodromy $\Del t = 8\pi G N$
        as a pure frame dragging effect:
        Sagnac interferometry
        \cite{Sagnac:1913ether,Sagnac:1913preuve}
        as gravitational AB effect
        \cite{Ashtekar:1975Sagnac,sakurai1980comments,Anandan:1981Sagnac}.
        Involves no conical deficit angle
        and thus is tensionless.
        Describes the very string defect
        of the Taub-NUT solution.
        See also \rrcite{LL:1975vol2,dowker1974nut,dowker1967gravitational,Ramaswamy:1981JMP,zee1985gravitomagnetic,samuel1986gravitational,zimmerman1989geodesics,maartens1998gravito,bunster2006monopoles,magnon1987mass,Alfonsi:2020lub,note-sdtn}.
\end{itemize}
A top-down angle for relativists
would be constructing
a one-form symmetry
that corresponds to each of these topologically gravitating defects.

Meanwhile,
we see that the universal template
established in \eqref{K1b}
rather immediately implies the existence of a one-form symmetry in tetradic Palatini gravity,
valued in the center of the local Lorentz group.
Amusingly,
this identifies a string defect
that is subtly different from the standard cosmic string:
\begin{itemize}[itemsep=3pt,topsep=5pt]
    \item 
        \textit{Tetradic Cosmic String}.
        A concentration of stress-energies
        that creates quantized spin precession angles.
        Detected by interferometry with 
        fermions,
        but
        \textit{cannot} be detected by 
        scalar particle geodesics.
        Hence differs from the standard cosmic string.
\end{itemize}
As a related, more elementary object,
we will also find the following:
\begin{itemize}[itemsep=3pt,topsep=5pt]
    \item 
        \textit{Chiral Tetradic Cosmic String}.
        A concentration of 
        a chiral (SD or ASD) stress-energy
        that creates quantized spin precession angles.
        Namely, a combined concentration of
        both electric and magnetic
        stress-energies
        in equal magnitudes.
        Detected by interferometry with 
        spinning particles, including fermions,
        but
        \textit{cannot} be detected by 
        scalar particle geodesics.
        Entails a distributional Ricci curvature
        as well as a distributional violation of the algebraic Bianchi identity.
\end{itemize}
Below, we investigate the gravitational one-form symmetry
whose vortex configuration
is the tetradic cosmic string,
although investigating other proposals
such as the Misner string
will be an intriguing research direction
(see \Sec{K2>Misner}).

\subsection{Gravity as a Lorentz Gauge Theory}

To recapitulate,
the universal construction in \Sec{K1b}
establishes 
the existence of a one-form symmetry in
nonabelian gauge theories
of a finite-dimensional matrix Lie group $\G$
by deriving
the Ward identity,
in which the Lagrangian top form, the charged operator, and the symmetry operator are
\begin{subequations}
    \label{recap.1form}
\begin{align}
    L[A,B]
    \,&=\,
        \frac{1}{g^2}\,\BB{
            B_a
            \wedge F^a[A]
            \,+\, f[B]
        }
    \,,\\
    W_\rho(\CC)
    \,&=\,
        \Tr_\rho
        \bP
        \exp\bb{
            \oint_\CC A
        }
    \,,\\
    U_\a(\S)
    \,&=\,
        \exp\bb{
            - \frac{i}{g^2}  \int_\M  
                B_a \wedge (\Omega_\S^{-1} dd\Omega_\S)^a
        }
    \,,
\end{align}
\end{subequations}
respectively.
The center element $\a \inn Z(\G)$
and the closed oriented submanifold support $\S \msubset \M$
of codimension two
provide the invariant characterization of 
the multivalued gauge parameter $\Omega_\S$.

For instance,
take YM theory defined on a background 
$d$-dimensional pseudo-Riemannian manifold $\M$.
In the first-order formulation,
the Lagrangian $d$-form reads
\begin{align}
    \label{eq:YM.L}
    L[A,B]
    \,=\,
        \frac{1}{g^2}\mem B_a \swedge F^a[A]
        - \frac{1}{2g^2} B_a \swedge {*}B^a
    \,,
\end{align}
where ${*}$ denotes the Hodge dual acting on spacetime indices.
The universal construction of \Sec{K1b}
immediately proves
the electric one-form symmetry,
identifies the Wilson loops as the charged operators,
and constructs the symmetry operator
explicitly in terms of the theory's local degrees of freedom.

\medskip

Now consider gravity in four spacetime dimensions.
The tetradic Palatini formulation of GR
describes a gauge theory of a Lorentz group $\G$,
i.e., a Lie group whose algebra
$\g = \Lie(\G)$ is either
$\so(1,3)$ or $\so(4)$ depending on the signature.
To be concrete,
we may presume $\g = \so(1,3)$
for the moment.

The Lagrangian four-form reads
\begin{align}
    \label{eq:eGrav.L.vec()}
    L[\gamma,e]
    \,=\, 
        \frac{1}{4\k^2}\mem 
        \e_{ABCD}\mem
        e^A \swedge e^B \swedge R^{CD}[\gamma]
        - \frac{\Lambda}{24\k^2}\mem
        \e_{ABCD}\mem e^A \swedge e^B \swedge e^C \swedge e^D
    \,,
\end{align}
where 
the internal indices
$A,B,\cdots \in \{0,1,2,3\}$
are the local Lorentz indices,
and $\e_{ABCD}$ is the permutation symbol
that is nondynamical.
We have assumed a generic cosmological constant $\L$
and identified the gravitational coupling as
$\k = \sqrt{8\pi\GN}$.

The coframe one-form $e^A$
provides local trivializations of the cotangent bundle $\Tstar\M$.
It defines the spacetime metric
as a \textit{composite} operator,
\begin{align}
    \label{eq:g[e]}
    g_{\m\n}[e]
    \,=\,
        \eta_{AB}\mem e^A{}_\m\mem e^B{}_\n
    \,,
\end{align}
via an invariant constant internal metric $\eta_{AB}$
that facilitates the raising and lowering of the internal indices.

The spin connection one-form
$\gamma^{AB} = -\gamma^{BA}$
refers to a Lorentz-valued one-form
that defines a connection
on the tangent bundle $T\M$.
Its covariant exterior derivative $D$ preserves the constant internal metric $\eta_{AB}$.
Its curvature two-form $R^A{}_B[\gamma]$
is defined as
\begin{subequations}
\label{eq:fsdef|GR}
\begin{align}
    \label{eq:Grav.fieldstrength}
    R^A{}_B[\gamma]
    \,=\,
        d\gamma^A{}_B
        + \gamma^A{}_C \wedge \gamma^C{}_B
    \,,
\end{align}
which precisely reproduces \eqref{eq:fsdef}.
In fact, by identifying the antisymmetric pair of fundamental indices $[AB]$ with an adjoint index $a$,
\eqref{eq:Grav.fieldstrength} can be written as
\begin{align}
    \label{eq:Grav.fsdef}
    R^a[\gamma]
    \,=\,
        d\gamma^a
        + \frac{1}{2}\, f^a{}_{bc}\mem \gamma^b \wedge \gamma^c
    \,,
\end{align}
\end{subequations}
if $f^a{}_{bc}$ denotes the structure constants
of the Lorentz algebra $\g$.

The Lagrangian four-form in \eqref{eq:eGrav.L.vec()}
is quadratic in the spin connection $\gamma^A{}_B$.
Its EoM give $D\bigbig{e^A \swedge e^B} = 0$,
which is algebraically\footnote{
    The torsion is non-propagating:
    any coupling of the spin connection to an external source
    generates nonzero torsion precisely only
    on the support of that source.
} equivalent to the vanishing of
the torsion two-form
$De^A = de^A + \gamma^A{}_B \swedge e^B$
provided $e^A$ is nondegenerate.
This analysis shows that
$\gamma^A{}_B = \gamma^A{}_{BC}\mem e^C$ 
can be viewed as an auxiliary field
that can be
eliminated at tree level 
by plugging its classical solution 
in terms of $e^A$
which the torsion-free condition stipulates.
This means to equate $\gamma^A{}_{BC}$ with
the connection coefficients of the Levi-Civita connection
of the metric in \eqref{eq:g[e]}
in the local Lorentz frame.
It is not difficult to show that
the resulting action
is equated with the Einstein-Hilbert action
for the metric in \eqref{eq:g[e]}.
Variation with respect to the coframe $e^A$ 
precisely reproduces Einstein's equations
with the cosmological constant $\L$.

In conclusion,
tetradic Palatini gravity is 
classically equivalent
to the metric theory of GR,
provided there are no sources that couple directly to the spin connection.
As noted earlier, quantum equivalence also follows
since the EFT of a massless spin-two particle is unique, modulo Wilson coefficients.

As emphasized earlier, we work in an EFT description of gravity which describes gravitons propagating over a fixed background. 
Hence, throughout our analysis 
the coframe and metric
are implicitly expanded as fluctuations about some choice of background values
$\bar e^A{}_\m$ and $\bar g_{\m\n}$,
respectively, though it will usually be simpler to manipulate the full field variables rather than their fluctuations.\footnote{
    In any sensible EFT, the background spacetime is nondegenerate and thus the background tetrad $\bar e^A{}_\m$ must be nonzero.
    This means the vacuum breaks diffeomorphism invariance and local Lorentz symmetry down to the diagonal, which naively hinders our analysis. 
    However, our calculation of the Ward identity for the symmetry and line operators utilizes the full tetrad field $e^A{}_\m$, which transforms covariantly, so there is no additional complication. 
    The very same phenomenon occurs in gauge theory, where expanding about a background gauge field $\bar A^a{}_\m$ technically breaks Lorentz invariance and color symmetry down to the diagonal, but of course with no effect on the Ward identities in the theory.  
}

\medskip

In consideration of \eqref{eq:Grav.fsdef},
we identify the two-form composite operator
\begin{align}
    \label{eq:pleb-def}
    B_{AB}[e]
    \,=\, 
        \frac{1}{2}\,
            \e_{ABCD}\mem e^C \swedge e^D 
    \,,
\end{align}
which we may want to refer to as 
the Pleba\'nski two-form
\cite{Plebanski:1977zz,Capovilla:1991qb}.
This is a two-form valued in the coadjoint of
the local Lorentz algebra $\g$.
\eqref{eq:pleb-def} facilitates a rewriting of
\eqref{eq:eGrav.L.vec()} as
\begin{align}
    \label{1form.L|GR}
    L[\gamma,e]
    \,=\,
        \frac{1}{\k^2}\mem 
        \bb{
            B_a[e] \wedge R^a[\gamma]
            - \frac{\Lambda}{6}\,
                B_a[e] \wedge {\star B}^a[e]
        }
    \,,
\end{align}
where $a,b,\cdots \in \{1,2,3,4,5,6\}$ are the adjoint indices.
Here, $\star$ 
denotes the Hodge star in the \textit{internal} space.
When the adjoint indices are all converted to
fundamental indices by the generators $(t_a)^{AB} = -(t_a)^{BA}$ of $\g$,
we have
\begin{align}
    B_{AB}[e]
    \,=\,  
        \frac{1}{2}\,
            \e_{ABCD}\mem e^C \swedge e^D
    \,,\quad
    {\star B}^{AB}[e]
    \,=\,
        e^A \wedge e^B
    \,.
\end{align}
Note that the internal Hodge dual $\star$ should be distinguished from the spacetime Hodge dual $\ast$ in \eqref{eq:YM.L}.

When presented as in \eqref{1form.L|GR},
it is evident that
tetradic Palatini gravity
conforms to the ``$BF\mem$'' grammar of 
the universal template in \Sec{K1b}:
compare it with \eqref{1form.L}.\footnote{
    It should be clear that 
    the statement here by no means implies that
    gravity is a topological $BF$ theory,
    even when the cosmological constant term is neglected.
    In \eqref{1form.L|GR},
    the $B$-field refers to the composite operator defined in \eqref{eq:pleb-def}
    and does not constrain all components of $R^{AB}[\gamma]$,
    which is precisely why Einstein's equations do not kill all Riemann components.
    The Pleba\'nski formulation \cite{Plebanski:1977zz}
    could only promote $B_a$ as 
    fundamental field variable
    by imposing a constraint on it
    (the simplicity constraint)
    so that not all components of $B_a$ are independent Lagrange multipliers.
}
The only difference
is that the $B$-field
is only \textit{composite}
for gravity,
via the formula in \eqref{eq:pleb-def}.
Importantly,
we will see shortly that
this does not raise any issue
in the path integral formalism
since 
the $B$-field
exhibits the correct behavior under
the multivalued gauge transformation
regardless of its compositeness.
Hence
the universal construction of \Sec{K1b}
precisely applies to tetradic Palatini gravity
up to a straightforward refinement.

The implication is that
dynamical gravity
exhibits a one-form symmetry
valued in the center $Z(\G)$\footnote{
    To be precise, this statement is
    up to self-screening:
    see the discussion in \Sec{sec:Breaking}.
    The symmetry group can technically 
    be smaller than $Z(\G)$
    for tetradic theories in Euclidean or complexified signatures,
    while, for the {\Pleb} formulation
    where the (co)adjoint-valued $B$-field is fundamental,
    no such subtlety arises in any signature.
    Below,
    we assume that this point
    on self-screening
    is understood.
}
as a gauge theory of the local Lorentz group $\G$.
The summary in \eqref{recap.1form} 
applies as
\begin{subequations}
    \label{recap.1form.GR}
\begin{align}
    L[\gamma,e]
    \,&=\,
        \frac{1}{\k^2}\mem 
        \bb{
            B_a[e] \wedge R^a[\gamma]
            - \frac{\Lambda}{6}\,
                B_a[e] \wedge {\star B}^a[e]
        }
    \,,\\
    W_\rho(\CC)
    \,&=\,
        \Tr_\rho
        \bP
        \exp\bb{
            \oint_\CC \gamma
        }
    \,,
\end{align}\begin{align}
    U_\a(\S)
    \,&=\,
        \exp\bb{
            - \frac{i}{\k^2}  \int_\M  
                B_a[e] \wedge (\Omega_\S^{-1} dd\Omega_\S)^a
        }
    \,,
\end{align}
\end{subequations}
where $W_\rho(\CC)$ is a functional of $\gamma$
and $U_\a(\S)$ is a functional of $e$.

This one-form symmetry
will be preserved by
the inclusion of higher-curvature corrections
for the EFT of gravity.
In any case,
the $B$-field will always describe
the conjugate field variable of the curvature in the EFT.%
\footnote{
    For instance,
    utilizing
    a Lagrangian such as
    $L[\gamma,e,r,B] =
        \k^{-2}\hhem
        \big(\mem
            B_a \wedge ( R^a[\gamma] - r^a )
            + f[e,r]
        \mem\big)
    $
    would incorporate the cases with algebraic higher-curvature operators.
}

\medskip
It remains to elaborate on the viable options for the symmetry group $Z(\G)$.
Note that actions and Lagrangian top forms
do not specify the global structure of the gauge group $\G$
but only specify the gauge algebra $\g$.

In Lorentzian signature,
the viable options for the gauge group $\G$ are\footnote{
    For simplicity, we restrict our attention to connected gauge groups.
}
\begin{align}
    \label{eq:GforL4}
        \Spin(1,3)
        \,\,\cong\,\,
        \SL(2,\mathbb{C}) 
    \,,\qquad
        \SO^+\nem(3,1)
        \,\,\cong\,\,
        \frac{\Spin(3,1)}{\mathbb{Z}_2}
    \,.
\end{align}
The latter is the 
restricted Lorentz group.
The former is the double cover of the latter.
The corresponding center subgroups are, respectively,
\begin{align}
    \label{Z2q}
        \mathbb{Z}_2 
    \,,\qquad
         \id  
    \,.
\end{align}
These are the allowed one-form symmetries of gravity
in Lorentzian signature.
Note that the zero-form symmetry associated with the former corresponds to \textit{net} parity on chiral and antichiral spinor indices, also known as fermion parity.

For completeness of our exposition,
we also discuss Euclidean signature,
in which case
$\g = \so(4)
    \cong
\su(2) {\mem\oplus\,} \su(2)
\cong \mathsf{spin}(4)$.
Hence we can choose 
the gauge group $\G$ 
(which we may refer to as the ``local Lorentz group'' by abuse of terminology)
to be either\footnote{
    A priori, one can also consider semispin groups such as  $\mathsf{SemiSpin}(4)\cong \SU(2) {\mem\times\,} \SO(3)$, which is yet a nonstandard quotient.
    Note also that, in the cases of 
    such semispin groups or
    $\G = \PSO(4) = \SO(4)/\Z_2$,
    the local Lagrangian will be realized in the {\Pleb} formulation
    (reviewed in \Secs{Jcx>COFRAME}{K1:app3})
    for not invoking fields in the vector representation
    such as the coframe $e^A$.
}
\begin{align}
\begin{split}
    \label{eq:Gfor4}
        \Spin(4)
    &\,\cong\,
        \SU(2) {\mem\times\,} \SU(2)
    \,,\\
        \SO(4)
    &\,\cong\,
        \frac{\Spin(4)}{\mathbb{Z}_2}
    \,,\\
        \frac{\SO(4)}{\mathbb{Z}_2}
    &\hem\cong\,
        \frac{\Spin(4)}{\mathbb{Z}_2 {\mem\times\,} \mathbb{Z}_2}
    \,,
\end{split}
\end{align}
whose center subgroups are given by 
\begin{align}
    \label{Z2Z2q}
        \mathbb{Z}_2 {\mem\times\,} \mathbb{Z}_2
    \,,\qquad
        \mathbb{Z}_2
    \,,\qquad
         \id  
    \,,
\end{align}
respectively.
The one-form charges will be valued in 
either of these center subgroups.
Note that the zero-form symmetry associated with 
the center $\mathbb{Z}_2 {\mem\times\,} \mathbb{Z}_2$ of
$\Spin(4)
\cong \SU(2) {\mem\times\,} \SU(2)$
acts as parity on chiral and antichiral spinor indices,
while
that of the center $\mathbb{Z}_2$
of
$\SO(4)$
acts as parity on \textit{vector} indices.\footnote{
    It is instructive to compare
    the cases when the gravitational one-form symmetry group is $\Z_2$.
    In the Lorentzian signature,
    the local Lorentz group should be
    $\G = \Spin(1,3)$,
    in which case vector holonomies are \textit{not charged}.
    In the Euclidean signature,
    we have
    $\G = \SO(4)$
    (for the standard option)
    so fermions do not exist
    while vector holonomies are \textit{charged}.
}

Lastly, as a formal consideration
one may be interested in the case of
the complexified gauge algebra
$\g = \so(4,\C) \cong \sl(2,\C) \moplus \sl(2,\C)$.
The viable center subgroups are again
$\Z_2 {\mem\times\,} \Z_2$
and its quotients
as in \eqref{Z2Z2q}.

\medskip
Below, we explicitly spell out
how the general construction in \Sec{K1b}
applies to
the tetradic Palatini formalism
to establish the one-form symmetry of gravity.
Without loss of generality,
we assume the context of Lorentzian signature
while being agnostic about the local Lorentz group $\G$.
Concretely,
we show that
the dynamical theory of a coframe $e^A$ and a spin connection $\gamma^{AB} = -\gamma^{BA}$,
defined on a four-manifold $\M$
by the Lagrangian four-form
\begin{align}
    \label{1form.L|Gen}
    L[\gamma,e]
    \,&=\,
        \frac{1}{\k^2}\mem 
        \BB{
            B_a[e] \wedge R^a[\gamma]
            + f[e]
        }
    \,,
\end{align}
exhibits a one-form symmetry valued in the center $Z(\G)$.
The definitions in \eqrefs{eq:pleb-def}{eq:Grav.fsdef}
state that $B_a[e]$ and $R^a[\gamma]$
respectively
transform in the coadjoint and adjoint of the local Lorentz group,
while $f[e]$ in \eqref{1form.L|Gen}
is an algebraic four-form functional
invariant under any local Lorentz transformations.

\subsection{Multivalued Local Lorentz Transformation}

The gauge transformations of tetradic Palatini gravity are
\begin{align}
    \label{1form.gt|GR}
    \gamma^A{}_B
    \,\,\mapsto\,\,
        (\Omega^{-1})^A{}_C\mem 
        \gamma^C{}_D\mem
        \Omega^D{}_B
        + (\Omega^{-1})^A{}_C\mem d\Omega^C{}_B
    \,,\quad
    e^A
    \,\,\mapsto\,\,
        (\Omega^{-1})^A{}_B\mem e^B
    \,,
\end{align}
where 
$\Omega$ is a single-valued zero-form parameter
valued in the local Lorentz group $\G$.
\eqref{1form.gt|GR}
leaves the Lagrangian four-form in 
\eqref{1form.L|Gen}
invariant.

The symmetry transformation
is a multivalued local Lorentz transformation.
For a multivalued parameter $\Omega_\S$
that exhibits a nontrivial winding around 
a closed oriented two-dimensional surface $\S \mem\msubset \M$,
it is
\begin{align}
\begin{split}
    \label{1form.tgt|GR}
    \gamma^A{}_B
    \,\,&\mapsto\,\,
        (\Omega^{-1}_\S)^A{}_C\mem 
        \gamma^C{}_D\mem
        (\Omega_\S)^D{}_B
        + (\Omega_\S^{-1})^A{}_C\mem d(\Omega_\S)^C{}_B
    \,,\\
    e^A
    \,\,&\mapsto\,\,
        (\Omega_\S^{-1})^A{}_B\mem e^B
    \,.
\end{split}
\end{align}
Just as in \eqref{eq:lambda-disc1.cen},
$\Omega_\S$ is stipulated to satisfy the center jump condition
\begin{align}
    \label{eq:lambda-disc1|GR}
    \lim_{
        \P_+ \to\, \P_-
    }\hem
        \Omega_\S(\P_{\hnem+})
        \mem 
        \Omega_\S^{-1}\hnem(\P_{\hnem-})
    \,=\,
        \cen
    \,\in\,
        Z(\G)
    \,,
\end{align}
across a branch sheet $\V$ such that $\S \eqq \partial\V$.
The relevant illustration has been given in \fref{fig:W'},
clarifying the meaning of the points $\P_{\hnem+}$ and $\P_{\hnem-}$ in \eqref{eq:lambda-disc1|GR}.

\subsection{Spin Holonomy as Charged Operator}
\label{K1b.W|GR}

The charged operator is
the spin holonomy $W_\rep(\CC)$
for a representation $\r$ of the local Lorentz group $\G$,
i.e., the Wilson loop for the spin connection
about a closed oriented one-dimensional contour in $\M$:
\begin{align}
    \label{1form.W|GR}
    W_\rho(\CC)
    \,=\,
        \Tr_\rho
        \bP
        \exp\bb{
            \oint_\CC \gamma
        }
    \,.
\end{align}

To show that $W_\rep(\CC)$
is charged under the symmetry transformation stated in \eqref{1form.tgt|GR},
it suffices to first consider the case
when $\CC$ links with $\S$ once.
With this assumption, we consider
an auxiliary open contour $\CC'$
defined previously in \fref{fig:W'}.
The Wilson line about this open contour
is the parallel propagator
with respect to the spin connection $\gamma$.
Its bilocal transformation behavior implies
\begin{align}
    W(\CC')
    \,\,\,\mapsto\,\,\,
    \Omega_\S^{-1}(\P_-)\mem W(\CC')\,\Omega_\S(\P_+)
    \,,
\end{align}
so the spin holonomy in \eqref{1form.W|GR} transforms as
\begin{align}
\begin{split}
    W_\rep(\CC)
    \,\,\,\mapsto\,\,\,
    {}&{}
    \Tr_\rep
    \lim_{\P_\pm \to \P} \BB{
        \Omega_\S^{-1}(\P_-)\mem W(\CC')\, \Omega_\S(\P_+)
    }
    \,,\\
    \,=\,
    {}&{}
        \Tr_\rep
        \lim_{\P_\pm \to \P} \BB{
            \Omega_\S(\P_+)\mem \Omega_\S^{-1}(\P_-)\, W(\CC')
        }
    \,,\\
    \,=\,
    {}&{}
        \Tr_\rep\nem
            \BB{
                \lim_{\P_\pm \to \P}
                    \a\mem
                    W(\CC')
            }
    \,=\,
        \rep(\cen)\, W_\rep(\CC)
    \,,
\end{split}
\end{align}
where $\rep(\cen)$ denotes the representation of the center element $\a$ as a complex phase.
Notably, this computation is independent of the branch sheet choice
by virtue of the fact that $\cen$ commutes with
any element in the local Lorentz group $\G$.
Finally, for a generic contour $\CC$,
it follows that
\begin{align}
    \label{1form.Wtransf|GR}
    W_\rep(\CC)
        \,\,\,\mapsto\,\,\,
    \rep(\cen)^{\link(\CC,\S)}\mem
    W_\rep(\CC)
    \,.
\end{align}

\subsection{Ward Identity}
\label{K1.Ward|GR}

The above computation boils down to
the following path integral identity:
\begin{align}
    \label{1form.wardderiv|GR}
    \nonumber
    &
    \int \D{\gamma}\D{e}\,\,
        \exp\bb{
            \frac{i}{\k^2} \int_\M  
                B_a[e] \wedge \BB{
                    R^a[\gamma] - (\Omega^{-1}_\S dd\Omega_\S)^a
                }
                + f[e]
        }
        \,
        W_\rep(\CC)
    \\
    &\,=\,
        \int \D{\gamma}\D{e}\,\,
            \exp\bb{
                \frac{i}{\k^2} \int_\M  
                    B_a[e] \wedge R^a[\gamma]
                    + f[e]
            }
            \,\mem
            \rep(\cen)^{\link(\CC,\S)}\mem
            W_\rep(\CC)
    \,.
\end{align}
This follows by using \eqref{1form.tgt|GR}
as a change of path integration variables.
Note that \eqrefs{eq:fsdef|GR}{1form.tgt|GR} imply
\begin{align}
\begin{split}
    \label{eq:GR.Ftransf}
    &
    R^A{}_B[\mem{
        \Omega_\S^{-1}\mem \gamma\mem \Omega_\S
        + \Omega_\S^{-1} d\Omega_\S
    }\hem]
    \\
    &
    =\,
    (\Omega_\S^{-1})^A{}_C\mem R^C{}_D[\gamma]\mem (\Omega_\S)^D{}_B
    \mem+\mem
        (\Omega_\S^{-1})^A{}_C\mem dd(\Omega_\S)^C{}_B
    \,,
\end{split}
\end{align}
where $dd\Omega_\S$
should not be equated to zero
due to the multivaluedness of $\Omega_\S$.

By definition,
the right-hand side of
\eqref{1form.wardderiv|GR}
is the expectation value
of the transformed spin holonomy.
The left-hand side, on the other hand,
describes
the expectation value of
$W_\rep(\CC)$
put together with the operator
\begin{align}
    \label{1form.Uderiv|GR}
    U_\a(\S)
    \,=\,
        \exp\bb{
            - \frac{i}{\k^2}  \int_\M  
                B_a[e] \wedge (\Omega_\S^{-1} dd\Omega_\S)^a
        }
    \,.
\end{align}
As explicated in detail in \Sec{K1mv},
the invariant geometrical data characterizing this operator are $\S$, the Dirac delta support, 
and $\a$, the multivaluedness of $\Omega_\S$.
Hence it is named $U_\a(\S)$.

To conclude,
the path integral identity in \eqref{1form.wardderiv|GR}
establishes
the one-form symmetry
of nonabelian gauge theories
in terms of the Ward identity
\begin{align}
    \label{1form.Ward|GR}
    \expval{
        W_\rep(\CC)\mem U_\cen(\S)
    }
    \,=\,
        \rho(\cen)^{\link(\CC,\S)}
    \,
        \expval{W_\rep(\CC)}
    \,.
\end{align}
The charged operators are the Wilson loops
$W_\rep(\CC)$
defined in \eqref{1form.W|GR},
which transform as \eqref{1form.Wtransf|GR}.
The symmetry operator 
is stipulated to be the operator $U_\cen(\S)$ defined in \eqref{1form.Uderiv|GR}.
This operator
is topological
because its action
on $W_\rep(\CC)$
does not change
under arbitrary smooth deformations of its support $\S$,
provided that the symmetry group is 
an abelian subgroup
of the local Lorentz group.

Again,
the symmetry operator $U_\cen(\S)$
is systematically deduced as
the operator that
induces the multivalued local Lorentz transformation in \eqref{1form.tgt|GR}
when inserted inside the path integral.
It should also be clear that
\begin{align}
\begin{split}
    \label{eq:arbWard|GR}
    \expval{
        U_\cen(\S)\mem \O
    }
    \,=\,
    \expval{
        \O_{\Omega_\S}
    }  
\end{split}
\end{align}
for any operator $\O$ in the theory,
if $\O_{\Omega_\S}$ denotes the image of $\O$
under the multivalued local Lorentz transformation in \eqref{1form.tgt|GR}.
For local operators
valued in the singlet of the local Lorentz group,
$\O_{\Omega_\S} = \O$.
It is worth emphasizing that
the metric $g_{\m\n}[e] = \eta_{AB}\mem e^A{}_\m\mem e^B{}_\n$ in \eqref{eq:g[e]}
is one such singlet operator.

As will be elaborated on in \Sec{K1.ccs},
it follows that $U_\cen(\S)$
is the creation operator for 
a two-dimensional vortex configuration,
i.e., a string defect,
that induces 
spin precession angles
for spinning particles
(as spin holonomies)
around $\S$.
This is the gravitational AB effect.

\medskip

Before moving on,
we may need a space 
to take stock
on some subtle aspects of the gravitational one-form symmetry.

Firstly,
we may have to
take stock on the thorny issue of diffeomorphism invariance.
Any local or quasilocal object
such as a point, curve, or surface in spacetime
is famously not diffeomorphism invariant
in the presence of dynamical gravitation,
and the charged operator $W_\rep(\CC)$
defined in \eqref{1form.W|GR}
is no exception.
Although it does not carry dangling indices,
it does specify a particular collection of spacetime points
in the same way
as a local operator $\O(x)$,
e.g., the Ricci scalar $\Ric(x)$,
picks a point $x \inn \M$.
In fact,
this annoyance has also been
present in any investigation of
Maxwell or YM theory Wilson loops
in the presence of dynamical gravitation,
which is central to 
many well-established
discussions of swampland conjectures.
A related discourse could be found in
\rrcite{Donnelly:2016rvo,Giddings:2022hba},
known as gravitational dressing.

However, 
we see that
the Ward identity in \eqref{1form.Ward|GR}
is a concrete result
in tetradic Palatini gravity.
Most importantly,
the linking number $\link(\CC,\S)$ is 
clearly diffeomorphism invariant.
The linking between $\CC$ and $\S$ is topological information
invariant under coordinate transformations,
encoded in their relative relationship.
We do not work in the topology-changing regime of gravity,
so there is no need to worry about
gravitational fluctuations
unlinking $\CC$ and $\S$
or changing $\link(\CC,\S)$.
The coordinate representations of
the submanifolds
$\CC$ and $\S$
will vary,
but $\link(\CC,\S)$ is always well-defined as the same value.
With this understanding, \eqref{1form.Ward|GR} stands as a perfectly diffeomorphism-covariant\footnote{
    Again, even a scalar operator such as $\Ric(x)$ is diffeomorphism covariant, not invariant.
} identity,
where the covariances of the left-hand side and the right-hand side match.
In this context,
diffeomorphisms pose no threat to
the well-definedness
of \eqref{1form.Ward|GR}.

Secondly,
the one-form symmetry of gravity 
will still be present in pure tetrad gravity.
While our derivations have made elaborate use of
the first-order (Palatini) formalism,
our final conclusions 
on the one-form symmetry
will remain valid 
also after integrating out the spin connection.

Certainly, it is generally said that
global symmetries shall be viewed as independent
of one's choice of formalism.
A much subtler question along this line is
whether there is a one-form symmetry
in the pure-metric or metric-affine theories of gravity,
where there is no tetrad, spin connection, or local Lorentz symmetry to speak of.
In particular,
there will be no such thing as
the one-form symmetry for spinor holonomy,
because of the simple reason that 
one cannot even write it down
since there is no tetrad field to characterize the gravitational interactions of fermions.

On the one hand,
it seems that
the various formulations of gravity
secretly had a subtle effect on changing
the declaration of admissible operators/observables.
An analogous situation is
studying YM theory with a stipulation that 
one can only ever use adjoint indices,
thus precluding the existence of the very fundamental Wilson loops which are charged under the one-form symmetry.
Conversely, 
it seems that
declaring a set of symmetries
constrains one's theory such that
certain formulations might become not available.

On the other hand,
one can also think of integrating out
field variables subject to the local Lorentz symmetry.
From this perspective,
the vector holonomy and its one-form symmetry should presumably have
a description in the metric-affine formulation
since the spin structure should not be necessary.

\subsection{Exponentiated Area as Symmetry Operator}
\label{sec:exparea}

\eqref{1form.Uderiv|GR}
is a concrete formula
expressing the gravitational one-form symmetry operator 
$U_\cen(\S)$
in terms of the local degrees of freedom of the theory.
A further explicit formula
can be obtained by choosing a specific representative for the multivalued gauge parameter $\Omega_\S$.
As shown in \Sec{K1mv},
there are many choices one can make.
One nice choice is
\begin{align}
    \label{Omega=exp(nphi)|GR}
    \Omega_\S
    \,=\,
        \exp\bigbig{
            \vn\mem \phi_\S
        }
    \transition{where}
    \mathe^{2\pi \vn}
    \,=\,
        \a
    \,\in\,
        Z(\G)
    \,,
\end{align}
where $\phi_\S$ is a multivalued function
that exhibits a unit winding from $0$ to $2\pi$
around $\S$,
and $\vn^a$ is an element of the local Lorentz algebra
$\g \eqq \Lie(\G)$
that exponentiates to the center element $\a$.
We take $\vn^a$ as a constant.
With this choice,
\eqref{1form.Uderiv|GR} boils down to
\begin{align}
    \label{1form.U|GR}
    U_\a(\S)
    \,=\,
        \exp\bb{
            - \frac{2\pi i}{\k^2}  \int_\S  
                B_a[e]\mem \vn^a
        }
    \,,
\end{align}
manifesting the surface support on $\S$.
Again, \eqref{1form.U|GR} should read as
one possible realization for $U_\cen(\S)$,
and our previous discussions
in \Sec{K1>1form>U}
on the spuriously breaking nature of $\vn^a$
applies in the same way.

Intriguingly,
this surface operator $U_\cen(\S)$
literally computes a certain area-like quantity
associated with $\S$.
By plugging in \eqref{eq:pleb-def}
and unpacking the gravitational coupling as
$\k = (8\pi\GN)^{1/2}$,
we find that
\eqref{1form.U|GR} describes
\begin{align}  
    \label{eq:Grav.U.alt}
    U_\cen(\S)
    \,=\,
    \exp\bigg(\mem{
        -\frac{i}{4\GN}
        \int_{\S}\mem
            \frac{1}{2}\:
            {\star\vn}_{AB}\mem
            (e^A \swedge e^B)
    }\mem\bigg)
    \,,
\end{align}
where 
${\star\lambda}_{AB}
= \tfrac{1}{2}\mem \e_{ABCD}\mem \lambda^{CD}$
is the Hodge dual of $\lambda^{AB}$.  Here we recognize
 $e^A \wedge e^B$ as
the infinitesimal area element
in the orthonormal frame,
so the exponent 
computes the area smeared with a reference $\star\lambda_{AB}$,
measured in Planck units;
$1/2$ is the canonical normalization factor
for contracting 
antisymmetric tensors.

Additionally, it is 
curious that the exponent in \eqref{eq:Grav.U.alt}
is tantalizingly similar to the
area operator in loop quantum gravity
which leads to the quantization of tetrahedral volume
\cite{Rovelli:1994ge,Ashtekar:1996eg}.
The only crucial difference is that here
the area is ``dotted'' with a Lorentz generator
$\lambda^{AB}$ that exponentiates to the center,
so in a sense it 
serves as a discrete and topological reincarnation of the area operator in loop quantum gravity.
On the other hand,
$U_\cen(\S)$ is also reminiscent of the Bekenstein-Hawking entropy formula \cite{Bekenstein:1973ur,Hawking:1975vcx}:
$(\text{Area})/4\GN$.
Of course,
these could easily be accidents of dimensional analysis on account of the $1/G_\mathrm{N}$ normalization in the exponent.
But in any case, 
it would be 
fascinating
to see if any of 
these superficial similarities
carry deeper significance. 

\medskip

It is also useful to define
the \textit{chiral surface operators} as
\begin{subequations}
    \label{Ublocks} 
\begin{align}
    \label{Ublocks.SD}
    \widetilde U(\S)
    \mem=\mem
    \exp\bb{
        \frac{1}{4G_\text{N}}
        \int_{\S}\mem
            \e_\wrap{\a\b}\mem \rambda_\wrap{\da\db}\hem
            (e^{\da\a} \swedge e^{\db\b})
    }
    &\,,\quad
    (\mathe^{2\pi\rambda})^\da{}_\db 
    \mem=\mem
        -\delta^\da{}_\db
    \,,\\
    \label{Ublocks.ASD}
    U(\S)
    \mem=\mem
    \exp\bb{
        -
        \frac{1}{4G_\text{N}}
        \int_{\S}\mem
            \lambda_\wrap{\a\b}\mem \te_\wrap{\da\db}\hem
            (e^{\da\a} \swedge e^{\db\b})
    }
    &\,,\quad
    (\mathe^{2\pi\l})_\a{}^\b 
    \mem=\mem
        -\delta_\a{}^\b
    \,,
\end{align}
\end{subequations}
which point to an intriguing idea of
measuring the chiral areas in Planck units
in terms of the chiral Pleba\'nski two-forms
\cite{Plebanski:1977zz,Capovilla:1991qb}.

In \eqref{Ublocks},
$\tilde\lambda^\da{}_\db$ and $\lambda_\a{}^\b$
belong to the SD and ASD sectors of the local Lorentz algebra $\g$, respectively.
Since the two sectors commute,
any element of $\g$
that exponentiates to a center element
splits into SD and ASD generators
that \textit{separately} exponentiate to center elements.
Therefore, it follows that 
the chiral surface operators in \eqref{Ublocks}
serve as primordial building blocks for
any symmetry operator.
The details of this decomposition yet
vary depending on the signature.

For example, 
suppose the complexified setup with
the largest group
$\G = \Spin(4,\C) \cong \SL(2,\C) \times \SL(2,\C)$.
Then 
each of
$\widetilde U(\S)$ and $ U(\S)$ in \eqref{Ublocks}
is separately a one-form symmetry operator,
implementing $\Z_2$ flips of the
holonomies of
the chiral spin connections
$\tilde\gamma^\da{}_\db$
and $\gamma_\a{}^\b$.
Note that, as zero-form symmetries, 
each of these $\Z_2$
acts as parity in the numbers of SD (dotted) and ASD (undotted) spinor indices, respectively.

For another example,
suppose Euclidean signature
with $\G = \Spin(4)$.
This case is similar to the above complexified case.
The chiral surface operators 
$\widetilde U(\S)$ and $ U(\S)$
are precisely the one-form symmetry operators corresponding to each $\Z_2$ factor of the center subgroup $Z(\G) = \mathbb{Z}_2\times \mathbb{Z}_2$. 
Again, as zero-form symmetries, each factor of this center acts as parity in the numbers of dotted and undotted spinor indices, respectively.

Finally,
in the Lorentzian signature
without complexification,
the one-form symmetry is only nontrivial if
$\G = \Spin(1,3) = \SL(2,\mathbb{C})$,
in which case the center is $Z(\G) = \mathbb{Z}_2$.  
In this case, only the \textit{real} combination of the chiral operators,
$\widetilde U(\S)\mem U(\S)$,
can qualify as a one-form symmetry operator.
Note that as a zero-form symmetry
this acts as \textit{net} parity on spinor indices,
drawing no distinction between the dotted and undotted ones.
For instance, it acts \textit{trivially} on vectors
but flips sign for fermions.

\subsection{Tetradic Cosmic String as a Local Lorentz Vortex}
\label{K1.ccs}

Finally, we explore the top corner of the symmetry triangle in \fref{SymTriangle}:
vortex geometry.
Our goal is to determine
the physical interpretation
of the vortex configuration
which
the symmetry operator $U_\a(\S)$
in \eqref{eq:Grav.U.alt}
generates
in GR.
The chiral surface operators in \eqref{Ublocks}
will also be considered in turn.

Recall how \Sec{K1mv} demonstrated that
the color vortex in gauge theory
is the Dirac string.
First, a flat configuration $\bar{A}^a$ was assumed
where $F^a[\bar{A}] = 0$.
Second, a multivalued gauge transformation
$\Omega_\S = \exp\bigbig{ \vn\mem \phi_\S }$
derived a new configuration
$A = \Ad{\Omega^{-1}_\S} \bar{A} + \Omega^{-1}_\S d\Omega_\S$
where
$F^a[A] = 2\pi\vn^a\mem \delta(\S)$.
Third,
this distributional field strength
was identified as implementing a thin tube of color flux.
Concrete demonstrations of this fact
were provided by
assuming various shapes for $\S$.
The background $\bar{A}^a$ could have
been chosen arbitrarily,
but
making it flat helps focus on the geometry and physics of the string itself.

To repeat this analysis for GR,
we suppose a flat spacetime
described by
a metric $\bar{g}_{\m\n} = \eta_{AB}\mem \bar{e}^A{}_\m\mem \bar{e}^B{}_\n$
and a spin connection $\bar{\gamma}^A{}_B$
such that $R^A{}_B[\bar{\gamma}] = 0$.
For instance, we may take
$\bar{e}^A{}_\m = \delta^A{}_\m$
and $\bar{\gamma}^A{}_B = 0$.
Next, we apply a multivalued local Lorentz transformation
represented as 
$\Omega_\S = \exp\bigbig{ \vn\mem \phi_\S }$.
The resulting geometry describes
\begin{align}
    \label{twisted-eR}
    e^A
    \,=\,
        \exp\bigbig{
            - \vn\mem \phi_\S
        }{}^A{}_B\mem \bar{e}^B
    \,,\quad
    R^A{}_B
    \,=\,
        2\pi\vn^A{}_B\mem \delta(\S)
    \,.
\end{align}
By analogy, \eqref{twisted-eR}
realizes a ``local Lorentz vortex.''
Notably, it encodes
a curvature singularity distributed over the two-dimensional surface $\S$,
reminiscently of the cosmic string.
To determine the precise physical interpretation of this string geometry,
we consider the following facts.

\paragraph{Invisibility to Scalar Particles}

First and foremost,
the metric is invariant under any local Lorentz transformation, 
whether multivalued or not.
The coframe given in \eqref{twisted-eR} implies
\begin{align}
    \label{g=bg}
    g_{\m\n}
    \,=\,
        \bar{g}_{\m\n}
    \,,
\end{align}
so the metric remains identically flat.
Physically, this implies that our string geometry
is \textit{completely invisible to scalar particles}.
The geodesics of scalar particles are left the same,
in both local and global senses.
Locally, the geodesic equation uses
the Christoffel symbols $\Gamma^\m{}_{\r\s}[g]$
defined by the metric.
Globally,
the scalar particle geodesics
will detect \textit{no angle deficit}
since no multivaluedness has been introduced
regarding the points of the base spacetime manifold.
The conclusion is that our string geometry
should somehow differ from the standard cosmic string.

\paragraph{Matter Content}

Secondly,
we reverse engineer the matter source
that generates our string geometry.
Certainly, this analysis will provide an explicit physical characterization.
With vanishing cosmological constant,
Einstein's equations describe
\begin{align}
    \label{eq:stress}
      -\frac{1}{\k^2}\mem
    {\star R}_{AB} \swedge e^B=
    \frac{1}{3!}\,
    T^\m{}_A\,
    \ve_{\m\n\r\s}\mem 
    dx^\n \swedge dx^\r \swedge dx^\s
    \,.
\end{align}
By plugging in \eqref{twisted-eR} to \eqref{eq:stress},
the stress-energy tensor $T^\m{}_A$ is determined as
\begin{align}
\begin{split}
    \label{eq:stress-string}
    T^\m{}_\k
    &=
        \frac{1}{8G_\text{N}}\,
        \ve^{\m\n\r\s}\mem
        {\star\vn}_{\n\k}\,
        \delta(\S)_{\r\s}
    \,,
\end{split}
\end{align}
which acts as a source for the coframe
in the first-order formulation.
Note that
the formula given in \eqref{eq:delta-param}
expresses this $T^\m{}_\k$ as
\begin{align}
        \frac{1}{4G_\text{N}}\mem
        \frac{1}{|e|}\,
        {\star\vn}_{\n\k}
    \int d^2\s\,\,
        \delta^{(4)}\hnem\bigbig{x{\mem-}X(\s^1,\s^2)\hnem}
        \bb{
            \frac{\partial X^{\m}}{\partial \s^1}
            \frac{\partial X^{\n}}{\partial \s^2}
            -
            \frac{\partial X^{\n}}{\partial \s^1}
            \frac{\partial X^{\m}}{\partial \s^2}
        }
    \,,
\end{align}
when the surface $\S$ is parameterized as $X^\m\hnem(\s^1,\s^2)$
in terms of worldsheet coordinates
$(\s^1,\s^2)$.
We see that the stress-energy tensor
is localized along a membrane $\S$
and describes a formula reminiscent of the cosmic string,
though it is not literally identical to the Nambu-Goto string.

Interestingly, we also see that the algebraic Bianchi identity, $R_{AB} \swedge e^B=0$, \textit{fails} for our string geometry, indicating the existence of \textit{magnetic} stress-energy.\footnote{
    This failure of algebraic Bianchi in the presence of magnetic matter
    is analogous to the failure of the Bianchi identity in Maxwell theory in the presence of
    magnetic charges.
}
The dual stress-energy tensor $T^\star{}^\m{}_A$ is defined as \cite{cho1991magnetic,bunster2006monopoles}
\begin{align}
    \label{eq:dualstress}
     -\frac{1}{\k^2}\mem
    R_{AB} \swedge e^B =
    \frac{1}{3!}\,
    T^\star{}^\m{}_A\,
    \ve_{\m\n\r\s}\mem 
    dx^\n \swedge dx^\r \swedge dx^\s
    \,.
\end{align}
By plugging in $R^A{}_B = \vn^A{}_B\mem \delta(\S)$,
we obtain
\begin{align}
\begin{split}
    \label{eq:dualstress-string}
    T^\star{}^\m{}_\k
    &=
        \frac{1}{8G_\text{N}}\,
        \ve^{\m\n\r\s}\mem
        {\vn}_{\n\k}\,
        \delta(\S)_{\r\s}
    \,.
\end{split}
\end{align}
It follows that 
\eqref{eq:dualstress-string} describes a line distribution of NUT charge,
just as \eqref{eq:stress-string} describes a line distribution of mass.

The conclusion here is that
the geometry in \eqref{twisted-eR}
represents a cousin of cosmic string,
sourced by a \textit{combined concentration of both
electric and magnetic matter}
localized on a two-dimensional surface $\S$.

\paragraph{No Misner String Component}

Lastly,
we shall clarify if 
a Misner string component 
is involved as well.
The answer turns out to be no.
First, 
the Misner string describes a stack of gravitomagnetic \textit{dipoles} \cite{Bonnor:1969ala,sackfield1971physical},
and as such
it can terminate on
gravitomagnetic monopoles
(NUT charges).
However, the magnetic stress-energy found in \eqref{eq:dualstress-string}
itself is
a line density of distributed gravitomagnetic \textit{monopole}.
Second,
the Misner string 
could also be characterized as
a localized source of torsion,
as it
induces a time monodromy
\cite{Misner:1963flatter,dowker1967gravitational,Alfonsi:2020lub}.
However,
it is easy to see that the torsion transforms covariantly and linearly even under the multivalued local Lorentz transformation,
so 
if the initial background exhibited vanishing torsion,
it stays identically zero after the insertion of the symmetry operator as well.
Concretely, one can establish that our string geometry has vanishing torsion
by plugging in the transformed tetrad and spin connection
into $De^A = de^A + \gamma^A{}_B \swedge e^B$,
which also reveals that the failure of algebraic Bianchi is solely due to the multivaluedness of the tetrad, $dde^A \neq 0$,
\textit{not} by a nonzero torsion.
The absence of torsion
is also clear
if one recalls the fact that
torsion is induced by local source for the spin connection
in the tetradic Palatini formalism,
while
the symmetry operator  only couples to the coframe.
For these reasons,
we conclude that
the geometry in \eqref{twisted-eR}
is of a purely cosmic kind,
with no Misner string component.

\paragraph{Tetradic Cosmic String}

\begin{figure}[htbp]
    \centering
    \includegraphics[width=0.8\linewidth]{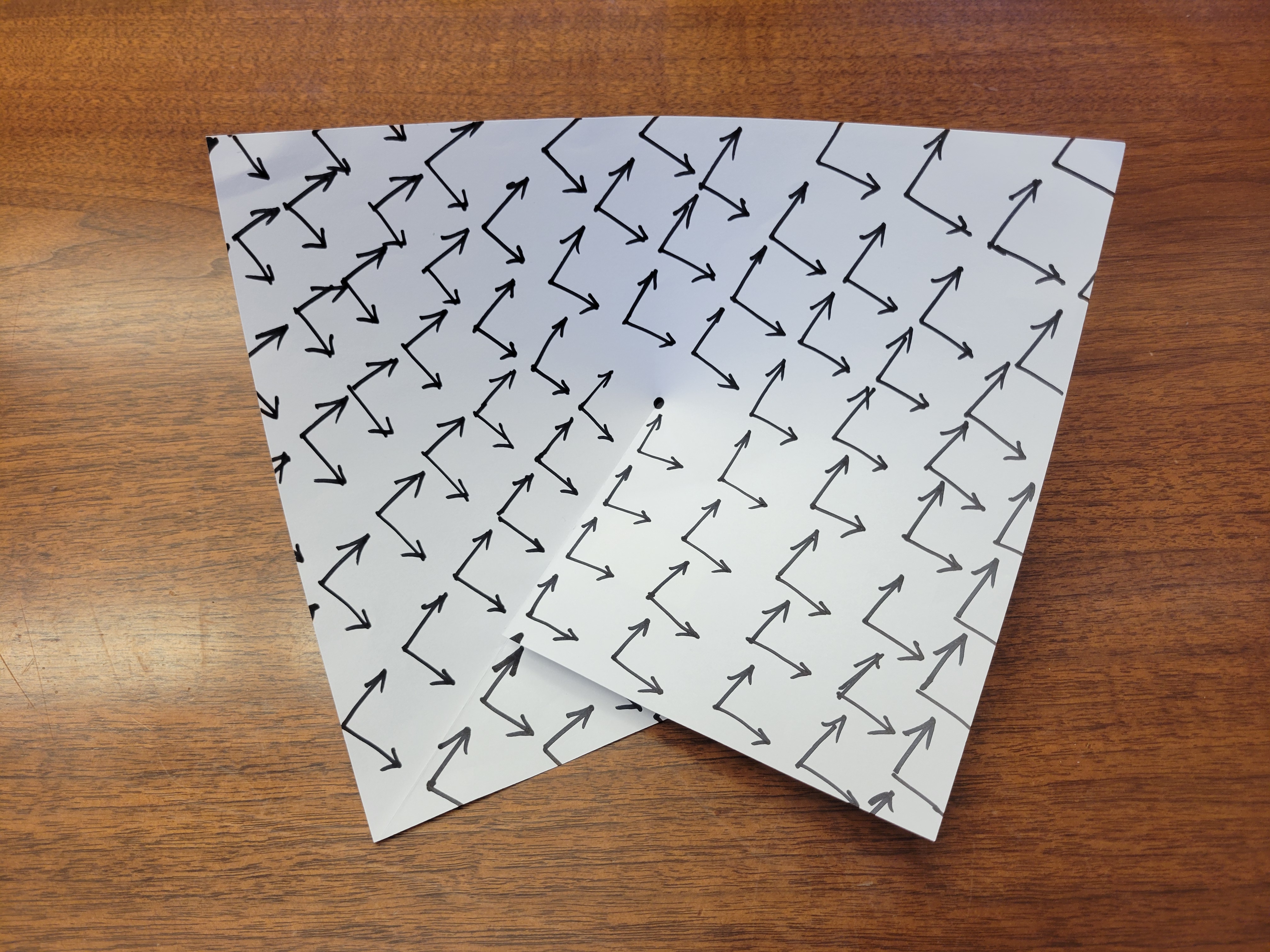}
    \\
    \adjustbox{scale=1.2,rotate=-90,valign=c}{$\,\,
        \xrightarrow[\quad]{}
    \,\,$}
    \\
    \includegraphics[width=0.8\linewidth]{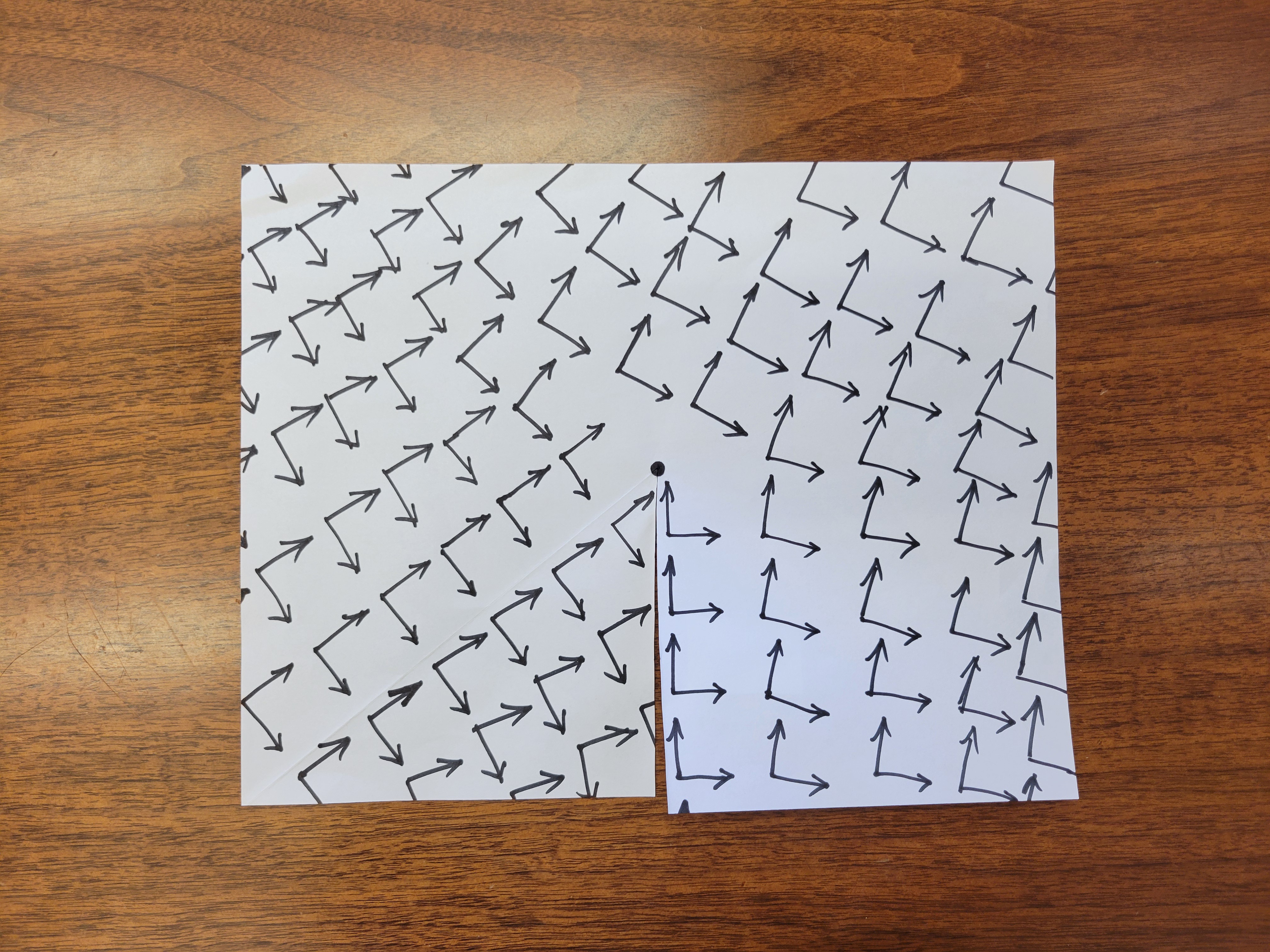}
    \vspace{0.6\baselineskip}
    \caption{%
        Construction of a tetradic cosmic string from the standard cosmic string.
        (\textit{Upper panel})
            A single-valued tetrad is installed on the spacetime of a standard cosmic string.
        (\textit{Lower panel}) 
            Unwrapping the spacetime by adding a wedge produces a multivalued tetrad on a spacetime without an angle deficit, realizing a tetradic cosmic string.
            The required center periodicity condition will be achieved by tuning the original string tension
            (figure depicts improper quantization for clarity of illustration).
        In this procedure,
        note the interplay between
        the multivalued geometries of the base
        and the fibration,
        manipulated by
        diffeomorphisms and
        local Lorentz transformations,
        respectively.
    }
    \label{fig:stringconstruct}
\end{figure}

Let us name
the geometry due to \eqref{twisted-eR}
a \textit{tetradic cosmic string}.
Our analysis above shows that
the tetradic cosmic string differs from 
the standard cosmic string
in terms of its invisibility to scalar particle geodesics,
involves both electric and magnetic stress-energies,
and carries no Misner string component.
It seems that the tetradic cosmic string
cannot be realized in non-tetradic
theories of gravity
such as pure-metric or metric-affine gravity.
The tetradic cosmic string will be detected by
the spin holonomy in \eqref{1form.W|GR},
i.e.,
interferometer experiments with spinning particles
to speak physically.
See \fref{fig:stringconstruct} for a visualization of the geometry of a tetradic cosmic string.

\paragraph{Concrete Construction and Relation to Standard Cosmic String}

To be further explicit about these points,
we demonstrate below
how a tetradic cosmic string
can be derived
by modifying the geometry of
the standard cosmic string
such that scalar particle geodesics get globally straightened out.
We wish to assume the complexified setup with
$\G = \Spin(4,\C)$.

We begin by reviewing the geometry of
the standard cosmic string with tension $\mu$.
In cylindrical coordinates,
a straight cosmic string inserted along the $z$-axis
describes the line element
\begin{align}
    \label{cosmic0.g}
    ds^2
    \,=\,
        - dt^2 
        + ds^2
        + \bigbig{ 1-4\hem\GN\m }^{\nem2}\mem
            s^2\mem d\phi^2
        + dz^2
    \,,\quad
    \phi \,\sim\, \phi + 2\pi
    \,.
\end{align}
The angle deficit is detected as
\begin{align}
    2\pi - \Del\theta
    \,=\,
        \oint 
            \bigbig{ 1-4\hem\GN\m }\mem d\phi
    \qiq
        \Del\theta \,=\, 8\pi\GN\mem \m
    \,.
\end{align}
A complexified orthonormal coframe for
the line element in \eqref{cosmic0.g} 
can be given as
\begin{align}
\begin{split}
    \label{cosmic0.coframe}
    e^0 \,&=\,
        \cos\phi\mem dt
        - i \sin\phi\mem dz
    \,,\\
    e^1 \,&=\,
        \cos\phi\mem ds
        - \bigbig{ 1-4\hem\GN\m }\hem 
            \sin\phi
            \:
            s\mem d\phi
    \,,\\
    e^2 \,&=\,
        \sin\phi\mem ds
        + \bigbig{ 1-4\hem\GN\m }\hem 
            \cos\phi
            \:
            s\mem d\phi
    \,,\\
    e^3 \,&=\,
        - i \sin\phi\mem dt
        + \cos\phi\mem dz
    \,,
\end{split}
\end{align}
which is \textit{single-valued}
over the domain 
$t \inn \R$, $s \geq 0$, $\phi \in [0,2\pi[$, $z \inn \R$.

Now consider plugging in $\phi = \psi/(1\mminus4\GN\m)$
and \textit{changing} the periodicity condition as
$\phi \sim \phi + 2\pi / (1\mminus4\GN\m)$:
\begin{align}
    \label{cosmic1.g}
    ds^2
    \,=\,
        - dt^2 
        + ds^2
        + s^2\mem d\psi^2
        + dz^2
    \,,\quad
    \psi \,\sim\, \psi + 2\pi
    \,.
\end{align}
Crucially, this derives a \textit{new} geometry
that globally differs from the previous one.
The line element in \eqref{cosmic1.g}
describes \textit{no angle deficit}
with the new periodicity condition:
\begin{align}
    2\pi - \Del\theta
    \,=\,
        \oint 
            \bigbig{ 1-4\hem\GN\m }\mem d\phi
    \,=\,
        \oint 
            d\psi
    \qiq
        \Del\theta \,=\, 0
    \,.
\end{align}
In the meantime,
the orthonormal coframe in \eqref{cosmic0.coframe}
is brought to
\begin{align}
\begin{split}
    \label{cosmic1.coframe}
    e^0 \,&=\,
        \cos\hnem\bb{
            \frac{\psi}{1\mminus4\GN\m}
        }\mem dt
        - i\hhem
        \sin\hnem\bb{
            \frac{\psi}{1\mminus4\GN\m}
        }
            \,
            dz
    \,,\\
    e^1 \,&=\,
        \cos\hnem\bb{
            \frac{\psi}{1\mminus4\GN\m}
        }\mem ds
        -
        \sin\hnem\bb{
            \frac{\psi}{1\mminus4\GN\m}
        }
            \,
            s\mem d\psi
    \,,\\
    e^2 \,&=\,
        \sin\hnem\bb{
            \frac{\psi}{1\mminus4\GN\m}
        }\mem ds
        + 
        \cos\hnem\bb{
            \frac{\psi}{1\mminus4\GN\m}
        }
            \,
            s\mem d\psi
    \,,\\
    e^3 \,&=\,
        - i\hhem
        \sin\hnem\bb{
            \frac{\psi}{1\mminus4\GN\m}
        }
            \,
            dt
        +
        \cos\hnem\bb{
            \frac{\psi}{1\mminus4\GN\m}
        }\mem dz
    \,,
\end{split}
\end{align}
which is \textit{multivalued}
over the domain 
$t \inn \R$, $s \geq 0$, $\psi \in [0,2\pi[$, $z \inn \R$
unless $(1\mminus4\GN\m)^{-1} \in \Z$.
Concretely,
let us take
\begin{align}
    \label{mu-k}
    \frac{1}{1\mminus4\GN\m}
    \,=\,
        k + \frac{1}{2}
    \qfq
    \m
    \,=\,
        \frac{1}{4\GN}\mem
        \frac{2k-1}{2k+1}
    \,,
\end{align}
where $k \in \Z$.

\newpage

Clearly,
\eqref{cosmic1.coframe} arises by
applying
a multivalued local Lorentz transformation
on the background coframe 
$(\bar{e}^0,\bar{e}^1,\bar{e}^2,\bar{e}^3) = 
    (dt,ds,s\mem d\psi,dz)
$:
\begin{align}
    \label{example-lambda-GN}
    \Omega_\S
    \,=\,
        \exp\bigbig{
            \lambda\hem \psi
        }
    \,,\quad
    \lambda^A{}_B
    \,=\,
        \bb{k + \frac{1}{2}}\nem
    \lrp{\,\,
    \begin{matrix}
        0 & 0 & 0 & \mathclap{i}
        \\
        0 &
            0
            &
            \mathclap{1}
        & 0
        \\
        0 &
            \mathclap{-1}
            &
            0
        & 0
        \\
        \mathclap{i} & 0 & 0 & 0
    \end{matrix}
    \,\,}
    \,.
\end{align}
As per \eqref{eq:lambda-disc1|GR},
the multivaluedness is characterized by the discontinuity,
\begin{align}
    \cen
    \,=\,
    \lim_{
        \e \to 0
    }
    \BB{
        \Omega_\S(\psi\eqq 2\pi\mminus\e)
        \,
        \Omega_\S^{-1}\hnem(\psi\eqq \e)
    }
    \,=\,
        \mathe^{2\pi\vn}
    \,=\,
        (-1)^{2k+1}\mem \id
    \,.
\end{align}
This shows how
a tetradic cosmic string
can be concretely constructed
by modifying the geometry of the standard cosmic string
whose tension $\mu$
is made to take
specific values
as stipulated in \eqref{mu-k}.


\paragraph{Chiral Tetradic Cosmic String}


It can also be seen that 
the Lorentz generator $\vn^A{}_B$
in \eqref{example-lambda-GN}
is SD:
$\star\vn^{AB} = +i\mem \vn^{AB}$.
Therefore,
the above example
has constructed
the geometry 
created by a SD surface operator, $\tU(\S)$, in \eqref{Ublocks.SD}.
Hence more precisely, it will be called a 
\textit{SD tetradic cosmic string}.
In general, we refer to the vortex configurations
generated by the chiral surface operators in \eqref{Ublocks}
as \textit{chiral tetradic cosmic strings}.

For chiral tetradic cosmic strings,
the electric and magnetic stress-energy tensors
in \eqrefs{eq:stress-string}{eq:dualstress-string}
satisfy the relation
\begin{align}
    T^\m{}_k
    \,=\,
        \pm\hem i\mem T^\star{}^\m{}_\k
    \,.
\end{align}
Hence they are concentrations of \textit{chiral matter}:
combinations of electric and magnetic stress-energies
in ``equal magnitudes.''

It is straightforward to compute
the holonomies
for various spin representations
in the background
of \eqref{cosmic1.coframe}:
\begin{align}
\begin{split}
    \label{shols.ex1}
    \Tr_\text{vector}
    \Pexp
    \bb{
        \vn
        \int_0^{2\pi}\kern-0.2em d\psi
    }
    \,&=\,
        4\hem (-1)^{2k+1}
    \,=\,
        -4
    \,,\\
    \Tr_{\text{spi}\mathclap{\widetilde{\phantom{iiiii}}}\text{nor}}
    \Pexp
    \bb{
        \vn
        \int_0^{2\pi}\kern-0.2em d\psi
    }
    \,&=\,
        2\hem (-1)^{2k+1}
    \,=\,
        -2
    \,,\\
    \Tr_\text{spinor}
    \Pexp
    \bb{
        \vn
        \int_0^{2\pi}\kern-0.2em d\psi
    }
    \,&=\,
        2
    \,.
\end{split}
\end{align}
Provided \eqref{mu-k} with $k\inn\Z$,
we see that the vector and SD spinor holonomies
flip their sign
while the ASD spinor holonomy is left the same.


With the imposition of the Lorentzian reality condition,
one considers the real combination $\tU(\S)\mem U(\S)$
as discussed in \Sec{sec:exparea}.
The SD tetradic cosmic string
described in \eqref{cosmic1.coframe}
can be superimposed with its ASD counterpart (``complex conjugate'')
in this sense,
in which case the coframe is\footnote{
    In principle,
    a real tetradic cosmic string
    will be able to
    involve both
    electric and magnetic stress-energies
    depending on the relative orientation
    between the surface support $\S$
    and the Lorentz generator $\vn$.
    For our specific example here,
    the magnetic stress-energy happens to vanish.
}
\begin{align}
\begin{split}
    \label{cosmic1.coframe.e}
    e^0 \,&=\,
        dt
    \,,\\
    e^1 \,&=\,
        \cos\hnem\BB{
            (2k\mem\mplus1)\mem \psi
        }\mem ds
        -
        \sin\hnem\BB{
            (2k\mem\mplus1)\mem \psi
        }
            \,
            s\mem d\psi
    \,,\\
    e^2 \,&=\,
        \sin\hnem\BB{
            (2k\mem\mplus1)\mem \psi
        }\mem ds
        + 
        \cos\hnem\BB{
            (2k\mem\mplus1)\mem \psi
        }
            \,
            s\mem d\psi
    \,,\\
    e^3 \,&=\,
        dz
    \,.
\end{split}
\end{align}
This arises by the local Lorentz transformation
\begin{align}
    \label{example-lambda-GN.e}
    \Omega_\S
    \,=\,
        \exp\bigbig{
            \lambda\hem \psi
        }
    \,,\quad
    \lambda^A{}_B
    \,=\,
        (2k\mem\mplus1)
    \lrp{\,\,
    \begin{matrix}
        0 & 0 & 0 & 0
        \\
        0 &
            0
            &
            \mathclap{1}
        & 0
        \\
        0 &
            \mathclap{-1}
            &
            0
        & 0
        \\
        0 & 0 & 0 & 0
    \end{matrix}
    \,\,}
    \,,
\end{align}
which is single-valued
in the vector representation.
The computation of the holonomies
in this background
gives
\begin{align}
\begin{split}
    \label{shols.ex2}
    \Tr_\text{vector}
    \Pexp
    \bb{
        \vn
        \int_0^{2\pi}\kern-0.2em d\psi
    }
    \,&=\,
        4\hem (-1)^{4k+2}
    \,=\,
        4
    \,,\\
    \Tr_{\text{spi}\mathclap{\widetilde{\phantom{iiiii}}}\text{nor}}
    \Pexp
    \bb{
        \vn
        \int_0^{2\pi}\kern-0.2em d\psi
    }
    \,&=\,
        2\hem (-1)^{2k+1}
    \,=\,
        -2
    \,,\\
    \Tr_\text{spinor}
    \Pexp
    \bb{
        \vn
        \int_0^{2\pi}\kern-0.2em d\psi
    }
    \,&=\,
        2\hem (-1)^{2k+1}
    \,=\,
        -2
    \,,
\end{split}
\end{align}
so the vector holonomy is not transformed.

\subsection{Linking and Spin Precession Angle}
\label{sec:ConDef}

In \Sec{K1.ccs},
we have established that 
the symmetry operator $U_\cen(\S)$ creates
a ``tetradic cosmic string''
that is invisible to scalar particle trajectories
but affects the spin holonomies $W_\rep(\CC)$
linking with it.

What is the physical interpretation
of this linking between
$U_\cen(\S)$ and $W_\rep(\CC)$,
leading to the Ward identity in \eqref{1form.Ward|GR}?
Notably,
this too has a simple physical interpretation
in terms of classical gravitation:
\textit{spin precession angles}
as a gravitational AB effect.

\begin{figure}[t]
\begin{align*}
    \text{
        Quantum corrected $U_\cen$
    }:&\qquad
        \includegraphics[scale=1.05,valign=c]{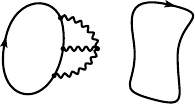}
        \quad\sim\quad
        \mathcal{O}(\k^{-2})
    \\[0.3\baselineskip]
    \text{
        Classical holonomy from $U_\cen$
    }:&\qquad
        \includegraphics[scale=1.05,valign=c]{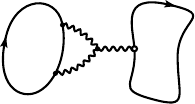}
        \quad\sim\quad
        \mathcal{O}(\k^{0})
    \\[0.3\baselineskip]
    \text{
        Quantum holonomy from $U_\cen$
    }:&\qquad
        \includegraphics[scale=1.05,valign=c]{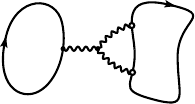}
        \quad\sim\quad
        \mathcal{O}(\k^{2})
    \\[0.3\baselineskip]
    \text{
        Quantum corrected $W_\rep$
    }:&\qquad
        \includegraphics[scale=1.05,valign=c]{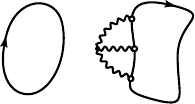}
        \quad\sim\quad
        \mathcal{O}(\k^{4})
    \\
    &\qquad
        \kern1.66em
        \mathclap{\adjustbox{scale=0.9}{$
            U_\cen %
        $}}
        \kern5.95em
        \mathclap{\adjustbox{scale=0.9}{$
            W_\rep %
        $}}
\end{align*}
    \caption{%
        A schematic depiction of
        Feynman diagrams contributing to the Ward identity
        at various perturbative orders in the gravitational coupling $\k$.
        The wavy lines denote gravitons, while the loops of solid lines on the left and right depict the symmetry operator $U_\cen$ and line operator $W_\rep$, respectively.   The first and fourth rows describe quantum corrections to $U_\cen$ and $W_\rep$ separately.  The second row is a contribution to the classical holonomy measured by $W_\rep$, treating $U_\cen$ as a source for the background spacetime.  
        The third row is the one-loop quantum correction to this quantity in that background.
        Since the linking number is dimensionless and topological, it arises purely from the classical holonomy.
    }
    \label{fig:duffs}
\end{figure}

It is easy to see that spin holonomies
arise from the parallel transport of spinning particles around a loop.
For example,
take the example calculation in \eqref{shols.ex1}.
It describes that
vector particles and SD fermions
exhibit spin precession angles
continuously accumulating from $0$ to $(2k\mem\mplus1)\mem\pi$
while traveling along a loop
that encloses the tetradic cosmic string.
This provides the physical interpretation of the Ward identity:
the SD surface operator $\tU(\S)$
induces
quantized spin precession angles for 
vector particles and SD fermions.

For the example calculation in \eqref{shols.ex2},
one learns that
fermions exhibit spin precession angles
continuously accumulating from $0$ to $(2k\mem\mplus1)\mem\pi$.
In the meantime,
vector particles exhibit
the total spin precession angle 
$(4k\mem\mplus2)\mem\pi$,
bringing it back to its initial state.

One may wonder
how this \textit{classical} computation
in the background geometries of
tetradic cosmic strings
captures the full content of the \textit{quantum} Ward identity.
Of course,
the exact path integral derivation of the Ward identity in \Sec{K1.Ward|GR}
clarifies that there is no mystery
at the mathematical level.
However, one may still want to revisit
the path integral derivation
from the perturbation theory point of view,
in which case
the linking number is computed by an infinite set of perturbative diagrams.
As sketched in \fref{fig:duffs},
a careful power counting argument
with the formulae in
\eqrefss{1form.W|GR}{1form.U|GR}{1form.L|GR}
shows that 
the linking number
receives contributions
solely from the diagrams
computing the classical holonomy
in the background of a $U_\cen(\S)$ insertion.
Note that $U_\cen(\S)$
couples gravitons to the surface $\S$
via the inverse factor $1/\k^2$.
Similarly, one can also argue that
the spin precession angles
are not renormalized.

Another remark is that
the computation of spin precession angles
under the insertion of a tetradic cosmic string
can be demonstrated in
general backgrounds,
such as the spacetime of an Anti-de-Sitter-Schwarzschild black hole.
First, identify the background coframe.
Second, apply a multivalued local Lorentz transformation.
Third, compute the spin holonomies directly
from the transformed spin connection
by using the path-ordered exponential formula.
By one's choice of the background spin connection
and the Lorentz generator $\vn$,
one will be able to
demonstrate a rather nontrivial calculation.

\section{ Physical Implications}
\label{sec:Breaking}

As is well-known, global higher-form symmetries can be broken, either explicitly or spontaneously.  
In analogy with YM theory, the one-form symmetry of gravity is explicitly broken in the presence of matter fields that transform nontrivially under the Lorentz group.  For example, if the theory includes a local operator in the spin representation $\rep$, then it is possible to define a Lorentz invariant line holonomy  $W_{\rep}(\CC)$ for a contour $\CC$ that is not closed, but rather terminates on this local operator.  The spin holonomy is then ``endable,'' so it can be unlinked topologically from the symmetry operator $U_\cen(\S)$, and the one-form symmetry is explicitly broken.  The physical interpretation of this phenomenon is that the spin holonomy is screened by spinning particles.

\begin{figure}[t]
    \centering
    \begin{align*}
        {\renewcommand{\arraystretch}{1.6}
        \renewcommand{\arraycolsep}{1.0em}
        \begin{array}{cc}
            \includegraphics[valign=c,scale=1.3]{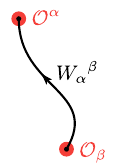}
            &
            \includegraphics[valign=c,scale=1.3]{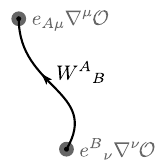}
        \end{array}
        }
    \end{align*}
    \caption{%
        Spinor holonomy is screened by fermions.
        Vector holonomy is screened by orbital angular momentum.
    }
    \label{fig:screen}
\end{figure}

Interestingly, this implies that the gravitational one-form symmetry is explicitly broken by particles with spin.  
For example, if the local Lorentz group is $\G = \Spin(4)$, then a holonomy in the spinor representation can only end on a local operator $\O_\alpha$ with a free spinor index.  Hence, the corresponding holonomy is endable if there exist fermions in the spectrum.
The case of $\G = \SO(4)$ is more subtle, however. 
A holonomy in the vector representation of the Lorentz group must end on an operator with a free Lorentz vector index,
which can be 
any operator carrying orbital angular momentum
dotted with a tetrad 
such as
$e^A{}_\m \nabla^\m\mathcal{O}$.\footnote{
    A more subtle question arises in formulations of gravity with differences in field content.
    For example, in the Pleba\'nski theory
    \cite{Plebanski:1977zz}
    described in \Sec{Jcx>COFRAME},
    the associated two-form field $B$ is a fundamental degree of freedom.
    The tetrad is sculpted from $B$ through a constraint 
    \cite{Buffenoir:2004vx,Celada:2016bf},
    while the metric can be expressed in terms of $B$ in closed form \cite{Urbantke:1984eb,Capovilla:1991qb}.
    Consequently, there is no field in the pure gravity sector that carries a single vector index, so it is unclear whether the vector spin holonomy is endable in this case.
        This illustrates a situation where 
        the one-form symmetry may be capable of distinguishing between
         theories
        that only differ nonperturbatively.
}
Consequently, the vector holonomy is analogous to the adjoint Wilson loop of gauge theory, which is automatically screened by dynamical gluons.  See \Fig{fig:screen} for a depiction of this phenomenon.

The above logic has implications for the standard model of physics.
Since fermions exist, we know that the local Lorentz group is $\G = \SL(2,\mathbb{C})$, which implies that there is a $\mathbb{Z}_2$ one-form gravitational symmetry under which spinor holonomies are charged.
Presuming the lightest neutrino is not massless, this one-form symmetry is unbroken below that scale.  This implies a new, albeit subtle, exact symmetry in the known laws of physics.

Notably, the explicit breaking of higher-form symmetry is strongly suggested by the so-called swampland conjectures.  In particular, there is strong evidence that all global symmetries are necessarily broken at some scale in a consistent theory of quantum gravity \cite{Misner:1957mt,Polchinski:2003bq,Banks:2010zn,Harlow:2018tng}.  A well-known avatar of this is the weak gravity conjecture \cite{Arkani-Hamed:2006emk,Harlow:2022ich}, which states that a $\U(1)$ gauge theory must exhibit a state whose charge exceeds its mass in Planck units.  The weak gravity conjecture quantitatively forbids the strict global symmetry limit of vanishing charge in a $\U(1)$ gauge theory.  

The swampland conjectures imply that any  higher-form symmetry should be either gauged or explicitly broken.  In the case of gauge theory coupled to gravity, the latter scenario requires the existence of a tower of charged states, as described  by the so-called completeness conjectures \cite{Heidenreich:2021xpr}.
Applying the same logic to dynamical gravity, we expect that something similar applies to the gravitational one-form symmetry.  One option is that this symmetry is gauged, for example as would occur if the Lorentz group is 
$\SO^+\nem(3,1) = \SL(2,\C) / \Z_2$.
Alternatively, if the Lorentz group is 
$\SL(2,\C)$,
then the one-form symmetry is not gauged and must be  explicitly broken, thus implying the existence of fermions in the spectrum.

Finally, let us speculate briefly on the possibility of phases in gravity.  Taking inspiration from gauge theory, it is natural to 
wonder whether the expectation value of the spin holonomy is an order parameter for symmetry breaking.  In the case of gauge theory, it is well-known that an area versus perimeter law scaling of the Wilson loop expectation value is a diagnostic of whether or not a theory is confining.  However, the analogous construction in gravity is far murkier.  
In particular, 
 the diffeomorphism invariance of dynamical gravity
 may suggest that the contour of the spin holonomy
 should be defined relationally with respect to some invariant boundary data.  So to be an order parameter, the spin holonomy may presumably be computed for a contour that circumscribes the boundary.

Even ignoring these subtleties, the fact that the EFT description of gravity is intrinsically weakly coupled suggests that confinement is not in play.
More generally it is very unclear whether a low-energy EFT of gravity on a fixed background could even access different phases, or what that would even mean.  One speculation is that this might have something to do with degenerate configurations of the metric and their corresponding domain walls \cite{Jacobson:1992xy,Romano:1993bj,Bengtsson:1997wr,Volovik:1999vb}.  Another possibility is that a putative gravitational phase diagram might delineate various choices of compactification or of asymptotic behavior of the metric.  Indeed, it is easy to see that the spin holonomy is highly sensitive to the cosmological constant.
For these reasons, it would be interesting to explicitly compute the expectation value of the spin holonomy in various examples. A number of existing works have calculated the spin holonomy in various contexts \cite{Modanese:1993zh,Modanese:1991nh,Fredsted:2001rt,Alawadhi:2021uie,Jacobson:1992ya,Donoghue:2016vck,Brandhuber:2008tf}. 

\section{ Future Directions}
\label{sec:Future}

In this work, we have initiated an exploration of generalized symmetry in the context of dynamical gravity.  Taking our cues from the one-form symmetries of YM theory, we have considered gravity in the tetradic Palatini formalism, which is a gauge theory of the local Lorentz group. We have argued that the gravitational one-form symmetry is defined by the center of the Lorentz group.  The  object which is charged under this symmetry is the spin holonomy $W_\rep(\CC)$.  The one-form symmetry transformation is implemented by an operator $U_\cen(\S)$, which has dual interpretations, both as a twisted Lorentz transformation but also as a chiral cosmic string defect carrying both electric and magnetic gravitational charge. The topological linking of the line and symmetry operators corresponds to the measurement of a quantized spin precession angle by the spin holonomy
as a gravitational AB effect.
In the standard model, this implies the existence of a new symmetry below the mass of the lightest neutrino.
The present work leaves numerous avenues for future exploration, which we now describe.

First and foremost are a number of very simple extensions of this work which should be relatively straightforward.  These include the question of generalization to higher spacetime dimensions, which offers a richer spectrum of one-form symmetry groups.
For example, the center group is maximally $Z(\Spin(4)\hnem) = \mathbb{Z}_2 {\mem\times\mem} \mathbb{Z}_2$ in four dimensions, but this grows to $Z(\Spin(6)\hnem) = \mathbb{Z}_4$ in six dimensions.  Another concrete direction is the inclusion of gravitational higher-curvature corrections.
As mentioned earlier,
these contributions will clearly preserve the one-form symmetry,
whose corresponding symmetry operator will be equal to the surface integral of the canonical conjugate of the curvature in the EFT.

Secondly, while the present work has focused on a gravitational one-form symmetry of electric type, it is natural to ask whether one can derive 
a magnetic counterpart.
For gauge theories,
the electric and magnetic symmetries are 
straightforwardly
related via Hodge duality
\cite{Schafer-Nameki:2023jdn}.
In contrast,
constructing the magnetic counterpart
of our gravitational one-form symmetry
stands as an open question.
One might possibly be considering 
the gravitational analog of the 't Hooft line
and attempting to identify the conjugate field
for the (internal) Hodge dual of the curvature two-form,
while an alternative pathway aiming to dualize the Pleba\'nski two-form
might also be a speculation.

A third topic of future study is
higher-group symmetry,
which describes a certain nonabelian structure built from the fusion of  multiple higher-form symmetries   \cite{Kapustin:2013uxa,Benini:2018reh,Cordova:2018cvg,Cordova:2020tij}.
A well-studied example of this is axion-YM theory, which exhibits a two-form symmetry for the axion, together with the one-form symmetry of the gauge theory \cite{Brennan:2020ehu}.
These symmetries fuse to yield a two-group structure, which in fact bounds the scale of axion strings from below by the lightest particle in the fundamental of color.
Acquainted with this remarkable fact, 
it would be interesting to investigate
if gravity can also exhibit a higher-group symmetry.
It seems quite likely that a similar two-group symmetry will appear in gravity coupled to an axion, in which case we should expect that the axion string scale is  bounded by the mass of the lightest fermion.

\begin{figure}[t]
    \centering
    \begin{align*}
        {\renewcommand{\arraystretch}{1.6}
        \renewcommand{\arraycolsep}{0.6em}
        \begin{array}{ccc}
            \includegraphics[valign=c,width=0.3\linewidth]{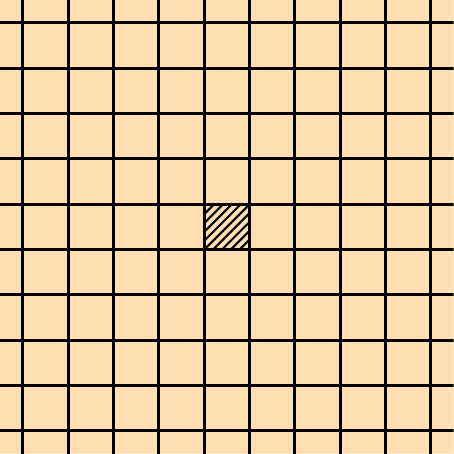}
            &
            \xrightarrow[\text{\footnotesize
                diffeomorphism
            }]{\text{\footnotesize
                \,
                multivalued
                \,
            }}
            &
            \includegraphics[valign=c,width=0.3\linewidth]{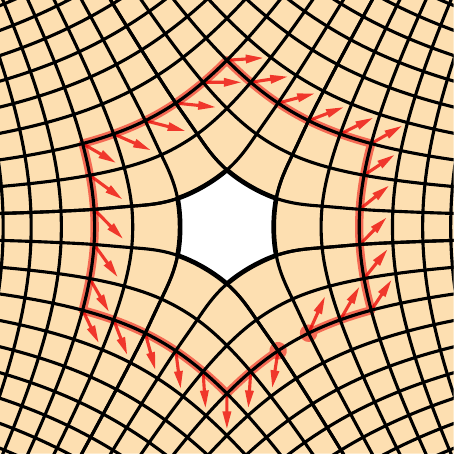}
            \\
            \adjustbox{scale=0.9,valign=c}{
                ``Empty Space''
            }
            &
            &
            \adjustbox{scale=0.9,valign=c}{
                ``Cosmic String''
            }
        \end{array}
        }
    \end{align*}
    \caption{%
        A diffeomorphism with a multivalued Jacobian
        creates a cosmic string from empty space,
        analogous to disinclinations in lattice systems.
    }
    \label{fig:cosmic}
\end{figure}

Fourthly, while our approach of formulating gravity into the ``$BF\mem$'' grammar yields a simple route to the one-form symmetry, there is the question of how this framework is explicitly realized in other physically equivalent formulations of gravity \cite{krasnov2020formulations} in the literature.
Another open question is the fate of the one-form symmetry in gravitational formulations with subtly different field content,
as referred to in a footnote in \Sec{sec:Breaking}.

A fifth area of study concerns the question of what topological symmetry can  teach us about classical gravitation.
As a theory of spacetime geometry,
gravity boasts a rich array of classical vacuum solutions,
each showcasing distinctive \textit{singularity structures} that are themselves a focal point of study
\cite{Penrose:1964wq}.
In particular, it could be very illuminating to 
initiate
a systematic analysis and classification of gravitational singularities 
from the perspective of topological operators and
their algebra.
For example, even in the absence of spin holonomy,
the one-form symmetry operator can link with  ``holes'' in spacetime.
It would be interesting to study whether there is
physical information
encoded in
such a linking, for example regarding the singularity of
a black hole.
Moreover, 
it is conceivable that known spacetime singularities in the literature have an alternative interpretation as symmetry operators, like we discovered for the chiral cosmic string.

A sixth question is 
whether there exists a gravitational one-form symmetry
that arises by the geometry of
\textit{multivalued diffeomorphisms}.
While our construction
has viewed gravity as a gauge theory of local Lorentz group,
the more fundamental redundancy endemic to dynamical gravitation
is diffeomorphisms.
As described below as well,
we anticipate that
this line of inquiry
will likely realize the Misner string proposal
suggested in \Sec{K1.grav}.
Difficulties are apparent,
however,
since the diffeomorphism group is not finite,
does not admit a nice center,
and even does not describe \textit{internal} transformations
in the strict sense of gauge theories.

\begin{figure}[t]
    \centering
    \begin{align*}
        {\renewcommand{\arraystretch}{1.6}
        \renewcommand{\arraycolsep}{0.6em}
        \begin{array}{ccc}
            \includegraphics[valign=c,width=0.36\linewidth]{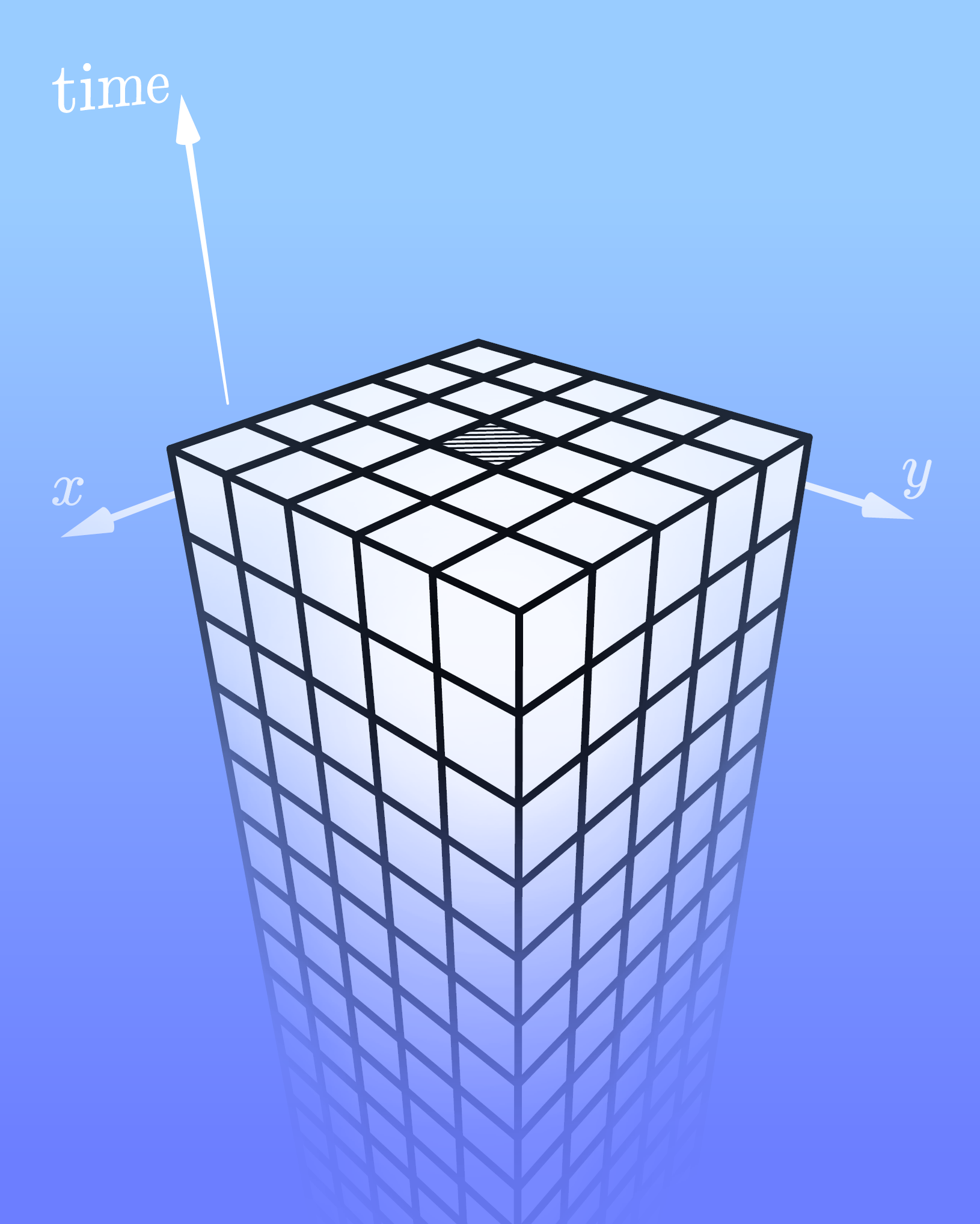}
            &
            \xrightarrow[\text{\footnotesize
                diffeomorphism
            }]{\text{\footnotesize
                \,
                multivalued
                \,
            }}
            &
            \includegraphics[valign=c,width=0.36\linewidth]{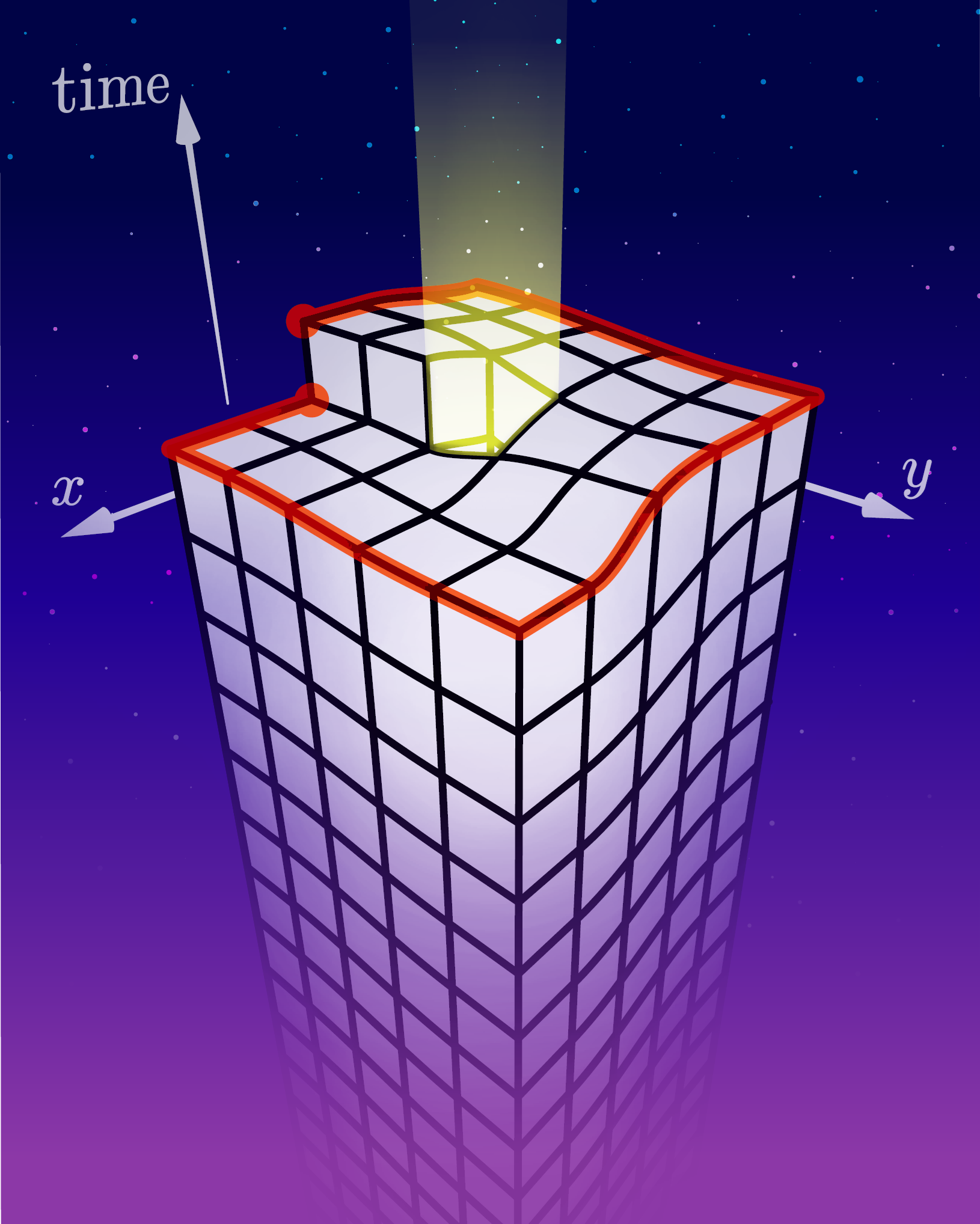}
            \\
            \adjustbox{scale=0.9,valign=c}{
                ``Empty Space''
            }
            &
            &
            \adjustbox{scale=0.9,valign=c}{
                ``Misner String''
            }
        \end{array}
        }
    \end{align*}
    \caption{%
        A multivalued diffeomorphism 
        creates a Misner string from empty space,
        whose time monodromy
        is analogous to the Burgers vector of dislocations in lattice systems.
    }
    \label{fig:misner}
\end{figure}

Last but not least,
it would be interesting to
see if the 
theoretical framework developed in this chapter
can be applied
to
lattice systems
such as crystals
with impurities
\cite{chaikin1995principles,Grozdanov:2018ewh,Pace:2023kyi}. Indeed, as described in
\cite{Fiziev:1995te,Kleinert:1996yi,Kleinert:2008zzb}, the physics of lattice systems
has a formulation that is strikingly
reminiscent of gravity.
In this picture,
lattice disinclinations
are analogous to cosmic strings, as depicted in \Fig{fig:cosmic}.
Furthermore,
the analog of ``translation holonomy'' is manifested by
the Burgers vector,
which measures
the net drift 
in the ``lattice frame''
per round trip about a defect.
Hence, as depicted in \Fig{fig:misner},
we observe that
lattice dislocations 
are analogous to
the Misner string%
---%
a Dirac string of 
time monodromy flux
endable on Taub-NUT charges
\cite{Misner:1963flatter,misner1967taub,Bonnor:1969ala,sackfield1971physical,dowker1967gravitational,cho1991magnetic,Alfonsi:2020lub}.
These two types of lattice defects
are described in terms of
\textit{multivalued} coordinate transformations
\cite{Fiziev:1995te,Kleinert:1996yi,Kleinert:2008zzb},
which are
the diffeomorphism analogs of the twisted gauge transformations
that played such a crucial role in our construction of a gravitational one-form symmetry.
It would be interesting
if 
this convergence between
gravity and lattice systems
could cross-pollinate new insights 
across these fields.

\begin{subappendices}

\section{ Canonical Formalism}
\label{sec:CanYM}

The one-form symmetry of gauge theories
can also be understood from the complementary point of view of the Hamiltonian formalism.
To this end,
we suppose a product spacetime
$\M = \M_3 {\,\times\mem} \mathbb{R}$
equipped with coordinates $x^\m = (x^i,x^4)$.
We perform a $3{\,+\,}1$ decomposition in which 
$x^i$ denotes coordinates of the spatial three-manifold $\M_3$ and $x^4=t$ defines equal-time slices.

\subsection{Phase Space}
\label{sec:CanYM.ps}

We suppose the first-order template Lagrangian top form in \eqref{1form.L}.
Writing out all indices explicitly, 
we obtain
\begin{align}
\begin{split}
    \label{eq:YM.compL}
    \frac{1}{g^2}\mem
    \bbsq{
         \frac{1}{2}\mem
            B_{a\m\n}
            \Big(\hem{
                \partial_\r A^a{}_\s
                +
                \minie\mem
                f^a{}_{bc}\mem A^b{}_\r\mem A^c{}_\s
            }\Big)\mem
            \e^{\m\n\r\s}
    }
    \mem d^4x
    \,+ f[B]
    \,,
\end{split}
\end{align}
where
$\e^{\m\n\r\s}$ denotes the permutation symbol.
Carrying out the $3{\,+\,}1$ decomposition
and integrating out $B_{ai4}$,
we immediately see that
the dynamical coordinates on the phase space are $A^a{}_i$ together with 
their canonical conjugates,
\begin{align}
    \label{eq:Edef}
    E^i{}_a
    \mem=\mem
        \frac{1}{2}\mem B_{a jk}\mem \e^{ijk}
    \,.
\end{align}
As a result, we identify the canonical commutation relations,
\begin{align}
\begin{split}
    \label{eq:YM.ccr}
    \comm{ A^a{}_i(x) }{ A^b{}_j(x') } &\,=\, 0 
    \,,\\ 
    \comm{ E^i{}_a(x) }{ A^b{}_j(x') } &\,=\, ig^2\, \delta^i{}_j\,\delta^b{}_a\, \delta^{(3)}\hnem(x{\,-\,}x') 
    \,,\\
    \comm{ E^i{}_a(x) }{ E^j{}_b(x') } &\,=\, 0
    \,,
\end{split}
\end{align}
where $x,x' \in \M_3$ are points in the spatial manifold.
The phase space is also equipped with the Gauss constraint and a Hamiltonian,
the details of which are not important for our 
purposes.

\subsection{Ward Identity}
\label{sec:CanYM.ward}

In the language of the path integral, a one-form symmetry transformation is implemented through the insertion of a symmetry operator which wraps the line operator.
In the operator formalism, however,
this corresponds to a conjugation of the latter by the former.
To see how 
this
works
in detail, consider a line operator $W_\rep(\CC)$, where $\CC$ is restricted to an equal-time slice, say at $t{\,=\,}0$. As before, we take the symmetry operator $U_\cen(\S)$ to be 
supported
on an exact surface $\S$ with an associated coboundary $\V$,
so $\S = \partial\V$.

\begin{figure}[t]
    \centering
    \includegraphics[scale=1.33]{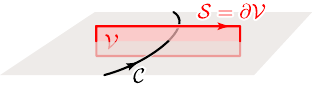}
    \\[1.25\baselineskip]
    \includegraphics[scale=1.33]{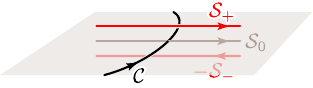}
    \\[1.25\baselineskip]
    \includegraphics[scale=1.33]{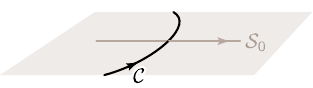}
    \caption{%
        Pancaking the symmetry operator
        on a purely spatial line operator $\CC$.
        The plane depicts an equal-time slice
        while time flows upwards.
        \textit{Top:}
            The equal-time slice bisects
            the 
            volume $\mathcal{V}$ 
            by a 
            surface $\S_0$.
        \textit{Middle:}
            The 
            surface $\S = \partial\mathcal{V}$
            splits into surfaces $\S_+$ and $-\S_-$
            at the infinitesimal future and past.
        \textit{Bottom:}
            $\S_+$ and $\S_-$ both
            project down to the surface $\S_0$.
    }
    \label{fig:pancake1}
\end{figure}

The geometric set-up is depicted in \Fig{fig:pancake1}.
We assume that
the surface
$\S$ \textit{links} once with 
the purely spatial loop $\CC$.  Consequently, 
the coboundary $\V$ is
\textit{intersected} by the loop $\CC$ and \textit{bisected} by the spatial slice.
Now imagine continuously squashing or pancaking 
the coboundary
$\V$ along the time direction such that  
its two-dimensional boundary $\S$ infinitesimally hugs the spatial slice.  In this limit,
$\S =\S_+ {\mem\cup\mem} (-\S_-)$ is the union
of two disjoint discs 
$\S_+$ and $-\S_-$
at the infinitesimal future and past
across $t{\,=\,}0$. Once $\V$ has completely collapsed into the spatial slice at $t {\,=\,} 0$, both discs
$\S_\pm$ approach the same surface, which we denote by $\S_0$.
From \Fig{fig:pancake1}, we see that this $\S_0$ 
will be the intersection between $\V$ and $t=0$.
As a result,
the
intersection of $\CC$ and $\V$ in the four-manifold $\M$ 
is equivalent to
intersection of
$\CC$ and $\S_0$
in the three-manifold $\M_3$ 
as the slice $t{\,=\,}0$.
Therefore,
we have
in general
\begin{align}
    \label{eq:4-to-3}
    \link(\CC,\S)
    \,=\,
    \intersect(\CC,\V)
    \,=\,
    \intersect_3(\CC,\S_0)
    \,,
\end{align}
where $\intersect_3$
denotes intersection number in $\M_3$.

Now,
we can describe how this
pancaking procedure
boils down the Ward identity
to an equal-time operator equation.
Since $\S =\S_+ {\mem\cup\mem} (-\S_-)$, we see that
the symmetry operator factorizes into
$U_\cen(\S) = 
U_\cen(\S_+)\mem U_\cen(-\S_-)$.\footnote{
    Note that 
    here we have allowed nonclosed surfaces
    for the support of symmetry operators,
    which might be a slight abuse of notation.
}
In turn,
the left-hand side of the Ward identity in \eqref{1form.ward}
becomes the time-ordered expression
$U_\cen(\S_+)\mem W_\rep(\CC)\mem U_\cen(-\S_-)$, which in the 
process of pancaking
limits to the equal-time operator product
$U_\cen(\S_0)\mem W_\rep(\CC)\mem \Uinv(\S_0)$.
Using \eqref{eq:4-to-3}, we then find that
the Ward identity 
translates to 
\begin{align}
    \label{eq:operatorWard}
    U_\cen(\S_0)\, W_\rep(\CC)\, \Uinv(\S_0)
    \,=\,
    \rho(\alpha)^{\intersect_3(\CC,\S_0)}\, W_\rep(\CC)
    \,,
\end{align}
which is an equal-time operator equation.

\eqref{eq:operatorWard}
is the avatar of the Ward identity in the operator formalism,
where all the relevant geometric objects and operations, including the disc $\S_0$ and the closed contour $\CC$, are defined within the three-manifold $\M_3$.
The key insight here is that
time ordering in the path integral formalism
turns into
operator ordering in the operator formalism.

\begin{figure}[t]
    \centering
    \includegraphics[
        scale=1.25
    ]{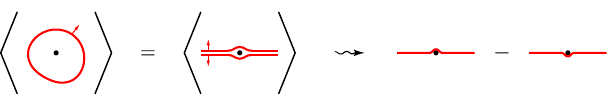}
    \caption{%
        Zero-form version of
        the pancaking procedure,
        which the reader may be familiar with.
        Wrapping an operator with the symmetry operator
        translates to an equal-time conjugation,
        where
        operator ordering
        traces back to
        time ordering.
    }
    \label{fig:pancake0}
\end{figure}

Finally,
it is straightforward to 
explicitly evaluate \eqref{eq:operatorWard}.
Applying the $3{\,+\,}1$ decomposition and using \eqref{eq:Edef},
the Hamiltonian formalism avatar of the symmetry operator,
supported on a 
disc
$\S_0 {\,\subset\,} \M_3$
in three dimensions,
is given as
\begin{align}
    \label{eq:YM.Q}
    U_\cen(\S_0)
    = \exp\bb{
        \frac{-i\pi}{g^2}
        \int_{\S_0} dx^j \swedge dx^k\, \e_{ijk}\mem E^i{}_a\mem \lambda^a
    }
    \,,\quad
    e^{2\pi\lambda} = \cen
    \,.
\end{align}
To compute 
$U_\cen(\S_0)\mem W_\rep(\CC)\mem \Uinv(\S_0)$,
let us first deduce how the conjugation acts on the phase space.
While $E^i{}_a(x)$ is left invariant
because $\comm{ E^i{}_a(x) }{ E^j{}_b(x') } = 0$,
the spatial gauge connection $A^a{}_k(x)$ has a nonvanishing commutator with the exponent of \eqref{eq:YM.Q},
\begin{align}
\begin{split}
    \label{eq:Qaction-calc}
    &
    \frac{-i\pi}{g^2}\mem \int_{\S_0} 
    dx'^i \swedge dx'^j\, \lambda^b(x')\,
        \e_{ijl}\mem
        \comm{ E^l{}_b(x') }{ A^a{}_k(x) }
    \,\\
    &\mem=\mem
    2\pi \int 
    d\s_1\hem d\s_2\,\,
        \frac{\partial X^i}{\partial\s_1} 
        \frac{\partial X^j}{\partial\s_2}
    \, \lambda^a(X)\,
        \e_{ijk}\mem
        \delta^{(3)}\hnem(x-X)
    \,,
\end{split}
\end{align}
where $(\s_1,\s_2) \mapsto X^i(\s_1,\s_2)$ is a parameterization of the surface $\S_0$.
According to \eqref{eq:delta-param},
this
describes the components of the one-form
$2\pi\lambda^a\mem \delta_3(\S_0)$,
where $\delta_3(\S_0)$ is the Dirac delta one-form 
of $\S_0$
defined
in the spatial three-manifold $\M_3$.
Therefore,
we can summarize the action of
the symmetry transformation 
on the phase space variables
as the following:
\begin{align}
\begin{split}
    \label{eq:Qaction}
    U_\cen(\S_0)\,{
        A^a{}_i
    }\,\Uinv(\S_0)
    &\,=\, \bigbig{
        A^a + 2\pi\lambda^a\mem \delta_3(\S_0)
    }_i
    \,,\\
    U_\cen(\S_0)\,{
        E^i{}_a
    }\,\Uinv(\S_0)
    &\,=\, E^i{}_a
    \,.
\end{split}
\end{align}
As a result, we find that the Wilson loop transforms as
\begin{align}
\begin{split}    
    \label{eq:QadjW}
    U_\cen(\S_0)\, W_\rep(\CC)\, \Uinv(\S_0)
    \,=\,
    \rho(\alpha)^{\intersect_3(\CC,\S_0)}\mem
    W_\rep(\mathcal{C})
    \,,
\end{split}
\end{align}
which 
proves 
the Hamiltonian counterpart of the
Ward identity.
We have used the fact that
the three-dimensional intersection number is
$\intersect_3(\CC,\S_0) = \oint_\CC \delta_3(\S_0) $.

\subsection{Pleba\'nski Gravity}
\label{K1:app3}

The above canonical analysis
for the Ward identity
straightforwardly applies
to gravity
with the same formulae in
\eqrefss{eq:Edef}{eq:YM.ccr}{eq:YM.Q}
(up to the $g \mapsto \k$ replacement),
by utilizing the
Pleba\'nski formulation \cite{Plebanski:1977zz} of gravity
in which the $B$-field is promoted to a fundamental field variable;\footnote{
    While the path originally taken by Ashtekar
    for deriving canonical gravity
    starts from
    the tetradic Palatini formulation
    (see \rcite{AshtekarBook}, for instance),
    the Pleba\'nski formulation offers a more straightforward path
    because it manifests the symplectic structure
    in the ``$BF$'' form
    with the $B$-field non-composite.
}
see \Sec{Jcx>COFRAME}.
The resulting Hamiltonian formulation of gravity
essentially describes a nonchiral version of
the Ashtekar \cite{Ashtekar:1986yd,AshtekarBook} formalism.
Namely,
it is known that
implementing the $3{\,+\,}1$ decomposition
and
integrating out the nondynamical degrees of freedom
from the Pleba\'nski Lagrangian
results in
a well-posed constrained Hamiltonian system
in the phase space $(\gamma^a{}_i , E^i{}_a)$
coordinatized in the Darboux basis
\cite{Buffenoir:2004vx,Alexandrov:2008fs,Celada:2012ua,Celada:2016bf}.
The constraints include
the Gauss constraints for local Lorentz transformations
as well as
the Hamiltonian and diffeomorphism constraints
familiar from the ADM \cite{Arnowitt:1959ah} analysis.
\nomenclature{ADM}{Arnowitt-Deser-Misner}

\end{subappendices}

\chapter[Double Copy and Diffeomorphism Gauge Theories]{Double Copy and\\[0.4\baselineskip] Diffeomorphism Gauge Theories}
\label{K2}

    We investigate the idea of diffeomorphism gauge theory
	as a strict implication of
    EoM level
    CK duality.
	In field theory,
	diffeomorphisms are external transformations shifting the base spacetime points
	and are considered disparate from the internal, fiberwise transformations
	of gauge theories
    such as YM theory.
	However,
    an EoM level formulation of
	manifest CK duality 
    is known to necessitate
	taking diffeomorphisms like the internal color algebra of a gauge theory,
    via the so-called 
    color-to-kinematics replacement
    (CKR).
    \nomenclature{CKR}{Color-to-Kinematics Replacement}
	  It is shown that CK duality
	systematically establishes a consistent definition of
	diffeomorphism gauge transformation,
	diffeomorphism gauge connection,
    diffeomorphism covariant derivative, 
    diffeomorphism field strength,
    and diffeomorphism Wilson line.
	Notably,
    such a diffeomorphism gauge theory
    can admit a simple and consistent geometrical construction
	as a field theory of a dynamical vielbein.
    In particular,
    this implies that classical Born-Infeld theory can be embedded in a teleparallel theory.
    For gravity,
    we note that
    diffeomorphism gauge theory formulations
    are viable
    in three dimensions or in the SD sector in four dimensions.
    For a classical solution of diffeomorphism gauge theories,
    we reinterpret the Misner string
    as a topological configuration
    generated by a multivalued diffeomorphism gauge transformation
    and view it as the 
    counterpart
    of the Dirac string.

\section{ Introduction}
\label{K2>INT}

YM theory is a gauge theory.
Gravity is a gauge theory.
What these statements precisely mean,
however,
is utterly different.

By the late twentieth century,
gauge theory acquired a systematic geometric interpretation,
establishing a paradigm in mathematical physics.
Yang and Mills had introduced 
the nonabelian gauge principle
as a local version of internal symmetry \cite{Yang:1954ek}.
Wu and Yang later studied the global aspects and emphasized the role of principal bundles \cite{Wu:1975es}.
As per the Wu-Yang dictionary \cite{Wu:1975es},
the gauge potential became a connection on a principal bundle;
the field strength became its curvature;
gauge transformations became changes in local trivializations;
and
Wilson loops, AB phases, monopoles, and instantons became manifestations of holonomy and global topology
\cite{Yang:1954ek,Wu:1975es,Yang:1977cq,Yang:1978fb,Trautman:1979gf,Trautman:1980fb,Eguchi:1980jx,Nakahara,tong2018gauge,SkinnerAQFTNotes}.

The success of this geometric ideal of Wu, Yang, and Mills was multifold.
To physicists,
it gave a common language to 
the three fundamental forces of our nature:
\textit{electromagnetism, the weak, and the strong}.
It also provided systematic tools for investigating topics such as 
monopoles, instantons, and anomalies
\cite{Eguchi:1980jx,Nakahara,tong2018gauge,SkinnerAQFTNotes}.
To mathematicians,
it gave the realization that
physics can be used as 
a tool for studying the topology of manifolds and moduli spaces of bundles
\cite{Atiyah:1983ah,Uhlenbeck:1982,Donaldson:1983wm,Donaldson:1983,Freed:1984ifm,Donaldson:1990gfm}.

A gauge theory, in this narrower context,
is a theory on a fixed base manifold $\M$,
equipped with an internal fiber modeled on a group $\G$.
The gauge transformations are vertical:
they act within each fiber while leaving the base point $x \inn \M$ completely fixed.
This fiberwise, \textit{internal} character is necessary.

\begin{Fullnote}
    YM theory is a gauge theory in the narrower,
    strict mathematical sense of Wu, Yang, and Mills,
    based on internal local transformations.
\end{Fullnote}

Interestingly, \textit{gravity}
has been both the most tempting and the most resistant candidate for inclusion in this class. 
On the one hand,
GR is already inherently geometric.
It features
the Levi-Civita connection and
the Riemannian curvature
as well as
parallel transport, holonomy, and topological invariants.
In particular,
the Riemannian curvature looks like a field strength of a gauge theory.
This proximity makes it natural to ask whether gravity, too, should be a gauge theory
in the same mathematical sense of Wu, Yang, and Mills.

Crucially,
the local symmetry of GR describes diffeomorphisms:
the freedom to choose coordinates.
Diffeomorphisms move spacetime points and pull back fields.
They act on the base manifold $\M$ itself.
They describe
\textit{external} transformations to say,
not internal.
In YM theory,
spacetime is the fixed arena on top of which 
the physics of gluons take place.
In GR,
the spacetime itself is dynamical.
Diffeomorphisms are not gauge transformations in the YM sense.

\begin{Fullnote}
    GR is sometimes said to be a gauge theory,
    but this is only in the broader sense 
    where the word ``gauge'' is used as a shorthand for ``redundancy.''
    The redundancies of GR are diffeomorphisms,
    which are external local transformations.
\end{Fullnote}

A related point is that
GR employs a connection on the tangent bundle $T\M$ of the spacetime $\M$,
and $T\M$ is not any bundle
(like the fiber bundles in YM theory)
but is intimately tied to $\M$.
Elements of each tangent fiber describe
actual directions in spacetime,
as
a vector
describes a pair of
infinitesimally neighboring points
on the very base space $\M$.
This is the soldering structure,
which can be represented as a coframe (vielbein).
Giving
a gauge theory style interpretation to the soldering structure
could be
subtle and tricky
\cite{Trautman:1979gf,Trautman:1980fb,Percacci:1984bq}.

Historically,
there had been 
aspirations toward reformulating GR 
as a gauge theory
\cite{Utiyama:1956sy,Kibble:1961ba,Sciama:1962,Hehl:1976kj,Hayashi:1967se,hayashi1977gauge,hayashi1981addendum,cho1976einstein,cho1976gauge,cho1992gravdiff,%
Grignani:1991nj,Blagojevic:2003cg,%
Percacci:1984bq,%
Blagojevic:2002grg,Blagojevic:2013xpa,Obukhov:2006pgt,Hehl:1994ue,MacDowell:1977jt,Wise:2010sm,Catren:2014mna,Ashtekar:1986yd},
envisioning a unified geometrization of
the four fundamental forces.
Early attempts began with gauging spacetime symmetries.
Utiyama \cite{Utiyama:1956sy} gauged the Lorentz group and obtained the spin connection,
but the vielbein had to be supplied as an additional structure.
To this end,
Kibble \cite{Kibble:1961ba} considered gauging the Poincaré group
so that the vielbein may be recast
as a kind of a ``translation gauge connection,''
arriving at
Einstein-Cartan theory.
Other notable or adjacent results include
the Poincar\'e gauge theory formulation of
three-dimensional gravity
\cite{Witten:1988hc,Witten:2007kt},
MacDowell–Mansouri formulation and its descendants
\cite{MacDowell:1977jt,Wise:2010sm,Catren:2014mna},
Ashtekar variables
\cite{Ashtekar:1986yd},
and
Pleba\'nski or $BF$ gravity
\cite{Plebanski:1977zz,Capovilla:1991qb,Urbantke:1984eb,DePietri:1998hnx,Buffenoir:2004vx,Alexandrov:2008fs,Celada:2012ua,Celada:2016bf}.
Despite these developments,
the gauge-theoretic formulation of GR
still appears as an open problem,
and
the practical outcomes are unclear.

\medskip

Notably, we remark that
modern explorations
through scattering amplitudes
bring
a new perspective
into this discourse.
In the scattering amplitudes program,
gauge redundancies
may be portrayed as pure evil,
hindering efficient computations
and obscuring physical structures.
Hence one seeks for more efficient representations of scattering amplitudes
instead of sticking to the Feynman diagram representations
arising from conventional Lagrangians.

Surprisingly,
deep computational explorations into
gluon amplitudes
have concluded that
the entire nonlinear content of YM theory
might be encapsulated in a single, universal \textit{cubic} interaction vertex,
given as the product of 
two structure constants.
The first are the very structure constants of the color gauge algebra, such as $\g = \su(N)$.
The second, however,
are the structure constants of a mysterious Lie algebra $\gk$.
This mystery Lie algebra $\gk$
is known as the kinematic algebra of YM theory,
as it would have to take the kinematic data (momenta and polarizations)
as ``adjoint indices.''
The fact that YM amplitudes
admit such remarkable cubic representations
is referred to as
CK duality
\cite{BCJ1,BCJ2}.
When extrapolated to a field theory statement
\cite{Cheung:2016prv,Cheung:2017ems,CCK},
CK duality 
provides striking evidence for 
a cubic reformulation of YM theory,
although
the Feynman rules of
the conventional YM Lagrangian
describe a quartic perturbation theory.
At this moment,
CK duality stands well-established 
at the level of on-shell scattering amplitudes.
See \cite{BCJReview} for a comprehensive review.

Further surprisingly,
deep computational explorations into graviton amplitudes
have concluded that
the entire nonlinear content of GR
might be encapsulated in a single, universal cubic interaction vertex
that simply squares the structure constants of 
the kinematic algebra $\gk$ of YM theory.\footnote{
    Technically, the precise double copy of YM theory
    is ``fat gravity,''
    containing not only the graviton
    but also a dilaton and a Kalb-Ramond $B$-field
    (the Neveu-Schwarz-Neveu-Schwarz (NS-NS) sector of type II supergravity).
    \nomenclature{NS}{Neveu-Schwarz}
    Yet, one can consistently truncate such extra modes.
}
This is the celebrated double copy correspondence
between YM theory and GR
\cite{BCJ1,BCJ2,BCJReview},
which has a string-theoretic precursor as well \cite{KLT}.
Again,
the double copy
is well-established 
at the level of on-shell scattering amplitudes.
When extrapolated to a field theory statement
\cite{Cheung:2016prv,Cheung:2017ems,CCK},
it gives striking evidence
for the conjecture that
GR could be reformulated 
as a YM theory
for a choice of a mysterious gauge algebra $\gk$,
at least at the EoM level:
\begin{align}
    \label{CKD:GR=YM}
    \GR
    \,\,\cong\,\,
    \YM(\gk)
    \,.
\end{align}
At this moment,
this conjecture stands verified
only within the SD sector
\cite{monteiro2011kinematic,MasonNewman:1989},
while its validity beyond the SD sector
has been inconclusive.

\begin{figure}[t]
    \centering
    \includegraphics[width=0.33\linewidth]{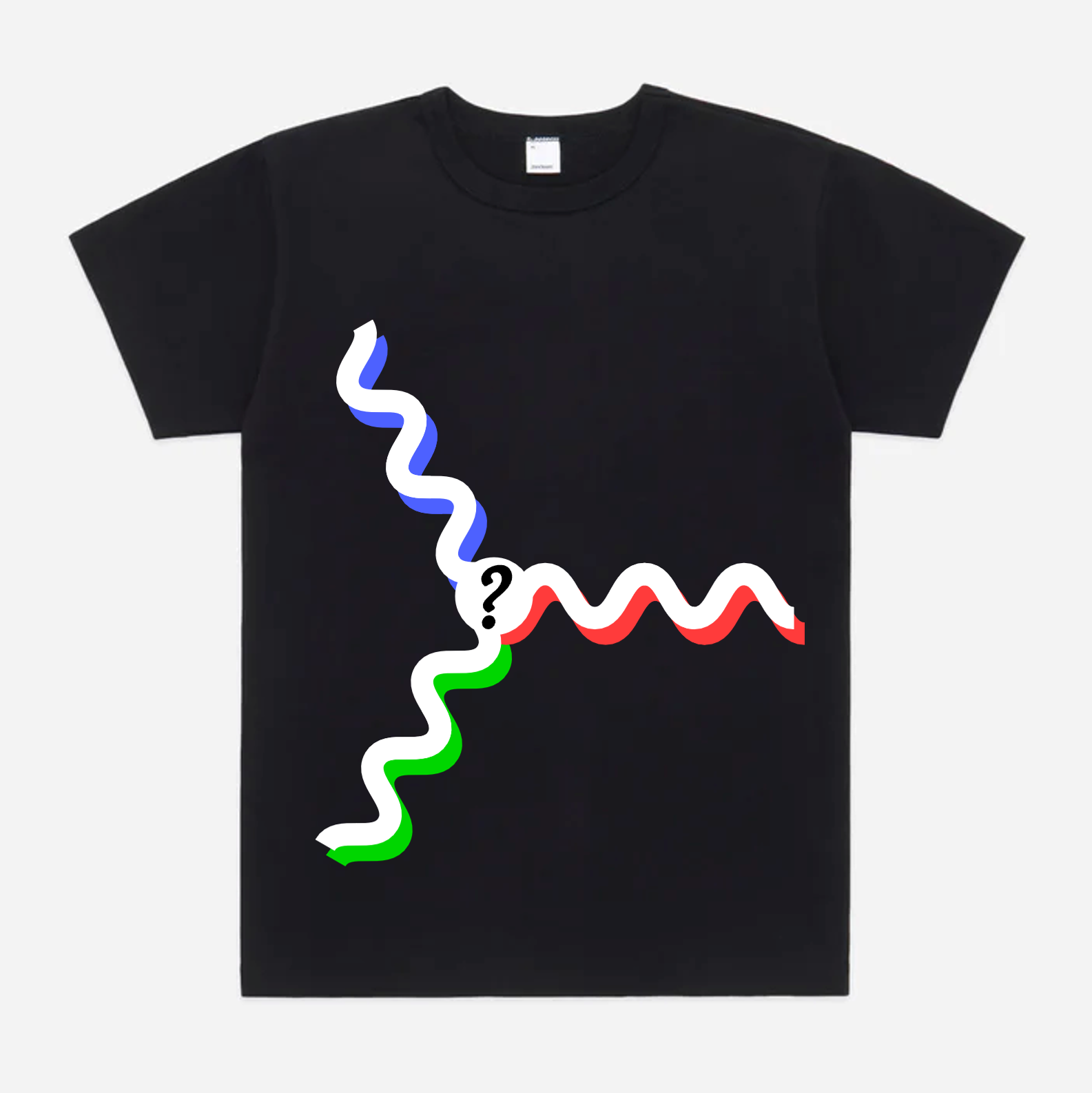}
    \medskip
    \caption{``What is the kinematic algebra of YM theory?''}
    \label{ck-tshirt}
\end{figure}

\nomenclature{NLSM}{Nonlinear Sigma Model}
\nomenclature{SG}{Special Galileon}

CK duality and double copy
are not properties that are exclusively enjoyed by YM theory or GR;
as well-reviewed in \rcite{BCJReview},
there exists a web of theories.
For example,
nonlinear sigma model (NLSM)
(as principal chiral model for a Lie group)
is known to exhibit CK duality,
whose field-theoretic explanation was established by
\rrcite{Cheung:2016prv,Cheung:2017ems,CCK}.
In particular,
Cheung and Mangan \cite{CCK}
explicitly identified 
the kinematic algebra of NLSM
at the EoM level
as the Lie algebra of
volume-preserving diffeomorphisms on spacetime,
$\sdiff(\M)$.
The double copy of NLSM
is a special instance of a Galileon
\cite{hinterbichler2015hidden,Cheung:2014dqa,Cheung:2015ota}
theory,
or ``special Galileon'' (SG) in short,
which arises by doubling the volume-preserving diffeomorphism algebra
\cite{Cachazo:2014xea,Cheung:2016prv,Cheung:2017ems,CCK,Mizera:2018jbh}.

It is also known that,
despite the mysterious status of the kinematic algebra $\gk$ of YM theory,
one can replace its color algebra
with the volume-preserving diffeomorphism algebra
to obtain a theory with
two kinematic algebras,
which turns out to be BI theory
\cite{BCJReview,CCK}:
\begin{align}
    \label{CKD:BI=YM}
    \BI
    \,\,\cong\,\,
        \YM(\sdiff(\M)\hnem)
    \,.
\end{align}
This action of replacing a color\footnote{
    Strictly speaking,
    ``color'' in the context of CK duality
    sometimes refers to a \textit{flavor} index,
    not necessarily gauged.
    This is a common abuse of terminology.
} (meaning finite-dimensional) Lie algebra
with a kinematic Lie algebra
is known as CKR
and is implemented in \rcite{CCK}.
The correspondence in \eqref{CKD:BI=YM}
precisely conforms to the grammar of \eqref{CKD:GR=YM},
and
BI theory serves as a concrete testing ground for
enhancing our understandings on kinematic algebras and double copy.

The implication of the double copy correspondence in \eqref{CKD:BI=YM} is the following.
The YM equations describe the vanishing of the covariant divergence of the nonabelian field strength $F^a{}_{\a\b}$,
whose explicit formula reads
\begin{align}
    \label{K2intro:fsdef}
    F^a{}_{\a\b}
    \,=\,
        \partial_\a A^a{}_\b
        - \partial_\b A^a{}_\a
        + f^a{}_{bc}\mem A^b{}_\a\mem A^c{}_\b
    \,.
\end{align}
Here, $a,b,c,\cdots$ are the adjoint indices of the color Lie algebra, say $\g = \su(N)$,
while greek indices are spacetime indices.
Also, $f^a{}_{bc}$ denote the structure constants of $\g$,
so $f^a{}_{bc}\mem A^b{}_\a\mem A^c{}_\b$
equals the Lie bracket
$\comm{A_\a}{A_\b}^a$.
Cheung and Mangan \cite{CCK}
showed that
the scattering amplitudes of
BI theory
are precisely reproduced by
applying the framework of BG recursion
\cite{BerendsGiele,Rosly:1996vr,Selivanov:1997an,Mizera:2018jbh}
to the YM equations 
with the color algebra
replaced with
the volume-preserving diffeomorphism algebra $\sdiff(\M)$.
What this means is 
to impose vanishing covariant divergence on
the ``diffeomorphism field strength,''
$F^\m{}_{\a\b}$.
In their Eq.\,(3.28),
Cheung and Mangan \cite{CCK}
explicitly write down the definition of this
$F^\m{}_{\a\b}$
as
\begin{align}
    \label{K2intro:Dfsdef}
    F^\m{}_{\a\b}
    \,=\,
        \partial_\a A^\m{}_\b
        - \partial_\b A^\m{}_\a
        + \BB{
            A^\r{}_\a\mem \partial_\r A^\m{}_\b
            -
            A^\r{}_\b\mem \partial_\r A^\m{}_\a
        }
    \,.
\end{align}
Here, $A^\m{}_\a$ is a ``diffeomorphism gauge connection,''
and the bracketed terms in \eqref{K2intro:Dfsdef}
evaluate $\comm{A_\a}{A_\b}^\m$,
the Lie bracket of $\sdiff(\M)$
that computes the commutator
between vector fields as differential operators:
$\comm{A^\m{}_\a\mem \partial_\m}{A^\n{}_\b\mem \partial_\n} = \comm{A_\a}{A_\b}^\m\mem \partial_\m$.
As 
Cheung and Mangan \cite{CCK}
explicitly state in their Eqs.\,(3.27)-(3.29)
the CKR means to
replace the color index $a$
with the vector index $\m$
while replacing the color Lie bracket
with the Lie bracket between vector fields.
Note that the elements of the diffeomorphism algebra
are vector fields.
The volume-preserving condition 
simply puts an additional restriction 
$\partial_\m A^\m{}_\a = 0$.

To reiterate,
this ``diffeomorphism YM theory''
formulation of BI theory
is an inevitable logical consequence
provided
the existence of
manifest CK duality
at the EoM level
and is also concretely verified
in terms of
explicit scattering amplitudes calculations,
as shown by Cheung and Mangan \cite{CCK}.

\medskip

If one 
questions
the mathematical basis of this construction,
however,
the very phrase
``diffeomorphism YM theory''
appears
seriously wrong
in the first place.
What does it mean to take \textit{diffeomorphisms as the internal symmetry of a YM theory}?
It
directly
violates 
the careful geometrical distinction
made at the beginning of this introduction,
where it is literally written that
``Diffeomorphisms are not gauge transformations in the YM sense.''
To reiterate, the CKR wishes to take
diffeomorphisms,
which are \textit{external} transformations,
as the \textit{internal} gauge transformations
of a YM theory.
From this point of view,
the CKR seems to be
a fundamentally forbidden operation.

Clearly,
physics tells otherwise.
The idea of ``diffeomorphism YM theory''
is valid in concrete senses
as clarified above,
and it is
indeed
even the very gist of CK duality.
The kinematic algebra describes spacetime transformations
because it is kinematic,
and CK duality
literally likens it to a color algebra.

\medskip

The purpose of this work
is to show that
these points of 
the mathematicians
and
the physicists 
can be
reconciled into a single consistent picture,
and
there is no strict illegality in thinking about
a ``diffeomorphism YM theory,''
yet with some crucial clarifications.
Arguably,
the insights gained 
here
would show us
some expectations onto
how
the unsolved problem of finding the kinematic algebra $\gk$ of GR
could unfold.

We begin by 
showing that
the CKR
provides
a working definition of diffeomorphism gauge theory.
We do so by concrete demonstrations.
We find that CKR mechanically constructs
	diffeomorphism gauge transformation,
	diffeomorphism gauge connection,
    diffeomorphism covariant derivative, 
    diffeomorphism field strength,
    and even diffeomorphism Wilson line.

In later parts,
we discover a consistent and remarkably simple
geometrical picture
that 
justifies diffeomorphism gauge theory
within
standard concepts of differential geometry or GR
in the textbook regime:
the diffeomorphism gauge connection $A^\m{}_\a$
is nothing but 
the vielbein perturbation.
Based on this fact,
we establish a novel geometric interpretation of 
Cheung and Mangan \cite{CCK}'s formulation of BI theory as a teleparallel theory.
For GR,
we remark that
three-dimensional gravity
and four-dimensional SD gravity
admit diffeomorphism gauge theory formulations
by the constructions due to
\rrcite{Ben-Shahar:2021zww,Maor3dGravity,%
MasonNewman:1989}.
Lastly,
we revisit the Misner string
\cite{Misner:1963flatter,Bonnor:1969ala,sackfield1971physical,Mazur:1986gb,GP_2009_Ch341}
geometry
as a classical solution of any diffeomorphism gauge theories
and view it as the 
classical double copy
of the Dirac string in ordinary gauge theories.

\section{ Color-Kinematics Duality from Equations of Motion}
\label{K2>CKREVIEW}

For a precise formulation of ideas,
we begin by a review of the EoM level manifest CK duality
\cite{CCK}.\footnote{
    The broader program
    has concerned
    manifesting CK duality 
    at the EoM level \cite{CCK,monteiro2011kinematic},
    at the Lagrangian level \cite{Cheung:2016prv,Cheung:2017ems,Tolotti:2013caa,Ben-Shahar:2022ixa},
    or 
    at the fully off-shell level
    \cite{Ben-Shahar:2021zww,Borsten:2021hua}.
    Here, we restrict our scope to the EoM level manifestation of \rcite{CCK},
    which makes the most conservative claim.
}

\subsection{Bi-Adjoint Scalar Theory as a Template}

A BAS field is a scalar field $\Phi^{a\ta}$
carrying
two adjoint indices $a \eqq {1,{\cdots\mem},\dim\g}$ and $\ta \eqq {1,{\cdots\mem},\dim\tg}$
for two Lie algebras $\g$ and $\tg$.
By definition,
their structure constants
$f^a{}_{bc}$ and $\tf^\ta{}_{\tb\tc}$
must satisfy the Jacobi identity.

Suppose a field theory is found to exhibit tree-level CK duality
at the scattering amplitudes level.
Cheung and Mangan \cite{CCK}
concretely established that
this tree-level CK duality
can be manifested and proved
to all multiplicities
if the perturbative EoM of the theory
can be formulated in the form of BAS equations
for a choice of the untilded and tilded Lie algebras $\g$ and $\tg$,
\begin{align}
    \label{K2:BAS}
    \BAS(\g,\tg):\quad
    \Box\mem \Phi^{a\ta}
    \,=\,
    \tfrac{1}{2}\,
        f^a{}_{bc}\mem 
        \tf^\ta{}_{\tb\tc}\,
            \Phi^{b\tb}\mem \Phi^{c\tc}
    \,,
\end{align}
provided that the relevant field redefinition
is invertible
on external scattering states
(asymptotically free states).

Cheung and Mangan \cite{CCK}
showed that
the EoM of
NLSM and SG
can be formulated
in the form of BAS equations
with the following choices of Lie algebras,
thus exhibiting manifest CK duality.
For NLSM, the algebras are
\begin{align}
    \label{CKstatement:NLSM}
    \NLSM(\g)
    \,\,&\cong\,\,
        \BAS(\g,\sdiff(\M)\hnem)
    \,,
\end{align}
where the color algebra is typically $\g = \su(N)$.
For SG, the algebras are
\begin{align}
    \label{CKstatement:SG}
    \SG
    \,\,&\cong\,\,
        \NLSM(\sdiff(\M)\hnem)
    \,\,\cong\,\,
        \BAS(\sdiff(\M),\sdiff(\M)\hnem)
    \,.
\end{align}
We will review the derivation of 
\eqrefs{CKstatement:NLSM}{CKstatement:SG}
shortly.
Here, $\sdiff(\M)$ denotes the volume-preserving diffeomorphism algebra of the spacetime, $\M$.

\begin{figure}[t]
\centering
    \adjustbox{valign=c}{\begin{tikzpicture}
        \node[empty] (O) at (0,0) {};
        \node[empty] (a) at (-1.2,1.2) {};
        \node[empty] (b) at ( 1.2,1.2) {};
        \node[w] (00) at ($(O)$) {\clap{\BAS}};
        \node[w] (1a) at ($(O)+1*(a)$) {\clap{\NLSM}};
        \node[w] (1b) at ($(O)+1*(b)$) {\clap{\YM}};
        \node[w] (2aa) at ($(O)+2*(a)$) {\clap{\SG}};
        \node[w] (2ab) at ($(O)+1*(a)+1*(b)$) {\clap{\BI}};
        \node[w] (2bb) at ($(O)+2*(b)$) {\clap{\GRf}};
        \node[w] (ph00) at ($(O)$) {};
        \node[w] (ph1a) at ($(O)+1*(a)$) {};
        \node[w] (ph1b) at ($(O)+1*(b)$) {};
        \node[w] (ph2aa) at ($(O)+2*(a)$) {};
        \node[w] (ph2ab) at ($(O)+1*(a)+1*(b)$) {};
        \node[w] (ph2bb) at ($(O)+2*(b)$) {};
        \draw[<-] (ph2aa)--(ph1a) node[] {};
        \draw[<-] (ph1a)--(ph00) node[] {};
        \draw[<-] (ph2ab)--(ph1b) node[] {};
        \draw[->] (ph1a)--(ph2ab) node[] {};
        \draw[->] (ph00)--(ph1b) node[] {};
        \draw[->] (ph1b)--(ph2bb) node[] {};
    \end{tikzpicture}}
    \medskip
	\caption{%
        A web of field theories
        via CK duality and double copy.
        Each arrow implements a CKR.
		The arrows pointing to the left replace
            $\su(N)$ with $\sdiff(\R^d)$.
		The arrows pointing to the right replace
            $\su(N)$ with $\protect\gk$,
            the mystery Lie algebra.
        Note how this web is structured in terms of
        spin (helicities $0,1,2$).
        It is remarkable that
        such an algebraic organization
        arises between forces in our nature.
	}
	\label{web}
\end{figure}
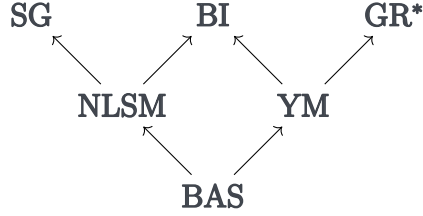

Similarly,
the CK duality of YM theory \cite{BCJ1}
is manifested at the EoM level
if one shows that
\begin{align}
    \label{CKstatement:YM}
    \YM(\g)
    \,\,\cong\,\,
        \BAS(\g,\gk)
    \,,
\end{align}
for a choice of a Lie algebra $\gk$.
The double copy 
property of gravity \cite{BCJ2}
is manifested at the EoM level
if one shows that
\begin{align}
    \label{CKstatement:GR}
    \GRf
    \,\,\cong\,\,
        \YM(\gk)
    \,\,\cong\,\,
        \BAS(\gk,\gk)
    \,,
\end{align}
for a choice of the same Lie algebra $\gk$
as in \eqref{CKstatement:YM}.
Here, $\GRf$ denotes
the NS-NS sector of type II supergravity,
sometimes called
``fat gravity.''
As remarked earlier in a footnote,
this theory contains
not only the graviton
but also a dilaton and a Kalb-Ramond $B$-field,
while a consistent truncation into GR is viable.

Note that these are ``if'' type statements.
Strictly speaking,
there 
might be
no guarantee that
CK duality or double copy
can always be manifested at the EoM level.
Still,
\eqrefs{CKstatement:YM}{CKstatement:GR}
may
be taken as
reasonable conjectures,
which are more conservative than Lagrangian or off-shell level claims,
in fact.

The reformulation statements in
\eqrefsss{CKstatement:NLSM}{CKstatement:SG}{CKstatement:YM}{CKstatement:GR}
are summarized in \fref{web},
which show a web of massless field theories.
To reiterate and clarify,
these reformulations
are at the perturbative EoM level
and presume invertibility
on external scattering states,
the meaning of which will become clear soon.

BI theory holds an interesting position in this web:
\begin{align}
    \label{CKstatement:BI}
    \NLSM(\gk)
    \,\,\cong\,\,
    \BI
    \,\,\cong\,\,
        \YM(\sdiff(\M)\hnem)
\end{align}
The statement here is the double copy property of BI theory,
which utilizes two different kinematic algebras,
is manifested at the EoM level if one shows
either one of the proposed equivalences in \eqref{CKstatement:BI}.
Examining this unique ``double duty'' of BI theory
might be holding a key toward identifying $\gk$.

The equivalence
$\BI \cong \YM(\sdiff(\M)\hnem)$
was established at the scattering amplitudes level
by Cheung and Mangan \cite{CCK}.
Showing this equivalence 
at the EoM level,
however,
remains as an open problem.
Concretely,
it demands identifying
a field redefinition
from an abelian gauge connection $a_\a$
to a diffeomorphism gauge connection $A^\m{}_\a$;
we will revisit this point in \Sec{K2>PGT}.

\subsection{Nonlinear Sigma Model}
\label{K2>REVIEW>NLSM}

\paragraph{Defining Formulation}

In the present context,
$\NLSM(\g)$ refers to
the principal chiral model $U: \M \to \G$
on flat spacetime $(\M,\eta)$
for a finite-dimensional matrix Lie group $\G$
whose Lie algebra is $\Lie(\G) \eqq \g$.
Here, $\eta$ is a Lorentzian-signature flat metric given to the spacetime manifold $\M$ with $\dim\M = d$,
while $U$ describes
a section of a principal $\G$-bundle over $\M$
which will be taken as trivial.
To define this as a perturbative field theory,
we presume a local coordinatization
$u: \g \to \G$
of the target manifold $\G$
centered around the vacuum expectation value $U = \id$,
while
recalling that
each tangent fiber of a Lie group $\G$ is isomorphic to its Lie algebra $\g$.
In this way
we obtain the pion field $\pi : \M \to \g$
as the composition $\pi = u^{-1} \circ U$
constructed locally.

The Lagrangian $d$-form is
\begin{align}
    \label{NLSM0.L}
    L[\pi]
    \,=\,
        -\frac{1}{2\k^2}\,
            \delta_{ab}\,
            j^a[\pi] \wedge {*} j^b[\pi]
    \,,
\end{align}
where $\delta_{ab}$ is a Killing form on $\g$,
$a,b,c,d,\cdots$ are the adjoint indices,
and $\k = 1/f_\pi$ is the coupling.
The left-invariant Maurer-Cartan form is
\begin{align}
    \label{j[pi]}
    j[\pi]
    \,=\,
        (u(\pi)\hnem)^{-1}\hem du(\pi)
    \,.
\end{align}

\begin{subequations}
From \eqref{j[pi]}, it is clear that
the pion field satisfies
the following equation off shell:
\label{NLSM0.em}
\begin{align}
    \label{NLSM0.m}
    dj^a[\pi] + \frac{1}{2}\, f^a{}_{bc}\, j^b[\pi] \swedge j^c[\pi]
    \,=\,
        0
    \,,
\end{align}
Variation of \eqref{NLSM0.L} shows that
the pion field satisfies
the following equation on shell:
\begin{align}
    \label{NLSM0.e}
    d\mem {*} j^a[\pi]
    \,=\,
        0
    \,.
\end{align}
\eqref{NLSM0.e} implies that
$j^a[\pi]$ is a conserved current:
the chiral current.
\end{subequations}

The arbitrariness in the coordinatization $u(\pi)$
describes a prime category of field redefinitions.
The only condition we demand on $u$ is that
it is ``proper,'' in the sense that the chiral current in \eqref{j[pi]} admits a well-behaved perturbative expansion
\begin{align}
    \label{j[pi]:proper}
    j^a[\pi]
    \,=\,
        d\pi^a
        + \O(\pi^2)
    \,.
\end{align}
Provided \eqref{j[pi]:proper},
any coordinatization of $\G$ is allowed.
For example,
take $u = \exp\mem(\pi)$:
normal coordinates via the exponential map.
In this case, the chiral current exhibits the expansion
\begin{subequations}
\begin{align}
    \label{j-pi.exp}
    j
    \,=\,
        \mathe^{-\pi}\mem d\mathe^{\pi}
    \,=\,
        d\pi
        - \frac{1}{2!}\, \comm{\pi}{d\pi}
        + \frac{1}{3!}\,
            \comm{\pi}{\comm{\pi}{d\pi}}
        + \O(\pi^4)
    \,.
\end{align}
For another example,
take $u = (\id\mplus\pi/2)(\id\mminus\pi/2)^{-1}$,
which is
known as Cayley basis.
In this case, the chiral current exhibits the expansion
\begin{align}
    \label{j-pi.cayley}
    j
    \,=\,
        d\pi - \frac{1}{2}\, \comm{\pi}{d\pi}
        + \frac{1}{4}\,
            \BB{
                \pi^2\mem d\pi
                - \pi\mem d\pi\hem \pi
                + d\pi\mem \pi^2
            }
        + \O(\pi^4)
    \,.
\end{align}
\end{subequations}
From \eqrefs{j-pi.exp}{j-pi.cayley},
we see that both coordinatizations are proper.

Plugging in either of \eqrefs{j-pi.exp}{j-pi.cayley}
to \eqref{j[pi]}
determines the EoM of the pion field
explicitly.
As a result, one obtains
a perturbation theory with
an infinite number of interaction vertices.
However, since both of \eqrefs{j-pi.exp}{j-pi.cayley}
exhibit the same leading behavior stipulated in \eqref{j[pi]:proper}
at the linearized level,
the S-matrix equivalence theorem
\cite{Chisholm:1961tha,Kamefuchi:1961sb,Arzt:1993gz,Cohen:2023ekv,Haag:1958vt}
applies and
ensures that
the scattering amplitudes are the same.

\paragraph{New Formulation}

An insight conceived by Cheung and Mangan \cite{CCK}
was that
one may have to focus on
objects that are invariant under 
the coordinate choice $u(\pi)$ on the target group manifold
to manifest CK duality,
a property observed in the field basis invariant S-matrix.
Again, coordinate transformations in the target space
are a prime example of field redefinitions.

\newpage

Motivated by this argument
(and in connection with an earlier creative exploration \cite{Cheung:2020NSE}),
Cheung and Mangan \cite{CCK}
envisioned taking
the chiral current as a fundamental field variable $j$.
This is viable by
imposing the flatness condition
as a constraint,
\begin{subequations}
\label{NLSMj.em}
\begin{align}
    \label{NLSMj.m}
    \partial_\m j^a{}_\n
    - \partial_\n j^a{}_\m
    + f^a{}_{bc}\mem j^b{}_\m\mem j^c{}_\n
    \,=\,
        0
    \,,
\end{align}
along with 
the conservation equation,
\begin{align}
    \label{NLSMj.e}
    \partial_\m j^{a\m}
    \,=\,
        0
    \,.
\end{align}
\end{subequations}
Combining \eqrefs{NLSMj.m}{NLSMj.e}
gives a second-order equation,
\begin{align}
    \label{NLSMj:eom}
    \Box\mem j^{a\m}
    \,=\,
        f^a{}_{bc}
        \,
        j^{b\r}\mem \partial_\r j^{c\m}
    \,=\,
        \frac{1}{2}\,
        f^a{}_{bc}
        \,
        \comm{j^b}{j^c}^\m
    \,,
\end{align}
where $\comm{j^b}{j^c}^\m = j^{b\r}\mem \partial_\r j^{c\m} - j^{c\r}\mem \partial_\r j^{b\m}$
describes the diffeomorphism Lie bracket.
Crucially,
\eqref{NLSMj:eom}
takes the form of BAS equations in \eqref{K2:BAS}
for the choice $\tg = \diff(\M)$.
More precisely,
the divergence-free condition in \eqref{NLSMj.e}
implies a restriction
to the volume-preserving subalgebra
$\tg = \sdiff(\M)$.
That is, the index-raised chiral current $j^{a\m}$
can be seen as a BAS field.

Let $\BAS(\g,\sdiff(\M)\hnem)$
denote the classical field theory of the field $j^{a\m}$
whose on-shell configurations satisfy \eqrefs{NLSMj.m}{NLSMj.e}.
The claim 
in \eqref{CKstatement:NLSM}
is that
this new formulation
is equivalent to
$\NLSM(\g)$
as a perturbative classical field theory
in specific senses.

\paragraph{Equivalence}

Firstly,
$\NLSM(\g)$ 
is embedded in
$\BAS(\g,\sdiff(\M)\hnem)$
by the map $\pi \mapsto j[\pi]$
given in \eqref{j[pi]},
for any proper target-space coordinatization $u(\pi)$.
Each on-shell configuration $\pi^a$
in $\NLSM(\g)$,
satisfying \eqref{NLSM0.em},
is mapped to
an on-shell configuration in $\BAS(\g,\sdiff(\M)\hnem)$,
satisfying \eqref{NLSMj.em}.

Secondly,
consider
\begin{align}
    \label{U[j]}
    U[j](x)
    \,=\,
        \Pexp\bb{
            \int_{\CC_x}
                j^a{}_\m\mem dx^\m
        }
    \,,
\end{align}
which 
computes the Wilson line
along a 
semi-infinite 
contour $\CC_x$
that ends at the bulk point $x$.
Then consider the map
$j \mapsto u^{-1}(U[j])$,
where $u^{-1} : \G \to \g$
is the inverse of $u : \g \to \G$ as a map.
It follows that
$\BAS(\g,\sdiff(\M)\hnem)$
is embedded in
$\NLSM(\g)$ 
by this map
for any choice of $\CC_x$.
Each on-shell configuration $j^a{}_\m$
in $\BAS(\g,\sdiff(\M)\hnem)$,
satisfying \eqref{NLSMj.em},
is mapped to
an on-shell configuration 
in $\NLSM(\g)$,
satisfying \eqref{NLSM0.em},
Especially, 
the flatness condition in \eqref{NLSMj.m}
ensures that 
\eqref{U[j]} evaluates to the same value
for each fixed $x \inn \M$,
regardless of one's choice of the contour $\CC_x$.

\newpage

Lastly,
the above two maps $\pi \mapsto j[\pi]$ and $j \mapsto u^{-1}(U[j])$
are invertible and are well-behaved in the perturbative regime.
At the linearized level,
in particular,
the former gives
$j^a{}_\m[\pi] = \partial_\m\pi^a + \O(\pi^2)$
as stipulated in \eqref{j[pi]:proper}
while
the latter gives,
e.g.,
$\pi^a[j] = (q\mdot\partial)^{-1} j^a{}_\m\mem q^\m + \O(j^2)$
when $\CC_x$ is realized as
a straight semi-infinite line
extending
along the direction of a constant reference vector $q^\m$.
It is easily checked that
these
are inverse to each other:
\begin{align}
    \label{j-pi.lin}
    j^a{}_\m
    \,=\,
        \partial_\m\pi^a 
        \mem+\mem \O(\pi^2)
    \qfq
    \pi^a
    \,=\,
        \frac{1}{q\mdot\partial}\:
            j^a{}_\m\mem q^\m 
        \mem+\mem \O(j^2)
    \,.
\end{align}

\begin{figure}[t]
    \centering
    \includegraphics[width=0.48\linewidth]{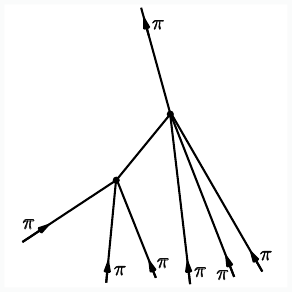}
    \hfill
    \includegraphics[width=0.48\linewidth]{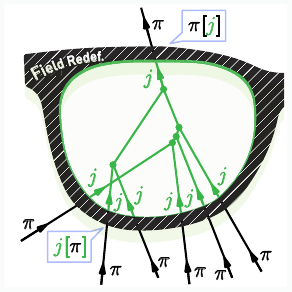}
    \medskip
    \caption{%
        Field redefinitions
        change one's description of 
        perturbative physics
        in the bulk.
        In one frame,
            one sees $\pi$-particles interacting via vertices of various valence.
        In another frame,
            one sees $j$-particles interacting via a universal cubic vertex.
        To obtain the same scattering amplitude,
        one simply needs to know
        how a single, \textit{free} $\pi$-particle turns into a free $j$-particle and vice versa.
        That is, it suffices to 
        perform the conversion between 
        the $\pi$- and $j$-frames
        at the boundary.
        When the amplitude is obtained
        via computing a one-point function
        by BG recursion,
        $j[\pi]$ is needed for the leaf legs
        while $\pi[j]$ is needed for the one last conversion for the root leg.
    }
    \label{nlsmfr}
\end{figure}

The linearized-level invertibility in \eqref{j-pi.lin}
implies that
there exists a prescription for
converting
a free $\pi$-state
to a unique free $j$-state
and vice versa.
In $\NLSM(\g)$,
a linearized on-shell configuration is given by
$\pi^a = c^a\mem \mathe^{ikx}$
for constant $c^a$ and
$k^2 = 0$.
\eqref{j-pi.lin} embeds this to
$\BAS(\g,\sdiff(\M)\hnem)$
by $j^a{}_\m = \partial_\m \pi^a = i\hhem k_\m\mem c^a\mem \mathe^{ikx}$,
which indeed solves the linear part of \eqref{NLSMj.em}.
Conversely,
in $\BAS(\g,\sdiff(\M)\hnem)$,
a linearized on-shell state is given by
$j^{a\m} = w^{a\m} \mem \mathe^{ikx}$
for $k^2 \eqq 0$
subject to
$w^{a\m} k_\m \eqq 0$
and $w^{a[\m} k^{\n]} \eqq 0$,
so
$j^{a\m} = i\hem c^a\hhem k^\m\mem \mathe^{ikx}$
for some $c^a$.
\eqref{j-pi.lin} embeds this to $\NLSM(\g)$
by
$\pi^a = (iq\mdot k)^{-1}\mem (
    i\hem c^a\hhem k_\m\mem \mathe^{ikx}
)\mem q^\m
= c^a\mem \mathe^{ikx}$,
which indeed solves the linear part of \eqref{NLSM0.em}.

\newpage

In sum,
an invertible field redefinition
is constructed
between
$\NLSM(\g)$ and $\BAS(\g,\sdiff(\M)\hnem)$
that is well-behaved in perturbation theory
and thus on free on-shell states.
This establishes the proposed perturbative equivalence
$\NLSM(\g) \cong \BAS(\g,\sdiff(\M)\hnem)$
stated in \eqref{CKstatement:NLSM}.

\medskip

To recapitulate,
Cheung and Mangan \cite{CCK}'s construction
can be taken as providing
the \textit{definition} of
EoM level manifest CK duality:
\begin{Fullnotew}
    A classical field theory
    exhibits \textit{EoM level manifest CK duality}
    if
    its EoM can be reformulated as BAS equations
    by a field redefinition
    that is invertible on asymptotic scattering states.
\end{Fullnotew}
To show that
this EoM level manifest CK duality
implies 
the manifest CK duality
of scattering amplitudes,
Cheung and Mangan \cite{CCK}
utilize the framework of
BG recursion
\cite{BerendsGiele}
(perturbiner method \cite{Rosly:1996vr,Selivanov:1997an,Mizera:2018jbh}).
Details are to be found in \rcite{CCK},
but nevertheless 
a quick review is provided in 
\Sec{app:BASBG}.

Note also that it is unnecessary to invoke Lagrangian formulations
in this discussion,
although the
Freedman-Townsend \cite{ft1981}
formulation of NLSM
may provide tempting pathways:
see \Sec{app:BZ-NLSM}.

\subsection{Born-Infeld Theory}

Given the identification of the kinematic algebra of NLSM as $\sdiff(\M)$,
and also the amplitudes-level fact that BI theory
arises by combining the
NLSM and YM kinematic algebras
(recall \fref{web} and \eqref{CKstatement:BI}),
Cheung and Mangan \cite{CCK}
established a novel formulation of BI theory
at the EoM level
in their Sec.\,3.4.3.
For the reader's sake,
let us reproduce their equations below
up to minor change in notation.

\begin{subequations}\label{CK3}
In their (3.27),
Cheung and Mangan
state the NLSM CKR rule
applied on YM gauge connection and field strength:
\begin{align}
    \label{CK3.27}
    \g
    \,\,\mapsto\,\,
    \sdiff(\M)
    \,:\quad
        A^a{}_\a
        \,\,\mapsto\,\,
        A^\m{}_\a
        \,,\quad
        F^a{}_{\a\b}
        \,\,\mapsto\,\,
        F^\m{}_{\a\b}
    \,.
\end{align}
This means to take the gauge algebra $\g$ of YM theory
as
the Lie algebra $\sdiff(\M)$
of volume-preserving diffeomorphisms.
In their (3.28),
Cheung and Mangan
expand out the precise mechanics of this replacement
for the definition of field strength:
\begin{align}
\begin{split}
    \label{CK3.28}
    &
    F^a{}_{\a\b}
    \,=\,
        \partial_\a A^a{}_\b
        - \partial_\b A^a{}_\a
        + f^a{}_{bc}\mem A^b{}_\a\mem A^c{}_\b
    \\
    &
    \mapsto\,\,
    F^\m{}_{\a\b}
    \,=\,
        \partial_\a A^\m{}_\b
        - \partial_\b A^\m{}_\a
        + A^\r{}_\a\mem \partial_\r A^\m{}_\b
        - A^\r{}_\b\mem \partial_\r A^\m{}_\a
    \,.
\end{split}
\end{align}
This clearly replaces
$f^a{}_{bc}\mem A^b{}_\a\mem A^c{}_\b
= \comm{A_\a}{A_\b}^a
$
with
$\comm{A_\a}{A_\b}^\m
= 
A^\r{}_\a\mem \partial_\r A^\m{}_\b
- A^\r{}_\b\mem \partial_\r A^\m{}_\a
$.
Lastly, 
in their (3.29),
Cheung and Mangan
explicate how the YM equations are mapped
by this replacement.
For the source-free case,
it is\footnote{
    The source term in their Eq.\,(3.29) is essentially used for
    specifying the external state prescription.
    It states $A^\m{}_\a = \partial^\m a_\a + \O(a^2)$
    if $a_\a$ is the conventional BI photon.
    Certainly,
    the linearized on-shell state is
    $A^\m{}_\a \sim p^\m e_\a\mem \mathe^{ipx}$
    when $a_\a \sim e_\a\mem \mathe^{ipx}$.
    As mentioned earlier,
    the exact field redefinition $A[a]$ is unknown
    for this moment
    (as well as its inverse $a[A]$).
}
\begin{align}
\begin{split}
    \label{CK3.29}
    &
    D^\a F^a{}_{\a\b}
    \,=\,
        \partial^\a F^a{}_{\a\b}
        +
        f^a{}_{bc}\mem A^{b\a}\mem F^c{}_{\a\b}
    \,=\,
        0
    \\
    &
    \mapsto\,\,
    D^\a F^\m{}_{\a\b}
    \,=\,
        \partial^\a F^\m{}_{\a\b}
        +
        A^{\r\a}\mem \partial_\r F^\m{}_{\a\b}
        -
        F^\r{}_{\a\b}\mem \partial_\r A^{\m\a}
    \,=\,
        0
    \,.
\end{split}
\end{align}
Cheung and Mangan
then state that their
``Eqs.\,(3.28) and (3.29) constitute a new first-order formulation of BI theory.''
Furthermore,
they report that they verified
``up to six-point scattering that the resulting BI
amplitudes agree with known expressions.''
\end{subequations}

For NLSM,
the CKR mapped a global Lie algebra index 
(flavor index)
to a kinematic index.
For BI theory,
the CKR maps the gauged index 
(color index)
of YM theory
to a kinematic index.
As a result,
one inevitably encounters the notion of
``diffeomorphism gauge transformation''
and
``diffeomorphism gauge covariance,''
which seem nonsensical at first sight.
What does it mean to take diffeomorphisms as the gauge algebra of a YM theory?
What does ``$\hem\YM(\sdiff(\M)\hnem)$'' mean?
Crucially,
the EoM level CK duality
strictly necessitates
such concepts,
as is asserted by Cheung and Mangan \cite{CCK}
and
as should be clear
from our review and exposition above.
To our understanding,
however,
this point
has not been properly clarified nor addressed
in the literature.

With this remark made,
we proceed to \Secs{K2>OGT}{K2>DGT}
to define
diffeomorphism gauge theory
as a strict implication of 
EoM level CK duality.

\section{ Ordinary Gauge Theory}
\label{K2>OGT}

To begin with,
we review basic concepts in ordinary gauge theory.
By an ordinary gauge theory, we mean a gauge theory
whose gauge group is
a finite-dimensional matrix Lie group $\G$.
For concreteness, we may take $\G = \mathrm{SU}(N)$.
The gauge algebra $\g = \su(N)$ describes
fundamental indices $i,j,k,l,\cdots$ and 
adjoint indices $a,b,c,d,\cdots$.
For $\psi \in \mathbb{C}^N$ and $u,v \in \mathsf{su}(N)$, we denote
\begin{align}
	\label{notations-su}
	(u\psi)^i \mem=\mem u^i{}_j\mem \psi^j \mem=\mem u^a\hem (t_a)^i{}_j\mem \psi^j
	\,,\quad
	\comm{u}{v}^a \mem=\mem f^a{}_{bc}\, u^b\mem v^c
	\,,
\end{align}
where $(t_a)^i{}_j$ and $f^a{}_{bc}$ are the generators and structure constants.
When working in the fundamental representation,
we may sometimes make indices implicit to avoid clutter.
We assume $d$-dimensional flat spacetime
and use greek letters for spacetime indices.

\subsection{Gauge Covariance}

In ordinary gauge theories such as YM theory,
local operators are subject to local $\G$-transformations,
geometrically formulated as transformations acting on the \textit{fibers} of a bundle over spacetime.
In this sense, gauge transformations are said to be \textit{internal} transformations:
the points in the base spacetime manifold are not altered at all.
It is important to understand that gauge transformations define equivalences, or redundancies, 
under a gauge orbit,
so physically meaningful quantities are gauge-invariant objects.

To describe gauge transformations in an explicit manner,
let $g(x)$ assign a group element in $\G$ for each spacetime point $x$.
Gauge-covariant operators,
such as those valued in the singlet, fundamental, or adjoint,
exhibit the following transformation behaviors:
\begin{align}
	\begin{split}
		\label{YM.gt}
		{\renewcommand{\arraystretch}{1.25}
			\begin{array}{rrl}
				\Sing:\quad&
				\phi(x) &\,\,\mapsto\,\,\, \phi(x)
				\,,\\
				\Fund:\quad&
				\phi^i(x) &\,\,\mapsto\,\,\, g^i{}_j(x)\, \phi^j(x)
				\,,\\
				\Adj:\quad&
				\phi^a(x) &\,\,\mapsto\,\,\, g^a{}_b(x)\, \phi^b(x)
				\,.
		\end{array}}
	\end{split}
\end{align}
Being valued in the singlet is a synonym for being gauge invariant.

Suppose a non-singlet field.
Due to the fact that the parameter $g(x)$ vary over spacetime,
the field values at different points transform differently.
As a result, their difference is not physically meaningful.
The implication is that the partial derivative of a non-singlet field is not a gauge-covariant object.

This provides the very motivation for introducing the gauge connection $A$,
which defines 
the gauge-covariant derivative $D$.
For each gauge-covariant operator,
the covariant derivative returns 
an operator
with one more spacetime index
that exhibits the \textit{same} gauge-covariant transformation behavior:
\begin{align}
	\begin{split}
		\label{YM.covD}
		{\renewcommand{\arraystretch}{1.25}
			\begin{array}{rrl}
				\Sing:\quad&
				D_\a \phi(x)
				&=\, \partial_\a \phi(x)
				\,,\\
				\Fund:\quad&
				D_\a \phi^i(x)
				&=\, \partial_\a \phi^i(x) + A^i{}_{j\a}(x)\, \phi^j(x)
				\,,\\
				\Adj:\quad&
				D_\a \phi^a(x)
				&=\, \partial_\a \phi^a(x) + \comm{ A_\a(x) }{ \phi(x) }^a
				\,.
		\end{array}}
	\end{split}
\end{align}
This is easily verified by direct computation,
provided that the connection transforms as
\begin{align}
	\label{YM.gtA}
	A_\a(x)
	\,\,\,\mapsto\,\,\,
	g(x)\mem \partial_\a g^{-1}\hnem(x) + g(x)\mem A_\a(x)\mem g^{-1}\hnem(x)
	\,.
\end{align}

\newpage

For reference, we also record the infinitesimal version of the gauge transformations in \eqref{YM.covD}:
\begin{align}
	\begin{split}
		\label{YM.igt}
		{\renewcommand{\arraystretch}{1.25}
			\begin{array}{rrl}
				\Sing:\quad&
				\phi(x) &\,\,\,\mapsto\,\,\, \phi(x)
				\,,\\
				\Fund:\quad&
				\phi^i(x) &\,\,\,\mapsto\,\,\, \phi^i(x) + (\ve(x)\mem \phi(x))^i
				\,,\\
				\Adj:\quad&
				\phi^a(x) &\,\,\,\mapsto\,\,\, \phi^i(x) + \comm{ \ve(x) }{ \phi(x) }^a
				\,.
		\end{array}}
	\end{split}
\end{align}
Here, we have taken $g(x) = \oldexp(\ve(x))$,
in which case \eqref{YM.gtA} becomes
\begin{align}
	\label{YM.igtA}
	A^a{}_\a(x)
	\,\,\,\mapsto\,\,\,
	A^a{}_\a(x)
	- D_\a \ve^a(x)
	\,.
\end{align}

The field strength $F$
is the curvature associated with the gauge connection $A$.
This describes the commutator between two covariant derivatives,
say $D_\a$ and $D_\b$.
Explicitly,
it is given as
\begin{align}
	\label{YM.fs}
	F^a{}_{\a\b}(x)
	\,=\, \partial_\a A^a{}_\b(x) - \partial_\b A^a{}_\a(x) + \comm{ A_\a(x) }{ A_\b(x) }^a
	\,.
\end{align}
The field strength is an adjoint-valued two-form.
Hence its gauge-covariant derivative is
\begin{align}
    \label{YM.DF}
    D_\c F^a{}_{\a\b}(x)
    \,=\,
        \partial_\c F^a{}_{\a\b}(x)
        + f^a{}_{bc}\mem A^b{}_\c(x)\mem F^c{}_{\a\b}(x)
    \,.
\end{align}
It is easy to establish 
either from \eqrefsor{YM.gtA}{YM.igtA}
that the field strength transforms covariantly.
Note also the Bianchi identity:
\begin{align}
	\label{YM.bianchi}
	D_\wrap{[\c} F^a{}_\wrap{\a\b]}(x) 
	\,=\, 
	D_\c F^a{}_{\a\b}(x) + D_\a F^a{}_{\b\c}(x) + D_\b F^a{}_{\c\a}(x)
	\,=\,
	0
	\,,
\end{align}
which holds 
by virtue
of the Jacobi identity of the gauge algebra $\g$.

\subsection{Wilson Line}
\label{sec:YM.wl}

Geometrically, the gauge connection $A$
defines parallel transport along contours in spacetime.
For a contour $\CC$
that stretches from point $\P$ to point $\P'$,
the parallel transport
establishes an isomorphism between
the gauge fibers at $\P$ and $\P'$.
This isomorphism 
describes a group element in $\G$ assigned for the contour $\CC$,
which is the Wilson line $W_\CC$.

To be concrete, suppose a finite contour $\CC$ parameterized as $s \mapsto x^\m + y^\m(s)$,
where $y^\m(0) = 0$ and $s \in [0,1]$.
The point $\P$ has coordinates $x^\m$, while the point $\P'$ has coordinates $x^\m + y^\m(1)$.
The parallel transport equation for this contour is given by
\begin{align}
	\label{YM.transpt.fund}
	\frac{d}{ds}\, \Psi^i(s)
	+ A^i{}_{j\a}(x{\,+\,}y(s))\, \dot{y}^\a(s)\, \Psi^j(s)
	\,=\,
	0
	\,,
\end{align}
where $\Psi^i(s)$ is a fundamental-valued variable restricted on the contour $\CC$.
As is well-known, the solution to this equation is available in terms of a path-ordered exponential.
For 
\begin{align}
	\label{YM.W.fund}
	W_\CC
	\,=\,
	\Pexp\bigg(\,{
		-\nem \int_0^1
		ds\,\,
		A_\a(x{\,+\,}y(s))\, \dot{y}^\a(s)
	}\,\bigg)
	\,,
\end{align}
it follows that
\begin{align}
	\label{YM.Wtransport.fund}
	\Psi^i(1)  =  (W_\CC)^i{}_j\, \Psi^j(0)
	\,.
\end{align}
\eqref{YM.W.fund} is the definition of the Wilson line $W_\CC$ about the contour $\CC$.

In \eqref{YM.Wtransport.fund},
$\Psi^i(0)$ and $\Psi^i(1)$ are covariant objects located at the points $\P$ and $\P'$
($x^\m$ and $x^\m + y^\m(1)$), respectively.
In particular, they transform with the gauge parameters at $\P$ and $\P'$.
This establishes that the gauge transformation behavior of the Wilson line is
\begin{align}
	\label{YM.bilocal}
	W_\CC
	\,\,\,\mapsto\,\,\,
	g(\P')\, W_\CC\, g^{-1}(\P)
	\,.
\end{align}
Namely, the Wilson line transforms bilocally.

When $\CC$ is a loop,
a singlet is formed by taking a trace of $W_\CC$ for a chosen representation.
This singlet is the Wilson \textit{loop},
which should be distinguished from the open Wilson \textit{line} described above.
The former is a number, whereas the latter is a matrix (group element).
The focus of this chapter is put on Wilson lines rather than Wilson loops.

An implication of the bilocal behavior in \eqref{YM.bilocal} is that
a gauge-covariant operator
at $\P$
can be alternatively covariantized at $\P'$
via the Wilson line $W_\CC$.
For a fundamental-valued operator,
one constructs
\begin{align}
	\label{YM.dressing*.fund}
		\Fund:\quad&
	W_\CC\, \phi(\P) \,\,\,\mapsto\,\,\, g(\P')\, \Big(\mem{ W_\CC\, \phi(\P) }\mem\Big)
	\,.
\end{align}
For adjoint-valued operators, one constructs
\begin{align}
	\label{YM.dressing*.adj}
		\Adj:\quad&
	W_\CC\, \phi(\P)\, W^{-1}_\CC \,\,\,\mapsto\,\,\, g(\P')\, \Big(\mem{ W_\CC\, \phi(\P)\, W^{-1}_\CC }\mem\Big)\mem g^{-1}\hnem(\P')
	\,.
\end{align}

Finally, it is instructive to note that
another explicit formula is available for the Wilson line in \eqref{YM.W.fund},
known as Magnus series \cite{Magnus:1954zz}.
The Magnus series computes the log of the path-ordered exponential.
For $W_\CC = \oldexp( \Omega_\CC )$,
it follows that
\begin{align}
	\label{YM.magnus}
	\Omega^a_\CC
	\,=\,
\begin{aligned}[t]
    \nonumber
	&
	-\nem \int_0^1 ds_1\,\,
		A^a{}_\a(x{\,+\,}y(s_1))\, \dot{y}^\a(s_1)
	\\
    \nonumber
	&
	+ \frac{1}{2}\mem \int_0^1 ds_1 \int_0^{s_1}\nem\nem ds_2\,\,
		\comm{ A_\a(x{\,+\,}y(s_1)) }{ A_\b(x{\,+\,}y(s_2)) }^a\,
		\mem 
		\dot{y}^\a(s_1)\, \dot{y}^\b(s_2)
	\\
	& + \cdots
	\,,
\end{aligned}
\end{align}
where higher-order terms describe integrals involving more Lie brackets.

The insight to be gained from this alternative representation
is that
\eqref{YM.magnus}
manifests the fact that the Wilson line defines a group element.
This is because 
each term in \eqref{YM.magnus}
is manifestly a Lie algebra element,
constructed in terms of nested Lie brackets.
Hence $\Omega_\CC \inn \g$,
which manifests that $W_\CC = \oldexp(\Omega_\CC) \in \G$.

It should be noted that the derivation of such nested Lie brackets 
counts on the fact that
the commutator of adjoint actions
is the adjoint action of the commutator.
This is implied by the Jacobi identity of the gauge algebra $\g$:
$[u,[v,w]] - [v,[u,w]] = [[u,v],w]$.

Note also that the sandwiched operator in \eqref{YM.dressing*.adj}
can now be described as
a sum of integrals of nested Lie brackets
rather than matrix products:
\begin{align}
\begin{split}
	(\Ad{W_\CC} \phi(\P))^a
	\,&=\,
	(e^{\ad{\Omega_\CC}} \phi(x))^a
    \,,\\
	\,&=\,
	-\nem \int_0^1 ds_1\,\,
		\comm{ A_\a(x{\,+\,}y(s_1)) }{ \phi(x) }^a\, \mem\dot{y}^\a(s_1)
	+ \cdots
	\,.
\end{split}
\end{align}
This expression manifests that 
the structure constants $f^a{}_{bc}$ are the only necessary algebraic information
needed to compute the adjoint-valued sandwiched operator,
crucially.

\subsection{Covariant Taylor Expansion}

In \eqrefs{YM.dressing*.fund}{YM.dressing*.adj},
we have seen that
the point of covariantization
can be shifted by attaching Wilson lines.
For instance,
$W_\CC\mem \phi(\P)$ in \eqref{YM.dressing*.fund},
which defines a nonlocal operator,
transforms like an operator at the point $\P'$.
In this light, one might wonder
if it is possible to explicitly express $W_\CC\mem \phi(\P)$
as an infinite sum of local operators at $\P'$.

This is indeed possible,
assuming that $\P'$ is sufficiently close to $\P$.
To understand this explicitly,
let us consider a straight-line contour 
$s \mapsto x^\m + q^\m s$,
where $q^\m$ describes a fixed vector.
The Wilson line for a part of this contour
will be denoted as
\begin{align}
	W(x{\,+\,}qs_2,x{\,+\,}qs_1)
	\,=\,
	\Pexp\bigg(\,{
		-\nem \int_{s_1}^{s_2}
		ds\,\,
		A_\a(x{\,+\,}qs)\, q^\a
	}\,\bigg)
	\,.
\end{align}
For any $s$, it holds that
\begin{align}
	\label{ids.WW}
	W(x{\,+\,}qs_2,x{\,+\,}qs)\, W(x{\,+\,}qs,x{\,+\,}qs_1)
	\,=\,
		W(x{\,+\,}qs_2,x{\,+\,}qs_1)
	\,.
\end{align}
Also, due to the transport equation in \eqref{YM.transpt.fund},
it follows that
\begin{align}
	\label{ids.dW}
	\dot{W}(x{\,+\,}qs,x{\,+\,}qs_1)
	\,=\,
		- A_\a(x{\,+\,}qs)\, q^\a\, W(s,s_1)
	\,.
\end{align}

Suppose a fundamental-valued field $\phi^i(x)$ in spacetime
and its restriction on the straight-line contour, $\phi(x{\,+\,}qs)$.
Direct computation shows that
\begin{align}
	\label{YM.ctr1}
	\frac{d}{ds}\mem
		\Big(\,{
				W(x,x{\,+\,}qs)\, \phi(x{\,+\,}qs)
		}\,\Big)
	\,=\,
		W(x,x{\,+\,}qs)\, 
		q^\a D_\a\phi(x{\,+\,}qs)
	\,,
\end{align}
where 
$D_\a\phi(x{\,+\,}qs) = \partial_\a\phi(x{\,+\,}qs) + A_\a(x{\,+\,}qs)\, \phi(x{\,+\,}qs)$.
Moreover, it also follows that
\begin{align}
	\label{YM.ctrl}
	\frac{d^\ell}{ds^\ell}\mem
	\Big(\,{
		W(x,x{\,+\,}qs)\, \phi(x{\,+\,}qs)
	}\,\Big)
	\,=\,
		W(x,x{\,+\,}qs)\, 
		(q\mdot D)^\ell \phi(x{\,+\,}qs)
	\,,
\end{align}
for any positive integer $\ell$.
As a result,
the Taylor expansion of 
$W(x,x{\,+\,}qs) $ $ \phi(x{\,+\,}qs)$
as a one-variable function of $s$
implies that
\begin{align}
	\label{YM.ctr}
	W(x,x{\,+\,}q)\, \phi(x{\,+\,}q)
	\,=\,
		\sum_{\ell=0}^\infty\,
			\frac{1}{\ell!}\,
				(q\mdot D)^\ell \phi(x)
	\,.
\end{align}
Clearly, \eqref{YM.ctr} shows that a nonlocal operator formed by attaching a Wilson line
is expressible in terms of an infinite sum of local operators.

The identity in \eqref{YM.ctr}
could be referred to as the covariant Taylor expansion.
It generalizes the usual Taylor expansion in spacetime
by replacing the partial derivative with the covariant derivative
while matching the covariances at two different points
via a Wilson line.
The similar identity holds for adjoint-valued fields as well.

Another insight that can be gained from \eqref{YM.ctr} is that
the covariant derivative has facilitated the comparison between field values at different points
via the Wilson line:
\begin{align}
	\label{YM.Dlimit}
	q^\a D_\a \phi(x)
	\,=\,
		\lim_{\e\to0}
		\,\frac{1}{\e}\mem
		\Big(\,{
			W(x,x{\,+\,}\e q)\, \phi(x{\,+\,}\e q)
			- \phi(x)
		}\,\Big)
	\,.
\end{align}
This faithfully expresses the very geometrical idea of gauge connection and covariant derivative.
In fact, 
we shall remark that
it is possible and may be desirable to view Wilson lines (holonomies, parallel propagators) as fundamental
and take \eqref{YM.Dlimit} as the definition of the covariant derivative derived from the Wilson lines.

\section{ Diffeomorphism Gauge Theory via Color-Kinematics Duality}
\label{K2>DGT}

In this section, we take the diffeomorphism algebra
in the sense that is implied by CK duality.
The strategy here is to
construct exact counterparts of
the well-established constructions of ordinary gauge theory
in \Sec{K2>OGT}
by mechanically implementing the CKR:
\begin{align}
	\label{ck}
	\g \,=\,
	\su(N)
	\quad\xrightarrow{\quad}\quad
	\g \,=\,	
	\diff(\mathbb{R}^d)
	\,.
\end{align}
This systematically deduces and defines what a diffeomorphism gauge theory is.
For instance,
every Lie bracket of $\su(N)$
will be replaced
with the Lie bracket between vector fields.

To elaborate,
the diffeomorphism Lie algebra in $d$ dimensions,
$\diff(\mathbb{R}^d)$,
is identified with the space $\Gamma(T\R^d)$ of vector fields on $\R^d$.
For 
a scalar field $\psi \in C^\infty(\mathbb{R}^d)$
and vector fields $u, v \in \mathsf{diff}(\mathbb{R}^d)$,
we denote
\begin{align}
\begin{split}
	\label{notations-diff}
	u\psi(x) \mem&=\mem u^\m(x)\, \partial_\m\psi(x)
	\,,\\
	\comm{ u(x) }{ v(x) }^\m \mem&=\mem u^\n(x)\, \partial_\n v^\m(x) - v^\n(x)\, \partial_\n u^\m(x)
	\,,
\end{split}
\end{align}
which defines the fundamental and adjoint actions
in parallel with \eqref{notations-su}.
Note that the Lie bracket between vector fields, $\comm{ u(x) }{ v(x) }^\m$,
is literally derived by taking the commutator between
$u^\m(x)\mem \partial_\m$ and $v^\m(x)\mem \partial_\m$
as \textit{differential operators}.
The mathematical basis of this view
is elaborated in \Sec{DYM>AG}.
The diffeomorphism Lie group is denoted as $\Diff(\mathbb{R}^d)$.

\subsection{Gauge Covariance}

In gravitational theories such as general relativity,
there exists the freedom to choose coordinates in spacetime.
This introduces the notion of \textit{general covariance},
which describes how local fields change under coordinate transformations.
It is important to understand that coordinate transformations in generally covariant theories
define equivalencies, or redundancies, under a gauge orbit,
on which physically meaningful quantities shall be invariant.

However, at the technical, precise mathematical level,
the gauge transformations for general covariance
work differently
than the gauge transformations described in \Sec{K2>OGT},
i.e., those in the YM theory, for instance.
Crucially, diffeomorphisms alter points in the \textit{base} spacetime manifold themselves.
Namely, they are \textit{external} transformations,
not internal.

Despite this crucial difference,
however,
we push the idea of having the diffeomorphism algebra as a gauge algebra in the YM sense,
as a necessary implication of CK duality.

We start with the transformation behavior of fields.
Based on \eqref{YM.igt},
the CKR in \eqref{ck} stipulates that
the singlet, fundamental, and adjoint transformation behaviors
are defined as
\begin{align}
    \label{DYM.igt}
    {\renewcommand{\arraystretch}{1.25}
    \begin{array}{rrl}
        \Sing:\quad&
        \phi(x) &\,\,\mapsto\,\,\, \phi(x)
        \,,\\
        \Fund:\quad&
            \phi(x) &\,\,\mapsto\,\,\, \phi(x) + \xi^\m(x)\, \partial_\m\phi(x)
        \,,\\
        \Adj:\quad&
        \phi^\m(x) &\,\,\mapsto\,\,\, \phi^\m(x) + \xi^\n(x)\, \partial_\n\phi^\m(x) - \phi^\n(x)\, \partial_\n\xi^\m(x)
        \,,
    \end{array}}
\end{align}
provided the fundamental and adjoint actions in \eqref{notations-diff}.
Here, $\xi = \xi^\m(x)\mem \partial_\m \in \diff(\mathbb{R}^d)$
is a diffeomorphism Lie algebra element
serving as an infinitesimal diffeomorphism gauge transformation parameter,
which is the counterpart of $\ve(x) \in \su(N)$ in \eqref{YM.igt}.

Suppose a non-singlet field,
such as the fundamental field in \eqref{DYM.igt}.
Due to the fact that the diffeomorphism parameter $\xi^\m(x)$ vary over spacetime,
the field values at different points transform differently.
As a result, their difference is not meaningful.
The implication is that the partial derivative of a non-singlet field is not a gauge-covariant object.
Explicitly, the partial derivative of the fundamental field, $\partial_\a\phi(x)$, transforms to
\begin{align}
\begin{split}
	\label{DYM.dphi}
    &
		\partial_\a\Big(\,{
			 \phi(x) + \xi^\m(x)\, \partial_\m\phi(x) 
		}\,\Big)
    \\
    &
	\,=\,
		\partial_\a\phi(x) + \xi^\m(x)\, \partial_\a\partial_\m\phi(x)  + \partial_\a\xi^\m(x)\, \partial_\m\phi(x)
	\,,
\end{split}
\end{align}
the behavior of
which is neither singlet, fundamental, or adjoint
according to \eqref{DYM.igt}.

In this light, 
we are invited to the idea of introducing
a diffeomorphism gauge connection $A$,
defining
the diffeomorphism-covariant derivative $D$.
To this end, one mechanically applies the CKR rule.
From \eqref{YM.covD}, one finds
\begin{align}
    \label{DYM.covD}
    \kern-0.8em
    \begin{adjustbox}{raise=-1.72em}$
    {\renewcommand{\arraystretch}{1.25}
        \begin{array}{rrl}
            \Sing:\quad&
            D_\a \phi(x)
            &=\, \partial_\a \phi(x)
            \,,\\
            \Fund:\quad&
            D_\a \phi(x)
            &=\, \partial_\a \phi(x) + A^\m{}_\a(x)\, \partial_\m\phi(x)
            \,,\\
            \Adj:\quad&
            D_\a \phi^\m(x)
            &=\, \partial_\a \phi^\m(x) + A^\n{}_\a(x)\, \partial_\n\phi^\m(x) - \phi^\n(x)\, \partial_\n A^\m{}_\a(x)
            \,,
    \end{array}}
    $\end{adjustbox}
    \kern-4em
\end{align}
while \eqref{YM.igtA} stipulates that the diffeomorphism gauge connection transforms as
\begin{align}
	\label{DYM.igtA}
	A^\m{}_\a(x)
	\,\,\,\mapsto\,\,\,
	A^\m{}_\a(x)
	- D_\a \xi^\m(x)
	\,.
\end{align}

Recall the description given around \eqref{YM.covD}.
Again,
for each diffeomorphism-covariant operator,
the diffeomorphism-covariant derivative returns 
an operator
with one more spacetime index
that exhibits the \textit{same} diffeomorphism-covariant transformation behavior.
That is,
the additional spacetime index is diffeomorphism-\textit{neutral}.

Let us verify this claim in a direct, explicit fashion.
For the singlet, 
\eqref{DYM.covD} stipulates that
its covariant derivative $D_\a\phi(x)$
is simply the partial derivative $\partial_\a\phi(x)$.
By \eqref{DYM.igt},
this transforms as $\partial_\a\phi(x) \mapsto \partial_\a\phi(x)$:
it is invariant.
Therefore, $D_\a\phi(x)$ is still a singlet,
where the index $\a$
is neutral under the diffeomorphism gauge transformations.

Next, 
consider a diffeomorphism-fundamental field $\phi(x)$,
whose infinitesimal gauge transformation 
is
$\delta_\xi \phi(x) = \xi^\m(x)\, \partial_\m\phi(x)$
according to \eqref{DYM.igtA}.
Using \eqrefs{DYM.dphi}{DYM.igtA},
the infinitesimal variation of
its covariant derivative $D_\a\phi(x)$
is found as the sum of the following three groups of terms:
\begin{align}
\begin{split}	
		\label{DYM.cdcheck.fund}
		&
		\Big(\,{
			\hlx{ \partial_\a\xi^\m\mem \partial_\m\phi} + \hlb{ \xi^\m\mem \partial_\a\partial_\m\phi  }
		}\,\Big)
		\,,\\
		&
		A^\m{}_\a\mem 
		\Big(\,{
			\hly{ \partial_\m\xi^\n\mem \partial_\n\phi }
			\hlb{ + \xi^\n\mem \partial_\m\partial_\n\phi }
		}\,\Big)
		\,,\\
		&
		\Big(\,{
			\hlx{{- \partial_\a\xi^\m}} \hly{ - A^\n{}_\a\mem \partial_\n\xi^\m } + \hlb{ \xi^\n\mem \partial_\n A^\m{}_\a }
		}\,\Big)\,
		\partial_\m\phi
		\,.
\end{split}
\end{align}
Here, we have temporarily omitted the argument ``$(x)$''
to avoid clutter
and employed a color coding to show cancellations/summations of terms efficiently.
As a result, one finds
\begin{align}
	\delta_\xi( D_\a\phi ) 
	\,=\,
		\hlb{ \xi^\m\mem \partial_\m( D_\a\phi ) }
	\,,
\end{align}
establishing that
$D_\a\phi(x)$ transforms exactly like a diffeomorphism-fundamental field \`a la \eqref{DYM.igt}.

Finally, consider a diffeomorphism-adjoint field $\phi^\m(x)$,
whose infinitesimal gauge transformation 
is given by
$\delta_\xi\phi^\m(x) = \comm{ \xi(x) }{ \phi(x) }^\m = \xi^\n(x)\, \partial_\n\phi^\m(x) - \phi^\n(x)\, \partial_\n\xi^\m(x)$
according to \eqref{DYM.igtA}.
The infinitesimal variation of 
its covariant derivative $D_\a\phi^\m(x)$,
given in \eqref{DYM.covD},
evaluates to the sum of five groups of terms:
\begin{align}
	\begin{split}	
		\label{DYM.cdcheck.adj1}
		&
		\partial_\a
		\Big(\,{
			\comm{ \xi }{ \phi }^\m
		}\,\Big)
		\,,\quad
		A^\r{}_\a\mem \partial_\r\mem
		\Big(\,{
			\comm{ \xi }{ \phi }^\m
		}\,\Big)
		\,,\quad
		\Big(\,{
			-\partial_\a \xi^\r - \comm{ A_\a }{ \xi }^\r
		}\,\Big)\,
		\partial_\r\phi^\m
		\,,\\
		&
		{-
		\Big(\,{
			\comm{ \xi }{ \phi }^\r
		}\,\Big)\,
		\partial_\r A^\m{}_\a}
		\,,\quad
		- \phi^\r\mem \partial_\r\mem
		\Big(\,{
			-\partial_\a \xi^\m - \comm{ A_\a }{ \xi }^\m
		}\,\Big)
		\,.
	\end{split}
\end{align}
Expanding each group, we find
\begin{align}
\begin{split}	
		\label{DYM.cdcheck.adj2}
		&
		\Big(\,{
			\hlx{ \partial_\a\xi^\n\mem \partial_\n\phi^\m }
			\hlb{ + \xi^\n\mem \partial_\a\partial_\n\phi^\m  }
			\hla{ - \partial_\a\phi^\n\mem \partial_\n\xi^\m }
			\hlz{ - \phi^\n\mem \partial_\a\partial_\n\xi^\m }
		}\,\Big)
		\,,\\
		&
		A^\r{}_\a\mem
		\Big(\,{
			\hly{ \partial_\r\xi^\n\mem \partial_\n\phi^\m }
			\hlb{ + \xi^\n\mem \partial_\r\partial_\n\phi^\m }
			\hla{ - \partial_\r\phi^\n\mem \partial_\n\xi^\m }
			\hlz{ - \phi^\n\mem \partial_\r\partial_\n\xi^\m }
		}\,\Big)
		\,,\\
		&
		\Big(\,{
			\hlx{{- \partial_\a\xi^\r}} \hly{ - A^\n{}_\a\mem \partial_\n\xi^\r } + \hlb{ \xi^\n\mem \partial_\n A^\r{}_\a }
		}\,\Big)\,
		\partial_\r\phi^\m
		\,,\\
		&
		- \Big(\,{
			\hlb{ \xi^\n\mem \partial_\n\phi^\r }
			\hlz{ - \phi^\n\mem \partial_\n\xi^\r }
		}\,\Big)\,
		\partial_\r A^\m{}_\a
		\,,\\
		&
		\phi^\r\mem
		\Big(\,{
			\hlz{ \partial_\r \partial_\a \xi^\m }
			\hla{ + \partial_\r A^\n{}_\a\mem \partial_\n \xi^\m}
			\hlz{ + A^\n{}_\a\mem \partial_\r \partial_\n \xi^\m }
			\hlz{ - \partial_\r \xi^\n\mem \partial_\n A^\m{}_\a }
			\hlb{ - \xi^\n\mem \partial_\r \partial_\n A^\m{}_\a }
		}\,\Big)\,
		\,.
\end{split}
\end{align}
As a result, it follows that
\begin{align}
\begin{split}
	\delta_\xi(D_\a\phi^\m)
	\,=\,
	{}
	&
	\hlb{
		\xi^\r\mem \partial_\r\mem
		\Big(\,{
			\partial_\a\phi^\m
			+ A^\n{}_\a\mem \partial_\n\phi^\m(x)
			- \phi^\nu\, \partial_\n A^\m{}_\a
		}\,\Big)
	} 
	\\
	&
	\hla{
		- \Big(\,{
			\partial_\a\phi^\r
			+ A^\r{}_\a\mem \partial_\r\phi^\r
			- \phi^\r(x)\mem \partial_\r A^\m{}_\a
		}\,\Big)\, \partial_\n\xi^\m
	}
	\,,
\end{split}
\end{align}
which simplifies to
\begin{align}
	\delta_\xi(D_\a\phi^\m)
	\,=\,
		\xi^\r\mem \partial_\r( D_\a\phi^\m )
		- ( D_\a\phi^\n )\mem \partial_\n\xi^\m
	\,=\, \comm{ \xi }{ D_\a\phi }^\m
	\,.
\end{align}
Therefore, $D_\a\phi^\m(x)$ transforms exactly like a diffeomorphism-adjoint field \`a la \eqref{DYM.igt}.

These calculations 
explicitly establish how the CKR rule
concretely realizes the notion of a diffeomorphism-covariant connection
and diffeomorphism-covariant derivative.
A crucial feature to be reiterated is that the covariant derivative does not change the transformation behavior.
If $\phi(x)$ is a fundamental,
its covariant descendants,
$D_\a \phi(x)$,
$D_\a D_\b \phi(x)$,
$D_\a D_\b D_\c \phi(x)$,
$\cdots$,
are all fundamentals as well.
If $\phi^\m(x)$ is an adjoint,
its covariant descendants,
$D_\a \phi^\m(x)$,
$D_\a D_\b \phi^\m(x)$,
$D_\a D_\b D_\c \phi^\m(x)$,
$\cdots$,
are all adjoints as well.

Consequently,
we discover that there naturally arise two types of spacetime indices:
diffeomorphism-neutral 
and diffeomorphism-charged.
The astute reader will point out that
we have been consistently using early greek letters $\a,\b,\c,\d,\cdots$
and late greek letters $\m,\n,\r,\s,\cdots$
for the former and the latter,
respectively.
A simple underlying geometric picture will be disclosed in \Sec{K2:tele}.

The diffeomorphism field strength is defined by
applying CKR on \eqref{YM.fs}:
\begin{align}
\begin{split}
	\label{DYM.fs}
	F^\m{}_{\a\b}(x)
	\,=\,
    {}
    &
		\partial_\a A^\m{}_\b(x) - \partial_\b A^\m{}_\a(x) 
    \\
    &
		+ A^\n{}_\a(x)\, \partial_\n A^\m{}_\b(x)
		- A^\n{}_\b(x)\, \partial_\n A^\m{}_\a(x)
	\,.
\end{split}
\end{align}
Again, this shall arise from the commutator between two diffeomorphism-covariant derivatives,
as is explicitly verified below.

First of all, suppose a fundamental field $\phi(x)$.
Computation shows that
\begin{align}
\begin{split}
	D_\a D_\b \phi
	\,&=\,
	\partial_\a\mem
		\Big(\,{
			\partial_\b \phi + A^\n{}_\b\mem \partial_\n\phi
		}\,\Big)
	+ A^\m{}_\a\mem \partial_\m\mem
		\Big(\,{
			\partial_\b \phi + A^\n{}_\b\mem \partial_\n\phi
		}\,\Big)
	\,,\\
	\,&=\
\begin{aligned}[t]
	&
		\Big(\,{
			\hlx{ \partial_\a\partial_\b \phi } 
			\hlb{ + \partial_\a A^\n{}_\b\mem \partial_\n\phi  }
			\hlx{ + A^\n{}_\b\mem \partial_\a\partial_\n\phi }
		}\,\Big)
	\\
	&
	+ A^\m{}_\a\mem 
		\Big(\,{
			\hlx{ \partial_\m\partial_\b \phi }
			\hla{ + \partial_\m A^\n{}_\b\mem \partial_\n\phi }
			\hlx{ + A^\n{}_\b\mem \partial_\m\partial_\n\phi }
		}\,\Big)
	\,,
\end{aligned}
	\\
\end{split}
\end{align}
where the sum of the terms marked in red 
does not survive after antisymmetrization in the indices $\a, \b$:
\begin{align}
	\comm{D_\a}{D_\b} \phi
	\,=\, 
	\Big(\,{
		\hlb{ \partial_\wrap{[\a} A^\n{}_\wrap{\b]} }
		\hla{ + A^\r{}_\wrap{[\a}\mem \partial_\r A^\n{}_\wrap{\b]} }
	}\,\Big)\,
		\partial_\n \phi
	\,=\,
		F^\n{}_{\a\b}\mem \partial_\n\phi
	\,=\,
		F_{\a\b}\phi
	\,.
\end{align}

Next, suppose an adjoint field $\phi^\m(x)$. Computation shows that
\begin{align}
		D_\a D_\b \phi^\m
		\,&=\,
		\partial_\a\mem
		\Big(\,{
			\partial_\b \phi^\m + [ A_\b , \phi ]^\m
		}\,\Big)
		+ \comm{ A_\a }{ \partial_\b\phi }^\m
		+ \comm{ A_\a }{ \comm{ A_\b }{ \phi } }^\m
		\,,\\
		\,&=\
		\begin{aligned}[t]
			&
			\Big(\,{
				\hlx{ \partial_\a\partial_\b \phi^\m } 
				\hlb{ + \comm{ \partial_\a A_\b }{ \phi }^\m  }
				\hlx{ + \comm{ A_\b }{ \partial_\a\phi }^\m }
			}\,\Big)
			\hlx{ + \comm{ A_\a }{ \partial_\b \phi }^\m }
			\hla{ + \comm{ A_\a }{ \comm{ A_\b }{ \phi } }^\m }
			\,,
		\end{aligned}
		\nonumber
\end{align}
where the sum of the terms marked in red 
does not survive after antisymmetrization in the indices $\a, \b$:
\begin{align}
\begin{split}
	\comm{ D_\a }{ D_\b } \phi^\m
	\,&=\, 
		\hlb{ \comm{ \partial_\wrap{[\a} A_\wrap{\b]} }{ \phi }^\m }
		\hla{ + \comm{ A_\a }{ \comm{ A_\b }{ \phi }}^\m - \comm{ A_\b }{ \comm{ A_\a }{ \phi }}^\m }
	\,,\\
	\,&=\,
		\hlb{ \comm{ \partial_\wrap{[\a} A_\wrap{\b]} }{ \phi }^\m }
		\hla{ + \comm{ \comm{ A_\a }{ A_\b } }{ \phi }^\m }
	\,,\\
	\,&=\,
		\comm{ F_{\a\b} }{ \phi }^\m
	\,=\,
		F^\n{}_{\a\b}\mem \partial_\n\phi^\m
		- \phi^\n\mem \partial_\n F^\m{}_{\a\b}
	\,.
\end{split}
\end{align}
Importantly, we have used the Jacobi identity of the diffeomorphism Lie bracket,
in the form of
$\comm{ A_\a }{ \comm{ A_\b }{ \phi }}^\m - \comm{ A_\b }{ \comm{ A_\a }{ \phi }}^\m = \comm{ \comm{ A_\a }{ A_\b } }{ \phi }^\m$.

These calculations show that
the diffeomorphism field strength
in \eqref{DYM.fs}
is covariantly constructed from
the commutator between two covariant derivatives,
so it serves as a diffeomorphism-covariant operator.
To verify this explicitly, we apply the transformation law for the gauge connection stipulated in \eqref{DYM.igtA},
$\delta_\xi A^\m{}_\a = -D_\a\xi^\m$,
to \eqref{DYM.fs}:
\begin{align}
\begin{split}
	\label{DYM.transform-F}
	\delta_\xi F^\m{}_{\a\b}
	\,&=\,
		\partial_\wrap{[\a} ( -D_\wrap{\b]} \xi^\m )
		+ \comm{ -D_\a\xi }{ A_\b }^\m
		+ \comm{ A_\a }{ -D_\b\xi }^\m
	\,,\\
	\,&=\,
		- \partial_\wrap{[\a} ( D_\wrap{\b]} \xi^\m )
		- \comm{ A_\wrap{[\a} }{ D_\wrap{\b]}\xi }^\m
	\,,\\
	\,&=\,
		- D_\wrap{[\a} D_\wrap{\b]} \xi^\m
	\,=\,
		- \comm{ F_{\a\b} }{ \xi }^\m
	\,=\, 
		\comm{ \xi }{ F_{\a\b} }^\m
    \,,\\
	\,&=\,
		\xi^\n\mem \partial_\n F^\m{}_{\a\b}
		- F^\n{}_{\a\b}\mem \partial_\n\xi^\m
	\,.
\end{split}
\end{align}
\eqref{DYM.transform-F} shows that
the diffeomorphism field strength
transforms as an adjoint
in the precise sense of \eqref{DYM.igt}.

The diffeomorphism field strength $F^\m{}_{\a\b}$
is a diffeomorphism adjoint.
Hence its diffeomorphism covariant derivative is defined as
\begin{align}
    \label{DYM.DF}
    D_\c F^\m{}_{\a\b}(x)
    \,=\,
        \partial_\c F^\m{}_{\a\b}(x)
        + A^\r{}_\c(x)\mem \partial_\r F^\m{}_{\a\b}(x)
        - F^\m{}_{\a\b}(x)\mem \partial_\r A^\m{}_\c(x)
    \,,
\end{align}
which arises by applying the CKR on
\eqref{YM.DF}.(x)

Finally,
the CKR
yields the diffeomorphism Bianchi identity
from \eqref{YM.bianchi}:
\begin{align}
	\label{DYM.bianchi}
	D_\wrap{[\c} F^\m{}_\wrap{\a\b]}(x) 
	\,=\, 
	D_\c F^\m{}_{\a\b}(x) + D_\a F^\m{}_{\b\c}(x) + D_\b F^\m{}_{\c\a}(x)
	\,=\,
	0
	\,,
\end{align}
Again, this is satisfied by the Jacobi identity of diffeomorphism Lie algebra.
The essential cancellation describes
$\comm{ A_\c }{ \comm{ A_\a }{ A_\b }}^\m + \comm{ A_\a }{ \comm{ A_\b }{ A_\c }}^\m + \comm{ A_\b }{ \comm{ A_\c }{ A_\a }}^\m = 0$.

These explicit and mechanical computations should convince the reader that
the idea 
a consistent working notion
of ``diffeomorphism gauge theory'' is 
not inconceivable
from the field theory perspective.

\subsection{Diffeomorphism Wilson Line}
\label{DYM>W}

Amusingly, the CKR
allows us to
construct the ``diffeomorphism Wilson line.''

Take the path-ordered exponential formula in \eqref{YM.W.fund},
given in the fundamental representation.
The CKR
maps it to
\begin{align}
	\label{DYM.W.fund}
	W_\CC
	\,=\,
	\Pexp\bigg(\,{
		-\nem \int_0^1
		ds\,\,
		A^\m{}_\a(x{\,+\,}y(s))\, \dot{y}^\a(s)\,
			\partial_\m
	}\,\bigg)
	\,,
\end{align}
where we remind ourselves that $\partial_\m$ denotes $\partial/\partial x^\m$.
Again, $s \mapsto y^\a(s)$ is a finite contour 
such that $y^\a(0) = 0$ and $s \in [0,1]$.
We propose that \eqref{DYM.W.fund}
defines the counterpart of Wilson line
in diffeomorphism gauge theory.
As will be further elaborated on in \Sec{DYM>AG},
we understand
\eqref{DYM.W.fund}
as a \textit{differential operator}.
That is,
\eqref{DYM.W.fund} states that
\begin{align}
	\label{DYM.W.fund.expand}
    {}&{}
	W_\CC \phi(x)
    \,=\,
	\phi(x)
	-\nem \int_0^1 ds_1\,\,
		A^\m{}_\a(x{\,+\,}y(s_1))\, \dot{y}^\a(s_1)\,
		\partial_\m\phi(x)
    \\
    \nonumber
    {}&{}
	+ \frac{1}{2}\mem \int_0^1 ds_1 \int_0^{s_1}\nem\nem ds_2\,\,
	A^\m{}_\a(x{\,+\,}y(s_1))\,
		\partial_\m A^\n{}_\b(x{\,+\,}y(s_2))\, \partial_\n\phi(x)
	\mem 
	\dot{y}^\a(s_1)\, \dot{y}^\b(s_2)
	\\
    \nonumber
    {}&{}
	+ \frac{1}{2}\mem \int_0^1 ds_1 \int_0^{s_1}\nem\nem ds_2\,\,
	A^\m{}_\a(x{\,+\,}y(s_1))\,
		A^\n{}_\b(x{\,+\,}y(s_2))\, \partial_\m\partial_\n\phi(x)
	\mem 
	\dot{y}^\a(s_1)\, \dot{y}^\b(s_2)
	\\
    \nonumber
    {}&{}
        + \cdots
	\,,
\end{align}
for any test function $\phi(x)$
valued in the fundamental.

From \eqref{DYM.W.fund.expand},
one can only deduce that $W_\CC$ is a generic differential operator,
collecting arbitrarily high orders of derivatives:
$\partial_\m$, $\partial_\m\partial_\n$, $\partial_\m\partial_\n\partial_\r$, $\cdots$.
Crucially, however, a closer look reveals that
it specifically describes
the transformation of test functions (pullback) induced by 
a diffeomorphism on spacetime,
establishing that
it is indeed valued in the diffeomorphism group:
$W_\CC \in \G = \Diff(\mathbb{R}^d)$.

Again, this will be established by
exploiting the Magnus series \cite{Magnus:1954zz}.
Applying
the CKR 
to \eqref{YM.magnus},
we find that
\begin{align}
	\label{DYM.Omega}
	W_\CC
	\,=\,
		\exp\Big(\,{ \Omega^\m_\CC(x)\mem \partial_\m }\mem\Big)
	\,,
\end{align}
where
\begin{align}
    \label{DYM.magnus}
    {}&{}
    \Omega_\CC^\m(x)
    \,=\,
        -\nem \int_0^1 ds_1\,\,
        A^\m{}_\a(x{\,+\,}y(s_1))\, \dot{y}^\a(s_1)
    \\
    {}&{}
        + \frac{1}{2}\mem \int_0^1 ds_1 \int_0^{s_1}\nem\nem ds_2\,\,
        \comm{ A_\a(x{\,+\,}y(s_1)) }{ A_\b(x{\,+\,}y(s_2)) }^\m\,
        \mem 
        \dot{y}^\a(s_1)\, \dot{y}^\b(s_2)
        + \cdots
        \,.
    \nonumber
\end{align}
Again, the derivation of this formula
makes uses of the Jacobi identity of the diffeomorphism Lie algebra,
from which the nested bracket structure arises.
To clarify, note that
the bracket $\comm{ A_\a(x{\,+\,}y(s_1)) }{ A_\b(x{\,+\,}y(s_2)) }^\m$ in \eqref{DYM.magnus}
is,
with $\partial_\n = \partial/\partial x^\n$,
\begin{align}
	A^\n{}_\a(x{\,+\,}y(s_1))\, \partial_\n A^\m{}_\b(x{\,+\,}y(s_2))
	-
	A^\n{}_\b(x{\,+\,}y(s_2))\, \partial_\n A^\m{}_\a(x{\,+\,}y(s_1))
	\,.
\end{align}

The formula in \eqref{DYM.magnus} describes that
a choice of a contour $\CC$ 
defines a vector field $\Omega^\m_\CC(x)$ on spacetime.
The formula in \eqref{DYM.Omega} manifests that
the Wilson line $W_\CC$
describes the unit-time flow under this vector field $\Omega_\CC^\m(x)$ in spacetime.
The crucial point here is that
$W_\CC$ is not just an any differential operator
but 
is the exponential of a \textit{first-order} derivative,
as \eqref{DYM.Omega} precisely shows.

\newpage

In sum, the diffeomorphism Wilson line defined in \eqref{DYM.W.fund}
can be understood as an element of the diffeomorphism Lie group
assigned for each contour $\CC$ in spacetime:
$W_\CC \in \G = \Diff(\mathbb{R}^d)$.
This is because it describes the exponentiation
of an element of the diffeomorphism Lie algebra 
as a vector field $\Omega_\CC \in \g = \diff(\mathbb{R}^d)$.

Another way to establish this fact is to study the transport equation.
From the path-ordered exponential formula described in \eqref{DYM.W.fund},
it follows that
\begin{align}
	&
	\frac{\partial}{\partial s}\,\bigg[\,\,{
		\Pexp\bigg(\,{
			-\nem \int_0^s
			ds'\,\,
			A^\m{}_\a(x{\,+\,}y(s'))\, \dot{y}^\a(s')\, \partial_\m
		}\,\bigg)\, \phi(x)
	}\mem\,\bigg]
	\\
	&=\,
		-A^\m{}_\a(x{\,+\,}y(s))\,
		\partial_\m
		\bigg[\,\,{
			\Pexp\bigg(\,{
				-\nem \int_0^s
				ds'\,\,
				A^\m{}_\a(x{\,+\,}y(s'))\, \dot{y}^\a(s')\, \partial_\m
			}\,\bigg)\, \phi(x)
		}\mem\,\bigg]
	\,,
    \nonumber
\end{align}
for $\phi(x)$ fundamental.
This implies that the solution to the transport equation
in the fundamental representation,
\begin{align}
	\label{DYM.transpt.fund}
	\frac{\partial}{\partial s}\, \Psi(x;s)
	+ A^\m{}_\a(x{\,+\,}y(s))\, \dot{y}^\a(s)\, \partial_\m\Psi(x;s)
	\,=\, 0
	\,,
\end{align}
with initial condition $\Psi(x;0) = \phi(x)$,
is given by $\Psi(x;1) = W_\CC\phi(x)$.
Namely,
\begin{align}
	\label{DYM.Wtransport.fund}
	\Psi(x;1)  = 
	W_\CC \Psi(x;0)
	\,.
\end{align}
Again, here $W_\CC$ acts on the function $\Psi(x;0)$ of $x^\m$ as a differential operator.

To establish the bilocal transformation behavior of the diffeomorphism Wilson line,
one can
study the transport equation in \eqref{DYM.transpt.fund}
in the fundamental.
Recalling \eqref{DYM.igt},
we compute three terms:
\begin{subequations}
\begin{align}
	\label{DYM.wtr.term1}
	&
	\frac{\partial}{\partial s}\BB{
		\xi^\m(x\mplus y(s))\mem \partial_\m \Psi(x;s)
	}
	\,,\\
	\label{DYM.wtr.term2}
	&
	{-\BB{
		\partial_\a \xi^\m(x\mplus y(s))
		+ \comm{ A_\a(x\mplus y(s)) }{ \xi(x\mplus y(s)) }^\m
	}}\,
		\dot{y}^\a(s)\, \partial_\m \Psi(x;s)
	\,,\\
	\label{DYM.wtr.term3}
	&
	A^\m{}_\a(x\mplus y(s))\, \dot{y}^\a(s)\, \partial_\m
	\BB{
		\xi^\r(x\mplus y(s))\mem \partial_\r \Psi(x;s)
	}
	\,.
\end{align}
\end{subequations}
\begin{subequations}
The first term evaluates to
\begin{align}
	\label{DYM.wtrev.term1}
	\hlx{
		\partial_\a\xi^\m(x\mplus y(s))\mem \dot{y}^\a(s)\,
			\partial_\m \Psi(x;s)
	}
	\hlb{ +
		\xi^\m(x\mplus y(s))\, \partial_\m\mem \frac{\partial}{\partial s}\mem \Psi(x;s)
	}
	\,,
\end{align}
while the sum of the second and third terms evaluates to
\begin{align}
\begin{split}
	\label{DYM.wtrev.term23}
	&
	\hlx{
		- \partial_\a \xi^\m(x\mplus y(s))\mem \dot{y}^\a(s)\,
			\partial_\m \Psi(x;s)
	}  
    \\
    &
	\hla{ +
		\xi^\r(x\mplus y(s))\, \partial_\r A^\m{}_\a(x\mplus y(s))
		\,
		\dot{y}^\a(s)\, \partial_\m \Psi(x;s)
	}
	\\
	&
	\hlb{ +
		A^\m{}_\a(x\mplus y(s))\, \dot{y}^\a(s)\,
		\xi^\r(x\mplus y(s))\mem \partial_\r
			\partial_\m \Psi(x;s)
	}
	\,.
\end{split}
\end{align}
\end{subequations}
As a result, the sum of \eqrefs{DYM.wtrev.term1}{DYM.wtrev.term23}
is
\begin{align}
	\xi^\r(x\mplus y(s))\mem \partial_\r
	\BB{
		\partial_\m \Psi(x;s)
		+ 
			A^\m{}_\a(x\mplus y(s))\, \dot{y}^\a(s)\,
				\partial_\m \Psi(x;s)
	}
	\,.
\end{align}

This shows that the transport equation in \eqref{DYM.transpt.fund}
is diffeomorphism gauge covariant at position $x \mplus y(s)$,
if $\Psi(x;s)$ is diffeomorphism gauge covariant at position $x \mplus y(s)$.
With this observation, one argues that
$\Psi(x;1) = W_\CC \Psi(x;0)$ in \eqref{DYM.Wtransport.fund}
is diffeomorphism gauge covariant at position $x \mplus y(1)$
if $\Psi(x;0)$ is diffeomorphism gauge covariant at position $x$.
Namely, if $\CC$ is a contour that starts at $\P$ and ends at $\P'$,
$W_\CC \phi(\P)$ transforms like a local operator at $\P'$
if $\phi(\P)$ is a local operator at $\P$
valued in the fundamental or adjoint.

\subsection{Covariant Taylor Expansion}

Again, one can express $W_\CC \phi(\P)$ as an infinite sum of local operators at $\P'$
provided $\P'$ is sufficiently close to $\P$.
To understand this explicitly,
let us consider a straight-line contour 
$s \mapsto x^\m + q^\m s$,
where $q^\m$ describes a fixed vector.
The diffeomorphism Wilson line for a part of this contour
will be denoted as
\begin{align}
	W(x{\,+\,}qs_2,x{\,+\,}qs_1)
	\,=\,
	\Pexp\bigg(\,{
		-\nem \int_{s_1}^{s_2}
		ds\,\,
		A^\m{}_\a(x{\,+\,}qs)\, q^\a
		\,\partial_\m
	}\,\bigg)
	\,.
\end{align}
For any $s$, the identities in \eqrefs{ids.WW}{ids.dW} hold,
with the understanding that
product of diffeomorphism Wilson lines
describes composition of differential operators.

For a fundamental-valued field $\phi(x)$,
computation shows that
the same identities as in \eqrefs{YM.ctr1}{YM.ctrl}
hold, but with 
$D_\a\phi(x{\,+\,}qs) = \partial_\a\phi(x{\,+\,}qs) + A^\m{}_\a(x{\,+\,}qs)\, \partial_\m \phi(x{\,+\,}qs)$.
Therefore the covariant Taylor expansion in \eqref{YM.ctr}
holds for the diffeomorphism gauge connection as well,
which also generalizes to adjoint-valued fields.

The insight that can be gained from this fact is that
the diffeomorphism-covariant derivative has facilitated the comparison between field values at different points
via the diffeomorphism Wilson line.
It is a nice exercise to verify that,
for a fundamental-valued scalar $\phi(x)$
and an adjoint-valued $\phi^\m(x)$,
\begin{align}
	\label{DYM.Dlimits}
	\Fund:\quad&
	q^\a D_\a \phi(x)
	\,=\,
		\lim_{\e\to0}
		\,\frac{1}{\e}\mem
		\Big(\,{
			W(x,x{\,+\,}\e q)\, \phi(x{\,+\,}\e q)
			- \phi(x)
		}\,\Big)
	\,,\\
	\Adj:\quad&
	q^\a D_\a \phi^\m(x)
	\,=\,
		\lim_{\e\to0}
		\,\frac{1}{\e}\mem
		\Big(\,{
			\mathe^{
				\comm{
					\Omega^\r(x,x{\,+\,}\e q)\mem \partial_\r
				}{\blank}
			}\, \phi^\m(x{\,+\,}\e q)
			- \phi^\m(x)
		}\,\Big)
	\,,
    \nonumber
\end{align}
where the second line
will compute the adjoint action 
with the help of the Magnus representation in \eqref{DYM.Omega}.
Notably, this faithfully expresses the very geometrical idea of a gauge connection and a gauge-covariant derivative,
in the strict gauge theory sense.
The diffeomorphism Wilson line provides a law of parallel transport
so that objects at different points can be compared.

\subsection{A Mathematical Perspective}
\label{DYM>AG}

In the above exposition, we regarded
vector fields as linear differential operators.
In pursuit of this differential operator approach,
we computed and defined
$W_\CC$ as a differential operator
and then identified it rather directly as an element of the diffeomorphism group as
$W_\CC \inn \Diff(\R^d)$,
although a diffeomorphism on $\R^d$ is usually thought of as a map $\R^d \too \R^d$.

We shall remark that
this identification is based on the differential operator representation of diffeomorphisms
\cite{vershikgelfand1975diffreps} by the pullback,
which traces back to
an algebraic perspective on geometry
\cite{vershikgelfand1975diffreps,Connes:1994yd,gracia2013elements,Szabo:2001kg,gelfand1943imbedding,segal1947irreducible,rmont-RingStructB}.
Below,
we elaborate on this mathematical viewpoint briefly.

The idea is that diffeomorphisms can be thought of as
maps that preserve the algebraic structure of smooth functions 
under pointwise addition and multiplication:
automorphisms of
the commutative ring $\Cinfty(\R^d)$ of smooth functions.
Namely, a differential operator $U : \Cinfty(\R^d) \to \Cinfty(\R^d)$
is a diffeomorphism $U \in \Diff(\R^d)$ iff
\begin{align}
	U\act{ \psi_1 + \psi_2 } \,=\, U\act{\psi_1} + U\act{\psi_2}
	\,,\quad
	U\act{ \psi_1 \psi_2 } \,=\, U\act{\psi_1} \, U\act{\psi_2}
\end{align}
for all $\psi_1,\psi_2 \in \Cinfty(\R^d)$.
In this precise mathematical sense, 
the differential operator in
\eqref{DYM.W.fund} has defined a diffeomorphism
$W_\CC \in \Diff(\R^d)$.

Similarly,
vector fields
on $\R^d$,
as infinitesimal diffeomorphisms,
are derivations on the commutative algebra $\Cinfty(\R^d)$:
they are differential operators that preserve the pointwise addition and multiplication.
That is, a differential operator $v : \Cinfty(\R^d) \to \Cinfty(\R^d)$
is 
a diffeomorphism generator $v \in \diff(\R^d)$
iff
\begin{align}
	v\act{ \psi_1 + \psi_2 } \,=\, v\act{ \psi_1 } + v\act{ \psi_2 }
	\,,\quad
	v\act{ \psi_1 \psi_2 } \,=\, v\act{\psi_1}\, \psi_2 + \psi_1\, v\act{\psi_2}
	\,.
\end{align}
This definition implies that $v$ is a linear differential operator, $v = v^\m \partial_\m$.
The Lie bracket between vector fields
is simply the commutator between linear differential operators.

In ordinary gauge theory,
the gauge connection
provides the isomorphism between
color spaces (finite-dimensional fibers) at two different points.
Importantly, this law of parallel transport preserves the 
color Lie algebra structure.
In diffeomorphism gauge theory,
the diffeomorphism gauge connection
may be thought of as
providing the isomorphism between
the infinite-dimensional algebra $\Cinfty(\R^d)$ of smooth functions
at two different points.

These mathematical formalizations
also conform to those of
noncommutative and Poisson gauge theories
\cite{Connes:1994yd,gracia2013elements,Szabo:2001kg,gelfand1943imbedding,segal1947irreducible,rmont-RingStructB,SW,Jurco:2000ja,Madore:2000en}.
For instance,
the diffeomorphism Wilson line in \eqref{DYM.W.fund}
can be deduced as a generalization of
the Poisson Wilson line from, e.g., \rrcite{Ishibashi:1999hs,Das:2000md}.

These mathematical justifications,
however,
may not be conceived as
providing the most clear
or practically useful
picture
for physicists.
Notably,
a simple geometrical interpretation
is found below.

\section{ Interpretation as Teleparallel Theory}
\label{K2:tele}

\subsection{Frame and Anholonomy}

As we have discussed in \Sec{Jb},
a \textit{teleparallel theory}
refers to a field theory of a dynamical frame field
\cite{weitzenbock1923invarianten,einstein1925einheitliche,einstein1928riemann,einstein1929einheitliche,einstein1956appendix,cartan1979lettres,Goenner:2004se,sauer2014einstein_unified_field_theory,Maluf:2013gaa,Aldrovandi:2013wha,Ferraro:2016wht,TrinityHeisenberg:2019}.
Strictly speaking, the name ``teleparallel theory''
suits
theories that describe geometries of absolute parallelism
in which the frame (diffeomorphism neutral) index is \textit{global},
as is discussed above \eqref{weit}.
In the case of GR,
we have seen in \eqref{tele.LLT}
that the frame index is, in fact, \textit{gauged}
due to local Lorentz symmetry.
Thus,
the frame formulation of GR
in \eqrefsor{FrameAction1}{FrameAction2}
is not a teleparallel field theory
in the strict sense.

A \textit{genuine teleparallel theory}
is a field theory of a dynamical frame $E^\m{}_\a$
whose diffeomorphism neutral index $\a$
is global,
i.e.,
not charged under any local symmetries.
It
governs
a dynamical geometry
of absolute parallelism.

Let us study a genuine teleparallel theory.
For clarity,
we use $\a,\b,\c,\d,\cdots$ for
the global diffeomorphism neutral index
(since $A,B,C,D,\cdots$ have been used and reserved as the local Lorentz index in our discussions on GR).
With this clarification in mind,
we recall the elements of frame geometry
from \Sec{Jb}.
The frame's
degree of anholonomy,
i.e., noncoordinateness,
is captured by 
the Lie bracket
$\comm{E_\a}{E_\b}$,
which we propose to denote as $F_{\a\b}$:
\begin{align}
    \label{F-E}
    F^\m{}_{\a\b}
    \,=\,
        \comm{E_\a}{E_\b}^\m
    \,=\,
        E^\r{}_\a\mem \partial_\r E^\m{}_\b
        -
        E^\r{}_\b\mem \partial_\r E^\m{}_\a
    \,.
\end{align}
For example,
the anholonomy coefficients 
defined in \eqref{anholonomy-def}
are
$\Omega^\c{}_{\a\b} = e^\c{}_\m\mem F^\m{}_{\a\b}$,
if the frame $E^\m{}_\a$
is invertible
to a coframe $e^\a{}_\m$.
The identity derived in \eqref{gym.1}
describes
$
    - E^\m{}_\c\mem de^\c
    = 
        \tfrac{1}{2}\,
            F^\m{}_{\a\b}\,
        e^\a \swedge e^\b
$.
Also notable is
\begin{align}
    \label{[E,F]}
    \comm{E_\c}{F_{\a\b}}^\m
    \,=\,
        E^\r{}_\c\mem \partial_\r F^\m{}_{\a\b}
        - F^\r{}_{\a\b}\mem \partial_\r E^\m{}_\c
    \,,
\end{align}
so the Bianchi identity is
\begin{align}
    \label{bianchiEF}
    \comm{E_\c}{F_{\a\b}}^\m
    + \comm{E_\a}{F_{\b\c}}^\m
    + \comm{E_\b}{F_{\c\a}}^\m
    \,=\,
        0
    \,.
\end{align}

Suppose the frame $E^\m{}_\a$ takes
the Kronecker delta $\delta^\m{}_\a$
as its vacuum expectation value.
The \textit{frame perturbation} $A^\m{}_\a$ is defined as
\begin{align}
    \label{vevE}
    E^\m{}_\a
    \,=\,
        \delta^\m{}_\a + A^\m{}_\a
    \,.
\end{align}
The background $\delta^\m{}_\a$ will spontaneously break diffeomorphisms.
Let us learn about
how the ``background independent'' formulae
of teleparallel geometry
boils down
when expanded around this vacuum expectation value.
By plugging in \eqref{vevE}
to
\eqref{F-E},
we learn that the anholonomy boils down to
\begin{align}
    F^\m{}_{\a\b}
    \,=\,
        \partial_\a A^\m{}_\b
        - \partial_\b A^\m{}_\a
        + A^\r{}_\a\mem \partial_\r A^\m{}_\b
        - A^\r{}_\b\mem \partial_\r A^\m{}_\a
    \,,
\end{align}
which is nothing but the diffeomorphism field strength
in \eqref{DYM.fs}.

The vacuum expectation value of the anholonomy $F^\m{}_{\a\b}$ is zero.
With this in mind,
the expansion of \eqref{[E,F]} around the Kronecker vacuum is
\begin{align}
    \comm{E_\c}{F_{\a\b}}^\m
    \,=\,
        \partial_\c F^\m{}_{\a\b}
        + A^\r{}_\c\mem \partial_\r F^\m{}_{\a\b}
        - F^\r{}_{\a\b}\mem \partial_\r A^\m{}_\c
    \,=\,
        D_\c F^\m{}_{\a\b}
    \,,
\end{align}
which is nothing but the
diffeomorphism covariant derivative
of the diffeomorphism field strength
in \eqref{YM.DF}.
Eventually, this leads to the realization that
the diffeomorphism covariant derivative
is nothing but the Lie derivative with respect to the frame vector field,
on various tensor fields:
\begin{align}
    \label{D=L}
    D_\a 
    \,=\,
        \pounds_{E_\a}
    \,.
\end{align}
In particular, \eqref{DYM.igt} is reproduced by
literally performing a spacetime diffeomorphism
while taking $\phi(x)$ and $\phi^\m(x)$ as scalar and vector fields,
while \eqref{DYM.igtA} arises by 
$E_\a \mapsto E_\a + \pounds_\xi E_\a$
realized around the Kronecker vacuum.

In sum,
we realize that
teleparallel theory
provides an alternative definition of
diffeomorphism gauge theory 
that is equivalent to the definition in \Sec{K2>DGT}
by the CKR,
provided
the crucial identification that\footnote{
    To our best understanding,
    this realization was first made in
    \rcite{dpb}, App.\,D.
    Later, it was also realized in \rcite{Maor3dGravity}.
    A historical precursor will be
    Mason and Newman \cite{MasonNewman:1989}.
}
\begin{Fullnote}
    The frame perturbation
    around the Kronecker vacuum
    is the diffeomorphism gauge connection.
\end{Fullnote}
\newpage\noindent
The diffeomorphism covariant derivative
is the Lie derivative
by the frame.

Lastly, consider a genuine teleparallel theory
provided with a constant metric $\eta^{\a\b}$
as an invariant global structure.
Suppose its classical EoM are given by
\begin{align}
    \label{dymE}
    \eta^{\c\a}\mem
    \comm{
        E_\c
    }{
        F_{\a\b}
    }^\m
    \,=\,
    \eta^{\c\a}\mem
    \comm{
        E_\c
    }{
        \comm{E_\a}{E_\b}
    }^\m
    \,=\,
        0
    \,.
\end{align}
Evidently, \eqref{dymE} is a background-independent equation.
Around the Kronecker vacuum,
plugging in \eqref{vevE} to \eqref{dymE} gives
\begin{align}
    \label{dymA}
    \eta^{\c\a}\mem
    \BB{
        \partial_\c F^\m{}_{\a\b}
        + A^\r{}_\c\mem \partial_\r F^\m{}_{\a\b}
        - F^\r{}_{\a\b}\mem \partial_\r A^\m{}_{\c}
    }
    \,=\,
        0
    \,,
\end{align}
while the vacuum expectation value of the anholonomy $F^\m{}_{\a\b}$ is zero.
\eqref{dymA}
is nothing but the diffeomorphism YM equations
of Cheung and Mangan \cite{CCK},
reproduced in \eqref{CK3.29}.

In conclusion,
we realize that 
Cheung and Mangan \cite{CCK}'s
formulation of BI theory,
reproduced in \eqref{CK3},
admits a remarkable geometric interpretation.
The diffeomorphism gauge connection 
introduced in \eqref{CK3.27}
is the vielbein perturbation
around the Kronecker vacuum,
$A^\m{}_\a = E^\m{}_\a - \delta^\m{}_\a$.
The diffeomorphism field strength in \eqref{CK3.28}
admits a background-independent rewriting
as the anholonomy $F^\m{}_{\a\b} = \comm{E_\a}{E_\b}^\m$.
The diffeomorphism YM equations in \eqref{CK3.29}
admit a background-independent presentation
as $\eta^{\c\a} \comm{E_\c}{F_{\a\b}} = 0$.
The diffeomorphism covariant derivative
means the Lie derivative:
$D_\a = \pounds_{E_\a}$.

\medskip

A few remarks are in order.
First and foremost,
not all diffeomorphism-invariant theories of the dynamical vielbein $E$
would turn out to be obtainable via 
applying a CKR on ordinary gauge theories.
That is,
a teleparallel theory
is not guaranteed to be
a diffeomorphism gauge theory of the strict sense.
This point is best demonstrated by examples;
we do so in \Secsto{K2>CaseBF}{K2>CaseGR}.
While the equations that we have explored just now
are certainly obtainable from the CKR,
there are also objects in teleparallel geometry
that do not straightforwardly admit
their counterparts
in ordinary gauge theory.
For example,
take the anholonomy \textit{coefficients}
$\Omega^\a{}_{\b\c}$,
which involve the inverse
$e^\a{}_\m = \delta^\a{}_\m - \delta^\a{}_\n\mem A^\n{}_\b\mem \delta^\b{}_\m + \O(A^2)$.
For another example,
take the determinant $|E| = 1 + \delta^\a{}_\m\mem A^\m{}_\a + \O(A^2)$
in general dimensions.
The appearance of bare connection $A$
may cause problems
when these objects are used in the Lagrangian,
for instance.

Secondly,
it is known that
teleparallel geometry
could admit an analogy with
lattice systems 
at
the level of mathematical frameworks
\cite{chaikin1995principles,Grozdanov:2018ewh,Pace:2023kyi,Fiziev:1995te,Kleinert:1996yi,Kleinert:2008zzb},
such as crystals with impurities.
In this case one can
import intuitions and ideas
such as
Burgers vector and dislocation defects,
as will be briefly touched upon in \Sec{K2>Misner}.
In this context
a diffeomorphism-invariant dressing
may be possible,
in connection with the diffeomorphism Wilson line in \Sec{DYM>W}.
The rough idea is to flow along the lattice frame
to find an anchor on the infinities.
While typical scalar fields are diffeomorphism fundamental,
a dressed field will be diffeomorphism singlet,
exhibiting the invariant behavior envisioned in
\eqref{DYM.igt}.

Thirdly,
it should be noted that
our diffeomorphism gauge theory program for the double copy
seeks to represent an approach
different from
previous constructions
such as
\rrcite{castro2004moyal,cho1992gravdiff}'s bundle construction
or
double field theory \cite{Diaz-Jaramillo:2021wtl,Lee:2018gxc,Siegel:1993th,Hull:2009mi,Hohm:2010pp,Hohm:2010jy,Hohm:2013bwa}
(cf.
generalized geometry \cite{hitchin2003generalized,gualtieri2004generalized,gualtieri2011generalized,courant1990dirac}).
In particular, our approach does not employ
any auxiliary or extra space
other than the spacetime manifold $\M$.
In this picture,
there is no such thing as 
two different species of coordinates 
$x^\m$, $\tilde{x}^{\tilde{\m}}$.
\rcite{CCK} writes down ``$A^{\tilde{\m}}{}_\m$'' as a notation,
but both indices
there describe the same spacetime kind of indices.

\subsection{Cofundamental, Coadjoint, and Densities}

The realization in \eqref{D=L}
facilitates a possible top-down definition of
cofundamental, coadjoint, or densities
in diffeomorphism gauge theory.\footnote{
    It also facilitates writing down
    noninfinitesimal diffeomorphism gauge transformations
    exactly, which we leave as an exercise.
}
There might be other versions
from representation theory points of view,
but
the identifications
below seem to provide one consistent version.

A cofundamental field $\psi(x)$ is a scalar density.
It is a field
that exhibits the infinitesimal transformation behavior
\begin{align}
    \label{igt.cofund}
    \Co\Fund:\quad
    \psi(x)
    \,\,\,\mapsto\,\,\,
    \psi(x)
    \mem+\mem
        \partial_\m\bigbig{
            \xi^\m(x)\mem \psi(x)
        }
    \,,
\end{align}
which can also be stated as
\begin{align}
    \psi(x)\mem d^dx
    \,\,\,\mapsto\,\,\,
    \psi(x)\mem d^dx
    \mem+\mem
    \pounds_\xi\bigbig{
        \psi(x)\mem d^dx
    }
    \,.
\end{align}

A coadjoint field $\psi_\m(x)$ is a one-form density.
It is a field 
that exhibits the infinitesimal transformation behavior
\begin{align}
    \label{igt.coadj}
    \kern-0.1em
    \Co\Adj:\quad
    \psi_\m(x)
    \,\,\,\mapsto\,\,\,
    \psi_\m(x)
    \mem+\mem
        \partial_\r\bigbig{
            \xi^\r(x)\mem \psi_\m(x)
        }
        + \psi_\r(x)\mem \partial_\m \xi^\r(x)
    \,.
    \kern-0.25em
\end{align}

In \eqrefs{YM.igt}{DYM.igt},
we see that
the infinitesimal transformation
$\delta_\xi \phi^\m = \comm{\xi}{\phi}^\m = (\ad{\xi}\hem \phi)^\m$
corresponds to
$\delta_\ve \phi^a = \comm{\ve}{\phi}^a = (\ad{\ve}\hem \phi)^a$
in ordinary gauge theory,
which identifies the adjoint action $\ad{\xi}$ for the diffeomorphism algebra.
Similarly,
comparing \eqref{igt.coadj}
with the infinitesimal transformation 
$\delta_\ve \psi_a = (\coad{\ve}\hem \psi)_a$
of ordinary gauge theory
identifies the coadjoint action for the diffeomorphism algebra as
\begin{align}
    \label{coad-diff}
    (\coad{\xi} \psi)_\m
    \,=\,
        \partial_\r\bigbig{
            \xi^\r\mem \psi_\m
        \hnem}
        + \psi_\r\mem \partial_\m \xi^\r
    \,.
\end{align}

Suppose $\phi(x)$ is fundamental and $\psi(x)$ is cofundamental.
A diffeomorphism-invariant pairing exists between the two:
\begin{align}
    \label{Dpairing.cofund}
    \int
        d^dx\,\,
        \psi(x)\mem
        \phi(x)
    \,\,\,\mapsto\,\,
    \int
        d^dx\,\,
        \psi(x)\mem
        \phi(x)
    \,.
\end{align}

Similarly,
suppose $\phi^\m(x)$ is adjoint and $\psi_\m(x)$ is coadjoint.
A diffeomorphism-invariant pairing exists between the two:
\begin{align}
    \label{Dpairing.coadj}
    \int
        d^dx\,\,
        \psi_\m(x)\mem
        \phi^\m(x)
    \,\,\,\mapsto\,\,
    \int
        d^dx\,\,
        \psi_\m(x)\mem
        \phi^\m(x)
    \,.
\end{align}
To show this, consider first 
how
$\psi_\m(x)\mem \phi^\m(x)$
transforms.
\eqrefs{DYM.igt}{igt.coadj} imply
that it changes by
\begin{align}
    \psi_\m\mem \BB{
        \xi^\r\mem \partial_\r \phi^\m
        - \phi^\r\mem \partial_\r \xi^\m
    }
    +
    \BB{
        \xi^\r\mem \partial_\r \psi_\m
        + \partial_\r \xi^\r\mem \psi_\m
        + \psi_\r\mem \partial_\m \xi^\r
    }\mem
    \phi^\m
    \,,
\end{align}
which simplifies to
$\partial_\r\bigbig{
    \xi^\r\mem \psi_\m \phi^\m
}$.
This is a total derivative.

It should be clear that
one can define various tensor or tensor density fields
in this way,
with any number of indices.

It will be interesting to reassess these
density representations
in the axiomatic point of view
demonstrated in \Sec{K2>DGT}.
Also, note that
the distinction between fundamental and cofundamental
vanishes
when the diffeomorphisms are restricted to be volume-preserving.

\subsection{Case Study I: Diffeomorphism \textit{BF} theory}
\label{K2>CaseBF}

Consider the action
\begin{align}
    \label{DBF.action}
    S[E,B]
    \,=\,
    \int 
        B^{\a\b}{}_\m\mem F^\m{}_{\a\b}[E]
        \,\, d^dx
    \,.
\end{align}
Here, $F^\m{}_{\a\b}[E]$ denotes the anholonomy
$\comm{E_\a}{E_\b}^\m$
as a functional of $E^\m{}_\a$.

\eqref{DBF.action}
is invariant under the infinitesimal local transformation
\begin{align}
\begin{split}
    \label{K2:tele-diff}
    E^\m{}_\a
    \,\,\,&\mapsto\,\,\,
    E^\m{}_\a
    \mem+\mem
        \xi^\r\mem \partial_\r E^\m{}_\a
        - E^\r{}_\a\mem \partial_\r \xi^\m
    \,,\\
    B^{\a\b}{}_\m
    \,\,\,&\mapsto\,\,\,
    B^{\a\b}{}_\m
    \mem+\mem
        \partial_\r\bigbig{
            \xi^\r
            B^{\a\b}{}_\m
        }
        + B^{\a\b}{}_\r\mem \partial_\m \xi^\r
    \,,
\end{split}
\end{align}
whose interpretation is an infinitesimal diffeomorphism.
We see that the $E$-field is a diffeomorphism adjoint
while the $B$-field is a diffeomorphism coadjoint,
while carrying diffeomorphism-neutral indices.

Note that 
we have already encountered
\eqref{K2:tele-diff}
previously in \eqref{tele.diff},
in fact.
\eqref{DBF.action}
is exactly the topological limit of GR
proposed in \Sec{J:topoGR}.

There is no local redundancy
acting on the frame index $\a$ of $E^\m{}_\a$,
so we identify \eqref{DBF.action}
as defining a genuine teleparallel field theory.

We stipulate that the frame $E^\m{}_\a$
gains the Kronecker vacuum expectation value, $\delta^\m{}_\a$.
By plugging in \eqref{vevE},
the action in \eqref{DBF.action} boils down to
\begin{align}
    \label{DBF.actionA}
    S[A,B]
    \,=\,
    \int 
        B^{\a\b}{}_\m\mem F^\m{}_{\a\b}[A]
        \,\, d^dx
    \,,
\end{align}
where $F^\m{}_{\a\b}[A]$ denotes the diffeomorphism field strength
as a functional of $E^\m{}_\a$.
It follows that \eqref{DBF.actionA}
is obtainable by
applying a CKR on an ordinary gauge theory,
\begin{align}
    \label{BF.actionA}
    S[A,B]
    \,=\,
    \int 
        B^{\a\b}{}_a\mem F^a{}_{\a\b}[A]
        \,\, d^dx
    \,,
\end{align}
which is $BF$ theory.
In other words,
\eqref{BF.actionA} is the image of \eqref{DBF.actionA}
under a 
kinematics-to-color replacement (KCR).
\nomenclature{KCR}{Kinematics-to-Color Replacement}

Here, we assumed that 
the KCR maps
a diffeomorphism coadjoint field
(with the definition in \eqref{igt.coadj})
in diffeomorphism gauge theory
to a coadjoint field
in ordinary gauge theory.
The rationale
is the invariance of the pairing in \eqref{Dpairing.coadj}.\footnote{
    If considered necessary, 
    the volume-preserving condition
    can be adopted to take $d^dx$ invariant.
}

In sum,
\eqref{DBF.action} admits a KCR into 
the ordinary $BF$ theory
after acquiring the Kronecker vacuum.
In the precise sense of CK duality,
the theory which \eqref{DBF.actionA} defines
can be called
diffeomorphism $BF$ theory:
\begin{align}
\begin{split}
	\label{KCR:BF}
    \text{KCR}:\quad\,\,\,
	\thy{BF}(\diff(\M)\hnem)
        \quad\mapsto\quad
	\thy{BF}(\g)
	\,.
\end{split}
\end{align}

The KCR is readily viable
at the EoM level.
The EoM of \eqref{DBF.actionA} impose
flatness on the diffeomorphism gauge connection,
\begin{subequations}\label{DBF.eomAB}
\begin{align}
    \label{DBF.eomA}
    F^\m{}_{\a\b}[A]
    \,=\,
        0
    \,,
\end{align}
as well as a diffeomorphism-covariant equation on the $B$-field,
\begin{align}
    \label{DBF.eomB}
    0
    \,=\,
        \partial_\a B^{\a\b}{}_\m
        + \partial_\r \bigbig{
            A^\r{}_\a\hem B^{\a\b}{}_\m
        }
        + B^{\a\b}{}_\m\mem \partial_\r A^\m{}_\a
    \,=\,
        D_\a B^{\a\b}{}_\m
    \,.
\end{align}
\end{subequations}
Here, we have used \eqref{coad-diff}
to define the
diffeomorphism covariant derivative
for coadjoint fields;
one may repeat the exercise in \Sec{K2>DGT}
for justification.
With this understanding,
\eqref{DBF.eomAB}
admits a KCR image (``single copy'') as
\begin{align}
    \label{BF.eomAB}
    F^a{}_{\a\b}[A]
    \,=\,
        0
    \,,\quad
    D_\a B^{\a\b}{}_a
    \,=\,
        0
    \,.
\end{align}
Certainly,
\eqref{BF.eomAB}
gives the EoM of 
the ordinary $BF$ theory in \eqref{BF.actionA}.

\subsection{Case Study II: Diffeomorphism Chern-Simons Theory}
\label{K2>CaseCS}

Another concrete instance
of a diffeomorphism gauge theory
is CS theory
with its color algebra replaced with the volume-preserving diffeomorphism algebra in three dimensions,
$\sdiff(\M_3)$:
\begin{align}
\begin{split}
	\label{CKR:CS}
    \text{CKR}:\quad\,\,\,
	\CS(\g)
        \quad\mapsto\quad
	\CS(\sdiff(\M_3)\hnem)
	\,.
\end{split}
\end{align}
The double copy of CS theory
was investigated in
\rcite{Ben-Shahar:2021zww}
with a construction of a Lagrangian.
A notable clarification has been given 
by a recent work \cite{Maor3dGravity}
at the Lagrangian level,
showing that
the double copy of CS theory
is on-shell equivalent to
three-dimensional Einstein gravity:
\begin{align}
    \label{CKstatement:3dGR}
    \CS(\sdiff(\M_3)\hnem)
    \,\,&\cong\,\,
        \GR_{\hem3}
    \,,
\end{align}
Here, we would like to briefly illuminate 
a key aspect of
\rcite{Maor3dGravity}'s construction from our perspective,
without reviewing every detail.

Ordinary CS theory
is defined by the action
that integrates the CS three-form 
on a three-manifold $\M_3$,
written in \eqref{CS3form}.
Crucially, the CS three-form requires the Killing form $\delta_{ab}$ on the gauge algebra $\g$.

In \Sec{K2>CaseBF},
we have seen that
diffeomorphism $BF$ theory 
admits a Lagrangian formulation,
facilitating the double copy
at both EoM and Lagrangian levels.
As is well-known,
$BF$ and CS Lagrangians are closely related;
the former provides the ``cotangent'' description of the latter.
Crucially, however,
the CS Lagrangian requires
the Killing form $\delta_{ab}$
unlike the $BF$ Lagrangian.
This raises an immediate problem
for the Lagrangian-level double copy:
one has to find
a diffeomorphism-invariant Killing form.

Interestingly, \rcite{Maor3dGravity}
shows that
a Killing form exists for
the volume-preserving diffeomorphism algebra
in three dimensions,
which is yet nonlocal.
The authors identify
an expression that can be compared to the CS three-form
with the aid of a nonlocal operator.
They then derive
the Euler-Lagrangian equations
from their proposed $\CS(\sdiff(\M_3)\hnem)$ action
and finds
$F^\m{}_{\a\b}[A] = 0$,
the flatness condition on the diffeomorphism gauge connection.
By KCR, this corresponds to $F^a{}_{\a\b}[A] = 0$,
the EoM of CS theory.

\subsection{Case Study III: Diffeomorphism Yang-Mills Theory}
\label{K2>CaseYM}

Diffeomorphism YM theory
can be defined as
a classical teleparallel field theory
around the Kronecker vacuum
whose EoM are
$\eta^{\c\a}\mem \comm{E_\c}{\comm{E_\a}{E_\b}} = 0$,
as discussed several times throughout our discussion.
Note that this theory requires the flat frame-indexed metric $\eta^{\c\a}$ as a background structure.

It seems unclear whether diffeomorphism YM theory could admit a Lagrangian formulation.
It will be interesting to see if
\rcite{Maor3dGravity}'s
Killing form can be utilized.
As a related subject,
it could also be interesting to think about the 
Pontryagin theory in four dimensions,
($F \wedge F$).

\rcite{CCK} established that
the volume-preserving diffeomorphism YM equations
provide a reformulation of BI theory.
Here, we realized that
they can be written
in terms of the $E$-variable as
$\eta^{\c\a}\mem \comm{E_\c}{\comm{E_\a}{E_\b}}^\m = 0$
with $\partial_\m E^\m{}_\a = 0$,
yet with the crucial clarification that
the Kronecker background $\delta^\m{}_\a$ has to be imposed.
Hence it can be said that
BI theory,
at the classical EoM level,
admits a formulation as a teleparallel theory.

A physical lesson that can be elicited from
this geometric rewriting
is that
expanding around
any background of the form
$\bar{E}^\m{}_\a(x) = X^\m{}_{,\n}(x)\mem \delta^\n{}_\a$
will also yield the same BI scattering amplitudes,
if $x^\m \mapsto X^\m(x)$ gives a well-behaved diffeomorphism
that is trivialized at the infinities
at a sufficiently fast speed.
This certainly changes many formulae
(e.g., the diffeomorphism field strength)
and the EoM
because the background
$\bar{E}^\m{}_\a(x)$ is position-dependent.
Verifying if the same tree amplitudes are obtained
as in the constant Kronecker vacuum
provides a nontrivial check and test of 
the teleparallel formulation of BI theory.

\subsection{Case Study IV: Frame formulation of General Relativity}
\label{K2>CaseGR}

Diffeomorphism YM theories
are genuine teleparallel theories,
and there is no local symmetry on the frame index.
In contrast,
the so-called teleparallel equivalent of GR
\cite{Maluf:2013gaa,Aldrovandi:2013wha,Ferraro:2016wht,TrinityHeisenberg:2019}
is not a genuine teleparallel theory
due to the local Lorentz symmetry.
This point applies to both action and EoM level formulations;
recall the discussion on the Mason-Newman \cite{MasonNewman:1989} idea in \eqref{MNE}.
An interesting perspective arises by 
the first-order action in \eqref{FrameAction1},
which understands GR
as a perturbation away from
diffeomorphism $BF$ theory
(suppose a rescaling $B \mapsto \k^2 B$).
Local Lorentz symmetry is not enjoyed in the topological limit
but emerges in the full theory.

Can the frame formulation of GR
still be viewed as a diffeomorphism gauge theory,
which is not necessarily YM?
Unfortunately,
neither of the actions in
\eqrefs{FrameAction1}{FrameAction2}
seems to allow for a KCR.
The anholonomy coefficients $\Omega^A{}_{BC}$
involves the coframe $e^A{}_\m$
inside which bare connection coefficients $A^\m{}_A$ are inserted.
The dynamical combination $\mathcal{G}^{\m\n}_{ABCD}[E]$ in \eqref{FrameAction1}
also suffers from this bare connection problem
around the Kronecker vacuum.
Hence, in general dimensions,
formulating GR as a genuine diffeomorphism gauge theory
still seems to be an open question.

\subsection{General Relativity as a Diffeomorphism Gauge Theory}

From the above case study,
a lesson emerges that
there exist different classes of
``teleparallel theories.''
\begin{itemize}
    \item 
        \textit{Teleparallel theory}:
        A field theory of a dynamical frame.
    \item
        \textit{Genuine teleparallel theory}:
        A field theory of a dynamical frame
        whose diffeomorphism-neutral index
        is not subject to any local symmetry.
        Governs geometries of absolute parallelism.
    \item 
        \textit{Nominal teleparallel theory}:
        A teleparallel theory that is not genuine.
    \item
        \textit{Diffeomorphism gauge theory}:
        A diffeomorphism-invariant teleparallel theory
        that is mapped to an ordinary gauge theory
        by a choice of a KCR rule
        when expanded around a choice of a vacuum.
\end{itemize}
According to this taxonomy,
the theories in
\Secsto{K2>CaseBF}{K2>CaseYM}
describe diffeomorphism gauge theories
at the EoM or Lagrangian levels,
while
the teleparallel equivalent of GR
is a nominal teleparallel theory
that seems to be also not diffeomorphism gauge theory
in general dimensions.
It shall be remarked that,
although
explorations on 
teleparallel gravity
may generally turn out to be fruitful
for envisioning a manifest double copy formulation of GR
or understanding CK duality,
emphasis must be put on the
viability of KCR
as a crucial criterion,
eventually.

\medskip
A few instances have been reported
in three and four dimensions,
however,
such that GR can be
exactly reformulated as
a diffeomorphism gauge theory
in the precise sense
(classical level, on-shell equivalence).
\begin{itemize}
    \item 
        As noted earlier in \Sec{K2>CaseCS},
        three-dimensional GR
        is shown to be on-shell equivalent to
        volume-preserving Chern-Simons theory:
        $\GR_3 \cong \CS(\sdiff(\M_3)\hnem)$.
        This refers to the result of
        \rcite{Maor3dGravity},
        which works in terms of Lagrangians.
    \item
        In four dimensions,
        the SD sector of GR
        is known to admit a formulation
        as a volume-preserving SDYM theory
        at the EoM level:
        $\SDGR \cong \SDYM(\sdiff(\M_4)\hnem)$.
        This refers to the result of
        Mason and Newman \cite{MasonNewman:1989},
        reviewed in \Sec{Jc.MN}.\footnote{
            As is reviewed in \Sec{Jc>SDG>HEAV},
            the Pleba\'nski second heavenly equation
            is also a notable formulation of SD gravity,
            in which case the kinematic algebra
            is a Poisson algebra.
            It seems that
            the Poisson gauge theory (cf. \Sec{K2>PGT})
            interpretation is yet subtle;
            see \App{COTANGENT}.
        }
        In this case
        the diffeomorphism gauge connection
        is the perturbation of
        a \textit{rescaled} tetrad $V$
        around the Kronecker vacuum.
\end{itemize}
    The three-dimensional GR construction of \rcite{Maor3dGravity}
    depends on several crucial features of three dimensions:
    first, the nonlocal Killing form construction,
    second, an interesting identity that rewrites the quadratic part of the determinant $|E|$
    via a nonlocal operator,
    third, the cubic nature of $|E|$
    around the Kronecker background,
    and fourth,
    the spin connection can be killed
    in the dreibein-Palatini formalism
    giving rise to a genuinely teleparallel description,
    owing to
    the celebrated fact that
    Einstein gravity describes flat Riemannian geometry
    on shell
    \cite{staruszkiewicz1963gravitation,deser1984three}.
It will be interesting to investigate
if these structures could be properly generalized
in $d>3$
and especially in $d=4$.\footnote{
    The determinant theory $|E|$ yields a quartic Lagrangian.
    Coincidentally or not,
    the YM and Pontryagin Lagrangians are also quartic.
}

Finally,
it is interesting to note that
the LL formulation of gravity,
reviewed in \Sec{Ja>LL},
points to suggestive formulae
in the context of the current discussion.\footnote{
    This observation is based on discussions with Clifford Cheung.
}
Suppose the de Donder gauge condition in \eqref{deDonder},
$\partial_\m \got^{\m\a} = 0$,
and the expansion
$\got^{\m\a} = \eta^{\m\a} + h^{\m\a}$
around the flat configuration.
A divergence of
the $H$-tensor in \eqref{LL:Htensor} gives
\begin{align}
\begin{split}
    \label{LL:dH}
    \partial_\m H^{\m\n\a\b}\nem
    \,&=\,
        \got^{\m\a}\mem \partial_\m \got^{\n\b}
        - \got^{\m\b}\mem \partial_\m \got^{\n\a}
    \,,\\
    \,&=\,
        \partial^\a h^{\n\b}
        - \partial^\b h^{\n\a}
        + h^{\m\a}\mem \partial_\m h^{\n\b}
        - h^{\m\b}\mem \partial_\m h^{\n\a}
    \,,
\end{split}
\end{align}
where $\partial^\a = \eta^{\m\a} \partial_\m$.
It is tempting to denote this as
$\partial_\m H^{\m\n\a\b} = F^{\n\a\b}$:
the anholonomy of the gothic metric
as a frame $\got^\a = \got^{\m\a}\mem \partial_\m$.
Then the diffeomorphism gauge-covariant derivative,
dressing the index $\n$,
is
\begin{align}
\begin{split}
    \label{LL:DF=}
    D_\a F^{\n\a\b}
    \,&=\,
        \partial_\a F^{\n\a\b}
        \mem+\mem 
        h^\r{}_\a\mem \partial_\r F^{\n\a\b}
        - F^{\r\a\b}\mem \partial_\r h^\n{}_\a
    \,,\\
    \,&=\,
        \t_\text{LL}^{\n\b}
        \mem+\mem
        h^\r{}_\a\mem \partial_\r F^{\n\a\b}
        - F^{\r\a\b}\mem \partial_\r h^\n{}_\a
    \,,
\end{split}
\end{align}
where the second line
imposes the EoM in \eqref{LL.eom},
and $h^\r{}_\a {\:=\:} h^{\r\s}\hem \eta_{\s\a}$.
Again, \eqref{LL:DF=} fails to be fully organized in the form of YM equations,
but it is suggestive that
an interpretation of the gothic metric geometry
may be possible in terms of
anholonomy or teleparallel structures.
In fact, on account of the index symmetry 
$\got^{\m\n} = \got^{\n\m}$,
it would be more relevant to
apply the double copy grammar of BAS theory
rather than the format of YM theory,
where one envisions a repeated insertion of the same kinematic algebra.

\section{ Misner String as a Diffeomorphism Vortex}
\label{K2>Misner}

So far, we have been studying diffeomorphism gauge theories
as classical field theories,
mainly at the level of EoM.
To be comprehensive,
we wish to describe
\textit{classical solutions}
of diffeomorphism gauge theories
to some explicit degree.
Generally speaking,
different classical solutions will arise
as each theory exhibits unique dynamics.
For instance, 
the on-shell configurations of
the diffeomorphism $BF$ theory 
in \Sec{K2>CaseBF}
are 
\textit{holonomic} frames,
i.e., coordinate vector fields.
In contrast,
the on-shell configurations of
the diffeomorphism YM theory 
in \Sec{K2>CaseYM}
are not necessarily holonomic:
the anholonomy dynamically evolves as the field strength.
There are propagating degrees of freedom,
and
the dynamics is far more complex.

The subject of finding and studying 
classical solutions of diffeomorphism gauge theories
seems to be an interesting program.
It will be exciting if
illuminations
can be drawn
on the classical double copy 
\cite{monteiro2014black,Luna:2015paa,Luna:2018dpt,Chacon:2021wbr,white2021twistorial,Luna:2022dxo}
program,
either
in general dimensions,
in three-dimensions,
or
for SD gravity.
The classical double copy program
investigates correspondences and maps between classical solutions
of theories in the double-copy web in \fref{web}.
In particular,
the KS classical double copy \cite{monteiro2014black}
has been mapping
stationary solutions in GR
to \textit{Maxwell} theory solutions,
which are then trivially embedded to YM theory
via a constant color factor.
Naively,
one could think of pursuing the same 
to embed Maxwell theory solutions
into a diffeomorphism YM theory,
thinking of a ``constant'' diffeomorphism generator.
However, this can be subtle
because
the diffeomorphism algebra involves derivatives.
It will be interesting
if this nontriviality
can lead to insightful illuminations
on the nature of classical double copy.

In this work,
we do not intend to
carry out
a comprehensive investigation
into classical solutions.
Yet, we provide a simple yet concrete example
that defines a classical solution
in \textit{any} diffeomorphism gauge theories:
the Misner string
\cite{Misner:1963flatter,Bonnor:1969ala,sackfield1971physical,Mazur:1986gb,GP_2009_Ch341},
reinterpreted in the teleparallel viewpoint.

As we have remarked in \Sec{K1.grav},
the Misner string
has been known as the gravitational analog
of the Dirac string,
in which case the NUT charge
is seen as the gravitomagnetic charge
\cite{demianski1966combined,Plebanski:1975xfb}.
Notably,
the teleparallel geometry interpretation
we have established in this chapter
facilitates
embedding the Misner string
to diffeomorphism gauge theories.
As a result,
we view the Misner string 
as a \textit{diffeomorphism vortex},
which exists in any diffeomorphism gauge theories
as a topological configuration.
Surely, the adequate mathematical framework
is \textit{multivalued} diffeomorphism gauge transformation,
which we have
established in detail in \Sec{K1mv}.

\subsection{Physicists' Approach}

Let us begin by reviewing the Misner string
in GR.
For simplicity, we assume $d=4$ spacetime dimensions for the moment.

\paragraph{Metric Description}

The metric of a typical straight Misner string reads
\cite{Misner:1963flatter,Bonnor:1969ala,sackfield1971physical,Mazur:1986gb,GP_2009_Ch341}
\begin{align}
    \label{Misner-straight}
    ds^2
    \,=\,
        - \bb{
            dt + 4\GN n\, \frac{x\mem dy \mminus y\mem dx}{x^2+y^2}
        }^{\nem\nem2}
        + dx^2 + dy^2 + dz^2
    \,,
\end{align}
where $n$ is the total NUT flux.
\eqref{Misner-straight} describes the image of
flat spacetime
under a \textit{multivalued diffeomorphism},
\begin{align}
    \label{mvdiff.t}
    t
    \,\,\,\mapsto\,\,\,
        t \,+\, 4\GN n\, \phi_\S
    \,.
\end{align}
Here, $\phi_\S(x,y)$ is the azimuthal angle:
a \textit{multivalued function} of $x$ and $y$
with periodicity $2\pi$,
such that $\tan\phi_\S(x,y) = y/x$.
\eqref{Misner-straight}
describes an infinite Misner string inserted along the $z$-axis.
By traveling around it,
one can meet oneself
$8\pi\GN n$ seconds ago,
$16\pi\GN n$ seconds ago,
$24\pi\GN n$ seconds ago,
and so on:
closed timelike curves.
This is the signature of \textit{multivalued time}:
\begin{align}
    \label{multi-time}
    {\Del}t
    \,=\,
    \oint
    \bb{
        dt + 4\GN n\, \frac{x\mem dy \mminus y\mem dx}{x^2+y^2}
    }
    \,=\,
        8\pi\GN n
    \,.
\end{align}
In the spacetime view, this Misner string
is inserted along the $t$-$z$ plane,
which we denote as $\S$.

The Misner string is a thin tube of 
gravitomagnetic flux,
or equivalently a thin solenoid of mass current
according to Bonnor and Sackfield
\cite{Bonnor:1969ala,sackfield1971physical}.
We have explained the key facts about the Misner string
in \Sec{K1mv},
and relevant references can be found there.
\eqref{multi-time}
is how the flux $n$ of the Misner string is measured.
Physically, this should describe an interferometry experiment
that serves as the gravitational AB effect.

\paragraph{Distributional Torsion}

Under a coordinate transformation 
from $x'^\m$ to $x^\m$,
the Christoffel symbols transform 
as
\begin{align}
    \label{ChrisTransform}
    \Gamma^\m{}_{\n\r}
    \,=\,
        \frac{\partial x^\m}{\partial x'^\k}\,
        \Gamma'^\k{}_{\l\zeta}\,
        \frac{\partial x'^\l}{\partial x^\n}\,
        \frac{\partial x'^\zeta}{\partial x^\r}
    +
        \frac{\partial x^\m}{\partial x'^\l}\,
        \frac{\partial}{\partial x^\r}\hem
        \bb{
            \frac{\partial x^\l}{\partial x^\n}
        }
    \,.
\end{align}
Note that
the ordering of the two derivatives in
the inhomogeneous term
is uniquely determined by
demanding that
$\nabla_\r V^\m = \partial_\r V^\m + \Gamma^\m{}_{\n\r}\mem V^\n$
transforms tensorially
with the aid of a test vector field $V^\m$,
for instance.
The torsion
$T^\m{}_{\n\r} = \Gamma^\m{}_{\r\n} - \Gamma^\m{}_{\n\r}$
transforms as
\begin{align}
    \label{TorsionTransform}
    T^\m{}_{\n\r}
    \,=\,
        \frac{\partial x^\m}{\partial x'^\k}\,
        T'^\k{}_{\l\s}\,
        \frac{\partial x'^\l}{\partial x^\n}\mem
        \frac{\partial x'^\s}{\partial x^\r}
    +
        \frac{\partial x^\m}{\partial x'^\l}\mem
        \bbsq{
            \frac{\partial}{\partial x^\n}
            ,
            \frac{\partial}{\partial x^\r}
        }\hem
        x'^\l
    \,.
\end{align}
The inhomogeneous part of \eqref{TorsionTransform}
vanishes for single-valued coordinate transformations.
However, for our case,
the multivalued coordinate transformation
$t' = t + 4\GN n\mem \phi_\S(x,y)$ gives
\begin{align}
    \label{Ttxy}
    T^t{}_{xy}
    \,=\,
        4\GN\mem
        \bbsq{
            \frac{\partial}{\partial x}
            ,\mem
            \frac{\partial}{\partial y}
        }\mem
            \phi_\S(x,y)
    \,=\,
        8\pi\GN n\mem \delta(x)\hem \delta(y)
    \,.
\end{align}
It can be seen that
this gives the only nonvanishing components of the torsion tensor.
We conclude that
the Misner string spacetime in \eqref{Misner-straight}
describes a distributional torsion
supported along the surface $\S$.
This torsion flux tube is a diffeomorphism vortex, to say,
inducing the time monodromy
in \eqref{multi-time}.

Similarly,
one can also examine if Riemannian curvature
gets generated by a multivalued coordinate transformation:
\begin{align}
    \label{RiemannTransform}
    R^\m{}_{\n\r\s}
    \,=\,
        \frac{\partial x^\m}{\partial x'^\k}\,
        R'^\k{}_{\l\zeta\xi}\,
        \frac{\partial x'^\l}{\partial x^\n}\mem
        \frac{\partial x'^\zeta}{\partial x^\r}\mem
        \frac{\partial x'^\xi}{\partial x^\s}
    +
        \frac{\partial x^\m}{\partial x'^\l}\,
        \bbsq{
            \frac{\partial}{\partial x^\r}
            ,
            \frac{\partial}{\partial x^\s}
        }\hem
            \frac{\partial x'^\l}{\partial x^\n}
    \,.
\end{align}
For our case,
the Jacobian is single-valued,
so the Misner string does not involve a Riemannian curvature
even in the distributional sense.

See also \frefs{fig:cosmic}{fig:misner}.
An insightful and inspiring reference relevant to the current discussion,
especially around \eqrefs{TorsionTransform}{RiemannTransform},
is \rcite{Kleinert:2008zzb}.

\paragraph{Coframe Description}

We can now switch to the tetrad formalism.
The coframe 
for the line element in \eqref{Misner-straight}
is
\begin{align}
    \label{edirac.A}
    e^0
    \mem=\mem
        dt \mem+\mem 4\GN n\, \frac{x\mem dy \mminus y\mem dx}{x^2+y^2}
    \,,\quad
    e^1 \mem=\mem dx
    \,,\quad
    e^2 \mem=\mem dy
    \,,\quad
    e^3 \mem=\mem dz
    \,,
\end{align}
which is certainly
the image of
the Kronecker coframe
under the multivalued coordinate transformation
$t' = t + 4\GN n\mem \phi_\S(x,y)$.

By formally treating \eqref{edirac.A} as 
a one-form field on the space $\R^4$
of $(t,x,y,z)$,
a physicist computes its curl as
\begin{align}
    \label{edirac.F}
    de^0
    \,=\,
        4\GN n\, dd\phi_\S
    \,=\,
        8\pi\GN n\,
            \delta(x)\hem \delta(y)
        \, dx \wedge dy
    \,.
\end{align}
Interestingly,
a suggestive parallel arises between
\eqrefs{adirac.A}{edirac.A}
as well as
\eqrefs{adirac.F}{edirac.F}.
The Misner string is indeed the gravitational analog
of the Dirac string.
\eqref{edirac.F},
which captures the torsion computed in \eqref{Ttxy},
will be viewed as
describing a ``gravitomagnetic field''
or ``gravitational field strength''
in the same sense as
$F$ in \eqref{adirac.F}
describes a magnetic field or field strength.

\paragraph{Frame Description}
To be more precise,
we have learned in \eqref{K2:tele}
the anholonomy $F_{\a\b} = \comm{E_\a}{E_\b}$
in teleparallel gravity
represents a field strength of diffeomorphisms
in a certain accurate sense.
Thus,
we consider the frame field 
that is dual to
the coframe field in \eqref{edirac.A}:
\begin{align}
\begin{split}
    \label{Edirac}
    E_0
    \mem=\mem
        \frac{\partial}{\partial t}
    &\,,\quad
    E_1
    \mem=\mem
        \frac{\partial}{\partial x}
        \mem+\mem 4\GN n\,
            \frac{y}{x^2 \mplus y^2}
        \mem
            \frac{\partial}{\partial t}
    \,,\\
    E_3
    \mem=\mem
        \frac{\partial}{\partial z}
    &\,,\quad
    E_2
    \mem=\mem
        \frac{\partial}{\partial y}
        \mem-\mem 4\GN n\,
            \frac{x}{x^2 \mplus y^2}
        \mem
            \frac{\partial}{\partial t}
    \,.
\end{split}
\end{align}
Around the Kronecker vacuum, \eqref{Edirac} describes
\begin{align}
\begin{split}
    \label{Edirac.A}
    A_0 \mem=\mem 0
    &\,,\quad
    A_1 \mem=\mem 
         +4\GN n\,
                \frac{y}{x^2 \mplus y^2}
            \mem
                \frac{\partial}{\partial t}
    \,,\\
    A_3 \mem=\mem 0
    &\,,\quad
    A_2 \mem=\mem 
        -4\GN n\,
            \frac{x}{x^2 \mplus y^2}
        \mem
            \frac{\partial}{\partial t}
    \,.
\end{split}
\end{align}
The nonvanishing component of the diffeomorphism field strength
arises by
\begin{align}
    \label{Edirac.Fcomp}
    F^t{}_{12}
    \,=\,
        \frac{\partial}{\partial x}\mem
            A^t{}_2
        -
        \frac{\partial}{\partial y}\mem
            A^t{}_1
    \,=\,
        - 8\pi\GN n\,
            \delta(x)\hem \delta(y)
    \,,
\end{align}
while the nonabelian part,
$\comm{A_1}{A_2}$,
vanishes.
\eqref{Edirac.Fcomp} is a direct characterization of the time monodromy,
representing flux measurement by
a tiny Amp\`erian square loop on the $x$-$y$ plane
around the origin, for instance.

\newpage

To clarify,
we now liberate the Misner string
from the GR context
and view \eqrefs{Edirac}{Edirac.A}
as describing a classical solution in 
a teleparallel or diffeomorphism gauge theory;\footnote{
    In this context,
    the Misner string geometry
    is defined by
    the frame in \eqref{Edirac}
    or perhaps even just by the multivalued diffeomorphism in \eqref{mvdiff.t},
    and
    the metric in \eqref{Misner-straight}
    is unnecessary.
}
accordingly, we put $8\pi\GN = 1$.
Since this is a topological configuration
generated by a multivalued \textit{gauge} transformation,
it is (locally) on-shell in any theory
regardless of the dynamics specified.

Consider the abelian Dirac string in \eqref{adirac.A}.
Its trivial embedding in YM theory
is given in \eqref{nadirac.A},
which uses a constant color factor $\l^a$.
This constant color factor trivializes the nonabelian part of the field strength $F^a$,
which happens to be
a common (and even characteristic) situation in classical double copy \cite{monteiro2014black}.
Interestingly, the very same kind of trivialization
happens in 
\eqref{Edirac.Fcomp},
which arises by 
applying the replacement 
$\l^a \mapsto -n\mem \delta^\m{}_t$
on \eqref{nadirac.F}.
Although 
the meaning of
``constant diffeomorphism generator''
would be unclear
in general,
it appears that
this effective constancy 
is achieved
by a collaboration between
the direction of the reference $\delta^\m{}_t$
and the orientation of the surface support
inducing monodromy for contours on the $x$-$y$ plane.

With these understandings,
we may identify
the Misner string
in \eqref{Edirac.A},
as a classical solution in any diffeomorphism gauge theory,
as the classical double copy
of the colored Dirac string
in \eqref{nadirac.A}.
In particular,
this identifies that
the colored Dirac string
in YM theory
classically double copies to
the Misner string
in the Cheung and Mangan \cite{CCK}'s formulation of BI theory.

\subsection{Mathematician's Approach}

Recalling \Sec{K1mv.math},
we also briefly remark on
a description
of the above Misner string geometry
that avoids using the Dirac delta.
In this case,
the surface $\S$
will be excised from the spacetime $\M$,
so one studies the manifold $\M{\setminus}\S$.

To begin with,
the multivalued diffeomorphism
may be formalized
in the basic framework of transition functions.
Recall that,
in manifold theory,
a transition function
relates the coordinates of one chart
to the coordinates of another.
Let $(U_i,\varphi_i)$ and $(U_j,\varphi_j)$ be 
two overlapping charts,
where $\varphi_i: U_i \to \R^d$
and $\varphi_j: U_j \to \R^d$.
The transition function
describes
the map
$\t_{ij} = \varphi_i \circ \varphi_j^{-1}:
\varphi_j(U_i \cap\hem U_j) \to \varphi_i(U_i \cap\hem U_j)
$.

Let $\M$ be a $d$-dimensional manifold and
$\S$ be a codimension two submanifold in $\M$.
Suppose an open cover $\{\CU_i\}$ of $\M{\setminus}\S$.
For each patch $\CU_i$,
we define coordinates by
$\varphi_i: U_i \to \R^d$.
This essentially means to choose
a nice set of $d$ elements from $\Cinfty(U_i,\R^d)$.
On each overlapping region $\CU_i \cap \CU_j$,
we demand that $\varphi_i$ and $\varphi_j$
are related by a transition function $\t_{ij}$
as follows:
\begin{align}
    \label{OO=z.diff}
    \varphi_i \circ \varphi_j^{-1}
    \,=\,
        \t_{ij}
    \,.
\end{align}
Note that
$\t_{ij}$ maps a region of $\R^d$ to a region of $\R^d$.
We choose the transition functions such that,
on each triple overlap $\CU_i \cap \CU_j \cap \CU_k$,
\begin{align}
    \t_{ij} \circ \t_{jk} \circ \t_{ki}
    \,=\,
        \id
    \,.
\end{align}
The topological nontriviality of the manifold
$\M{\setminus}\S$,
seen as the atlas $\{(U_i,\varphi_i)\}$,
will arise
by stipulating that
the collection
$\{ \t_{ij} \}$ 
of our transition functions
represents
a nontrivial {\Cech} cohomology element,\footnote{
    Here we have posited 
    the maximally general
    $\Diff(\R^d)$
    while sweeping subtleties under the rug.
    One expects that
    an appropriate restriction to a subgroup
    would be adequate.
    For instance,
    our example in \eqref{mvdiff.t}
    will concern the group of constant translations.
    In the teleparallel theory context,
    the notion of absolute parallelism could
    facilitate a nice diffeomorphism-invariant
    incarnation of ``constant translations,''
    ensuring consistency
    in identifying $\a$.
}
\begin{align}
    \label{cechalpha.diff}
    [\{\t_{ij}\}]
    \,=\,
        \cen
    \,\in\,
        \cH^1(\M{\setminus}\S,\Diff(\R^d))
    \cong
        \Diff(\R^d)
    \,.
\end{align}
This characterizes the multivaluedness of the diffeomorphism entailed in the Misner string
as an obstruction to a single-valued lift.
Namely,
each time one wraps around $\S$,
one's coordinates are transformed by $\a$.
Note also the parallel between
\eqrefsto{OO=z}{cechalpha}
and
\eqrefsto{OO=z.diff}{cechalpha.diff}.

We should talk in terms of an explicit example.
Concretely,
suppose $\M = \R^4$ be the space of $(t,x,y,z)$
and take $\S$ to be the $t$-$z$ plane.
Install four patches on $\M{\setminus}\S$
by taking the product of
$\R = \{ t \mem|\mem t \inn \R \}$ and 
the four patches in \eqref{fourpatches}.
Choose the transition functions as
\begin{align}
\begin{split}
\label{clocks4}
    \t_{12}:
    &\quad
    (t,x,y,z)
    \,\,\,\mapsto\,\,\,
    (t-3{\Del}t,x,y,z)
    \,,\\
    \t_{23}:
    &\quad
    (t,x,y,z)
    \,\,\,\mapsto\,\,\,
    (t+7{\Del}t,x,y,z)
    \,,\\
    \t_{34}:
    &\quad
    (t,x,y,z)
    \,\,\,\mapsto\,\,\,
    (t-2{\Del}t,x,y,z)
    \,,\\
    \t_{41}:
    &\quad
    (t,x,y,z)
    \,\,\,\mapsto\,\,\,
    (t-{\Del}t,x,y,z)
    \,,
\end{split}
\end{align}
where ${\Del}t = 8\pi\GN n$.
Physically, \eqref{clocks4}
means that
four \textit{time zones}
are installed on a spacetime.
When hopping from zone 3 to zone 2,
the reading of the clock jumps by $+7{\Del}t$:
$t_2 = t_3 + 7{\Del}t$.
When hopping from zone 2 to zone 1,
the reading of the clock jumps by $-3{\Del}t$:
$t_1 = t_2 - 3{\Del}t$.
In this way, one travels through the four time zones
and find that
one's clock
is shifted by the total of
$+7{\Del}t - 3{\Del}t - {\Del}t - 2{\Del}t = {\Del}t$.
This is exactly the content of
the multivalued diffeomorphism
in \eqref{mvdiff.t}.
Also, it should be clear that
different choices for the clock jumps
can be made
while keeping the total jump ${\Del}t$
the same.
This is what is implied by the equivalence class in \eqref{cechalpha.diff}.
The choice in \eqref{clocks4}
corresponds to
the last example in \eqref{cech.nlist},
$(-3,7,-2,-1)$,
which is listed with other choices such as
$(0,0,0,1)$ and $(0,-1,0,2)$.

This provides a mathematical formalization for 
a multivalued diffeomorphism.
The Misner string
as a frame configuration
will arise by
preparing a single-valued set of vector fields on $\M{\setminus}\S$
and then 
applying the pushforward of this multivalued diffeomorphism.
\subsection{Measurement of Torsion Flux}
\label{INTF1}

We end with a brief physical discussion on
nonlocal functionals
in the teleparallel theory
that can characterize the torsion flux of Misner string.

\paragraph{Translation Holonomy}

Consider a Misner string in a generic $d$-dimensional spacetime $\M$,
supported on a codimension two submanifold $\S$.
As a generalization of \eqref{edirac.A},
this Misner string
will be described by the coframe
\begin{align}
    \label{e-n}
    e^\a
    \,=\,
        \delta^\a{}_\m\mem dx^\m
        \mem+\mem
        \frac{n^\a}{2\pi}\,
            d\phi_\S
    \,,
\end{align}
where $n^\a$ is a constant factor
characterizing the total flux of torsion.

We may take insights from GR,
where
it has been pointed out that
is that the action of a scalar particle
gives a version of gravitational AB phase.
See, e.g.,
\rrcite{LL:1975vol2,sakurai1980comments,dowker1974nut,dowker1967gravitational,Ramaswamy:1981JMP,zee1985gravitomagnetic,samuel1986gravitational,zimmerman1989geodesics,maartens1998gravito,Alfonsi:2020lub}.
The time monodromy in \eqref{multi-time}
also admits an interpretation as
Sagnac interferometry
\cite{Ashtekar:1975Sagnac,Anandan:1981Sagnac,sakurai1980comments,Sagnac:1913ether,Sagnac:1913preuve}.
The first-order action of a scalar particle
is given by
a line integral of the one-form $m_\a\hem e^\a{}_\m\mem dx^\m$
plus a Lagrange multiplier term imposing the mass-shell constraint on the momentum $m_\a$.
Inspired by this fact,
one integrates the one-form
$m_\a\hem e^\a{}_\m\mem dx^\m$
over a loop $\CC$
in the background in \eqref{e-n} and finds
\begin{align}
    \label{mnlink}
    \oint_\CC\mem m_\a\hem e^\a
    \,=\,
        m_\a\hem n^\a
        \,
        \link(\CC,\S)
    \,.
\end{align}
Although
this calculation
describes the basis of the
argument for NUT charge quantization
in, e.g.,
\rrcite{Misner:1963flatter,dowker1974nut,zee1985gravitomagnetic,samuel1986gravitational,bunster2006monopoles,Alfonsi:2020lub},
a caveat is that
$m_\a$ is treated as being constant
and nondynamical.

A related idea that has also existed for a long time
is ``translation holonomy,''
which naturally arises in the Poincar\'e gauge theory
\cite{Utiyama:1956sy,Kibble:1961ba,Sciama:1962,Hehl:1976kj,Hayashi:1967se,hayashi1977gauge,hayashi1981addendum,cho1976einstein,cho1976gauge,%
Grignani:1991nj,Blagojevic:2003cg,%
Percacci:1984bq}
context.
See, e.g.,
\rrcite{Anandan:1993ri,Anandan:1992rf,Mielke:2017nwt}.
In this case one considers the combination
$\mathbf{P}_\a\mem e^\a + \tfrac{1}{2}\mem \mathbf{S}_{\a\b}\mem \gamma^{\a\b}$,
where $\mathbf{P}_\a$ and $\mathbf{S}_{\a\b}$ are the generators of the Poincar\'e group,
and constructs a ``Wilson loop'' valued in the Poincar\'e group.
When the translation mode is only activated,
one obtains $\mathbf{P}_\a \oint_\CC e^\a$.
This essentially amounts to \eqref{mnlink}
if one replaces $\mathbf{P}_\a$
with the constant frame momentum $m_\a$.

Although insightful,
these interpretations
can be subtle in GR
because the frame index is gauged.
Also, the translation algebra is abelian
while gravitational interactions are nonlinear.
Crucially, it seems unclear
what would be the trace
for the translation group,
which may be needed to achieve a Wilson \textit{loop}.
In genuine teleparallel theories,
however,
we may safely take
$\oint_\CC e^\a$
as a valid nonlocal operator
since the frame index $\a$ is global.

It is interesting to note that
$\oint_\CC e^\a$ resembles the Burgers vector
in lattice systems,
which characterizes disinclination defects
\cite{chaikin1995principles,Grozdanov:2018ewh,Pace:2023kyi,Fiziev:1995te,Kleinert:1996yi,Kleinert:2008zzb,Lin:2022rvx}.
In this case the index $\a$ describes directions in the so-called \textit{lattice frame}
and thus is considered global.

\paragraph{Diffeomorphism Wilson Line}

In the meantime,
the very object
that would directly characterize the multivalued diffeomorphism
is arguably
the diffeomorphism Wilson line
defined in \Sec{DYM>W}.
Concretely,
take the example configuration in \eqref{Edirac.A}.
With the Magnus representation in \eqref{DYM.Omega},
the logarithm of
the diffeomorphism Wilson line $W_\CC = \exp\bigbig{
    \Omega^\m_\CC(x)\mem \partial_\m
}$
for a closed contour $\CC$
that wraps around the Misner string once
is\footnote{
    Note that the plus sign, $+n\mem \delta^\m{}_t$,
    will be obtained in the convention of \Chap{K1}.
}
\begin{align}
\begin{split}
    \Omega_\CC^\m(x)
    \,&=\,
        -\nem \int
            A^\m{}_1(x\mplus y(s)\hnem)\mem dy^1(s)
            +
            A^\m{}_2(x\mplus y(s)\hnem)\mem dy^2(s)
    \,,\\
    \,&=\,
        -n\mem \delta^\m{}_t
    \,,
\end{split}
\end{align}
regardless of one's choice of the base point $x$.
Certainly,
this gives the direct computation of the diffeomorphism group element $\a$ in \eqref{cechalpha.diff}.

This proposal,
however,
may be considered less robust than
the translation holonomy.
To have a genuine gauge-invariant
Wilson \textit{loop},
one will have to take a trace.
It is unclear
what would be the trace for the diffeomorphism group.

In sum,
we have established the Misner string
as a classical solution in 
any teleparallel or diffeomorphism gauge theory,
which arises by applying a multivalued diffeomorphism.
In a certain sense, this Misner string
serves as the classical double copy of the Dirac string.

Note that it could be natural to ask if
this exploration on the Misner string
gives rise to a one-form symmetry
in (genuine) teleparallel theories,
as per the symmetry triangle in \fref{SymTriangle}.
Namely, a one-form symmetry
that arises by multivalued diffeomorphisms.\footnote{
    Investigations into this direction
    are beyond the scope of the present work,
    but several speculative remarks could be made.
    First, recall that
    the explicit breaking of
    the one-form symmetry due to
    multivalued local Lorentz symmetry
    has mandated the existence of fermions
    as spinning particles
    (\Chap{K1}).
    Similarly,
    the explicit breaking of
    the one-form symmetry due to
    multivalued diffeomorphisms,
    if exists,
    may mandate the existence of massive particles.
    In this case we encounter a reincarnation of the Poincar\'e charges (mass, spin)
    in GR
    via the higher-form (topological surface operator) framework.
    Second, the Misner string is famously endable on NUT charges.
    Hence, if a magnetic picture exists
    for this hypothetical one-form symmetry,
    then the explicit breaking of it
    might mandate the existence of a NUT-charged object
    in our universe.
    Indeed, we will see
    a concrete configuration representing a finite Misner string
    ending on opposite NUT charges
    in \Chap{K3:NJA}.
    Third, 
    it will again be interesting to note similarities with lattice systems (cf. \rcite{Armas:2023tyx}).
    Lastly,
    it will also be natural to consider multivalued 
    volume-preserving diffeomorphisms
    for SD gravity
    in the frameworks of \Secsor{Jc.MN}{Jcx>FRAME}.
    Note that this seems to be what is
    envisioned or suggested in \rcite{Freidel:2026iia},
    although 
    the symmetry transformations
    were not explicitly discussed.
}
We had briefly remarked about this possibility in \Chap{K1},
and
if it is indeed viable,
it will describe an interesting
interaction between the two ideas,
generalized symmetries and CK duality.
In this context,
a question one can ask is
``Do symmetries double copy?''

\section{ Poisson Gauge Theory}
\label{K2>PGT}

We end with a brief remark on Poisson gauge theories,
which arises by
replacing the gauge algebra
of an ordinary gauge theory
with a Poisson algebra.
This is implemented in, e.g., \rcite{Cheung:2022mix}.

In a Poisson gauge theory,
the gauge potential $A_\a$ carries just one index.
The field strength is given by
$F_{\a\b} = 
    \partial_\a A_\b
    - \partial_\b A_\a
    + \pb{A_\a}{A_\b}
$,
where the nonlinear term is the Poisson bracket
between $A_\a$ and $A_\b$.
Infinitesimal Poisson gauge transformations
act on the Poisson gauge potential as
$A_\a \mapsto A_\a - D_\a\xi$,
where $D_\a\xi = \partial_\a\xi + \pb{A_\a}{\xi}$.
On the Poisson field strength,
we have
$F_{\a\b} \mapsto F_{\a\b} + \pb{\xi}{F_{\a\b}}$.
It is left as an exercise to define Poisson gauge theory
via CKR
and explicitly establish
proper gauge transformation behaviors
as in \Sec{K2>DGT},
while also constructing Poisson Wilson line\footnote{
    Note that, in noncommutative gauge theories,
    a gauge-invariant Wilson \textit{loop} is viable;
    see, e.g., \rrcite{Ishibashi:1999hs,Das:2000md}.
}
and its Magnus formula.

Although 
similar in spirit of CK duality,
the status of Poisson gauge theory seems
more established than diffeomorphism gauge theory;
see, e.g., \rcite{Jurco:2000fs,Jurco:2001my,Kurkov:2023mrz,Kupriyanov:2021aet,Abla:2022wfz}.
Part of this status 
may owe to the fact that
Poisson gauge theory can be obtained as
the semi-commutative limit of
noncommutative gauge theory,
which is a well-studied subject \cite{SW}.
Note also that diffeomorphism gauge theories
on $\M$
are automatically Poisson gauge theories on $\Tstar\M$
by implementing the diffeomorphism generator as the cotangent ``momentum''
\cite{Kupriyanov:2021aet,dpb}.

The semi-commutative limit of
the Seiberg-Witten map \cite{SW}
provides a field redefinition
from an ordinary abelian gauge connection $a_\a$
to a Poisson gauge connection $A_\a$,
which is based on the Moser lemma 
\cite{moser1965volume,Jurco:2000fs,Jurco:2001my}.
It will be interesting to investigate if
the semi-commutative limit of Seiberg-Witten map
(or its adequate volume-preserving diffeomorphism generalization, if exists)
can be useful
in explicitly constructing
the field redefinition from
the BI photon field
to the manifest CK dual field basis,
explaining the equivalence $\BI \cong \YM(\sdiff(\M)\hnem)$
in \eqref{CKstatement:BI}
in two (or $d>2$) dimensions.
One lesson that can be gained from this investigation
will be the fate of gauge symmetries
in the problem of double copy.
A key philosophy and feature of the Seiberg-Witten map
is that
gauge orbits match
between the two sides.
One of the perennial questions of double copy
has been that
whether
one has to
pursue 
an explicitly gauge-fixed description
to manifest double copy
or
aim to formulate everything in terms of gauge-covariant or gauge-invariant objects
(cf. the desire for field basis independence during the NLSM discourse in \Sec{K2>REVIEW>NLSM}).

\begin{subappendices}

\section{ From Equations of Motion to Amplitudes}
\label{app:BASBG}

To see why formulating a theory 
as a BAS theory
at the EoM level
manifests CK duality
or double copy,
consider 
the derivation of
the scattering amplitudes of BAS theory
via BG recursion
\cite{BerendsGiele}
(perturbiner method \cite{Rosly:1996vr,Selivanov:1997an,Mizera:2018jbh}),
which performs LSZ reduction on a one-point function
evaluated on a background
\cite{CCK}.
This represents BAS amplitudes
as a sum over cubic rooted tree graphs $\Gamma$
with momentum labels:
\begin{align}
    \label{BASamp}
    \A
    \,=\,
    \sum_{\Gamma}\,
        \frac{
            C(\Gamma)\mem \tC(\Gamma)
        }{D(\Gamma)}
    \,.
\end{align}
Here, $C(\Gamma)$, $\tC(\Gamma)$, and $D(\Gamma)$
are graph functions.
$D(\Gamma)$ encodes the propagators,
while $C(\Gamma)$ and $\tC(\Gamma)$
are color factors associated with $\g$ and $\tg$,
respectively.
By the very structure of BAS equations,
$C(\Gamma)$ and $\tC(\Gamma)$
describe the same tensor structure.
Hence,
for each triplet of cubic graphs
$(\Gamma_i,\Gamma_j,\Gamma_k)$
whose color factors 
are related by the Jacobi identity
$C(\Gamma_i) + C(\Gamma_j) + C(\Gamma_k) = 0$,
the same relation holds for the tilded color factors as
$\tilde C(\Gamma_i) + \tilde C(\Gamma_j) + \tilde C(\Gamma_k) = 0$
and vice versa.
As a result,
the BCJ definition of CK duality \cite{BCJ1,BCJ2,BCJReview}
is satisfied
if a choice for the Lie algebra $\tg$
exists such that
$\tilde{C}(\Gamma)$ can define
a tree graph function of
the external momenta
(and also polarizations for spinning theories).
For NLSM, one takes $\tg = \sdiff(\M)$.
Subtleties may lie in 
converting the external scattering states
back and forth
between $\NLSM(\g)$ and $\BAS(\g,\sdiff(\M)\hnem)$,
but
it can be shown that
the invertibility on free on-shell configurations
in \eqref{j-pi.lin}
ensures obtaining the same amplitude.
The embedding $j^a{}_\m = \partial_\m \pi^a$
is relevant for
the leaf-leg conversion of $\pi$-particles to $j$-particles;
the inverse mapping $\pi^a = (q\mdot\partial)^{-1} j^a{}_\m\mem q^\m$
is relevant for
the root-leg conversion of a $j$-particle to a $\pi$-particle.
Details are to be found in \rcite{CCK}.

\section{ Lagrangian Formulations}
\label{app:BZ-NLSM}

BAS theory admits
a Lagrangian formulation by
the Lagrangian $d$-form
\begin{align}
    \label{BAS.L}
    L[\Phi]
    \,=\,
        \frac{1}{\k^2}\,
        \delta_{ad}\mem \delta_{\ta\td}\mem
        \bb{\nem{
        - \frac{1}{2}\,
            \Phi^{a\ta}\mem
            \Box\mem
            \Phi^{d\td}
        + \frac{1}{6}\,
            f^a{}_{bc}\mem 
            \tf^\ta{}_{\tb\tc}\,
                \Phi^{b\tb}\mem \Phi^{c\tc}
                \mem
                \Phi^{d\td}
        }}
    \mem d^dx
    \,,
\end{align}
if $\k$ denotes the coupling.

Although not necessary,
it is natural to ask if
the chiral current formulation of NLSM,
$\BAS(\g,\sdiff(\M)\hnem)$,
admits a Lagrangian formulation,
and whether manifest CK duality is exhibited in that description.\footnote{
    This exploration is based on
    discussions with Clifford Cheung and Nabha Shah.
}
This will primarily refer to the
Freedman-Townsend \cite{ft1981}
formulation of NLSM.
The idea is to
promote the chiral current 
in \eqref{NLSM0.L}
to a fundamental field variable
by imposing the flatness condition as a constraint:
\begin{align}
    \label{NLSMft.L}
    L[j,B]
    \,=\,
        \Cont{B_a}{
            dj^a + \frac{1}{2}\mem
                f^a{}_{bc}\mem j^b \swedge j^c
        }
        -\frac{1}{2\k^2}\,
            \delta_{ab}\mem
            \cont{j^a}{j^b}
    \,.
\end{align}
Here, $\cont{\a}{\b} := \a \wedge\hnem {*}\b$.
The coadjoint two-form $B_a$ is a Lagrange multiplier.

The Freedman-Townsend formulation,
however,
does not seem to establish CK duality
at the Lagrangian level:
\eqref{NLSMft.L}
does not arise by implementing a CKR on \eqref{BAS.L}.
To this end,
one might be offered with a suggestion that
the volume-preserving condition of the $j$-field
may be imposed directly
by an explicit parametrization,
\begin{align}
    \label{j[Z]}
    j^a
    \,=\,
        d^\dagger Z^a
    \qfq
    j^{a\m}
    \,=\,
        \partial_\r Z^{a\m\r}
    \,,
\end{align}
for an adjoint-valued two-form field $Z^a$.
Here, $d^\dagger$ is the Hodge codifferential operator.
By analogy with fluid dynamics,
a proper name for this $Z$-field seems to be
\textit{streamfunction}.\footnote{
    The $j$-field could be interpreted as the velocity field for an incompressible fluid
    (with flavors \cite{Cheung:2020NSE}).
    The streamfunction is a useful
    tool in describing two-dimensional incompressible flows,
    whose contours
    represent streamlines.
    The $Z$-field here is essentially
    a generalization of this concept
    in $d$ dimensions.
}
By plugging in \eqref{j[Z]} to \eqref{NLSMft.L}
while performing a gauge-fixing $dZ^a = 0$,
one finds\footnote{
    \eqref{NLSMBZ.L} is classically equivalent to
    \eqref{NLSMft.L}.
    In fact, a careful consideration of the one-loop determinant shows that
    they are also one-loop equivalent.
}
\begin{align}
\begin{split}
    \label{NLSMBZ.L}
    L[Z,B]
    \,=\,
    {}&{}
        \cont{B_a}{
            \Box\hem Z^a
        }
        - \frac{1}{2}\,
        \cont{B_a}{
            f^a{}_{bc}\mem
            \comm{Z^b}{Z^c}
        }
        - \frac{1}{2\k^2}\,
            \delta_{ab}\mem
            \cont{Z^a}{\Box\hem Z^b}
    \,,
\end{split}
\end{align}
where $\Box$ describes the Hodge-de Rham Laplacian
$d^\dagger d + d\mem d^\dagger$.
We have also defined,
for any two-forms $Z_1,Z_2 \in \Omega^2(\M)$,
\begin{align}
    \label{Zbr}
    \comm{Z_1}{Z_2}
    \,:=\,
        -d^\dagger Z_1 \wedge d^\dagger Z_2
    \,.
\end{align}
Explicitly,
$
    \comm{Z_1}{Z_2}^{\m\n}
    = 
        - 
            (\partial_\r Z_1^{\m\r})
            (\partial_\s Z_2^{\n\s})
        +
            (\partial_\r Z_2^{\m\r})
            (\partial_\s Z_1^{\n\s})
$.

As the notation has suggested,
\eqref{Zbr} defines a Lie bracket on the space $\Omega^2(\M)$ of two-forms on $\M$.
It is not difficult to verify this directly.
In fact,
$\Omega^2(\M)$ equipped with this Lie bracket
realizes a two-form version of
the volume-preserving diffeomorphism algebra $\sdiff(\M)$
(yet up to the equivalence
$Z^a \sim Z^a + d^\dagger\lambda^a$).

\eqref{NLSMBZ.L}
is an interesting action,
formulated solely in terms of \textit{two-forms}.\footnote{
    Cf. Pleba\'nski gravity in \Sec{Jcx>COFRAME}.
}  
It also beautifully exhibits the invariant tensors of the Lie algebra,
$\delta^a{}_b$, $f^a{}_{bc}$, and $\delta_{ab}$.
Diagrammatically,
the $B$-field and the $Z$-field
(as species)
serve as ``up and down indices''
in the Feynman rules.

It is tempting to
view \eqref{NLSMBZ.L}
as arising from
performing the CKR on the following theory
via replacing $\tg$ with 
the two-form version of $\sdiff(\M)$:
\begin{align}
    \label{BAS.L1}
    &
    L[\Phi,\Psi]
    \\
    &
    =\,
        \cont{
            \Psi_{a\ta}
        }{
            \Box\hem \Phi^{a\ta}
        }
        - \frac{1}{2}\,
        \cont{
            \Psi_{a\ta}
        }{
            f^a{}_{bc}\mem 
            \tf^\ta{}_{\tb\tc}\,
            \Phi^{b\tb}\mem \Phi^{c\tc}
        }
        - \frac{1}{2\k^2}\,
            \delta_{ab}\mem \delta_{\ta\tb}\,
            \cont{
                \Phi^{a\ta}
            }{
                \Box\hem \Phi^{b\tb}
            }
    \,.
    \nonumber
\end{align}
It can be said that
\eqref{BAS.L1} embeds BAS theory
as a subsector
(cf. the EoM for $\delta\Psi_{a\ta}$),
but it is unclear 
whether such a two-field setup
could be genuinely 
relevant for BCJ duality.

\end{subappendices}

\chapter{Simplicity of Self-Dual Black Hole}
\label{K3:HYDROGEN}

In his classic monograph \cite{chandrasekhar1983mathematical},
Chandrasekhar described black holes as
``the most perfect macroscopic objects'' in the universe.
With the first direct detections of gravitational waves
\cite{LIGOScientific:2016aoc,LIGOScientific:2017vwq}
and the iconic horizon-scale image produced by the Event Horizon Telescope
\cite{EventHorizonTelescope:2019dse},
these theoretical objects have now turned into
precision experimental probes of strong-field gravity.
It is in this sense that the black hole may be viewed as
``the atom of the twenty-first century'' \cite{Dijkgraaf:2019bhatom}:
simple enough to be universal, yet rich enough to reveal profound structures.

In the meantime,
recent theoretical studies
\cite{Adamo:2023fbj,Adamo:2024xpc,Adamo:2025fqt,Crawley:2021auj,Crawley:2023brz,Guevara:2023wlr,Guevara:2024edh,Guevara:2025psg,note-sdtn,nja,AAS,Skvortsov:2025ohi,Doran:2026bng}
have provided deeper investigations into
\textit{SD black holes},
realizing their exceptional simplicity.
This line of study
may be traced back to
the classic works
\cite{hawking1977gravitational,Gibbons:1978tef,Gibbons:1979xm,Gibbons:1986hz,Gibbons:1987sp}
of 
Gibbons, Hawking, and Ruback
on gravitational instantons
and hidden symmetries.
The SD Taub-NUT solution,
which is an instance of a Gibbons-Hawking
gravitational instanton
\cite{hawking1977gravitational,Gibbons:1978tef},
has been investigated as a prime example of a SD black hole
(see, e.g., \rrcite{Adamo:2023fbj,Adamo:2025fqt,AAS}
for justifications of this terminology).
In light of its associated hidden symmetry,
it has been suggested that
the SD Taub-NUT solution
is
``the analog of the hydrogen atom for black
holes''
\cite{Guevara:2023wlr}.
Moreover,
the classical integrability of SD gravity \cite{Penrose:1976js} renders its perturbation theory tractable \cite{Adamo:2025fqt};
for instance,
it has been shown that
the SD Taub-NUT solution
exhibits a graviton helicity selection rule
up to multiplicity two
\cite{Adamo:2023fbj,Adamo:2025fqt}.

Certainly, SD black holes could be viewed as purely mathematical:
the self-duality condition inevitably makes their gravitational fields complex-valued.
See the discussions on complex spacetimes in \rrcite{penrose1976nonlinear,newman1988remarkable,Penrose:1976js,shaviv1975general,plebanski1975some,Plebanski:1977zz,grg207flaherty},
for instance.
One may at least
regard them as formal idealizations
leading to theoretical curiosities,
while keeping in mind that
the ultimate desire will be to restore
the missing ASD part
either perturbatively or nonperturbatively
(recall \Secs{Jc}{Jcx}).

To give the reader at least a concrete glimpse of why the SD black hole has been compared to the hydrogen atom \cite{Guevara:2023wlr}
and why they are of an active theoretical interest
\cite{Adamo:2023fbj,Adamo:2024xpc,Adamo:2025fqt,Crawley:2021auj,Crawley:2023brz,Guevara:2023wlr,Guevara:2024edh,Guevara:2025psg,note-sdtn,nja,AAS,Skvortsov:2025ohi,Doran:2026bng},
this chapter aims to analyze a simpler yet analogous system in electromagnetism: the SD dyon.
As is established by \rcite{note-sdtn}
and will be reproduced in \Chap{K3:SDTN} of this thesis,
the SD dyon has been shown to describe
the exact gauge theory counterpart of
the SD Taub-NUT solution
via the Kerr-Schild double copy.
See also works
\cite{Adamo:2023fbj,Adamo:2024xpc,Adamo:2025fqt}
and discussions therein.
The SD dyon has served as a simplified testing ground for questions related to SD black holes
\cite{Adamo:2024xpc,Garner:2024tis}.

Our conclusion in this chapter
will be that the motion of a charged test particle
in the SD dyon background
is \textit{maximally superintegrable}.
Note that
later parts of this thesis
will extend
this superintegrability
to SD black hole backgrounds
and to the all-orders-in-spin dynamics of
spinning probe particles.

\section{ Dyon Motion in Dyon Background}

Consider flat spacetime with Cartesian coordinates
$x^\m = (t,x^i)$.
A dyon is a particle that carries both electric and magnetic charges,
say $Q$ and $\Qm$.
The electromagnetic field strength
of a static dyon at $\vex = 0$
is given by
\begin{align}
    \label{SDDeq:dyon-EB}
    \vec{E} = \frac{Q}{4\pi}\mem \frac{\rhat}{r^2}
    \,,\quad
    \vec{B} = \frac{Q^\star}{4\pi}\mem \frac{\rhat}{r^2}
    \,,
\end{align}
where $r := |\vex|$ and $\rhat := \vex/r$.
The reader may consult to
\cite{Shnir:2011zz}
for more information on the physics of dyons.

We investigate the special-relativistic dynamics of a charged massive test particle in this background.
For full generality,
let $q$ and $q^\star$
be its electric and magnetic charges,
respectively.
In a constant einbein parameterization,
the classical EoM read
\begin{subequations}
    \label{SDDeq:dyon-eom}
\begin{align}
    \label{SDDeq:dyon-eom-0}
    m\mem \frac{dt}{d\t}
        \,=\, p^0
    &\,,\quad
    m\mem \frac{dp^0}{d\t}
        \,=\, 
        \vep\cdot (q \vec{E} {\,+\,} q^\star \vec{B})
    \,,\\
    \label{SDDeq:dyon-eom-i}
    m\mem \frac{d\vex}{d\t}
        \,=\, \vep
    &\,,\quad
    m\mem \frac{d\vep}{d\t}
        \,=\,
        p^0\mem (q \vec{E} {\,+\,} q^\star \vec{B})
        + \vep \mtimes\hnem (q\vec{B} {\,-\,} q^\star \vec{E})
    \,,
\end{align}
\end{subequations}
where $m$ is the rest mass.
Plugging in \eqref{SDDeq:dyon-EB},
these boil down to
\begin{align}
    \label{SDDeq:eom}
    \frac{dp^0}{d\t}
        \,=\, -\mem \frac{d}{d\t}\bigg(\mem{
            \frac{\k}{r}
        }\mem\bigg)
    \,,\quad
    m\mem \frac{d\vep}{d\t}
        \,=\, \k\mem p^0\mem \frac{\rhat}{r^2}
        + \tk\mem \vep\mt \frac{\rhat}{r^2}
    \,,
\end{align}
where we have denoted
\begin{align}
    \k \,:=\, \frac{ 
        q Q {\,+\,} q^\star Q^\star 
    }{4\pi}
    \,,\quad
    \tk \,:=\, \frac{
        q Q^\star\nem {\,-\,} q^\star Q 
    }{4\pi}
    \,.
\end{align}
The EoM preserve the mass-shell condition,
\begin{align}
    \label{SDDeq:mass-shell}
    (p^0)^2 \,=\, \vep^2 + m^2
    \,.
\end{align}

\newpage

As the background in \eqref{SDDeq:dyon-EB} is static and spherically symmetric,
we expect
the conservation of energy and angular momentum.
Indeed, the conserved energy 
can be immediately identified
from the first equation in \eqref{SDDeq:eom}:
\begin{align}
    \label{SDDeq:Edef}
    \E \,=\, p^0 + \frac{\k}{r}
    \,.
\end{align}
Meanwhile,
the conserved angular momentum
is given by
\begin{align}
    \label{SDDeq:Jdef}
    \vec{\J} \,=\, \vex\mtimes\vep - \tk \rhat
    \,,
\end{align}
which follows from using 
the second equation in \eqref{SDDeq:eom}
and also an identity
\begin{align}
    \label{SDDeq:rhat-id}
    \frac{1}{r}\, 
        \rhat \mtimes (\vep \mtimes \rhat)
    \,=\,
    \frac{1}{r}\mem
        \Big(\mem\hem{
            \vep - \rhat\, \rhat\mdot\vep
        }\,\Big)
    \,=\, \vep\mdot\vp\, \rhat
    \,.
\end{align}
Famously, the term $\tk \rhat$
in \eqref{SDDeq:Jdef}
describes the field angular momentum
due to the interaction between
the dyon and the test particle.

Consequently, we find
three independent conserved quantities in involution:
$\E$, $\vec{\J}^2$, and a component of $\vec{\J}$
(say $\J_z$).
As the particle has exactly three physical degrees of freedom,
the system is Liouville integrable.
Hence the EoM are
solvable by quadratures.

Let us elaborate on 
the concrete steps of solving the EoM.
First of all,
it follows from
\eqrefs{SDDeq:mass-shell}{SDDeq:Jdef}
that
\begin{align}
    \label{SDDeq:Veff}
    \frac{m}{2}\mem \bigg({
        \frac{dr}{d\t}
    }\bigg)^{\nem\hnem2}
    + \frac{ \vec{\J}^2 {\mem-\,} \c }{2mr^2}
    + \frac{\E\mem\k}{mr}
    \,=\,
        \frac{\E^2 {\mem-\,} m^2}{2m}
    \,,
\end{align}
where we have denoted
\begin{align}
    \label{SDDeq:cdef}
    \c 
    \,:=\,
        \k^2 + \tk^2
    \,=\,
        \bigbig{
            Q^2 \mplus {Q^\star}^2
        }
        \bigbig{
            q^2 \mplus {q^\star}^2
        }
    \,.
\end{align}
Clearly, 
\eqref{SDDeq:Veff} describes an effective potential governing the radial motion,
from which 
one can find $r$ as a function of the proper time $\t$.
One can then find $\rhat$ as a function of $\t$
by integrating the equation
\begin{align}
    \label{SDDeq:Omega}
    \frac{d\rhat}{d\t}
    \,=\,
        \frac{\vec{\J}}{mr^2} \mtimes \rhat
    \,,
\end{align}
which follows from \eqrefs{SDDeq:Jdef}{SDDeq:rhat-id}.
This completely determines the position $\vex = r \rhat$
of the particle as a function of time.

It is worth understanding 
the geometry
behind \eqref{SDDeq:Omega}.
Crucially,
the particle's trajectory lies on a cone
defined by $\vec{\J} \mdot \rhat = -\tk$,
known as Poincar\'e cone \cite{poincare1896remarques}.
The cone's half-angle
from the $\hat{\J}$-axis
is given by
$\pi/2 + \xi$, where
\begin{align}
    \xi := \sin^{-1}\hnem\bigg(\mem{
        \frac{\tk}{|\vec{\J}|}
    }\mem\bigg)
    \,.
\end{align}
Note that $\xi$ approaches zero when the limit of vanishing interaction
so that the cone degenerates to a flat plane.
Now the interpretation of \eqref{SDDeq:Omega} is obvious:
the particle revolves around 
the axis of the Poincar\'e cone
with angular velocity
$|\vec{\J}|/mr^2$.
In consequence,
determining
the radial motion
from the effective potential
fixes the entire trajectory.

\section{ Self-Dual Dyon as the Hydrogen Atom}
\label{K3a:SDD>H}

A SD dyon is a dyon whose electric and magnetic charges satisfy the relation $Q = i\Qm$.
It sources a SD electromagnetic field:
$\vec{E} = i\vec{B}$.

An ASD dyon is a dyon whose electric and magnetic charges satisfy the relation $Q = -i\Qm$.
It sources an ASD electromagnetic field:
$\vec{E} = -i\vec{B}$.

Together, they are referred to as chiral dyons.
Clearly, these chiral limits are the extremal cases
in which the parameter $\c$ in \eqref{SDDeq:cdef}
becomes zero.

From now on,
we study the motion of the test particle
in the background of a chiral dyon.
As mentioned earlier,
this may be seen as a rather bizarre pathway
from a conservative point of view.
Due to the imaginary magnetic charge,
the particle experiences a complex-valued Lorentz force.
Hence its velocity becomes complex,
and even its position becomes complex;
otherwise there is no solution to the classical EoM.
However, this complexified limit
turns out to be intriguing enough
to pursue,
at least as a formal curiosity.

First of all,
upon imposing the chiral limit $\gamma \too 0$,
\eqref{SDDeq:Veff} boils down to
\begin{align}
    \label{SDDeq:Veff.SD}
    \gamma \mem=\mem 0
    \qiq
    \frac{m}{2}\mem \bigg({
        \frac{dr}{d\t}
    }\bigg)^{\nem\hnem2}
    + \frac{ \vec{\J}^2 }{2mr^2}
    + \frac{\E\mem\k}{mr}
    \,=\,
        \frac{\E^2 {\mem-\,} m^2}{2m}
    \,.
\end{align}
Remarkably, this
is isomorphic to 
the effective potential equation
for the \textit{nonrelativistic} hydrogen atom
(i.e., Kepler problem).

The nonrelativistic hydrogen atom
is famously known to exhibit
the Laplace-Runge-Lenz (LRL) vector 
\nomenclature{LRL}{Laplace-Runge-Lenz}
as a conserved quantity.
Motivated by this fact,
we perform
brute-force computation
to find that
\eqref{SDDeq:eom} implies
\begin{align}
    \frac{d}{d\t}\mem \BB{
        \vep \mtimes \vec{\J}
    }
    \,&=\,
        \frac{
            \gamma - \E\mem\k\mem r
        }{mr^3}\mem
        \frac{d\hat{r}}{d\t}
    \,.
\end{align}
As a result, it follows that
\begin{align}
    \label{SDDeq:dK}
    \vec{\KA}
    \,:=\,
        \vep \mtimes \vec{\J} \mem+\mem \E\mem\k\mem \hat{r}
    \qiq
    \frac{d\mem\vec{\KA}}{d\t}
    \,=\,
        \frac{\gamma}{mr^3}\mem
        \BB{
            \vep r - m\dot{r}\hem \vex
        }
    \,.
\end{align}
Therefore, 
a new set of conserved quantities emerges
in the chiral limit $\gamma \too 0$:
\begin{align}
    \label{SDDeq:K}
    \gamma \mem=\mem 0
    \qiq
    \frac{d\mem\vec{\KA}}{d\t}
    \,=\,
        0
    \,.
\end{align}
It is amusing how the LRL vector
admits a relativistic reincarnation
via complexifications and the formal chiral limits.

In sum,
we identify seven conserved quantities:
$\E$, $\vec{\J}$, $\vec{\KA}$.
There are two algebraic relations relating them.
On $\gamma = 0$,
we find
\begin{align}
    \vec{\J} \mdot \vec{\KA}
    \,=\,
        -\E\mem \k\mem \tk
    \,,\quad
    \vec{\KA}^2
    \,=\,
        \bigbig{
            \E^2 \mminus m^2
        }
        \bigbig{
            \vec{\J}^2 \mplus \k^2
        }
        + \E^2\hem \k^2
    \,.
\end{align}
Therefore the number of
independent conserved quantities is five,
which is $2\times3 -1$
in a six-dimensional phase space.
Hence the system is \textit{maximally superintegrable}.
A further computation shows that
the dynamical symmetry algebra is
nicely closed,
describing 
$\so(4)$ for $\E^2 < m^2$,
$\so(1,3)$ for $\E^2 > m^2$
(or $\so(4,\C)$ in the complexified category),
and $\iso(3)$ for $\E^2 = m^2$.

In these concrete senses,
we conclude that
the dynamics of a relativistic test particle
in the background of a chiral dyon
is isomorphic to
that of the nonrelativistic Hydrogen atom
or Kepler problem.

\section{ Maximal Superintegrability from Killing-Yano Tensors}

In the above exposition,
we have intentionally implemented the $1+3$ split
to be pedagogical;
after all, the background stipulates a preferred time direction.
However, 
it should be pointed out that
there is also a more elegant approach
that retains the four-vector notation.

Without loss of generality,
we now assume the background of a SD dyon
with electric charge $Q/2$ and magnetic charge $-iQ/2$.
Let $u^\m$ be the timelike Killing vector
encoding the stationary direction.
In the relativistic notation,
the electromagnetic field in \eqref{SDDeq:dyon-EB} 
describes
\begin{align}
    \label{SDD.F}
    F_{\m\n}(x)
    \,=\,
        \frac{Q}{8\pi|\vex|^3}\:
            {*}Y^+_{\m\n}(x)
    \,.
\end{align}
where we have defined
SD and ASD two-forms $Y^\pm_{\m\n}$ as
\begin{align}
    {*}Y^\pm_{\m\n}(x)
    \,=\,
        \bigbig{
            u \wedge x
        }\hnem{}^\pm_{\m\n}
    \,=\,
        \tfrac{1}{2}\,
        \bigbig{
            u_\m x_\n - x_\m u_\n
        }
        \mp \tfrac{i}{2}\, 
            \ve_{\m\n\r\s}\hhem u^\r x^\s
    \,.
\end{align}
The $\pm$ superscript
denotes the SD/ASD projection.
These essentially encode the position three-vector $\vex$
via each $\sl(2,\C)$ sector of
$\so(4,\C) \cong \sl(2,\C) \oplus \sl(2,\C)$:
$(\vex\mdot\vec{\s})^\da{}_\db$
and
$(\vex\mdot\vec{\s})_\a{}^\b$.
If $Y_{\m\n}(x) = Y^+_{\m\n}(x) + Y^-_{\m\n}(x)$,
we find
\begin{align}
    \label{KYT0}
    \partial_\r Y_{\m\n}(x)
    \,=\,
        \ve_{\m\n\r\s}\hem u^\s
    \,=\,
        \partial_{[\r} Y_{\m\n]}(x)
    \,.
\end{align}
This is the defining equation for
a Killing-Yano \cite{yano1952some} tensor.
Moreover, it is easy to see that
it ``commutes'' with the field strength in \eqref{SDD.F} as
\begin{align}
    \label{FY=YF}
    F^\m{}_\r(x)\mem Y^\r{}_\n(x)
    \,=\,
        Y^\m{}_\r(x)\mem F^\r{}_\n(x)
    \,.
\end{align}

In generic electromagnetic fields,
the dyonic test particle
with charges $(q,q^\star)$ 
exhibits the EoM
\begin{align}
\begin{split}
    p^\m \,=\, m\mem \frac{dx^\m}{d\t}
    \,,\quad
    m\mem \frac{dp_\m}{d\t}
    \,&=\,
        q F_{\m\n}(x)\mem p^\n
        -
        q^\star\mem {*}F_{\m\n}(x)\mem p^\n
    \,,\\
    \,&=\,
        q_+ F^+_{\m\n}(x)\mem p^\n
        +
        q_- F^-_{\m\n}(x)\mem p^\n
    \,,
\end{split}
\end{align}
where $p^2 + m^2 = 0$
and $q_\pm = q \mp iq^\star$.
In SD backgrounds, this reduces to
\begin{align}
    \label{lorforce-in-sd}
    p^\m \,=\, m\mem \frac{dx^\m}{d\t}
    \,,\quad
    \frac{dp^\m}{d\t}
    \,=\,
        \frac{q_+}{m}\mem
        F^\m{}_\n(x)\mem p^\n
    \,.
\end{align}
Provided the two conditions in \eqrefs{KYT0}{FY=YF},
\eqref{lorforce-in-sd} implies
\begin{align}
    \frac{d}{d\t}\mem\BB{
        Y^\m{}_\r(x)\mem p^\r
    }
    \,=\,
        \frac{q_+}{m}\mem F^\m{}_\n(x)\mem 
        \BB{
            Y^\n{}_\r(x)\mem p^\r
        }
    \,.
\end{align}
This means that $p^\m$ and $Y^\m{}_\r(x)\mem p^\r$
exhibit the same cyclotron precession angular frequency,
$q_+ F^\m{}_\n(x)/m$.
See also 
\rrcite{Carter:1968ks,yano1952some,Walker:1970un,Hughston:1972qf,hughston1973spacetimes,jeffryes1984space,Jezierski:2005cg,Nozawa:2015qea,hansen2014killing,Penrose:1973naked,floyd1973dynamics,marck1983solution}.

Meanwhile, consider the ASD two-forms of Pleba\'nski \cite{Plebanski:1977zz,Capovilla:1991qb}:
\begin{align}
\begin{split}
    \Sigma_1 \,&=\,
        i\mem\bigbig{
            dx \wedge dt + i\mem dy \wedge dz
        }
    \,,\\
    \Sigma_2 \,&=\,
        i\mem\bigbig{
            dy \wedge dt + i\mem dz \wedge dx
        }
    \,,\\
    \Sigma_3 \,&=\,
        i\mem\bigbig{
            dz \wedge dt + i\mem dx \wedge dy
        }
    \,.
\end{split}
\end{align}
These describe a trio of two-forms,
$(\Sigma_a)_{\m\n}$ for $a=1,2,3$,
and correspond to $(\s_a)_\a{}^\b$ in the spinor notation,
so ${*}(\Sigma_a)_{\m\n} \eqq {-i}\mem (\Sigma_a)_{\m\n}$.
Notably, these
trivially
serve as Killing-Yano tensors
\cite{Gibbons:1987sp,Nozawa:2015qea}
and define a quaternionic structure
as 
\cite{Gibbons:1987sp,Atiyah:1978wi,Nozawa:2015qea}
\begin{align}
    \label{quaternion}
    (\Sigma_a)^\m{}_\r\hem (\Sigma_b)^\r{}_\n
    \,=\,
        -\delta_{ab}\, \delta^\m{}_\n
        + \ve^c{}_{ab}\mem (\Sigma_c)^\m{}_\n
    \,.
\end{align}
Moreover, they also commute with the field strength in \eqref{SDD.F} as
\begin{align}
    (\Sigma_a)^\m{}_\r\mem F^\r{}_\n(x)
    \,=\,
        F^\m{}_\r(x)\mem (\Sigma_a)^\r{}_\n
    \,,
\end{align}
as any ASD element commutes with a SD element
in the Lorentz algebra.

\newpage

As a result,
the classical EoM in
\eqref{lorforce-in-sd} also imply
\begin{align}
    \frac{d}{d\t}\mem\BB{
        (\Sigma_a)^\m{}_\r\mem p^\r
    }
    \,=\,
        \frac{q_+}{m}\mem F^\m{}_\n(x)\mem 
        \BB{
            (\Sigma_a)^\n{}_\r\mem p^\r
        }
    \,,
\end{align}
so we now find that
a triple of four-vectors,
$p^\m$, $Y^\m{}_\r(x)\mem p^\r$, and $(\Sigma_a)^\m{}_\r\mem p^\r$
exhibit a synchronized precession behavior.
Clearly the describe three axes of a common Lorentz frame
attached to the particle.

From this geometric realization,
it is clear that the following scalar quantities are dynamically conserved:
\begin{align}
    \label{SDDeq:Cart}
    \Cart
    \,&=\,
        - p_\m\mem Y^\m{}_\r(x)\mem Y^\r{}_\n(x)\mem p^\n
    \,=\,
        (\vex \mtimes \vep)^2
    \,,\\
    \label{SDDeq:Ka}
    \K_a
    \,&=\,
        - p_\m\mem (\Sigma_a)^\m{}_\r\mem Y^\r{}_\n(x)\mem p^\n
    \,=\,
        \vep \mtimes (\vex \mtimes \vep)
        + i\hem p^0\mem \vex \mtimes \vep
    \,.
\end{align}

Of course, there are also conserved quantities due to the Killing vectors.
It is left as an exercise to check that
they reproduce the energy $\E$
and the angular momentum $\vec{\J}$
in \eqrefs{SDDeq:Edef}{SDDeq:Jdef},
for which it will be helpful to adopt the approach in \rcite{Hughston:1972qf}.
With this understanding,
we plug in $p^0 = \E - \k/r$ and $\vex \mtimes \vep = \vec{\J} + \tk\rhat$
according to \eqrefs{SDDeq:Edef}{SDDeq:Jdef}.
Then \eqrefs{SDDeq:Cart}{SDDeq:Ka} 
boil down to
\begin{align}
    \Cart
    \,=\,
        \vec{\J}^2 + \k^2
    \,,\quad
    \K_a
    \,=\,
        \KA_a + i\mem \E\mem \J_a
    \,,
\end{align}
on the support of $\k = i\mem\tk$.
While $\Cart$---%
the Carter \cite{Carter:1968ks} constant---%
does not yield a new conserved quantity,
$\K_a$ does:
the LRL vector identified earlier.

In conclusion,
we have reproduced the conserved charges found in \Sec{K3a:SDD>H}
in the Lorentz-covariant notation.
The underlying geometric structures are identified as
the Killing-Yano tensor $Y_{\m\n}(x)$
and the ASD Pleba\`nski two-forms $(\Sigma_a)_{\m\n}$.
It is left as an exercise to understand how
the quaternionic structure in \eqref{quaternion}
boils down to the symmetry algebra.

Notably, 
the same geometric structures arise
for the SD Taub-NUT black hole as well.
This forms the basis of the statement
\cite{Guevara:2023wlr}
that the SD Taub-NUT black hole
is secretly the hydrogen atom.

More details can be found in the next chapter,
where it is shown that
the SD Taub-NUT black hole
is the double copy of the SD dyon.
Specifically,
we discover a new metric for
the SD Taub-NUT solution
by double copying the Kerr-Schild gauge potential of the SD dyon.
\chapter{Self-Dual Black Hole from Double Copy}
\label{K3:SDTN}

We derive a previously unknown KS metric for the SD Taub-NUT solution
by classical double copy.
We show its equivalence with the
well-known Gibbons-Hawking instanton metric
by means of an explicit complex coordinate transformation.
Our KS metric develops
a semi-infinite string defect
as the Misner string,
much like the original metric.
It also establishes two more
classical double copy
correspondences to the SD dyon in electromagnetism,
one of which is a novel nonlocal kind
based on propagators between dyonic matter.

\begin{fullnote}
    Contents of \Secsto{NOTE-SDTN>INTRO}{NOTE-SDTN>CONC} 
    are adapted from \rcite{note-sdtn}, \fullcite{note-sdtn}.
\end{fullnote}
\medskip

\section{ Introduction}%
\label{NOTE-SDTN>INTRO}

Recently,
an exciting program dubbed
classical double copy
\cite{Berman:2018hwd,bahjat2020monopoles,note-sdtn,nja,%
Carrillo-Gonzalez:2017iyj,CarrilloGonzalez:2019gof,Bahjat-Abbas:2017htu,Banerjee:2019saj,elor2020newman,bah2020kerr,Monteiro:2020plf,Alfonsi:2020lub,Monteiro:2021ztt,gabriel1,%
Luna:2015paa,Luna:2018dpt,Chacon:2021wbr,white2021twistorial,Luna:2022dxo}
has refined analogies between gravity and electromagnetism
into
concrete
mappings
between classical solutions
while being rooted in the study of scattering amplitudes
\cite{KLT,BCJ1,BCJ2,BCJReview,monteiro2011kinematic}.
The prime example is the
KS double copy,
established
in the seminal work \cite{monteiro2014black}.\footnote{
    KS metrics give an abelian sector of GR
    in the sense of
    the effective linearization property
    \cite{xanthopoulos1978exact,Harte:2016vwo}.
    In this light,
    the KS double copy
    may be viewed as probing 
    the double copy correspondence between
    the abelian sectors of YM theory and GR.
}
It is shown that
a stationary KS solution
$g = \eta + \Phi\mem \ell \otimes \ell$
in GR
defines
a stationary solution in Maxwell theory
by the gauge potential
$A = \Phi\mem \ell$,
which takes a ``single copy'' of the null one-form $\ell$.
Conversely, one envisions
the gravitational counterpart
$g = \eta + \Phi\mem \ell \otimes \ell$
of a stationary solution
$A = \Phi\mem \ell$
in electromagnetism,
by taking a ``double copy'' of $\ell$.
Well-known
instances
of this correspondence
are
Schwarzschild to point charge
and
Kerr to
a rotating solution dubbed ``{\kerr}\mem''
\cite{monteiro2014black,aho2020,Lynden-Bell:2002dvr,Newman:1965tw-janis}.

From this angle,
it is natural to expect that
the Taub-NUT solution
\cite{Taub:1950ez,Newman:1963yy}
could correspond to
a dyon,
given the interpretation of the NUT charge
as the gravitomagnetic monopole
\cite{Misner:1963flatter,Bonnor:1969ala,sackfield1971physical,cho1991magnetic,dowker1967gravitational,samuel1986gravitational,maartens1998gravito,Alfonsi:2020lub}.
Unfortunately,
only a \textit{double} KS metric
$g = \eta + \Phi_1\hem \ell_1 \hnem\otimes\hnem \ell_1 + \Phi_2\mem \ell_2 \hnem\otimes\hnem \ell_2$
has been known to exist
for the Taub-NUT solution:
a classic result due to Pleba\'nski and Demia\'nski
\cite{Plebanski:1975xfb,PD,Chong:2004hw}.
Consequently,
the anticipated correspondence between
the Taub-NUT solution
and the electromagnetic dyon
had to be
achieved only by
an alternative construction
using the double KS ansatz
\cite{Luna:2015paa}.

Remarkably,
the present letter
points out that
a direct application of the original KS double copy is viable for the \textit{SD} Taub-NUT solution.
This is the extremal case of the Taub-NUT solution
that
has recently gained
a revived interest \cite{Adamo:2023fbj,Guevara:2023wlr,Crawley:2021auj,Crawley:2023brz,Adamo:2024xpc}.

Crucially, we discover that
the SD Taub-NUT solution admits a \textit{single} KS metric:
$g = \eta + \Phi\mem \ell \otimes \ell$.
First, a KS metric
is obtained
by KS double copying the gauge potential of
the SD dyon 
in electromagnetism.
We then show that
it describes the SD Taub-NUT solution
by explicitly constructing a coordinate transformation to
the well-known Gibbons-Hawking
\cite{hawking1977gravitational,Gibbons:1978tef}
instanton metric.
To our understanding,
this gives the first instance
of the classical double copy program
yielding a new result in GR (a previously unknown metric).

Next, the physical interpretation of the metric
as a gravitational dyon
is established
within the KS description.
Most importantly,
we confirm the characteristic
\textit{Misner string} geometry:
the gravitational analog
\cite{Misner:1963flatter,Bonnor:1969ala}
of the Dirac string
\cite{Dirac:1931kp,Dirac:1948um}.
Notably,
finding
its distributional source
\cite{Bonnor:1969ala,sackfield1971physical}
becomes a significantly simpler problem
in KS coordinates,
thanks to the effective linearization property
\cite{Harte:2016vwo,xanthopoulos1978exact,vines2018scattering}.

Lastly,
the classical double copy
is established
in two more ways.
Our KS description
concretely validates the Weyl double copy
proposed by \rrcite{Luna:2018dpt,white2021twistorial,Chacon:2021wbr},
which has been
stated without referencing an explicit metric.
We then propose
a new exotic
form of classical double copy
employing
a \textit{nonlocal} operator,
which arises from
propagators between dyonic matter
\cite{Zwanziger:1970hk,Gubarev:1998ss,Shnir:2011zz,Terning:2020dzg,Moynihan:2020gxj}.

In this chapter,
we use the field theorists' natural unit
in which $8\pi G = 1$.

\section{ Derivation of Self-Dual Taub-NUT Solution from Double Copy}%

In electromagnetism, a dyon is a point particle
carrying both electric and magnetic charges
(recall \eqref{SDDeq:dyon-EB}).
The SD dyon is a dyon with electric and magnetic charges, say, 
$Q$ and $-iQ$.
Its gauge potential can be given as
\begin{align}
\begin{split}
    \label{eq:A0}
    A
    \,=\,
    \frac{Q}{4\pi R}\mem
    \Big(\hem{
        - dt
        - i\hem R\mem
        (1{\mem-\mem}\cos\theta)\mem d\phi
    }\mem\Big)
    \,,
\end{split}
\end{align}
where
we have set up spherical coordinates $(R,\theta,\phi)$
and
oriented the Dirac string
\cite{Dirac:1931kp,Dirac:1948um}
along the negative $z$-axis (south pole).
This describes a complexified solution
where
the field strength $F=dA$ is SD:
$*F = +i\hem F$.

Interestingly,
adding a total derivative
$(Q/4\pi)\mem d\log(R{\,+\,}z)$
gauge transforms
\eqref{eq:A0}
to a KS form:
\begin{align}
    \label{eq:A1}
    A
    =
    \frac{Q}{4\pi R}\mem
    \bigg({
        -dt
        + \frac{x{\,+\,}iy}{R{\,+\,}z}\mem dx
        + \frac{y{\,-\,}ix}{R{\,+\,}z}\mem dy
        + dz
    }\bigg)
    \,.
\end{align}
Here, $(x,y,z) = (R\sin\theta\cos\phi, R\sin\theta\sin\phi, R\cos\theta)$
are the Cartesian coordinates.
Importantly,
the terms inside the bracket comprise a null one-form.

With this understanding,
the KS double copy \cite{monteiro2014black}
associates \eqref{eq:A1}
to the following line element,
by literally taking ``two copies'' of the null one-form:
\begin{align}
\begin{split}
    \label{eq:g1}
    ds^2
    \,=\,
    {}&{}{
        -dt^2 + dx^2 + dy^2 + dz^2
    }
    \\
    {}&{}
        +
        \frac{M}{4\pi R}\mem
        \bigg({
            -dt
            + \frac{x{\,+\,}iy}{R{\,+\,}z}\mem dx
            + \frac{y{\,-\,}ix}{R{\,+\,}z}\mem dy
            + dz
        }\mem\bigg)^{\nem\hnem2}
    \,.
\end{split}
\end{align}
Here, the charge $Q$ is mapped to a mass parameter $M$.

Remarkably,
\eqref{eq:g1}
derives a new metric for
the SD Taub-NUT solution,
whose mass and NUT charge are
$M$ and $-iM$.
To prove this,
we simply construct
an explicit coordinate transformation
to a previously known metric.

\section{ The Diffeomorphism}%

\subsection{Direct Construction}

It is helpful to employ
complex coordinates
$\zeta = (x{\,+\,}iy)/(R{\,+\,}z)$
and
$\smash{\tzeta} = (x{\,-\,}iy)/(R{\,+\,}z)$
that arise
from stereographically projecting
the two-sphere
with respect to the south pole,
so
\eqref{eq:g1} translates to
\begin{align}
\begin{split}
    \label{eq:g1-stereographic}
    ds^2
    \,=\,
    {}&{}{
        -dt^2 + dR^2 + R^2\mem \frac{4\hem d\smash{\tzeta}\hem d\zeta}{(1{\,+\,}\tzeta\zeta)^2}
    }
        +
        \frac{M}{4\pi R}\mem
        \bigg({
            -dt
            + dR
            - R\mem \frac{2\tzeta\hem d\smash{\zeta}}{1{\,+\,}\tzeta\zeta}
        }\mem\bigg)^{\nem\hnem2}
        \,.
\end{split}
\end{align}
By writing down an ansatz at each order in the expansion in $M/4\pi$,
we could perturbatively construct a diffeomorphism
whose resummation gives
\begin{align}
\begin{split}
    \label{eq:diff}
    t
        \,\,\mapsto\mem\,
    t + \frac{M}{4\pi}
        &\,,\quad
    \zeta
        \,\,\mapsto\mem\,
    \zeta
    \,,\\[-0.1\baselineskip]
    R
        \,\,\mapsto\mem\,
    R + \frac{M}{4\pi}
        &\,,\quad
    \tzeta
        \,\,\mapsto\mem\,
    \frac{1}{
        1 + (M/4\pi R) (1{\,+\,}\tzeta\zeta)
    }\mem
    \tzeta
    \,.
\end{split}
\end{align}
Applying \eqref{eq:diff} to \eqref{eq:g1-stereographic} gives
\begin{align}
    \label{eq:g0}
    ds^2
    \,=\,
    \lrp{
    \begin{aligned}[c]
    {}&{}
        {-
        \bigg(\hem{
            1 + \frac{M}{4\pi R}
        }\hem\bigg)^{\nem\nem-1}\hnem
        \bigg(\mem{
            dt
            + \frac{M}{4\pi R}\mem
            \bigg(\mem{
                dR
                - R\mem \frac{2\tzeta\hem d\smash{\zeta}}{1{\,+\,}\tzeta\zeta}
            }\mem\bigg)\nem\nem
        }\mem\bigg)^{\nem\hnem2}
        }
    \\
    &
        +
        \bigg(\hem{
            1 + \frac{M}{4\pi R}
        }\hem\bigg)
        \bigg(\mem{
            dR^2
            + R^2\hem
            \frac{4\mem d\tzeta d\zeta}{(1{\,+\,}\tzeta\zeta)^2}
        }\mem\bigg)
    \end{aligned}
    \,}
    \,,
\end{align}
which is a Gibbons-Hawking metric
\cite{hawking1977gravitational,Gibbons:1978tef}
for the SD Taub-NUT solution
with mass $M$.
This diffeomorphism
is surely invertible;
it becomes self-evident soon that
the inverse map
is simply
\eqref{eq:diff}
with $M$ replaced with $-M$.

\medskip
As a reminder,
the Gibbons-Hawking line element reads \cite{hawking1977gravitational,Gibbons:1978tef}
\begin{align}
    \label{gh}
    ds^2
    \,=\,
	- \frac{1}{V}\mem
    	\BB{
    		dt + a
    	}^{\nem2}
	+ V\mem
        \BB{
            dx^2
            + dy^2
            + dz^2
        }
    \,.
\end{align}
\eqref{gh} defines a SD, Ricci-flat metric if
$V$ and $a$
are scalar and one-form fields
on the three-dimensional space of $(x,y,z)$ such that
\begin{align}
    \label{gh-sdcond}
    -dV \,=\, i\, {*}_3\mem da
    \,.
\end{align}
Here, ${*}_3$ denotes the Hodge dual
of the three-dimensional flat metric $dx^2 + dy^2 + dz^2$.
\eqref{gh-sdcond} means that
$V$ and $a$ can be interpreted as 
describing
a stationary SD electromagnetic field.
Diffeomorphisms of the form
$t \mapsto t + \chi(x,y,z)$
correspond to
the gauge transformations
$a \sim a + d\chi$.

The SD Taub-NUT solution
is included in the Gibbons-Hawking class
as a single-centered, asymptotically locally flat instance.
When the spherical coordinates are used
for the flat three-space,
it arises by taking
\begin{align}
    \label{gh-sdtn}
    V
    \mem=\mem
        1 + \frac{M}{4\pi R}
    \,\,\,\,\implies\,\,\,\mem
    {-dV}
    \mem=\mem
        \frac{M}{4\pi R^2}\mem dR
    \,,\,\,\,\,
    da
    \mem=\mem
        \frac{-iM}{4\pi}\mem
            \sin\theta\mem d\theta \swedge d\phi
    \,.
\end{align}
It is easy to see that \eqref{eq:g0}
precisely encodes \eqref{gh-sdtn}.

\medskip

To sum up,
we have provided a \textit{constructive} proof that
the KS and Gibbons-Hawking
metrics in
\eqrefs{eq:g1}{eq:g0}
are diffeomorphic.
Consequently,
we establish that the SD Taub-NUT solution admits
a construction from the KS double copy.
Previously,
these facts
were only recently \textit{conjectured}
by \rcite{gabriel1}
in split signature.

An important remark is in order.
The astute reader may have noticed
an unusual feature of
the diffeomorphism in \eqref{eq:diff}:
it
describes a \textit{complexified diffeomorphism},\footnote{
    Such a complexification is exactly what is practiced in
    Pleba\'nski \cite{Plebanski:1975xfb}, Sec.\,7,
    and 
    Pleban\'nski and Demia\'nski \cite{PD}, Sec.\,13.
    Mathematically,
    this means to work in the holomorphic category of manifolds,
    where the ``barred'' coordinates play no role.
    See the clarification which Penrose makes in \rcite{penrose1976nonlinear}.
    See also 
    \rrcite{Penrose:1976js,shaviv1975general,plebanski1975some,Plebanski:1977zz,newman1988remarkable,grg207flaherty}
    for related discussions and reviews.
}
since $\zeta$ is frozen
while $\smash{\tzeta}$ gets scrambled:
hence the notation $\smash{\tzeta}$
instead of
$\smash{\bar{\zeta}}$.
This is justified as
a legitimate operation
in the current context
because
the Lorentzian-signature SD Taub-NUT solution
is inherently a complex saddle
as a SD solution
\cite{Adamo:2023fbj}.
Just as
the SD dyon gauge potential
described in
\eqref{eq:A1}
is a solution to
complexified Maxwell's equations
due to the imaginary magnetic charge $-iQ$,
the SD Taub-NUT metric in \eqref{eq:g1}
solves
complexified Einstein's equations
governing holomorphic metrics
\cite{plebanski1975some,Plebanski:1977zz,shaviv1975general,Newman:1976gc,Adamo:2023fbj}.
As a consequence
of this interpretation,
$\zeta$ and $\smash{\tzeta}$ can be transformed independently as \textit{holomorphic} coordinates.

\subsection{Kerr-Schild to Gibbons-Hawking via Null Geodesic Flow}

In fact,
the explicit diffeomorphism
in \eqref{eq:diff}
simply turns out to be
a flow along the null congruence
associated with the KS metric:
\begin{align}
    \label{ks-to-gh}
	\bigbig{
		t
		,\mem
		x
		,\mem
		y
		,\mem
		z
	}
	\,\,\mapsto\,\,
	\bb{
		t
		+ \frac{M}{4\pi}
		\mem,\,
		x + \frac{M}{4\pi}\mem \frac{x+iy}{|\vex|+z}
		\mem,\,
		y + \frac{M}{4\pi}\mem \frac{y-ix}{|\vex|+z}
		\mem,\,
		z + \frac{M}{4\pi}
	}
    \,.
\end{align}

This fact can be explained as follows.
Firstly,
suppose a vector field $\ell^\m$ in complexified Minkowski space
defines a geodesic congruence:
$\ell^\m{}_{,\n}\mem \ell^\n = 0$.
The time-$\e$ flow generated by this vector field
is simply
$x^\m \mapsto
x^\m + \e\mem \ell^\m(x)
$,
as
geodesics are straight lines in flat space.
The pull-back of the flat metric
$\eta =
    \eta_{\m\n}\mem dx^\m {\mem\otimes\mem} dx^\n$
is
$
    \eta +
        2\e\mem \ell_\wrap{(\m,\n)}\mem
    dx^\m {\mem\otimes\mem} dx^\n
        + \e^2\mem \eta_{\r\s}\mem \ell^\r{}_{,\m}\mem \ell^\s{}_{,\n}
    dx^\m {\mem\otimes\mem} dx^\n
$.

The null vector field (NVF) associated with the KS metric of the SD Taub-NUT solution,
\nomenclature{NVF}{Null Vector Field}
$\ell^\m = (1 , \zeta, -i\zeta , 1)$,
is stationary, geodesic, and shear-free.
With the self-duality condition,
they together imply that
$2\hem \partial_\n \ell^\m = \rho\mem \tilde{m}^\m m_\n$
when described with a null tetrad $(\ell^\m,n^\m = 2u^\m {\mem-\,} \ell^\m,m^\m,\tm^\m)$
where $u^\m := \delta^\m{}_0$.
The complex expansion computes to $\rho = 1/r$.

Thus,
it follows that
the flat metric transforms to
$\eta + \e\mem \rho\mem \tm {\mem\odot\mem} m$
if the geodesic flow $x^\m \mapsto x^\m + \e\mem \ell^\m(x)$
is used
as a diffeomorphism,
where $\odot$ denotes the symmetrized tensor product.
In turn,
the KS metric
$\eta + \phi\mem \ell^2$
transforms to
\begin{align}
    \eta \,+\, \e\mem \rho\mem \tm {\mem\odot\mem} m
    \,+\, \phi'\mem \ell^2
    \,,
\end{align}
where $\phi'(x) := \phi\bigbig{
    x {\,+\,} \e\mem \ell(x)\hnem
}$.

Meanwhile,
using $\eta = -2u {\mem\odot\mem} \ell + \ell^2
+ \tm {\mem\odot\mem} m$,
the Gibbons-Hawking metric
$-(1{\,+\,}\phi)^{-1}\mem
(\hem{
    u - \phi\mem (\ell{\,-\,}u)
}\hem)^2
+ (1{\,+\,}\phi)\mem (u^2 {\,+\,} \eta)$
boils down to
\begin{align}
\begin{split}
        \eta \,+\, \phi\mem \tm {\mem\odot\mem} m
        \,+\, (\phi^{-1} {\,+\,} 1)^{-1}\mem \ell^2
    \,.
\end{split}
\end{align}
Therefore, 
the KS metric
is mapped to the Gibbons-Hawking metric
if we can flow along the geodesic congruence
such that $\phi^{-1} {\,+\,} 1 = \phi'^{-1}$
and $\e\mem \rho = \phi$.
Clearly, $\e = M/4\pi$
satisfies both of these conditions.
\newpage

\section{ Physical Interpretation}%
We should now examine the physical validity of the
KS metric in \eqref{eq:g1}.
To begin with,
we point out that
\eqref{eq:g1}
\textit{itself}
is quite unusual
as a KS metric.
While the KS potential is simply
$\Phi = M/4\pi r$,
the null one-form
$\ell$
involves a \textit{semi-infinite string defect} along the negative $z$-axis,
which
is exactly how
the gravitomagnetic charge
is encoded
in this metric:
\begin{align}
    \label{eq:ell-zag}
    \ell
    \,=\,
        -dt
        + \frac{x{\,+\,}iy}{r{\,+\,}z}\mem dx
        + \frac{y{\,-\,}ix}{r{\,+\,}z}\mem dy
        + dz
    \,.
\end{align}
This is intriguing,
as the null one-form is usually regular
wherever the KS potential is regular---%
which is especially true for
the double KS construction of Taub-NUT \cite{Luna:2015paa}.

We expect that
the interpretation
of this string defect
should be the \textit{Misner string}
\cite{Misner:1963flatter,Bonnor:1969ala},
just as in the original metric in \eqref{eq:g0}:
the gravitational analog of the Dirac string,
inputting a gravitomagnetic flux of $-im$
at the origin.
To show this, we recall that
there are at least three senses in which
the Taub-NUT metric
\cite{Misner:1963flatter,Bonnor:1969ala}
describes
a gravitational analog of a dyon:
(a)
    the Misner string is
    invisible as
    a coordinate artifact
    if
    the gravitational analog of the Dirac quantization condition holds \cite{Misner:1963flatter},
(b)
    the Misner string implements
    a flux tube of
    time monodromy
    \cite{cho1991magnetic,dowker1967gravitational,Alfonsi:2020lub},
and
(c)
    the distributional stress-energy of the string
    describes
    an energy-less thin solenoid of mass current \cite{Bonnor:1969ala,sackfield1971physical}.
Our goal now is to
show that all of these properties
hold
in the KS description as well.

Firstly, we derive the Dirac quantization condition.
For the SD dyon gauge potential in \eqref{eq:A1},
adding a term
\smash{$(Q/2\pi)\mem d\log\zeta$}
repositions its Dirac string
along the positive $z$-axis (north pole).
This encodes
the complexified group-valued
transition function
\smash{$\zeta{}^{\hem ieQ/2\pi}$},
where $e$ is the electromagnetic coupling.
Demanding its single-valuedness
on the ``equator''
\cite{Wu:1975es}
implies
$ieQ \in 2\pi\mathbb{Z}$.

The gravitational Dirac quantization
can be derived
in a similar fashion,
i.e., by
hiding the strings with overlapping coordinate patches
\cite{Misner:1963flatter,dowker1967gravitational,cho1991magnetic,Alfonsi:2020lub}.
In the KS description,
we find that
the diffeomorphism
\begin{align}
\begin{split}
    \label{eq:reposition}
    t
        \,\,\mapsto\mem\,
    t
    + \frac{M}{2\pi} \log \zeta
    \,,\,\,\,\,
    \tzeta
        \,\,\mapsto\mem\,
    \frac{
        1 - (M/4\pi r) (1{\,+\,}\tzeta\zeta)/\tzeta\zeta
    }{
        1 + (M/4\pi r) (1{\,+\,}\tzeta\zeta)
    }\mem \tzeta
\end{split}
\end{align}
with $r$ and $\zeta$ fixed
transforms
the KS metric in
\eqref{eq:g1-stereographic} to
another KS metric:
\begin{align}
\begin{split}
    \label{eq:north-g1-stereographic}
    ds^2
    =
    {}&{}{
        -dt^2 + dr^2 + r^2\mem \frac{4\hem d\smash{\tzeta}\hem d\zeta}{(1{\,+\,}\tzeta\zeta)^2}
    }
        +
        \frac{M}{4\pi r}\mem
        \bigg({
            -dt
            + dr
            + r\mem \frac{2\hem d\smash{\zeta}}{\zeta(1{\,+\,}\tzeta\zeta)}
        }\mem\bigg)^{\nem\hnem2}
        \,,
\end{split}
\end{align}
which develops a string defect along the positive $z$-axis
(north pole).
For this diffeomorphism
to be single-valued within the overlapping region,
the time coordinate
should be periodic as $t \sim t + iM$,
given the $2\pi i$ periodicity of $\log \zeta$.
Hence
the energy $e$
of a test particle
is quantized as
$ieM \in 2\pi\mathbb{Z}$.
This is precisely the gravitational
Dirac quantization condition
due to the NUT charge $-iM$.

Secondly, we study the near-string geometries
in an explicit manner.
Let
$\xi := x+iy$,
$\smash{\txi} := x-iy$
be complex coordinates
for the (complexified) $x$-$y$ plane,
and suppose a positive infinitesimal $\e$.
Applying the replacement $z \mapsto -1/\e + z$
to the gauge potential in \eqref{eq:A1}
and then taking the limit $\e \to 0$,
we obtain
$A = (Q/2\pi)\mem d\log\smash{\txi}$.
This precisely describes an infinite Dirac string
generated by a large gauge transformation
\cite{tong2018gauge}.

\begin{figure}[t]
    \centering
    \includegraphics[valign=c,width=0.4\linewidth]{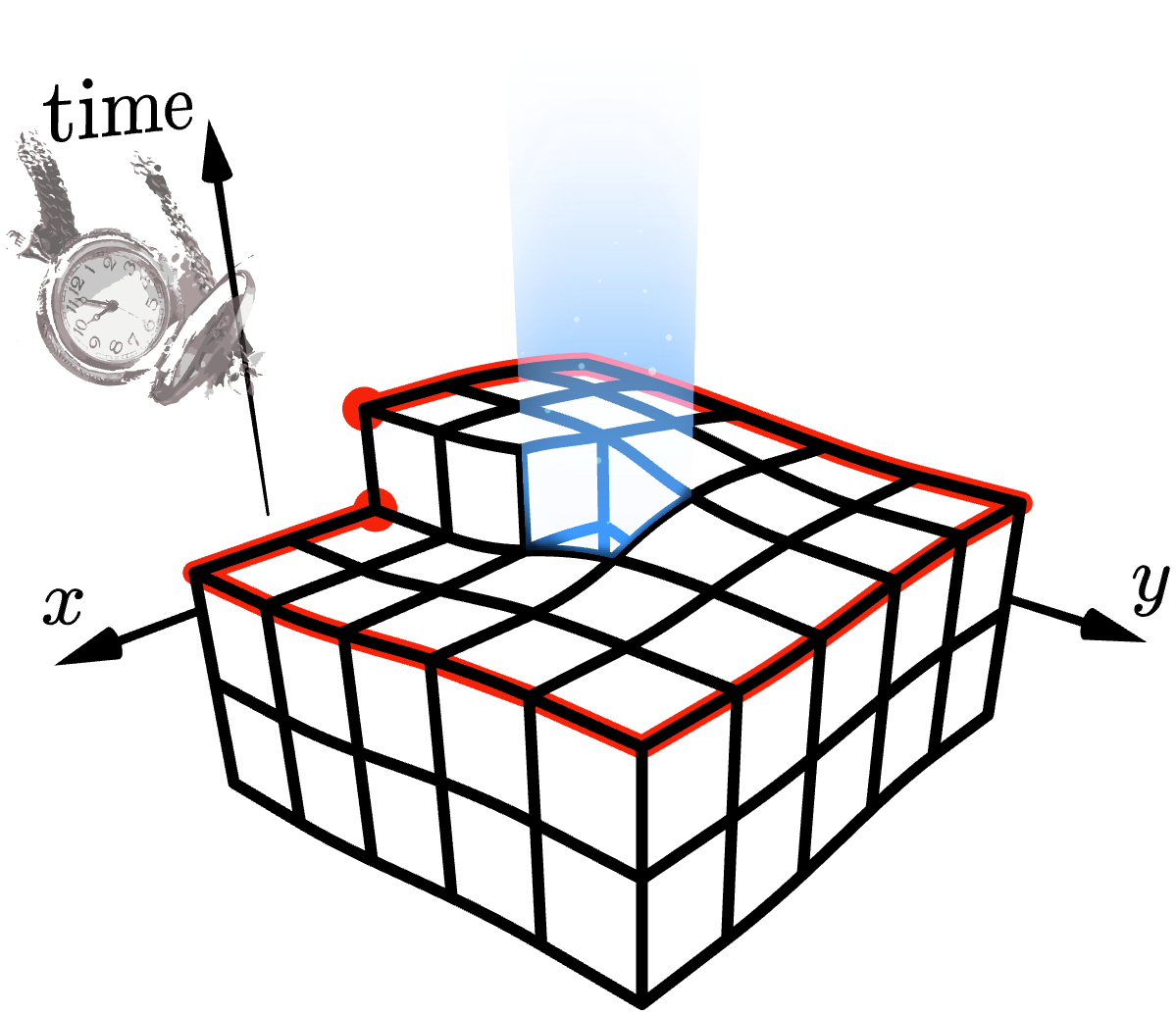}
    \smallskip
    \caption{%
        The Misner string is a ``dislocation''
        in spacetime,
        inducing time monodromy as an AB phase.
    }
\end{figure}

For the line element in
\eqref{eq:g1},
the same limiting procedure
leaves out a divergent part
that
can be compensated by
a shift
$\smash{\xi} \mapsto \smash{\xi} + (
    (M/\pi)\mem (1/\e{\,-\,}z)
    - 3M^2 \nem/4\pi^2
) /\smash{\txi}$.
Amusingly,
the resulting line element is
simply the Minkowski line element,
but with a multivalued time:
\begin{align}
    t' 
    \,=\,
        t \,+\, \frac{M}{2\pi}\mem \log\txi
    \,.
\end{align}
Namely,
in the vicinity of the string,
the geometry approaches the image of empty spacetime
under
a large diffeomorphism.
Here,
a Sagnac interferometer
measures a monodromy
in time as
the gravitational AB phase
\cite{sakurai1980comments,dowker1967gravitational,samuel1986gravitational,zimmerman1989geodesics},
localized along the $z$-axis (as $\smash{\txi} = 0$).
This precisely describes
an infinite Misner string
as a thin tube of gravitomagnetic flux $-iM$.\footnote{
    One might make a parallel between the field strength
    $F = dA$ and the torsion two-form $ddt'$
    (cf.\,\rrcite{Fiziev:1995te,Kleinert:1996yi,Kleinert:2008zzb})
    associated with these two strings,
    which are distributional.
}

Thirdly,
we derive the distributional stress-energy tensor
of this string.
Among various approaches
\cite{israel1970source,Balasin:1993kf,Balasin:1993fn},
we particularly
consider
the ``harmonic-gauge linearization''
of \rcite{vines2018scattering}
which
stationary KS metrics enjoy
as a luxury.
Provided
the congruence of $\ell^\m\mem \partial_\m$
is
geodesic and shear-free (GSF),
\nomenclature{GSF}{Geodesic and Shear-Free}
this approach
identifies
\smash{$T^{\m\n} {\,=\,} u^{(\m} J^{\n)}$}
as the source
for the double copy solution
\cite{vines2018scattering}.
Here
$J^\m$ is the single copy source,
while $u^\m {\,=\,} \delta^\m{}_0$
is the unit timelike Killing vector.
For our case,
the GSF property of the congruence associated with the null one-form in \eqref{eq:ell-zag}
can be established from an identity
\begin{align}
    \partial_\m \ell_\n
    =
    \frac{1}{2R}\mem
    \Big(\,{
        \eta_{\m\n}
        + \ell_\wrap{(\m} (2u{\,-\,}\ell)_\wrap{\n)}
        - i\mem \ve_{\m\n\r\s} u^\r \ell^\s
    }\mem\Big)
    \,,
\end{align}
while
the single copy source $J^\m$
easily follows by
acting on the Laplacian $\square := -\eta^{\m\n}\partial_\m\partial_{\n}$
to \eqref{eq:A0}.
Applying the formula
\smash{$T^{\m\n} {\,=\,} u^{(\m} J^{\n)}$}
then immediately reproduces
Bonnor's stress-energy tensor
\cite{Bonnor:1969ala,sackfield1971physical},
describing a point mass $m$
attached to
a massless
semi-infinite
rotating rod.

To sum up,
we have convinced ourselves
from multiple angles
that the KS metric in \eqref{eq:g1} indeed
exhibits the geometrical and physical
characteristics of
a Taub-NUT spacetime
arising from the Misner string.
This analysis
demonstrates
that
the KS description of the solution
is not merely a formal construct
but physical.
Moreover,
the quick derivation of the distributional source
demonstrates its usefulness
as well.

\section{ More Double Copy Correspondences}

Finally, 
for interested readers,
we discuss alternative classical double copy relations
that can be drawn
between SD Taub-NUT and SD dyon.
The possible options are
Weyl double copy
\cite{Luna:2018dpt,Chacon:2021wbr,white2021twistorial},
operator KS double copy
\cite{monteiro2014black,Berman:2018hwd,elor2020newman},
and SD double copy
\cite{monteiro2014black,monteiro2011kinematic}.

To this end, we first note that
our KS metric for the SD Taub-NUT solution
describes an elegant formula
in the spinor notation:
\eqref{eq:g1} translates to
\begin{align}
    ds^2
    \,=\,
        ds^2_\text{flat}
        \,+\,
            \frac{M}{4\pi R}\mem
            \BB{
                \eta_\a\mem \tdo_\da\mem
                dx^{\da\a}
            }^{\nem2}
    \,,
\end{align}
where the null one-form splits into two spinors as
\begin{align}
    \label{eq:lada}
    \ell^\m
    = (\s^\m)_{\a\da}\mem
        \tdo^\da \eta^\a
    \,,\quad
    \tdo^\da = \lrp{\begin{matrix}
        1 \\ \zeta
    \end{matrix}}
    \,,\quad
    \eta^\a = \lrp{\begin{matrix}
        1 \\ 0
    \end{matrix}}
    \,.
\end{align}
Here, the SD spinor $\tdo^\da$ 
describes
a ``square root'' of the position three-vector $\vex$
in essence,
while the ASD spinor $\eta^\a$
is a constant reference
controlling the direction of the Misner string.
It can be seen that the former encodes physical information,
whereas the latter encodes pure gauge information.

To elaborate,
computation shows that
the principal spinors
for the KS metric in \eqref{eq:g1}
are
$\tdo^\da = (1,\zeta)$
and
$\ti^\da {\,=\,} ( -\smash{\tzeta} , 1 ) /(1{\,+\,}\smash{\tzeta}\zeta)$,
the former of which is precisely the SD spinor in \eqref{eq:lada}.
It is interesting to note that
the \textit{real} combinations
$[\tdo_\da]^*\hem \tdo_\db$,
$[\ti_\da]^*\hem \ti_\db$
of these principal spinors
exactly reproduce
the null one-forms $-dt \pm dr$
associated with the
KS metrics of the
Schwarzschild solution
as envisioned in
\rrcite{white2021twistorial,Chacon:2021wbr,gabriel1}.\footnote{
    Cf.
    the GSF spinor equation
    \cite{penrose:1986spinors2,huggett1994introduction}
    is chiral.
}

\subsection{Weyl Double Copy}%

The \textit{Weyl double copy}
\cite{Luna:2018dpt,Chacon:2021wbr,white2021twistorial}
seeks to directly construct the Weyl curvature
of four-dimensional algebraically special spacetimes
by ``squaring''
electromagnetic field strengths
via a choice of a scalar field (``zeroth copy field'')
while not
referencing
a metric nor a gauge potential.
It has been formulated in both
position \cite{Luna:2018dpt} and dual twistor \cite{Chacon:2021wbr,white2021twistorial} spaces.

Below, we establish that
the SD Taub-NUT solution
is the Weyl double copy of the SD dyon
in both position space and dual twistor space senses.

First of all,
direct computation based on \eqrefs{eq:A1}{eq:g1}
shows that
the SD curvature tensors
(field strength and Weyl)
of SD dyon and SD Taub-NUT solutions
are given by
\begin{align}
\begin{split}
    \label{eq:NPformalism}
    F_\wrap{\da\db}
    \,\propto\,
    \frac{Q}{4\pi R^2}\,
        \tdo_\wrap{(\da} \ti_\wrap{\db)}
    &\,,\quad
    C_\wrap{\da\db\dc\dd}
    \,\propto\,
    \frac{M}{4\pi R^3}\,
        \tdo_\wrap{(\da} \tdo_\wrap{\db}
        \hem
        \ti_\wrap{\dc} \ti_\wrap{\dd)}
    \,,
\end{split}
\end{align}
respectively.
Of course, the ASD curvature tensors vanish.
The Weyl double copy in position space refers to
the squaring relation \cite{Luna:2018dpt}
\begin{align}
    \label{weyldc-pos}
    C_\wrap{\da\db\dc\dd} 
    \,=\,
        F_\wrap{(\da\db} F_\wrap{\dc\dd)}
        \big/S
    \,.
\end{align}
We see that \eqref{weyldc-pos} holds for \eqref{eq:NPformalism}
with the choice $S \propto 1/R$.

Second of all,
we see that
the above
$S$, $F_\wrap{\da\db}$, and $C_\wrap{\da\db\dc\dd}$
satisfy the spin-$h$ zero-rest-mass equations
in flat background
for $h=0,1,2$, respectively.
As such, they arise from
the Penrose transform \cite{Penrose:1969ae}
as
\begin{align}
    \label{eq:Penrose}
    \phi_{\da_1\cdots\da_{2h}}\nem(x)
    \,=\mem
    \oint \frac{
        \pi_\dc\hem d\pi^\dc
    }{2\pi i}\,
        \frac{
            \pi_{\da_1} \nem\cdots \pi_{\da_{2h}}
        }{
            Q(\pi,\pi x)^{h+1}
            \vphantom{\tilde{Z}}
        }
    \,.
\end{align}
Here, $Q(\pi,\pi x)$ evaluates
the following dual twistor quadric
\cite{penrose-TN1-10,sparling-TN1-14,hughston1979advances,white2021twistorial}
on the dual twistor line:
\begin{align}
    \label{eq:F}
    Q(W)
    \,=\,
        \omega^\a\mem \delta_{\a\da}\mem \pi^\da
    \,,\quad
    \delta_{\a\da} \,\propto\, u_\m (\s^\m)_{\a\da}
    \,,
\end{align}
where $W^\rmA = (\pi_\da,\omega^\a)$.
Note that the zeros of $Q(\pi,\pi x)$
encode the principal spinors
$\tdo_\da,\ti_\da$
as per
the Kerr theorem
\cite{penrose1967twistoralgebra,penrose:1986spinors2,Newman:2004ba}.
From this analysis,
we identify a relation
between dual twistor space data,\footnote{
    See \rcite{Guevara:2021yud} for 
    a scattering amplitudes interpretation,
    where
    the dual twistor data via the Penrose transform
    is identified as the three-point scattering amplitude for absorbing one negative-helicity massless quantum,
    half-Fourier transformed to the dual twistor space.
}
\begin{align}
    \label{weyldc-twistor}
    \mathcal{M}_h(W)
    \,=\,
        Q(W)^{-h-1}
    \qiq
    \mathcal{M}_2(W)
    \,=\,
        \bigbig{
            \mathcal{M}_1(W)
        \hnem}^{\nem2}
        \big/
        \mathcal{M}_0(W)
    \,.
\end{align}
\eqref{weyldc-twistor}
is the squaring relation for
the Weyl double copy in dual twistor space
\cite{white2021twistorial,Chacon:2021wbr}.

Previously,
the above squaring relations
for the SD Taub-NUT solution
had been speculated on in
\rrcite{Chacon:2021wbr,white2021twistorial}.
The argument was that
the SD Taub-NUT solution would describe the ``SD part'' of Schwarzschild,
so its SD Weyl curvature
and principal spinors
might be exactly equated to
those of the Schwarzschild solution.
Although physically plausible,
it should be pointed out that
this claim is
strictly speaking
a statement whose validity is restricted within the linearized level.
Otherwise, it is a highly nontrivial assertion
in light of the nonlinearity of GR
demanding an extra effort for justification.

The clarification which
our explicit KS metric in \eqref{eq:g1} brings
is that
\eqref{eq:NPformalism}
has in fact described the \textit{exact} Weyl tensor
of the SD Taub-NUT solution at the fully nonlinear level.
The Cartesian coordinates $x^{\da\a}$ of the flat background
employed in the Penrose transform in \eqref{eq:Penrose}
can be identified with the KS coordinates for the exact curved SD Taub-NUT geometry.\footnote{
    Note that
    the ``position vector'' $\vex$
    which one identifies from the Killing spinor
    as
    \smash{$
        \tchi^\da{}_\db
        = \minie\mem (\vex {\mem\cdot\mem} \vec{\sigma})^\da{}_\db
    $}
    \cite{Adamo:2023fbj}
    also describes the (spatial part of) our KS coordinates.
}
Consequently,
the Weyl double copy construction
of the SD Taub-NUT solution
is validated
at the fully nonlinear orders.

Generally speaking,
it seems that the validity of the Weyl double copy
could only be strictly ensured
at the nonlinear orders
when 
it can be assisted by
an underlying
KS structure at the metric level,
in which case one can appeal to
the effective linearization property \cite{xanthopoulos1978exact,Harte:2016vwo}.

\subsection{Nonlocal Operator Kerr-Schild Double Copy}%

Another known instance of a classical double copy construction
is the operator KS double copy
\cite{monteiro2014black,Berman:2018hwd,elor2020newman}.
The operator KS double copy
promotes
the null one-form $\ell_\m$
in the KS ansatz
to a differential operator $\smash{\hat{k}_\m}$,
which
may factorize as
$-\minie\mem (\s^\m)_{\a\da}\mem \smash{\hat{k}_\m} = \o_\a \smash{\hat{k}_\da}$
so that the fields are
manifestly
put in the lightcone gauge
\cite{monteiro2014black,Berman:2018hwd,elor2020newman}.

Intriguingly,
we find that
an operator KS double copy
for the SD Taub-NUT solution
is viable
only in an exotic \textit{nonlocal} form.
\eqrefs{eq:A1}{eq:g1}
arise precisely if
the operator
\smash{$\hat{k}_\da$} is given by
\begin{align}
    \label{eq:khat}
    \hat{k}_\da
    \mem=\hem
    \bigg(\mem{
        \hhnem-
        \frac{
            \partial_x {\mem+\,} i\partial_y
        }{
            \partial_z
        }
        \mem,\mem
        1
    }\mem\bigg)
    \,,
\end{align}
which is nonlocal.
From
$\smash{\hat{k}_\da} (1/r) = (1/r)\mem \tdo_\da$
and
$\smash{\hat{k}_\da}\smash{\hat{k}_\db} (1/r)
= \smash{(1/r)\mem \tdo_\da \tdo_\db}$,
it follows
that
a double copy correspondence
between
SD dyon gauge potential
and 
the KS graviton field $h_{\m\n} = g_{\m\n} - \eta_{\m\n}$ of
SD Taub-NUT
is realized as
\begin{align}
    \label{eq:operator-dc}
    A_{\a\da} =
    \o_\a \smash{\hat{k}_\da}
    \mem
        \frac{q}{4\pi r}
    \,,\quad
    h_\wrap{\a\da\b\db} =
    \o_\a\o_\b\mem \smash{\hat{k}_\da}\smash{\hat{k}_\db}
    \mem
        \frac{m}{4\pi r}
    \,.
\end{align}

Notably,
this new form of classical double copy
admits a nice physical explanation
from an electric-magnetic dual perturbation theory.
The key fact is that
the SD dyon can be viewed as an object in the ``extended'' Maxwell theory
where the violation of Bianchi identity
by \textit{magnetic matter} $J^\star$
is allowed by the introduction of the dual gauge potential $A^\star$
\cite{Dirac:1948um}:
$F = dA + \ast dA^\star$.
In this picture, the distributional source for the SD dyon is
\begin{align}
\begin{split}
    \label{eq:Jdual}
    J^\m
    =
        \partial_\n F^{\m\n}
    &=
    qu^\m\mem\delta^{(3)}\hnem(\vex)
    \,,\\
    J^{\star \m}
    =
        -\partial_\n {*}F^{\m\n}
    &=
    -iqu^\m\mem\delta^{(3)}\hnem(\vex)
    \,.
\end{split}
\end{align}
We may redescribe these in the ``helicity basis'' as
$J_+^\m = J^\m - i\mem J^\star{}^\m = 0$
and
$J_-^\m = J^\m + i\mem J^\star{}^\m = 2qu^\m\mem \smash{\delta^{(3)}(\vex)}$,
while having
\smash{$A^\pm_\m := \minie\mem (A^{\vphantom{+}}_\m {\,\pm\,} iA^\star_\m)$}
for the potentials.

As a dynamical field theory,
a formulation of this extended Maxwell theory
has been given by Zwanziger
\cite{Zwanziger:1970hk,Gubarev:1998ss,Shnir:2011zz,Terning:2020dzg,Moynihan:2020gxj}.
We note that
the propagators of this theory
can be summarized into
a single formula in the spinor notation:
\begin{align}
\begin{split}
   \label{eq:zprop}
   \frac{
    \hat{\Delta}^{+-}_{\wrap{\a\da\b\db}}
   }{\Box}
   =
   \frac{
       n_{\wrap{\a\db}}\hem \partial_{\wrap{\b\da}}
   }{-n\mdot \partial}
   \frac{1}{\square}
   \,,
\end{split}
\end{align}
where $n$ is an auxiliary four-vector
that traces back to the Dirac string.
\eqref{eq:zprop}
describes the position-space propagator
from \smash{$A^-_\wrap{\b\db}$} to \smash{$A^+_\wrap{\a\da}$},
while the $\langle A^- A^- \rangle$ and $\langle A^+ A^+ \rangle$
propagators vanish.
A simple calculation then shows that the SD dyon gauge potential in \eqref{eq:A1}
is precisely reproduced by this propagator,
if one takes
$n_\wrap{\a\da} = { -\o_\a \bo_\da }$
for the auxiliary vector
so that
$\smash{\hat{k}}_\da = {- \partial_\wrap{1\da} / \partial_\wrap{0\dot{0}} }$.

Amusingly,
the graviton field
arises in the same fashion by
using the linearized gravity equivalent of Zwanziger propagator,
which is given by doubling \eqref{eq:zprop}
as is advocated in \rcite{Moynihan:2020gxj}.
In this case,
the sources in the two-potential theory
are
$T_{+}^{\m\n} = 0$ and
$T_{-}^{\m\n} = 2m u^{\m} u^{\n}\mem \smash{\delta^{(3)}\hnem(\vex)}$.
The resulting linearized metric perturbation
happens to be
also the full metric perturbation
by taking the KS form
\cite{Harte:2016vwo,xanthopoulos1978exact,vines2018scattering}.

In summary,
the double copy structure proposed in \eqref{eq:operator-dc}
can be traced back to
propagators between dyonic matter
in
electrodynamics and linearized gravity:
\begin{align}
    A^+_\wrap{\a\da}
    =
        \frac{
            \hat{\Delta}^{+-}_\wrap{\a\da\b\db}
        }{\Box}\,
            J_-^{\db\b}\mem
    ,\,\,\,\,
    h^+_\wrap{\a\da\b\db}
    =
        \frac{
            \hat{\Delta}^{+-}_\wrap{\a\da\c\dc}\hem
            \hat{\Delta}^{+-}_\wrap{\b\db\d\dd}
        }{\Box}\,
            T_-^{\dc\c\dd\d}\mem
    .
\end{align}
A few comments are in order.
Firstly,
it is instructive to translate the propagator in \eqref{eq:zprop}
into the vector notation,
where one can identify a combination of operators
\smash{$(\vec{n}\mtimes\vp/\vec{n}\mdot\vp)$}
connecting between the Gilbertian and Amp\`erian implementations \cite{griffiths3ed}
of multipoles.
This implies that the nonlocality of
the operator in \eqref{eq:khat}
has encoded the Dirac string structures.
Secondly,
our construction
could be justified
from the on-shell perspective.
The ``numerator'' of
the off-shell propagator in \eqref{eq:zprop}
correctly factorizes into the null polarization vectors
so that
the double copy of sources from
$\smash{J_-^\m} = \smash{2qu^\m\mem \delta^{(3)}\hnem(\vex)}$
to
$\smash{T_-^{\m\n}} = \smash{2mu^\m u^\n\mem \delta^{(3)}\hnem(\vex)}$
encodes
the squaring relation of the massive on-shell amplitudes
\cite{Huang:2019cja,Emond:2020lwi,ahh2017,Moynihan:2020gxj}.
In this perspective,
the nonlocality traces back to that of the $x$-factor.
Lastly,
it is interesting to
apply this approach to construct
generic Taub-NUT spacetimes,
in which case one naturally obtains a double KS metric perturbation.

\subsection{Self-Dual Double Copy}%
\label{NOTE-SDTN>SD dyonC}

Computation shows that
our KS coordinates are also
the Pleba\'nski coordinates
\cite{plebanski1975some,Adamo:2021bej}
for the second heavenly equation,
on which the
SD double copy
\cite{monteiro2011kinematic,monteiro2014black,Berman:2018hwd}
is based.
The SD double copy
could draw fascinating conclusions about
the SD dyon and Taub-NUT solutions
in the context of
Moyal deformation of the heavenly equation
\cite{Plebanski:1994qi,Park:1990fp,park19922d,Husain:1993dp,Cheung:2022mix,Armstrong-Williams:2022apo}.
Strangely,
however,
the SD Taub-NUT metric
is not straightforwardly
the SD double copy
of SD dyon
in terms of the Pleba\'nski scalar field:
$A_{\a\da} = \o_\a \partial_{0\da}
(\hem{
    q\zeta/4\pi
}\hem)$,
$h_\wrap{\a\da\b\db}
=
    \o_\a \o_\b\mem \partial_\wrap{0\da} \partial_\wrap{0\db}
    (\hem{
        m (2r{\,+\,}z)\mem \zeta^2
        \hnem/24\pi
    }\hem)
$.
This puzzle deserves further investigation,
in comparison with the
Eguchi-Hanson case \cite{Berman:2018hwd}.

\section{ Conclusions}%
\label{NOTE-SDTN>CONC}

In this work, we derived a KS metric for the SD Taub-NUT solution
by classical double copy
and validated it by an explicit complexified coordinate transformation.
The KS metric exhibits the Misner string defect,
supporting its physical interpretation.
Overall,
we established
three concrete senses
in which
the SD Taub-NUT solution
is the gravitational counterpart of
the SD dyon in electromagnetism:
KS,
Weyl,
and nonlocal operator KS
classical double copies.
The last one
adds to
the inventory of
classical double copies
as a new entry,
whose physical origin
is propagators between dyonic matter.

The distinctive charm of this work
lies in the exciting development
along the very first three equations.
The usual punchline of
classical double copy
has been
starting with a gravitational solution
and then
deducing the electromagnetic counterpart.
This work,
in contrast,
takes a reverse path:
we constructed an electromagnetic solution
and then applied the double copy to yield a new statement in GR.

It is also insightful to frame this work
in the angle of
``classical double copy of gravitational instantons.''
The only known instance has been
\rcite{Berman:2018hwd}'s work on
the Eguchi-Hanson instanton solution
\cite{Eguchi:1978xp,Eguchi:1978gw,Eguchi:1979yx},
which traces back to
\rrcite{Tod:1982mmp,Sparling:1981nk}.
This work
adds another instance to the discourse,
namely
the Taub-NUT instanton.
It is amusing that
the correspondence between
SD objects in electromagnetism and gravity
which the Gibbons-Hawking
ansatz \cite{hawking1977gravitational,Gibbons:1978tef}
had suggested a long time ago
becomes strengthened from new perspectives
by the advent of the classical double copy program.

The KS metric
of the SD Taub-NUT solution
conveys
an aesthetic elegance
in terms of
its
minimalist
and
spinorial nature.
We anticipate that
this new metric
will contribute to
future studies on the SD Taub-NUT solution,
both conceptually and practically.
For further consequences or applications,
we suggest the following directions:
    asymptotic symmetries \cite{Godazgar:2019dkh,Godazgar:2019ikr},
    integrability of
    test-particle motion
    \cite{Gibbons:1986hz,Gibbons:1987sp},
    black hole
    perturbation theory
    \cite{Guevara:2023wlr,Adamo:2023fbj,Teukolsky:1973ha,Press:1973zz},
    exact graviton propagators
    \cite{Sasaki:2003xr,Mano:1996vt,Bonelli:2021uvf},
    or self-force calculations
    \cite{Poisson:2011nh,DeWitt:1960fc,Barack:1999wf,Cheung:2023lnj}
    in an idealized setup.
Especially,
explorations in the
scattering context
are likely to
shed insights on
the Kerr black hole scattering
for an integrable subsector,
on account of
the Newman-Janis shift
\cite{Newman:1965tw-janis}:
the ``SD part'' of Kerr
is a SD Taub-NUT solution
\cite{Newman:1973yu,Newman:2002mk,Ghezelbash:2007kw,Crawley:2021auj,gabriel1,Adamo:2023fbj}.

\begin{subappendices}

\section{ Double Kerr-Schild Metric}
\label{NOTE-SDTN>DKS}

Consider the following double KS metric:
\begin{align}
\begin{split} 
    \label{tndks}
    ds^2
    \,=\,
    \lrp{
    \begin{aligned}[c]
        &
            - dt^2 + dx^2 + dy^2 + dz^2
        \\
        &
        +
            \frac{M_1}{4\pi|\vex|}\,
            \bb{
                - dt + \frac{x+iy}{|\vex|+z}\mem dx + \frac{y-ix}{|\vex|+z}\mem dy + dz
            }^{\nem\nem2}
        \\
        &
        +
            \frac{M_2}{4\pi|\vex|}\,
            \bb{
                - dt + \frac{x-iy}{|\vex|-z}\mem dx + \frac{y+ix}{|\vex|-z}\mem y - dz
            }^{\nem\nem2}
    \end{aligned}
    }
    \,.
\end{split}
\end{align}
\eqref{tndks} is a double KS metric for the SD Taub-NUT solution with mass $M_1 + M_2$
according to Pleba\'nski and Demia\'nski \cite{PD}
(which arises by a zooming-in procedure
taking $a \to \infty$).

It can be explicitly shown that
\eqref{tndks}
is diffeomorphic to a Gibbons-Hawking metric
with mass $M_1 \mplus M_2$.
Explicitly,
apply the diffeomorphism
\begin{align}
\begin{split}
    \label{tndks.diff}
    t
	&\,\,\,\mapsto\,\,\,
    t + \frac{M_1}{4\pi} + \frac{M_2}{4\pi}
    \,,\\
    x
	&\,\,\,\mapsto\,\,\,
		x + \frac{M_1}{4\pi}\, \frac{x+iy}{|\vex|+z} + \frac{M_2}{4\pi}\, \frac{x-iy}{|\vex|-z}
    \,,\\
    y
	&\,\,\,\mapsto\,\,\,
		y + \frac{M_1}{4\pi}\, \frac{y-ix}{|\vex|+z} + \frac{M_2}{4\pi}\, \frac{y+ix}{|\vex|-z}
    \,,\\
    z
	&\,\,\,\mapsto\,\,\,
		z + \frac{M_1}{4\pi} - \frac{M_2}{4\pi}
	\,,
\end{split}
\end{align}
which induces $|\vex| \mapsto |\vex| + (M_1 + M_2)/4\pi$.
Evidently, \eqref{tndks.diff} arises by
taking the sum of the two NVFs involved in \eqref{tndks}.
Computation shows that \eqref{tndks.diff} maps \eqref{tndks} to the Gibbons-Hawking metric for
\begin{align}
    V \,=\,
        1 + \frac{(M_1 \mplus M_2)}{4\pi|\vex|}
    \,,\quad
    a \,=\,
        a_1 + a_2
    \,,
\end{align}
where
\begin{subequations}
\label{a1a2.cart}
\begin{align}
	a_1
	\,&=\,
        \frac{M_1}{4\pi|\vex|}\mem\bb{
            \frac{x+iy}{|\vex|+z}\mem dx 
            + \frac{y-ix}{|\vex|+z}\mem dy
            + dz
        }
	\,,\\
	a_2
	\,&=\,
        \frac{M_2}{4\pi|\vex|}\,
        \bb{
            \frac{x-iy}{|\vex|-z}\mem dx + \frac{y+ix}{|\vex|-z}\mem dy - dz
        }
	\,.
\end{align}
\end{subequations}
In spherical coordinates, \eqref{a1a2.cart} boils down to
\begin{subequations}
\label{a1a2.sph}
\begin{align}
	a_1
	\,&=\,
        \frac{M_1}{4\pi}\mem
        \bb{
    		\frac{1}{R}\, dR
    		- \tan \frac{\theta}{2}\, d\theta
    		- i\mem (1\mminus\nem\cos\theta)\, d\phi
        }
	\,,\\
	a_2
	\,&=\,
        \frac{M_2}{4\pi}\mem
        \bb{
    		\frac{1}{R}\, dR
    		+ \cot \frac{\theta}{2}\, d\theta
    		+ i\mem (1\mplus\nem\cos\theta)\, d\phi
        }
	\,,
\end{align}
\end{subequations}
from which it is clear that
$da = d(a_1 \mplus a_2) = -i\hem ((M_1\mplus M_2)/4\pi) \sin\theta\, d\theta \swedge d\phi$.

A corollary is that \eqref{tndks} for $M_2 = -M_1$
provides a double KS representation of empty spacetime.

\section{ A Puzzle Regarding Two Classical Double Copies}
\label{NOTE-SDTN>TWOROOTS}

The amplitudes-level double copy
of BCJ \cite{BCJ1}
implies that
the perturbiner (Duff \cite{Duff:1973zz}) constructions
of classical solutions in YM theory and gravity
can exhibit a correspondence
through replacing the color Lie algebra with
a kinematic Lie algebra,
whose identity is known in the SD sector \cite{monteiro2011kinematic}
but not beyond.
In the meantime,
the classical double copy
originally
proposed by Monteiro and O'Connell \cite{monteiro2014black}
has resorted to the effective linearization property \cite{xanthopoulos1978exact,Harte:2016vwo} of KS metrics.

To our best understanding,
an explicit demonstration of
the former implying the latter
has been not given so far.
First, the KS double copy program has not discussed
a faithful diagram-by-diagram mapping
at the nonlinear orders.
Second, such a mapping is
unlikely to be identified easily after all,
due to the mysterious status of the kinematic algebra beyond the SD sector.
Hence to address the question of whether
the amplitudes double copy implies the classical double copy,
it could be more promising to first focus on the SD sector
(cf. \rcite{Raeymaekers:2025akr}).

In this context,
\rcite{note-sdtn}
identified a puzzle that
there seems to be a disagreement between
the amplitudes double copy
and
the classical double copy.
On the one hand,
\rcite{note-sdtn}'s KS metric
shows that
the SD Taub-NUT solution
is the classical double copy \cite{monteiro2014black}
of the SD dyon,
which describes a correspondence between
type [1,1] and type D solutions.
On the other hand,
the amplitudes double copy 
via the second heavenly equation
\cite{monteiro2014black}
works by taking the same
Pleba\'nski scalar
for both gauge theory and gravity
up to attaching a constant color factor.
\rcite{note-sdtn} explicitly constructs
the Pleba\'nski scalars
for the SD dyon and the SD Taub-NUT solutions
and finds mismatch.

Taking lessons from a recent work
\cite{AAS},
we realize that
the mapping
stipulated by the amplitudes double copy
generically
describes a correspondence between
type [2] (null Maxwell field) and type D solutions.
This is because Tod \cite{Tod:1982mmp}'s construction
essentially concerns
the linearized case of the second heavenly equation.
This validates the mismatch
pointed out in this chapter \cite{note-sdtn}.
Consequently,
we learn that
all SD black holes seem to have ``two square roots.''

\end{subappendices}
\chapter[
    Newman-Janis Algorithm from Taub-NUT Instantons
]{
    Newman-Janis Algorithm from
    \\[0.4\baselineskip]
    Taub-NUT Instantons
}
\label{K3:NJA}

It is shown that the Kerr metric represents the nonlinear superposition of self-dual and anti-self-dual Taub-NUT instantons.
This promotes the Newman-Janis algorithm to a rigorous derivation of the Kerr metric with a definite physical origin.
In the same way, the Kerr-Newman and charged Kerr-Taub-NUT solutions are systems of Taub-NUT instantons and chiral dyons.

\begin{fullnote}
    This chapter reproduces the contents of \rcite{nja}, \fullcite{nja}.
\end{fullnote}

\section{ Introduction}%
The NJA \cite{Newman:1965tw-janis} 
is
a mystery in 
the history of relativity.
Discovered
about sixty years ago,
the NJA
aims to
derive the
rotating black hole solution of Kerr \cite{Kerr:1963ud}
by applying a ``complex coordinate transformation''
to the 
static black hole solution of Schwarzschild.
Apparently, this procedure
is a sequence of ad hoc manipulations.
Each element in the Schwarzschild metric
must be
promoted to 
specific combinations of complex variables,
which are then transformed by a rule
of unknown physical origin.
Hence
the NJA has stood merely as
a formal mathematical trick.

Such a lack of clear understanding
is unsettling
because the NJA
holds the historical significance
as the very method that led to the discovery of
the Kerr-Newman metric
\cite{Newman:1965my-kerrmetric}:
the charged rotating black hole solution.

Despite early skepticism---%
epitomized by characterizations such as
a ``fluke'' \cite{Drake:1998gf}
or ``methods which transcend logic''  \cite{ernst1968new2}---%
research efforts have persisted over decades
toward a satisfactory justification
\cite{%
	Erbin:2014aja,Erbin:2014aya,Erbin:2016lzq,Brauer:2014wwa,Keane:2014sta,Ferraro:2013oua,%
	Whisker:2008kk,Erbin:2015pla,Erbin:2014lwa,Mirzaiyan:2017adt,Tavakoli:2020uzr,Yazadjiev:1999ce,CiriloLombardo:2004qw,Roberts:1988un,bambi2013rotating,%
	Azreg-Ainou:2014pra,ernst1968new1,ernst1968new2,%
	Rajan:2015ffs,Rajan:2016qiy,Rajan:2016zmq,giampieri1990introducing,quevedo1992complex,harvey1989complex,santos1975newtonian,flaherty1976hermitian,Aksteiner:2022bwr,%
	Ayon-Beato:2015nvz,Lan:2024wfo,kamenshchik2023newman,beltracchi2021physical1,beltracchi2021physical2,Canonico:2011lba,canonico2010theoretical,%
	Drake:1997hh,herrera1982complexification,ibohal2005rotating,viaggiu2006interior,%
	Talbot:1969bpa,schiffer1973kerr,Gurses:1975vu,Finkelstein:1974nr,demianski1972new,demianski1966combined,Carter:1968rr,Drake:1998gf,%
	Kerr:2007dk,Adamo:2014baa,Adamo:2009vu,%
	newman1988remarkable,newman1974curiosity,newman1974collection,%
	Newman:1973yu,Newman:1973afx,Newman:2002mk,Newman:2004ba,%
	Newman:1976gc,ko1981theory,grg207flaherty,%
	sst-asym,ambikerr1,gmoov,%
	note-sdtn,%
	Crawley:2021auj,%
	gabriel1,%
	Guevara:2018wpp,Guevara:2019fsj,chkl2019,aho2020%
}.
These works may be classified into three categories, as outlined below.

The first category could be named
\textit{classical explanations}.
Here, the goal is to justify the NJA
from geometrical perspectives.
This means to
resolve inherent ambiguities,
identify hidden assumptions,
and explore simplifications or generalizations.
Notable contributions 
in this direction
include works by
Talbot \cite{Talbot:1969bpa},
Drake and Szekeres \cite{Drake:1998gf},
G\"urses and G\"ursey \cite{Gurses:1975vu},
Flaherty \cite{flaherty1976hermitian,grg207flaherty},
Rajan and Visser \cite{Rajan:2016zmq},
and
Giampieri \cite{giampieri1990introducing}.
The original article \cite{Newman:1965tw-janis}
also provides a brief justification
along this direction:
an allowed possibility in Kerr theorem \cite{penrose1967twistoralgebra}.
These appr\-oaches could be agnostic about physical origins, however.

The second category describes
\textit{the works of Newman himself}
\cite{%
	newman1988remarkable,newman1974curiosity,newman1974collection,%
	Newman:1973yu,Newman:1973afx,Newman:2002mk,Newman:2004ba,%
	Newman:1976gc,ko1981theory,grg207flaherty%
},
which portray a more serious stance.
Emphasis is put on the fact that
the NJA realizes the spin of Kerr black hole
via an imaginary displacement into complexified spacetime.
The NJA is treated not as a mere mathematical curiosity
but a signal of a deeper physical structure,
such as
a Hodge duality
on angular momenta as
mass dipole moments
\cite{newman1974curiosity}.
A fundamental unification of spin and spacetime
is envisioned from
the construction of ``complex center of mass''
\cite{newman1974curiosity,newman1974collection,Newman:1973yu,Newman:1973afx,Newman:2002mk,Newman:2004ba,Newman:1976gc,ko1981theory,grg207flaherty,sst-asym}.

Lastly, the third category
describes the \textit{modern explanation} emerged through works
\cite{ahh2017,Guevara:2018wpp,Guevara:2019fsj,chkl2019,aho2020}.
Here, one adopts
a particle physicist's perspective
and imagines observing the Kerr black hole
by throwing gravitons
(i.e., through gravitational waves)
from far away,
as if one detects the internal structure of a proton
by firing electrons 
in a particle collider.
In terms of scattering cross-sections,
it is revealed
that
the Kerr black hole
behaves like 
an ideally pointlike object
when the incident gravitons are all
circularly polarized
in either right or left handedness,
i.e.,
self-dual (SD) or anti-self-dual (ASD)
\cite{Guevara:2018wpp,Guevara:2019fsj,chkl2019,aho2020,Johansson:2019dnu,Aoude:2020onz,Lazopoulos:2021mna,zihan23,fabian2}.
Thus,
despite the black hole's
extended geometrical features
such as the ring singularity or the horizon,
its ``graviton X-ray image''
displays just two objects:
a point that
absorbs SD gravitons only 
and another 
point
that
absorbs ASD gravitons only.
This stunning simplicity
is identified as the basis of NJA \cite{aho2020},
as
the coordinates of such points
precisely predict Newman's complex centers of mass
by taking complex values.

\begin{figure}[t]\centering
	\centering
	\includegraphics[scale=3.0,valign=c,
	clip= true,
	trim= 2.5pt 2.5pt 2.5pt 2.5pt
	]{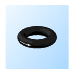}
	\quad
	\adjustbox{scale=1.5,valign=c}{$=$}
	\quad
	\includegraphics[scale=3.0,valign=c,
	clip= true,
	trim= 2.5pt 2.5pt 2.5pt 2.5pt
	]{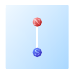}
    \medskip
	\caption[%
		In electromagnetism,
		a current loop
		behaves like
		a pair of
		opposite
		monopoles:
		Amp\`erian and Gilbertian dipoles.
		This letter reveals
		a gravitational version of this duality.
    ]{%
		In electromagnetism,
		a current loop
		behaves like
		a pair of
		opposite
		monopoles:
		Amp\`erian and Gilbertian dipoles
		\cite{griffiths3ed}.
		This letter reveals
		a gravitational version of this duality.
	}
	\label{fig:gad}
\end{figure}

The modern explanation shows that
the NJA is not 
a mere mathematical construct
but rather a representation of some physical structure:
the NJA is recast as a statement about scattering amplitudes,
a gauge-invariant physical observable.
That said,
it did not decipher every aspect of the NJA
in its original, relativist form
as a statement about bulk metrics.
The conversion of the scattering amplitudes to bulk metrics
is based on
perturbative methods
\cite{Duff:1973zz,Neill:2013wsa,xanthopoulos1978exact,Harte:2016vwo,vines2018scattering},
so a resummed picture is lacking.

A similar caveat applies to the fact
attributed to
Crawley, Guevara, Miller, and Strominger \cite{Crawley:2021auj}
or
Ghezelbash, Mann, and Sorkin \cite{Ghezelbash:2007kw},
which is limited to the SD sector
and thus fall short of 
fully explaining the nonlinear aspect of NJA
with both SD and ASD components.

In this letter,
we provide 
a principled derivation of the Kerr metric
that faithfully reproduces the NJA
in its original form.
This not only identifies 
a definite physical origin
for structures observed in the classical explanations
and Newman's works
but also reveals a fully nonlinear, nonperturbative picture
completing the modern explanation.

The key new finding is that
the Kerr metric describes
a pair of SD and ASD Taub-Newman-Unti-Tamburino (NUT) instantons,
which are known to be gravitational analogs of chiral dyons.
These instantons
are exactly the ideally pointlike objects
in the field theorists' 
graviton X-ray image.
The Kerr metric is derived by
establishing a mathematical theorem
about nonlinear superposition of 
Kerr-Schild (KS) spacetimes.
This builds upon
a crucial recent finding \cite{note-sdtn}
and
a modern framework
known as
KS double copy
\cite{monteiro2014black}.
Our construction generalizes to
charged Kerr-Taub-NUT solutions,
thus validating the historical derivation \cite{Newman:1965my-kerrmetric} of the Kerr-Newman metric
as well.

\section{ Intuition and Overview}%
The prototype of this chapter
grew out of
a moment of pure imagination 
that struck the present author
some years ago.
In electromagnetism,
a magnetic dipole arises from
not only
an Amp\`erian current loop
but also
a static monopole pair,
like the N and S poles of a bar magnet
(\fref{fig:gad}).
By analogy,
the question
was whether
the ring singularity of Kerr black hole
can be dualized into a pair of gravitomagnetic monopoles.

It is known
\cite{Misner:1963flatter,Bonnor:1969ala,sackfield1971physical,demianski1966combined,Plebanski:1975xfb,dowker1974nut,Griffiths:2009dfa}
that the gravitational analog of the monopole
is the Taub-NUT \cite{Taub:1950ez,Newman:1963yy} solution.
This
is a stationary vacuum solution characterized by two parameters:
mass $M$ and magnetic mass $N$,
which are gravitoelectric and gravitomagnetic charges, respectively.
The metric of the Taub-NUT solution develops a semi-infinite line defect 
as the gravitational Dirac string,
dubbed Misner string
\cite{Misner:1963flatter,Bonnor:1969ala}.

A modern framework 
known as KS double copy \cite{monteiro2014black}
has elevated
such analogies
to exact mathematical correspondences
between
stationary vacuum solutions in Maxwell theory and GR.
For example,
the Coulomb solution
is the counterpart of
the Schwarzschild solution.
It has been recently shown \cite{note-sdtn} that
this correspondence applies to the Taub-NUT solution as well,
yet only in chiral (SD or ASD) limits
in which
$M = \pm i\hem N$.

The chiral limits of the Taub-NUT solution
are referred to as Taub-NUT instantons.
Historically, they were introduced
by the seminal works of Gibbons and Hawking
\cite{hawking1977gravitational,Gibbons:1978tef}
as paradigmatic instances of gravitational instantons;
hence the name.

\begin{figure}[t]
    \centering
    \includegraphics[width=0.9\linewidth]{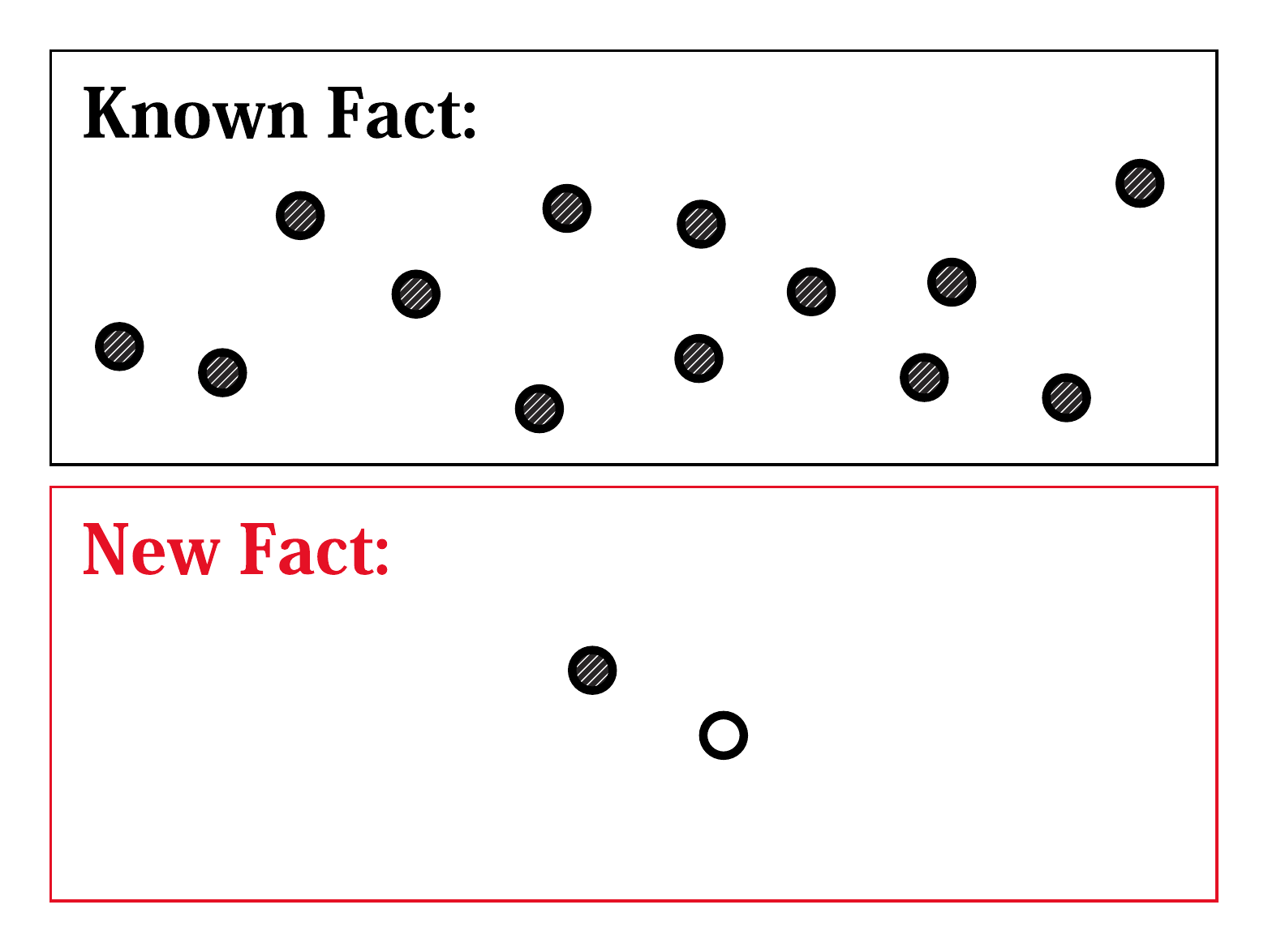}
    \vspace{-5pt}
    \caption[%
        The established fact due to Gibbons and Hawking 
        concerns
        Taub-NUT instantons of the same chirality.
        The new fact proved in this work
        concerns 
        two Taub-NUT instantons of the opposite chirality.
    ]{%
        The established fact due to Gibbons and Hawking \cite{hawking1977gravitational,Gibbons:1978tef}
        concerns
        the nonlinear superposition of
        Taub-NUT instantons of the same chirality.
        The new fact proved in this work
        concerns 
        the nonlinear superposition of
        two Taub-NUT instantons of the opposite chirality.
    }
    \label{newold}
\end{figure}

A previously \textit{known fact}
is that 
any number of
Taub-NUT instantons of the same chirality
can be put together
to yield a multi-centered exact solution
\cite{hawking1977gravitational,Gibbons:1978tef}.
The \textit{new fact}
we prove
is that
SD and ASD Taub-NUT instantons can 
form
a two-centered exact solution:
opposite chiralities.
Remarkably, 
this two-centered solution is nothing other than the Kerr solution.

\section{ Factorization of Kerr}%
As an introduction,
we begin with an innocuous mathematical fact
about a quartic polynomial.
Consider the zero locus
\begin{align}
	\label{eq:variety}
	(x^2{\,+\,}y^2{\,+\,}z^2 - a^2)^2
	+ (2az)^2 = 0
	\,,
\end{align}
where $a$ is a constant.
\eqref{eq:variety} describes a \textit{ring} of radius $a$
in the three-dimensional space $(x,y,z) \in \R^3$.
Intriguingly,
this ring can also appear as a set of
\textit{two disjoint points}.
Suppose we allow $x,y,z$ in \eqref{eq:variety}
to take complex values
so that
we obtain
a complex algebraic variety
in the affine space $\mathbb{C}^3$.
Then the ring is the slice of this variety
by the real section $x,y,z \in \mathbb{R}$.
However,
consider the slice
$x,y,z \in i\hem\mathbb{R}$.
Wick-rotating $x,y,z$
in \eqref{eq:variety} gives
\begin{align}
	\label{eq:variety-wick}
	(x^2{\,+\,}y^2{\,+\,}z^2 + a^2)^2
	- (2az)^2 = 0
	\,,
\end{align}
which factorizes to
$(x^2 + y^2 + (z+a)^2)(x^2 + y^2 + (z-a)^2) = 0$.
This admits just two points as solutions:
$(0,0,\pm a)$.

\begin{figure}[t]\centering
	\centering
	\adjustbox{valign=c}{
		\includegraphics[scale=1.2
		,clip=true,trim=7pt 7pt 7pt 7pt
		]{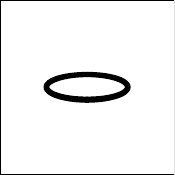}
	}
	\,\,$\xrightarrow[]{\,\,\textsc{\scriptsize{Wick}}\,\,}$\,\,
	\adjustbox{valign=c}{
		\includegraphics[scale=1.2
		,clip=true,trim=7pt 7pt 7pt 7pt
		]{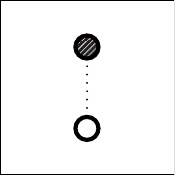}    
	}
    \smallskip
	\caption{%
		The ring singularity of Kerr black hole
		factorizes into
		a pair of
		SD and ASD
		Taub-NUT instantons.
	}
    \medskip
	\label{fig:factorization}
\end{figure}

The astute reader will point out that
\eqref{eq:variety}
is the very equation of the ring singularity
of the Kerr black hole.
The Kerr metric is characterized by 
mass $M$ and ring radius $a$.
In KS coordinates, for instance, it reads
\begin{align}
	\label{eq:line-in-real}
	&
	{-dt^2} {\,+\,} dx^2 {\,+\,} dy^2 {\,+\,} dz^2
	\\
	&{
		+\mem
		\frac{2M \robl^3}{\robl^4{\,+\,}a^2z^2}
		\bigg({
			-dt 
			+ \frac{\robl x{\,-\,}ay}{\robl^2{\,+\,}a^2}\mem dx
			+ \frac{\robl y{\,+\,}ax}{\robl^2{\,+\,}a^2}\mem dy
			+ \frac{z}{\robl}\mem dz
		}\mem\bigg)^{\hnem\nem2}
	}
	\,.
	\nonumber
\end{align}
Here, $\robl$ is a function of $x,y,z$
implicitly defined by
\begin{align}
	\label{eq:r-oblate}
	\frac{x^2{\,+\,}y^2}{\robl^2{\,+\,}a^2}
	+ \frac{z^2}{\robl^2}
	= 1
	\,.
\end{align}
The contour surfaces of $\robl$
are oblate spheroids,
which degenerate to a disc 
of radius $a$
at $\robl = 0$.
The boundary of this disc
is exactly the ring singularity
shown in \eqref{eq:variety}.

By analytic continuation,
we regard
\eqref{eq:line-in-real}
as a holomorphic metric
on a complex four-manifold
solving
complexified vacuum Einstein's equations
(cf. \rrcite{plebanski1975some,Plebanski:1977zz,Plebanski:1975xfb,PD}).
Then the total Wick rotation,
$x^\m \mapsto ix^\m$,
is an allowed coordinate change.
The resulting line element is
\begin{align}
	\label{eq:line-in-wick}
	&
	{dt^2} {\,-\,} dx^2 {\,-\,} dy^2 {\,-\,} dz^2
	\\
	&{
		+\mem
		\frac{2iM \rprol^3}{\rprol^4{\,-\,}a^2z^2}
		\bigg(\mem{
			dt 
			- \frac{\rprol x{\,+\,}iay}{\rprol^2{\,-\,}a^2}\mem dx
			- \frac{\rprol y{\,-\,}iax}{\rprol^2{\,-\,}a^2}\mem dy
			- \frac{z}{\rprol}\mem dz
		}\mem\bigg)^{\hnem\nem2}
	}
	\,,
	\nonumber
\end{align}
where $\rprol$ is
a function of 
the transformed
$x,y,z$:
\begin{align}
	\label{eq:r-prolate}
	\frac{x^2{\,+\,}y^2}{\rprol^2{\,-\,}a^2}
	+ \frac{z^2}{\rprol^2}
	= 1
	\,.
\end{align}
The level sets of $\rprol$ are
now prolate spheroids.
The line element in \eqref{eq:line-in-wick}
is ill-defined
over a ``needle'' along
$x,y=0$ and $z \in [-a,a]$,
which is the degenerate prolate spheroid at $\rprol = a$.
Curvature invariants imply that 
the boundary of this needle is only truly singular,
which exactly describes the set of two points 
specified by \eqref{eq:variety-wick}.

Direct computation shows that
these point singularities are
Taub-NUT instantons of opposite chiralities.
By taking $z \mapsto z \pm a$ and sending $a \to \infty$,
one finds that
the metric in the vicinity of the upper/lower tip of the needle
is exactly the SD/ASD Taub-NUT metric of \rcite{note-sdtn}.

Direct computation also shows that
the needle is a Misner string
that transports the magnetic mass flux
from one instanton to another.
The metric near any point on the needle
can be represented as
$
(\hhem
dt {\,-\,} 2iM\, d\log (x{\,-\,}iy)
)^2 $ $
- dx^2 - dy^2 - dz^2
$
\footnote{
	The line element diverges
	when zooming to an arbitrary point
	$z {\,=\,} \eta\hem a \in (-a,a)$.
	However, it
	can be compensated by a diffeomorphism 
	$
	(x{\,+\,}iy)
	\mapsto 
	(x{\,+\,}iy)
	-
	2iM
	(
	(1{\hem-\mem}\eta^2)\hem a {\,-\,} 2iM {\,-\,} 2\eta z
	)/(x{\,-\,}iy)
	$,
	$
	(x{\,-\,}iy)
	\mapsto
	(x{\,-\,}iy)
	$.
},
which
describes the image of
flat spacetime
under a large diffeomorphism.
Such a line defect of time monodromy
is the very notion of Misner string as a gravitomagnetic flux tube
\cite{dowker1967gravitational,deser1984three,Griffiths:2009dfa,misner1967taub,Alfonsi:2020lub,note-sdtn}.

In sum, we have explicitly shown that
the Kerr metric, as a \textit{whole},
represents a pair of SD and ASD Taub-NUT instantons.
In \fref{fig:factorization},
the instantons are visualized as blobs
while the Misner string is drawn as a dotted line:
\begin{align}
	\label{story1}
	\text{Kerr}
	\,\,\,\,=\,\,\,\,
	\zag 
	\:\:+\:\:
	\zig
	\,.
\end{align}

It is easily verified that their
masses, magnetic masses, and center positions
are
given as below
(if measured with respect to the slice of mostly-positive metric signature):
\begin{align}
	\label{eq:kerr-summary}
	\kern-0.3em
	\bigg\{\,
	\begin{aligned}
		\text{%
			SD instanton,
			\zag
		}
		\,&:\,
		(M/2,-iM/2)
		\,\,\,\text{at}\,\,\,
		\vex = +i\vea
		\,,\\
		\text{%
			ASD instanton,
			\zig
		}
		\,&:\,
		(M/2,+iM/2)
		\,\,\,\text{at}\,\,\,
		\vex = -i\vea
		\,.
	\end{aligned}
	\kern-0.1em
\end{align}
Note that the magnetic mass dipole moment computes to
{$\smash{\vec{J}} = (-iM/2)(2i\vea) = M\vea$},
which is precisely the angular momentum of the Kerr black hole.
In fact,
the static configuration in \eqref{eq:kerr-summary} reproduces
all the
black hole's
spin-induced 
mass multipole moments
for finite $\vea$,
which are known to take the same magnitude with alternating signs
\cite{janis1965structure,Newman:1965tw-janis,Newman:1973yu,Hernandez:1967zza,Geroch:1970cd,Hansen:1974zz}
\footnote{
	$(M/2)\mem (ia)^{2k} + (M/2)\mem (ia)^{2k} = (-1)^k Ma^{2k}$ for electric $2^{2k}$-pole moments and
	$(-iM/2)\mem (ia)^{2k+1} + (iM/2)\mem (-ia)^{2k+1} = (-1)^k Ma^{2k+1}$ for magnetic $2^{2k+1}$-pole moments.
}.

The separation $2i\vea$ has to be pure-imaginary
for $\smash{\vec{J}}$ to be real,
since Taub-NUT instantons carry imaginary magnetic mass in Lorentzian signature.
In fact, 
we will learn that
the imaginary coordinates $\pm i\vea$
are exactly the imaginary displacements of the NJA.
Therefore,
we realize that
the NJA is a statement about
the curvature singularity of Kerr solution
when taken
as a complex saddle.

It should be stressed that our analysis above
does not count on any methods of linearization or perturbative expansions.
It is readily performed in other coordinate systems as well,
such as the Boyer-Lindquist system.

To our knowledge,
the above explicit proof of the exact equivalence between
Kerr and a dipolar configuration of Taub-NUT instantons
is new,
although
an argument appears in
\rcite{Gross:1983hb}.
Our result is also disparate from
the ``black dihole'' of
\rrcite{%
	Manko:2009xx,Clement:2018ynm,Clement:2021ukb,%
	bonnor1966exact,Emparan:1999au,Liang:2001sp,Emparan:2001bb,Teo:2003ug,Manko:2009zza,Chen:2012dr,Manko:2013iva,Cabrera-Munguia:2015fha,Manko:2017avt%
}.

\section{ Nonlinear Superposition Theorem}%
Next, we regard the Taub-NUT instantons as \textit{parts}
and show that
they can be assembled together
to form
the Kerr black hole:
\begin{align}
	\label{story2}
	\zag 
	\:\:+\:\:
	\zig
	\,\,\,\,=\,\,\,\,
	\text{Kerr}
	\,.
\end{align}
A crucial feature of Einstein's gravity theory,
however,
is nonlinearity of field equations.
Hence generally speaking,
it is virtually impossible to assemble two vacuum solutions together to obtain another
by a simple process.
The $+$ symbol in \eqref{story2}
signifies a \textit{nonlinear} sum.

Yet remarkably,
the Taub-NUT instanton solution
admits a KS metric \cite{note-sdtn}
as mentioned earlier.
For instance, the KS metric of a SD Taub-NUT instanton
is given in an elegant formula in the spinor notation:
\begin{align}
	\label{eq:tnks}
	ds^2 \,=\, ds^2_\text{flat} + 
	\frac{M}{|\vex|}\mem \Big(\,{
		\eta_\a\hhem \tdo_\da\mem dx^{\da\a}
	}\mem\Big)^2
	\,.
\end{align}
Here, the ASD spinor
$\eta_\a$ is an arbitrary constant reference
that controls the direction of the Misner string
(as an ASD null plane),
whereas
the SD spinor $\tdo_\da$
describes a ``square root'' of the position three-vector $\vex$
\footnote{
	Concretely, $\protect\tdo^{\protect\da}$ is the eigenspinor of 
	the Killing spinor
	\smash{$\protect\tchi^{\protect\da}{}_{\protect\db}  = \frac{1}{2}\mem (\protect\vex\mdot\protect\vec{\sigma})^{\protect\da}{}_{\protect\db}$}
	whose eigenvalue is $+|\protect\vex|/2$.
}.
That is,
the gauge and physical data
are respectively stored in the ASD and SD spinors.
The Weyl tensor is purely SD.
See \rcite{note-sdtn}
for the derivation of
\eqref{eq:tnks}
from the well-known
Gibbons-Hawking metric
\cite{hawking1977gravitational,Gibbons:1978tef}.

Notably,
KS metrics
are known to
enjoy an \textit{effective linearization} property:
the exact vacuum Einstein's equations
are solved if
the linearized equations
are solved
\cite{xanthopoulos1978exact,Harte:2016vwo,vines2018scattering}.
Via this effective linearization,
two vacuum KS metrics
can be assembled
into another
as shown below.

First of all,
a NVF
$\ell$ in complexified Minkowski space is \textit{canonical}
if it is stationary, GSF, and unit-normalized as $\ell^0 = 1$
with respect to the time direction.
One can treat
canonical NVFs as projective objects,
due to the fixed normalization $\ell^0 = 1$.

Each canonical NVF
is associated with two solutions.
The \textit{single copy solution} is the electromagnetic gauge potential $A_\m = \theta\mem \ell_\m$.
The \textit{double copy solution} is the metric $g_{\m\n} = \eta_{\m\n} + \e\, \theta\mem \ell_\m \ell_\n$,
where $\e$ is an auxiliary parameter.
Provided that $\theta = \partial_\m \ell^\m$ is the expansion of the congruence of $\ell$,
$A_\m$ solves vacuum Maxwell's equations in flat spacetime
while $g_{\m\n}$ solves vacuum Einstein's equations.

The above set of definitions is
our formulation of the framework known as KS double copy \cite{monteiro2014black}.

Next, we present what we call
the \textit{superposition lemma}.
Suppose two canonical NVFs $\ell_1$ and $\ell_2$
with expansions $\theta_1$ and $\theta_2$.
Then
\begin{align}
	\label{eq:elformula-spinor}
	\ell^{\da\a}
	\,\propto\,\mem
	\ell_1^{\da\b} \delta_{\b\db}\mem \ell_2^{\db\a}
	\,,\quad
	\tell^{\da\a}
	\,\propto\,\mem
	\ell_2^{\da\b} \delta_{\b\db}\mem \ell_1^{\db\a}
\end{align}
are canonical NVFs
with expansions $\theta$ and $\tilde{\theta}$
such that
\begin{align}
	\label{eq:linearized-1}
	\theta_1\mem \ell_1{}_\m + \theta_2\mem \ell_2{}_\m
	\,&=\, \theta\mem \ell_\m + \tilde{\theta}\mem \tell_\m
	+ \partial_\m \chi
	\,,\\
	\label{eq:linearized-2}
	\theta_1\mem \ell_1{}_\m \ell_1{}_\n + \theta_2\mem \ell_2{}_\m \ell_2{}_\n
	\,&=\, \theta\mem \ell_\m \ell_\n + \tilde{\theta}\mem \tell_\m \tell_\n
	+ 2\mem \partial_\wrap{(\m} \xi_\wrap{\n)}
	\,,
\end{align}
and $\theta_1 + \theta_2 = \theta + \tilde{\theta}$.
Here,
$\delta_{\b\db} = (\s^0)_{\b\db}$
is the stationary direction boiled down into the spinor notation.
$\chi$ and $\xi_\n$ are some scalar and one-form fields.

Suppose $\tilde{\theta}$ happens to vanish.
Then \eqref{eq:linearized-1} implies that
the single copy solution of $\ell$
describes the superposition of the single copy solutions of $\ell_1$ and $\ell_2$.
Similarly,
\eqref{eq:linearized-2} implies that
the double copy solution of $\ell$
describes the nonlinear superposition of the double copy solutions of $\ell_1$ and $\ell_2$.
First,
the linear sum $\theta_1\mem \ell_1{}_\m \ell_1{}_\n + \theta_2\mem \ell_2{}_\m \ell_2{}_\n$
is gauge-equivalent to $\theta\mem \ell_\m \ell_\n$
in linearized gravity.
Second, $\theta\mem \ell_\m \ell_\n$ uplifts to an exact solution
by the effective linearization property \cite{xanthopoulos1978exact,Harte:2016vwo,vines2018scattering} of KS metrics.

The \textit{nonlinear superposition theorem} is the following:
if two canonical NVFs $\ell_1$ and $\ell_2$ 
are respectively associated with SD and ASD single copy solutions,
then
the double copy solution of 
$\ell$
in \eqref{eq:elformula-spinor}
describes the nonlinear superposition of the double copy solutions of $\ell_1$ and $\ell_2$.
This is because $\tilde{\theta}$ vanishes.

The proofs are detailed in the supplemental material\footnote{
	The supplemental material includes \rrcite{plebanski1998linear,robinson-TN44-03,robinson1987some,penrose:1986spinors2,huggett1994introduction}.
},
but we shall remark here that
the chiral nature of \eqref{eq:elformula-spinor} plays an important role:
the SD (ASD) spinor index $\da$ ($\a$) is controlled by the SD (ASD) solution's NVF $\ell_1$ ($\ell_2$).
Also, \eqref{eq:elformula-spinor}
translates into the vector notation as
\begin{align}
	\label{eq:elformula}
	\ell^\m
	\,=\,
	\frac{\mem{
			\ell_1^\m {\,+\,} \ell_2^\m 
			\mem+\mem 
			\ell_1\nem\mdot\ell_2\mem u^\m
			\mem+\mem
			i\mem \ve^\m{}_{\n\r\s}\mem u^\n\hem \ell_1^\r\mem \ell_2^\s
		}\mem}{2+\ell_1\mdot\ell_2}
	\,,
\end{align}
which describes
a highly nonlinear formula.

\section{ Derivation of Kerr Metric}%
The nonlinear superposition theorem provides
a \textit{derivation} of Kerr metric
(without quotation marks).

First,
note that
the SD Taub-NUT instanton metric in \eqref{eq:tnks}
is the double copy solution
of the canonical NVF
$\ell^{\da\a} \propto \tdo^\da\hem \eta^\a$,
whose single copy solution is SD
\cite{note-sdtn}.

Second,
prepare
a SD Taub-NUT instanton centered at $\vex_1$
and an ASD Taub-NUT instanton centered at $\vex_2$.
Their canonical NVFs are
$\ell_1^{\da\a} \propto \tdo_1^\da\mem \eta^\a$
and $\ell_2^{\da\a} \propto \teta^\da\mem o_2^\a$,
where
$\tdo^\da_1(\vex) = \tdo^\da(\vex{\,-\,}\vex_1)$
and
$o^\a_2(\vex) = o^\a(\vex{\,-\,}\vex_2)$
are ``square roots'' of
$\vex{\,-\,}\vex_{1,2}$.
The expansions are
$\theta_{1,2} = |\vex{\,-\,}\vex_{1,2}|^{-1}$,
as should be clear from \eqref{eq:tnks}.

Third,
construct their nonlinear superposition
via the nonlinear superposition theorem.
\eqref{eq:elformula-spinor} gives
\begin{align}
	\label{eq:o1o2}
	\ell^{\da\a}
	\,\,\propto\,\,\mem
	\tdo^\da_1\, \eta^\b
	\delta_{\smash{\b\db}\vphantom{\b}}\mem 
	\teta^\db\mem o^\a_2
	\,\,\propto\,\,\mem
	\tdo^\da_1\mem o^\a_2
	\,,
\end{align}
which simply transvects
the physical spinors 
$\tdo^\da_1$
and
$o^\a_2$.
The expansion is $\theta = \theta_1 + \theta_2
= |\vex{\,-\,}\vex_1|^{-1} + |\vex{\,-\,}\vex_2|^{-1}
$.

Finally,
the new vacuum spacetime
is obtained as
\begin{align}
	\label{eq:ig-kerr}
	\partial^2_\text{flat}
	- M\mem
	\bigg({
		\frac{1}{|\vex{\,-\,}\vex_1|}
		{\,+\,} 
		\frac{1}{|\vex{\,-\,}\vex_2|}
	}\bigg)
	\bigg(\mem{
		\frac{\tdo^\da_1 o^\a_2}{ 
			o_2^\c\mem \delta_{\c\dc}\mem \tdo_1^\dc
		}\, \partial_{\a\da}
	}\mem\bigg)^{\hnem\nem2}
	\,,
\end{align}
where we have written down the inverse metric
for a direct comparison with \rcite{Newman:1965tw-janis}
while setting $\e = M$.

Remarkably, the vacuum solution in \eqref{eq:ig-kerr}
is exactly the Kerr solution
with mass $M$ and angular momentum \smash{$\vec{J} = -iM\mem (\vex_1{\,-\,}\vex_2)/2$},
centered around $(\vex_1{\,+\,}\vex_2)/2$.

In particular, take 
$\vex_1 = +i\vea$ and $\vex_2 = -i\vea$ for $\vea = (0,0,a)$.
Then \eqref{eq:ig-kerr}
exactly coincides with the Kerr inverse metric
obtained in the very article \cite{Newman:1965tw-janis}.

This establishes a mathematically rigorous derivation of the Kerr solution
as the nonlinear superposition of SD and ASD Taub-NUT instantons.

\section{ Newman-Janis Algorithm Deciphered}%
\begin{figure}[t]\centering
	{
		\centering
		\adjustbox{valign=c}{
			\includegraphics[scale=1.2
			,clip=true,trim=5pt 0pt 2pt 0pt
			]{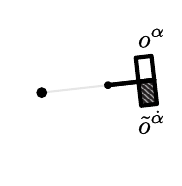}
		}
		$\xrightarrow[]{\textsc{\scriptsize{separate}}}$
		\adjustbox{valign=c}{
			\includegraphics[scale=1.2
			,clip=true,trim=5pt 0pt 2pt 0pt
			]{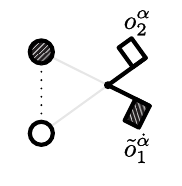}    
		}
	}
	\vspace{-0.5\baselineskip}
	\caption{%
		In an accurate sense,
		the NJA
		first decomposed a Schwarzschild black hole
		into SD\,(\zag)
		and ASD\,(\zig) 
		Taub-NUT instantons
		and then moved them independently.
	}
	\label{fig:NJA}
\end{figure}

It remains to explicate that
the above derivation faithfully reproduces the NJA.
Consider the inverse metric of the Schwarzschild solution,
identified as the nonlinear superposition of
SD and ASD Taub-NUT instantons overlapped at the same point:
\begin{align}
	\label{eq:ig-sch}
	\partial^2_\text{flat}
	- M\mem
	\bigg({
		\frac{1}{|\vex|}
		{\,+\,} 
		\frac{1}{|\vex|}
	}\bigg)
	\bigg(\mem{
		\frac{\hem \tdo^\da o^\a}{
			o^\c\mem \delta_{\c\dc}\mem \tdo^\dc
		}\, \partial_{\a\da}
	}\mem\bigg)^{\hnem\nem2}
	\,.
\end{align}
Each term here acquires a unique meaning
as per the nonlinear superposition theorem,
so
there exists a unique way of
deforming \eqref{eq:ig-sch}
such that the instantons get displaced to 
$\vex_1 = +i\vea$ and $\vex_2 = -i\vea$.
\eqref{eq:o1o2}
dictates the transformation of the spinors:
the SD (ASD) spinor shifts by $+i\vea$ ($-i\vea$).
The KS potential is mandated to be the total expansion:
$\theta_1 + \theta_2
= |\vex{\,-\,}i\vea|^{-1} + |\vex{\,+\,}i\vea|^{-1}
$.
Thus,
one
unambiguously
generates 
the Kerr inverse metric in
\eqref{eq:ig-kerr}
from 
the Schwarzschild inverse metric in
\eqref{eq:ig-sch}.
See \fref{fig:NJA} for a visualization.

This operation precisely reproduces the NJA in its very original formulation \cite{Newman:1965tw-janis},
as elaborated in the supplemental material.
In particular,
Eqs.\:(4) and (7)
of \rcite{Newman:1965tw-janis}
are \textit{literally} reproduced
upon using spheroidal coordinates.

In \rcite{Newman:1965tw-janis},
the $\pm i\vea$ imaginary shifts of the spinors (null tetrad)
and the ``complexification'' of the KS potential 
$2/|\vex| \to 2\mem\Re\mem |\vex{\,-\,}i\vea|^{-1}$
are posited as ad hoc rules.
The nonlinear superposition theorem
shows that this set of replacements, i.e., the NJA, is
a strict implication of the fact that
the Schwarzschild-Kerr family of solutions
are systems of SD and ASD Taub-NUT instantons.

This refined understanding implies the following
clarifications on the NJA.

First,
the NJA is \textit{not} a coordinate transformation
as originally conceived 
\cite{Newman:1965tw-janis}.
The instanton pairs with zero and nonzero separati\-ons,
i.e., Schwarzschild and Kerr,
are never diffeomorphic.

Second,
the NJA is not about one object
as originally conceived
\cite{Newman:1973yu,Newman:2002mk},
but \textit{two}.
It does not merely shifts
a Schwarzschild black hole
but instead \textit{splits} it into SD and ASD parts
and then \textit{separates} them,
as is illustrated in \fref{fig:NJA}.
The displacements $\vex_1, \vex_2$ are also \textit{independent},
although a reality condition can be imposed a posteriori.

Third,
it is important to explicitly address the nonlinear gravitational interaction
between the SD and ASD instantons.
Otherwise the NJA is not fully explained nor justified
as a method that generates an exact solution in Einstein gravity.
Newman's complex center of mass analysis
\cite{newman1974curiosity,Newman:1973yu,Newman:2002mk},
for instance,
identifies one worldline
from the SD sector (in linearized gravity)
and does not elaborate much on the validity of
``plus complex conjugate''
in the nonlinear theory.
Works \cite{Ghezelbash:2007kw,Crawley:2021auj}
have shown that
the SD Kerr-Taub-NUT solution
describes a SD Taub-NUT instanton,
which is not a statement about the Kerr solution which the NJA concerns.
To reiterate,
the NJA is about \textit{two} Taub-NUT instantons,
not just one SD instanton.

The technical structure of the formula in
\eqref{eq:elformula}
may have been effectively observed in
works \cite{schiffer1973kerr,Gurses:1975vu,Rajan:2016zmq},
but the instanton pair interpretation is not realized there.

\section{ Newman-Janis Algorithm for Charged Kerr-Taub-NUT Solution}%
Finally, we construct more general black holes.
First of all,
the electric-magnetic dual version of Kerr solution
arises as
\begin{align}
	\label{story2*}
	\:\:-\:\:
	\zag 
	\:\:+\:\:
	\zig
	\,\,\,\,=\,\,\,\,
	\text{Kerr}^\star
	\,.
\end{align}
The nonlinear superposition theorem
admits no freedom for 
relative coefficients,
but there is a binary choice for the NVF
per each instanton.
In particular,
the KS metric of SD instanton
can be realized in either
outgoing ($\ell^{\da\a} \propto \tdo^\da \eta^\a$) or ingoing ($\ell^{\da\a} \propto  \ti^\da \eta^\a$)
conventions.
Switching to ingoing flips the sign of the expansion,
and in this sense
the instantons can be subtracted as in \eqref{story2*}.

\begin{figure}[t]\centering
	\adjustbox{valign=c}{
		\includegraphics[scale=1.2]{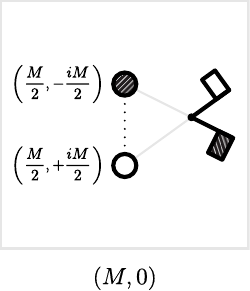}
	}
	\,
	\adjustbox{valign=c}{
		\includegraphics[scale=1.2]{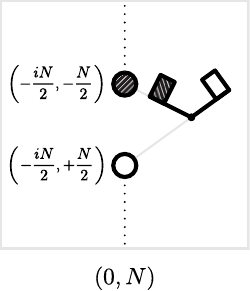} 
	}
	\caption{%
		Implementing pure
		mass
		(\textit{left})
		and magnetic mass
		(\textit{right})
		with Taub-NUT instantons.
	}
	\label{fig:MN}
\end{figure}

Explicitly, the metric for
$\ell^{\da\a} \propto \ti^\da_1\hem o^\a_2$ is
\begin{align}
	\label{eq:Kerr*}
	&
	{dt^2} {\,-\,} dx^2 {\,-\,} dy^2 {\,-\,} dz^2
	\\
	&{
		-
		\frac{2 N a\rprol z}{\rprol^4{\hem-\,}a^2z^2}
		\bigg(\mem{
			dt 
			{\,+\,} \frac{\rprol}{a}
			\bigg({
				\frac{zx{\,+\,}i\rprol y}{\rprol^2{-\,}z^2}\mem dx
				{\,+\,} \frac{zy{\,-\,}i\rprol x}{\rprol^2{-\,}z^2}\mem dy
				{\,-\,} dz
			}\bigg)\nem\nem
		}\mem\bigg)^{\hnem\nem2}
	}
	\,,
	\nonumber
\end{align}
which describes 
a massless black hole with 
magnetic mass $N$
and gravitoelectric dipole moment ${-\hnem N a}$.
The exercise
of zooming to each part of this metric
can be repeated
to explicitly verify the nonlinear superposition;
see \fref{fig:MN}.

By adding up the two KS perturbations in 
\eqrefs{eq:line-in-wick}{eq:Kerr*},
one obtains the generic three-parameter Kerr-Taub-NUT solution
in the double KS form
\cite{Plebanski:1975xfb,Chong:2004hw,Farnsworth:2023mff,Luna:2015paa}
\footnote{
	In the SD limit, $N \to -iM$,
	this double KS metric
	describes a SD Taub-NUT instanton,
	consistently with \rrcite{Ghezelbash:2007kw,Crawley:2021auj}.
}.
The two canonical NVFs
$\tdo^\da_1\hem o^\a_2$
and
$\ti^\da_1\hem o^\a_2$
share a common spinor $o^\a_2$,
making them mutually orthogonal.

Meanwhile,
the Maxwell stress-energy tensor
$\smash{T^{\da\a\db\b}} = \smash{\tilde{F}^{\da\db} F^{\a\b}}$
\cite{penrose:1985spinors1}
vanishes for SD field strength (\smash{$F^{\a\b} = 0$}).
Hence the single copy and double copy solutions of
$\ell^{\da\a} \propto \tdo^\da\hem \eta^\a$
together define a solution in Einstein-Maxwell theory:
a SD Taub-NUT instanton
endowed with SD electromagnetic charge.
By using such Einstein-Maxwell configurations
as building blocks,
a generalization of the nonlinear superposition theorem
given in the supplemental material
derives
the Kerr-Newman solution,
stipulating that the KS potential exactly needs to be modified as
\begin{align}
	\label{eq:KN-potential}
	\frac{M}{|\vex{\,-\,}\vex_1|} + \frac{M}{|\vex{\,-\,}\vex_2|}
	- \frac{Q^2}{ |\vex{\,-\,}\vex_1| |\vex{\,-\,}\vex_2| }
	\,.
\end{align}
This reproduces
the complexification prescription
posited in
\rcite{Newman:1965my-kerrmetric}.
The $Q^2$ term in \eqref{eq:KN-potential}
reflects
static electromagnetic interaction between
SD and ASD dyons.
Working in the similar fashion,
one derives
the five-parameter subclass of
Pleba\'nski-Demia\'nski \cite{PD,debever1971type,Plebanski:1975xfb} solutions
with mass, magnetic mass, ring radius, electric charge, and magnetic charge,
in the double KS form.

\section{ Summary}%
This chapter elevates the NJA to a rigorous derivation of rotating black hole solutions
with a definite origin:
factorizations into SD and ASD Taub-NUT instantons
and chiral dyons.

This resolution of an old mystery
will spark new perspectives on
modern theoretical studies of rotating black holes:
hidden symmetries \cite{Carter:1968ks,Penrose:1973naked,Compere:2023alp,Guevara:2023wlr}
\footnote{
	The Killing-Yano tensor of Kerr
	decomposes into
	the conformal Killing-Yano tensors
	of the instantons.
},
perturbation theory
\cite{Teukolsky:1973ha,Adamo:2023fbj},
scattering amplitudes
\cite{ahh2017,Guevara:2018wpp,Guevara:2019fsj,chkl2019,aho2020,Johansson:2019dnu,Aoude:2020onz,Lazopoulos:2021mna,zihan23,fabian2},
and
effective dynamics
\cite{ShR065,Porto:2016pyg,Kalin:2020mvi,vines2018scattering,Guevara:2018wpp,Guevara:2019fsj,chkl2019,Gibbons:1986df,atiyah2014geometry,ferrell1987slow}.

\begin{subappendices}

\section{ Nonlinear Superposition Theorem}

The idea of patching SD and ASD spacetimes together
has been a long-standing aspiration
\cite{plebanski1998linear,robinson-TN44-03,robinson1987some}.\footnote{
    The author thanks Maciej Dunajski for this anecdote.
}
Here,
we tackle this problem
in a narrow setup:
\textit{stationary KS spacetimes}.

The complexified Minkowski space $\mflat$
is the linear space $\C^4$ equipped with a holomorphic flat metric $\eta$
with a signature-$(-,+,+,+)$ real section.
Let $x^\m$ be Cartesian coordinates.
Let $u^\m$ be a unit vector such that $u^2 = \eta_{\m\n}\mem u^\m u^\n = -1$;
this is the characterization of the stationary direction $u^\m = \delta^\m{}_0$ in the main article.
All indices are raised and lowered with the flat metric $\eta$.

\smallskip
\smallskip
\noindent{\bfseries Definition 1.1}|%
A null vector field (NVF) $\ell$ on $\mflat$
is a vector field $\ell \in \Gamma(T\mflat)$
such that $\ell^2 = \eta_{\m\n}\mem \ell^\m \ell^\n = 0$.

\smallskip
\noindent{\bfseries Definition 1.2}|%
A NVF $\ell$ on $\mflat$
is \textbf{canonical} iff it is
stationary, geodesic, shear-free, 
and unit-normalized as $-u\mdot\ell = 1$.

\smallskip
\noindent{\bfseries Proposition 1.3}|%
The first-order derivative $\partial_\m \ell_\n$ 
of a canonical NVF $\ell$
takes the following general form:
\begin{align}
	\label{eq:canonical}
	\partial_\m \ell_\n
	\,=\,
	\frac{\theta}{2}\, 
	\Big(\,{
		\eta_{\m\n}
		+ \ell_\wrap{(\m} (2u{\,-\,}\ell)_\wrap{\n)}
	}\mem\Big)
	+ \frac{\psi}{2}\, \ve_{\m\n\r\s}\hem u^\r\hem \ell^\s
	\,.
\end{align}
Here, the scalar fields $\theta$ and $\psi$ are referred to as 
the \textbf{expansion} and \textbf{twist} of $\ell$,
defined as
$\theta = \partial_\m \ell^\m$
and 
$\psi = 
	u_\m\mem
	\ve^{\m\n\r\s}
	\ell_\n\mem \partial_\r \ell_\s
$.
The \textbf{complex expansions} of $\ell$ refer to
$\rho^\pm = (\theta \pm i\psi)/2$.

(Proposition 1.3)
is easily verified.
For instance, the shear-free property
arises from the fact that
$\eta_{\m\n}
+ \ell_\m u_\n
+ u_\m \ell_\n
- \ell_\m \ell_\n
$ and
$\ve_{\m\n\r\s}\hem u^\r \ell^\s$
in \eqref{eq:canonical}
are tensors lying within the two-dimensional plane
transverse to both $u$ and $\ell$,
spanning the trace-only and rotation modes.
\rcite{vines2018scattering}
makes a fruitful use of the formula in \eqref{eq:canonical}.

\smallskip
\noindent{\bfseries Proposition 1.4}|%
From \eqref{eq:canonical}
and its integrability $\partial_\m(\partial_\n\ell_\r) = \partial_\n(\partial_\m\ell_\r)$, it follows that
$\ell^\m\mem \partial_\m \rho^\pm = - (\rho^\pm)^2$,
$\ell^\m\mem \ell^\n\mem \partial_\m \partial_\n \rho^\pm = 2\hem (\rho^\pm)^3$,
and
$\partial^2 \rho^\pm = 0$.
The first equation is
the Sachs equation \cite{penrose:1986spinors2,huggett1994introduction} in flat background.

\smallskip
\smallskip
\noindent{\bfseries Definition 2.1}|%
For each canonical NVF $\ell$ on $\mflat$,
the \textbf{single copy solution} is the abelian gauge connection $A_\m = \theta\mem \ell_\m$
while the \textbf{double copy solution} is the metric $g_{\m\n} = \eta_{\m\n} + \e\, \theta\mem \ell_\m \ell_\n$,
where $\e$ is a constant parameter.

\smallskip
\noindent{\bfseries Proposition 2.2}|%
The single copy and double copy solutions of a canonical NVF
solve vacuum flat-background Maxwell's equations and vacuum Einstein's equations
in the holomorphic category,
respectively.

\smallskip
\noindent{\bfseries Definition 2.3}|%
A canonical NVF $\ell$ is \textbf{self-dual} (SD) iff
its expansion is nonvanishing 
and
the field strength $F_{\m\n} = \partial_\m (\theta\mem \ell_\n) - \partial_\n (\theta\mem \ell_\m)$
of its single copy solution is SD.
A canonical NVF $\ell$ is \textbf{anti-self-dual} (ASD) iff
its expansion is nonvanishing 
and
its field strength is ASD.

\smallskip
\noindent{\bfseries Proposition 2.4}|%
A canonical NVF is SD iff
$\rho^+ \neq 0$ and $\rho^- = 0$.
A canonical NVF is ASD iff
$\rho^+ = 0$ and $\rho^- \neq 0$.

\smallskip
\noindent{\bfseries Proposition 2.5}|%
The double copy solution of a SD (ASD) canonical NVF
exhibits SD (ASD) curvature:
${*}R_{\m\n\r\s} = \pm i\mem R_{\m\n\r\s}$.

(Proposition 2.5) can be shown by
employing the orthonormal coframe
$e^A{}_\m = \delta^A{}_\m\mem (\delta^\m{}_\n + \e\, \theta\mem \ell^\m \ell_\n/2)$,
in which case the spin connection $\gamma_{AB\r}$ is
\begin{align}
\begin{split}
	\label{sc}
	- \frac{\e}{2}\,
    \lrp{
        \begin{aligned}[c]
        &
    		F_{\m\n}\mem \delta^\m{}_A\mem \delta^\n{}_B\,
    		\ell_\r
        \\
        &
        +
    	\BB{
    		\rho^+\mem \Pi^+_{ABCD}
    		+
    		\rho^-\mem \Pi^-_{ABCD}
    		\nem}
    	\BB{
    		\delta^C{}_\r
    		+ \delta^C{}_\k\mem u^\k \ell_\r
    	}
    	\BB{
    		\delta^D{}_\s\mem A^\s
    	}
        \end{aligned}
    }
	\,,
\end{split}
\end{align}
where the SD and ASD projectors are
$\Pi^\pm_{ABCD} = \tfrac{1}{2}\mem(
	\eta_{AC} \eta_{BD}
	- \eta_{AD} \eta_{BC}
	\mp i\mem \ve_{ABCD}
)$.

\smallskip
\smallskip
\noindent{\bfseries Lemma 3.1} (\textit{Little Superposition Lemma})|%
Suppose a collection of canonical NVFs
$\ell_i$ ($i=1,2,\cdots n$)
such that 
$\sum_i c_i\mem \theta_i = 0$
for a set of constants $c_i$.
If there exists a scalar function $\chi$ such that
$\sum_i c_i\, \theta_i \ell_i{}_\m = \partial_\m \chi$,
then there exists a one-form field $\xi_\m$ such that
$\sum_i c_i\, \theta_i \ell_i{}_\m \ell_i{}_\n = 2\mem \partial_\wrap{(\m} \xi_\wrap{\n)}$.
Explicitly, $\xi_\m = \chi\mem u_\m - \protect\sum_i c_i\, \ell_i{}_\m$.

\smallskip
\noindent{\bfseries Lemma 3.2} (\textit{Superposition Lemma})|%
If $\ell_{1,2}$ are canonical NVFs
with complex expansions $\rho_{1,2}^\pm$,
then
\begin{align}
	\label{eq:elformulas}
	\ell^{\da\a}
	\,=\, \frac{
		\ell_1^{\da\b} \delta_{\b\db}\mem \ell_2^{\db\a}
	}{1{\,+\,}\ell_1\nem\mdot\ell_2/2}
	\,,\quad
	\tell^{\da\a}
	\,=\, \frac{
		\ell_2^{\da\b} \delta_{\b\db}\mem \ell_1^{\db\a}
	}{1{\,+\,}\ell_1\nem\mdot\ell_2/2}
\end{align}
are canonical NVFs with respective complex expansions
$(\rho^+,\rho^-) = (\rho^+_1,\rho^-_2)$
and
$(\trho^+,\trho^-) = (\trho^+_2,\trho^-_1)$,
satisfying
\begin{align}
\begin{split}
	\label{eq:chi12}
    &
	\theta\mem \ell_\m
	+ \ttheta\mem \tell_\m
	\,=\,
	\theta_1\mem \ell_1{}_\m
	+ \theta_2\mem \ell_2{}_\m
	+ \partial_\m\chi
    \\
	&
    \iq
	\theta\mem \ell_\m \ell_\n
	+ \ttheta\mem \tell_\m \tell_\n
	\,=\,
	\theta_1\mem \ell_1{}_\m \ell_1{}_\n
	+ \theta_2\mem \ell_2{}_\m \ell_2{}_\n
	+ 2\mem \partial_\wrap{(\m} \xi_\wrap{\n)}
\end{split}
\end{align}
with $\chi = \log( (1 {\mem+\,} \ell_1\nem\mdot\ell_2/2)^2 )$
and 
$\xi_\m = \chi\mem u_\m - \ell_\m - \tell_\m + \ell_1{}_\m + \ell_2{}_\m$.
Here, \smash{$\delta_{\a\da} = -u_\m (\s^\m)_{\a\da}$}.

The GSF condition plays an important role in (Lemma 3.1).
(Lemma 3.2) follows by direct computation
and (Lemma 3.1),
where the chiral nature of \eqref{eq:elformulas} is crucial.

\smallskip
\noindent{\bfseries Theorem 3.3} (\textit{Nonlinear Superposition Theorem})|%
A SD canonical NVF $\ell_1$
and an ASD canonical NVF $\ell_2$
together define
an exact solution 
$g_{\m\n} = \eta_{\m\n} + \e\mem \big(\mem{ \theta_1 + \theta_2 }\mem\big)\mem \ell_\m\ell_\n$
to Einstein's equations
via 
\smash{$
\ell^{\da\a}
=
	\ell_1^{\da\b} \delta_\wrap{\b\db}\mem \ell_2^{\db\a}
	/({1{\,+\,}\ell_1\nem\mdot\ell_2/2})
$}.

Proof is straightforward from (Lemma 3.2)
and (Definition 2.3):
$(\trho^+,\trho^-) = (\trho^+_2,\trho^-_1)$ vanishes.
This describes the nonlinear superposition of SD and ASD vacuum metrics in the sense that
the Riemann tensor of
$\eta_{\m\n} + \e\, \theta\mem \ell_\m \ell_\n$
equals the sum of the Riemann tensors of
$\eta_{\m\n} + \e\, \theta_1\mem \ell_1{}_\m \ell_1{}_\n$
and
$\eta_{\m\n} + \e\, \theta_2\mem \ell_2{}_\m \ell_2{}_\n$
at leading order in the parameter $\e$.

\smallskip
\noindent{\bfseries Lemma 4.1} (\textit{Maxwell Stress-Energy via Modifying the KS Potential})|%
If $\ell$ is a canonical NVF on $\mflat$
with complex expansions $\rho^\pm$,
then
the gauge potential
$A_\m = \theta\mem \ell_\m$
and 
the metric
\smash{$g_{\m\n} = \eta_{\m\n} + \e\mem \big(\mem{ \theta -2\alpha \mem \rho^+ \rho^- }\mem\big)\mem \ell_\m\ell_\n$}
together define a solution to 
$G^\m{}_\n[g] = \e\mem\alpha\, T^\m{}_\n[A,g]$
and $\nabla^\n F_{\m\n} = 0$:
Einstein-Maxwell theory
in the holomorphic category.
Here,
$G^\m{}_\n[g]$ is the Einstein tensor associated with the metric $g$,
$T^\m{}_\n[A,g]$ is the Maxwell stress-energy tensor,
$\nabla$ is the Levi-Civita connection of the metric $g$,
$F_{\m\n} = 2\mem \nabla_\wrap{[\m} A_\wrap{\n]} = 2\mem \partial_\wrap{[\m} A_\wrap{\n]}$
is the field strength,
and $\a$ is a constant parameter.

\smallskip
\noindent{\bfseries Theorem 4.2} (\textit{Nonlinear Superposition for Einstein-Maxwell Solutions})|%
A SD canonical NVF $\ell_1$
and an ASD canonical NVF $\ell_2$
together define an exact solution
$A_\m = \big(\mem{ \theta_1 + \theta_2 }\mem\big)\mem \ell_\m$,
$g_{\m\n} = \eta_{\m\n} + \e\mem \big(\mem{ \theta_1 + \theta_2 -2\alpha \mem \theta_1 \theta_2 }\mem\big)\mem \ell_\m\ell_\n$
to Einstein-Maxwell theory
via 
\smash{$
	\ell^{\da\a}
	=
	\ell_1^{\da\b} \delta_\wrap{\b\db}\mem \ell_2^{\db\a}
	/({1{\,+\,}\ell_1\nem\mdot\ell_2/2})
$}.
Here, $Q$ is a constant parameter such that
$\e\hem\alpha = Q^2/2$.

For deriving the Kerr-Newman solution,
one rescales the gauge potential
by $Q/8\pi\varepsilon_0$
so that the electromagnetic charges of the SD and ASD solutions
become $(Q/2, \mp\hem i\hem Q/2)$.
Consequently,
the constant parameters are taken as
$\e = GM/c^2$
and 
$\e\hem \alpha 
= (8\pi G/c^4) \mdot (1/\mu_0) \mdot (Q/8\pi\ve_0)^2 / c^2
= Gk_\text{e} Q^2 \nem/2c^4
$.
Hence,
in the relativists' natural unit
($c{\,=\,}1$, $G{\,=\,}1$, $k_\text{e}{\,=\,}1$),
one finds
$\e = M$ and $\e\hem \alpha = Q^2 \hnem/2$
as stated in the main article.

\smallskip
\smallskip
\noindent{\bfseries Theorem 5.1} (\textit{Effective Linearization})|%
A metric of the form $g_{\m\n} = \eta_{\m\n} + \phi\mem \ell_\m\hem \ell_\n$
with $\ell^2 = 0$
solves
the exact vacuum Einstein's equations
if it solves
the linearized vacuum Einstein's equations
\cite{xanthopoulos1978exact,Harte:2016vwo,vines2018scattering}.

\section{ Spheroidal Coordinates}

The main article describes the inverse metrics in 
\eqrefs{eq:ig-kerr}{eq:ig-sch}
in the spinor language.
The original paper \cite{Newman:1965tw-janis},
on the other hand,
uses
the null tetrad language
in spheroidal coordinates.
We show that they are equivalent.

\smallskip\noindent
\textbf{Null Tetrad}.|%
Given two spinors
$\tdo_1^\da$ and $o_2^\a$
and $\delta_{\a\da} = -u_\m\mem (\s^\m)_{\a\da}$,
a null tetrad in flat spacetime is formed by
\begin{align}
	\label{eq:tetrad-flat}
    &
	\ell^{\da\a} =
	\frac{\tdo^\da_1 o^\a_2}{ 
		\langle o_2 | \delta | \tdo_1 \rsq
	}
	\,,\,\,\,
	n^\circ{}^{\da\a} =
	\frac{\ti^\da_2 \iota^\a_1}{ 
		\langle o_2 | \delta | \tdo_1 \rsq
	}
	\,,\,\,\,
	m^{\da\a} =
	\frac{\tdo^\da_1 \iota^\a_1}{ 
		\langle o_2 | \delta | \tdo_1 \rsq
	}
	\,,\,\,\,
	\tm^{\da\a} =
	\frac{\ti^\da_2 o^\a_2}{ 
		\langle o_2 | \delta | \tdo_1 \rsq
	}
    \nonumber\\
    &
    \iq
	\eta^{\m\n} = -\ell^{(\m} n^\circ{}^{\n)} + m^{(\m} \tm^{\n)}
	\,,
\end{align}
where
$\ti_2{}_\da = - o^\a_2\mem \delta_{\a\da}$
and
$\iota_1{}_\a = - \delta_{\a\da}\mem \tdo^\da_1$
so that
$\lsq \tdo_1 \ti_2 \rsq = \langle o_2 | \delta | \tdo_1 \rsq$
and
$\langle \iota_1 o_2 \rangle = \langle o_2 | \delta | \tdo_1 \rsq$.

For example,
the principal spinors of the Schwarzschild solution are given by
$\tdo^\da = (1,(x{\,+\,}iy)/(|\vex|{\,+\,}z))$
and
$o^\a = (1,(x{\,-\,}iy)/(|\vex|{\,+\,}z))$.
Taking
$\tdo^\da_1 = \tdo^\da$
and
$o^\a_2 = o^\a$
in \eqref{eq:tetrad-flat}
gives
$\ell^\m = (1,x/|\vex|,y/|\vex|,z/|\vex|)$,
which is the outgoing canonical NVF 
for the KS metric of the Schwarzschild solution.

The physical spinor of SD/ASD Taub-NUT instanton
equals the SD/ASD principal spinor of Schwarzschild
\cite{note-sdtn}.

The main article sets
$\tdo_1^\da$ and $o_2^\a$
as the physical spinors of the SD and ASD Taub-NUT instantons
at $\vex_1 = +i\vea$ and $\vex_2 = -i\vea$
with $\vea = (0,0,a)$.
Explicitly, this means to take
\smash{$\tdo^\da_1 = (1,\tzeta)$},
\smash{$o^\a_2 = (1,\zeta)$},
\smash{$\ti^\da_2 = (-\tzeta,1)$},
\smash{$\iota^\a_1 = (-\zeta,1)$},
and
$1/\langle o_2|\delta|\tdo_1\rsq = (\robl{\,+\,}z)/2\robl$,
where
\begin{align}
\begin{split}
	\tzeta
	\,&=\, \frac{x+iy}{|\vex{\,-\,}i\vea|+(z{\,-\,}ia)}
	\,=\, \frac{x+iy}{\robl+z}\mem \frac{\robl}{\robl-ia}
	\,,\\
	\zeta
	\,&=\, \frac{x-iy}{|\vex{\,+\,}i\vea|+(z{\,+\,}ia)}
	\,=\, \frac{x-iy}{\robl+z}\mem \frac{\robl}{\robl+ia}
	\,.
\end{split}
\end{align}
Here, $\robl$ denotes the function of $x,y,z$ defined in \eqref{eq:r-oblate}.
A branch cut prescription
\cite{Newman:2002mk,aho2020}
sets $|\vex{\,\mp\,}i\vea| = \sqrt{(\vex{\,\mp\,}i\vea)^2} = \robl \mp iaz/\robl$.
The computation of the resulting flat null tetrad due to \eqref{eq:tetrad-flat}
is straightforward.

\smallskip\noindent
\textbf{Spheroidal Coordinates}.|%
A nice choice of angle variables $(\theta,\varphi)$
simplifies the result as
$\ell^\m = (1, \sin\theta \cos\varphi,$ $ \sin\theta\sin\varphi, \cos\theta)$:
$x \pm iy = (\robl \mp ia)\mem \sin\theta\mem e^{\pm i\varphi}$,
$z = \robl\mem \cos\theta$.
In fact, $(r,\theta,\varphi)$ is known as
the twisted oblate spheroidal coordinate system.
As a result,
the components of $\ell$ are simply $(1,1,0,0)$
in the $(t,\robl,\theta,\varphi)$ coordinate system.
Thus, by introducing the retarded time $u = t-\robl$,
the null tetrad is
\begin{align}
\begin{split}
	\label{eq:njatetrad-flat}
	\ell \,=\, \partial_\robl
	&\,,\quad
	m \,=\, \frac{\mathe^{i\varphi}}{\robl{\,+\,}ia\cos\theta}\mem
	\bigg(\,{
		ia\sin\theta\mem \Big(\mem{
			\partial_u - \partial_\robl
		}\mem\Big)
		+ \partial_\theta
		+ \frac{i}{\sin\theta}\mem \partial_\varphi
	}\,\bigg)
	\,,\\
	n^\circ \,=\, 2\partial_u - \partial_\robl
	&\,,\quad
	\tm \,=\,
    \frac{\mathe^{-i\varphi}}{\robl{\,-\,}ia\cos\theta}\mem
	\bigg(\,{
		-ia\sin\theta\mem \Big(\mem{
			\partial_u - \partial_\robl
		}\mem\Big)
		+ \partial_\theta
		- \frac{i}{\sin\theta}\mem \partial_\varphi
	}\,\bigg)
	\,.
\end{split}
\end{align}

The null tetrad $(\ell,n,m,\tm)$ for
the curved inverse metric in \eqref{eq:ig-kerr} 
follows by
perturbing $n^\circ \to n = n^\circ + M\mem \theta\mem \ell$:
\begin{align}
	\label{eq:njatetrad-n}
	n
	\,=\,
	2\partial_u - \partial_\robl
	+ M\mem 
	\bb{
		\frac{1}{\robl{\,-\,}ia\cos\theta}
		{\,+\,} 
		\frac{1}{\robl{\,+\,}ia\cos\theta}
    }
	\, \partial_\robl
	\,.
\end{align}
\eqrefs{eq:njatetrad-flat}{eq:njatetrad-n}
are literally Eq.\,(7) of \rcite{Newman:1965tw-janis}.
Taking the limit $a\to0$ exactly reproduces Eq.\,(4) of \rcite{Newman:1965tw-janis}.

The NJA inverts this limit.
\rcite{Newman:1965tw-janis} views it as
a ``complex coordinate transformation''
due to the suggestiveness of the spheroidal coordinates.
However, the main article clarifies that
the NJA is not a coordinate transformation.

By repeating the above exercise,
one systematically derives the NJA for
Kerr-Newman, $\text{Kerr}^\star$, Kerr-Taub-NUT, and charged Kerr-Taub-NUT
(five-parameter Pleba\'nski-Demia\'nski)
solutions.
We emphasize that
what facilitates such a systematic and universal derivation of NJAs
is the very instanton pair structure
elucidated in the main article.
In other words,
the origin of the NJA is the factorization of black holes into Taub-NUT instantons and chiral dyons.

Note that this derivation of the NJA works in any signature:
Euclidean, Lorentzian, Kleinian, mostly-plus, mostly-minus, etc.
For instance,
one can employ twisted prolate hyperboloidal coordinate system
$x \pm iy = (\rprol \pm ja) $ $\sin\theta\, \mathe^{\pm j\varphi}$,
$z = \rprol \cos\theta$,
where $j$ denotes the unit split complex number ($j^2 = 1$).

\end{subappendices}
\chapter[
    Geodesic Deviation to All Orders
]{
    Geodesic Deviation to All Orders
}
\label{K3:GDE}

    We establish an in-in formalism for geodesic deviation
    as an alternative to Synge calculus,
	based on a covariant calculus of differential forms in tangent bundle.
    This derives
    the exact Lagrangian and equations
    governing the finite geodesic deviation between
    a free-falling test particle
    and an arbitrary observer,
    in terms of
    infinite sums
    whose coefficients are products of binomial coefficients.
	Explicit expressions are provided up to tenth order,
	finding agreements with the previous fourth-order result.
    
\begin{fullnote}
    This section reproduces the contents of \rcite{gde}, \fullcite{gde}.
\end{fullnote}

\section{
Introduction}%
The 
GDE
is a foundational topic in GR,
commonly covered in textbooks
\cite{Carroll:2004st,mtw,Hawking:1973uf,Wald:1984rg,PoissonWill:2014,synge1952tensor,synge1960general}.
First derived in 1927
by Levi-Civita and Synge
\cite{levicivita1927ecart,synge1927ii,synge1928geodesics,Synge:1935zz,synge1934deviation},
this equation describes the evolution of the \textit{infinitesimal} separation between two nearby free-falling test particles.
Explicitly, it reads
\begin{align}
    \label{eq:GDE2}
    \frac{D^2 y^\m}{d\t^2}
    \,=\,
        - R^\m{}_{\n\r\s}\mem u^\n y^\r u^\s
    \,,
\end{align}
where $y^\m$ describes the infinitesimal separation as a vector
and $u^\m$ describes the unit-normalized four-velocity.

The generalization of the GDE to the case of \textit{finite} separations
has been a topic of research in the literature
\cite{hodgkinson1972modified,bazanski1977kinematics,bazanski1977dynamics,aleksandrov1979geodesic,li1979coupled,mashhoon1975tidal,mashhoon1977tidal,ciufolini1986generalized,ciufolini1986measure,Chicone:2002kb,Chicone:2006rm,Perlick:2007ux,Vines:2014oba,Puetzfeld:2015uxi,Obukhov:2018nnt,Flanagan:2018yzh,Waldstein:2021olw}.
Here, the objective is to find corrections to the GDE
perturbatively in the orders of the separation vector $y$,
covariantly defined as a tangent to the geodesic segment
joining the two particles.
Namely,
the right-hand side of \eqref{eq:GDE2}
is augmented
with terms involving
nonlinear powers of $y$,
coupled to
the Riemann tensor $R^\m{}_{\n\r\s}$ and its derivatives.
Physically, this captures
the higher-order tidal effects
due to the non-infinitesimal size of $y$.

The higher-order extensions of the GDE
have found fruitful physical applications
\cite{Vines:2014oba,Waldstein:2021olw}:
gravitational wave detectors,
astrophysical jets,
measurement of spacetime curvature,
and
an analytical treatment of
eccentric relativistic orbits
\cite{li1979coupled,mashhoon1975tidal,mashhoon1977tidal,ciufolini1986generalized,ciufolini1986measure,Chicone:2002kb,Chicone:2006rm,Perlick:2007ux,%
Kerner:2001cw,vanHolten:2001ea,Colistete:2002ka,colistete2002higher,Baskaran:2003bx,Koekoek:2010pv,Koekoek:2011mm,tammelo1984physical,tammelo2006pressure}.

The explicit GDEs through
$\mathcal{O}(y^3)$ and $\mathcal{O}(y^4)$
were obtained by
works \cite{hodgkinson1972modified,bazanski1977kinematics,bazanski1977dynamics,aleksandrov1979geodesic}
in the 1970s
and Vines \cite{Vines:2014oba}
in 2014,
respectively.
Moreover,
Vines \cite{Vines:2014oba}
also provided
formulae for
the all-orders extension of the GDE
and its Lagrangian formulation,
though
in terms of
tensor expressions
that are not fully expanded out in terms of
Riemann curvature and its derivatives.
To find the explicit equation or Lagrangian at an order,
one has to solve
a group of interrelated recursion relations
in Appendix 3 of \rcite{Vines:2014oba},
which arise
in the context of
various and intricate identities of
the Synge bitensor formalism
\cite{Ruse:1931ht,Synge:1931zz,Synge:1960ueh,Poisson:2011nh}.

In this work,
we intend to revisit 
this problem
from a different framework.
The achievements are the following.
Firstly,
we develop an alternative 
to the Synge formalism 
in which
the relevant tensor expressions
are
computed 
from
a covariant calculus of differential forms.
Secondly, we provide
the exact all-orders formula
for the so-called Jacobi propagators
\cite{Dixon:1970zza,Dixon:1974xoz,Harte:2008xt,Harte:2012uw},
fully expanded out in terms of Riemann tensor and its derivatives.
Our exact expressions are infinite sums
whose coefficients are products of binomial coefficients,
showing
agreements with Vines \cite{Vines:2014oba}.
Consequently,
we provide
the explicit GDE and its Lagrangian up to $\O(y^{10})$

The key idea of our approach
is to formulate 
geodesic deviation 
as an initial-value problem
like in
works
\cite{Thiemann:1995ug,Thiemann:2000bw,Thiemann:2002vj,Hall:2001jq,Hall:2002du,hall2011adapted}.
Namely, we specify an \textit{in-in} boundary condition for the geodesic:
point $x$ and a tangent vector $y$ at $x$.
This is to be contrasted with
the Synge bitensor formalism
\cite{Ruse:1931ht,Synge:1931zz,Synge:1960ueh,Poisson:2011nh}
where a geodesic segment is characterized by
its two endpoints
$x$ and $z$
as an \textit{in-out} boundary condition.

\section{
Geodesic Deviation in Tangent Bundle}%
Let $(\M,g)$ be
a $d$-dimensional real-analytic manifold
with local coordinates $x^\m$
and metric $g_{\m\n}$,
which we call spacetime.
Its tangent bundle, $T\M$,
can be viewed as a $2d$-dimensional manifold
with local coordinates $(x^\m,y^\m)$,
based on
its local trivialization
by the coordinate vector fields.

Coordinate transformations of $T\M$
are restricted in the form
$(x^\m,y^\m) \mapsto (f^\m(x),f^\m{}_{,\n}(x)\mem y^\n)$,
where $f^\m(x)$ describes a set of real-analytic functions in a local patch of $\M$.
Importantly,
the notion of
covariance in $T\M$ is defined with respect to
these coordinate transformations.

In particular,
the following vector field 
in $T\M$
is \textit{invariant} under 
such coordinate transformations:
\begin{align}
    \label{eq:N}
    N
    \,=\,
        y^\m \frac{\partial}{\partial x^\m}
        - \Gamma^\m{}_{\r\s}(x)\mem y^\r\mem y^\s\mem \frac{\partial}{\partial y^\m}
    \,,
\end{align}
where $\Gamma^\m{}_{\r\s}(x)$ denote the Christoffel symbols.
This can be easily seen by considering its interior products with
a one-form basis in $T\M$ that transform covariantly:
\begin{align}
    \label{eq:iN}
    \i_N\hem dx^\m = y^\m
    \,,\quad
    \i_N Dy^\m = 0
    \,.
\end{align}
Here, $D$ denotes the covariant exterior derivative:
$Dy^\m = dy^\m + \Gamma^\m{}_{\n\r}(x)\mem y^\n\mem dx^\r$.
In light of its invariance,
$N$ defines a structure
characteristic of the tangent bundle.
In the mathematical jargon,
it describes a horizontal vector field 
due to the Ehresmann notion of a connection
\cite{ehresmann1948connexions,Mason:2013sva}.

\begin{figure}\centering[t]
    \centering
    \includegraphics[scale=1.5]{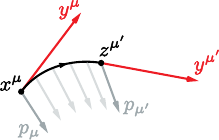}
    \smallskip
    \caption{%
        A geodesic segment joins two spacetime points $x$ (observer) and $z$ (test particle).
        The tangent vectors are respectively denoted as $y^\m$ and $y^{\m\sprime}$, 
        which are normalized such that
        $x$ and $z$ are related by unit-time geodesic flows.
    }
    \label{fig:GDprime}
\end{figure}

Crucially,
$N$ 
holds the very significance as
the \textit{generator of geodesic deviation}.
To see this, consider
the first-order formulation of the geodesic equation:
\begin{align}
\begin{split}
    \label{eq:diffeq-XY}
    \frac{d}{d\eta}\mem X^\m(\eta) 
    &\mem=\mem 
        Y^\m(\eta)
    \,,\\
    \frac{d}{d\eta}\mem Y^\m(\eta)
    &\mem=\mem
        -\Gamma^\m{}_{\r\s}(X(\eta))\,
        Y^\r(\eta)\, Y^\s(\eta)
    \,.
\end{split}
\end{align}
Evidently,
the explicit components of $N$ in \eqref{eq:N}
encode \eqref{eq:diffeq-XY}.
This implies that
the power series solution of \eqref{eq:diffeq-XY}
is given by
$X^\m(\eta) = \mathe^{\eta N} x^\m$
and
$Y^\m(\eta) = \mathe^{\eta N} y^\m$
for initial conditions
$X^\m(0) = x^\m$ and $Y^\m(0) = y^\m$,
where
$N$ is understood as a differential operator.
Namely,
the time-$\eta$ flow of $N$ solves the geodesic equation.

In particular, consider the unit-time flow:
\begin{align}
    \label{eq:XY(1)}
    X^\m(1) \mem=\mem \mathe^{N} x^\m
    \,,\quad
    Y^\m(1) \mem=\mem \mathe^{N} y^\m
    \,.
\end{align}
Borrowing the notation in the Synge bitensor formalism
\cite{Ruse:1931ht,Synge:1931zz,Synge:1960ueh,Poisson:2011nh},
we rephrase \eqref{eq:XY(1)} as the following:
\begin{align}
    \label{eq:zy}
    z^{\m\sprime}
    \mem=\mem
        \delta^{\m\sprime}{}_\m\mem
            (\mathe^{N} x^\m)
    \,,\quad
    y^{\m\sprime}
    \mem=\mem
        \delta^{\m\sprime}{}_\m\mem
            (\mathe^{N} y^\m)
    \,,
\end{align}
where $\delta^{\m\sprime}{}_\m$ is Kronecker delta.
Namely, we assign unprimed and primed indices
to objects at the original ($x$) and deviated ($z$) points,
respectively.
The rationale might be that indices represent the behavior under coordinate transformations,
and 
tensors at $x$ and $z$ transform with different Jacobian factors.
\fref{fig:GDprime} provides a spacetime picture
that visualizes \eqref{eq:zy}.

The unit-time flow of $N$
can be regarded as a diffeomorphism in $T\M$.
By the very definition of the Lie derivative,
the pullback of 
a tensor
$\a$ in $T\M$
is given by $\mathe^{\pounds_N} \a$.
Practically, this pullback
replaces
every $x$ with $\mathe^N x$ and $y$ with $\mathe^N y$.
For instance, the following identity holds:
\begin{align}
    \label{eq:ELN.dx}
    dz^{\m\sprime}
    &\mem=\mem
        \delta^{\m\sprime}{}_\m\mem
            (\mathe^{\pounds_N} dx^\m)
    \,.
\end{align}
Notably,
for differential forms,
Lie derivatives are neatly computed
by using the Cartan magic formula:
\begin{align}
    \label{eq:magic}
    \pounds_N
    \mem=\mem
        d\mem \i_N
        + \i_N d
    \,.
\end{align}
We leave it 
as an exercise to
check \eqref{eq:ELN.dx}
at each order in $y$
by using \eqref{eq:magic}.

\section{
Geodesic Deviation in A Direct Sum Bundle}%
For pedagogical reasons, let us also consider
a slight generalization of the above construction.
Suppose the direct sum of
the tangent and cotangent bundles:
$\P = T\M \oplus T^*\nem\M$,
where local coordinates are $(x^\m,y^\m,p_\m)$.
Eventually, this $3d$-dimensional space
will serve as our geometrical arena
for a first-order formulation of GDE.
In this larger bundle,
the vector field $N$ is uniquely defined by
$\i_N\hem dx^\m = y^\m$,
$\i_N Dy^\m = 0$,
and
$\i_N Dp_\m = 0$.

It is not difficult to see that this $N$ is the generator of geodesic deviation and parallel transport.
In particular,
consider the analog of the first-order differential equations in \eqref{eq:diffeq-XY}
to find that
$p_{\m\sprime} = (\mathe^{N} p_\m)\mem \delta^\m{}_{\m'}$
is the covector $p_\m$
parallel-transported to the point $z$
along the geodesic;
see \fref{fig:GDprime}.
This finding is succinctly stated as
\begin{align}
    \label{eq:W.p}
    p_{\m\sprime}
    \,=\, 
        p_\m\mem W^\m{}_{\m'}
    \,,
\end{align}
where $W^\m{}_{\m\sprime}$ denotes the \textit{parallel propagator}:
the Wilson line of the Levi-Civita connection
computed about the geodesic between $x$ and $z$.
Namely,
$p_\m \mapsto p_{\m\sprime} = p_\m\mem W^\m{}_{\m'}$
describes the isomorphism between the cotangent spaces
at $x$ and $z$
facilitated by the geodesic parallel transport.

Again,
$(x,y,p) \mapsto \mathe^{\eta N}(x,y,p)$
describes the time-$\eta$ flow in $\P$ generated by the vector field $N$.
Due to
this geometrical interpretation,
the unit-time geodesic deviation and parallel transport 
is given by the
exponentiated Lie derivative $\mathe^{\pounds_N}$.

For a concrete example,
consider a one-form $p_\m\hem dx^\m$,
i.e., the trivial extension of the cotangent bundle's tautological one-form
to $\P$.
Then we have
\begin{align}
    \label{eq:pdz0}
    p_{\m'}\hem dz^{\m\sprime}
    \mem=\mem
        \mathe^{\pounds_N} (p_\m\mem dx^\m)
    \,,
\end{align}
the right-hand side of
which can be computed order-by-order
by series-expanding $\mathe^{\pounds_N}$.
We leave it as an exercise to carry out this computation
and check consistency between the left-hand and right-hand sides.

\section{
Covariant Lie Derivative and Dressing Identity}%
The computations utilizing $\pounds_N$,
however,
are not very efficient.
One encounters unoccupied connection coefficients
arising from the components of $N$,
so covariance at the original point $x$
is not manifested
in intermediate steps.

In this light, we introduce a shorthand notation,
\begin{align}
    \label{eq:cov-magic}
    \pounds_N^D
    \mem=\mem
        D\mem \i_N
        + \i_N D
    \,,
\end{align}
defined on any tensor-valued differential form.
This describes a covariantized analog of the Lie derivative;
more information might be found in
\rrcite{Jackiw:2001jb,Ashtekar:2020xll,de2015covariant}.

Crucially,
since $p_\m\hem dx^\m$ 
is a scalar-valued one-form carrying no free indices,
using $D$ instead of $d$ makes no difference.
Hence we can equivalently use $\exp\,({\pounds_N^D})$
in \eqref{eq:pdz0}.
It is a privilege of exponentiated operators
that they are distributable as
\begin{align}
    \label{eq:pdz-dist-cov}
    \mathe^{\pounds_N^D}(p_\m\hem dx^\m)
    \hem=\hem
    (\mathe^{\pounds_N^D} p_\m)\hem
    (\mathe^{\pounds_N^D} dx^\m)
    \hem=\hem
    p_\m\hem
    (\mathe^{\pounds_N^D} dx^\m)
    \,,
\end{align}
where
the last equality follows from
$\pounds_N^D p_\m = \i_N Dp_\m = 0$.
Therefore, from
\eqrefss{eq:W.p}{eq:pdz0}{eq:pdz-dist-cov},
it follows that
\begin{align}
    \label{eq:W.dz}
    W^\m{}_{\m'}\mem dz^{\m\sprime}
    \mem=\mem
        \mathe^{\pounds_N^D} dx^\m
    \,.
\end{align}

In the same way,
it can be shown that
the $\exp\,(\pounds_N^D)$
of a tensor-valued differential form
computes its value at the deviated point,
followed by the
parallel-transportation back to the original point
via dressing by the Wilson lines.
This fact will be referred to as
the \textit{dressing identity}.

\section{
Recursion for Jacobi Propagators}%
Having set up the foundations of our formalism,
we now concern explicit evaluations.
In particular,
it follows from \eqref{eq:cov-magic} that
the right-hand side of \eqref{eq:W.dz} 
evaluates as
\begin{align}
    \label{eq:pdz1}
        dx^\m
	+
        Dy^\m
	+
        \sum_{\ell=2}^\infty 
            \frac{1}{\ell!}\mem
                (\hhnem{
                    (\i_ND)^{\ell-2} \i_N R^\m{}_\n
                }\hhem)\mem y^\n
    \,,
\end{align}
where $R^\m{}_\n$ denotes the Riemann curvature two-form such that $D^2 y^\m = R^\m{}_\n\mem y^\n$.
This computation is illustrated in Fig.\,\ref{fig:liecd-e}
as a curved deformation 
of the sequence of differential forms due to the Cartan magic formula.
As a tensor at the point $x$,
\eqref{eq:pdz1} eventually boils down to 
the following form:
\begin{align}
\begin{split}
    \label{eq:jacXY}
    W^\m{}_{\m'}\mem dz^{\m\sprime}
    \mem=\mem
    \mathe^{\pounds_N^D} dx^\m
   	\mem&=\mem
        X^\m{}_\s\mem dx^\s
        +
        Y^\m{}_\s\mem Dy^\s
    \,.
\end{split}
\end{align}
The objective now is to find the tensors
$X^\m{}_\s$ and $Y^\m{}_\s$ explicitly.
Note that
$W^{\m\sprime}{}_\m\mem X^\m{}_\s$
and
$W^{\m\sprime}{}_\m\mem Y^\m{}_\s$
are
exactly what are known as Jacobi propagators
(denoted as 
$K^{\m\sprime}{}_\s$ and $H^{\m\sprime}{}_\s$
in \rrcite{Vines:2014oba,dixon1979isolated}%
),
where $W^{\m\sprime}{}_\m$ is the inverse of $W^\m{}_{\m'}$.
We find it preferable
to peel off the Wilson line,
for which the path-ordered exponential formula is well-known and amenable.
\begin{figure}\centering
	{\begin{tikzpicture}
			\node[empty] (O) at (0,0) {};
			\node[empty] (X) at (3.0, 0) {};
			\node[empty] (Y) at (0, -0.9) {};
			\node[w] (a00) at ($(O)$) {$dx^\m$};
			\node[w] (a01) at ($(O)+1*(X)$) {$Dy^\m$};
			\node[w] (a02) at ($(O)+2.0*(X)$) {$0$};
			\node[w] (a10) at ($(O)+1*(Y)$) {$0$};
			\node[w] (a11) at ($(O)+1*(Y)+1*(X)$) {$(\i_NR^\m{}_\n)\hem y^\n$};
			\node[w] (a12) at ($(O)+1*(Y)+2*(X)$) {$0$};
			\node[w] (a21) at ($(O)+2*(Y)+1*(X)$) {$(\i_ND\mem \i_NR^\m{}_\n)\hem y^\n$};
			\node[w] (a22) at ($(O)+2*(Y)+2*(X)$) {$0$};
			\node[w] (a31) at ($(O)+3*(Y)+1*(X)$) {$\vdots$};
			\node[w] (phantom-a00) at ($(O)$) {};
			\node[w] (phantom-a01) at ($(O)+1*(X)$) {};
			\node[w] (phantom-a02) at ($(O)+2.0*(X)$) {};
			\node[w] (phantom-a10) at ($(O)+1*(Y)$) {};
			\node[w] (phantom-a11) at ($(O)+1*(Y)+1*(X)$) {};
			\node[w] (phantom-a12) at ($(O)+1*(Y)+2*(X)$) {};
			\node[w] (phantom-a10) at ($(O)+1*(Y)$) {};
			\node[w] (phantom-a11) at ($(O)+1*(Y)+1*(X)$) {};
			\node[w] (phantom-a12) at ($(O)+2*(Y)+2*(X)$) {};
			\node[w] (phantom-a21) at ($(O)+2*(Y)+1*(X)$) {};
			\node[w] (phantom-a31) at ($(O)+3*(Y)+1*(X)$) {};
			\draw[->] (a00)--(a01) node[midway,above] {\scriptsize $D\mem\i_N$};
			\draw[->] (a01)--(a02) node[] {};
			\draw[->] (a11)--(a12) node[] {};
			\draw[->] (a21)--(a22) node[] {};
			\draw[->] (phantom-a00)--(phantom-a10) node[midway,left] {\scriptsize $\i_ND$};
			\draw[->] (phantom-a01)--(phantom-a11) node[] {};
			\draw[->] (phantom-a11)--(phantom-a21) node[] {};
			\draw[->] (phantom-a21)--(phantom-a31) node[] {};
	\end{tikzpicture}}
	\caption{%
		A sequence of one-forms originating from $dx^\m$
		via the covariant Lie derivative $\pounds_N^D = D\mem \i_N + \i_ND$.
		Note that $(D\i_N)(\i_ND) = 0$,
		while $D\hem dx^\m$
		vanishes for zero torsion.
	}
	\label{fig:liecd-e}
\end{figure}

To this end, one needs to compute
$(\i_ND)^{\ell-2} \i_NR^\m{}_\n$
for $\ell {\,\geq\,} 2$.
For $\ell {\,=\,} 3$, one finds
$\i_N D\mem \i_N R^\m{}_\n
=
	y^\k
	R^\m{}_{\n\r\s;\k}(x) $ $
		y^\r\mem dx^\s
	+
		R^\m{}_{\n\r\s}(x)\mem
			y^\r\hem Dy^\s
$.
When one hits this with yet another $\i_ND$,
the covariant exterior derivative $D$ can
act on $Dy^\s$ to generate another Riemann tensor.
As a result,
one finds
two single-Riemann terms
and one double-Riemann term
at $\ell {\,=\,} 4$
(see \Sec{organic-chemistry}).
In the same fashion, higher concatenations of Riemann tensors
arise at higher orders.

A recursive structure can be identified 
in these calculations.
First of all, it follows that
\begin{align}
	\begin{split}
		\label{eq:boldQs}
		({\i_ND})^{\ell-2}\mem \i_N R^\m{}_\n\mem y^\n
        \,=\,
        {}&{}
		\sum_{p=1}^{\lfloor{\ell/2}\rfloor}\kern-0.2em
			\sum_{\a \in \Omega_p(\ell)}\kern-0.2em
				\co{\a_1,\cdots,\a_p}\mem
				(Q_{\a_1} {\nem\cdots\mem} Q_{\a_p})^\m{}_\s
			\mem dx^\s		
        \\
        {}&{}
		+
		\sum_{p=1}^{\lfloor{\hnem\tfrac{\ell-1}{2}\hnem}\rfloor}\kern-0.3em
			\sum_{\a \in \Omega_p(\ell-1)}\kern-0.2em
				\cop{\a_1,\cdots,\a_p}\mem
				(Q_{\a_1} {\nem\cdots\mem} Q_{\a_p})^\m{}_\s
			\mem Dy^\s
		\,,
	\end{split}
\end{align}
where $\a$ runs over ordered partitions
such that
\begin{align}
\begin{split}
    &
	\a = (\a_1,\a_2,\cdots,\a_p)
		\,\in\,
		\Omega_p(\ell)
    \\
    &
    \iff\quad
	\a_1 + \a_2 + \cdots + \a_p 
	\mem=\mem \ell
	\,,\quad
	\a_i \geq 2
	\,.
\end{split}
\end{align}
Note that the total number of such partitions
at each $\ell$, i.e.,
$\sum_{p=1}^{\lfloor\ell/2\rfloor}\nem |\mem{ \Omega_p(\ell) }\hem|$,
is the $(\ell{\,-\mem}1)$\textsuperscript{th} Fibonacci number.
In \eqref{eq:boldQs},
the
``$Q$-tensors'' are defined as
\begin{align}
\begin{split}
    \label{eq:def-Qten}
    &
    (Q_\ell)^\m{}_\s
    \,:=\,
	  y^{\k_1}{\cdots}y^{\k_\ell}\mem
	  R^\m{}_{\k_1\k_2\s;\k_3;\cdots;\k_\ell}\hnem(x)
    \\
    &\iq
    (Q_\ell)_{\m\s} = (Q_\ell)_{\s\m}
    \,,\quad
    (Q_\ell)^\m{}_\s\hem y^\s = 0
    \,.
\end{split}
\end{align}
The recursion relations
for the coefficients in \eqref{eq:boldQs}
are easily found from
identifying the following action of $\i_ND$:
\begin{subequations}
\begin{align}
		(Q_{\a_1} {\nem\cdots\mem} Q_{\a_p})^\m{}_\s
		\mem dx^\s
	\,\,\mapsto\,\,
        &
		\bbsq{\mem
			\sum_{i=1}^{p}\,
				(Q_{\a_1} {\nem\cdots\mem} Q_{\a_{i-1}} Q_{\a_i +1} Q_{\a_{i+1}} {\nem\cdots\mem} Q_{\a_p})^\m{}_\s
				\mem dx^\s
		}
        \nonumber
        \\
        &
		+
		(Q_{\a_1} {\nem\cdots\mem} Q_{\a_p})^\m{}_\s
			\mem Dy^\s
    \,,
	\\[-0.1\baselineskip]
		(Q_{\a_1} {\nem\cdots\mem} Q_{\a_p})^\m{}_\s
		\mem Dy^\s
	\,\,\mapsto\,\,
        &
		\bbsq{\mem
			\sum_{i=1}^{p}\,
				(Q_{\a_1} {\nem\cdots\mem} Q_{\a_{i-1}} Q_{\a_i +1} Q_{\a_{i+1}} {\nem\cdots\mem} Q_{\a_p})^\m{}_\s
				\mem Dy^\s
		}
        \nonumber
        \\
        &
		+
		(Q_{\a_1} {\nem\cdots\mem} Q_{\a_p} Q_2)^\m{}_\s
			\mem dx^\s
	\,.
\end{align}
\end{subequations}
With the proper identification of the boundary conditions \footnote{
	(a)
	$\co{\a_1,\a_2,\cdots,\a_p} = 0$,
	$\cop{\a_1,\a_2,\cdots,\a_p} = 0$
	if any of $\a_1,\a_2,\cdots,\a_p$ equals one.
	(b) $\co{2} = 1$, $\cop{2} = 0$.
},
the solution is determined as
\footnote{
	The generating function
	$G(\zeta_1,\cdots,\zeta_p) = \Sigma_\alpha c^{\alpha_1,\cdots,\alpha_p}\mem (\zeta_1)^{\alpha_1} \cdots (\zeta_p)^{\alpha_p}$
	is determined as
	$G(\zeta_1,\cdots,\zeta_p) = ((\zeta_p)^2/(\zeta_1 \mplus\cdots\mplus \zeta_p - 1)) \cdot (\zeta_1/(\zeta_1-1))^2 \cdot (\zeta_2/(\zeta_1\mplus\zeta_2 -1))^2 \cdots (\zeta_{p-1}/(\zeta_1\mplus\cdots\zeta_{p-1}-1))^2$.
}
\begin{align}
	\label{c-sols}
	c^{\a_1,\cdots,\a_p}
	\mem=\mem
		\prod_{i=1}^p
		\binom{
			\bigbig{
				\sum_{j=i}^p \a_j
			} \mminus 2
			\hem
		}{
			\a_i \mminus 2
		}
	\,,\quad
	c'^{\a_1,\cdots,\a_p}
	\mem=\mem
		\prod_{i=1}^p
		\binom{
			\bigbig{
				\sum_{j=i}^p \a_j
			} \mminus 1
			\hem
		}{
			\a_i \mminus 2
		}
	\,,
\end{align}
which describes products of binomial coefficients.
Finally, plugging in \eqref{c-sols} to \eqref{eq:boldQs},
we arrive at the following formula
for $\mathe^{\pounds_N^D} dx^\m$
from which
the tensors
$X$ and $Y$ in \eqref{eq:jacXY}
are readily read off:
\begin{align}
	\label{sol}
	\mathe^{\pounds_N^D} dx^\m 
    \,=\,
&
	 	dx^\m
	 	+ Dy^\m
	 	\\[-0.2\baselineskip]
&
	 	+\mem \sum_{\ell=2}^\infty
	 		\frac{1}{\ell!}
			\sum_{p=1}^{\lfloor{\ell/2}\rfloor}\kern-0.2em
			\sum_{\a \in \Omega_p(\ell)}\kern-0.2em
	 			\bbsq{
	 				\prod_{i=1}^p
	 				\binom{
	 					\bigbig{
	 						\sum_{j=i}^p \a_j
	 					} \mminus 2
	 					\hem
	 				}{
	 					\a_i \mminus 2
	 				}
	 				\nem\nem
	 			}\mem
        \nonumber\\
        &\qquad\quad
	 		\BB{
	 				(Q_{\a_1} {\nem\cdots\mem} Q_{\a_p})^\m{}_\s\, dx^\s
 				+ (\a_p\mminus2)\mem
	 				(Q_{\a_1} {\nem\cdots\mem} Q_{\a_{p-1}} Q_{\a_p - 1})^\m{}_\s\, Dy^\s
	 		}
	 \,.
	 \nonumber
\end{align}
See
\Sec{APP-XY}
for explicit enumerations.

The explicit solution
for the Jacobi propagators
in \eqref{sol}
has not been spelled out in the literature
to our best knowledge,
though
Appendix 3 of Vines \cite{Vines:2014oba}
has identified a set of relevant recursion relations
by building upon
Ottewill and Wardell \cite{Ottewill:2009uj}
and Dixon \cite{dixon1979isolated}.

\section{
The Lagrangians}%
We are now ready to derive
the all-orders GDE and its Lagrangian.
When described with the primed variables,
the first-order action of a free-falling test particle is
\begin{align}
	\label{eq:L0}
	\int\hhnem d\s\,
	\bigg[\,\,{
		p_{\m'}\hem \frac{d{z}^{\m\sprime}}{d\s}
		- \frac{e}{2}\mem \Big(\,{
			g^{\m\sprime\mem\n\sprime}\hhnem(z)\mem p_{\m'} p_{\n\sprime}
			+ m^2
		}\,\Big)
	}\mem\,\bigg]
	\,,
\end{align}
where $e$ is the einbein (a Lagrange multiplier),
and $m$ is the rest mass.
To describe
this particle from 
an observer's worldline, $\s \mapsto x^\m(\s)$,
we identify that
the first term
in \eqref{eq:L0}
originates from
the one-form $p_{\m'}\hem dz^{\m\sprime} = p_\m\mem (\hem {\exp\,(\pounds_N^D)\mem dx^\m} \hhem)$.
Consequently,
\eqref{eq:L0} boils down to
\begin{align}
	\label{eq:L1}
	\kern-0.4em
	\int\hhnem d\s\,
	\bigg[\,\,{
		p_\m\mem \bigg(\hem{
			X^\m{}_\n\hem 
			\frac{dx^\n}{d\s}
			{\mem+\,} Y^\m{}_\n\hem
			\frac{Dy^\n}{d\s}
		}\mem\bigg)\hnem
		- \frac{e}{2}\mem (p^2 {\mem+\,} m^2)
	}\mem\,\bigg]
	\,,
	\kern-0.5em
\end{align}
where $p^2 = g^{\m\n}\hhnem(x)\mem p_\m\hem p_\n$
since
Wilson lines due to the Levi-Civita connection respect the metric.
By integrating out $p_\m$,
we also find
a second-order Lagrangian:
\begin{align}
	\label{eq:L2}
	\int\hhnem d\s\,
	\bigg[\,\,{
		\frac{1}{2e}\mem
		\bigg(\mem{
			X\mem \frac{dx}{d\s}
			{\mem+\,} 
			Y\mem \frac{Dy}{d\s}
		}\mem\bigg)^{\nem2}
		- \frac{m^2e}{2}
	}\mem\,\bigg]
	\,.
\end{align}
Adopting an invariant measure of time $d\t = me\mem d\s$
for $m\neq0$
reproduces Vines \cite{Vines:2014oba}'s action
for affinely parameterized worldlines (isochronous correspondence \cite{aleksandrov1979geodesic,Vines:2014oba}):
\begin{align}
	\label{eq:L2-tau}
	m\mem
	\int\hnem d\t\,\,
	\frac{1}{2}\mem
	\bigg(\mem{
		\bigbig{\nem
			X\mem u {\mem+\,} Y\mem v
		}^{\nem2}
		- 1
	}\,\bigg)
	\,.
\end{align}
Here, we have denoted $u^\m := dx^\m/d\t$ and $v^\m := Dy^\m/d\t$.

Eqs.\,(\ref{eq:L1})-(\ref{eq:L2-tau})
provide exact first-order and second-order Lagrangian formulations of the all-orders geodesic deviation,
where the deviation $y^\m$ is defined as a vector attached to the observer's worldline.
We clarify that the observer's worldline is introduced as a nondynamical reference \cite{Vines:2014oba},
while $y^\m$ and $p_\m$ are dynamical variables \footnote{
	It might be interesting to understand
	the transformations to other correspondences 
	as gauge transformations in the first-order setup.
}.

The Lagrangian in \eqref{eq:L2-tau}
is explicitly given
up to $\O(y^{10})$
in \Sec{APP-L2},
showing perfect agreement with
the previous $\O(y^5)$ result
due to 
Vines \cite{Vines:2014oba}.

Note that the second-order actions in \eqrefs{eq:L2}{eq:L2-tau}
could have been directly obtained
by implementing our formalism simply in the tangent bundle $T\M$,
instead of employing the extended bundle $\P$.

\section{
The All-Orders Geodesic Deviation Equation}%
The all-orders GDE follows by varying the above Lagrangians.
Otherwise, it can also be derived 
from our formalism
in the following way.

Consider 
the first-order formulation of the free-falling EoM,
associated with \eqref{eq:L0}:
\begin{align}
	\label{ff'}
	\frac{dz^{\m\sprime}}{d\s}
	\,=\,
		e\mem p^{\m\sprime}
	\,,\quad
	\frac{Dp^{\m\sprime}}{d\s}
	\,=\,
		0
	\,.
\end{align}
The idea is to re-covariantize \eqref{ff'}
at the observer's position, $x^\m$,
via the geodesic Wilson line dressing.
Earlier, we have obtained
$W^\m{}_{\m'}\hem dz^{\m\sprime} = X^\m{}_\n\mem dx^\n + Y^\m{}_\n Dy^\n$ 
in \eqref{eq:jacXY}.
This applies to the 
first equation in \eqref{ff'}.
Taking a similar approach for the second equation as well,
we obtain a first-order formulation of the all-orders GDE:\hnem
\begin{subequations}
\label{1GDE}
\begin{align}
	\label{1GDE.x}
	e\mem p^\m
	\,&=\,
		X^\m{}_\n\mem \frac{dx^\n}{d\s}
		+ Y^\m{}_\n\mem \frac{Dy^\n}{d\s}
	\,,\\
	\label{1GDE.p}
	\frac{Dp^\m}{d\s}
	\,&=\,
		-\bb{
			\acute{X}^\m{}_{\n\s}\mem \frac{dx^\s}{d\s}
			+ \acute{Y}^\m{}_{\n\s}\mem \frac{Dy^\s}{d\s}
		}\mem p^\n
	\,.
\end{align}
\end{subequations}

For obtaining \eqref{1GDE.p}, we have 
applied the dressing identity to the one-form $Dp^\m$:
\begin{align}
\begin{split}
	\label{dressing-Dp}
	W^\m{}_{\m'}
	Dp^{\m'}
	\,&=\,
		\mathe^{\pounds_N^D} Dp^\m
	\,=\,
		Dp^\m
		+
			\acute{X}^\m{}_{\n\s}\mem p^\n\mem dx^\s
		+
			\acute{Y}^\m{}_{\n\s}\mem p^\n\mem Dy^\s
	\,.
\end{split}
\end{align}
The tensors $\acute{X}^\m{}_{\n\s}$ and $\acute{Y}^\m{}_{\n\s}$ 
are given as
\begin{subequations}
	\label{sol-acute}
\begin{align}
	\label{sol-acute.X}
		\acute{X}^\m{}_{\n\s}
		\mem&=\mem
	 	\sum_{\ell=2}^\infty
	 		\frac{1}{(\ell-1)!}
			\sum_{p=1}^{\lfloor{\ell/2}\rfloor}\kern-0.2em
			\sum_{\a \in \Omega_p(\ell)}\kern-0.2em
	 			\bbsq{
	 				\prod_{i=1}^p
	 				\binom{
	 					\bigbig{
	 						\sum_{j=i}^p \a_j
	 					} \mminus 2
	 					\hem
	 				}{
	 					\a_i \mminus 2
	 				}
	 				\nem\nem
	 			}\,
	 				(\acute{Q}_{\a_1})^\m{}_{\n\k}
	 				\mem
	 				(Q_{\a_2} \cdots Q_{\a_p})^\k{}_\s
	 \,,\\
 	\label{sol-acute.Y}
 		\acute{Y}^\m{}_{\n\s}
 		\mem&=\mem
	 	\sum_{\ell=2}^\infty
	 		\frac{1}{(\ell-1)!}	
			\sum_{p=1}^{\lfloor{\ell/2}\rfloor}\kern-0.2em
			\sum_{\a \in \Omega_p(\ell)}\kern-0.2em
	 			\bbsq{
	 				\prod_{i=1}^p
	 				\binom{
	 					\bigbig{
	 						\sum_{j=i}^p \a_j
	 					} \mminus 2
	 					\hem
	 				}{
	 					\a_i \mminus 2
	 				}
	 				\nem\nem
	 			}\,
	 				(\a_p\mminus2)\mem
	 				(\acute{Q}_{\a_1})^\m{}_{\n\k}
	 				\mem
	 				(Q_{\a_2} \cdots Q_{\a_{p-1}} Q_{\a_p - 1})^\k{}_\s
	\,,
\end{align}
\end{subequations}
which simply replaces $\ell!$ in \eqref{sol}
to $(\ell{\mem-\,}1)!$
regarding the combinatorial factors.
Here, we have defined, for $\ell \geq 2$,
\begin{align}
\begin{split}
    &
	(\acute{Q}_\ell)^\m{}_{\n\s}
	:= 
	    y^{\k_1}{\cdots}y^{\k_{\ell-1}}
	    R^\m{}_{\n\k_1\s;\k_2\cdots\k_{\ell-1}}\hnem(x)
	\\
    &
    \iq
\nem\left\{\,
\begin{aligned}[c]
	&
	(\acute{Q}_\ell)_{\m\n\s} = -(\acute{Q}_\ell)_{\n\m\s}
	\,,\quad
	\\
	&
	(\acute{Q}_\ell)_\wrap{[\m\n\s]} = 0
	\,,\quad
\end{aligned}
\begin{aligned}[c]
	&
	(\acute{Q}_\ell)_{\m\n\s} y^\s = 0
	\,,\\
	&
	(\acute{Q}_\ell)^\m{}_{\n\s}\mem y^\n = (Q_\ell)^\m{}_\s
	\,.
\end{aligned}
\right.
\end{split}
\end{align}

Combining \eqrefs{1GDE.x}{1GDE.p},
the second-order formulation
of the all-orders GDE 
in the isochronous correspondence
is found as
\begin{align}
\begin{split}
	\label{2GDE}
	- Y^\m{}_\n\mem \frac{Dv^\n}{d\tau}
	\,=\,{}
    &
		\BB{				
			\nabla_\r X^\m{}_\s
			{\,+\,} \acute{X}^\m{}_{\n\r}\mem X^\n{}_\s
		\nem}\mem u^\r u^\s
    \\&
		+ 2\mem
		\BB{					
			\nabla_\r Y^\m{}_\s
			{\,+\,} \acute{X}^\m{}_{\n\r}\mem Y^\n{}_\s
		\nem}\mem u^\r v^\s
    \\&
		+
		\bb{
			\frac{\partial}{\partial y^\r}\mem Y^\m{}_\s
			{\,+\,} \acute{Y}^\m{}_{\n\r}\mem Y^\n{}_\s
		\nem}\mem v^\r v^\s
	\,,
\end{split}
\end{align}
where
it should be understood that $\nabla_\r$ will be acted only on the Riemann tensors inside the $Q$-tensors.

By inverting the matrix $Y^\m{}_\n$ on the left-hand side
and using \eqref{2GDE},
we obtained
the explicit GDE in the form $-Dv^\m \nem/d\t = \cdots$
up to $\O(y^{10})$.
Moreover,
we also verified that 
the GDE obtained in this way
is exactly reproduced from varying the second-order Lagrangian in \eqref{eq:L2-tau},
up to $\O(y^5)$.
Here, we spell out the explicit GDE up to $\O(y^5)$
due to limited space:\footnote{
	The $\O(y^4)$ GDE provided by Vines \cite{Vines:2014oba}
	describes
	$-1$ (instead of $4$) for the coefficient of the term
	\smash{$\protect\acQ_2(Q_3u,u)$}
	and
	$-8$ (instead of $8$) for the coefficient of the terms
	\smash{$\protect\acQ_2(v,Q_2u)$} and \smash{$\protect\acQ_2(Q_2v,u)$}.
	As our $\O(y^5)$ Lagrangian in \oldeqref{eq:GDEL-5} agrees flawlessly with Eq.\,(5) in Vines,
	and since it is explicitly verified in the ancillary file \texttt{Low.nb}  
        provided with \rcite{gde}
	that the variation of the $\O(y^5)$ Lagrangian
	implies our $\O(y^4)$ GDE,
	we suppose the above mismatched coefficients are typos.
}
\begin{align}
	\label{GDEX5}
	&
	{- \frac{Dv}{d\t}}
	{}={}
	\acQ_2(u,u)
	+
	\frac{1}{2!}\mem
	\bbsq{
		\BB{
			\acQ_3(u,u)
			+
			\nabla_{\nem u} (Q_2u)
		}
		+ 4\mem \acQ_2(v,u)
	\mem}
	\\\nonumber
	&
	+ \frac{1}{3!}
	\lrsq{
    \begin{aligned}[c]
        &
		\BB{
			\acQ_4(u,u)
			+
			\nabla_{\nem u} (Q_3u)
			+
			\acQ_2(u,Q_2u)
			+
			3\mem \acQ_2(Q_2u,u)
		\hls{
			-
			Q_2\mem \acQ_2(u,u)
		}
		}
        \\
        &
		+
		\BB{
			2\mem \nabla_{\nem v} (Q_2u)
			+ 6\mem \acQ_3(v,u)
		}
		+
			4\mem \acQ_2(v,v)
    \end{aligned}
	}
	\\\nonumber
	&
	+ \frac{1}{4!} \lrsq{\begin{aligned}[c]
		&
		\lrp{
		\begin{aligned}[c]
		&
			\acQ_5(u,u)
			+
			\nabla_{\nem u} (Q_4u)
			+
			(\nabla_{\nem u} Q_2)\mem (Q_2u)
			+
			Q_2\mem \nabla_{\nem u} (Q_2u)
		\hls{
			- 2\mem Q_2\mem \nabla_{\nem u} (Q_2u)
		}
		\\
		&
			+
			6\mem \acQ_3(Q_2u,u)
			+
			3\mem \acQ_3(u,Q_2u)
			+
			\acQ_2(u,Q_3u)
			+
			4\mem \acQ_2(Q_3u,u)
        \\
        &
		\hls{
			- 2\mem Q_2\mem \acQ_3(u,u)
			- 2\mem Q_3\mem \acQ_2(u,u)
		}
		\end{aligned}
		}
		\\
		&
		+
		\BB{
			4\mem \nabla_{\nem u} (Q_3v)
			+ 8\mem \acQ_4(v,u)
			+ 8\mem \acQ_2(v,Q_2u)
			+ 8\mem \acQ_2(Q_2v,u)
		\hls{
			- 8\mem Q_2\mem \acQ_2(v,u)
		}
		}
		\\
		&
		+
		\BB{
			2\mem \nabla_{\nem v} (Q_2v)
			+
			10\mem \acQ_3(v,v)
		}
	\end{aligned}}
	\\\nonumber
	&
	+ \frac{1}{5!} \lrsq{\begin{aligned}[c]
		&
		\lrp{\,
		\begin{aligned}[c]
		&
			\acQ_6(u,u)
			+
			\nabla_{\nem u} (Q_5u)
			+
			10\mem \acQ_4(Q_2u,u)
			+
			10\mem \acQ_3(Q_3u,u)
			+
			5\mem \acQ_2(Q_4u,u)
        \\&
			+
			5\mem \acQ_2(Q_2Q_2u,u)
			+
			Q_2\mem \nabla_{\nem u} (Q_3u)
			+
			3\mem Q_3\mem \nabla_{\nem u} (Q_2u)
        \\&
			+
			(\nabla_{\nem u} Q_2)\mem Q_3
			+
			3\mem (\nabla_{\nem u} Q_3)\mem Q_2
		\hls{
			- \tfrac{10}{3}\mem Q_2\mem \nabla_{\nem u} (Q_3u)
			- 5\mem Q_3\mem \nabla_{\nem u} (Q_2u)
		}
		\\
		&
			+
			\acQ_2(u,Q_4u)
			+
			4\mem \acQ_3(u,Q_3u)
			+
			6\mem \acQ_4(u,Q_2u)
			+
			10\mem \acQ_2(Q_2u,Q_2u)
        \\&
			+
			\acQ_2(u,Q_2Q_2u)
		\hls{
			- \tfrac{10}{3}\mem Q_2\mem \acQ_4(u,u)
		}
		\hls{
			- 10\mem Q_2\mem \acQ_2(Q_2u,u)
			- 5\mem Q_3\mem \acQ_3(u,u)
        }
        \\&
        \hls{
			- 3\mem Q_4\mem \acQ_2(u,u)
			+ \tfrac{7}{3}\mem Q_2 Q_2\mem \acQ_2(u,u)
			- \tfrac{10}{3}\mem Q_2 \acQ_2(u,Q_2u)
		}
		\end{aligned}
		}
		\\\nonumber
		&
		+
		\lrp{
		\begin{aligned}[c]
		&
			10\mem \acQ_5(v,u)
			+
			6\mem \nabla_{\nem u} (Q_4v)
			+
			2\mem Q_2\mem \nabla_{\nem u} (Q_2v)
			+
			2\mem (\nabla_{\nem u} Q_2)\mem (Q_2v)
        \\&
		\hls{
			- \tfrac{20}{3}\mem Q_2\mem \nabla_{\nem u} (Q_2v)
		}
			+
			20\mem \acQ_3(Q_2v,u)
			+
			20\mem \acQ_2(Q_3v,u)
			+
			10\mem \acQ_2(v,Q_3u)
        \\&
			+
			30\mem \acQ_3(v,Q_2u)
		\hls{
			-20\mem Q_2\mem \acQ_3(v,u)
			-20\mem Q_3\mem \acQ_2(v,u)
		}
		\end{aligned}
		}
		\\\nonumber
		&
		+
		\lrp{
        \begin{aligned}[c]
        &
			6\mem \nabla_{\nem v} (Q_3v)
			+
			18\mem \acQ_4(v,v)
			+
			6\mem \acQ_2(Q_2v,v)
			- 2\mem Q_2\mem \acQ_2(v,v)
        \\&
			+ 10\mem \acQ_2(v,Q_2v)
		\hls{
			- \tfrac{40}{3}\mem Q_2\mem \acQ_2(v,v)
		}
        \end{aligned}
		}
	\end{aligned}}
	\\\nonumber
	&
	+ \O(y^6)
	\,.
\end{align}
We have
color-coded terms due to $Y^{-1}$
and
adopted a condensed notation:
$Dv/d\t \to Dv^\m \nem/d\t$,
$\acQ_3(v,u) \to (\smash{\acQ_3})^\m{}_{\r\s} $ $  v^\r u^\s$,
$Q_2\mem \smash{\acQ_2}(u,u) \to (Q_2)^\m{}_\n\mem (\smash{\acQ_2})^\n{}_{\r\s}\mem u^\r u^\s$,
$\nabla_{\nem v} (Q_2u) \to v^\r\mem (\nabla_{\nem \r} (Q_2)^\m{}_\s)\mem u^\s$,
$(\nabla_{\nem u} Q_2)\hem (Q_2u) \to u^\r\mem (\nabla_{\nem \r}(Q_2)^\m{}_\s)\mem (Q_2)^\s{}_\k\mem u^\k$,
etc.
The sources of
the minus signs on the right-hand side
are
either $Y^{-1}$ or 
the Riemann tensor identities
used for simplifying the quadratic-in-$v$ part.

\section{
Zero-Torsion Identities}%
One may notice that 
the behavior of
\smash{$\acX^\m{}_{\n\r}$} and \smash{$\acY^\m{}_{\n\r}$}
in \eqref{2GDE}
is suggestive of 
connection coefficients.
In fact, 
they arise from
the conjugation
$
	\smash{\mathe^{\pounds^D_N}\nem D\mem \mathe^{-\pounds^D_N}}
	=
	\smash{\mathe^{[\pounds^D_N,\blank]}\nem D}
$
of the covariant exterior derivative.
In turn,
it can be seen that
they encode the covariant derivative
at the deviated point:
see \eqref{avatar-gravity}.

On a related note,
the torsion-free condition 
for the Levi-Civita connection,
as $Ddx^\m = \Gamma^\m{}_{\r\s}\mem dx^\s \swedge dx^\r = 0$,
implies
$0 
= \smash{
	(\mathe^{\pounds_N^D} D\mem \mathe^{-\pounds_N^D})
	(\mathe^{\pounds_N^D} dx^\m) 
}
= 
	D(X^\m{}_\s dx^\s {+\hem} $ $ Y^\m{}_\s Dy^\s)
	+ 
		(\smash{\acX}^\m{}_{\n\r} dx^\r {\hem+\hem} \smash{\acY}^\m{}_{\n\r} Dy^\r)
		\wedge 
		(X^\n{}_\s dx^\s {+\hem} Y^\n{}_\s Dy^\s)
$,
which unpacks into three identities:
\begin{subequations}
\label{torid}
\begin{align}
	\label{torid-a}
	\nabla_\wrap{[\r} X^\m{}_\wrap{\s]}
	{\mem+\mem} \acX^\m{}_\wrap{\n[\r}\mem X^\n{}_\wrap{\s]}
	\mem&=\mem
		- \tfrac{1}{2}\mem Y^\m{}_\l\mem R^\l{}_{\n\r\s}
	\,,\\
	\label{torid-b}
	\nabla_\r Y^\m{}_\s
		{\mem+\mem} \acute{X}^\m{}_{\n\r}\mem Y^\n{}_\s
	\mem&=\mem
	\frac{\partial}{\partial y^\r}\mem X^\m{}_\s
		{\mem+\mem} \acute{Y}^\m{}_{\n\r}\mem X^\n{}_\s
	\,,\\
	\label{torid-c}
	\frac{\partial}{\partial y^{[\r}}\mem Y^\m{}_\wrap{\s]}
	{\mem+\mem} \acY^\m{}_\wrap{\n[\r}\mem Y^\n{}_\wrap{\s]}
	\mem&=\mem
		0
	\,.
\end{align}
\end{subequations}
Especially,
we have made use of \eqref{torid-b}
in \eqref{2GDE}
to simplify the computation of the term linear in both $u$ and $v$.
The ancillary file \texttt{Low.nb}
    provided with \rcite{gde}
provides an explicit check of \eqrefs{torid-a}{torid-c} up to $\O(y^5)$ or $\O(y^4)$,
which exploits various identities about the Riemann tensor.
Also, note that more identities follow 
in a similar fashion
by conjugating $D^2 = R$, $[D, R] = 0$, etc.

\section{
Summary and Outlook}%
In this work, we revisited the problem of finding the all-orders-exact GDE for finite separations
by formulating
geodesic deviation and transport
as a flow along a vector field in tangent bundle.
The technique of covariant Lie derivative
then systematically defines and generates
various bitensors with the parallel propagators peeled off,
directly producing
manifestly covariantized expressions at the original point.
This achieves an in-in formalism for geodesic deviation
that serves as an alternative to the Synge calculus.

The explicit outcomes are the all-orders formula for the Jacobi propagators in \eqref{sol}
as well as
the second-order GDE and its Lagrangian
obtained up to $\O(y^{10})$.

Our tangent-bundle formalism for geodesic deviation
is a versatile tool
and can find applications in various areas of physics,
besides classical gravitation:
covariant worldline perturbation theory
\cite{Feynman:1948ur,Feynman:1950ir,Schwinger:1951nm,vH,polchinski1985worldlineformalism,Bern:1990cu,Bern:1991aq,Strassler:1992zr,Schubert:2001he,schubert2012lectures,schwartz2014qft,witten2015every,townsend2004string,wlf-ps,csg},
quantum field theory \cite{Cheung:2021yog}, and sigma models \cite{Alvarez-Gaume:1980zra,Alvarez-Gaume:1981exa,Callan:1985ia,dpb,csg},
for instance.
\Sec{gaugetheory} 
implements our formalism
for gauge-covariant translations
in nonabelian gauge theories.

Furthermore, we realize that
our innocuous attempt to remaster the fundamental subject of 
all-orders-in-deviation GDE
surprisingly connects to 
a persistent and seemingly disparate problem in 
modern gravitational physics:
finding the all-orders-in-spin EoM
for the Kerr black hole 
in its effective point-particle description
\cite{Levi:2015msa,Levi:2018nxp,Porto:2016pyg,Guevara:2018wpp,Guevara:2019fsj,chkl2019,aho2020,gmoov}.
By implementing a probe counterpart of the 
Newman-Janis algorithm \cite{Newman:1965tw-janis},
we have found that
the all-orders GDE for an imaginary deviation
can deduce a part of the black hole's
all-orders-in-spin EoM,
yielding a consistent nonlinear completion of the 
Mathisson-Papapetrou-Dixon  \cite{Mathisson:1937zz,Papapetrou:1951pa,Dixon:1970zza}
equations.
The details have been given in
\rrcite{probe-nj,njmagic.1,njmagic.11}.

\begin{subappendices}
\section{
The ``Organic Chemistry'' of Covariant Lie Derivative Calculus}%
\label{organic-chemistry}

In this appendix, we devise
a variant of
the Penrose graphical notation \cite{penrose1956tensor,penrose1971applications,penr04-tensor}
with the following rules.
\begin{align}
	\label{chem-legend}
	\adjustbox{valign=c}{\includegraphics[scale=0.4,
		trim={0 120pt 0 0},clip
	]{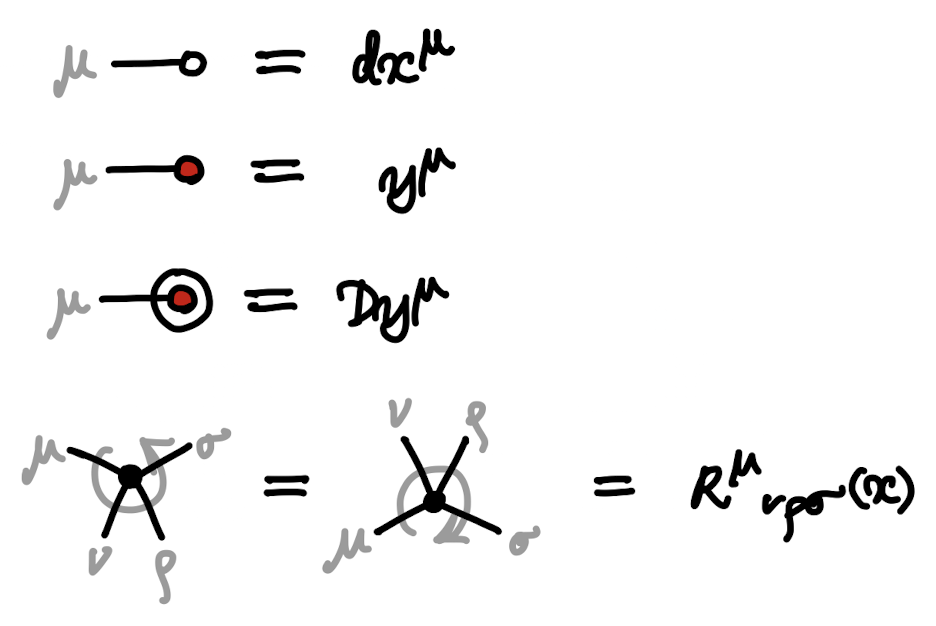}}
	\kern-4em
	\adjustbox{valign=c}{\includegraphics[scale=0.4,
		trim={0 0 0 200pt},clip
	]{figs/GDE/legend}}
\end{align}
The higher covariant derivatives of the Riemann tensor will be denoted as the following.
\begin{align}
	\label{chem-carbon}
	\adjustbox{valign=c}{\includegraphics[scale=0.4]{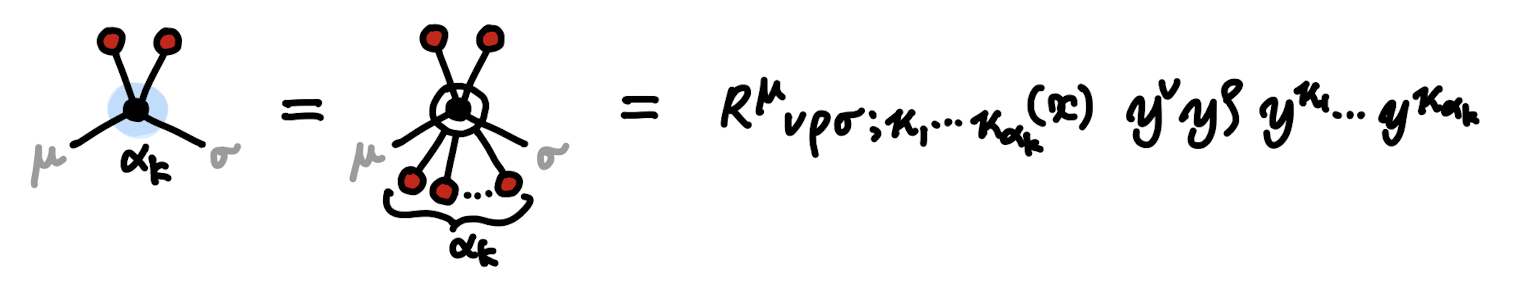}}
\end{align}
The relevant ``molecular backbones'' are the ``carbon'' (as Riemann tensor) chains of the following form.
\begin{align}
	\label{chem-chain}
	\adjustbox{valign=c}{\includegraphics[scale=0.4]{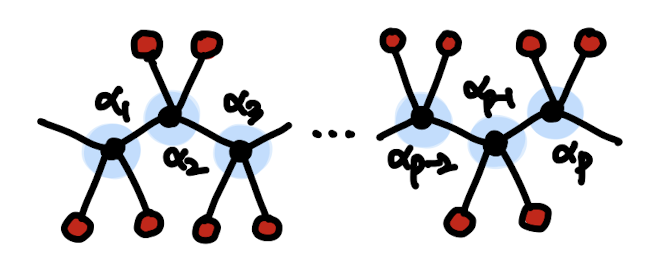}}
\end{align}
The computation of
$\i_N D\mem (\i_N R^\m{}_\n)$
and
$(\i_N D)^2\mem (\i_N R^\m{}_\n)$
proceeds as the following.
\begin{align}
	\label{chem-reaction}
	\adjustbox{valign=c}{\includegraphics[scale=0.33]{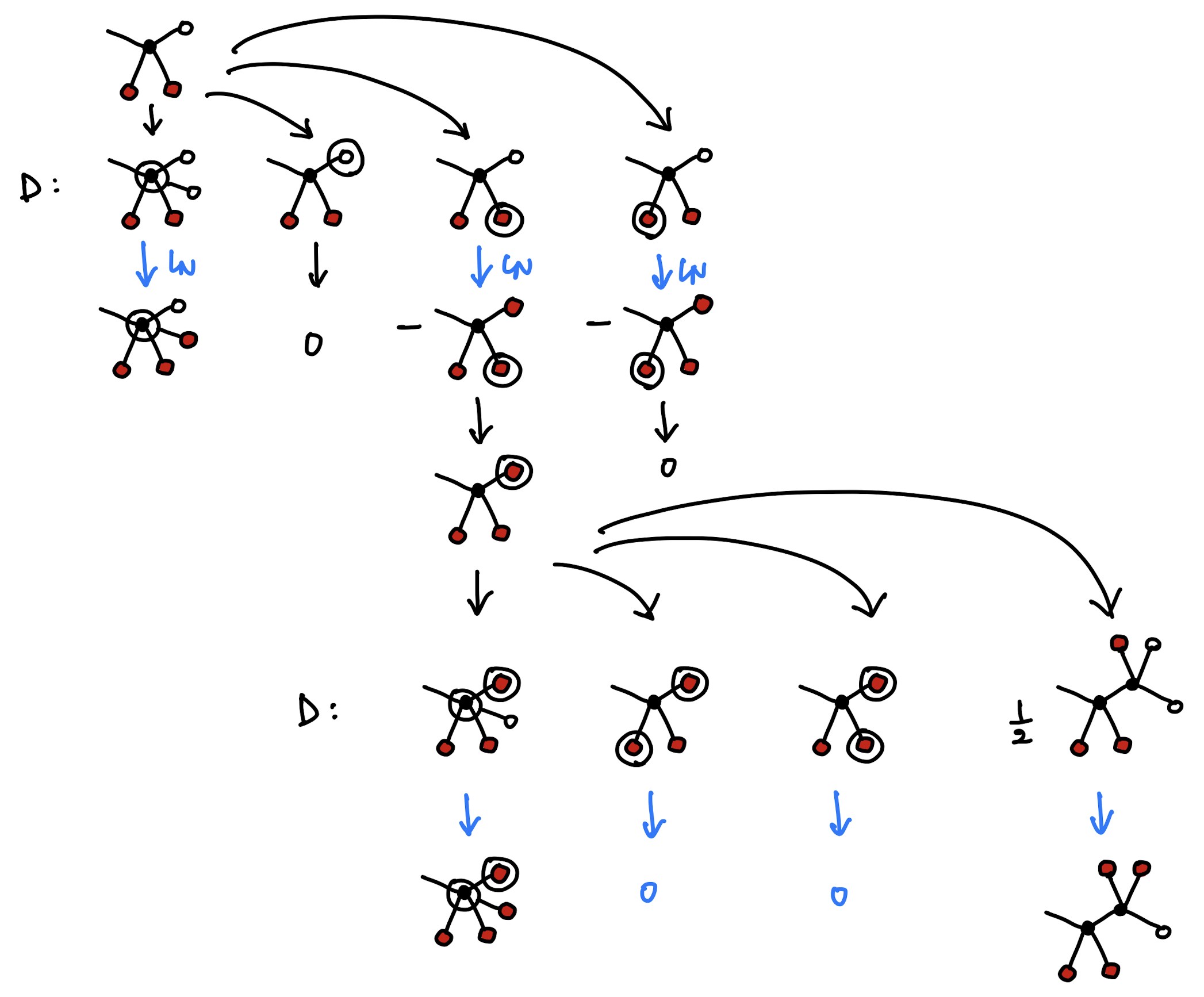}}
\end{align}
The ordering for the differential forms should be clear
from the orientations designated in \eqref{chem-legend}.
The interior product $\i_N$ 
is a ``reduction reaction''
that substitutes a ``hydroxyl group'' (as $dx$) with a ``hydrogen'' (as $y$).
Observe the mechanism for the ``carbon polymerization'' (as $Q$-tensor concatenation).

\section{
Explicit Results}%
\label{10th}

\subsection{The Jacobi Propagators}%
\label{APP-XY}%
The Jacobi propagators 
with the parallel propagator peeled off
are given in \eqref{sol}.
It is then trivial to enumerate them explicitly up to any desired order,
say $\O(y^{9})$.
With
the definition of the $Q$-tensors
in \eqref{eq:def-Qten},
they are
\begin{subequations}
\begin{align}
    \label{eq:Qaction-1}
	Dy^\m
	= {}\phantom{+}{}&
		Dy^\m
	\,,\\
    \label{eq:Qaction-2}
    (\i_N R^\m{}_\n)\hem y^\n
    = {}\phantom{+}{}&
        (Q_2)^\m{}_\s \mem{dx^\s}
    \,,\\
    \label{eq:Qaction-3}
    ({\i_ND}\mem \i_N R^\m{}_\n)\hem y^\n
    = {}\phantom{+}{}&
        (Q_3)^\m{}_\s \mem{dx^\s}
        + (Q_2)^\m{}_\s \mem{Dy^\s}
    \,,\\
    \label{eq:Qaction-4}
    (\hhnem{({\i_ND})^2\mem \i_N R^\m{}_\n}\hhhem)\hem y^\n
    = {}\phantom{+}{}&
        (Q_4 + Q_2 Q_2)^\m{}_\s \mem{dx^\s}
        + (2 Q_3)^\m{}_\s \mem{Dy^\s}
    \,,\\
    \label{eq:Qaction-5}
    (\hhnem{({\i_ND})^3\mem \i_N R^\m{}_\n}\hhhem)\hem y^\n
    = {}\phantom{+}{}&
        (Q_5 + 3 Q_3 Q_2 + Q_2 Q_3)^\m{}_\s \mem{dx^\s}
    \nonumber\\ {}+{}&
        (3 Q_4 + Q_2 Q_2)^\m{}_\s \mem{Dy^\s}
    \,,\\
    \label{eq:Qaction-6}
    (\hhnem{({\i_ND})^4\mem \i_N R^\m{}_\n}\hhhem)\hem y^\n
    = {}\phantom{+}{}&
        (Q_6 + 6 Q_4 Q_2 + 4 Q_3 Q_3 + \phantom{2}Q_2 Q_4)^\m{}_\s \mem{dx^\s}
    \nonumber\\ {}+{}&
        (4 Q_5 \phantom{{}+Q_4Q_2{}} + 4 Q_3 Q_2 + 2 Q_2 Q_3)^\m{}_\s \mem{Dy^\s}
    \nonumber\\ {}+{}&
        (Q_2 Q_2 Q_2)^\m{}_\s \mem{dx^\s}
    \,,\\
    \label{eq:Qaction-7}
    (\hhnem{({\i_ND})^5\mem \i_N R^\m{}_\n}\hhhem)\hem y^\n
    = {}\phantom{+}{}&
        (Q_7 
            + 10 Q_5 Q_2 + 10 Q_4 Q_3 + \phantom{1}5 Q_3 Q_4 + \phantom{3}Q_2 Q_5
        )^\m{}_\s \mem{dx^\s}
    \nonumber\\ {}+{}&
        (5 Q_6 \phantom{{}+1Q_5Q_2{}} 
            + 10 Q_4 Q_2 + 10 Q_3 Q_3 + 3 Q_2 Q_4
        )^\m{}_\s \mem{Dy^\s}
    \nonumber\\ {}+{}&
        (
            5Q_3 Q_2 Q_2
            + 3Q_2 Q_3 Q_2
            + Q_2 Q_2 Q_3
        )^\m{}_\s \mem{dx^\s}
    \nonumber\\ {}+{}&
        (Q_2 Q_2 Q_2)^\m{}_\s \mem{Dy^\s}
    \,,\\
    \label{eq:Qaction-8}
    (\hhnem{({\i_ND})^6\mem \i_N R^\m{}_\n}\hhhem)\hem y^\n
    = {}\phantom{+}{}&
        (Q_8
            + 15 Q_6 Q_2 + 20 Q_5 Q_3 + 15 Q_4 Q_4 + \phantom{1}6 Q_3 Q_5 + \phantom{4}Q_2 Q_6
        )^\m{}_\s \mem{dx^\s}
    \nonumber\\ {}+{}&
        (6Q_7 \phantom{{}+1Q_5Q_2{}} 
            + 20 Q_5 Q_2 + 30 Q_4 Q_3 + 18 Q_3 Q_4 + 4 Q_2 Q_5
        )^\m{}_\s \mem{Dy^\s}
    \nonumber\\ {}+{}&
        \bigg(
        \begin{array}{l}
            \phantom{+}{}
            15 Q_4 Q_2 Q_2
            + 18 Q_3 Q_3 Q_2
            + 6 Q_2 Q_4 Q_2
            \\
            + \phantom{00}\mathllap{6} Q_3 Q_2 Q_3
            + \phantom{00}\mathllap{4}  Q_2 Q_3 Q_3 
            + \phantom{0} Q_2 Q_2 Q_4
        \end{array}
        \bigg){\vphantom{\Big)}}^\m{\vphantom{\Big)}}_\s \,{dx^\s}
    \nonumber\\ {}+{}&
        (
            6 Q_3 Q_2 Q_2 
            + 4 Q_2 Q_3 Q_2
            + 2 Q_2 Q_2 Q_3
        )^\m{}_\s \mem{Dy^\s}
    \nonumber\\ {}+{}&
        (Q_2 Q_2 Q_2 Q_2)^\m{}_\s \mem{dx^\s}
    \,,\\
    \label{eq:Qaction-9}
    (\hhnem{({\i_ND})^7\mem \i_N R^\m{}_\n}\hhhem)\hem y^\n
    = {}\phantom{+}{}&
        (Q_9
            + 21 Q_7 Q_2 + 35 Q_6 Q_3 + 35 Q_5 Q_4 + 21 Q_4 Q_5 + \phantom{2}7 Q_3 Q_6 + \phantom{5}Q_2 Q_7
        )^\m{}_\s \mem{dx^\s}
    \nonumber\\ {}+{}&
        (7Q_8 \phantom{{}+1Q_5Q_2{}} 
            + 35 Q_6 Q_2 + 70 Q_5 Q_3 + 63 Q_4 Q_4 + 28 Q_3 Q_5 + 5 Q_2 Q_6
        )^\m{}_\s \mem{Dy^\s}
    \nonumber\\ {}+{}&
        \left(
        \begin{array}{l}
            \phantom{+}{} 35 Q_5 Q_2 Q_2 
            + 63 Q_4 Q_3 Q_2 
            + 42 Q_3 Q_4 Q_2
            + 10 Q_2 Q_5 Q_2
            \\
            + 21 Q_4 Q_2 Q_3
            + 28 Q_3 Q_3 Q_3
            + 10 Q_2 Q_4 Q_3
            \\
            + \phantom{00}\mathllap{7} Q_3 Q_2 Q_4
            + \phantom{00}\mathllap{5} Q_2 Q_3 Q_4
            + \phantom{00} Q_2 Q_2 Q_5
        \end{array}
        \right){\vphantom{\Big)}}\kern-0.2em{\vphantom{\Big)}}^\m{\vphantom{\Big)}}_\s \,{dx^\s}
    \nonumber\\ {}+{}&
        \bigg(
        \begin{array}{l}
            \phantom{+}{} 
            21 Q_4 Q_2 Q_2 
            + 28 Q_3 Q_3 Q_2
            + 10 Q_2 Q_4 Q_2 
            \\
            + 14 Q_3 Q_2 Q_3
            + 10 Q_2 Q_3 Q_3
            + \phantom{00}\mathllap{3} Q_2 Q_2 Q_4 
        \end{array}
        \bigg){\vphantom{\Big)}}^\m{\vphantom{\Big)}}_\s \,{Dy^\s}
    \nonumber\\ {}+{}&
        (7 Q_3 Q_2 Q_2 Q_2 + 5 Q_2 Q_3 Q_2 Q_2 + 3 Q_2 Q_2 Q_3 Q_2 + Q_2 Q_2 Q_2 Q_3)^\m{}_\s \mem{dx^\s}
    \nonumber\\ {}+{}&
        (Q_2 Q_2 Q_2 Q_2)^\m{}_\s \mem{Dy^\s}
    \,.
\end{align}
\end{subequations}
Eqs.~(\ref{eq:Qaction-1})-(\ref{eq:Qaction-5})
agree exactly
with Eq.\,(83) of \rcite{Vines:2014oba}.

\subsection{The Lagrangian}
\label{APP-L2}

The Lagrangian for the all-orders GDE 
in isochronous correspondence
is given in \eqref{eq:L2-tau}:
\begin{align}
\begin{split}
	L
	\,&=\,
		\frac{m}{2}\mem (Xu+Yv)^2
		- \frac{m}{2}
    \,,\\
	\,&=\,
		m\mem
		\bb{
			\frac{u^2{\mem-\,}1}{2}
			+ u\mdot v
		}
		+ \frac{1}{2}\mem mv^2
		+ m\mem \sum_{\ell=2}^\infty \frac{1}{\ell!}\mem \mathcal{L}_\ell
	\,.
\end{split}
\end{align}
In the last expression, the bracketed terms can be discarded
as a constant plus a total derivative.
The term $\tfrac{1}{2}\mem m v^2$, on the other hand, is the standard kinetic energy.
Hence it remains to spell out the ``interaction Lagrangian'' $\mathcal{L}_\ell$ at each order $\ell$,
which follows from the explicit Jacobi propagators by straightforward algebra:
\begin{subequations}
\begin{align}
    \label{eq:GDEL-2}
	\mathcal{L}_2
	= {}\phantom{+}{}&
			u\mem Q_2 u
	\,,\\
	\label{eq:GDEL-3}
	\mathcal{L}_3
	= {}\phantom{+}{}&
			u\mem Q_3 u
		+
			v\mem \BB{
				4 Q_2
			}\hhem u
	\,,\\
	\label{eq:GDEL-4}
	\mathcal{L}_4
	= {}\phantom{+}{}&
			u\mem \BB{
				Q_4 + 4 Q_2 Q_2
			}\hem u
		+
			v\mem \BB{
				6 Q_3
			}\hhem u
		+
			v\mem \BB{
				4 Q_2
			}\hhem v
	\,,\\
	\label{eq:GDEL-5}
	\mathcal{L}_5
	= {}\phantom{+}{}&
			u\mem \BB{
				Q_5 + 14 Q_3 Q_2
			}\hem u
		+
			v\mem \BB{
				8 Q_4 + 16 Q_2 Q_2
			}\hhem u
		+
			v\mem \BB{
				10 Q_3
			}\hhem v
	\,,\\
	\label{eq:GDEL-6}
	\mathcal{L}_6
	= {}\phantom{+}{}&
			u\mem \BB{
				Q_6 + 22 Q_4 Q_2 + 14 Q_3 Q_3 + 16 Q_2 Q_2 Q_2
			}\hem u
		\nonumber\\
		{}+{}&
			v\mem \BB{
				10 Q_5 + 50 Q_3 Q_2 + 30 Q_2 Q_3
			}\hhem u
		+
			v\mem \BB{
				18 Q_4 + 16 Q_2 Q_2
			}\hhem v
	\,,\\
	\label{eq:GDEL-7}
	\mathcal{L}_7
	= {}\phantom{+}{}&
			u\mem \BB{
				Q_7 
				+ 32 Q_5 Q_2 + 50 Q_4 Q_3 
				+ 62 Q_3 Q_2 Q_2 + 66 Q_2 Q_3 Q_2 
			}\hem u
		\nonumber\\
		{}+{}&
			v\mem \BB{
				12 Q_6 
				+ 108 Q_4 Q_2
				+ 108 Q_3 Q_3 
				+ 52 Q_2 Q_4
				+ 64 Q_2 Q_2 Q_2
			}\hhem u
		\nonumber\\
		{}+{}&
			v\mem \BB{
				28 Q_5 + 112 Q_3 Q_2
			}\hhem v
	\,,\\
	\label{eq:GDEL-8}
	\mathcal{L}_8
	= {}\phantom{+}{}&
			u\mem \lrp{
			\begin{aligned}[c]
				&
				Q_8
				+ 44 Q_6 Q_2 + 82 Q_5 Q_3 + 50 Q_4 Q_4
				\\
				&
				+ 114 Q_4 Q_2 Q_2 + 302 Q_3 Q_3 Q_2 + 62 Q_3 Q_2 Q_3
                \nem\nem
				\\
				&
                + 174 Q_ 2 Q_4 Q_2
				+ 64 Q_2 Q_2 Q_2 Q_2
			\end{aligned}
			}\hem u
		\nonumber\\
		{}+{}&
			v\mem \bb{
			\begin{aligned}[c]
				&
				14 Q_7 
				+ 196 Q_5 Q_2 
				+ 266 Q_4 Q_3 
				+ 210 Q_3 Q_4
				+ 84 Q_2 Q_5 
				\\
				&
				+ 238 Q_3 Q_2 Q_2
				+ 308 Q_2 Q_3 Q_2
				+ 126 Q_2 Q_2 Q_3
			\end{aligned}
			}\hhem u
		\nonumber\\
		{}+{}&
			v\mem \BB{
				40 Q_6 + 220 Q_3 Q_3 + 272 Q_4 Q_2 + 64 Q_2 Q_2 Q_2
			}\hhem v
	\,.
\end{align}
\end{subequations}
Eqs.~(\ref{eq:GDEL-2})-(\ref{eq:GDEL-5})
agree flawlessly with Eq.\,(5) in Vines \cite{Vines:2014oba}.

\subsection{The Geodesic Deviation Equation}
\label{APP-2GDE}
The GDE, in the second-order formulation, is given in \eqref{2GDE}.
The definition of $X^\m{}_\s$ and $Y^\m{}_\s$
is given in \eqref{sol}.
The definition of $\acute{X}^\m{}_{\n\r}$ and $\acute{Y}^\m{}_{\n\r}$
is given in \eqref{sol-acute}.
For the reader's sake, we explicitly spell them out at low orders:\nem
\begin{subequations}
\label{AllXYs5}
\begin{align}
	\label{AllX5}
	\hphantom{\acute{X}^\m{}_{\n\s}{}}
	\mathllap{X^\m{}_\s}
	\mem&=\mem{}
    \begin{aligned}[t]
    &
		\delta^\m{}_\s
		+ \frac{1}{2!}\mem (Q_2)^\m{}_\s
		+ \frac{1}{3!}\mem (Q_3)^\m{}_\s
		+ \frac{1}{4!}\mem (Q_4 + Q_2 Q_2)^\m{}_\s
    \\&
		+ \frac{1}{5!}\mem (Q_5 + 3 Q_3 Q_2 + Q_2 Q_3)^\m{}_\s
		+ \O(y^6)
	\,,
    \end{aligned}\\
	\label{AllY5}
	\hphantom{\acute{X}^\m{}_{\n\s}{}}
	\mathllap{Y^\m{}_\s}
	\mem&=\mem{}
    \begin{aligned}[t]
    &
		\delta^\m{}_\s
		+ \frac{1}{3!}\mem (Q_2)^\m{}_\s
		+ \frac{1}{4!}\mem (2 Q_3)^\m{}_\s
		+ \frac{1}{5!}\mem (3 Q_4 + Q_2 Q_2)^\m{}_\s
    \\
    &
		+ \frac{1}{6!}\mem (4 Q_5 + 4 Q_3 Q_2 + 2 Q_2 Q_3)^\m{}_\s
		+ \O(y^6)
    \end{aligned}
	\,,
\\
	\label{AllacX5}
	\hphantom{\acute{X}^\m{}_{\n\s}{}}
	\mathllap{\acute{X}^\m{}_{\n\s}}
	\mem&=\mem{}
		  \frac{1}{1!}\mem (\acute{Q}_2)^\m{}_{\n\s}
		+ \frac{1}{2!}\mem (\acute{Q}_3)^\m{}_{\n\s}
		+ \frac{1}{3!}\mem \BB{
			(\acute{Q}_4)^\m{}_{\n\s} + (\acute{Q}_2)^\m{}_{\n\l}\mem (Q_2)^\l{}_\s
		}\\
    &\hphantom{{}=\mem{}}{}
		+ \frac{1}{4!}\mem \BB{
			(\acute{Q}_5)^\m{}_{\n\s} 
			+ 3 (\acute{Q}_3)^\m{}_{\n\l}\mem (Q_2)^\l{}_\s
			+ (\acute{Q}_2)^\m{}_{\n\l}\mem (Q_3)^\l{}_\s
		}
		+ \O(y^5)
		\,,\nonumber
	\\
	\label{AllacY5}
	\hphantom{\acute{X}^\m{}_{\n\s}{}}
	\mathllap{\acute{Y}^\m{}_{\n\s}}
	\mem&=\mem{}
		  \frac{1}{2!}\mem (\acute{Q}_2)^\m{}_{\n\s}
		+ \frac{1}{3!}\mem (2 \acute{Q}_3)^\m{}_{\n\s}
		+ \frac{1}{4!}\mem \BB{
			3 (\acute{Q}_4)^\m{}_{\n\s} + (\acute{Q}_2)^\m{}_{\n\l}\mem (Q_2)^\l{}_\s
		}\\
    &\hphantom{{}=\mem{}}{}
	    + \frac{1}{5!}\mem \BB{
			4 (\acute{Q}_5)^\m{}_{\n\s} 
			+ 4 (\acute{Q}_3)^\m{}_{\n\l}\mem (Q_2)^\l{}_\s
			+ 2 (\acute{Q}_2)^\m{}_{\n\l}\mem (Q_3)^\l{}_\s
		}
		+ \O(y^5)
		\,.
        \nonumber
\end{align}
\end{subequations}

Next, we need to compute 
the inverse $(Y^{-1})^\m{}_\n$ of the Jacobi propagator $Y^\m{}_\n$
in \eqref{AllY5},
which is viable by geometric series expansion around $\delta^\m{}_\n$:
\begin{align}
	Y^{-1}
	\mem=\mem
    {}
		&
		\id
		- \frac{1}{3!}\mem \BB{
			Q_2
		}
		- \frac{1}{4!}\mem \BB{
			2\mem Q_3
		}
		- \frac{1}{5!}\mem \bb{
			3\mem Q_4 - \frac{7}{3}\mem Q_2 Q_2
		}
    \\\nonumber
        &
		- \frac{1}{6!}\mem \bb{
			4\mem Q_5 - 8\mem Q_2 Q_3 - 6\mem Q_3 Q_2
		}
    \\\nonumber
		&
		- \frac{1}{7!}\mem \bb{
			5\mem Q_6 
			- 11\mem Q_4 Q_2 - 25\mem Q_3 Q_3 - 18\mem Q_2 Q_4 
			+ \frac{31}{3}\mem Q_2 Q_2 Q_2
		}
		\\\nonumber
		&
		- \frac{1}{8!}\mem \lrp{
        \begin{aligned}[c]
            &
			6\mem Q_7 
			- \frac{52}{3}\mem Q_5 Q_2
			- 54\mem Q_4 Q_3 
			- 66\mem Q_3 Q_4
			- \frac{100}{3}\mem Q_2 Q_5
            \\
            &
			+ 34\mem Q_3 Q_2 Q_2
			+ \frac{124}{3}\mem Q_2 Q_3 Q_2 
			+ \frac{146}{3}\mem Q_2 Q_2 Q_3 
        \end{aligned}
		}
	\,.
\end{align}

Finally,
the right-hand side of \eqref{2GDE}
should be computed.
The algebraic terms, 
\smash{$\acX^\m{}_{\n\r}\mem X^\n{}_\s$},
\smash{$\acX^\m{}_{\n\r}\mem Y^\n{}_\s$},
\smash{$\acY^\m{}_{\n\r}\mem X^\n{}_\s$},
and
\smash{$\acY^\m{}_{\n\r}\mem Y^\n{}_\s$},
are straightforward to evaluate
from \eqref{AllXYs5}.
The evaluation of the differential terms,
however,
is relatively less trivial.
The issue in particular is the evaluation of the $y$-derivative.
When
a $y$-derivative hits a $Q$-tensor,
it inserts $v$ in all possible positions in the string of $y$-vectors.
Such terms can be gathered and simplified
by commuting covariant derivatives
and using Bianchi identities of the Riemann tensor
in various ways.

For the part linear in both $u$ and $v$,
this issue is simply avoided 
by virtue of
the zero-torsion identity in \eqref{torid-b}.
Hence it suffices to massage
the part quadratic in $v$,
which is the third term in the right-hand side of \eqref{2GDE}.

As the result of this simplification,
we obtain the final expression
as a sum of terms
that strictly conform to the following particular tensor structures:
\begin{subequations}
\label{structures}
\begin{align}
	\label{structure-D}
	\text{Derivative}:\quad
	&
	(Q_{\c_1} \cdots Q_{\c_r})^\m{}_\n
	\,
	(Q_\ell)^\n{}_{\k|\r}
	\,
	(Q_{\a_1} \cdots Q_{\a_p})^\k{}_\s
	\,,\\
	\label{structure-B}
	\text{Branching}:\quad
	&
	(Q_{\c_1} \cdots Q_{\c_r})^\m{}_\n
	\,
	(\acQ_\ell)^\n{}_{\l\k}
	\,
	(Q_{\a_1} \cdots Q_{\a_p})^\k{}_\s
	\,
	(Q_{\b_1} \cdots Q_{\b_q})^\l{}_\r
	\,.
\end{align}
\end{subequations}
Here, we have denoted
\begin{align}
	\label{DQ-def}
	(Q_\ell)^\m{}_{\s|\r}
	\,:=\,
		y^{\k_1}{\cdots}y^{\k_{\ell-1}}\mem
	    R^\m{}_{\k_1\k_2\s;\k_3;\cdots;\k_{\ell-1};\r}(x)
	\quad\text{for}\quad
	\ell \geq 3
	\,.
\end{align}
Crucially, 
the free index $\r$ in \eqref{DQ-def},
which can be contracted with $u$ or $v$, for instance,
is made to
describe the last (outmost) covariant derivative
acting on the Riemann tensor.
This is always possible by commuting covariant derivatives,
while
employing the Bianchi identities
facilitates
rewriting the remaining terms 
strictly in terms of the $Q$- and $\acQ$-tensors.

The final results are contained in the ancillary Mathematica notebook \texttt{All.nb}
provided with \rcite{gde}.

\section{
Gauge-Covariant Translations in Nonabelian Gauge Theory}
\label{gaugetheory}

It should be remarked that
our framework applies to 
not only gravity 
but also
nonabelian gauge theories.
Suppose a nonabelian gauge theory in a $d$-dimensional flat spacetime $\mflat$
with gauge group $G$.
Let $A^i{}_j = A^i{}_{j\m}(x)\mem dx^\m$ be the gauge connection
and let $F^i{}_j = dA^i{}_j + A^i{}_k \swedge A^k{}_j$
be its curvature,
where $i,j,\cdots = 1,2,\cdots,N$ are the fundamental indices.
For example, suppose a vector bundle $E$ over $\mflat$
whose typical fiber is $\mathbb{C}^N$.
This bundle is coordinatized by
$x^\m$ and $\psi^i$,
associated respectively with the base and fiber.

Our tangent bundle formalism
then considers the larger bundle
$\P = T\mflat \oplus E$,
which is locally isomorphic to
$\mathbb{R}^{2d} \times \mathbb{C}^N$.
Local trivialization equips $\P$ with
coordinates $x^\m$, $y^\m$, and $\psi^i$.
In $\P$, the generator of gauge-covariant translations 
is given by the vector field
\begin{align}
	N
	\,=\,
		y^\r\,
		\bb{
			\frac{\partial}{\partial x^\r}
			- A^i{}_{j\r}(x)\mem \psi^j
			\,\frac{\partial}{\partial \psi^i}
		}
	\,.
\end{align}
Again, this describes a horizontal vector field
due to the Ehresmann \cite{ehresmann1948connexions}
notion of the connection $A$.
Recalling \eqref{eq:diffeq-XY},
one derives the set of first-order differential equations
it encodes,
which leads to the conclusion that
it generates gauge-covariant translations:
$x^\m \mapsto x^\m + y^\m$,
$y^\m \mapsto y^\m$,
and
$\psi^i \mapsto \psi^{i'} = W^{i\sprime}{}_i\mem \psi^i$.
Here, the Wilson line $W^{i\sprime}{}_i$ is given by the path-ordered exponential,
\begin{align}
	\label{wilsonline}
	\mathrm{P} \exp \bb{\nem
		{-\nem \int_0^1 ds\,\,
			A_\r(x + ys)\,
				y^\r
		}
	}
	\,,
\end{align}
which describes the parallel transport
along the straight path
from $x$ to $x+y$.
Under gauge transformations, \eqref{wilsonline} transforms bilocally.

Now consider the ``$\pounds_N^D$ sequence''
of the one-form $D\psi^i = d\psi^i + A^i{}_{j\r}(x)\mem \psi^j\mem dx^\r$.
\begin{align}
    \adjustbox{valign=c}{\begin{tikzpicture}
			\node[empty] (O) at (0,0) {};
			\node[empty] (X) at (3.0, 0) {};
			\node[empty] (Y) at (0, -0.9) {};
			\node[w] (a00) at ($(O)$) {$D\psi^i$};
			\node[w] (a01) at ($(O)+1*(X)$) {$0$};
			\node[w] (a10) at ($(O)+1*(Y)$) {$\i_N F^i{}_j\mem \psi^j$};
			\node[w] (a11) at ($(O)+1*(Y)+1*(X)$) {$0$};
			\node[w] (a20) at ($(O)+2*(Y)$) {$\i_N D \i_N F^i{}_j\mem \psi^j$};
			\node[w] (a21) at ($(O)+2*(Y)+1*(X)$) {$0$};
			\node[w] (a30) at ($(O)+3*(Y)$) {$\vdots$};
			\node[w] (phantom-a00) at ($(O)$) {};
			\node[w] (phantom-a01) at ($(O)+1*(X)$) {};
			\node[w] (phantom-a10) at ($(O)+1*(Y)$) {};
			\node[w] (phantom-a11) at ($(O)+1*(Y)+1*(X)$) {};
			\node[w] (phantom-a20) at ($(O)+2*(Y)$) {};
			\node[w] (phantom-a21) at ($(O)+2*(Y)+1*(X)$) {};
			\node[w] (phantom-a30) at ($(O)+3*(Y)$) {};
			\draw[->] (a00)--(a01) node[midway,above] {\scriptsize $D\mem\i_N$};
			\draw[->] (a10)--(a11) node[] {};
			\draw[->] (a20)--(a21) node[] {};
			\draw[->] (phantom-a00)--(phantom-a10) node[midway,left] {\scriptsize $\i_ND$};
			\draw[->] (phantom-a10)--(phantom-a20) node[] {};
			\draw[->] (phantom-a20)--(phantom-a30) node[] {};
	\end{tikzpicture}}
\end{align}
This derives 
\begin{subequations}
\label{Dpsi}
\begin{align}
	\label{Dpsi-series}
	\mathe^{\pounds_N^D} D\psi^i
	\,&=\,
		D\psi^i
		+ \sum_{\ell=1}^\infty\,
			\frac{1}{\ell!}\,
			(\i_ND)^{\ell-1} \i_N F^i{}_j\mem \psi^j
	\,,\\
	\label{Dpsi-series-expanded}
	\,&=\,
		D\psi^i
		+ \sum_{\ell=1}^\infty\,
			\frac{1}{\ell!}\,
			\BB{
				(P_\ell)^i{}_{j\s}\mem dx^\s
				+ (\ell {\,-\mem} 1)\mem (P_{\ell-1})^i{}_{j\s}\mem dy^\s
			}
			\mem \psi^j
	\,,
\end{align}
\end{subequations}
where the ``$P$-tensors'' are defined for $\ell \geq 1$ as
\begin{align}
    \label{eq:def-Pten}
    (P_\ell)^i{}_{j\s}
    {}:={}\mem&
		y^{\k_1}{\cdots}y^{\k_\ell}\mem
		F^i{}_{j\k_1\s;\k_2;\cdots;\k_\ell}\hnem(x)
    \qiq
    (P_\ell)^i{}_{j\s}\hem y^\s = 0
    \,.
\end{align}
The ancillary file \texttt{P.nb} 
    provided with \rcite{gde}
verifies \eqref{Dpsi-series-expanded} in a direct fashion up to $\O(y^{10})$.
For a physical application,
\eqref{Dpsi} can be used for deriving the Wong equations \cite{wong1970field}
of a color-charged particle
as seen by an arbitrary observer.

To be further concrete, the dressing identity
$\delta^{i\sprime}{}_i\mem \mathe^{\pounds_N}( D\psi^i ) = W^{i\sprime}{}_i\mem (\mathe^{\pounds_N^D} D\psi^i)$ 
unpacks the content of \eqref{Dpsi-series-expanded} as
\begin{align}  
\begin{split}
    &
	d\psi^{i\sprime}
	+ 
		A^{i\sprime}{}_{j\sprime\r}(z)\mem \psi^{j\sprime}\mem dz^\r
    \\
	&=\,
		W^{i\sprime}{}_i\mem
		\BB{
			d\psi^i 
			+ A^i{}_{j\r}(x)\mem \psi^j\mem dx^\r
		}
    \\
    &\quad\,
		\mem+\,
		W^{i\sprime}{}_i\mem
		\sum_{\ell=1}^\infty\,
			\frac{1}{\ell!}\,
			\BB{
				(P_\ell)^i{}_{j\s}\mem dx^\s
				+ (\ell {\mem-\,} 1)\, (P_{\ell-1})^i{}_{j\s}\mem dy^\s
			}
		\mem \psi^j
	\,,
\end{split}
\end{align}
where the deviated coordinates $z^\m$ are understood as $x^\m + y^\m$.
This reveals that \eqref{Dpsi-series-expanded} encodes
\begin{subequations}
\label{avatar}
\begin{align}
	\label{avatar-a}
	W^i{}_{i'}\hem A^{i\sprime}{}_{j\sprime\mem\r}(z)\mem W^{j\sprime}{}_j
	+
	W^i{}_{i'}\mem \frac{\partial}{\partial x^\r}\mem W^{i\sprime}{}_j
	\,&=\,
		A^i{}_{j\r}(x) + 
		\sum_{\ell=1}^\infty\,
		\frac{1}{\ell!}\,
			(P_\ell)^i{}_{j\r}
	\,,\\
	\label{avatar-b}	
	W^i{}_{i'}\hem A^{i\sprime}{}_{j\sprime\mem\r}(z)\mem W^{j\sprime}{}_j
	+
	W^i{}_{i'}\mem \frac{\partial}{\partial y^\r}\mem W^{i\sprime}{}_j
	\,&=\,
		\sum_{\ell=2}^\infty\,
		\frac{1}{\ell!}\,
			(\ell {\,-\mem} 1)\mem
			(P_{\ell-1})^i{}_{j\r}
	\,.
\end{align}
\end{subequations}
Especially, \eqref{avatar-a} describes
a ``gauge transformation'' of the connection $A^{i\sprime}{}_{j\sprime\mem\r}(z)$
at the deviated point $z$,
via the Wilson line
to the original point $x$.
In this sense, its right-hand side describes an avatar of the gauge connection.
In fact, the sum of $P$-tensors in \eqrefs{avatar-a}{avatar-b},
which computes the difference between 
the connection at $z$ dragged back to $x$
and the connection at $x$,
can be used for studying the Fock-Schwinger gauge \cite{triyanta1991fsgauge}.

Note that all variables in \eqrefs{avatar-a}{avatar-b} are understood as functions of $x$ and $y$, by the very construction of our formalism.
Also, \eqrefs{avatar-a}{avatar-b} can be reproduced
from the following formula
that describes generic variations of the Wilson line
$W(s_2,s_1) = \mathrm{P}\exp\bigbig{ -\hnem\nem\int_{s_1}^{s_2} ds\, A_\r(\gamma(s)\hnem)\mem \dot{\gamma}^\r(s) }$
about an arbitrary contour $s \mapsto \c^\m(s)$:
\begin{align}
\begin{split}
	W(0,1)\mem \delta W(1,0)
	\mem&=\mem
	\begin{aligned}[t]
		&
		A_\s(\gamma(0)\hnem)\mem \delta\gamma^\s(0)
		- W(0,1)\mem 
			A_\s(\gamma(1)\hnem)\mem \delta\gamma^\s(1)
		\hem W(1,0)
		\\
		&
		+ \int_0^1 ds\,\,
			\dot{\gamma}^\r(s)\,
			W(0,s)\mem F_{\r\s}(\gamma(s)\hnem)\mem W(s,0)
			\,\delta\gamma^\s(s)
		\,.
	\end{aligned}
\end{split}
\end{align}

Note that the identities in \eqref{avatar}
are the consequences of the conjugation 
$\mathe^{\pounds_N^D} D\mem \mathe^{-\pounds_N^D}$;
recall the discussion around \eqref{torid}.
In the same fashion,
more identities 
follow by conjugating
$D^2 = F$, $[D , F] = 0$, {etc.}
by \smash{$\mathe^{\pounds_N^D}$}.

Lastly,
the identities analogous to \eqref{avatar}
in Riemannian geometry
are
\begin{subequations}
\label{avatar-gravity}
\begin{align}
	\label{avatar-gravity-a}
\begin{split}
    &{}
		W^\m{}_{\m'}\mem
		\Gamma^{\m\sprime}{}_{\n\sprime\r'}(z)\mem
		W^{\n\sprime}{}_\n
		\, W^{\r\sprime}{}_\k\mem X^\k{}_\r
	+
		W^\m{}_{\m'}\mem
		\bb{
			\frac{\partial}{\partial x^\r}
			- \Gamma^\k{}_{\l\r}(x)\mem 
				y^\l
				\frac{\partial}{\partial y^\k}
		}\mem 
			W^{\m\sprime}{}_\n
    \\[-0.1\baselineskip]
    &{}
    \,=\,
		\Gamma^\m{}_{\n\r}(x)
		+ \acX^\m{}_{\n\r}
	\,,
\end{split}\\[0.125\baselineskip]
	\label{avatar-gravity-b}
    &{}
		W^\m{}_{\m'}\mem
		\Gamma^{\m\sprime}{}_{\n\sprime\r'}(z)\mem
		W^{\n\sprime}{}_\n
		\, W^{\r\sprime}{}_\k\mem Y^\k{}_\r
	+
		W^\m{}_{\m'}\mem
		\frac{\partial}{\partial y^\r}\mem 
			W^{\m\sprime}{}_\n
    \,=\,
		\acY^\m{}_{\n\r}
	\,,
\end{align}
\end{subequations}
which can be deduced from \eqref{dressing-Dp}.
Notice the presence of the Jacobi propagators $X^\k{}_\r$ and $Y^\k{}_\r$
on the left-hand sides,
which reflects the fact that
the curve for the Wilson lines
depends on the gravitational fields
unlike in gauge theory.

\end{subappendices}
\chapter{Probe Newman-Janis Algorithm}
\label{K3:PROBENJ}

We implement a probe counterpart of NJA,
which Wick rotates the all-orders geodesic deviation equation 
into a part of exact spinning-particle EoM.
Consequently,
the gravitational dynamics of the Kerr black hole in its point-particle effective theory
is completely constrained in the SD sector
for a hidden symmetry,
implying
the spin exponentiation of same-helicity gravitational Compton amplitudes to all multiplicities.

\begin{fullnote}
    This section reproduces the contents of \rcite{probe-nj}, \fullcite{probe-nj}.
\end{fullnote}

\section{
Introduction}%
Suppose a free-falling probe particle in space,
followed by an observer.
Their separation
is famously described by 
the geodesic deviation equation.
Yet, typical textbooks
describe this equation only in
the limit of infinitesimal separation.
The exact equation
valid to all orders in \textit{separation} and \textit{curvature}
was investigated by Vines \cite{Vines:2014oba}.
Recently,
we have revisited this topic
from an alternative formalism \cite{gde}.

It seems that
this foundational subject of geodesic deviation
is intimately tied to
a pressing problem
in contemporary gravitational physics:
the exact dynamics of the Kerr black hole as a probe object.

Suppose a Kerr black hole
in the presence of mild gravitational perturbations;
it may travel through macroscopic gravitational waves
or constitute an inspiral-phase binary system.
It is well-established
\cite{ShR065,Porto:2005ac,Levi:2015msa,Porto:2016pyg,Levi:2018nxp,Kalin:2020mvi}
that
the black hole
can be treated as a relativistic spinning point particle
in the \textit{effective theory} sense,
provided separation of scales.
In this 
effective description,
the persistent problem
has been
determining
the exact EoM for the Kerr black hole
in the probe limit,
to all orders in the so-called \textit{spin length}
(spin per mass, encoding the ring radius)
and \textit{curvature}.
At the linear order in curvature,
the EoM
have been completely fixed
to all orders in spin length
\cite{ambikerr1,gmoov}.
Frustratingly,
ambiguities arise
from the very quadratic order in curvature:
in layman's terms,
physicists in this 21\textsuperscript{st} century
do not know the \smash{``$\vec{F} = m\vec{a}$\mem''} of
spinning black holes.
This agony has gained a considerable attention
lately
and has been scrutinized
in terms of scattering amplitudes
\cite{ahh2017,%
Guevara:2018wpp,Guevara:2019fsj,chkl2019,aho2020,%
Johansson:2019dnu,Aoude:2020onz,Lazopoulos:2021mna,%
Bern:2022kto,Aoude:2022trd,%
Falkowski:2020aso,Chiodaroli:2021eug,%
Ochirov:2022nqz,Cangemi:2022bew,Cangemi:2023bpe,%
fabian1,fabian2,zihan23,Saketh:2023bul,%
Cangemi:2022abk,Alessio:2025nzd%
}.

The widespread expectation is that 
the Kerr black hole will exhibit the \textit{simplest} EoM
among all massive spinning objects
in four dimensions,
in light of its simplicity
uncovered in a modern analysis \cite{ahh2017,Guevara:2018wpp,Guevara:2019fsj,chkl2019,aho2020%
}
that traces back to the NJA \cite{Newman:1965tw-janis}.
The NJA,
introduced by the very article
``Note on the Kerr Spinning-Particle Metric'' in 1965 \cite{Newman:1965tw-janis},
is a method that
derives the Kerr solution
by displacing the Schwarzschild solution
along a complexified direction
in a sense.
Crucially, this
reinterprets spin as an imaginary deviation \cite{newman1974curiosity,newman1974collection}.
Due to \rcite{nja},
the NJA now holds the status of 
a rigorous derivation of the Kerr solution
with a definite origin.

This article shows that
a unique answer exists for the simplest spinning-particle EoM
to all orders in spin and curvature,
in SD spacetimes.
This Wick rotates the \textit{deviation vector} in the exact geodesic deviation equation
to the \textit{spin length pseudovector} in exact spinning-particle EoM,
as an incarnation of NJA
at the probe level.

Our proposal is distinguished in terms of a hidden symmetry
in the presence of Killing-Yano tensors.
In perturbation theory,
it implies
the property known as
spin exponentiation \cite{ahh2017,Guevara:2018wpp,Guevara:2019fsj,chkl2019,aho2020,Johansson:2019dnu,Aoude:2020onz,Lazopoulos:2021mna}
at all graviton multiplicities.

\section{
Schwarzschild Equations of Motion}%
The free-fall EoM of a scalar probe
is given in the first-order formulation as
\begin{align}
	\label{GDE.z}
	p^{\m\sprime}
	\mem=\mem
		\frac{dz^{\m\sprime}}{d\t}
	\,,\quad
	\frac{Dp^{\m\sprime}}{d\t}
	=\,
		0
	\,,
\end{align}
where $D$ denotes the covariant differential.
$z^{\m\sprime}$ describes the position,
while $p^{\m\sprime}$ describes the momentum.
We have put primes on the indices
with hindsight.

Now introduce an arbitrary observer at position $x^\m$.
\rcite{gde} shows that
\eqref{GDE.z} is equivalent to
\begin{align}
	\label{GDE}
	p^\m
	=\,
		\a^\m{}_\n\mem \frac{dx^\n}{d\t}
		+ \b^\m{}_\n\mem \frac{Dy^\n}{d\t}
	\,,\quad
	\frac{Dp^\m}{d\t}
	=\,
		\Omega^\m{}_\n\mem p^\n
	\,,
\end{align}
where
$\a^\m{}_\n$,
$\b^\m{}_\n$,
and
$\Omega^\m{}_\n$
are tensors
composed of
the Riemann tensor $R^\m{}_{\n\r\s}(x)$, its covariant derivatives,
and the vector $y^\m$,
etc.,
all transforming tensorially at $x$.
Also, $\Omega_{\m\n} = -\Omega_{\n\m}$
so that $p^2$ is conserved.

The idea is to construct a geodesic joining between $x$ and $z$,
whose tangent at $x$ is $y^\m$:
\begin{subequations}
\label{gds}
\begin{align}
	\label{z}
	z^{\m\sprime}
	\mem=\mem
		\delta^{\m\sprime}{}_\m\mem
		\BB{
			x^\m + y^\m - \tfrac{1}{2}\mem \Gamma^\m{}_{\r\s}(x)\mem y^\r y^\s 
			+ \O(y^3)
		}
	\,.
\end{align}
In the well-known Synge formalism
\cite{Ruse:1931ht,Synge:1931zz,Synge:1960ueh,Poisson:2011nh},
the parallel propagator
about the geodesic path
is given by
\begin{align}
	\label{W}
	W^{\m\sprime}{}_\n
	\mem=\mem
		\delta^{\m\sprime}{}_\n
		-
			\delta^{\m\sprime}{}_\m\mem
			\Gamma^\m{}_{\n\r}(x)\mem y^\r
		+ \O(y^2)
	\,,
\end{align}
\end{subequations}
so a vector $p^\m$ based at $x$ is parallel-transported to $z$ as
$p^{\m\sprime} =$ $ W^{\m\sprime}{}_\n\mem p^\n$.
Especially,
the geodesic's tangent vector at $z$ is
$y^{\m\sprime} = W^{\m\sprime}{}_\n\mem y^\n$.
Multiplying 
\eqref{GDE.z}
by
the inverse
$W^\m{}_{\m'}$ of $W^{\m\sprime}{}_\m$
re-covariantizes tensors at $z$
with respect to the point $x$,
hence deriving \eqref{GDE}.

\rcite{gde} provides all-orders formulae for
$\a^\m{}_\n$, $\b^\m{}_\n$, and $\Omega^\m{}_\n$
explicitly as power series in $y$:
\begin{align}
\begin{split}
	\label{alpha}
	\kern-0.1em
	\a^\m{}_\n ={}
	&
		\delta^\m{}_\n 
		+ \tfrac{1}{2}\mem y^\r y^\s R^\m{}_{\r\s\n}(x)
		+ \tfrac{1}{6}\mem y^\r 
		y^\s y^\k R^\m{}_{\r\s\n;\k}(x)
	\\
	&
		+ \tfrac{1}{24}\mem y^\r y^\s y^\k y^\l R^\m{}_{\r\s\n;\k\l}(x)
	\\
	&
		+ \tfrac{1}{24}\mem y^\r y^\s y^\k y^\l R^\m{}_{\r\s\xi}(x)\hem R^\xi{}_{\k\l\n}(x)
		+ \O(y^5) 
	\,,
	\kern-0.1em
\end{split}
\end{align}
for instance.
Note the nonlinearity in curvature.

\eqref{GDE}
is the first-order EoM of
a free-falling probe
as seen by an arbitrary observer.
The dynamical variables are
$y^\m$ and $p^\m$,
while $x^\m$ is a nondynamical reference.
The second-order version is worked out in \rcite{gde}.

\eqref{GDE}
characterizes the exact dynamics of a minimally coupled scalar probe,
which can model a Schwarzschild black hole in its point-particle effective theory.

\section{
``Derivation'' of Kerr Equations of Motion}%
Now we implement a three-step procedure that derives
a distinct class of exact spinning-particle EoM
in four dimensions.

Firstly,
take the Schwarzschild EoM in \eqref{GDE}
for an imaginary (Wick-rotated) deviation:
\begin{align}
	\label{wick}
	y^\m 
		\,\,\mapsto\,\,
	iy^\m
	\,.
\end{align}
The result reads
\begin{subequations}
\label{heaven}
\begin{align}
	\label{heaven.x}
	p^\m
	&=\,
		\a^\m{}_\n\mem \frac{dx^\n}{d\t}
		+ i \b^\m{}_\n\mem \frac{Dy^\n}{d\t}
	\,,\\
	\label{heaven.p}
	\frac{Dp^\m}{d\t}
	&=\,
		\Omega^\m{}_\n\mem p^\n
	\,,
\end{align}
with the understanding that
the expressions for
$\a^\m{}_\n$,
$\b^\m{}_\n$,
and $\Omega^\m{}_\n$
should change accordingly
by analytic continuation of power series:
for instance,
$\a^\m{}_\n$ in \eqref{heaven.x}
equals
$
	\delta^\m{}_\n 
	- \frac{1}{2}\mem y^\r y^\s R^\m{}_{\r\s\n}(x) 
	- \frac{i}{6}\mem y^\r y^\s y^\k R^\m{}_{\r\s\n;\k}(x) 
	+ \O(y^4)
$.

Secondly,
stipulate an additional pair of equations,
\begin{align}
	\label{heaven.y}
	\frac{Dy^\m}{d\t}
	\,&=\,
		\Omega^\m{}_\n\mem y^\n
	\,,
\end{align}
\end{subequations}
which demands
that the precession of $y^\m$
is synchronized with that of
the momentum
$p^\m$ in \eqref{heaven.p}.

Let us pause for a second to clarify the interpretations.
The Wick rotation in \eqref{wick}
implements the idea of NJA
at the level of probe EoM,
so $y^\m$ is now interpreted as the spin length pseudovector.
Its definition reads
\begin{align}
	\label{ydef}
	y^\m 
	\,=\,
		-{*}S^{\m\n} p_\n / p^2
	\,,
\end{align}
where $S^{\m\n}$ is the spin angular momentum tensor.
\eqref{ydef} is equivalent to imposing
$S^{\m\n} = \varepsilon^{\m\n\r\s} y_\r p_\s$
and $p \mdot\hnem y = 0$.

Consequently,
all of $x^\m$, $p^\m$, and $y^\m$
had to be interpreted as dynamical variables
of a spinning particle,
requiring a total of $12$ equations.
The additional stipulation in \eqref{heaven.y}
was exactly for this purpose.
Physically, such a synchronization of orbital ($p^\m$)
and spin ($y^\m$) precessions
will encode
a spinning/complexified analog of equivalence principle 
proposed in \rcite{sst-asym}.

Finally,
we enter the last step of our procedure.
The $12$ equations in \eqrefss{heaven.x}{heaven.p}{heaven.y}
together define well-posed spinning-particle EoM,
only if the variables $x^\m, p^\m, y^\m$ are complexified.
This is because the Wick rotation in \eqref{wick}
induces imaginary units inside the tensors
$\a^\m{}_\n$, $\b^\m{}_\n$, $\Omega^\m{}_\n$.
Therefore,
\eqref{heaven} describes
the EoM of a spinning particle in \textit{complexified spacetime}.

Thus, we propose that
each imaginary unit in \eqref{heaven}
is eliminated by trading off with
a Hodge star $*$ on a Riemann tensor,
which is justified if the complexified spacetime is taken as \textit{SD}:
$i \mapsto *$ via
$i\hem R_{\m\n\r\s} = {*}R_{\m\n\r\s}$.
In this manner, \eqref{heaven}
is uplifted to a set of completely real equations
that dictates the time evolution of $x^\m,p^\m,y^\m$ as real variables.
Details will be elaborated later.

Parity property of $y^\m$ implies that
the linear-in-curvature part of \eqref{heaven}
reproduces the unity 
\cite{Newman:1965tw-janis,Newman:1973yu,janis1965structure}
of spin-induced multipole moments
of the Kerr black hole,
if uplifted in this precise fashion.
This provides a justification.
Intuitively, we have turned the localized orbital momentum
in geodesic deviation
(yielding unity of mass multipoles)
into a spin angular momentum.

In sum, we have proposed a probe counterpart of NJA
that uniquely derives
a well-posed instance of spinning-particle EoM
in \textit{heaven},
which will then be readily uplifted to \textit{earth}.
Here, 
we have used technical terminologies coined by 
Newman \cite{shaviv1975general,Newman:1976gc} and Pleba\'nski \cite{plebanski1975some,Plebanski:1977zz}:
\begin{align}
\begin{split}
	\label{newman-plebanski}
	\textsc{Heaven}
	\,&=\,
	\textsc{SD Spacetime}
	\,,\\
	\textsc{Earth}
	\,&=\,
	\textsc{Real Spacetime}
	\,.
\end{split}
\end{align}
To clarify,
heaven is a complex-analytic four-manifold with a holomorphic metric,
while
earth is a real-analytic four-manifold with a pseudo-Riemannian metric.

Remarkably,
the following hold for \eqref{heaven} in heaven:
\begin{enumerate}[label={}]
\item 
	(\textsc{a})
		The linear-in-curvature part of
		\eqref{heaven} describes the established linear coupling
		for Kerr.
\item 
	(\textsc{b})
		The dynamics due to \eqref{heaven}
		features exact (all-orders) conserved quantities
		in the presence of Killing vectors and Killing-Yano tensors.
\item 
	(\textsc{c})
		\eqref{heaven} admits a Lagrangian formulation.
\end{enumerate}
In these senses,
\eqref{heaven}
is an exact and consistent nonlinear completion
of the linearized Kerr spinning-particle EoM,
characterized by hidden symmetry.
Hence we propose that
\eqref{heaven}
is \textit{the} exact Kerr spinning-particle EoM
in SD spacetimes.
Below, we refer to \eqref{heaven}
as the \textit{heavenly Kerr spinning-particle EoM}.

\section{
Hidden Symmetry}%
Let us begin with the proof of
proposition (\textsc{b}).
First of all,
recall the equivalence between \eqrefs{GDE.z}{GDE}
via \eqref{gds}.
Due to the Wick rotation in \eqref{wick},
\eqref{gds} is now redefined as
\begin{subequations}
\label{igds}
\begin{align}
	\label{z.wick}
	z^{\m\sprime}
	&=\mem
		\delta^{\m\sprime}{}_\m\mem
		\BB{
			x^\m + iy^\m + \tfrac{1}{2}\mem \Gamma^\m{}_{\r\s}(x)\mem y^\r y^\s 
			+ \O(y^3)
			\nem
		}
	\,,
	\kern-0.3em\\
	\label{W.wick}
	\kern-0.2em
	W^{\m\sprime}{}_\n
	&=\mem
		\delta^{\m\sprime}{}_\n
		- i\mem
			\delta^{\m\sprime}{}_\m\mem
			\Gamma^\m{}_{\n\r}(x)\mem y^\r
		+ \O(y^2)
	\,.
\end{align}
\end{subequations}
\eqref{z.wick}
describes a complexified geodesic
for which the complexified parallel propagator is \eqref{W.wick}.
These are well-established ideas in differential geometry:
adapted complex structure
\cite{guillemin1991grauert,guillemin1992grauert,lempert1991global,szHoke1991complex,halverscheid2002complexifications,aguilar2001symplectic,burns2000symplectic,hall2011adapted}.
See \rcite{gmoov} for a reinvention.

Via \eqref{igds},
\eqref{heaven} is equivalent to
\begin{align}
	\label{heaven.z}
	p^{\m\sprime}
	\mem=\mem
		\frac{dz^{\m\sprime}}{d\t}
	\,,\quad
	\frac{Dp^{\m\sprime}}{d\t}
	\mem=\mem
		0
	\,,\quad
	\frac{Dy^{\m\sprime}}{d\t}
	\mem=\mem
		0
	\,,
\end{align}
where $p^{\m\sprime} = W^{\m\sprime}{}_\m\mem p^\m$
and
$y^{\m\sprime} = W^{\m\sprime}{}_\m\mem y^\m$
via \eqref{W.wick}.
Again, multiplying by $W^{\m\sprime}{}_\m$
yields \eqref{heaven.z}
from \eqref{heaven}.

By the very construction of the probe NJA,
the first two equations in
\eqref{heaven.z}
are simply 
the free-fall equations in \eqref{GDE.z}
with the revised definitions 
of $z^{\m\sprime}$ and $p^{\m\sprime}$
due to \eqref{igds}.
Consequently,
a part of 
the heavenly Kerr spinning-particle EoM
is literally
the Schwarzschild scalar-particle EoM,
albeit imaginary-shifted.

The last equation in \eqref{heaven.z},
on the other hand,
is a remarkable restatement of the orbit-spin synchronization.
It is amusing to see 
how the black hole's
spin precession parameters
such as the
gravimagnetic ratio $\k {\,=\,} 1$ \cite{Yee:1993ya,khriplovich1989particle,Khriplovich:1997ni}
are precisely reformulated 
as parallel transport (covariant constancy) of $p^{\m\sprime}, y^{\m\sprime}$
along the complex worldline $z^{\m\sprime}$.

Next,
recall the following well-known facts
\cite{Carter:1968ks,yano1952some,Walker:1970un,Hughston:1972qf,hughston1973spacetimes,jeffryes1984space,Jezierski:2005cg,Nozawa:2015qea,hansen2014killing,Penrose:1973naked,floyd1973dynamics,marck1983solution}
about free-fall motion
with position $x^\m$ and momentum $p^\m$.
First, a vector field $X^\m(x)$
is Killing if $X^{\m;\r}(x) = X^{[\m;\r]}(x)$,
which implies constancy of
$p_\m\hem X^\m(x)$.
Second, a two-form $Y_{\m\n}(x)$ 
is Killing-Yano 
if $Y_{\m\n;\r}(x) = Y_\wrap{[\m\n;\r]}(x)$,
which implies covariant constancy of
$Y^\m{}_\n(x)\mem p^\n$.
In turn, the scalar combination
$(\hhem Y(x)\hem p)^2 = -p_\m Y^\m{}_\r(x)\mem Y^\r{}_\n(x)\mem p^\n$
is strictly constant
along the worldline.

Via complexification, these facts are immediately 
inherited to the Kerr spinning particle:
\begin{align}
	\label{pX}
	\frac{d}{d\t}\mem
	\BB{
		p_{\m'} X^{\m\sprime}(z)
	}
	\mem&=\,
		p_{\m'} X^{\m\sprime}{}_{;\r'}(z)\, p^{\r\sprime}
	\,=\,
		0
	\,,\\
	\label{Yp}
	\frac{D}{d\t}\mem
	\BB{
		Y^{\m\sprime}{}_{\n'}\hnem(z)\mem p^{\n\sprime}
	\mem}
	\mem&=\,
		Y^{\m\sprime}{}_{\n';\r'}\hnem(z)\mem p^{\n\sprime} p^{\r\sprime}
	\,=\,
		0
	\,.
\end{align}
Meanwhile, the spinning particle features
yet another variable
that is covariantly constant:
$y^{\m\sprime}$ in \eqref{heaven.z}.
Consequently,
$p^{\m\sprime}$,
$Y^{\m\sprime}{}_{\n'}\hnem(z)\mem p^{\n\sprime}$,
and
$y^{\m\sprime}$
are
parallel-transported altogether along the complex worldline $z^{\m\sprime}$,
so
any scalar product between them
is strictly constant.

In this way, it holds true that
the following
are exact, meaning all-orders in spin and curvature,
conserved quantities
of the Kerr spinning particle:
\begin{align}
	\label{Q}
	\mathds{Q}
	\,&=\,
		p_{\m'} X^{\m\sprime}(z)
	\,,\\
	\label{R}
	\mathds{R}
	\,&=\,
		p_{\m'} Y^{\m\sprime}{}_{\n'}\hnem(z)\mem y^{\n\sprime}
	\,,\\
	\label{C}
	\mathds{C}
	\,&=\,
		- p_{\m'} Y^{\m\sprime}{}_{\r'}\hnem(z)\mem Y^{\r\sprime}{}_{\n'}\hnem(z)\mem p^{\n\sprime}
	\,.
\end{align}
\eqrefs{R}{C} are
all-orders completions of the R\"udiger \cite{rudiger1981conserved,rudiger1983conserved} and Carter \cite{Carter:1968ks} constants,
respectively.

Explicitly,
\eqref{Q} can be re-covariantized 
at $x$ as
\begin{align}
\begin{split}
	\label{pX.split}
	p_\m\hem W^\m{}_{\m\sprime}\, X^{\m\sprime}(z)
	\mem=\,
		\sum_{\ell=0}^\infty\,
			\frac{i^\ell}{\ell!}\,
				p_\m\mem 
				X^\m{}_{;\r_1\cdots\r_\ell}(x)\mem
				y^{\r_1} {\cdots\hem} y^{\r_\ell}
	\,,
	\nem
\end{split}
\end{align}
yielding an infinite expansion in the spin length.
When written in this form,
the conservation
becomes quite nontrivial to check:
the use of identities regarding $\a^\m{}_\n$, $\b^\m{}_\n$, $\Omega^\m{}_\n$
is mandated
at each order.
This highlights the power of the spin-resummed formulation
facilitated by \eqref{heaven.z}.
The same analysis applies to \eqrefs{R}{C} as well.

Furthermore,
heaven is stipulated to be SD,
so its ASD spinor bundle is flat.
As a result, the ASD Pleba\'nski two-forms
\cite{Plebanski:1977zz,Capovilla:1991qb}
are covariantly constant:
$(\Sigma^a)_{\m\n}$ for $a {\,=\,} 1,2,3$.
These serve as Killing-Yano tensors
\cite{Gibbons:1987sp,Nozawa:2015qea}
and define a quaternionic structure
as
$
    (\Sigma_a)^\m{}_\r\hem (\Sigma_b)^\r{}_\n
    = -\delta_{ab}\mem \delta^\m{}_\n
    + \ve^c{}_{ab}\mem (\Sigma_c)^\m{}_\n
$
\cite{Gibbons:1987sp,Atiyah:1978wi,Nozawa:2015qea}.
Consequently,
we now identify
$p^{\m\sprime}$,
$Y^{\m\sprime}{}_{\n'}\hnem(z)\mem p^{\n\sprime}$,
$(\Sigma_a)^{\m\sprime}{}_{\n'}\hnem(z)\mem p^{\n\sprime}$,
$y^{\m\sprime}$
being covariantly constant altogether,
so any scalar product between them is a constant of motion,
such as
\begin{align}
	\label{K}
	\mathds{K}_a
	\,=\,
		p_{\m'} (\Sigma_a)^{\m\sprime}{}_{\r'}\hnem(z)\mem Y^{\r\sprime}{}_{\n'}\hnem(z)\mem p^{\n\sprime}
	\,.
\end{align}

A physically relevant example
is the SD Taub-Newman-Unti-Tamburino (NUT) background,
which describes a gravitational instanton
\cite{hawking1977gravitational,Gibbons:1978tef}.
It is now well-understood 
\cite{nja,note-sdtn,Newman:1973yu,Newman:2002mk,Gross:1983hb,Ghezelbash:2007kw,Crawley:2021auj,gabriel1,gabriel2,Adamo:2023fbj}
to an explicit extent \cite{nja,note-sdtn}
that this background
is the SD sector of the Kerr background,
such that the same Killing-Yano tensor is shared \cite{note-sdtn,nja}.
By using the Kerr-Schild metric provided by \rcite{note-sdtn},
\eqref{K} evaluates to
$\vec{\mathds{K}} = \vep \mt (\vex\mtimes\vep) - ip_0\mem (\vex\mt\vep)$,\footnote{
	Also, the three-vector index of $\mathds{K}_a$ is global
	since we gauge-fixed the ASD spinor bundle
	by zeroing the spin connection coefficients.
}
which
amusingly
reincarnates
the LRL vector
(cf.\:\rrcite{Gibbons:1986df,Gibbons:1986hz,Gibbons:1987sp,Feher:1986ib,Feher:1988th,Cordani:1989ip,Iwai:1992zu,Guevara:2023wlr}).

By incorporating time translation and rotation invariances as well,
we establish that
the motion of the Kerr spinning-particle
in the SD sector of the Kerr background
is \textit{superintegrable},
with hidden symmetry $\mathrm{so}(4)$:\footnote{
	with suitable complexification
}
that of
Kepler problem.
This is a new result on the all-orders dynamics of
the Kerr-Kerr system.
Previous works have truncated the probe's spin order \cite{Compere:2023alp,Jakobsen:2021susysky,Gibbons:1993ap,Akpinar:2025tct}
or NJ (Newman-Janis) shifted only the background \cite{Guevara:2023wlr,Guevara:2024edh}.
\nomenclature{NJ}{Newman-Janis}

To summarize,
we have shown that
exact symmetries arise
for our proposed Kerr spinning particle
by the very structure of the probe NJA.
The conserved quantities of the spinning probe
can arise as direct NJ shifts of
the conserved quantities of the scalar probe.
For the explicit example of physical relevance,
we have found a sufficient number of functionally independent conserved quantities,
establishing superintegrability.

\section{
Lagrangian Formulation}%
Next, we prove proposition (\textsc{c}).
We find an explicit first-order worldline action,
\begin{align}
	\label{Lz}
	\int\mem
		p_{\m'}\mem \BB{
			dz^{\m\sprime} 
			- i\mem Dy^{\m'}
			+ {*}\Theta^{\m\sprime}{}_{\n'}\mem y^{\n\sprime}
		\,}
		- \mathscr{C}
	\,,
\end{align}
where we have omitted constraint terms ($\mathscr{C}$)
suitably
encoding the conditions such as $p^2 = -m^2$ and $p \mdot\hnem y = 0$.
Computation shows that
the saddle of \eqref{Lz}
is precisely \eqref{heaven.z};
see \Chap{K3:OSD1} for details.

In \eqref{Lz},
$\Theta^{\m\sprime}{}_{\n'}$
describes the ``infinitesimal angle''
as a one-form,
coupled to $S^{\n\sprime}{}_{\m'\nem} = {*}(y \mwedge p)^{\n\sprime}{}_{\m\sprime}$
to realize the kinetic term for spin.
The covariant exterior derivative $D$ acts as
$D\Theta^{\m\sprime}{}_{\n'}
= \tfrac{1}{2}\mem R^{\m\sprime}{}_{\n'\r'\s'}(z)\mem dz^{\r\sprime} \swedge dz^{\s\sprime}
$,
implying
\begin{align}
	\label{D=0}
	D\mem
	\BB{
		- i\mem Dy^{\m'}
		+ {*}\Theta^{\m\sprime}{}_{\n'}\mem y^{\n\sprime}
	\,}
	\,=\,
		0
\end{align}
for the SDity.
Via \eqref{D=0},
\eqref{Lz} is equivalent to
\begin{align}
	\label{L}
	\int\mem
		p_\m\mem \BB{\hem
			\a^\m{}_\n\mem dx^\n
			+ i\mem \b^\m{}_\n\mem Dy^\n
			- i\mem Dy^\m
			+ {*}\Theta^\m{}_\n\mem y^\n
		}
		- \mathscr{C}
	\,,
\end{align}
which re-covariantizes a one-form
at the point $x$ \cite{gde}.
Both
\eqrefs{Lz}{L}
retrieve the action of a free massive spinning particle
\cite{Hanson:1974qy,ambikerr0}
in the flat limit.

\eqref{Lz} is a complexified action
dictating the evolution of
complexified variables $z^{\m\sprime}$, $p^{\m\sprime}$, $y^{\m\sprime}$.
When converted to \eqref{L},
it can be uplifted to a purely real action
dictating the evolution of real variables
$x^\m$, $p^\m$, $y^\m$.
\section{
Linearized Coupling}%
Finally, we prove proposition (\textsc{a})
by showing that
dropping all terms nonlinear in the curvature
in \eqref{L}
reproduces the well-known 
Levi-Steinhoff action \cite{Levi:2015msa}
up to iterating EoM.
The only important term is $p_\m\mem \a^\m{}_\n\mem dx^\n$,
as $\beta^\m{}_\n - \delta^\m{}_\n = \O(R^1y^2)$ \cite{gde} and
$Dy^\m\hnem/d\t = \Omega^\m{}_\n\mem y^\n = \O(R^1y^2)$
on EoM.
As is glimpsed in \eqref{alpha},
the $\O(R^1)$ part of $\a^\m{}_\n$
in heaven
is given by \cite{gde}
\begin{align}
	\label{alphaR1}
	\delta^\m{}_\n
	\,+\,
	\sum_{\ell=2}^\infty\,
		\frac{1}{\ell!}\,
			{*}^\ell R^\m{}_{\k_1\k_2\n;\k_3;\cdots;\k_\ell}(x)\,
			y^{\k_1} \cdots y^{\k_\ell}
	\,,
\end{align}
after the Wick rotation in \eqref{wick}.
Via $dx^\m\hnem/d\t = p^\m + \O(R^1y^2)$
and $S^{\m\n} = \ve^{\m\n\r\s} y_\r p_\s$,
the relevant parts of the action boil down to
$\int p_\m\mem \a^\m{}_\n\mem p^\n - \tfrac{1}{2}\mem S^{\m\n} {*}\Theta_{\m\n} - \mathscr{C}$,
reproducing the
Levi-Steinhoff action \cite{Levi:2015msa}
for the Kerr multipole coefficients $C_\ell = 1$
\cite{ahh2017,Guevara:2018wpp,Guevara:2019fsj,chkl2019,aho2020}
upon
plugging in \eqref{alphaR1}.

In fact, the unity $C_\ell {\:=\,} 1$ of multipole coefficients
can be shown directly
from the EoM
by matching  
$\Omega^\m{}_\n$ in \eqref{heaven}
with
the universal template for spin precession EoM \cite{ambikerr1}
extending the Mathisson-Papapetrou-Dixon  \cite{Mathisson:1937zz,Papapetrou:1951pa,Dixon:1970zza}
equations.
Invariantly, it is verified by the spin exponentiation for one graviton
(see \eqref{spin-exp}).

\section{
Spin Exponentiation at All Multiplicities}%
Finally, let us explore all orders in perturbation theory:
\begin{enumerate}[label={}]
\item 
	(\textsc{d})
		Positive-helicity graviton Compton amplitudes
		due to the action in \eqref{Lz}
		exhibit spin exponentiation at all multiplicities.
\end{enumerate}
No physical explanation 
has existed
for this property
to the best of our knowledge,
although its consistency at the amplitudes level has been known \cite{Johansson:2019dnu,Aoude:2020onz,Lazopoulos:2021mna}.
In our case,
the structure of the probe NJA
is highly constrained for the hidden symmetries in the SD sector,
whose implication is spin exponentiation at all multiplicities.

To prove proposition (\textsc{d}),
we expand the gravitational fields around Minkowski background,
yet after switching to the tetrad formalism.
In tetrad formalism, the constraints and Hamiltonian (put in $\mathscr{C}$ in \eqref{Lz}) take the same form as in the free theory.
Also, the spin connection describes antisymmetric indices as 
$\gamma^{AB}{}_\r = - \gamma^{BA}{}_\r$,
on which the Hodge star 
acts to yield $+i$ in heaven.

Crucially,
this shows that the coupling to the spin connection
identically vanish in \eqref{Lz}---%
consistently with \eqref{D=0}---%
so that the only graviton coupling is via the coframe perturbation,
$h^A{}_\m = e^A{}_\m - \delta^A{}_\m$:
the interaction Lagrangian describes a single term,
$p_{A'}\mem h^{A\sprime}{}_{\m'}(z)\mem \dot{z}^{\m\sprime}$.
This is identical to the interaction Lagrangian 
$p_A\mem h^A{}_\m(x) \dot{x}^\m$
of
the minimally coupled scalar particle
in tetrad formalism,
via complexifications
$p_A \mapsto p_{A'}$,
$x^\m \mapsto z^{\m\sprime}$.
Consequently, it can be diagrammatically proven that
all Feynman diagrams
for the Kerr spinning-particle
are mere complexifications of those of the Schwarzschild scalar particle.
This implies 
that the amplitudes for $(n{\mem-\,}2)$ gravitons satisfy
\begin{align}
	\label{spin-exp}
	\M_\text{Kerr}(3^+{\cdots\mem}n^+)
	\,=\,
		\M_\text{Sch}(3^+{\cdots\mem}n^+)\,
			\mathe^{(\sum k) \cdot a}
	\,,
\end{align}
where $(\sum k) = k_3 + \cdots + k_n$ is the total momentum.
Here, the asymptotic free trajectory 
is $z^{\m\sprime} = -ia^{\m\sprime} + mu^{\m\sprime}\mem \t$
for constant spin $a^{\m\sprime}$
and velocity $u^{\m\sprime}$ parameters
\footnote{
	We are sorry that our convention for the spin length pseudovector
	flips sign from the standard convention:
	$y^\m = -a^\m$,
	so $S_{12} = \varepsilon_{1203} p^0 a^3 = (+1)\mem (+m)\mem a^3$
	in the rest frame.
}.

It suffices to work in heaven for showing \eqref{spin-exp}
because heaven consists of positive-helicity gravitons
in the incoming convention
\cite{bialynicki1981note,ashtekar1986note}.
Also, it should be clear that
any real uplift of the action in \eqref{L}
implies
the all-multiplicity
spin exponentiation
for both positive- and 
negative-helicity gravitons.
One should be able to reproduce the all-multiplicity spin exponentiation with any such real uplift,
as scattering amplitudes are (worldline) field redefinition invariant.

\section{
Uplift to Earth}%
Eventually,
we elaborate on the uplifting procedure,
which provides the extension to generic backgrounds
with \textit{both SD and ASD modes}.
For instance, recall $\a^\m{}_\n$ in \eqref{alpha}
and suppose the Wick rotation in \eqref{wick}
is applied.
On the support of SDity,
each imaginary unit $i$ 
is absorbed to a Riemann tensor.
Clearly, ambiguities arise from $\O(y^4)$
due to the nonlinear-in-Riemann term:
which Riemann tensor
are we absorbing $i$ into,
$R^\m{}_{\r\s\xi}(x)$ or  $R^\xi{}_{\k\l\n}(x)$?

We dub this ambiguity, in distributing the Hodge stars over nonlinear strings of Riemann tensors,
the \textit{star-shifting ambiguity}.
Its invariant content at 
$\O(R^{n-2})$
in the Lagrangian
can be substantiated as contact deformations of the
$n$-graviton Compton amplitude.

Take $n {\:=\:} 4$, for instance.
By plugging in the straight-line trajectory,
we learn that
it suffices to compare between
the following two terms
in the plane-wave background 
$R_{\m\n\r\s}(x) =
	\e_3\mem \varphi^+_{3\m\n}\mem \varphi^+_{3\r\s}\mem \mathe^{ik_3x}
	+ \e_4\mem \varphi^-_{4\m\n}\mem \varphi^-_{4\r\s}\mem \mathe^{ik_4x}
$,
where
\smash{$\varphi^+_3$} and \smash{$\varphi^-_4$} are gauge-invariant polarizations for the positive- and negative-helicity gravitons:
\begin{subequations}
\label{starshift}
\begin{align}
	\label{starshiftL}
	S^{\m\n}\mem {
		((a{\mem\cdot\mem}\partial)^{j_1}\mem
			{*}R_{\m\n}{}_{a \k})\mem 
		((a{\mem\cdot\mem}\partial)^{j_2}\mem
			R^{\k}{}_{a \r\s})
	}\mem S^{\r\s}
	\,,\\
	\label{starshiftR}
	S^{\m\n}\mem {
		((a{\mem\cdot\mem}\partial)^{j_1}\mem
			R_{\m\n}{}_{a \k})\mem 
		((a{\mem\cdot\mem}\partial)^{j_2}\mem
			{*}R^{\k}{}_{a \r\s})
	}\mem S^{\r\s}
	\,.
\end{align}
\end{subequations}
Upon extracting the part multilinear in $\e_3$ and $\e_4$,
\eqrefs{starshiftL}{starshiftR}
are equal in magnitude but \textit{opposite in sign}:
$
	\pm\mem
	(\varphi^+_3 S)
	(\varphi^-_4 S)
	(
		(k_3a)^{j_1}
		(k_4a)^{j_2}
		-
		(k_4a)^{j_1}
		(k_3a)^{j_2}
	)
$ $
	(a\hem\varphi^+_3\varphi^-_4a)
$.
Thus, each time one shifts the Hodge star from one Riemann tensor to another,
one deforms the Compton amplitude
by a contact term
if $j_1 \neq j_2$.

In fact, one can simply add an explicit worldline operator
in the Lagrangian
that involves at least one SD and one ASD curvatures.
Yet, such an explicit deformation may
seem more artificial than star-shifting.

Most generally,
we conclude that the uplift
of the heavenly Kerr EoM
and its Lagrangian
to earth
is ambiguous
due to the possibility of adding such mixed-chirality terms:
\textit{earthly deformations}, so to speak.\footnote{
    Despite this ambiguity
    and despite the possible mismatches with black hole perturbation theory computations,
    our explicit result on an earthly uplift
    is at least practically useful:
    it provides the ansatz that automatically guarantees the same-helicity spin exponentiation to all multiplicities,
    on top of which one can add mixed-chirality operators systematically.
}
Star-shifting is 
a natural instance of earthly deformations.

\section{
Compton Amplitude}%
Note that
star-shifting ambiguities
begin at $\O(y^4)$.
For instance,
we had obtained the spurious-pole-free mixed-helicity gravitational Compton amplitude
from an earthly Kerr Lagrangian:
\begin{align}
\begin{split}
\label{MKerr}
	\frac{
		\M_\text{Kerr}(3^+4^-)
	}{
		\M_\text{Sch}(3^+4^-)\mem
		\mathe^{-\a/2}\mem \mathe^{\b/2}
	}
	\,=\,
		1 + \gamma + \gamma^2\mem
		\bb{
			\frac{\gamma+\beta}{\alpha+\beta}\mem
			E_2(\a)
			-
			\frac{\gamma-\alpha}{\alpha+\beta}\mem
			E_2(-\b)
		}
	\,,
\end{split}
\end{align}
where $\a,\b,\c$ are linear-in-spin parameters such that
the trivial spin exponentiation is
$
	\M_\text{Sch}(3^+4^-)\mem
	\mathe^{-\a/2}\mem \mathe^{\b/2}
$
and the 
AHH
amplitude 
with the spurious pole 
\cite{ahh2017}
is 
$
	\M_\text{Sch}(3^+4^-)\mem
	\mathe^{-\a/2}\mem \mathe^{\b/2}\mem
	\mathe^{\gamma}
$,
and $E_2(\xi) := (\mathe^\xi {\,-\,} \xi {\,-\,} 1)/\xi^2$.
Notably,
\eqref{MKerr}
deviates from AHH
already from $\O(\gamma^4)$,
although the strict necessity of contact deformations 
arises from $\O(\gamma^5)$
\cite{ahh2017}.\footnote{
	A similar phenomenon had occurred in the electromagnetic analog
	\cite{Kim:2024grz}.
}

\section{
Interpretation}%
The NJA had been in a mysterious status
for decades.
Despite initial criticisms
\cite{ernst1968new2,Drake:1998gf},
constant attempt was made for explanations
\cite{Talbot:1969bpa,Drake:1998gf,Gurses:1975vu,flaherty1976hermitian,grg207flaherty,Rajan:2016zmq,giampieri1990introducing,Newman:1965tw-janis,penrose1967twistoralgebra}.
Especially,
Newman himself
envisioned a fundamental unification of spin and spacetime
into a complex geometry,
in which the spinning particle
truly becomes a scalar particle
\cite{%
	newman1988remarkable,Newman:1973afx,Newman:2002mk,Newman:1973yu,newman1974curiosity,newman1974collection,Newman:1973afx,Newman:2004ba,
	Newman:1976gc,ko1981theory,grg207flaherty%
}.
This idea was not only pursued at the level of sourced fields
but also at the level of probe particles:
``complex center of mass.''
An attempt for EoM
appears in \rcite{ko1981theory}.

This chapter
vividly realizes Newman's dream
in terms of 
explicit and rigorous frameworks
of geodesic deviation
\cite{gde}
and
adapted complex structure
\cite{guillemin1991grauert,guillemin1992grauert,lempert1991global,szHoke1991complex,halverscheid2002complexifications,aguilar2001symplectic,burns2000symplectic,hall2011adapted}.
The holomorphic worldline $z^{\m\sprime}$
is Newman's complex center of mass,
realized in fully nonlinear gravity.
However,
our analysis on earthly deformations
clarifies that
Newman's proposal
of the Kerr spinning particle
being equivalent to a scalar particle
is valid only in the SD sector.
To reiterate,
the probe NJA
pinpoints a unique dynamics
\textit{within} heaven.
This point is made further evident in \Chap{K3:OSD1},
where 
an explicit ASD group of terms
ruins
the equivalence between \eqrefs{Lz}{L} in earth.

A similar realization has been critical in
grasping the full content of the NJA \cite{nja}.
The SD limit of
the Kerr-Taub-NUT solution
is diffeomorphic to the SD Taub-NUT solution \cite{Crawley:2021auj}.
In Kerr-Schild coordinates \cite{note-sdtn,nja},
this diffeomorphism is just a translation by $+i\vea$.
Crucially, however,
the non-SD Kerr solution
describes \textit{two} centers at $\pm i\vea$
in the complexified description:
a pair of SD \textit{and ASD} Taub-NUT solutions
\cite{nja}.
Thus, the equivalence between
Kerr and a pointlike object
is lost
beyond the SD sector.

Based on this repertoire
that applies for both our probe NJA and the NJA,
it seems that a natural interpretation for the
holomorphic worldline $z^{\m\sprime}$ in \eqref{z}
is the worldline of the ASD Taub-NUT instanton
constituting the Kerr black hole.
Surely,
the anti-holomorphic worldline could have also been
obtained by
geodesically deviating in the direction of $-iy^\m$,
in which case the SD Taub-NUT instanton
is reached.

This interpretation is nicely consistent with spin exponentiation,
disclosing an underlying physical picture:
spin exponentiation is the helicity selection rule \cite{Adamo:2023fbj,Adamo:2024xpc,Adamo:2025fqt} of Taub-NUT instantons.
It also provides a physical interpretation for the worldsheet structure speculated in \rcite{gmoov}:
the Misner string
(gravitational Dirac string)
as a flux tube of NUT charge
joining the SD and ASD instantons.

Due to the nonlinearity of gravity
inherent in \eqref{z},
there is a difficulty in
simultaneously manifesting the nonlinear NJ shifts
(all-multiplicity spin exponentiations, spin resummations)
for both SD (positive-helicity) and ASD (negative-helicity) sectors.
Still, we shall make
maximum use of the superintegrability of the SD sector
shown in \eqref{K}
by perturbing around the holomorphic variable $z^{\m\sprime}$:
a \textit{googly}
\cite{Penrose:2015lla:Palatial,penr04-googly,Witten:2003nn,Adamo:2017qyl}
approach
to Kerr spinning-particle effective theory.

\section{
Conclusion}%
We proposed
the probe counterpart of NJA
to uniquely pinpoint
a remarkable instance of a massive spinning particle
in SD spacetimes,
which exhibits exact symmetries
in the presence of Killing vectors and Killing-Yano tensors.
It is shown that this distinguished point
in the effective theory landscape of all massive spinning particles
enjoys
the privileges of
superintegrability in SD Taub-NUT background
and
spin exponentiation at every graviton multiplicity.
It is also determined that the uplift to real spacetimes
is ambiguous 
precisely by mixed-chirality deformations,
corresponding to contact deformations for Compton amplitudes.

This work provides a partial answer to the persistent question,
``What principle in the effective theory
characterizes Kerr among all massive spinning objects?''
Previously, 
the speculations had described
a \textit{simplicity} of linearized interactions
\cite{ahh2017,Guevara:2018wpp,Guevara:2019fsj,chkl2019,aho2020}
or its continuation \cite{Johansson:2019dnu,Aoude:2020onz,Lazopoulos:2021mna},
a \textit{pattern} suggested at low orders \cite{Bern:2022kto,Aoude:2022trd},
or the \textit{gauge redundancy} in the higher-spin formalism \cite{Ochirov:2022nqz,Cangemi:2022bew}.
However, we identified 
an exact hidden \textit{symmetry}
embedded in nonlinear interactions,
thus demonstrating that
a physical principle can directly address the question.

\chapter{The Kerr Effective Action}
\label{K3:OSD1}

\begin{fullnote}
    Contents of
    this chapter are adapted from
    \rcite{njmagic.1}, \fullcite{njmagic.1}.
\end{fullnote}

\section{
Introduction}%
The twistor particle program
\cite{Penrose:1974di,Perjes:1974ra,%
tod1975dissertation,%
hughston1980programme,
perjes1982introduction,%
perjes1982internal,%
tod1977some,%
tod1976two,%
Perjes:1976sy,penrose1977twistor,
perjes1976evidence,hughston1976twistor,hughston1979twistors,perjes1979unitary%
}
was a movement in the '70s to '80s
that applied twistor theory to particle physics.
In particular, massive particles were implemented as systems of two or more twistors,
whose internal symmetries were associated with color or flavor.

In a modern reboot of this program \cite{ambikerr0,ambikerr1},
the two-twistor implementation is adopted
such that
the $\SU(2)$ internal symmetry
describes the massive little group instead,
in accordance with \rcite{ahh2017}.
This revival has also expanded the scope
from particle physics to
astrophysics,
based on the point-particle effective theory of
macroscopic bodies
\cite{ShR065,Porto:2005ac,Levi:2015msa,Porto:2016pyg,Levi:2018nxp,Kalin:2020mvi}
such as neutron stars or black holes.

The two-twistor system should be able to realize
interacting particles;
see \rrcite{tod1976two,Bette:1989zt,Bette:2004ip,Fedoruk:2007dd,Deguchi:2015iuw,ambikerr1}
for previous attempts.
The problem of direct relevance to the current PM gravity
community
is to implement
the Kerr black hole
coupled to curved spacetime,
capturing all its $2^\ell$-pole moments for $\ell = 0,1,2,\cdots$
\cite{janis1965structure,Newman:1965tw-janis,Newman:1973yu,vines2018scattering,Guevara:2018wpp,Guevara:2019fsj,chkl2019,Levi:2015msa}
as well as further nonlinear-in-curvature couplings.

In this chapter,
we describe how
an attempt toward curved massive twistor theory
had resulted in
a candidate effective point-particle Lagrangian for the Kerr black hole
to all orders in spin and curvature.
This construction was originally obtained in 2022
by the present author.

This derivation implements
an intrinsic feature of twistor particle theory
that spin is literally an imaginary deviation
in terms of complexified incidence relation
\cite{Shirafuji:1983zd,penrose:maccallum,newman1974curiosity}.
As a result,
the NJA \cite{Newman:1965tw-janis}
is realized
for
dynamical particles.
A differential-geometric formulation
arises 
in terms of
the generator of geodesic deviation, $N$, 
and an almost-complex structure, $J$.

\section{
Flat}

\subsection{Free Particle}%
The twistor space $\tflat$ is the K\"ahler vector space $\mathbb{C}^4$
with $(2,2)$-signature metric;
we consider the product $\tflat {\,\times\,} \tflat$
\cite{penrose:maccallum,tod1976two}.
Linear coordinates on $\tflat {\,\times\,} \tflat$ are $Z_\rmA{}^I$.
The index
$\rmA = 0,1,2,3$ is a Dirac spinor
while $I = 0,1$ is an $\SU(2)$ spinor.
The Weyl blocks are
\begin{align}
	Z_\rmA{}^I
	\,=\,
		\Bigg({
		\begin{aligned}
			\lambda_\a{}^I
			\\
			i\hem \mu^{\da I}
		\end{aligned}
		}\Bigg)
	\,,\quad
	\bZ_I{}^\rmA
	\,=\,
	\BB{
		-i\hem \bmu_I{}^\a	
		\,\,\,
		\bar{\lambda}_{I\da}
	}
	\,,
\end{align}
where $\bZ_I{}^\rmA = [Z_\rmA{}^I]^*$.
The symplectic form is
\begin{align}
	\label{omega0}
	\omega^\circ \,=\, i\mem d\bZ_I{}^\rmA \swedge dZ_\rmA{}^I
	\,=\,
		d\mu^{\da I}\nem \swedge d\bar{\lambda}_{I\da}
		+
		d\bmu_I{}^\a\nem \swedge d\lambda_\a{}^I
	\,.
\end{align}
The conformal group $\SU(2,2)$ acts from the left
while the massive little group $\SU(2)$ acts from the right.
The $\mathrm{U}(1)$
which $\bZ_I{}^\rmA Z_\rmA{}^I$ generates
is merely gauge
\cite{ambikerr0,ambikerr1}.

Consider the quadric
$-\tfrac{1}{2}\, I^{\rmA\rmB}\mem Z_\rmA{}^I Z_\rmB{}^J\mem \e_{IJ} = \det(\lambda) = c$
in $\tflat {\,\times\,} \tflat$,
which we denote as $\Delta_c$.
Here, \smash{$I^{\rmA\rmB}$} is the infinity twistor \cite{penrose:maccallum}.
For mathematical precision, one may define
the massive twistor space as
$\mtflat = (\tflat {\,\times\,} \tflat) {\,\setminus\,} \Delta_0$.
\eqref{omega0} shows that
$\mtflat$ is
the cotangent bundle
of the space $\lambda_\a{}^I \in \GL(2,\C)$
of \rcite{ahh2017}'s massive spinor-helicity variables.

Symplectic reduction derives
a $12$-dimensional constrained phase space
$\mt_* = \Delta_m / \C$,
realizing the $3$ translational and $3$ rotational degrees of freedom
of a massive spinning body
with rest mass $m>0$
\cite{ambikerr0,ambikerr1}
\footnote{
	It is easy to incorporate the Regge trajectory \cite{Hanson:1974qy};
	see \rrcite{ambikerr0,ambikerr1}.
}.

The Poincar\'e group is included as a subgroup of the $\SU(2,2)$.
The Poincar\'e charges are found as
\begin{align}
\begin{split}
	\label{NJMAGIC|poincare}
	p_{\a\da}
	\,&=\,
		-\lambda_\a{}^I\hem \bar{\lambda}_{I\da}
	\,,\\
	j^\da{}_\db
	\,&=\, 
		\mu^{\da I}\hem \bar{\lambda}_\wrap{I\db}
		- \tfrac{1}{2}\mem \delta^\da{}_\db\,
			\mu^{\dc I}\hem \bar{\lambda}_\wrap{I\dc}
	\,.
\end{split}
\end{align}
The Poincar\'e-invariant symplectic potential is
\begin{align}
	\label{theta0Z}
	\theta^\circ \,=\,
	\frac{i}{2}\,\BB{
		\bZ_I{}^\rmA\mem dZ_\rmA{}^I
		- d\bZ_I{}^\rmA\mem Z_\rmA{}^I
	}
	\,,
\end{align}
such that $\omega^\circ = d\theta^\circ$.

\subsection{Twistor Magic}%
A remarkable feature
of the twistor particle framework
is that
an imaginary shift of the two-twistor's co-incident position
derives a spinning particle
from a scalar particle.

As is well-known,
the relation between twistor space and 
spacetime $\mflat = (\R^4,\eta)$ is
established by the incidence relation
\cite{penrose:maccallum,penrose1967twistoralgebra},
the massive adaptation of which reads
\begin{align}
	\label{incidence}
	\mu^{\da I} 
	\,=\,
		x^{\da\a}\mem \lambda_\a{}^I
	\,.
\end{align}
Geometrically, \eqref{incidence} means that the two twistors $Z_\rmA{}^{I=0,1}$
are co-incident at the same spacetime point $x \in \mflat$.
By plugging in \eqref{incidence} to \eqref{theta0Z},
one obtains
\begin{align}
	\label{free.theta}
	\theta^\circ
	\,=\,
	p_{\a\da}\mem dx^{\da\a}
	\,,
\end{align}
which is the symplectic potential
of a scalar particle.

Now consider the complexification of
\eqref{incidence}:
\begin{align}
	\label{incidencec}
	\mu^{\da I} 
	\,=\,
	\BB{
		x^{\da\a} + i\hem y^{\da\a}
	}\mem \lambda_\a{}^I
	\,.
\end{align}
This shifts the co-incident point of the twistors $Z_\rmA{}^{I=0,1}$ to
a point
in complexified Minkowski space
$\mhat = (\C^4,\widehat{\eta})$,
where $\widehat{\eta}$ is the holomorphic extension of the Lorentzian metric $\eta$.
By plugging in \eqref{incidencec} to \eqref{theta0Z},
one obtains
\begin{align}
\label{free.thetac}
	\theta^\circ
	\,=\,
	p_{\a\da}\mem dx^{\da\a}
	+ i\mem y^{\da\a}\mem
	\BB{
		\lambda_\a{}^I\hem d\bar{\lambda}_{I\da}
		{\,-\,}
		d\lambda_\a{}^I\mem \bar{\lambda}_{I\da}
	}
	\,,
\end{align}
which is precisely 
the symplectic potential
of a massive spinning particle
\cite{ambikerr0,ambikerr1,tod1976two}.
Inserting \eqref{incidencec}
in \eqref{NJMAGIC|poincare}
shows that
$y^{\da\a}$ describes
the spin (Pauli-Luba\'nski) pseudovector normalized in the unit of length.
Hence, to put in a well-recognized notation \cite{Hanson:1974qy,Levi:2015msa}
for the reader's sake,
the transition from
\eqref{free.theta} to \eqref{free.thetac} describes
\footnote{
	\label{py}
	Note that the $\mathrm{U}(1)$ gauge generator evaluates to
	$\protect\bZ_I{}^\rmA Z_\rmA{}^I = -p_{\a\protect\da}\mem y^{\protect\da\a}$
	on \protect\eqref{incidencec}.
	In \eqref{rough}, we have imposed $-p_{\a\protect\da}\mem y^{\protect\da\a} = 0$.
}
\begin{align}
\label{rough}
	p\mem dx
	\mem\,\,\,\,\mapsto\,\,\,\,
	p\mem dx
	+ \tfrac{1}{2}\mem S\mem \Lambda\mem d\Lambda
	\,.
\end{align}

The insight that the complexified Minkowski space in twistor theory
unifies spacetime $x^\m$ and spin $y^\m$
as real and imaginary parts
traces back to
\rrcite{penrose:maccallum,newman1974curiosity},
whose explicit demonstration for dynamical massless particles
is given in \rcite{Shirafuji:1983zd}.
Here, we have provided the explicit demonstration for dynamical massive particles.

\subsection{Spinspacetime}%
The complexified Minkowski space
in this context
is dubbed
spinspacetime \cite{sst-asym}:
\begin{align}
	\label{tau-flat}
	\mhat
	\,\,\cong\,\,
	T\mflat
	\,.
\end{align}
As a real manifold,
spinspacetime is the tangent bundle $T\mflat$ of the spacetime $\mflat$,
with base coordinates $x^{\da\a}$ and fiber coordinates $y^{\da\a}$.
As a complex manifold,
spinspacetime is equipped with complex coordinates
\begin{align}
	\label{NJMAGIC|z}
	z^{\da\a} \,=\, x^{\da\a} + iy^{\da\a}
	\,,
\end{align}
which are holomorphic
due to the almost-complex structure
$dx^{\da\a} \mapsto dy^{\da\a}$,
$dy^{\da\a} \mapsto -dx^{\da\a}$.
(In this chapter, we wish to formulate complex structures as
isomorphisms of the cotangent bundle
rather than of the tangent bundle.)

Smooth tensor fields on spacetime
are analytically continued to
smooth tensor fields on spinspacetime.
Their Cauchy-Riemann flow
arises by the exponentiated Lie derivative
$\exp\mem(\mem i\pounds_N)$,
where $N \in \Gamma(T\mflat)$:
\begin{align}
	\label{N-flat}
	N \,=\, y^{\da\a}\mem \frac{\partial}{\partial x^{\da\a}}
	\,.
\end{align}
In particular,
the complexified metric is
$\widehat{\eta} = \exp\mem(\mem i\pounds_N)\mem \eta$.

As explicated in \rrcite{newman1974curiosity,sst-asym},
the physical origin of \eqref{NJMAGIC|z} is
a Hodge duality in
the decomposition of angular momentum
for massive particles,
\begin{align}
	\label{Jsplit}
	j \,=\, (x \mwedge p) + {*}(y \mwedge p)
	\,.
\end{align}
The SD part of \eqref{Jsplit}
is the SD part of $(z \mwedge p)$,
as $*$ is sent to $+i$.
The complexification in \eqref{incidencec}
precisely arises in this way
via $j^\da{}_\db$ in \eqref{NJMAGIC|poincare}.

\subsection{Correspondence Space}%
In massless twistor theory, the incidence relation is mathematically formalized as the double fibration.
Let us achieve its massive equivalent.

Let $S_2\mhat$ be a trivial $\GL(2,\C)$ bundle over $\mhat$.
\fref{fibration-flat}
concerns two maps
$\varphi: (z^{\da\a};\lambda_\a{}^I) \mapsto (\lambda_\a{}^I , i\hem z^{\da\a}\lambda_\a{}^I)$
and
$\pi_2 : (z^{\da\a};\lambda_\a{}^I) \mapsto z^{\da\a}$
from it.
The first is an invertible diffeomorphism.
The second is the bundle projection.
This provides a mathematical formalization of
the massive incidence relation in \eqref{incidencec},
establishing the relation between
massive twistor space $\mtflat$ and spinspacetime $\mhat$.

When $\mhat$ is regarded as a real manifold
as per \eqref{tau-flat},
$S_2\mhat$
is viewed as
the direct sum bundle $(T {\,\oplus\,} S_2)\hem\mflat$
over spacetime $\mflat$.
This is represented as a map 
$\gamma: (x^{\da\a};y^{\da\a},\lambda_\a{}^I) \mapsto (x^{\da\a} \mplus iy^{\da\a};\lambda_\a{}^I)$
in \fref{fibration-flat}.

The space
\begin{align}
	\label{K-flat}
	\kflat
	\mem=\mem
		(T {\,\oplus\,} S_2)\hem\mflat
	\,\,\cong\,\,
		S_2\mhat
\end{align}
serves as
the massive analog of the correspondence space,
which has the same dimension as $\mtflat$.
Crucially, $\kflat$ inherits the K\"ahler geometry of $\mtflat$
through $\varphi$ and $\gamma$.

First,
$\kflat$ is equipped with an almost-complex structure
\begin{align}\begin{split}
	\label{J-flat}
	\kern-0.2em
	J
	\,:\,
	(dx^{\da\a},dy^{\da\a},d\lambda_\a{}^I)
	&\mem\mapsto
	(dy^{\da\a},-dx^{\da\a},-i\mem d\lambda_\a{}^I)
	\mem.
	\kern-0.2em
\end{split}\end{align}
This corresponds to the complex structure
$dZ_\rmA{}^I \mapsto -i\mem dZ_\rmA{}^I$ 
of $\mtflat$.

Second,
$\kflat$ is equipped with the symplectic potential in \eqref{free.thetac}.
This is the pullback of the symplectic potential of $\mtflat$ given in \eqref{theta0Z}
by $\varphi \circ \gamma$.
(By abuse of notation, we have denoted both symplectic potentials as $\theta^\circ$.)

As a result, $\kflat$ is a K\"ahler manifold.
It is left as an exercise to 
derive
the pullback of 
the metric
by $\varphi \circ \gamma$
and
check the K\"ahler triple relation within $\kflat$.

We also equip $\kflat$
with a vector field $y^{\da\a} \partial/\partial x^{\da\a} \in \Gamma(T\kflat)$,
which is again denoted as $N$ by abuse of notation.

\begin{figure}[t]
	\centering
	\adjustbox{valign=c}{\begin{tikzpicture}
		\node[empty] (o) at (0,0) {};
		\node[empty] (i) at (-1.35, -1.35) {};
		\node[empty] (j) at ( 1.35, -1.35) {};
		\node[empty] (x) at ( 2.5, 0) {};
		\node[b] (A) at ($(o)$) {${
			S_2\mhat
		}\phantom{|}$};
		\node[b] (Ai) at ($(A)+(i)$) {$\mathclap{
			\mt
		}\phantom{|}$};
		\node[b] (Aj) at ($(A)+(j)$) {$\mathclap{
			\mhat
		}\phantom{|}$};
		\node[b] (B) at ($(o)+(x)$) {${
			(T {\,\oplus\,} S_2)\hem\mflat
		}\phantom{|}$};
		\node[b] (Bj) at ($(B)+(j)$) {$\mathclap{
			\mflat
		}\phantom{|}$};
		\draw[->] (A)--(Ai) node[midway,left, pos=0.35] 
		{\scriptsize $\varphi$\,\hnem};
		\draw[->] (A)--(Aj) node[midway,right,pos=0.35] {\scriptsize $\pi_2$};
		\draw[->] (B)--(Bj) node[midway,right,pos=0.35] {\scriptsize $\pi_1$};
		\draw[->] (B)--(A) node[midway,above,pos=0.45] {\scriptsize $\gamma$};
	\end{tikzpicture}}
	\caption{%
		Flat massive twistor theory.
	}
	\label{fibration-flat}
\end{figure}

Note that
working in the massive correspondence space $\kflat$
is what \rrcite{Shirafuji:1983zd,Fedoruk:2007dd}
refer to the so-called ``hybrid'' description.
Namely, $\kflat$ is the phase space of a free massive spinning body
described in terms of
spacetime position $x^{\da\a}$, spin length pseudovector $y^{\da\a}$,
and the spin frame $\lambda_\a{}^I$ 
encoding momentum and Eulerian angles.

\subsection{\textit{N} and \textit{J} for Newman-Janis}%
The twistor magic can be implemented in the massive correspondence space $\kflat$.
Namely, the spinning-particle symplectic potential
arises from the scalar-particle symplectic potential as
\begin{align}
\label{magic-in-K}
	\theta^\circ
	\,=\,
	\BB{
		1 + J\mem \i_N\hem d
	}\bigbig{
		p_{\a\da}\mem dx^{\da\a}
	}
	\,.
\end{align}
Here, $\i_N$ denotes the interior product
with respect to $N$
while $J$ is the almost-complex structure in \eqref{J-flat}.

\eqref{magic-in-K}
precisely
transcribes
the complexification of the co-incidence relation in \eqref{incidencec}.
Firstly,
$\i_N\hem d$ implements 
a deviation $x \mapsto x + y$
in the manner that conforms to the polarization choice.
Then
$J$ implements $\pm i$ factors
such that the twistors $Z_\rmA{}^I$ are co-incident at $x+iy$
and the dual twistors $\bZ_I{}^\rmA$ are co-incident at $x-iy$.

Note that $J$ implements
Hodge duality on the massive spinor-helicity fibers:
$+i$ for the right-handed $\bar{\lambda}_{I\da}$
and
$-i$ for the left-handed $\lambda_\a{}^I$.
Physically,
this converts the electric mass dipole
of a deviated scalar particle
to
the magnetic mass dipole,
i.e., spin,
of the spinning particle:
\begin{align}
	\label{dipoles}
	J\mem\bigbig{
		- dp_{\a\da}\mem y^{\da\a}
	}
	\,=\,
	i\mem y^{\da\a}\mem
		\BB{
			\lambda_\a{}^I\hem d\bar{\lambda}_{I\da}
			{\,-\,}
			d\lambda_\a{}^I\mem \bar{\lambda}_{I\da}
		}
	\,.
\end{align}
\eqref{dipoles} is an incarnation of the Hodge duality between orbital and spin angular momenta in \eqref{Jsplit}.

\section{
Earth}

The desire of this chapter is to
implement the twistor magic in curved backgrounds.
However,
we do not have
a definition of
``curved massive twistor space'' 
yet.

Recall the scaffolding of curved twistor theory
\cite{Penrose:1976js,GravityMHVTwistors,MasonNewman:1989}
(cf. \App{REVIEW>T}).
One may start by discussing the geometry of the correspondence space
within the conventional mathematical languages of general relativity.
Then the ``radical''
twistor-space picture
will  arise afterwards
via the self-duality condition
as
integrability along $\a$-surfaces.

We shall follow the same approach.
First of all,
we have to realize
the curved massive correspondence space $\mathcal{K}$
as
the typical phase space of a massive spinning particle
in textbook general relativity.
After establishing this ``conservative'' description,
we may then envision how 
a notion of
curved massive twistor space $\mtflat$ could emerge.

\subsection{Curved Correspondence Space}%
In textbook general relativity,
spacetime $(\M,g)$ is a real-analytic pseudo-Riemannian four-manifold.
Local trivialization of $T\M$
describes the 
frame $E_m = E^\m{}_m(x)\mem \partial_\m$
and
coframe $e^m = e^m{}_\m(x)\mem dx^\m$
such that
$\langle e^m , E_n \rangle = \delta^m{}_n$.
Here, $m,n,\cdots = 0,1,2,3$ are local Lorentz indices
while $\m,\n,\cdots = 0,1,2,3$ are spacetime indices.

The decomposition
$T\M \cong (S^+ \mtensor S^-)\M$
into SD and ASD spinor bundles
describes that
a local Lorentz index is a pair of spinor indices
as
$e^m \leftrightarrow e^{\da\a}$.
The Levi-Civita connection $\nabla$ of $g$
splits to $\SL(2,\C)$-valued connections $\bgamma$ and $\gamma$ on
$S^\pm\hnem\M$.
The structure equations are
\smash{$
	0 = 
		De^{\da\a}
	=
		de^{\da\a} - \bgamma^\da{}_\db \swedge e^{\db\a} + e^{\da\b} \swedge \gamma_\b{}^\a
$}
and
\smash{$
	R_\a{}^\b =
		d\gamma_\a{}^\b + 
		\gamma_\a{}^\c \wedge \gamma_\c{}^\b
$},
where
$D$ denotes the covariant exterior derivative
with respect to $\nabla$.

Physically,
the degrees of freedom of
a massive spinning particle in $(\M,g)$
are described by
$x^\m$, $y^m$, and $\lambda_\a{}^I$.
Here, the index form declares the behaviors
under coordinate and local Lorentz transformations.
The position variable $x^\m$ is supposed to describe coordinates on $\M$.
The spin length pseudovector $y^m$ 
and 
the spin frame $\lambda_\a{}^I$
are supposed to be local Lorentz degrees of freedom.

Therefore, the curved massive correspondence space
should be defined as the direct sum bundle
\begin{align}
	\label{K-curved}
	\mathcal{K} 
	\,=\,
		(T {\,\oplus\,} S_2)\hem\M
	\,,
\end{align}
equipped
with base coordinates $x^\m {\,\in\,} \R^4$
and fiber coordinates 
$y^m {\,\in\,} \R^4$,
$\lambda_\a{}^I {\,\in\,} \GL(2,\C)$.
Here,
$S_2\M$ is a principal $\GL(2,\C)$-bundle over $\M$
which is isomorphic to an open subbundle of $(S^- \moplus S^-)\M$.

The transition functions for $\mathcal{K}$ 
as a manifold
are restricted to the following specific forms:
\begin{align}\begin{split}
\label{transfs}
	(x^\m,y^m,\lambda_\a{}^I)
	&\,\,\mapsto\,\mem
	(f^\m(x),y^m,\lambda_\a{}^I)
	\,,\\
	(x^\m,y^m,\lambda_\a{}^I)
	&\,\,\mapsto\,\mem
	(x^\m,
		\Omega^m{}_n(x)\, y^n,
		\Omega_\a{}^\b\hnem(x)\, \lambda_\a{}^I
	)
	\,.
\end{split}\end{align}
These encode
spacetime diffeomorphisms
and local Lorentz transformations,
respectively.
Here,
$\Omega^m{}_n(x) \in \SO(1,3)$
arises from
$\Omega_\a{}^\b(x) \in \SL(2,\C)$.

The curved massive correspondence space
is equipped with 
a triple of geometrical structures:
\begin{align}
	(\mathcal{K},N,J,\omega)
	\,.
\end{align}
$N {\,\in\,} \Gamma(T\mathcal{K})$ is a vector field,
$J : T^*\mathcal{K} {\,\to\,} T^*\mathcal{K}$ is an almost-complex structure,
and $\omega {\,\in\,} Z^2(\mathcal{K}) {\:\subset\:} \Omega^2(\mathcal{K})$ is a symplectic structure.
$N$ and $J$
are invariant under the restricted coordinate transformations of $\mathcal{K}$
stipulated in \eqref{transfs}.

First,
$N$ is
uniquely defined by the interior products
\begin{align}
	\label{Ndef}
	\i_N
	\,\,\,:\,\,\,
	(e^m,Dy^m,D\lambda_\a{}^I)
	\,\,\mapsto\,\,
	(y^m,0,0)
	\,.
\end{align}
It is easy that
$N = y^m \tilde{E}_m$,
where $\tilde{E}_m \in \Gamma(T\mathcal{K})$
is
the horizontal lift \cite{ehresmann1948connexions,Mason:2013sva} of
the frame vector field
$E_m \in \Gamma(T\M)$
with respect to $\nabla$.
It follows that $N$ is the generator of geodesic deviation
and transport
\cite{gde}.

Second, $J$ is uniquely defined as the map such that
\begin{align}\begin{split}
	\label{NJMAGIC|J}
	J
	\,\,:\,\,
	(e^m,Dy^m,D\lambda_\a{}^I)
	&\,\hhem\mapsto\mem
	(Dy^m,-e^m,-i\mem D\lambda_\a{}^I)
	\mem.
\end{split}\end{align}

Third,
$\omega \in Z^2(\mathcal{K})$ is defined as a symplectic form
that approaches the following two-form $\omega^\bullet \in \Omega^2(\mathcal{K})$
in the limit of vanishing curvature of $\nabla$,
namely $\lim_{R\to0} \omega = \omega^\bullet$:
\begin{align}\begin{split}
\label{cov.omega}
	\omega^\bullet
	\,=\,
	{}&{}
	\bar{\lambda}_{I\da}\mem \bigbig{
		e^{\da\a} \mminus i\mem Dy^{\da\a}
	} \wedge D\lambda_\a{}^I
	- D\bar{\lambda}_{I\da} \wedge \bigbig{
		e^{\da\a} \mplus i\mem Dy^{\da\a}
	}\mem \lambda_\a{}^I
	\\ 
	{}&{}
	+ 2i\mem y^{\da\a}\mem
		D\lambda_\a{}^I \swedge D\bar{\lambda}_{I\da}
	\,.
\end{split}\end{align}
Here, we define
$p_{\a\da} {\:=\:} {-\lambda_\a{}^I\hem \bar{\lambda}_{I\da}}$.

It should be clear that
$N$ in \eqref{Ndef},
$J$ in \eqref{NJMAGIC|J},
and $\omega^\bullet$ in \eqref{cov.omega}
respectively
reduce to
$N$ in \eqref{N-flat},
$J$ in \eqref{J-flat},
and
the exterior derivative of \eqref{free.thetac}
in the limit of vanishing curvature of $\nabla$,
in which case $\mathcal{K} \cong \kflat$.
They are the very covariantizations of
the free theory's
$N$, $J$, and $\omega^\circ = d\theta^\circ$
due to $\nabla$.

When $\nabla$ is curved,
$\omega$ must differ from $\omega^\bullet$
since $d\omega^\bullet \neq 0$.
In this sense the particle's symplectic structure
necessarily develops a curvature correction.

The structures $N$ and $J$ 
could also develop curvature corrections in principle,
but
it will suffice to deform just the symplectic structure
for our physical purposes.

The particle's $12$-dimensional physical phase space
arises by
symplectic reduction of $\mathcal{K}$ 
by
gauge generators
$-p_m\hhem y^m \approx 0$ 
and $p^2 \mplus m^2 \approx 0$.

\subsection{Traditional Minimal Coupling}%
By choosing the symplectic structure $\omega$ differently,
one realizes various gravitational couplings of the massive spinning particle
in the phase space $\mathcal{K}$.
In particular, the choice of $\omega$ 
amounts to specifying the particle's action.

For example, the choice
dubbed ``minimal coupling''
in the traditional sense
defines the symplectic potential on $\mathcal{K}$
by directly covariantizing \eqref{free.thetac}:
\begin{align}
\label{cov.theta}
	\theta_{(0)}
	\,=\,
	p_m\hhem e^m
	+ i\mem y^{\da\a}\mem
	\BB{
		\lambda_\a{}^I\hem D\bar{\lambda}_{I\da}
		{\,-\,}
		D\lambda_\a{}^I\mem \bar{\lambda}_{I\da}
	}
	\,.
\end{align}
In this case, 
the symplectic form is $\omega_{(0)} = d\theta_{(0)} = \omega^\bullet + \smash{\omega'_{(0)}}$,
where 
\smash{$\omega'_{(0)} = p_m\hem {\star\hnem}R^m{}_n\hem y^n$}.
$\star$ denotes the Hodge star acting on the internal (local Lorentz) indices.
Note that it could be instructive to rewrite \eqref{cov.theta} as
\begin{align}
	\theta_{(0)}
	\,=\,
		p_m\hhem e^m
		+ p_m {\hem\star\Theta}^m{}_n\hem y^n + W_0\mem d\psi
	\,,
\end{align}
where $\Theta^m{}_n$ is a Lorentz-valued Maurer-Cartan form
such that $D\Theta^m{}_n {\,=\,} R^m{}_n$,
$\psi$ is a $\mathrm{U}(1)$ angle,
and $W_0 {\,=\,} {-p_m\hhem y^m}$.

\begin{figure*}[t]
	\centering
	\begin{tikzpicture}
	    \node[empty] (O) at (0,0) {};
	    \node[empty] (X) at (4.4, 0) {};
	    \node[empty] (x) at (3.9, 0) {};
	    \node[empty] (Y) at (0, -0.92) {};
	    \node[w] (a00) at ($(O)$) {$p_me^m$};
	    \node[w] (a01) at ($(O)+1*(X)$) {${d(p_my^m)}$};
	    \node[w] (a02) at ($(O)+2.0*(X)$) {$0$};
	    \node[w] (a10) at ($(O)+1*(Y)$) {$-Dp_m\hem y^m$};
	    \node[w] (a11) at ($(O)+1*(Y)+1*(X)$) {$0$};
	    \node[w] (a20) at ($(O)+2*(Y)$) {$p_m (\i_NR^m{}_n)\hem y^n$};
	    \node[w] (a21) at ($(O)+2*(Y)+1*(X)$) {$0$};
	    \node[w] (a30) at ($(O)+3*(Y)$) {$p_m (\i_ND\mem \i_NR^m{}_n)\hem y^n$};
	    \node[w] (a31) at ($(O)+3*(Y)+1*(X)$) {$0$};
	    \node[w] (a40) at ($(O)+4*(Y)$) {$\vdots$};
	    \node[w] (a2K) at ($(O)+2*(Y)-1*(X)$) {$p_m {\hem\star\Theta}^m{}_n\hem y^n {\mem+\mem} W_0\mem d\psi$};
	    \node[w] (a3K) at ($(O)+3*(Y)-1*(X)$) {$p_m (\i_N{\star R}^m{}_n)\hem y^n$};
	    \node[w] (a4K) at ($(O)+4*(Y)-1*(X)$) {$p_m (\i_ND\mem \i_N{\star R}^m{}_n)\hem y^n$};
	    \node[w] (a5K) at ($(O)+5*(Y)-1*(X)$) {$\vdots$};
	    \node[w] (a2k) at ($(O)+2*(Y)-1*(X)+(x)$) {$0$};
	    \node[w] (a3k) at ($(O)+3*(Y)-1*(X)+(x)$) {$0$};
	    \node[w] (a4k) at ($(O)+4*(Y)-1*(X)+(x)$) {$0$};
	    \node[w] (phantom-a00) at ($(O)$) {};
	    \node[w] (phantom-a01) at ($(O)+1*(X)$) {};
	    \node[w] (phantom-a02) at ($(O)+1.5*(X)$) {};
	    \node[w] (phantom-a10) at ($(O)+1*(Y)$) {};
	    \node[w] (phantom-a11) at ($(O)+1*(Y)+1*(X)$) {};
	    \node[w] (phantom-a20) at ($(O)+2*(Y)$) {};
	    \node[w] (phantom-a21) at ($(O)+2*(Y)+1*(X)$) {};
	    \node[w] (phantom-a30) at ($(O)+3*(Y)$) {};
	    \node[w] (phantom-a31) at ($(O)+3*(Y)+1*(X)$) {};
	    \node[w] (phantom-a40) at ($(O)+4*(Y)$) {};
	    \node[w] (phantom-a2K) at ($(O)+2*(Y)-1*(X)$) {};
	    \node[w] (phantom-a3K) at ($(O)+3*(Y)-1*(X)$) {};
	    \node[w] (phantom-a4K) at ($(O)+4*(Y)-1*(X)$) {};
	    \node[w] (phantom-a5K) at ($(O)+5*(Y)-1*(X)$) {};
	    \node[w] (phantom-a2k) at ($(O)+2*(Y)-1*(X)+(x)$) {};
	    \node[w] (phantom-a3k) at ($(O)+3*(Y)-1*(X)+(x)$) {};
	    \node[w] (phantom-a4k) at ($(O)+4*(Y)-1*(X)+(x)$) {};
	    \draw[->] (a00)--(a01) node[midway,above] {\scriptsize $d\mem\i_N$};
	    \draw[->] (a01)--(a02) node[] {};
	    \draw[->] (a10)--(a11) node[] {};
	    \draw[->] (a20)--(a21) node[] {};
	    \draw[->] (a30)--(a31) node[] {};
	    \draw[->] (phantom-a00)--(phantom-a10) node[midway,left] {\scriptsize $\i_N d$};
	    \draw[->] (phantom-a10)--(phantom-a20) node[] {};
	    \draw[->] (phantom-a20)--(phantom-a30) node[] {};
	    \draw[->] (phantom-a01)--(phantom-a11) node[] {};
	    \draw[->] (phantom-a10)--(a2K) node[midway,above] {\scriptsize $J$};
	    \draw[->] (phantom-a2K)--(phantom-a3K) node[] {};
	    \draw[->] (phantom-a3K)--(phantom-a4K) node[] {};
	    \draw[->] (a2K)--(a2k) node[] {};
	    \draw[->] (a3K)--(a3k) node[] {};
	    \draw[->] (a4K)--(a4k) node[] {};
	    \draw[->] (phantom-a30)--(phantom-a40) node[] {};
	    \draw[->] (phantom-a4K)--(phantom-a5K) node[] {};
	\end{tikzpicture}
	\caption{%
		The ``$\i_N d / J$ sequence'' for Kerr.
		A tree of one-forms 
		emanates from
		the scalar-particle symplectic potential, $p_m\hhem e^m$.
	}
	\label{tree-of-life}
\end{figure*}

The resulting Hamiltonian EoM are
immediate by
the technique of covariant symplectic perturbations.
They reproduce the Mathisson-Papapetrou-Dixon \cite{Mathisson:1937zz,Papapetrou:1951pa,Dixon:1970zza}
equations,
which describe the quadrupolar coupling
\cite{Steinhoff:2009tk,Harte:2011ku,Vines:2016unv,Compere:2023alp,Ramond:2026fpi,%
khriplovich1989particle,Yee:1993ya,Khriplovich:1997ni,Thorne:1984mz}
$C_2 {\,=\,} 0$.
However,
it is well-known that
the Kerr black hole carries $C_2 {\,=\,} 1$
\cite{janis1965structure,Newman:1965tw-janis,Newman:1973yu,vines2018scattering,Guevara:2018wpp,Guevara:2019fsj,chkl2019}.
Therefore,
the symplectic potential in \eqref{cov.theta}
fails to describe the Kerr black hole.

\subsection{\textit{N} and \textit{J} for Newman-Janis, with Curvature}%
In spirit of the NJA,
we ask whether
the Kerr symplectic potential
can be constructed in $\mathcal{K}$
solely from
the Schwarzschild symplectic potential $p_m\hem e^m$
and
the invariant structures $N$ and $J$.

\eqref{magic-in-K} has implemented the twistor magic
in the flat correspondence space $\kflat$.
Its covariantization is
\begin{align}
\label{theta0}
	\theta_{(0)}
	\,=\,
	\BB{
		1 + J\mem \i_N\hem d
	}\bigbig{
		p_m\hhem e^m
	}
	\,,
\end{align}
which unfortunately evaluates to \eqref{cov.theta}.
A natural modification is
\begin{align}
\label{theta1}
	\theta_{(1)}
	\,=\,
	\BB{
		\cos(\pounds_N)
		+
		\sinc(\pounds_N)\,
		J\mem \i_N\hem d
	}\bigbig{
		p_m\hhem e^m
	}
	\,.
\end{align}

\eqref{theta1}
faithfully implements the concept of NJA:
imaginary shifts along spin direction.
Since $N$ is the generator of geodesic deviation,
the imaginary deviations arise by
$\exp\mem(\pm i\pounds_N)
= \cos(\pounds_N) \pm i \sinc(\pounds_N)\hem \pounds_N
$.
\eqref{theta1} implements
the $\pm i$ factors with the almost-complex structure $J$
while replacing a $\pounds_N$ with $\i_N\hem d$
as a direct generalization of
the flat construction in \eqref{magic-in-K}.

Crucially, the repeated action of $\pounds_N$
does not truncate at a finite order
in the presence of curvature.
As per the Cartan magic formula,
the repeated action of $\pounds_N = \i_N\hem d + d\hem \i_N$
is diagrammatically represented as
a planar grid
where each node is a differential form
and each edge is an arrow representing
the action of either
$\i_N\hem d$ or $d\hem \i_N$,
such that $(\i_N\hem d)(d\hem \i_N) = 0$
and $(d\hem \i_N)(\i_N\hem d) = 0$.
As shown in \fref{tree-of-life},
this grid does not truncate
if $R^m{}_n \neq 0$.

Hence we shall regard that $\cos(\pounds_N)$ and $\sinc(\pounds_N)$
were secretly present in \eqref{magic-in-K} as well,
albeit trivialized.
In curved backgrounds,
they ensure the unity $C_\ell {\:=\:} 1$ of even and odd multipoles coefficients.
That is,
\eqref{theta1}
correctly captures the linear gravitational coupling of Kerr
established through
\rrcite{janis1965structure,Newman:1965tw-janis,Newman:1973yu,vines2018scattering,Guevara:2018wpp,Guevara:2019fsj,chkl2019}.

In fact, \eqref{theta1}
pinpoints a unique gravitational coupling
to all orders in spin and curvature.
From \fref{tree-of-life},
it can be seen that
\eqref{theta1} can be represented as
\begin{align}
	\label{sum1}
	\theta_{(1)}
	\mem&=\,
	\theta_{(0)}
	+\mem\hhem
		\sum_{\ell=2}^\infty
			\frac{1}{\ell!}\,
				p_m
				\bigbig{\hnem
					(\i_N\hem D)^{\ell-2} \i_N\mem {\star^\ell\hnem} R^m{}_n
				\hhnem}\hem y^n
	\,,
\end{align}
where ${\star^\ell}$ means to act on the internal Hodge star $\ell$ times.
By methods in differential geometry \cite{gde},
\eqref{sum1} is explicitly computed as
\begin{align}
\begin{split}
\label{theta-earth}
	\theta_{(1)}
    	\mem=\,
    {}
    		\theta_{(0)}
&
    	 	+\mem\hhem \sum_{\ell=2}^\infty
    	 		\frac{1}{\ell!}
    			\sum_{p=1}^{\lfloor{\ell/2}\rfloor}\kern-0.2em
    			\sum_{\a \in \Omega_p(\ell)}\kern-0.2em
    	 			\bbsq{
    	 				\prod_{i=1}^p
    	 				\binom{
    	 					\bigbig{
    	 						\sum_{j=i}^p \a_j
    	 					} \mminus 2
    	 					\hem
    	 				}{
    	 					\a_i \mminus 2
    	 				}
    	 				\nem\nem
    	 			}\mem
            \\
            &
    	 		\lrp{
    	 		\begin{aligned}[c]
    	 			&
				\bigbig{\hnem
					({\star}^\ell Q_{\a_1}\hnem)\mem Q_{\a_2}\, {\nem\cdots\mem} Q_{\a_p}
				\hnem}
				{}^m{}_s\, e^s
				\\
				&
				+ (\a_p\mminus2)\mem
				\bigbig{\hnem
					({\star}^\ell Q_{\a_1}\hnem)\mem Q_{\a_2}\, {\nem\cdots\mem} Q_{\a_{p-1}} Q_{\a_p - 1}
				\hnem}
				{}^m{}_s\, Dy^s
			\hnem
                \end{aligned}
    	 		}
  	 \,.
\end{split}
\end{align}
where $\a = (\a_1,\a_2,\cdots,\a_p) \in \Omega_p(\ell)$
runs over ordered partitions
such that
$\a_1 + \a_2 + \cdots + \a_p  = \ell$
and
$\a_i \geq 2$.
The so-called $Q$-tensors \cite{gde} are defined as
\begin{align}
    \label{Qtensor}
    ({\star}^\ell Q_j)^m{}_s
    \,=\,
	  {\star}^\ell R^m{}_{r_1r_2s;r_3;\cdots;r_j}\hnem(x)
	  \, y^{r_1}{\cdots}y^{r_j}
    \,.
\end{align}

The nonlinear-in-Riemann terms
in \eqref{theta-earth}
are fine-tuned to yield
remarkable physical consequences
in SD backgrounds,
as we have explained in \rcite{probe-nj}.

\section{
Heaven}

\subsection{Heavenly Portal to Curved Spinspacetime}%
\eqref{theta1} and its explicit evaluation in \eqref{theta-earth}
provide a manifestly covariant, gauge-invariant, and reparametrization-invariant
construction of the Kerr action
in generic, real spacetimes.
It is based on two invariant geometric structures in the spinning-particle phase space:
$N$ and $J$.

However, it is based on the typical languages of general relativity.
Apparently,
the spinspacetime and massive twistor descriptions are lost.
This means that the curved analogs of the left part of the diagram in \fref{fibration-flat}
are yet to be constructed.

The first step toward the ``radical'' reformulations
is to identify the map $\gamma$.
It can be discovered by
traveling to SD spacetimes,
i.e., \textit{heaven}
in the terminologies of
Newman \cite{shaviv1975general,Newman:1976gc} and Pleba\'nski \cite{plebanski1975some,Plebanski:1977zz}.

The spacetime curvature $R^m{}_n$
splits to SD and ASD parts as
\smash{$\bR^\da{}_\db$} and \smash{$R_\a{}^\b$}.
Consider the (formal) complexified limit in which the latter vanishes,
so $\star R^m{}_n = +i\hem R^m{}_n$.
Then \eqref{sum1} equals
\begin{align}
	\label{kerr.heaven}
	\theta_{(1)}
	\mem&=\,
	\theta_{(0)}
	+\mem\hhem
		\sum_{\ell=2}^\infty
			\frac{i^\ell}{\ell!}\,
				p_m
				\bigbig{\hnem
					(\i_N\hem D)^{\ell-2} \i_N R^m{}_n
				\hhnem}\hem y^n
	\,.
\end{align}
Notably, \eqref{kerr.heaven} is nearly equal to
the symplectic potential of a scalar particle,
geodesically deviated in a pure-imaginary direction $iy$
(cf. \rcite{gde}):
\begin{align}
	\label{sch+ia}
	\mathe^{i\pounds_N}\bigbig{
		p_m\hhem e^m
	}
    \,=\,
		p_m\hhem e^m
		+ i\mem p_m Dy^m
		+\mem\hhem
			\sum_{\ell=2}^\infty
				\frac{i^\ell}{\ell!}\,
					p_m
					\bigbig{\hnem
						(\i_N\hem D)^{\ell-2} \i_N R^m{}_n
					\hhnem}\hem y^n
	\,.
\end{align}
That is, \eqrefs{kerr.heaven}{sch+ia}
differ only by the combination
\begin{align}
	\label{NJMAGIC|g2a}
	\theta_\ga
	\,=\,
		p_m\mem {\star}\Theta^m{}_n\hem y^n
		- i\mem p_m Dy^m
		+ W_0\mem d\psi
	\,,
\end{align}
precisely since
the SD limit facilitates
replacing $J$ in \eqref{theta1}
with $+i$
except at the dipolar spin order $\ell = 1$.
\newpage

Therefore,
our Kerr particle
is indistinguishable from
an imaginary-deviated Schwarzschild particle
in SD backgrounds,
up to a slight subtlety at the dipolar order.

\begin{figure}[t]
	\centering
	\includegraphics[scale=1.25]{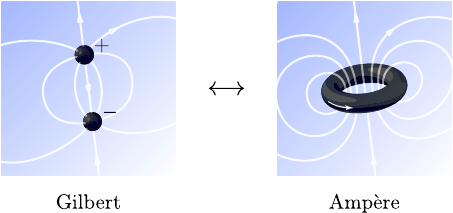}
    \vspace{0.6\baselineskip}
	\caption{%
		Gilbert-Amp\`ere duality.
	}
	\label{fig:G2A}
\end{figure}

To understand 
the origin of
this subtlety
from a physicist's intuition,
note that \eqref{NJMAGIC|g2a} replaces
an imaginary \textit{electric mass dipole} term $i\mem p_m Dy^m$
with 
the \textit{magnetic mass dipole} (i.e., spin) term $p_m\mem {\star}\Theta^m{}_n\hem y^n$.
This reflects a kinematics-level difference between
spinning and non-spinning particles.
Specifically,
there is a physical sense in which
\eqref{NJMAGIC|g2a}
replaces a Gilbertian dipole
(displaced charges)
with an Amp\`erian dipole
(current loop),
as illustrated in \fref{fig:G2A}.
In SD backgrounds,
magnetic charge is indistinguishable from $i$ times electric charge.
Thus the Amp\`erian dipole
$p_m\mem {\star}\Theta^m{}_n\hem y^n$
gets replaced with $i$ times the Gilbertian dipole $p_m Dy^m$,
and upon this very mechanism
spin becomes $i$ times deviation---%
which is the gist of the NJA, in fact \cite{nja}.

Hence,
physically speaking, 
the almost-complex structure $J$
has implemented a duality between Gilbertian and Amp\`erian multipoles.
This observation can be rephrased as
electric-magnetic or Hodge duality;
recall the discussions around \eqrefs{magic-in-K}{dipoles}.

With this insight,
we rewrite the Kerr symplectic potential
in the SD limit as
\begin{align}\begin{split}
	\label{kerr.g2a}
	\theta_{(1)}
	\mem&=\,
	\mathe^{i\pounds_N}\bigbig{
		p_m\hhem e^m
	}
	\hem+\mem\hhem \theta_\ga
    \,=\,
	\mathe^{i\pounds_N}\bigbig{
		p_m\hhem e^m
		\hem+\mem\hhem \theta_\ga
	}
	\,,
\end{split}\end{align}
where the second line follows from the fact that
$\theta_\ga$ in \eqref{NJMAGIC|g2a}
carries no gravitational charge
in the SD limit
such that
$d\theta_\ga = 0$;
it describes a sort of a topological defect.

Remarkably, \eqref{kerr.g2a} 
shows that
the Kerr symplectic potential
in SD spacetime
is 
a one-form
localized at a ``complex spacetime point''
whose coordinates are
\begin{align}
	\label{zcurved}
	\mathe^{iN} x^\m
	\,=\,
		x^\m 
		+ i\mem y^\m
		+ \tfrac{1}{2}\, \Gamma^\m{}_{\r\s}(x)\mem y^\r y^\s
		+ \O(y^3)
	\,.
\end{align}
\eqref{zcurved} generalizes \eqref{NJMAGIC|z}
in a nonlinear and generally covariant fashion:
\textit{curved spinspacetime}.

\begin{figure}[t]
	\centering
	\adjustbox{valign=c}{\begin{tikzpicture}
		\node[empty] (o) at (0,0) {};
		\node[empty] (i) at (-1.35, -1.35) {};
		\node[empty] (j) at ( 1.35, -1.35) {};
		\node[empty] (x) at ( 3.0, 0) {};
		\node[b] (A) at ($(o)$) {${
			\mathcal{K}
		}\phantom{|}$};
		\node[b] (Ai) at ($(A)+(i)$) {$\mathclap{
			\MT
		}\phantom{|}$};
		\node[b] (Aj) at ($(A)+(j)$) {$\mathclap{
			\M_z
		}\phantom{|}$};
		\node[b] (B) at ($(o)+(x)$) {${
			(T^{1,0} {\,\oplus\,} S_2^{1,0} {\,\oplus\,} S_2^{1,0})\hem\M
		}\phantom{|}$};
		\node[b] (Bj) at ($(B)+(j)$) {$\mathclap{
			\M_x
		}\phantom{|}$};
		\draw[->] (A)--(Ai) node[midway,left, pos=0.35] 
		{\scriptsize $\varphi$\,};
		\draw[->] (A)--(Aj) node[midway,right,pos=0.35] {\scriptsize $\pi_2$};
		\draw[->] (B)--(Bj) node[midway,right,pos=0.35] {\scriptsize $\pi_1$};
		\draw[->] (B)--(A) node[midway,above,pos=0.44] {\scriptsize $\gamma$};
	\end{tikzpicture}}
	\caption{%
		Curved massive twistor theory.
	}
	\label{fibration-curved}
\end{figure}

\subsection{Curved Spinspacetime}%
As shown in \Sec{ACS},
spinspacetime exists for any
real-analytic spacetime.
To expedite our discussion, however,
we may want to directly jump to a complexified setup
regarding its relevance to the SD limit and twistorial ideas.

From now on, spacetime 
is a complex-analytic four-manifold $(\M,g)$
equipped with a holomorphic metric $g$.

Spinspacetime is the holomorphic tangent bundle $\Mhat = T^{1,0}\M$,
which has base coordinates $x^\m$ and fiber coordinates $y^\m$.
This means that coordinate transformations on $\M$
induce
$(x^\m,y^\m) \mapsto (f^\m(x),f^\m{}_{,\n}(x)\mem y^\n)$,
where $f^\m(x)$ are holomorphic functions.
Spinspacetime is equipped with the holomorphic vector field
$N \in T^{1,0}\Mhat$
that is horizontal with respect to the Levi-Civita connection $\nabla$ of $g$
and satisfies $\i_N dx^\m = y^\m$.

The time-$i$ flow of $N$ is
a map $\gamma : \Mhat \to \Mhat$:
\begin{align}\begin{split}
	z^{\m\sprime}
	\,=\,
		\delta^{\m\sprime}{}_\m\mem
		(\mathe^{iN} x^\m)
	\,,\quad
	y^{\m\sprime}
	\,=\,
		\delta^{\m\sprime}{}_\m\mem
		(\mathe^{iN} y^\m)
	\,.
\end{split}\end{align}
Here, we are adopting the primed index notation
of Synge calculus
\cite{Ruse:1931ht,Synge:1931zz,Synge:1960ueh,Poisson:2011nh,gde}.
$z^{\m\sprime}$ are the coordinates of the point in $\M$
that is reached by geodesically deviating from $x$ along the direction of $y$,
so
$\Exp : T^{1,0}\M \too \M : (x,y) \mapsto z$
is the holomorphic exponential map;
$y^{\m\sprime} {\:=\:} W^{\m\sprime}{}_\m\mem y^\m$ describe the tangent to the geodesic segment
at the endpoint $z$;
$W^{\m\sprime}{}_\m$ is the parallel propagator from $x$ to $z$.

In sum, we identify two coordinate charts on $\Mhat = T^{1,0}\M$.
The ``defining'' chart is $(x^\m;y^\m)$;
the other chart \smash{$(z^{\m\sprime};y^{\m'})$}
will be referred to as a ``googly'' chart.

\subsection{Curved Correspondence Space}%
For the complex spacetime $(\M,g)$,
the curved massive correspondence space
is the holomorphic direct sum bundle
\begin{align}
	\label{K-curved-holic}
	\mathcal{K} 
	\,=\,
		(T^{1,0} {\,\oplus\,} S_2^{1,0} {\,\oplus\,} S_2^{1,0})\hem\M
	\,,
\end{align}
equipped
with base coordinates $x^\m {\,\in\,} \C^4$
and fiber coordinates 
$y^m {\,\in\,} \C^4$,
$\lambda_\a{}^I, \trambda_{I\da} \in \GL(2,\C)$.
Its transition functions are restricted 
in terms of holomorphic functions
$f^\m(x) {\:\in\:} \C^4$
and
$\Omega_\a{}^\b(x), \tilde{\Omega}^\da{}_\db(x) {\:\in\:} \SL(2,\C)$
(cf. \eqref{transfs}).

The curved correspondence space is equipped with
the holomorphic vector field
$N {\:\in\:} \Gamma(T^{1,0}\mathcal{K})$
that is horizontal with respect to $\nabla$
and satisfies $\i_N dx^\m {\:=\:} y^\m$,
which we again denote as $N$ by abuse of notation.

Similarly,
the time-$i$ flow of $N$ is denoted as $\gamma$.
This is
the map $\gamma : \mathcal{K} \to \mathcal{K}$
such that
$x^\m \mapsto z^{\m\sprime} = (\exp\mem(x,y))^{\m\sprime}$,
$y^\m \mapsto y^{\m\sprime} {\:=\:} W^{\m\sprime}{}_\m\mem y^\m$,
and
$\lambda_\a{}^I \mapsto \lambda_{\a'}{}^I {\:=\:} W_{\a'}{}^\a\mem \lambda_\a{}^I$,
$\trambda_{I\da} \mapsto \trambda_{I\da\sprime} {\:=\:} \trambda_{I\da}\mem W^\da{}_{\da'}$.
Here,
the parallel propagator \smash{$W^{\m\sprime}{}_\m$}
decomposes into $\SL(2,\C)$ blocks.
Note that it
is metric-preserving as
$g_{\m\sprime\n'}(z)$ $ W^{\m\sprime}{}_\m\mem W^{\n\sprime}{}_\n
{\:=\:} g_{\m\n}(x)$.

The defining chart on $\mathcal{K}$ is
$(x^\m;y^\m,\lambda_\a{}^I,\trambda_{I\da})$.
The googly chart on $\mathcal{K}$ is
$(z^{\m\sprime};y^{\m\sprime},\lambda_{\a'}{}^I,\trambda_{I\da'}\hnem)$.
The former chart establishes the projection
$\pi_1 : \mathcal{K} \to \M : (x^\m;y^\m,\lambda_\a{}^I,\trambda_{I\da}) \mapsto x^\m$.
The latter chart establishes the projection
$\pi_2 : \mathcal{K} {\:\to\:} \M : (z^{\m\sprime};y^{\m\sprime},\lambda_{\a'}{}^I,\trambda_{I\da'}\hnem)$ $ \mapsto z^{\m\sprime}$.
As a result, $\mathcal{K}$ fibers on $\M$ in two different ways.
If a disambiguation is needed, we denote the former and latter versions of $\M$ as 
$\M_x$ and $\M_z$, respectively.

Note that $\Mhat = T^{1,0}\M_x \cong \M_z \mtimes \M_\tz$,
where $z$ and $\tz$ arise by $\mathe^{+iN}x$ and $\mathe^{-iN}x$.
In the flat limit $\M_x {\,\to\:} \mflat$,
$\M_z$ and $\M_\tz$
are precisely the holomorphic and anti-holomorphic
subspaces of $\mhat = \mflat^\C$,
$\mhat^{1,0}$ and $\mhat^{0,1}$.

Lastly,
$\mathcal{K}$ is not only equipped with the horizontal vector field
$N {\,\in\,} \Gamma(T^{1,0}\mathcal{K})$
but also
\smash{$J : T^*_{1,0}\mathcal{K} {\,\to\,} T^*_{1,0}\mathcal{K}$}
such that $J^2 {\:=\:} {-\mathrm{id}}$
and $\omega {\,\in\,} \Omega^{2,0}(\mathcal{K})$
such that $d\omega {\:=\:} 0$:
$(\mathcal{K},N,J,\omega)$.
$N$ and $J$
are invariant under the restricted coordinate transformations of $\mathcal{K}$
stipulated above.

The Kerr symplectic potential 
arises by the same formula in \eqref{theta1},
yet now $\theta^{(1)} \in \Omega^{1,0}(\mathcal{K})$.

\subsection{Chiral Action and Motion in Heaven}%
\label{OSD1>CHIRAL}
If the holomorphic metric $g$ is SD,
then \eqref{kerr.g2a} implies that
the Kerr symplectic potential is
$\theta_{(1)} = \gamma^* \theta_{(1)}^+ + d( iW_0 )$,
where
\begin{align}
	\label{chiral.theta}
	\theta_{(1)}^+
	\,=\,
		- \trambda_{I\da'}\hem e^{\da\sprime\a}\hem \lambda_\a{}^I
		- 2i\mem \trambda_{I\da'}\mem y^{\da\sprime\a}\mem
			d\lambda_\a{}^I
	\,.
\end{align}
For simplicity, we have set the ASD spin connection coefficients to zero in a local patch
so that the ASD Wilson line $W_{\a'}{}^\b$ is trivial;
hence the primes on left-handed spinor indices have been dropped.

\eqref{chiral.theta} provides a chiral formulation of the Kerr particle,
treating $\lambda_\a{}^I$ and $\trambda_{I\da'}$ asymmetrically.
This formulation makes it evident that the symplectic form
is isomorphic to the free theory's symplectic form $\omega^\circ$
via the map
$\varrho: 
(dz^{\da\a}, d\bz^{\da\a}, d\lambda_\a{}^I , d\bar{\lambda}_{I\da})
\mapsto
(e^{\da\sprime\a} , e^{\da\sprime\a} - 2iDy^{\da\sprime\a}, d\lambda_\a{}^I, D\trambda_{I\da'})
$.
Hence
there is no covariant symplectic perturbation,
and the resulting Hamiltonian EoM
are just isomorphic to free theory:
\begin{align}\begin{split}
	\label{heavenly-eom}
	e^{m\sprime}{}_{\m'}(z)\mem \dot{z}^{\m\sprime} \mem=\mem p^{m\sprime}
	&\,,\quad
	\dot{\lambda}_\a{}^I \mem=\mem 0
	\,,\\
	Dy^{m\sprime}\nem/d\t
		\mem=\mem 0
	&\,,\quad
	D\trambda_{I\da\sprime}\mem/d\t
		\mem=\mem 0
	\,.
\end{split}\end{align}
Remarkably,
the dynamics of the Kerr black hole in heaven
is thus simply geodesic motion
with parallel transportation of the local Lorentz degrees of freedom
$y^{m\sprime}, \trambda_{I\da'}, \lambda_\a{}^I$
on the complex worldline $z^{\m\sprime}$
\footnote{
	A similar idea seems to had been conceived in \rcite{ko1981theory}
	in the context of $\mathscr{H}$-space.
}.

\eqref{heavenly-eom} concretely verifies the spinning, complexified analog of equivalence principle
that uniquely characterizes spinning black holes in SD external fields,
which the current author proposed through works \cite{sst-asym,ambikerr1}:
the left-handed spin frame $\lambda_\a{}^I$ is frozen, i.e., does not precess at all,
in SD backgrounds.

When transcribed to the defining chart
(covariantized at $x^\m$ in spacetime),
\eqref{heavenly-eom}
yields
an all-orders-in-spin extension of 
Mathisson-Papapetrou-Dixon equations
with $C_\ell {\:=\:} 1$ for all $\ell$.
Moreover,
\eqref{heavenly-eom}
implies exact hidden symmetries
in the presence of Killing and Killing-Yano tensors
by dynamically NJ shifting the conserved quantities of the Schwarzschild probe
\cite{probe-nj}.

It is also easily seen that the holomorphic coordinates $z^{\m\sprime}$ of 
the Kerr black hole's
curved spinspacetime $\Mhat$ remain Poisson-commutative
in SD gravity.

\subsection{Curved Massive Twistor Space}%
Lastly, we expand the holomorphic coframe 
around a flat background as
$e^{\da\sprime\a} = dz^{\da\sprime\a} + h^{\da\sprime\a}$.
The physical relevance of this setup is perturbation theory, for instance.
Concretely, employ the second Pleba\'nski coordinates
\cite{plebanski1975some,Adamo:2021bej}
to let $\a$ remain as the ``rigid'' $\SL(2,\C)$ index.
Then \eqref{chiral.theta} becomes
\begin{align}
	\label{chiral.theta.h}
	\kern-0.5em
	{- \trambda_{I\da'}\hem d\bigbig{
		z^{\da\sprime\a}\hem \lambda_\a{}^I
	}}
	+ \trambda_{I\da'}
	\bigbig{
		z^{\da\sprime\a} {\hem-\mem} 2iy^{\da\sprime\a}
	}\hem d\lambda_\a{}^I
	+ \theta'_{(1)}
	\,,
	\kern-0.6em
\end{align}
where 
$\theta' = - \trambda_{I\da'}\hem h^{\da\sprime\a}\hem \lambda_\a{}^I
	= p_{m'} h^{m\sprime}
	= p_{m'} h^{m\sprime}{}_{\m'}(z)\mem dz^{\m\sprime}
$.
This suggests to identify the ``mu variables'' as
\begin{align}\begin{split}
	\label{NJMAGIC|muvariables}
	\mu^{\da\sprime\mem I}
	\,=\,
		z^{\da\sprime\a}\hem\hhem \lambda_\a{}^I
	\,,\quad
	\tmu_I{}^\a
	\,&=\,
		\trambda_{I\da'}
		\bigbig{
			z^{\da\sprime\a} {\hem-\mem} 2iy^{\da\sprime\a}
		}
	\,,
\end{split}\end{align}
in which case \eqref{chiral.theta.h} 
and its exterior derivative become
\begin{align}\begin{split}
	\label{NJMAGIC|twistor.h}
	\theta^+_{(1)}
	\,&=\,
	- \trambda_{I\da'}\hem d\mu^{\da\sprime\mem I}
	+ \tmu_I{}^\a\hem d\lambda_\a{}^I
	+ \theta'_{(1)}
	\,,\\
	\omega_{(1)}
	\,&=\,
	d\mu^{\da\sprime\mem I} \swedge d\trambda_{I\da'}
	+ d\tmu_I{}^\a \swedge d\lambda_\a{}^I
	+ d\theta'_{(1)}
	\,.
\end{split}\end{align}

The free-theory part of \eqref{NJMAGIC|twistor.h}
is identical to $\omega^\circ$ in \eqref{omega0}
upon unpriming the right-handed spinor indices.
\eqref{NJMAGIC|twistor.h}---eventually---verifies the Kerr symplectic structure in SD spacetime we derived in \rcite{ambikerr1}
from a bootstrap approach based on the K\"ahler geometry (zig-zag structure) of massive twistor space
and the Souriau insight
\cite{souriau1970structure,dyson1990feynman}
that
deformations of particle symplectic structure
implement couplings to external fields.

\eqref{NJMAGIC|twistor.h}
establishes that 
the coupling of the Kerr particle to SD gravity
can be formulated as
a deformation of the symplectic structure on massive twistor space:
\begin{align}
	(\mtflat,\omega^\circ)
	\,\,\,\xrightarrow{\,\,\,\,\,\,}\,\,\,
	\MT
    \,\cong\,
        \bigbig{
            \mtflat,
            \omega^\circ \mplus\mem\hhem d\theta'_{(1)}
        }
	\,.
\end{align}
Crucially,
this deformation is via a $(1,0)$-perturbation $\theta'_{(1)}$
at the level of symplectic potential
such that the Poisson commutativity of holomorphic spinspacetime coordinates \cite{sst-asym} is preserved:
$\pb{z^{\m\sprime}}{z^{\n\sprime}} = 0$.
Physically,
this encodes that
the gravitoelectromagnetic interaction action
is supported on the NJ shifted worldline $z^{\m\sprime}$
for the SD (incoming positive-helicity \cite{bialynicki1981note,ashtekar1986note}) sector.

In this context,
\eqref{NJMAGIC|muvariables} is identified
as the deformed massive incidence relation,
from which the scaffolding of curved massive twistor theory arises.
Namely, \eqref{NJMAGIC|muvariables} defines a diffeomorphism
$\varphi: \mathcal{K} \to \MT$
and completes the map of our journey shown in \fref{fibration-curved}.

We may note that
$\varphi$ seems to encode a deformation of complex structure as well
when approached from the holomorphic complex structure of curved spinspacetime $\Mhat$.
The original complex coordinates
of $\Mhat \cong \M_z {\,\times\,} \M_\tz$
are $z^{\m\sprime}$ and \smash{$\tz^{\m'\nem\sprime}$}
that describe $\mathe^{\pm iN}x$.
However, the deformed massive incidence relation in \eqref{NJMAGIC|muvariables}
identifies
$z^{\m\sprime} \mminus 2iy^{\m\sprime}$
as the ``$z$-bar'' variable:
\begin{align}
	\label{tz}
	w^{\m\sprime}
	\,=\,
		z^{\m\sprime} - 2iy^{\m\sprime}	
	\,=\,
		z^{\m\sprime} - 2i\mem \s^{\m\sprime}(z,\tz)
	\,.
\end{align}
Here, $\s(z,\tz)$ describes the complexified Synge world function 
\cite{Ruse:1931ht,Synge:1931zz,Synge:1960ueh,Poisson:2011nh},
$\s : \M_z \mtimes \M_\tz \to \C$.
Crucially,
$w^{\m\sprime}$ differs from \smash{$\tz^{\m'\nem\sprime}$}
for nonzero curvature.

In massless curved twistor theory,
the celebrated nonlinear graviton construction of Penrose \cite{Penrose:1976js}
formulates SD gravity
as deformations of the complex structure in twistor space.
This describes deformations of the twistor lines
in terms of the deformed incidence relation.

Our proposal above
envisions curved massive twistor theory
from a deformed massive incidence relation
relevant to perturbation theory,
which is reminiscent yet not fully equivalent to massless curved twistor theory.
It may be also possible to consider
a formulation
that follows the ideas of massless curved twistor theory
more closely,
in which case 
the symplectic structure might be fixed.

\subsection{Googly Kerr Action}%
In \eqref{theta-earth},
we have already obtained the Kerr action
in generic spacetimes with both SD and ASD modes,
i.e., \textit{earth}
in the terminologies of
Newman \cite{shaviv1975general,Newman:1976gc} and Pleba\'nski \cite{plebanski1975some,Plebanski:1977zz}.

We then took an excursion to SD backgrounds, i.e., \textit{heaven},
to discover the nonlinear NJ shift in \eqref{zcurved}.
The key punchline is that
the \textit{NJ shift manifests in heaven}
as the localization of the Kerr action on the holomorphic worldline $z^{\m\sprime}$.

It should be stressed that
there is an obstruction in simultaneously manifesting
the NJ shifts in earth for both SD and ASD sectors.
This obstruction is due to the very nonlinearity of gravity
and poses a massive analog of the googly 
\cite{Penrose:2015lla:Palatial,penr04-googly,Witten:2003nn,Adamo:2017qyl} problem.
Concretely, it is impossible to write the earthly Kerr action
as a sum of two separate worldline actions localized respectively at
$z^{\m\sprime}$ and $\tz^{\m'\nem\sprime}$,
since there exist nonlinear mixing between SD and ASD curvatures.

See also \Sec{NJMAGIC>RKERR},
where the same problem is found for the case of nonabelian gauge theory.

The best we can do is to perturb away from the SD sector,
which is a typical twistor-theorist move.
Assuming the complexified setup,
the Kerr symplectic potential in 
\eqref{theta1}
can be written as
\begin{align}
\label{wsform}
	\theta_{(1)}
	\,=\,
		p_m\mem e^m
	+
		\frac{
			\mathe^{i\pounds_N} \mminus 1
		}{\pounds_N}
		\,\theta^+_\ga
	+
		\frac{
			\mathe^{-i\pounds_N} \mminus 1
		}{\pounds_N}
		\,\theta^-_\ga
	\,,
\end{align}
where we have defined
\begin{align}
\label{strings.kerr}
	\theta^\pm_\ga
	\mem:=\mem
		\frac{1}{2}\,
		\BB{
			p_m Dy^m
			\mp i\mem
			\bigbig{
				p_m\mem {\star}\Theta^m{}_n\hem y^n
				\mplus W_0\mem d\psi
			}
		\nem}
	\,.
\end{align}
Meanwhile, the symplectic potential
$\mathe^{i\pounds_N}\bigbig{p_m\hhem e^m}$
of the imaginary-deviated Schwarzschild particle is
\begin{align}
\label{strings.sch+ia}
	p_m\hhem e^m
	+
		\frac{
			\mathe^{i\pounds_N} \mminus 1
		}{\pounds_N}
		\,\theta^+_\ga
	+
		\frac{
			\mathe^{i\pounds_N} \mminus 1
		}{\pounds_N}
		\,\theta^-_\ga
	\,.
\end{align}
The difference between \eqrefs{strings.kerr}{strings.sch+ia} derives
\begin{align}
\label{kerr.googly}
	\theta_{(1)}
	\mem&=\,
	\mathe^{i\pounds_N}\mem
	\bb{
		p_m\hhem e^m
		+ 
		\frac{
			\mathe^{-2i\pounds_N} \mminus 1
		}{\pounds_N}
		\,\theta^-_\ga
	}
    \,,\\
	\,&=\,
		\mathe^{i\pounds_N}\mem
		\bb{
			\theta_{(0)}
			- i\mem p_m Dy^m
			+\mem\hhem
				\sum_{\ell=2}^\infty
					\frac{(-2i)^\ell}{\ell!}\mem
						p_m
						\bigbig{\hnem
							(\i_N\hem D)^{\ell-2} \i_N\mem R^-{}^m{}_n
						\hhnem}\hem y^n
		}
	\nonumber
	\,,\\
    	\mem&=\,
\begin{aligned}[t]
		&
		p_{m'}\hhem e^{m\sprime}
		+ p_{m'}\mem {\star}\Theta^{m\sprime}{}_{n'}\hem y^{n\sprime}
		- i\mem p_{m'} Dy^{m\sprime}
		+ W_0\mem d\psi
    		\\
    		&
    	 	+\mem\hhem \sum_{\ell=2}^\infty
\begin{aligned}[t]
{}&{}
    	 		\frac{(-2i)^\ell}{\ell!}
    			\sum_{p=1}^{\lfloor{\ell/2}\rfloor}\kern-0.2em
    			\sum_{\a \in \Omega_p(\ell)}\kern-0.2em
    	 			\bbsq{
    	 				\prod_{i=1}^p
    	 				\binom{
    	 					\bigbig{
    	 						\sum_{j=i}^p \a_j
    	 					} \mminus 2
    	 					\hem
    	 				}{
    	 					\a_i \mminus 2
    	 				}
    	 				\nem\nem
    	 			}\mem
\\
{}&{}
    	 		\lrp{
    	 		\begin{aligned}[c]
    	 			&
				\bigbig{\hnem
					(Q^-_{\a_1}\hnem)\mem Q_{\a_2}\, {\nem\cdots\mem} Q_{\a_p}
				\hnem}
				{}^{m\sprime}{}_{s'}\mem e^{s\sprime}
				\\
				&
				+ (\a_p\mminus2)\mem
				\bigbig{\hnem
					(Q^-_{\a_1}\hnem)\mem Q_{\a_2}\, {\nem\cdots\mem} Q_{\a_{p-1}} Q_{\a_p - 1}
				\hnem}
				{}^{m\sprime}{}_{s'}\mem Dy^{s\sprime}
    	 		\end{aligned}}
 	 \,,
\end{aligned}
\nonumber
\end{aligned}
\end{align}
where 
$R^\pm := \minie\mem (R {\,\mp\,} i\mem {\star} R)$
and
$(Q_j^\pm) := \minie\mem (Q_j {\,\mp\,} i\mem {\star} Q_j)$.
To be precise, the first line is the pullback of the second line by the diffeomorphism $\gamma : \mathcal{K} \to \mathcal{K}$.

\eqref{kerr.googly} provides
a googly formulation of the Kerr action.
It describes
\textit{both} SD and ASD couplings
as an action localized on the \textit{holomorphic} worldline.
It reduces to \eqref{kerr.g2a} in the SD limit.
It treats SD and ASD modes,
and the $z^{\m\sprime}$ and \smash{$\tz^{\m'\nem\sprime}$} worldlines,
asymmetrically.

In actual applications,
\eqref{kerr.googly} facilitates
a perturbation theory
in which the NJ shift (as spin exponentiation  \cite{ahh2017,Guevara:2018wpp,Guevara:2019fsj,chkl2019})
of scattering amplitudes
is manifestly guaranteed in the SD sector
to all graviton multiplicities
\cite{probe-nj},
on top of which
the ASD perturbations are systematically added on.

\section{
Interpretation}%
In \eqref{kerr.googly},
$(\mathe^{-2i\pounds_N} \mminus 1)/(-2i\pounds_N) = 
\smash{\int_0^1\mem d\eta\, \mathe^{-2i\eta\pounds_N}}
$
inserts the NJ shifts of $\theta^-_\ga$
along the geodesic segment
joining the two points $z$ and $\tz$.
Hence \eqref{kerr.googly} states that
the Kerr black hole 
describes 
a dangling open string attached to
a point mass at $z$.
This string is charged under ASD gravitons
($d\theta^-_\ga {\:=\:} p_m R^-{}^m{}_n\hem y^n$)
but is invisible to SD fields
as remarked earlier.
In fact,
recalling our earlier discussions on Gilbertian and Amp\`erian dipoles,
it precisely describes the ASD combination
of electric and magnetic mass dipole moments.
Thus, we identify its interpretation as
an ASD Misner string.

Similarly,
it can be seen that
\eqref{theta1}
describes two point masses
joined by a thin gravitomagnetic flux tube:
an ordinary Misner string
\footnote{
	This string structure was noticed in
	\rcite{gmoov},
	but the physical interpretations and googly formulation
	are not provided there.
	\rcite{gmoov} also does not provide explicit expressions 
	such as Eqs.\:(\ref{theta-earth}) and (\ref{kerr.googly})
	that expand out the action fully in terms of covariant worldline operators.
}.
The two pictures,
googly and real,
are equivalent
via identifying a pair of opposite-sign masses
as a line defect of electric mass dipole.

The above analysis elicits a provocative idea that
the Kerr black hole describes a system of SD and ASD dyons.
At least in the effective theory of Kerr black holes,
\eqrefs{kerr.g2a}{kerr.googly} show that
this statement holds true as the physical origin of the NJ shift.
The question is whether it holds in the ultraviolet theory,
i.e., general relativity,
as a nonperturbative fact.

The persistent endeavor of the current author around this question
had resulted in works \cite{note-sdtn,nja},
which established that the Kerr metric
represents the exact nonlinear superposition of
SD and ASD Taub-NUT solutions
as a holomorphic saddle,
from which the NJA \cite{Newman:1965tw-janis}
is faithfully reproduced.

The worldlines
$z^{\m\sprime}$ and \smash{$\tz^{\m'\nem\sprime}$} 
are the infrared avatars of the Taub-NUT instantons.
The worldline actions due to
\eqrefs{theta1}{kerr.googly}
insert the Misner string along the geodesic
between the two instantons.
As long as the geodesicity of the Misner string
(or locality in worldline proper time)
can be assumed,
\eqref{theta1},
equivalently \eqref{kerr.googly},
defines the unique answer
for the effective action 
that describes
a dynamical pair of SD and ASD Taub-NUT instantons.
Regarding this point,
it will be helpful to investigate if
the Misner string
serves as a genuinely topological surface defect
in dynamical, full (non-SD) gravity
(cf. \Sec{K1.grav}),
in which case
the Misner worldsheet can be realized in any shape
to predict the same physical observables.

\section{
Conclusions}%
In this chapter,
we derived an explicit all-orders worldline effective action for the Kerr black hole
by generalizing an intrinsic feature of twistor particle theory
in curved spacetime.
This implements
the NJA at the level of test-particle actions,
in terms of differential-geometric elements $N$ and $J$.
Nonlinear, dynamical NJ shift manifests in SD backgrounds
as localizations on the holomorphic worldline.
The full gravitational interactions
are systematically understood by
perturbing around the SD sector.
Clear physical interpretations are given
and are substantiated by 
the ultraviolet description
of the Kerr black hole.

This chapter is written in the geometrical language
and wishes to deliver itself to the twistor theory community as well.
This chapter attempts to make
a small yet ambitious step towards
``curved massive twistor theory,''
which shall exist
in light of the histories of massless twistor theory 
and twistor particle program.
Some tentative identifications and definitions are given for
the deformed massive incidence relation
and curved massive twistor space,
based on the particular case of the Kerr black hole
as a massive twistor.

\begin{subappendices}

\section{
Spinspacetime from Adapted Complex Structure}%
\label{ACS}
This appendix elaborates on the definition of spinspacetime
for real spacetime.

In \eqref{tau-flat},
we discussed
two perspectives toward
spinspacetime:
real and complex manifolds.
In the case of curved real-analytic spacetime $(\M,g)$,
however,
the latter picture
is seemingly lost.
The formula in \eqref{NJMAGIC|z}
becomes utterly nonsensical,
since
$x^\m$ are \textit{coordinates}
while $y^\m$ are components of a \textit{vector}.
This represents a  clash between
two necessary features
of ``curved spinspacetime'':
general covariance and holomorphy.

Luckily,
the mathematical framework
known as adap\-ted complex structure
\cite{guillemin1991grauert,guillemin1992grauert,lempert1991global,szHoke1991complex,halverscheid2002complexifications,aguilar2001symplectic,burns2000symplectic,hall2011adapted}
provides a nice resolution,
showing that
a sufficiently small neighborhood
of the zero section of
the tangent bundle $T\M$
of a real-analytic $(\M,g)$
can be embedded into a complex manifold $\Mhat$
via complexifying the exponential map
(geodesic flow):
\begin{align}
	\label{tau}
	\Pi_i
	\,\,:\,\,
	T'\hnem\M
	\,\,\to\,\,
	\Mhat
	\,.
\end{align}
Here, we have supposed 
a sufficiently narrow neighborhood $T'\hnem\M$ of the zero section in $T\M$.

The idea is simple.
The time-$\eta$ exponential map
sends the pair $(x,y) {\:\in\:} T'\hnem\M$ to 
the point in $\M$
whose coordinates are
$w^\m(x,y;\eta)
= x^\m
	+ \eta\mem y^\m
	- \smash{\frac{\eta^2}{2}}\, \Gamma^\m{}_{\r\s}(x)\mem y^\r y^\s
	+ \O(y^3)
$.
This is a
power series solution
which converges in $T'\hnem\M$.
Upon a real-analytic coordinate transformation $x^\m \mapsto f^\m(x)$ on $\M$,
the coordinates
of the geodesically deviated point
transform as 
$w^\m(x,y;\eta)
\mapsto f^\m(w(x,y;\eta))
$
for any $\eta$,
by construction.

By analytic continuation,
$z^\m(x,y) {\,:=\,} w^\m(x,y;i)$
is well-defined within $T'\hnem\M$,
whose explicit formula
is exactly \eqref{zcurved}.
By analyticity,
the coordinate transformation on $\M$
induces
$z^\m(x,y)
\mapsto f^\C{}^\m(z(x,y))
$,
where $f^\C$ is the analytic continuation of $f$.
This shows that
$z^\m(x,y)$
defines a holomorphic coordinate chart on $T'\hnem\M$.
Therefore, $T'\hnem\M$ becomes a complex manifold.

Physically speaking,
this construction ensures that
holomorphy is an observer-independent notion.
This is important since holomorphy will be linked with self-duality
by the NJ shift.

Now let us unravel the precise mathematical details.
The time-$\eta$ exponential map can be formalized as
\begin{align}
	\Pi_\eta
	\,=\,
		\pi \circ \Phi_\eta
	\,\,:\,\,
	T'\hnem\M \,\to\, \M
	\,,
\end{align}
where $\Phi_\eta : T'\hnem\M \to T\M$ denotes the time-$\eta$ flow
by the horizontal vector field $N \in \Gamma(T\M)$
such that $\i_N dx^\m = y^\m$,
and $\pi : T\M \to \M$ is the bundle projection.
The pullback ${\Pi_\eta}^*$
maps
analytic functions on $\M$
to
analytic functions on $T'\hnem\M$,
the explicit power series solutions for which
arise by
the exponentiated Lie derivative
$\mathe^{\eta\pounds_N}$.

By analytic continuation to $\eta \in \C$,
one defines ${\Pi_i}^*$ as a map from
real-analytic functions on $\M$
to complex-analytic functions on $T'\hnem\M$,
the application of which on coordinate functions
establishes 
the map $\Pi_i$ in \eqref{tau}
such that $\M$, as the zero section of $T\M$,
is embedded in the complex-analytic manifold $\Mhat = \M^\C$
as a totally real submanifold of maximal dimension
\cite{whitney1959quelques,grauert1958levi,hall2011adapted}.

\begin{figure}[t]
	\centering
	\adjustbox{valign=c}{\begin{tikzpicture}
		\node[empty] (o) at (0,0) {};
		\node[empty] (i) at (-1.4, -1.4) {};
		\node[empty] (j) at ( 1.4, -1.4) {};
		\node[b] (A) at ($(o)$) {${
			\a^\C \nem\in \Omega^p(\Mhat)
		}\phantom{|}$};
		\node[b] (Ai) at ($(A)+(i)$) {$\mathclap{
			\widehat{\a} \in \Omega^p(T'\hnem\M)
		}\phantom{|}$};
		\node[b] (Aj) at ($(A)+(j)$) {$\mathclap{
			\a \in \Omega^p(\M)
		}\phantom{|}$};
		\draw[->] (A)--(Ai) node[midway,left,pos=0.4] 
		{\scriptsize ${\Pi_i}^*$\nem};
		\draw[<-] (A)--(Aj) node[midway,right, pos=0.4] 
		{\scriptsize \,\text{a.c.}};
	\end{tikzpicture}}
	\caption{%
		The ``central dogma'' of 
		adapted complex structure.
	}
	\label{dogma}
\end{figure}

The practical use of this formalism
is facilitated by the fact that
the exponentiated Lie derivative $\mathe^{i\pounds_N}$
computes the pullback ${\Pi_i}^*$.
Any differential $p$-form $\a {\:\in\:} \Omega^p(\M)$
is transcribed to
$\a^\C \nem{\:\in\:} \smash{\Omega^p(\Mhat)}$
by analytic continuation,
which translates to
$\widehat{\a} = {\Pi_i}^*[\mem{ \a^\C }\mem] = \mathe^{i\pounds_N} \a \in \Omega^p(T'\hnem\M)$.
This is the ``central dogma'' of adapted complex structure,
depicted in \fref{dogma}.
For instance,
a zero-form
$\phi \in \Cinfty(\M)$
exhibits
the covariant Taylor expansion
\begin{align}\begin{split}
	\label{taylor0}
	\widehat{\phi}(x,y)
	\,&=\,
		\phi^\C(z(x,y))
	\,=\,
		\mathe^{i\pounds_N} \phi(x)
	\,,\\
	\,&=\,
		\sum_{\ell=0}^\infty\,
			\frac{i^\ell}{\ell!}\,
				\phi_{;\r_1;\cdots;\r_\ell}(x)\,
			y^{\r_1} {\cdots} y^{\r_\ell}
	\,.
\end{split}\end{align}
Note how general covariances coexist 
at both points $x$ and $z$
\eqref{taylor0}.

Curved spinspacetime 
is the complex-analytic four-manifold $\Mhat$
that recasts the tangent bundle of spacetime
via adapted complex structure,
in the sense of \eqref{tau}.

Spacetime fields,
such as $\phi(x)$ in \eqref{taylor0},
permeate into spinspacetime
via analytic continuation:
$\phi^\C(z)$.
The spinspacetime field $\phi^\C(z)$
is represented in the ``spacetime $+$ spin'' form,
\smash{$\widehat{\phi}(x,y)$},
by means of the operator
$\mathe^{i\pounds_N}$
\footnote{
	The latter description is perturbative in spin
	and may break down beyond 
	the region
	$T'\hnem\M$
	in $T\M$.
	However, 
	it suffices to consider small enough spin
	for the physical purposes of this chapter.
}.

Historically, 
it was Newman who first envisioned
the concept of spinspacetime
and its physical applications
\cite{%
	newman1988remarkable,newman1974curiosity,newman1974collection,%
	Newman:1973yu,Newman:1973afx,Newman:2002mk,Newman:2004ba,%
	Newman:1976gc,ko1981theory,grg207flaherty%
}.
The curved spinspacetime constructions of
this chapter transparently realizes Newman's provisional insights
in terms of concrete differential-geometric constructions.

\section{
Nonabelian Root-Kerr Action
}
\label{NJMAGIC>RKERR}

Here, we construct the {\kerr} symplectic structure coupled to nonabelian gauge theory.
Suppose gauge group $G$
and gauge algebra $\g$.
Let $a,b,\cdots$ be the adjoint indices.
Let $\V$ be a representation space
assigned with indices $i,j,\cdots$,
on which the generators are $(t_a)^i{}_j$.

The color degrees of freedom
can be implemented by phase spaces such as
$T^*G$, $T^*\V$, etc.
Take $T^*\V$ with complex coordinates
$\psi^i$, for instance.
The correspondence space is enlarged as $\kflat_\text{col} = (T{\:\oplus\:}S_2)\mflat \oplus E$,
where $E$ is a vector bundle over $\mflat$
whose typical fiber is $T^*\V$.

Let $A {\:\in\:} \Omega^1(\mflat;\g)$ be the nonabelian gauge connection.
A complete, gauge-covariant basis of one-forms on $\kflat_\text{col}$
is $(dx^\m,dy^\m,d\lambda_\a{}^I,D\psi^i)$,
where $D$ is the gauge-covariant exterior derivative.
The horizontal vector field $N \in \Gamma(\kflat_\text{col})$
is uniquely defined by 
the interior products
$\i_N : (dx^\m,dy^\m,d\lambda_\a{}^I,D\psi^i) \mapsto (y^\m,0,0,0)$.
It follows that $N$ is the generator of gauge-covariant translations \cite{gde}.

The Coulomb particle is described by
the color-sector symplectic potential
\begin{align}
	\label{stheta0}
	\stheta_{(0)}
	\mem=\,
		i\mem \bpsi_i\mem D\psi^i
	\,=\,
		i\mem \bpsi_i\mem d\psi^i
		+ q_a A^a
	\,,
\end{align}
where $q_a {\:=\:} i\mem \bpsi_i\mem (t_a)^i{}_j\hem \psi^j$
is the adjoint color charge.
A deviated Coulomb particle is described by
$\mathe^{\pounds_N} \stheta_{(0)}$:
\begin{align}
	\label{stheta0.dev}
		\cos(\pounds_N)\bigbig{
			i\mem \bpsi_i\mem D\psi^i
		}
		+ 
		\sinc(\pounds_N)\bigbig{
			q_a\mem \i_N F^a
		}
	\,.
\end{align}

In \eqref{stheta0.dev}, the $\cos(\pounds_N)$ term
implements two electric charges separated apart
at $x\pm iy$.
The $\sinc(\pounds_N)$, on the other hand,
inserts a thin electric flux tube between these electric charges.
We apply electric-magnetic duality
on the $\sinc(\pounds_N)$ term
via Hodge dual on the field strength:
\begin{align}
	\label{stheta1}
	\stheta_{(1)}
	\mem=\,
		\cos(\pounds_N)\bigbig{
			i\mem \bpsi_i\mem D\psi^i
		}
		+ 
		\sinc(\pounds_N)\bigbig{
			q_a\mem \i_N {*}F^a
		}
	\,.
	\kern-0.25em
\end{align}
As a result, we obtain the symplectic potential
describing two electric charges at $x\pm iy$
joined by a magnetic flux tube, i.e., a Dirac string
(cf. \rcite{GenSymGrav}).
Upon Gilbert-Amp\`ere duality, this describes a pair of SD and ASD dyons.
This pair is nothing other than 
the 
(dynamical version of)
{\kerr} solution:
repeat the exercise in \rcite{nja}.

\begin{figure}[t]
	\centering
	\label{eq:Liecd-rKerr}
	\adjustbox{valign=c}{\begin{tikzpicture}
	    \node[empty] (O) at (0,0) {};
	    \node[empty] (X) at (3.75, 0) {};
	    \node[empty] (x) at (3.0, 0) {};
	    \node[empty] (Y) at (0, -0.95) {};
	    \node[w] (a00) at ($(O)$) {$i\mem \bpsi_i\hem D\psi^i$};
	    \node[w] (a01) at ($(O)+1*(x)$) {$0$};
	    \node[w] (a10) at ($(O)+1*(Y)$) {$q_a (\i_N F^a)$};
	    \node[w] (a11) at ($(O)+1*(Y)+1*(x)$) {$0$};
	    \node[w] (a20) at ($(O)+2*(Y)$) {$q_a (\i_ND\mem \i_N F^a)$};
	    \node[w] (a21) at ($(O)+2*(Y)+1*(x)$) {$0$};
	    \node[w] (a30) at ($(O)+3*(Y)$) {$\vdots$};
	    \node[w] (a2K) at ($(O)+2*(Y)-1*(X)$) {$q_a (\i_N {\ast F}^a)$};
	    \node[w] (a3K) at ($(O)+3*(Y)-1*(X)$) {$q_a (\i_ND\mem \i_N {\ast F^a})$};
	    \node[w] (a4K) at ($(O)+4*(Y)-1*(X)$) {$\vdots$};
	    \node[w] (a2k) at ($(O)+2*(Y)-1*(X)+(x)$) {$0$};
	    \node[w] (a3k) at ($(O)+3*(Y)-1*(X)+(x)$) {$0$};
	    \node[w] (phantom-a00) at ($(O)$) {$\phantom{\big|}$};
	    \node[w] (phantom-a01) at ($(O)+1*(x)$) {};
	    \node[w] (phantom-a10) at ($(O)+1*(Y)$) {};
	    \node[w] (phantom-a11) at ($(O)+1*(Y)+1*(x)$) {};
	    \node[w] (phantom-a20) at ($(O)+2*(Y)$) {};
	    \node[w] (phantom-a21) at ($(O)+2*(Y)+1*(x)$) {};
	    \node[w] (phantom-a30) at ($(O)+3*(Y)$) {};
	    \node[w] (phantom-a31) at ($(O)+3*(Y)+1*(x)$) {};
	    \node[w] (phantom-a2K) at ($(O)+2*(Y)-1*(X)$) {};
	    \node[w] (phantom-a3K) at ($(O)+3*(Y)-1*(X)$) {};
	    \node[w] (phantom-a4K) at ($(O)+4*(Y)-1*(X)$) {};
	    \node[w] (phantom-a2k) at ($(O)+2*(Y)-1*(X)+(x)$) {};
	    \node[w] (phantom-a3k) at ($(O)+3*(Y)-1*(X)+(x)$) {};
	    \draw[->] (a00)--(a01) node[midway,above] {\scriptsize \smash{${d}\mem\i_N$}\vphantom{d}};
	    \draw[->] (a10)--(a11) node[] {};
	    \draw[->] (a20)--(a21) node[] {};
	    \draw[->] (phantom-a00)--(phantom-a10) node[midway,left] {\scriptsize $\i_N{d}$};
	    \draw[->] (phantom-a10)--(phantom-a20) node[] {};
	    \draw[->] (phantom-a20)--(phantom-a30) node[] {};
	    \draw[<->] (phantom-a10)--(a2K) node[midway,above] {\adjustbox{raise=1.5pt}{\scriptsize dual}};
	    \draw[->] (phantom-a2K)--(phantom-a3K) node[] {};
	    \draw[->] (phantom-a3K)--(phantom-a4K) node[] {};
	    \draw[->] (a2K)--(a2k) node[] {};
	    \draw[->] (a3K)--(a3k) node[] {};
	\end{tikzpicture}}
	\caption{%
		The ``$\i_N d / {*}$ sequence'' for {\kerr}.
		A tree of one-forms
		emanates from
		the Coulomb particle's
		color-sector symplectic potential,
		$i\mem \bpsi_i\hem D\psi^i$.
	}
	\label{stree-of-life}
\end{figure}

\eqref{stheta1}
pinpoints a unique gauge theory interaction
to all orders in spin and gauge coupling.
From \fref{stree-of-life},
it can be seen that
\eqref{stheta1} can be represented as
\begin{align}
	\label{ssum1}
	\stheta_{(1)}
	\mem&=\,
	\stheta_{(0)}
	+\mem\hhem
		\sum_{\ell=1}^\infty
			\frac{1}{\ell!}\,
				q_a\mem
				\bigbig{\hnem
					(\i_N\hem D)^{\ell-1} \i_N\mem {*^\ell\hnem} F^a
				\hhnem}
	\,,
\end{align}
where ${*^\ell}$ means to act on the internal Hodge star $\ell$ times.
\eqref{ssum1} parallels \eqref{sum1}
by identifying spin angular momentum as ``the gravitational charge''
and Riemann curvature two-form as the ``gravitational field strength'':
gravity as a gauge theory of Lorentz group.

By methods in differential geometry \cite{gde},
\eqref{ssum1} evaluates to
\begin{align}
	\label{stheta-earth}
	\stheta_{(1)}
	\mem&=\,
	i\mem\bpsi_i D\psi^i
	+\mem\hhem
		\sum_{\ell=1}^\infty
			\frac{1}{\ell!}\,
				q_a
	    	 		\bb{\hnem
					\bigbig{\hnem
						{*}^\ell P_\ell
					\hnem}
					{}^a{}_\s\, dx^\s
					+
					(\ell\mminus1)\mem
					\bigbig{\hnem
						{*}^\ell P_{\ell-1}
					\hnem}
					{}^a{}_\s\, dy^\s
	    	 		}
	\,,
\end{align}
where
the so-called $P$-tensors \cite{gde} are defined as
\begin{align}
    \label{Ptensor}
    ({*}^\ell P_j)^a{}_\s
    \,=\,
	  {*}^\ell F^a{}_{\r_1\s;\r_2;\cdots;\r_j}\hnem(x)
	  \, y^{\r_1}{\cdots}y^{\r_j}
    \,.
\end{align}

\eqref{stheta-earth} provides the action of the {\kerr} particle
coupled to generic nonabelian gauge field configurations (earth).
It is manifestly real and parity-symmetric
and is based on the conventional ``spacetime $+$ spin'' picture.

NJ shift manifests in heaven.
On the support of self-duality ${*}F^a = +i\mem F^a$,
\eqref{ssum1} becomes
\begin{align}\begin{split}
	\label{rkerr.heaven}
	\stheta_{(1)}
	\mem&=\,
		\mathe^{i\pounds_N} \stheta_{(0)}
	\,=\,
		i\mem \bpsi_{i'} D\psi^{i\sprime}
	\,,
\end{split}\end{align}
which localizes on $z^\m = x^\m + iy^\m$.
Here, we have complexified $\V$
(as a real vector space)
to $\V^\C$.

There is an obstruction in simultaneously manifesting
the NJ shifts in earth for both SD and ASD sectors.
This obstruction is due to the very nonlinearity of nonabelian gauge theory
and poses a massive analog of the googly problem.
Concretely, it is impossible to write the earthly {\kerr} action
as a sum of two separate worldline actions localized respectively at
$z^\m$ and $\bz^\m$,
since
$P_\ell$ inevitably exhibits
nonlinear mixing between SD and ASD modes
from $\ell \geq 2$.

Note that there is no such googly problem in abelian gauge theory,
as the gauge potential cleanly splits into SD and ASD parts
and the {\kerr} symplectic potential is linear in fields.

Still, we can manifest the spinspacetime localization
on either $z^\m$ or $\bz^\m$.
By mimicking \eqref{wsform}
in terms of $\bigbig{(\mathe^{i\pounds_N} \mminus 1)/\pounds_N}\mem q_a\hem (\i_N F^\pm{}^a)$,
it follows that
\begin{align}
\label{rkerr.googly}
	\stheta_{(1)}
	\mem&=\,
	\mathe^{i\pounds_N}\mem
	\bb{
		i\mem \bpsi_i D\psi^i
		+ 
		\frac{
			\mathe^{-2i\pounds_N} \mminus 1
		}{\pounds_N}
		\,q_a\mem (\i_NF^-{}^a)
	}
	\,,\\
    	\mem&=\,
		i\mem \bpsi_{i'} D\psi^{i\sprime}
		+\mem\hhem
			\sum_{\ell=1}^\infty
				\frac{(-2i)^\ell}{\ell!}\,
					q_{a'}
		    	 		\bb{\hnem
						\bigbig{\hnem
							P^-_\ell
						\hnem}
						{}^{a\sprime}{}_\s\, dz^\s
						+
						(\ell\mminus1)\mem
						\bigbig{\hnem
							P^-_{\ell-1}
						\hnem}
						{}^{a\sprime}{}_\s\, dy^\s
		    	 		}
 	 \,,
     \nonumber
\end{align}
where 
$F^\pm{}^a := \tfrac{1}{2}\mem (F^a {\,\mp\,} i\mem {*}F^a)$
and
$(P_\ell^\pm) := \minie\mem (P_\ell {\,\mp\,} i\mem {*} P_\ell)$.

\newpage

\eqref{rkerr.googly} provides
the ``radical'' spinspacetime formulation of the {\kerr} action,
i.e., a googly formulation.
It describes
\textit{both} SD and ASD couplings
as an action localized on the \textit{holomorphic} worldline $z^\m$.
It reduces to \eqref{rkerr.heaven} in the SD limit.
It treats SD and ASD modes,
and the $z^\m$ and $\bz^\m$ worldlines,
asymmetrically.
In perturbation theory, it manifests spin exponentiation of the positive-helicity (but not negative-helicity) Compton amplitudes.

\end{subappendices}
\chapter{The Kerr-Newman Effective Action}\label{K3:OSD11}

	An all-orders worldline effective action
	for Kerr-Newman black hole
	is achieved in
	twistor particle theory.
	Exact hidden symmetries are identified
	in SD backgrounds.

\begin{fullnote}
    This chapter reproduces the contents of \rcite{njmagic.11}, \fullcite{njmagic.11}.
\end{fullnote}

\section{
Introduction}%
An important topic in the current gravitational physics literature has been
the construction of worldline effective actions
\cite{ShR065,Porto:2005ac,Levi:2015msa,Porto:2016pyg,Levi:2018nxp,Kalin:2020mvi}
of spinning black holes.
Recently, \rcite{njmagic.1}
has derived an exact first-order action of the Kerr black hole
and provided its explicit all-orders formula,
by using
massive twistor theory
\cite{Penrose:1974di,Perjes:1974ra,tod1977some}
and the tangent bundle formalism for geodesic deviation \cite{gde}.

\rcite{njmagic.1}'s derivation
of the Kerr action
provides the test-particle counterpart
of the NJA
\cite{njmagic.1,probe-nj}.
The NJA is
a solution-generating technique
that rederived the Kerr metric
by performing a complex-geometrical transformation on
the Schwarzschild metric
\cite{Newman:1965tw-janis}.
This trick then
facilitated the very historical derivation of the Kerr-Newman solution \cite{Newman:1965my-kerrmetric}.
According to a modern view \cite{nja},
the NJA encodes
that the Kerr solution
is diffeomorphic to
a pair of SD and ASD Taub-NUT instantons
when taken as a holomorphic saddle.

The present paper
generalizes
\rcite{njmagic.1}'s construction
to derive the first-order action of the Kerr-Newman black hole
coupled to background gravitational and electromagnetic fields.
Again, the systematic formalism of \rcite{gde} proves instrumental in obtaining explicit formulae for the action.
We identify a Hamiltonian system
that can represent the exact, all-orders dynamics of the Kerr-Newman black hole
in its point-particle effective theory.

It is found that the conclusions of \rcite{njmagic.1} are preserved.
Nonlinear NJ shift manifests in SD backgrounds.
A chiral (``googly'') formulation of the Kerr-Newman action is viable
for generic, non-SD external fields.
In perturbation theory,
the spin exponentiation
of same-helicity Compton amplitudes
\cite{Guevara:2018wpp,Guevara:2019fsj,aho2020,Johansson:2019dnu,Aoude:2020onz,Lazopoulos:2021mna}
is manifested to all graviphoton multiplicities.

We also derive the classical EoM
in SD backgrounds.
We confirm the complexified, spinning analog of equivalence principle proposed by \rcite{sst-asym}.
This implies
dynamical hidden symmetries
in the presence of Killing-Yano tensors,
generalizing \rcite{probe-nj}'s previous result
about the Kerr particle.

\section{
Geodesic Deviation in Einstein-Maxwell Geometry}%
First, we present the new application of \rcite{gde}'s tangent bundle formalism
for geodesic deviation.

Let $(\M,g)$ be a real-analytic pseudo-Riemannian four-manifold,
and let $F {\:\in\:} \Omega^2(\M)$ be a two-form.
Let $\nabla$ be the Levi-Civita connection.
Let $D$ be the covariant exterior derivative due to $\nabla$.

Let $N \in T(T\M)$ be the horizontal lift
of the tautological vector field of the tangent bundle $T\M$
with respect to $\nabla$.
This is the geodesic spray, i.e., the generator of geodesic deviation in $T\M$.

We explicitly evaluate 
$(\i_Nd)^{\ell-1}\mem \i_NF = (\i_ND)^{\ell-1}\mem \i_NF$
for all integers $\ell \geq 1$.
The result is
\begin{align}
\label{PQseq}
    &{}
   (\i_ND)^{\ell-1}\mem \i_N F
   \,=\,
           (P_\ell)_\m\mem
           dx^\m
           +
           (\ell\mminus1)\mem
           (P_{\ell-1})_\m\mem
           \, Dy^\m
\\\nonumber
&\quad
       +
       \sum_{p=2}^{\lfloor{(\ell+1)/2}\rfloor}\kern-0.4em
       \sum_{\a \in \Omega_p\hnem(\ell+1)}\kern-0.6em
           \mathscr{K}_\a
       \lrp{\hem
       \begin{aligned}[c]
       &
           \bigbig{\hnem
               P_{\a_1-1}
               Q_{\a_2}\, {\nem\cdots\mem} Q_{\a_p}
           \hnem}
           _{\nem\m}
           \, dx^\m
       \\
       &
           +
           (\a_p\mminus2)
           \bigbig{\hnem
               P_{\a_1-1}
               Q_{\a_2}\, {\nem\cdots\mem} Q_{\a_{p-1}} Q_{\a_p - 1}
           \hnem}
           _{\nem\m}
           \, Dy^\m
           \nem
       \end{aligned}
       }
   \,.
\end{align}

This is a new calculation,
not presented in our previous works
\cite{gde,njmagic.1}.
Yet, our notations are the same.
Each $\a = (\a_1,\a_2,\cdots,\a_p) \in \Omega_p(\ell)$
is an ordered partition
such that
$\a_1 + \a_2 + \cdots + \a_p = \ell$
and $\a_i \geq 2$.
The coefficient $\mathscr{K}_\a$
for $\a \in \Omega_p(\ell)$
is a product of binomial coefficients:
\begin{align}
\label{coeffK}
   \mathscr{K}_\a
   \,=\,
   \bbsq{
       \prod_{i=1}^p
       \binom{
           \bigbig{
               \sum_{j=i}^p \a_j
           } \mminus 2
           \hem
       }{
           \a_i \mminus 2
       }
       \nem\nem
   }
   \,.
\end{align}
For any $\a = (\a_1,{\cdots},\a_p)$
such that $\min\{\a_1,{\cdots},\a_p\} < 2$,
we set $\mathscr{K}_\a = 0$.
The $P$- and $Q$-tensors \cite{gde} are
\begin{subequations}
\begin{align}
   \label{P,Q}
     (P_\ell)_\m
     &\,=\,
         F_{\r_1\m;\r_2;\cdots;\r_\ell}(x)\,
         y^{\r_1} y^{\r_2} {\hem\cdots\mem} y^{\r_\ell}
     \,,\\
	(Q_\ell)^\m{}_\s
	&\,=\,
		R^\m{}_{\r_1\r_2\s;\r_3;\cdots;\r_\ell}\hnem(x)\,
		y^{\r_1} y^{\r_2} y^{\r_3} {\hem\cdots\mem} y^{\r_\ell}
	\,,
\end{align}
\end{subequations}
while
$P_\ell$ and $Q_\ell$
are set to zero
for $\ell = 0$ and $\ell = 0,1$, respectively.

\eqref{PQseq}
is readily proven by mathematical induction,
based on the recursive structure illustrated in \fref{chemistry-A}.
This applies the ``organic chemistry'' intuition of \rcite{gde}.

Amusingly, \eqref{PQseq}
can be quickly seen by
a ``covariant color-kinematics'' \cite{CCK} mapping,
which views gravity as a gauge theory of the Lorentz group.
Namely, \eqref{PQseq} can be deduced from
the formula for
$(\i_ND)^{\ell-1}\mem \i_N R^m{}_n$
given in \rcite{gde,njmagic.1}.
By viewing the antisymmetric pair of local Lorentz indices
as the adjoint index of the Lorentz group as a gauge group,
$(\i_ND)^{\ell-1}\mem \i_N R^m{}_n$
corresponds to $(\i_ND)^{\ell-1}\mem \i_N F^a$
where $F^a$ is a nonabelian field strength.
Taking the abelian limit then reproduces \eqref{PQseq}.

\section{
The Kerr-Newman Action}%
Second, we implement
the probe counterpart of NJA,
in the version of either
\rcite{njmagic.1} (twistorial, top-down)
or
\rcite{probe-nj} (relatively bottom-up),
to derive the exact effective action of the Kerr-Newman black hole
in the presence of external gravitational and electromagnetic fields.

The gravitational part has been already provided
by \rcite{njmagic.1}:
the Kerr two-twistor particle.
In this chapter, we simply need to add the electromagnetic interaction action.
Let $A \in \Omega^1(\M)$ be the electromagnetic gauge potential,
and let $F = dA \in Z^2(\M) \subset \Omega^2(\M)$ now denote the electromagnetic field strength.

\begin{figure}[t]
	\centering
	\adjustbox{valign=c}{\begin{tikzpicture}
	    \node[empty] (O) at (0,0) {};
	    \node[empty] (X) at (3.75, 0) {};
	    \node[empty] (x) at (3.0, 0) {};
	    \node[empty] (Y) at (0, -0.85) {};
	    \node[w] (a00) at ($(O)$) {$A$};
	    \node[w] (a01) at ($(O)+1*(x)$) {$0$};
	    \node[w] (a10) at ($(O)+1*(Y)$) {$\i_NF$};
	    \node[w] (a11) at ($(O)+1*(Y)+1*(x)$) {$0$};
	    \node[w] (a20) at ($(O)+2*(Y)$) {$\i_ND\mem \i_NF$};
	    \node[w] (a21) at ($(O)+2*(Y)+1*(x)$) {$0$};
	    \node[w] (a30) at ($(O)+3*(Y)$) {$\vdots$};
	    \node[w] (a2K) at ($(O)+2*(Y)-1*(X)$) {$\i_N {\ast F}$};
	    \node[w] (a3K) at ($(O)+3*(Y)-1*(X)$) {$\i_ND\mem \i_N {\ast F}$};
	    \node[w] (a4K) at ($(O)+4*(Y)-1*(X)$) {$\vdots$};
	    \node[w] (a2k) at ($(O)+2*(Y)-1*(X)+(x)$) {$0$};
	    \node[w] (a3k) at ($(O)+3*(Y)-1*(X)+(x)$) {$0$};
	    \node[w] (phantom-a00) at ($(O)$) {$\phantom{\big|}$};
	    \node[w] (phantom-a01) at ($(O)+1*(x)$) {};
	    \node[w] (phantom-a10) at ($(O)+1*(Y)$) {};
	    \node[w] (phantom-a11) at ($(O)+1*(Y)+1*(x)$) {};
	    \node[w] (phantom-a20) at ($(O)+2*(Y)$) {};
	    \node[w] (phantom-a21) at ($(O)+2*(Y)+1*(x)$) {};
	    \node[w] (phantom-a30) at ($(O)+3*(Y)$) {};
	    \node[w] (phantom-a31) at ($(O)+3*(Y)+1*(x)$) {};
	    \node[w] (phantom-a2K) at ($(O)+2*(Y)-1*(X)$) {};
	    \node[w] (phantom-a3K) at ($(O)+3*(Y)-1*(X)$) {};
	    \node[w] (phantom-a4K) at ($(O)+4*(Y)-1*(X)$) {};
	    \node[w] (phantom-a2k) at ($(O)+2*(Y)-1*(X)+(x)$) {};
	    \node[w] (phantom-a3k) at ($(O)+3*(Y)-1*(X)+(x)$) {};
	    \draw[->] (a00)--(a01) node[midway,above] {\scriptsize \smash{${d}\mem\i_N$}\vphantom{d}};
	    \draw[->] (a10)--(a11) node[] {};
	    \draw[->] (a20)--(a21) node[] {};
	    \draw[->] (phantom-a00)--(phantom-a10) node[midway,left] {\scriptsize $\i_N{d}$};
	    \draw[->] (phantom-a10)--(phantom-a20) node[] {};
	    \draw[->] (phantom-a20)--(phantom-a30) node[] {};
	    \draw[<->] (phantom-a10)--(a2K) node[midway,above] {\adjustbox{raise=1.5pt}{\scriptsize dual}};
	    \draw[->] (phantom-a2K)--(phantom-a3K) node[] {};
	    \draw[->] (phantom-a3K)--(phantom-a4K) node[] {};
	    \draw[->] (a2K)--(a2k) node[] {};
	    \draw[->] (a3K)--(a3k) node[] {};
	\end{tikzpicture}}
	\caption{%
		The ``$\pounds_N$ sequence'' for Einstein-Maxwell geometry.
	}
	\label{liecd-A}
\end{figure}

The ``$\pounds_N$ sequence'' for $A$ is shown in \fref{liecd-A}.
The recipe of \rrcite{njmagic.1,probe-nj} applies Hodge duality $\ell$ times on the $2^\ell$-pole moment term in this sequence.
This determines the electromagnetic interaction symplectic potential.
Consequently, the full symplectic potential
of the Kerr-Newman particle
is
\begin{align}
	\label{NJA}
	\theta_\text{KN}
	\,=\,
		&
	   	\BB{
	   		\cos(\pounds_N)
	   		+
	   		\sinc(\pounds_N)\,
	   		J\mem \i_N\hem d
	   	}\bigbig{
	   		p_\m\hhem dx^\m
	   	}
	   	\\
		&
		+
	   	\BB{
	   		\cos(\pounds_N)
	   		+
	   		\sinc(\pounds_N)\,
	   		\i_N\mem {*}\mem d
	   	}\bigbig{
	   		A_\m(x)\mem dx^\m
	   	}
	   	\,.
\nonumber
\end{align}
For simplicity, we have set the electric charge $q$ to a unit value;
to reinstate it, rescale $A \mapsto qA$.
We are now working in 
the massive correspondence space $\mathcal{K}$
defined in \rcite{njmagic.1}.
Due to limited space, we 
leave the details to \rcite{njmagic.1}.
Recall that $\mathcal{K}$ is equipped with the geometric data
$N \in \Gamma(T\mathcal{K})$ and $J : T^*\mathcal{K} \to T^*\mathcal{K}$,
where $J^2 = -\mathrm{id}$.

\eqref{PQseq} facilitates the explicit evaluation of \eqref{NJA}:
\begin{align}
\label{KN}
	\theta_\text{KN}
    	\mem=\,
    	{}&{}
    		\theta_{(0)}
   		   	 	+ A_\m(x)\mem dx^\m
	\\
\nonumber
    	 	{}&{}
    	 	+\mem\hhem \sum_{\ell=2}^\infty
    	 		\frac{1}{\ell!}
    			\sum_{p=1}^{\lfloor{\ell/2}\rfloor}\kern-0.2em
    			\sum_{\a \in \Omega_p(\ell)}\kern-0.2em
    	 			\mathscr{K}_\a\mem
    	 		\,p_\m
    	 		\lrp{
    	 		\begin{aligned}[c]
    	 			&
				\bigbig{\hnem
					({*}^\ell Q_{\a_1}\hnem)\mem Q_{\a_2}\, {\nem\cdots\mem} Q_{\a_p}
				\hnem}
				{}^\m{}_\s\, dx^\s
				\\
				&
				+ (\a_p\mminus2)\mem
				\bigbig{\hnem
					({*}^\ell Q_{\a_1}\hnem)\mem Q_{\a_2}\, {\nem\cdots\mem} Q_{\a_{p-1}} Q_{\a_p - 1}
				\hnem}
				{}^\m{}_\s\, Dy^\s
			\hnem
    	 		\end{aligned}
                }
    	 \\
\nonumber
   	 	{}&{}
		+\mem\hhem
			\sum_{\ell=1}^\infty
				\frac{1}{\ell!}\,
		    	 		\bb{\hnem
						\bigbig{\hnem
							{*}^\ell P_\ell
						\hnem}
						{}_\m\, dx^\m
						+
						(\ell\mminus1)\mem
						\bigbig{\hnem
							{*}^\ell P_{\ell-1}
						\hnem}
						{}_\m\, Dy^\m
		    	 		}
    	\\
\nonumber
    	 	{}&{}
    	 	+\mem\hhem \sum_{\ell=2}^\infty
    	 		\frac{1}{\ell!}
		   \sum_{p=2}^{\lfloor{(\ell+1)/2}\rfloor}\kern-0.4em
		   \sum_{\a \in \Omega_p\hnem(\ell+1)}\kern-0.6em
		       \mathscr{K}_\a
		   \lrp{
		   \begin{aligned}[c]
		   &
		       \bigbig{\hnem
		           ({*}^\ell P_{\a_1-1})\mem
		           Q_{\a_2}\, {\nem\cdots\mem} Q_{\a_p}
		       \hnem}
		       _{\nem\m}
		       \, dx^\m
		   \\
		   &
		       +
		       (\a_p\mminus2)
		       \bigbig{\hnem
		           ({*}^\ell P_{\a_1-1})\mem
		           Q_{\a_2}\, {\nem\cdots\mem} Q_{\a_{p-1}} Q_{\a_p - 1}
		       \hnem}
		       _{\nem\m}
		       \, Dy^\m
		   \end{aligned}
		   \hnem}
 	 \,,
\end{align}
Here, ${*}^\ell P_j$ and ${*}^\ell Q_j$
means to act on the Hodge star $\ell$ times
on the electromagnetic field strength $F$ and the Riemann tensor $R$
inside $P_j$ and $Q_j$.

Again, $\theta_{(0)}$ is defined in \rcite{njmagic.1}:
the covariantization of the free-theory symplectic potential
of the two-twistor particle.
For the reader's sake, we reproduce it below:
\begin{align}
\label{OSD11:theta0}
	\theta_{(0)}
	\,=\,
	p_\m\hhem dx^\m
	+ i\mem y^{\da\a}\mem
	\BB{
		\lambda_\a{}^I\hem D\rambda_{I\da}
		{\,-\,}
		D\lambda_\a{}^I\mem \rambda_{I\da}
	}
	\,.
\end{align}

By incorporating the first-class constraints $\phi_0 = \frac{1}{2}\mem (p^2 {\,+\,} m^2)$ and $\phi_1 = -p\mdot y$
via Lagrange multipliers,
\eqref{KN}
concretely defines an all-orders first-order worldline action 
in which each term 
is fully and explicitly evaluated
as a manifestly covariant worldline operator.

It should be clear that
\eqref{KN} nicely reduces to the Kerr and {\kerr} symplectic potentials
in \rcite{njmagic.1}.
Also, it is straightforward that
the nonabelian Kerr-Newman particle
is obtained from \eqref{KN} by simple replacements
$A_\m(x)\mem dx^\m \mapsto i\mem \bpsi_i\mem D\psi^i$
and
$(P_\ell)_\s \mapsto q_a\hem (P_\ell)^a{}_\s$
(see the appendix of \rcite{njmagic.1} for the relevant setup).

\section{
Nonlinear Newman-Janis Shift}%
Third, we impose SDity condition on both
electromagnetic and Riemannian curvatures
to reveal the dynamical, nonlinear NJ shift:
${*}F_{\m\n} = i\hem F_{\m\n}$,
${*}R^\m{}_{\n\r\s} = i\hem R^\m{}_{\n\r\s}$.

To this end, we simply replace ${*}^\ell$ with $i^\ell$ in \eqref{KN}.
Recalling the explanations given in \rcite{njmagic.1},
it is seen that this replacement results in
\begin{align}
	\label{NJA.exp}
	\theta_\text{KN}
	\,\sim\,
		\mathe^{i\pounds_N}\BB{
			p_\m\hhem dx^\m
			\hem+\mem\hhem A_\m(x)\mem dx^\m
			\hem+\mem\hhem \theta_\ga
		}
   	\,,
\end{align}
where $\sim$ signifies equivalence modulo total derivative.
Specifically,
we have discarded
$i\mem \sinc(\pounds_N)\mem d(A_\m(x)\mem y^\m)$.

\eqref{NJA.exp}
is nothing but 
a minimally coupled charged scalar particle---%
i.e.,
the Reissner-Nordstr\"om particle---%
geodesically deviated into complex spacetime,
up to the ``Gilbert-to-Amp\`ere'' replacement \cite{njmagic.1}
implemented by $\theta_\ga$
which reflects a kinematical detail:
\begin{align}
	\label{g2a}
	\theta_\ga
	\,=\,
		i\mem y^{\da\a}\mem
			\BB{
				\lambda_\a{}^I\hem D\trambda_{I\da}
				{\,-\,}
				D\lambda_\a{}^I\mem \trambda_{I\da}
			}
		- i\mem p_\m Dy^\m
	\,.
\end{align}

Clearly, this incarnates the spirit of the NJA
\cite{Newman:1965tw-janis}
and its very historical application
\cite{Newman:1965my-kerrmetric}
to the Reissner-Nordstr\"om solution
for the derivation of the Kerr-Newman solution.

The exponentiated Lie derivative $\mathe^{i\pounds_N}$ in \eqref{NJA.exp}
describes the map $\gamma$ defined in \rcite{njmagic.1}
that brings us to the ``primed chart''
$(z^{\m\sprime},y^{\m\sprime},\lambda_{\a'}{}^I)$
on the (complexified) massive correspondence space $\mathcal{K}$.
With this understanding, \eqref{NJA.exp} boils down to
\begin{align}
	\label{KN+}
	\theta_\text{KN}^+
	\,=\,
		p_{\m'}\hhem dz^{\m\sprime}
		- 2i\mem \trambda_{I\da'}\mem y^{\da\sprime\a\sprime}\mem
			D\lambda_{\a'}{}^I
		+ A_{\m'}(z)\mem dz^{\m\sprime}
   	\,,
\end{align}
where we have discarded a total derivative
just as in \rcite{njmagic.1}.
Here, $z^{\m\sprime} = \delta^{\m\sprime}{}_\m\mem (
	x^\m + iy^\m + \tfrac{1}{2}\mem \Gamma^\m{}_{\r\s}(x)\mem y^\r y^\s + \O(y^3)
)$
are the holomorphic coordinates
of curved spinspacetime.

Reviving the analysis in \rcite{probe-nj},
one sees that the worldline perturbation theory due to \eqref{KN+}
manifests the same-helicity (positive-helicity, in particular) spin exponentiation
of the graviphotonic Compton amplitudes
at all multiplicities
and any massless species.

\section{
Heavenly Equations of Motion}%
The classical EoM due to
the symplectic potential in \eqref{KN+}
and the first-class constraints $\phi_0$ and $\phi_1$
are easy to derive
by the technique of covariant symplectic perturbations
\cite{csg}.
By fixing their Lagrange multipliers respectively as
$\k^0(\t) = 1$ and $\k^1(\t) = 0$,
we find
\begin{align}
\begin{split}
	\label{heom}
	\dot{z}^{\m\sprime} 
		\mem=\mem p^{\m\sprime}
	&\,,\,\,\,
	\frac{Dy^{\m\sprime}}{d\t}
		\mem=\mem 
			F^{\m\sprime}{}_{\n'\hnem}(z)\, y^{\n\sprime}
	\,,\\
	\frac{D\lambda_{\a'}{}^I}{d\t}
		\mem=\mem 
			0
	&\,,\,\,\,
	\frac{D\trambda_{I\a\sprime}}{d\t}
		\mem=\mem \trambda_{I\b\sprime}\,
			\tF^{\db\sprime}{}_{\da'\hnem}(z)
	\,.
\end{split}
\end{align}
This shows that
the motion of
our Kerr-Newman particle
describes
a complex geodesic worldline $z^{\m\sprime}$
in spinspacetime
on which the left-handed spin frame
$\lambda_{\a'}{}^I$
is covariantly constant
(parallel-transported),
while the right-handed spin frame $\trambda_{I\da'}$
and the spin length pseudovector $y^{\m\sprime}$
are transported according to the Lorentz force law
due to the dynamically NJ shifted field strength.

Based on the fact that
the left-handed spinor bundle is flat for the moment,
$D\lambda_{\a'}{}^I\nem/d\t = 0$
explicitly verifies 
the spinning, complexified analog of equivalence principle,
which was proposed by \rcite{sst-asym}
as the property characterizing spinning black holes.
Namely,
there is no ASD component in the spin precession,
in SD backgrounds.

When re-covariantized at $x^\m$ in spacetime
(i.e.,
transcribed to the defining chart of $\mathcal{K}$),
\eqref{heom}
yields
an all-orders extension of the
Bargmann-Michel-Telegdi (BMT) \cite{Bargmann:1959gz}
\nomenclature{BMT}{Bargmann-Michel-Telegdi}
and 
MPD 
\cite{Mathisson:1937zz,Papapetrou:1951pa,Dixon:1970zza}
equations
with all the multipole coefficients, both electromagnetic and gravitational, set to unity.
This is indeed the desired behavior for the Kerr-Newman particle.
Crucial point here is that
the unity of the gravitational multipole coefficients
is achieved because we have utilized geodesic deviation
in generating the spin-curvature couplings in the action.

Similarly,
it is also easily seen that the holomorphic coordinates $z^{\m\sprime}$ of 
the Kerr black hole's
curved spinspacetime are Poisson-commutative
in this SD limit:
\begin{align}
	\label{OSD11:zz=0}
	\pb{\hem z^{\m\sprime}\hem}{z^{\n'}} \,=\, 0
	\,.
\end{align}

\section{
Hidden Symmetry}%
Fourth, let us suppose
the presence of
a Killing vector $X \in \Gamma(T\M)$
as well as
a Killing-Yano tensor $Y \in \Omega^2(\M)$.
These shall satisfy 
\begin{align}
	\label{KYT}
	\pounds_X g = 0
	\,,\quad
	\pounds_X F = 0
	\,,\quad
	Y^\m{}_\r\hem F^\r{}_\n
	\,=\,
		F^\m{}_\r\hem Y^\r{}_\n
	\,,
\end{align}
and $Y_{\m\n;\r} {\:=\:} Y_{[\m\n;\r]}$.
The last condition
describes a commuting property,
which holds
for the Kerr-Newman background
by the type-D property \cite{Hughston:1972qf}
and also for
its SD sector:
the SD charged Taub-Newman-Unti-Tamburino (NUT) solution
in which $F^\m{}_\n \propto Y^+{}^\m{}_\n / |\vex|^3$.
See \rcite{nja} and also \rcite{Adamo:2023fbj};
recall \Chap{K3:HYDROGEN}.

As is well-known \cite{Hughston:1972qf},
the Reissner-Nordstr\"om particle,
governed by the EoM
$\dot{x}^\m {\:=\:} p^\m$, $Dp^\m\nem/d\t = F^\m{}_\n(x)\mem p^\n$,
enjoys
the conserved quantities
\begin{subequations}
    \label{charges0}
\begin{align}
	\label{charge0.Q}
	\mathds{Q} \,&=\,
		p_\m\hem X^\m(x)
		+ \a(x)
	\,,\\
	\label{charge0.C}
	\mathds{C} \,&=\,
		- p_\m\mem Y^\m{}_\r(x)\mem Y^\r{}_\n(x)\, p^\n
	\,.
\end{align}
\end{subequations}
In \eqref{charge0.Q},
$\a \in \Cinfty(\M)$ is a scalar field such that
$d\a = -\i_X F$ \cite{Hughston:1972qf},
whose local existence is guaranteed
because $-d\i_X F = \i_X dF = 0$.
Concretely, one can take $\a = \i_X A$
if one is willing to make an explicit gauge choice such that $\pounds_X A = 0$.
To derive \eqref{charge0.C},
one notes that
the covariant precession of the vector $Y^\m{}_\n(x)\mem p^\n$
is the same as $p^\m$
by virtue of the commuting property in \eqref{KYT}.

\rcite{probe-nj}
has established the bold proposal that
exact (all-orders in spin and coupling) conserved charges
of the spinning black hole probe
directly arise by
dynamically NJ shifting
the conserved quantities of the non-spinning black hole probe,
in the SD sector.
By appreciating this idea,
we identify
the following exact conserved charges
for our Kerr-Newman probe:
\begin{subequations}
\label{charges}
\begin{align}
	\label{charge.Q}
	\mathds{Q} \,&=\,
		p_{\m'} X^{\m\sprime}(z)
		+ \a(z)
	\,,\\
	\label{charge.R}
	\mathds{R} \,&=\,
		p_{\m'} Y^{\m\sprime}{}_{\n'\hnem}(z)\mem y^{\n\sprime}
	\,,\\
	\label{charge.C}
	\mathds{C} \,&=\,
		- p_{\m'} Y^{\m\sprime}{}_{\r'\hnem}(z)\mem Y^{\r\sprime}{}_{\n'\hnem}(z)\mem \mem p^{\n\sprime}
	\,.
\end{align}
\end{subequations}
In the explicit gauge choice for the electromagnetic gauge potential,
one finds $\a(z) {\:=\:} A_{\m'\hnem}(z)\mem X^{\m\sprime}(z)$.
It is easy to verify that $\mathds{Q}$, $\mathds{R}$, $\mathds{C}$ in \eqref{charges}
are all conserved by the EoM in \eqref{heom}.
Note that $Dp^{\m\sprime}\nem/d\t = F^{\m\sprime}{}_{\n'\hnem}(z)\mem p^{\n\sprime}$.
The covariant precession behaviors 
exhibited by
$p^{\m\sprime}$,
$y^{\m\sprime}$,
and
$Y^{\m\sprime}{}_{\n'\hnem}(z)\mem p^{\n\sprime}$
are all identical.

The conserved charges in \eqref{charges}
are sufficient to establish
\textit{integrability} of the Kerr-Newman probe dynamics
in SD Einstein-Maxwell backgrounds
admitting at least two Killing vectors and a Killing-Yano tensor.
In \rcite{probe-nj},
such integrability
has been identified as
the property that characterizes black holes
among all possible massive spinning objects.
While \rcite{probe-nj}
has substantiated this proposal
by the Kerr particle,
this work reports the same conclusion
for the Kerr-Newman particle.

In fact, one can establish \textit{superintegrability}
by incorporating the ASD Pleba\`nski two-forms
\cite{Plebanski:1977zz},
just as in \rcite{probe-nj}:
\begin{align}
	\label{charge.K}
	\mathds{K}_a \,&=\,
		p_{\m'} (\Sigma_a)^{\m\sprime}{}_{\r'\hnem}(z)\mem Y^{\r\sprime}{}_{\n'\hnem}(z)\mem \mem p^{\n\sprime}
	\,.
\end{align}
It should also be highlighted that
the symmetry algebra 
formed by $\mathds{Q}$, $\mathds{C}$, and $\mathds{K}_a$
is \textit{identical}
to the non-spinning case
up to complexification,
by virtue of the very important property 
that $\pb{z^{\m\sprime}}{z^{\n\sprime}} = 0$
(\eqref{OSD11:zz=0})
which is characteristic of black hole spinspacetime.

In short, we identify superintegrability
in the dynamics of the Kerr-Newman black hole binary system
at the zeroth self-force order,
yet in the SD sector.

\section{
Curved Massive Twistor Space}%
Fifth,
we identify the ``mu variables''
to return to the massive twistor space.

To this end, we trivialize the ASD spin connection coefficients,
employ tetrad formalism,
and expand the holomorphic coframe 
around a flat background as
$e^{\da\sprime\a} = dz^{\da\sprime\a} + h^{\da\sprime\a}$
which may utilize the second Pleba\'nski coordinates \cite{plebanski1975some}.
Then \eqref{KN+} becomes
\begin{align}
	\label{NJA'}
	\theta_\text{KN}^+
	\,=\,
		- \trambda_{I\da'}\hem dz^{\da\sprime\a}\hem \lambda_\a{}^I
		- 2i\mem \trambda_{I\da'}\mem y^{\da\sprime\a}\mem
			d\lambda_\a{}^I
		+ \theta'_\text{KN}
   	\,.
\end{align}
We have discarded a total derivative, while
\begin{align}
	\theta'_\text{KN}
	\,=\,
		\BB{
			- \trambda_{I\da'}\hem h^{\da\sprime\a}{}_{\m'\hnem}(z)\hem \lambda_\a{}^I
			+ A_{\m'\hnem}(z)
		}\,
			dz^{\m\sprime}
	\,.
\end{align}
As a result, the same identification for the mu variables 
as in \rcite{njmagic.1}
are made:
\begin{align}
\begin{split}
	\label{muvariables}
	\mu^{\da\sprime\mem I}
	\,=\,
		z^{\da\sprime\a}\hem\hhem \lambda_\a{}^I
	\,,\quad
	\tmu_I{}^\a
	\,&=\,
		\trambda_{I\da'}
		\bigbig{
			z^{\da\sprime\a} {\hem-\mem} 2iy^{\da\sprime\a}
		}
	\,.
\end{split}
\end{align}
Eventually, we find
\begin{align}
\begin{split}
	\label{twistor.h}
	\theta^+_\text{KN}
	\,&=\,
	- \trambda_{I\da'}\hem d\mu^{\da\sprime\mem I}
	+ \tmu_I{}^\a\hem d\lambda_\a{}^I
	+ \theta'_\text{KN}
	\,,\\
	\omega_\text{KN}
	\,&=\,
	d\mu^{\da\sprime\mem I} \swedge d\trambda_{I\da'}
	+ d\tmu_I{}^\a \swedge d\lambda_\a{}^I
	+ d\theta'_\text{KN}
	\,.
\end{split}
\end{align}
\eqref{twistor.h} identifies the notion of curved massive twistor space
in terms of
a deformation of the symplectic structure on massive twistor space:
\begin{align}
	\bigbig{
		\mathbb{MT}
		,\mem
		\omega^\circ
	}
	\,\,\,\xrightarrow{\,\,\,\,\,\,}\,\,\,
	\bigbig{
		\MT
		,\mem
		\omega^\circ \mplus\mem\hhem d\theta'_\text{KN}
	}
	\,.
\end{align}
This deformation is via a $(1,0)$-perturbation $\theta'_\text{KN}$
at the level of symplectic potential
such that the Poisson commutativity of holomorphic spinspacetime coordinates \cite{sst-asym} is preserved as \eqref{OSD11:zz=0}.
Physically,
this encodes that
the gravitoelectromagnetic interaction action
is supported on the NJ shifted worldline $z^{\m\sprime}$
for the SD (incoming positive-helicity \cite{bialynicki1981note,ashtekar1986note}) sector.

As remarked in \rcite{njmagic.1},
the above recipes are motivated by the practical use of our framework
in perturbation theory,
so other pathways toward the definition of curved massive twistor space
might exist.

\section{
Googly Formulation}%
Lastly,
our final task is to derive the chiral (``googly'') formulation of the Kerr-Newman action.
This computes the generic, non-SD action
described in \eqref{KN}
in terms of the complex, primed variables
$(z^{\m\sprime},y^{\m\sprime},\lambda_{\a'}{}^I)$
by considering the map $\gamma$
of \rcite{njmagic.1}.

To this end,
we revive the gymnastics in \rcite{njmagic.1}.
Define
\begin{align}
\label{strings}
	&
	\theta^\pm_1
	\,=\,
		\frac{1}{2}\,
		\BB{
			p_\m Dy^\m
			\pm
				i\mem y^{\da\a}\mem
				\BB{
					\lambda_\a{}^I\hem D\trambda_{I\da}
					{\,-\,}
					D\lambda_\a{}^I\mem \trambda_{I\da}
				}
		\nem}
	+
		\i_N F^\pm
	\,,
\end{align}
where the $\pm$ superscripts describe
the projections by $\tfrac{1}{2}\mem (1 \mp i\mem {*})$.
Then it follows that
\begin{subequations}
    \label{wsforms}
\begin{align}
\label{wsform.KN}
&
	\theta_\text{KN}
	\,\,=\,\,
		p_\m\hhem dx^\m + A_\m(x)\mem dx^\m
	+
		\frac{
			\mathe^{i\pounds_N} \mminus 1
		}{\pounds_N}
		\,\theta^+_1
	+
		\frac{
			\mathe^{-i\pounds_N} \mminus 1
		}{\pounds_N}
		\,\theta^-_1
	\,,\\
\label{wsform.RN+ia}
&
	\mathe^{i\pounds_N}\BB{
		p_\m\hhem dx^\m + A_\m(x)\mem dx^\m\hnem
	}
	\,\,\sim\,\,
		p_\m\hhem dx^\m + A_\m(x)\mem dx^\m
	+
		\frac{
			\mathe^{i\pounds_N} \mminus 1
		}{\pounds_N}
		\,\theta^+_1
	+
		\frac{
			\mathe^{i\pounds_N} \mminus 1
		}{\pounds_N}
		\,\theta^-_1
	\,.
\end{align}
\end{subequations}
The difference between \eqrefs{wsform.KN}{wsform.RN+ia}---%
Kerr-Newman and ``Reissner-Nordstr\"om${}+{} iy$''---%
derives
\begin{align}
\begin{split}
	\theta_\text{KN}
	\,\mem&\sim\,\,
	\mathe^{i\pounds_N}\mem
	\bb{
		p_\m\hhem dx^\m
		+ 
		\frac{
			\mathe^{-2i\pounds_N} \mminus 1
		}{\pounds_N}
		\,\theta^-_1
	}
	\,,\\
	\,\,&=\,\,
		\mathe^{i\pounds_N}\mem
		\lrp{
        \begin{aligned}[c]
        &
			\theta_{(0)}
			- i\mem p_\m Dy^\m
        \\&
			+\mem\hhem
				\sum_{\ell=2}^\infty
					\frac{(-2i)^\ell}{\ell!}\mem
						p_\m
						\bigbig{\hnem
							(\i_N\hem D)^{\ell-2} \i_N\mem R^-{}^\m{}_\n
						\hhnem}\hem y^\n
        \\&
			+\mem\hhem
				\sum_{\ell=1}^\infty
					\frac{(-2i)^\ell}{\ell!}\mem
						(\i_N\hem d)^{\ell-1} \i_N\mem F^-
        \end{aligned}
		}
	\,,
\end{split}
\end{align}
which leads to
\begin{align}
\label{KN.googly}
	\theta_\text{KN}
    	\mem\,\sim\,\,
    	{}&{}
   			p_{\m'}\hhem dz^{\m\sprime}
			- 2i\mem \trambda_{I\da'}\mem y^{\da\sprime\a\sprime}\mem
				D\lambda_{\a'}{}^I
	   	 	+ A_{\m'\hnem}(z)\mem dz^{\m\sprime}
	\\
    	 	{}&{}
    	 	+\mem\hhem \sum_{\ell=2}^\infty
    	 		\frac{(-2i)^\ell}{\ell!}
    			\sum_{p=1}^{\lfloor{\ell/2}\rfloor}\kern-0.2em
    			\sum_{\a \in \Omega_p(\ell)}\kern-0.2em
    	 			\mathscr{K}_\a\mem
   	 		\,p_{\m\sprime}\hem
    	 		\lrp{
    	 		\begin{aligned}[c]
    	 			&
				\bigbig{\hnem
					Q_{\a_1}^-\hem Q_{\a_2}\, {\nem\cdots\mem} Q_{\a_p}
				\hnem}
				{}^{\m\sprime}{}_{\s'}\mem dz^{\s\sprime}
				\\
				&
				+ (\a_p\mminus2)\mem
				\bigbig{\hnem
					Q_{\a_1}^-\hem Q_{\a_2}\, {\nem\cdots\mem} Q_{\a_{p-1}} Q_{\a_p - 1}
				\hnem}
				{}^{\m\sprime}{}_{\s'}\hhem Dy^{\s\sprime}\mem
			\hnem
    	 		\end{aligned}}
\nonumber
    	 \\
   	 	{}&{}
		+\mem\hhem
			\sum_{\ell=1}^\infty
				\frac{(-2i)^\ell}{\ell!}\,
		    	 		\bb{\hnem
						\bigbig{\hnem
							P^-_\ell
						\hnem}
						{}_{\s'}\mem dz^{\s\sprime}
						+
						(\ell\mminus1)\mem
						\bigbig{\hnem
							P^-_{\ell-1}
						\hnem}
						{}_{\s'}\hhem Dy^{\s\sprime}
		    	 		}
\nonumber
    	\\
    	 	{}&{}
    	 	+\mem\hhem \sum_{\ell=2}^\infty
    	 		\frac{(-2i)^\ell}{\ell!}
		   \sum_{p=2}^{\lfloor{(\ell+1)/2}\rfloor}\kern-0.4em
		   \sum_{\a \in \Omega_p\hnem(\ell+1)}\kern-0.6em
		       \mathscr{K}_\a
		   \lrp{
		   \begin{aligned}[c]
		   &
		       \bigbig{\hnem
		           P^-_{\a_1-1}\mem
		           Q_{\a_2}\, {\nem\cdots\mem} Q_{\a_p}
		       \hnem}
		       _{\nem\m'}
		       \mem dz^{\m\sprime}
\nonumber
		   \\
		   &
		       +
		       (\a_p\mminus2)
		       \bigbig{\hnem
		           P^-_{\a_1-1}\mem
		           Q_{\a_2}\, {\nem\cdots\mem} Q_{\a_{p-1}} Q_{\a_p - 1}
		       \hnem}
		       _{\nem\m'}
		       \hhem Dy^{\m\sprime}
		   \end{aligned}
		   }
 	 \,.
\nonumber
\end{align}
Again, we find the structure that
the first entry in each string of $P$ or $Q$ tensors
develops the ASD projection.
\eqref{KN.googly} reproduces \eqref{KN+}
in the SD limit.
As a friendly reminder, our index notation \cite{gde} implies
\begin{subequations}
\begin{align}
   \label{P,Q(z)}
     (P_\ell)_{\m'}
     &\,=\,
         F_{\r\sprime_1\m\sprime\hem;\r\sprime_2;\cdots;\r\sprime_\ell}(z)\,
         y^{\r\sprime_1} y^{\r\sprime_2} {\hem\cdots\mem} y^{\r\sprime_\ell}
     \,,\\
	(Q_\ell)^{\m\sprime}{}_{\s'}
	&\,=\,
		R^\m{}_{\r\sprime_1\r\sprime_2\s\sprime\mem;\r\sprime_3;\cdots;\r\sprime_\ell}\hnem(z)\,
		y^{\r\sprime_1} y^{\r\sprime_2} y^{\r\sprime_3} {\hem\cdots\mem} y^{\r\sprime_\ell}
	\,.
\end{align}
\end{subequations}

\eqref{KN.googly} provides
the googly formulation of the Kerr-Newman action.
To reiterate,
it describes the completely general case of
non-SD
gravitoelectromagnetic field configurations
but formulates the SD and ASD modes
on an unequal footing.
It provides a worldline action localized on the holomorphic worldline $z^{\m\sprime}$,
the perturbation theory of which
manifests the same-helicity spin exponentiation
for incoming positive-helicity photons and gravitons.
The photon coupling is always linear.
Graviphoton contact vertices arise iff the incoming photon carries negative helicity.
When examined in the spin length expansion,
the first graviphoton contact vertex
appears at $\O(y^3)$.

\section{
Interpretation}%
We end with quick remarks on the physical interpretation
by recalling the comments given in \rcite{njmagic.1}.
\eqref{KN} represents two charged masses
joined by 
a Dirac-Misner string
(line defect of magnetic and NUT fluxes).
\eqref{KN.googly} represents 
a dangling ASD Dirac-Misner string
attached to
a charged mass.
The shape of the Dirac-Misner string
happens to be geodetic
for each slice of simultaneity defined by the worldline proper time.
For the Dirac string part,
this is completely justified
based on its well-established status
as a topological surface operator.

By virtue of the transparent decoding of the NJA
revealed by \rcite{nja},
we readily realize that
this string structure
is the precise IR counterpart
of the actual Dirac-Misner string
that exists in the UV description of the Kerr-Newman solution.\footnote{
	Here,
	IR refers to the point-particle effective theory,
	while
	UV refers to GR.
}
Namely, it is shown in \rcite{nja}
that the Kerr-Newman solution
as a holomorphic saddle in Einstein-Maxwell theory
represents a pair of SD and ASD
charged Taub-NUT instantons.
In conclusion,
the holomorphic worldline $z^{\m\sprime}$
is the IR avatar of the worldline of the ASD charged Taub-NUT instanton
that constitutes the Kerr-Newman black hole.

The googly action in \eqref{KN.googly}
perturbs around the SD sector---%
the sector 
in which we, at least, have a full understanding on the dynamics
in terms of remarkable properties
such as the superintegrability due to \eqrefs{charges}{charge.K}
or the same-helicity spin exponentiation.
The googly agenda means to approach the complicated dynamics of the Kerr-Newman probe
from its exactly solvable part,
which should sound reasonable.

Note that \rcite{gmoov} had speculated on a worldsheet structure,
although no clear physical interpretation
nor explicit formulae such as \eqrefs{KN}{KN.googly}
were provided.

\begin{figure}[t]
\centering
	\includegraphics[width=1.0\linewidth]{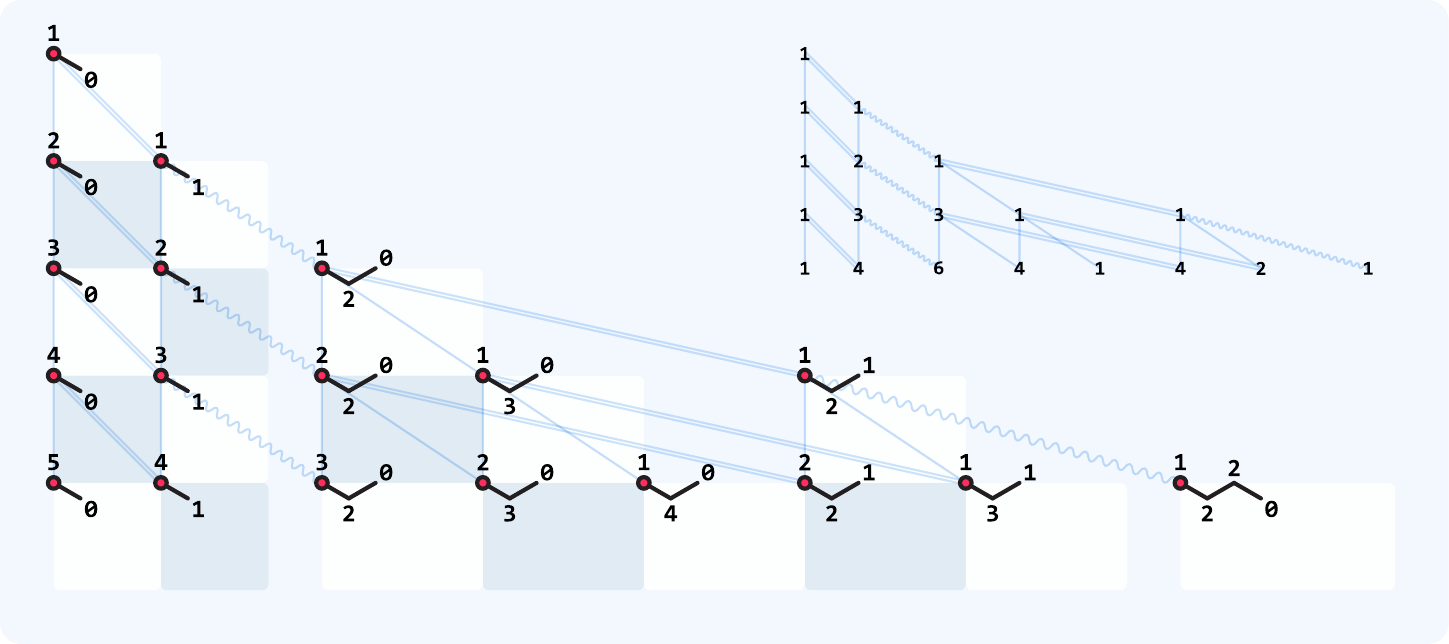}
    \medskip
	\caption[%
		The ``organic chemistry'' of Einstein-Maxwell geometry.
		Molecules at each row follow from those on the former row by the action of $\i_ND$.
		Single lines represent the reaction
		$(P_{\a_1}Q_{\a_2}{\cdots\mem}Q_{\a_p})\mem dx \mapsto \sum_{\b_1+\cdots+\b_p = 1} (P_{\a_1+\b_1} Q_{\a_2+\b_2}{\cdots\mem}Q_{\a_p+\b_p})\mem dx$
		with $\b_1,{\cdots},\b_p {\:\geq\:} 0$.
		Double lines represent the reaction
		$(PQ{\mem\cdots\mem}Q)\mem dx \mapsto (PQ{\mem\cdots\mem}Q)\mem Dy$.
		Squiggly lines represent the ``polymerization'' reaction
		$(PQ{\mem\cdots\mem}Q)$ $Dy \mapsto (PQ{\mem\cdots\mem}Q)\mem (Q_2\hem dx)$.
		See \fref{molecules} and also \Chap{K3:GDE}.
		The smaller diagram in the upper right corner
		displays the coefficient of each molecule.
    ]{%
		The ``organic chemistry'' of Einstein-Maxwell geometry.
		Molecules at each row follow from those on the former row by the action of $\i_ND$.
		Single lines represent the reaction
		$(P_{\a_1}Q_{\a_2}{\cdots\mem}Q_{\a_p})\mem dx \mapsto \sum_{\b_1+\cdots+\b_p = 1} (P_{\a_1+\b_1} Q_{\a_2+\b_2}{\cdots\mem}Q_{\a_p+\b_p})\mem dx$
		with $\b_1,{\cdots},\b_p {\:\geq\:} 0$.
		Double lines represent the reaction
		$(PQ{\mem\cdots\mem}Q)\mem dx \mapsto (PQ{\mem\cdots\mem}Q)\mem Dy$.
		Squiggly lines represent the ``polymerization'' reaction
		$(PQ{\mem\cdots\mem}Q)\mem Dy \mapsto (PQ{\mem\cdots\mem}Q)\mem (Q_2\hem dx)$.
		See \fref{molecules} and also \Chap{K3:GDE}.
		The smaller diagram in the upper right corner
		displays the coefficient of each molecule.
	}
	\label{chemistry-A}
\end{figure}

\begin{figure}[t]
\centering
	\includegraphics[scale=0.85]{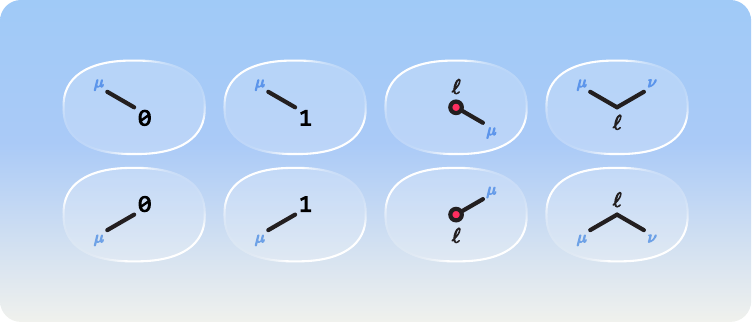}
    \medskip
	\caption{%
		The graphical notation used in \fref{chemistry-A}.
		Elements in the first row denote
		$dx^\m$, $Dy^\m$, $(P_\ell)_\m$, and $(Q_\ell)^{\m\n}$,
		respectively.
		Elements in the second row denote
		$dx_\m$, $Dy_\m$, $(P_\ell)^\m$, and $(Q_\ell)_{\m\n}$,
		respectively.
	}
	\label{molecules}
\end{figure}

\chapter{Spinspacetime}\label{K3:SST}

    We show that 
    Poincar\'e invariance directly implies the existence of
    a complexified Minkowski space whose real and imaginary directions unify spacetime and spin, which we dub spinspacetime.
    Despite the intrinsic noncommutativity of spin,
    spinspacetime exhibits
    mutually commuting
    holomorphic coordinates.

    \noindent
    Its twistorial construction can derive the NJ shift property of spinning black holes by massive half-Fourier transforming complexified on-shell kinematics, which encode a spinning analog of equivalence principle. 
    
\begin{fullnote}
    This section reproduces the contents of
    \rcite{sst-asym}, \fullcite{sst-asym}.
\end{fullnote}

\section{
Introduction}%
In the 70s,
Ezra Ted Newman
investigated
the curious idea of
unifying spacetime and spin into a complex geometry
\cite{newman1974curiosity,newman1974collection,Newman:1973afx,Newman:2004ba,Newman:1973yu,Newman:2002mk,Newman:1976gc,ko1981theory}.
While offering
a unique perspective
on relativistic angular momenta,
it 
provided
a suggestive 
argument for
the NJ \cite{Newman:1965tw-janis,Newman:1965my-kerrmetric} derivation of spinning black hole solutions
where spin turns into an imaginary deviation.
However, it remains uncertain if
the NJ shift really follows solely from this complex geometry,
since Newman's argument implicitly counts on an assumption
that the black hole would be
localized in the complex 
for an unknown reason.
On a related note,
it has also remained unclear
whether this concept is necessarily limited to black holes or not.
Unfortunately,
this line of ideas
appears to be
no longer
actively revived in the current literature.

In this chapter,
we aim to give a dedicated spotlight on this ``forgotten lore''
and unveil its relevance to
the modern program of
describing spinning black holes
in terms of scattering amplitudes
\cite{ahh2017,aho2020,Guevara:2018wpp,Guevara:2019fsj,chkl2019,Johansson:2019dnu,Lazopoulos:2021mna,Aoude:2020onz,Aoude:2022trd,Cangemi:2022bew,Cangemi:2023bpe,gmoov,ambikerr1,Kim:2024grz}.
As the very first step towards this modern resurrection,
we endow it with a name: ``spinspacetime.''
We provide
a complete definition of spinspacetime
that unifies various perspectives
\cite{newman1974curiosity,newman1974collection,Newman:1973afx,Newman:2004ba,Newman:1973yu,Newman:2002mk,Newman:1976gc,ko1981theory,grg207flaherty,almond1973time,casalbuoni1976classical,hughston1979twistors,bette1984pointlike,bette1996directly,bette1997extended,bette2000twistor,bette2004massive04,Bette:2004ip,bette2005twistors-0402150,bette2005two-0503134,lukierski2014noncommutative,Filippas:2022soduality}.
Especially, we highlight its characteristic commutator structure,
which, to our knowledge, was not investigated in Newman's works
\cite{newman1974curiosity,newman1974collection,Newman:1973afx,Newman:2004ba,Newman:1973yu,Newman:2002mk,Newman:1976gc,ko1981theory}.
Consequently,
we refine Newman's ideas
to precise statements.
In particular,
we envision
a new derivation of the NJ shift from scattering amplitudes,
which may trigger
an extension of the on-shell/twistor diagrams
\cite{hodges1980twistor,hodges1990feynman,hodges2005a,hodges2005b,arkani2010s-matrix,arkani2012positivegrassmannian,Atiyah:2017erd}
to massive states.

Our central findings are twofold.
Firstly, we show that spinspacetime \textit{universally} arises in
any massive system
exhibiting global Poincar\'e symmetry:
particles and fields in asymptotically flat spacetimes,
or even systems that may lack a spacetime formulation.
We construct real variables $x^\m$, $y^\m$
directly from the Poincar\'e charges.
Then
Poincar\'e symmetry dictates
their commutators to be
\begin{align}
\begin{split}
    \label{eq:prop.zigzag}
    \comm{z}{z} = 0
    \,,\quad
    \comm{z}{\bz} \neq 0
    \,,\quad
    \comm{\bz}{\bz} = 0
    \,,
\end{split}
\end{align}
if $z^\m = x^\m + iy^\m$
and $\bz^\m = x^\m - iy^\m$.
These complex coordinates unify spacetime and spin as real and imaginary parts,
which can be seen from the following split of the total angular momentum into orbital and spin parts:
\begin{align}
    \label{eq:prop.J}
    J = (x\wedge p) + *(y\wedge p)
    \qiq
    J^+ = (z\wedge p)^+
    \,.
\end{align}
Here,
an intrinsic association between chirality and holomorphy
is implied,
as the SD part of the total angular momentum, $J^+$, arises solely from the holomorphic coordinates $z$
as $* \to +i$.
Therefore, a complexified Minkowski space 
exhibiting a fascinating geometry
arises,
whose complex coordinates $z^\m$, $\bz^\m$
(1) realize a particular ``off-diagonal'' form of commutators as in \eqref{eq:prop.zigzag}
and (2) are inherently linked with chirality
as in \eqref{eq:prop.J}.
This provides the complete characterization of spinspacetime.

Secondly,
we specialize this framework
in the scattering problem of massive spinning states
from asymptotically free regions.
To this end,
we adopt a twistorial implementation of spinspacetime,
in which case the off-diagonal form of \eqref{eq:prop.zigzag}
originates from the oscillator commutator algebra
(K\"ahler geometry)
\cite{Penrose:1968me,penrose:maccallum,penrose1975aims}
of twistor space:
\begin{align}
    \label{eq:ZZpb}
    \comm{Z}{Z} = 0
    \,,\quad
    \comm{Z}{\bZ} \neq 0
    \,,\quad
    \comm{\bZ}{\bZ} = 0
    \,.
\end{align}
In this description,
the NJ shift
(as localizations of scattering amplitudes at complex positions
\cite{ahh2017,aho2020,Guevara:2018wpp,Guevara:2019fsj,chkl2019,Johansson:2019dnu,Lazopoulos:2021mna,Aoude:2020onz,Aoude:2022trd})
is directly derived from
complexified on-shell kinematics
in \fref{fig:hkins}
via \textit{massive} half-Fourier transforms.
As they
provide 
a massive counterpart of the maximally helicity violating (MHV) kinematics
\nomenclature{MHV}{Maximally Helicity Violating}
\cite{arkani2012positivegrassmannian}
and describe a spinning analog of equivalence principle,
a novel perspective
is provided for
the minimality 
\cite{ahh2017,aho2020,Guevara:2018wpp,Guevara:2019fsj,chkl2019}
of spinning black holes.

In conclusion, spinspacetime is a universal framework 
for any massive Poincar\'e-invariant system, 
in which the minimality of spinning black holes is strikingly manifest through its distinctive complex-geometrical structures.
This refined understanding fully resolves the questions left 
by Newman's works
and allows us to extend and generalize its curious story
with exciting future applications.

\section{
Reconstructing Spacetime from Poincar\'e Symmetry}%
In any system enjoying global Poincar\'e symmetry,
there shall exist Poincar\'e charges:
\begin{align}
\begin{split}
    \label{eq:pa}
    \comm{ J^{\m\n} }{ J^{\r\s} }
    &= i\mem (-4 \delta^{[\m}{}_{[\k} \eta^{\n][\r} \delta^{\s]}{}_{\l]})\hem J^{\k\l}
    \,,\\
    \comm{ J^{\m\n} }{ p_\r }
    &= i\mem p_\s\hem (-2\eta^{\s[\m} \delta^{\n]}{}_\r)
    \,,\quad
    \comm{ p_\m }{ p_\n }
    = 0
    \vphantom{(-4 \delta^{[\m}{}_{[\k} \eta^{\n][\r} \delta^{\s]}{}_{\l]})\hem J^{\k\l}}
    \,.
\end{split}
\end{align}
Additionally,
the notion of mass dimension is accompanied
in any physical situation.
This amounts to having an operator $D$ such that
\begin{align}
    \label{eq:D|Jp}
    \comm{ J^{\m\n} }{ D } = 0
    \,,\quad
    \comm{ p_\m }{ D } = +i\mem p_\m
    \,,
\end{align}
assigning dimensions $0$ and $+1$ to $J$ and $p$, respectively.

Suppose the system is massive so that
its configurations 
do not realize $p^2 = 0$.
Then, define
\begin{align}
    \label{eq:xdef}
    x^\m
    = 
    J^{\m\n} p_\n /p^2
    -
    D\hem p^\m \nem/p^2
    \,,
\end{align}
as a function of $J$, $p$, and $D$.
By straightforward calculations,
it follows that
$x^\m$ exactly transforms like a position vector in Minkowski space
under the Poincar\'e group action:
$\comm{ x^\m }{ J^{\r\s} } = 
    i\mem
    (-2\eta^{\m[\r} \delta^{\s]}{}_\n)\mem
    x^\n
$,
$\comm{ x^\m }{ p_\r } = 
    i\mem
    \delta^\m{}_\r
$.
In this manner,
a Minkowski space
emerges
from Poincar\'e symmetry.

It should be clear that this construction applies to 
\textit{any} massive system with global Poincar\'e symmetry:
a particle, a group of particles, 
a field, a group of fields,
or even a system not prescribed in a spacetime formulation.
The prime example is
a massive system in an asymptotically flat spacetime,
in which case
our argument asserts that 
a \textit{flat bulk}
emerges
from the Poincar\'e charges at asymptotic infinity.

A few remarks are in order.
First, suppose we have only assumed the strict Poincar\'e algebra in \eqref{eq:pa}.
Then $J^{\m\n}p_\n/p^2$ behaves almost, but not perfectly, like a position vector.
Amusingly,
taking \eqref{eq:xdef} then as an ansatz,
imposing the desired transformation behavior of $x^\m$
mandates the commutation relations in \eqref{eq:D|Jp},
so the same construction is reproduced.
Next,
at the technical level,
the above calculations
will closely parallel
those carried out in \rrcite{brooke1985relativistic,bacry1967space}.
However,
\rrcite{brooke1985relativistic,bacry1967space}
specifically restricted
their scope
to a free particle,
whereas
our very point here is the universality of this construction.

\section{
Spin-induced Spacetime Fuzziness}%
Next, we observe
a puzzling feature
in this notion of an emergent flat spacetime:
it is noncommutative as
\begin{align}
    \label{eq:[xx]}
    [x^\m , x^\n]
    = -\frac{i}{p^2}\mem S^{\m\n}
    \,,
\end{align}
which directly follows from 
\eqrefs{eq:pa}{eq:xdef}.
Here
$S^{\m\n}$
denotes the transverse projection of $J^{\m\n}$
such that
$p_\m S^{\m\n} = 0$.
Namely, it is
the spin angular momentum
which by definition is
the part of
the Lorentz charge
$J^{\m\n}$
unexplainable by the orbital angular momentum 
computed with spacetime coordinates:
$J^{\m\n} = 2x^{[\m}p^{\n]} + S^{\m\n}$.

This peculiarity is inevitable in the sense that
\eqref{eq:xdef} is the only Poincar\'e-covariant and real spacetime coordinates
that can be directly constructed from $J^{\m\n}$, $p_\m$, and $D$,
as implied by the analysis provided in the supplemental material.
Indeed, the noncommutativity in \eqref{eq:[xx]}
has been observed in
a considerable amount of work
\cite{Born:1935ap,pryce1948mass,fleming1965covariant,Hanson:1974qy,Casalbuoni:1975hx,casalbuoni1976classical,Barut:1980mv,brooke1985relativistic,zakrzewski1995extended,GuzmanRamirez:2013ynp,ramirez2015lagrangian,Deriglazov:2017jub,Deriglazov:2019vcj,hughston1979twistors,bette1984pointlike,bette1996directly,bette1997extended,bette2000twistor,bette2004massive04,Bette:2004ip,bette2005twistors-0402150,bette2005two-0503134,lukierski2014noncommutative,Filippas:2022soduality}
from various angles
and
from numerous spinning particle models.

\section{
Unification of Spacetime and Spin}%
\label{sec:spinspacetime}%
In the co-moving frame of the momentum,
\eqref{eq:[xx]} 
boils down to
$\comm{x^i}{x^j} = (i\hbar / m^2c^2)\mem \ve^{ij}{}_k\mem S^k$,
which implies a noncommutativity scale of
$\mathit{\Delta}x \sim s^{1/2}\mem (\hbar/mc)$ for spin $s$.
Here, we have 
temporarily
restored 
the fundamental constants.
Meanwhile, recall the famous commutator,
$\comm{S^i}{S^j} = i\hbar\mem \ve^{ij}{}_k\mem S^k$.
Amusingly, these right-hand sides 
can be canceled by 
composing complex combinations $x^i \mp i\mem S^i\nem/mc$,
precisely because the imaginary unit squares to $-1$.

Let us implement this solution to the noncommutativity puzzle
in a fully Poincar\'e-covariant way.
Define the spin length pseudovector,
which describes
the Pauli-Lubanski pseudovector normalized in the units of length:
\begin{align}
    \label{eq:ydef}
    y^\m = - {*J}^{\m\n} p_\n /p^2
    \qiq
    S^{\m\n} = \ve^{\m\n\r\s} y_r\hhem p_\s
    \,.
\end{align}
Then \eqref{eq:pa} implies
that $y^\m$ transforms like a tangent vector:
$
    \comm{ y^\m }{ J^{\r\s} }
    = 
    i\mem
    (-2\eta^{\m[\r} \delta^{\s]}{}_\n)\mem
    y^\n
$, $
    \comm{ y^\m }{ p_\r }
    =
    0
$.
In turn,
for $z^\m = x^\m + iy^\m$,
\eqrefs{eq:pa}{eq:D|Jp} imply
\begin{align}
\begin{split}
    \label{eq:zzpb}
    \comm{ z^\m }{ z^\n }
    &= \comm{ \bz^\m }{\bz^\n }
    = 0
    \,,\\
    \comm{ z^\m }{ \bz^\n }
    &
    = 
    -\frac{2}{p^2}\mem
    \Big(\mem{
        2y^{(\m} p^{\n)}
        + i\mem \ve^{\m\n\r\s}\hhnem y_\r\hhem p_\s
    }\mem\Big)
    \,.
\end{split}
\end{align}
Remarkably, $z^\m$ themselves are
\textit{commutative}.

In conclusion,
a complexified Minkowski space
\cite{penrose1967twistoralgebra},\footnote{
    Mathematically, this is the space $\mathbb{C}^4$
    equipped with the flat metric $\eta^\mathbb{C}$ whose signature reduces to Lorentzian on the real section $z^\m {\,=\,}\protect\bz^\m$
    \cite{Adamo:2017qyl}.
    One may regard it as the tangent bundle of Minkowski space
    equipped with the adapted complex structure\cite{guillemin1991grauert,guillemin1992grauert,lempert1991global,szHoke1991complex,hall2011adapted}.
}
equipped with 
the commutators in \eqref{eq:zzpb}
featuring
commutative holomorphic coordinates,
can be constructed in any massive system with global Poincar\'e symmetry.
This complexified Minkowski space is the \textit{spinspacetime}.

One might wonder if the use of the imaginary unit,
seemingly a bit radical,
is really necessary.
However, as we have shown in a bootstrap analysis in
the supplemental material,
this is \textit{the only Poincar\'e-covariant solution} to the noncommutativity puzzle.
This uniqueness seems to have not been realized explicitly,
although proposals for the complex-valued coordinates have existed
\cite{almond1973time,casalbuoni1976classical}.

Thus,
it appears that Poincar\'e invariance intrinsically entails a curious complex geometry,
which we then must pay attention to.
Especially,
it is easily checked that
the Poincar\'e group action preserves the complex structure:
it does not mix up holomorphic and anti-holomorphic variables.
Hence there is a possibility for
this complex structure, or holomorphy, of spinspacetime 
acquiring a physical significance
as a frame-independent notion.

\section{
Newman's Derivation of Spinspacetime}%
Indeed,
holomorphy in spinspacetime
is intrinsically associated with
\textit{chirality}.\footnote{
    The easiest way of seeing this is to realize that
    $y^\m$ is a pseudovector:
    parity flip
    induces complex conjugation, $z^\m {\,\leftrightarrow\,} \protect\bz^\m$.
}
This point is emphasized in Newman's original derivation of spinspacetime \cite{newman1974curiosity},
sketched in \eqref{eq:prop.J}.
To recapitulate,
Newman 
derives the complex coordinates by
observing
that
the SD and ASD parts of 
the angular momentum
are
$J^\pm = ((x{\,\pm\,}iy) \mwedge p)^\pm$,
where 
$(\a \mwedge \b)^\pm$ denotes the SD/ASD projection of the bivector $(\a \mwedge \b)^{\m\n} = \a^\m \b^\n - \a^\n \b^\m$.
Crucially, 
the SD part $J^+$ depends solely on the holomorphic position,
while
the \textit{A}SD part $J^-$ depends solely on the \textit{anti}-holomorphic position.
In this light,
the imaginary unit in $x {\,\pm\,} iy$
has originated from
a Hodge (electric-magnetic \cite{Filippas:2022soduality}) duality between
orbital and spin angular momenta,
$(x \mwedge p)$
and
${*}(y \mwedge p)$.

Newman \cite{newman1974curiosity}'s approach to spinspacetime
is rather independent from 
the approach in the previous sections
where the focus is put on the commutator structure
\cite{almond1973time,casalbuoni1976classical}.
However,
we emphasize that
both perspectives capture essential defining features of spinspacetime:
the geometry of spinspacetime
is completely characterized only if \textit{both}
(1)~%
the commutator structure in \eqref{eq:prop.zigzag}
(more precisely \eqref{eq:zzpb})
and
(2)~%
the association between holomorphy and chirality
tracing back to \eqref{eq:prop.J}
are understood.

These two unique geometrical features 
lead to remarkable physical implications.
In fact,
they can be demonstrated nicely in a twistorial implementation of spinspacetime,
which we now elaborate on.

\section{
Spinspacetime from Massive Twistor Space}%
\label{K3:SST>MT}
The astute reader
will realize that
the 
development
so far strongly resonates with the very spirit of twistor theory.
First of all,
spacetime is regarded as a secondary construct
\cite{penrose1987origins,penrose1975aims,penrose:maccallum}.
And not to mention,
a complex-geometrical structure is disclosed.
Further,
the $\comm{z^\m}{\bz^\n}$ bracket
takes a simple form in the spinor notation,
as will be revealed shortly in \eqref{eq:z|bz}.
In addition, $D = -p_\m\hem x^\m$ in \eqref{eq:D|Jp} and\:(\ref{eq:xdef})
is the dilatation charge.
These all point to an intimate connection to twistor theory
where complex geometry, spinors, and conformal algebra
play central roles.

A definition of twistor space is
the vector space $\mathbb{C}^4$
equipped with a $(2,2)$-signature Hermitian form,
serving as
the linear representation space of $\mathrm{U}(2,2)$
\cite{penrose:maccallum,penrose1975aims}.
Geometrically,
a twistor represents a null ray in complexified Minkowski space \cite{penrose:maccallum,penrose1975aims,penrose1967twistoralgebra}.
To implement the mass,
one considers
two copies of twistor space,
$\mathbb{C}^8 \cong \mathbb{C}^4 {\mem\times\mem} \mathbb{C}^4$
\cite{Penrose:1974di,Perjes:1974ra,tod1976two},
where $\mathrm{U}(2,2)$
acts from the left
and $\mathrm{U}(2)$
acts from the right
with a shared $\mathrm{U}(1)$.
In essence, this realizes 
a massive momentum as a sum of two null momenta.
This space $\mathbb{C}^8$ is the \textit{massive twistor space}.
According to
a modern view 
developed in
\rrcite{ambikerr0,ambikerr1},
which differs a bit from the original interpretation taken in
the 70's twistor particle program
\cite{Perjes:1974ra,Penrose:1974di},
the right $\mathrm{SU}(2)$ is the massive little group
while the shared $\mathrm{U}(1)$ is gauge.

Having defined the massive twistor space,
the next step is to
construct its spinspacetime coordinates explicitly.
Let $Z_\rmA{}^I = (\lambda_\a{}^I, i\mu^{\da I})$
and $\bZ_I{}^\rmA = (-i\bmu_I{}^\a , \blambda_{I\da})$
be holomorphic and anti-holomorphic coordinates
of the massive twistor space.
Here
$\rmA,\rmB,\cdots$ 
are $\mathrm{SU}(2,2)$ (Dirac spinor) indices,
while
$I,J,\cdots$
are $\mathrm{SU}(2)$ indices.
This space is a K\"ahler vector space
\cite{Penrose:1968me,penrose:maccallum,penrose1975aims},
so
the commutation relations are given by the oscillator algebra
as sketched in \eqref{eq:ZZpb}:
$
    [ Z_\rmA{}^I , \bZ_J{}^\rmB ]
    = \delta_\rmA{}^\rmB\mem \delta_J{}^I
$,
to be explicit.

The $\mathrm{U}(2,2)$ generators
are given by 
$G_\rmA{}^\rmB = Z_\rmA{}^I \bZ_I{}^\rmA $,
the Weyl block decomposition of which gives
\begin{align}
\begin{split}
    \label{eq:pG-twistor}
    p_{\a\da}
    = -\lambda_\a{}^I \blambda_{I\da}
    \,,\quad
    G^\da{}_\db
    = \mu^{\da I} \blambda_{I\db}
    \,.
\end{split}
\end{align}
The first equation is the famous
``massive spinor-helicity'' decomposition
of the massive momentum
\cite{Penrose:1974di,Perjes:1974ra,ahh2017,conde2016spinor,conde2016lorentz}.
The second equation 
unifies dilatation and SD angular momentum:
\smash{$
    \smash{G^\da{}_\db}
    = \smash{J^\da{}_\db}
    + \minie\mem \smash{\delta^\da{}_\db}\mem 
    (D {\,+\,} i\tilde{D})
$}.
A new ingredient is $\tilde{D}$,
which generates the $\mathrm{U}(1)$ gauge group.
It has zero commutator with any other generators
and induces
a central extension from $\mathrm{SU}(2,2)$ to $\mathrm{U}(2,2)$.

With this understanding,
we translate
our earlier formula for
spinspacetime coordinates
into the spinor notation.
Notably, it
takes a strikingly simple form:
\begin{align}
    \label{eq:z=Gp}
    z^{\da\a} = -G^\da{}_\db\hem (p^{-1})^{\db\a}
    \,.
\end{align}
Here,
\smash{$(p^{-1})^{\da\a} = \smash{\be^{\da\db} \e^{\a\b}} p_{\wrap{\b\db}} / \det(p)$}
denotes the inverse of $p_{\a\da}$ as a square matrix,
defined for $p^2 \neq 0$.
In turn,
the conformal algebra
directly implies that
\begin{align}
\begin{split}
    \label{eq:z|bz}
    \comm{ z^{\da\a} }{ z^{\db\b} } 
    = 0
    \,,\quad
    \comm{ z^{\da\a} }{ \bz^{\db\b} }
    &= i\mem (z {\,-\,} \bz)^{\da\b} (p^{-1})^{\db\a}
    \,,
\end{split}
\end{align}
reproducing \eqref{eq:zzpb}
up to the extension $y^\m \mapsto y^\m - \tilde{D}\mem p^\m \nem/p^2$.\footnote{
    \eqrefs{eq:z=Gp}{eq:z|bz} apply
    generally to any
    Hamiltonian system in asymptotically flat spacetime,
    where
    one can find conformal generators
    from the conformal killing vectors
    (even though the 
    Hamiltonian
    may not enjoy
    the full conformal symmetry).
}
Finally,
applying \eqref{eq:z=Gp}
to
the charges in
\eqref{eq:pG-twistor},
the spinspacetime coordinates 
for the massive twistor space
are found as
\begin{align}
    \label{eq:incidence}
    z^{\da\a}
    = \mu^{\da I} (\lambda^{-1})_I{}^\a
    \qiq
    \mu^{\da I} = z^{\da\a} \lambda_\a{}^I
    \,.
\end{align}
Interestingly,
\eqref{eq:incidence}
reinvents the incidence relation of twistor theory
\cite{penrose:maccallum,penrose1967twistoralgebra,newman1974curiosity}
for the massive case.
Its interpretation
is that
the two complexified null rays represented by
the twistors
$Z_\rmA{}^{I=0,1}$
are ``co-incident''
at a point $z^{\da\a}$
in complexified Minkowski space.

\begin{figure}[t]
    \centering
    \includegraphics[width=0.7\linewidth,trim=0 20pt 0 10pt]{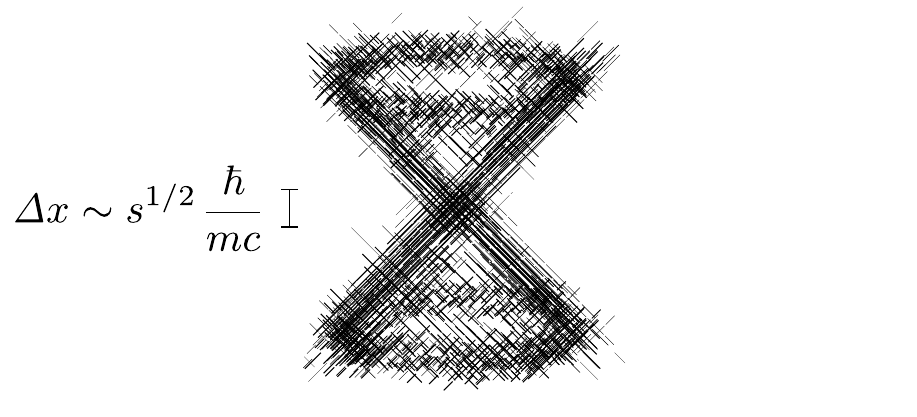}
    \medskip
    \caption[%
        Lightcone in twistor theory.
        Penrose advocates a point of view that
        there is an absurdity inherent in the notion of spacetime
        already at the Compton scale of elementary particles.
        Accordingly,
        the very philosophy of
        twistor theory
        is to
        take 
        spacetime points as
        secondary constructs
        emerging from intersections between null rays.
    ]{%
        Lightcone in twistor theory 
        according to Penrose
        \cite{penrose1975aims,penr04}.
        Penrose advocates a point of view that
        there is an absurdity inherent in the notion of spacetime
        already at the Compton scale of elementary particles \cite{Penrose:1974di,Penrose:1980hi,penrose1987origins}.
        Accordingly,
        the very philosophy of
        twistor theory
        is to
        take 
        spacetime points as
        secondary constructs
        emerging from intersections between null rays.
    }
    \label{fig:LCdx}
\end{figure}

In conclusion,
the spinspacetime of massive twistor space
is simply the complexified Minkowski space 
which its incidence relation in \eqref{eq:incidence} defines.\footnote{
    Note that a double fibration picture could be pursued to formulate this relationship in a more mathematically precise sense,
    analogously to the massless case.
}
Intriguingly, this
implies that
the noncommutativity in \eqref{eq:[xx]}
could be reinterpreted
as the fuzziness which spacetime points acquires in twistor theory
as light rays become fundamental objects
\cite{penrose1987origins,penrose1975aims,penrose:maccallum,hughston1979twistors,bette1984pointlike,bette1996directly,bette1997extended,bette2000twistor,bette2004massive04,Bette:2004ip,bette2005twistors-0402150,bette2005two-0503134,lukierski2014noncommutative}:
see Fig.\,\ref{fig:LCdx}.

The twistorial picture
associates fascinating geometric origins to the defining features of spinspacetime.
First,
the off-diagonal commutators
(\eqref{eq:prop.zigzag})
have signaled
the K\"ahler geometry of twistor space
(\eqref{eq:ZZpb}).
Second,
holomorphy in spinspacetime
corresponds to
the holomorphy of twistors.
Third,
the link between holomorphy and chirality in spinspacetime
(\eqref{eq:prop.J})
traces back to
that of
twistor space.
Note that this link is deeply endemic in twistor space:
the \textit{holomorphic} $Z_\rmA{}^I$
contains the \textit{left-handed} spinor-helicity variable $\lambda_\a{}^I$.

\section{
S-Matrix in Massive Twistor Space}%
\label{K3:SST>FOURIER}%
Eventually,
we study
scattering in asymptotically flat spacetimes.
This setup concerns
the spinspacetime 
which
Poincar\'e charges 
in asymptotically free regions
reconstruct,
together with
an asymptotic massive twistor space as a \textit{phase space}.

As well-known,
a massive on-shell scattering state
is represented by
the massive spinor-helicity variables
$\lambda_\a{}^I$, $\rambda_{I\da}$
\cite{ahh2017}.
The massive twistor space
can be viewed as the cotangent bundle
of the space $\mathbb{C}^4$ of massive spinor-helicity variables,
in light of the commutators
$
    \comm{ \lambda_\a{}^I }{ \bmu_J{}^\b }
    = i\mem \delta_\a{}^\b\mem \delta_J{}^I
$,
$
    \comm{ \rambda_{I\da} }{ \mu^{\db J} }
    = i\mem \delta_I{}^J\mem \delta^\db{}_\da
$.

The massive twistor space admits various polarizations,
offering
different bases for the scattering states.
For instance,
the scattering states can be realized in the coherent state basis
(K\"ahler polarization),
\begin{align}
\begin{split}
    Z \Ket{Z_1}
    = \Ket{Z_1}\mem Z_1
    \,,\quad
    \Bra{\bZ_2} \bZ
    = \bZ_2\hem\hhem \Bra{\bZ_2} 
    \,,
\end{split}
\end{align}
or in the spinor-helicity basis,
\begin{align}
    \lambda \mem\Ket{\lambda_1\rambda_1}
    = \Ket{\lambda_1\rambda_1}\mem \lambda_1
    \,,\quad
    \rambda \mem\Ket{\lambda_1\rambda_1}
    = \Ket{\lambda_1\rambda_1}\mem \rambda_1
    \,,
\end{align}
where we have started to omit indices to avoid clutter.
The overlaps are given by
$\langle \bZ_2 | Z_1 \rangle = \smash{e^{\bZ_2 Z_1}}$
and
$\langle \lambda_2\rambda_2 | \lambda_1\rambda_1 \rangle = \delta^{(4)}({\lambda_1}{\mem-\,}{\lambda_2}) \, \delta^{(4)}({\rambda_1}{\mem-\,}{\rambda_2})$,
where $\bZ_2 Z_1$ abbreviates $(\bZ_2)_I{}^\rmA (Z_1)_\rmA{}^I$.
The transformations between these two bases
are given by the half-Fourier transforms,
which one may define concretely
by analytically continuing to the $(2,2)$-signature
\cite{Witten:2003nn,arkani2010s-matrix}:
\begin{align}
\begin{split}
    \label{eq:half-Fourier}
    \Ket{Z_1}
    = \Ket{\lambda_1\mu_1}
    &= \int\, [d^4\rambda]\,\,
        \mathe^{i\rambda\mu_1}
        \,\, \Ket{\lambda_1\rambda}
    \,,\\
    \Bra{\bZ_2}
    = \Bra{\rambda_2\bmu_2}
    &= \int\, [d^4\lambda]\,\,
        \mathe^{-i\bmu_2\lambda\nem}
        \,\, \Bra{\lambda\rambda_2}
    \,.
\end{split}
\end{align}
In turn,
S-matrix elements in the spinor-helicity basis
can be converted to the massive twistor basis:
$\langle \lambda_2\rambda_2 | S | \lambda_1\rambda_1 \rangle$ $\to$ $\langle \bZ_2 | S | Z_1 \rangle$.
To our best knowledge,
such half-Fourier transforms
have been actively pursued only
for massless legs
\cite{Witten:2003nn,arkani2010s-matrix,Guevara:2021yud}.

\section{
What Defines Spinning Black Holes}%
\begin{figure}[t]
    \centering
    \begin{align*}
        \includegraphics[valign=c,scale=1.2]{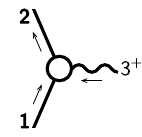}
        &
        \adjustbox{scale=1.1}{$
            \quad\sim\quad
            \del{
                \lambda_1
            }{
                \lambda_2
            }
        $}
        \\
        \includegraphics[valign=c,scale=1.2]{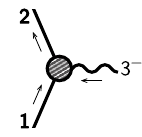}
        &
        \adjustbox{scale=1.1}{$
            \quad\sim\quad
            \del{
                \blambda_1
            }{
                \blambda_2
            }
        $}
    \end{align*}
    \caption{%
        The building blocks of 
        ``massive on-shell diagrams,''
        describing extremal cases of complexified on-shell kinematics.
    }
    \label{fig:hkins}
\end{figure}

Now we ask the following question:
What is the simplest S-matrix in massive twistor space?
For the case of massive-massive-massless three-particle scattering with equal masses,
the on-shell kinematics exhibit two \textit{extremal} cases
that freeze
either of the left-handed or right-handed massive spinor-helicity variables.
Let us call these \textit{zig} and \textit{zag}\footnote{
    This terminology is inspired by Penrose
    \cite{penr04},
    Chapter 25.2: The zigzag picture of the electron, pp. 628-633.
} kinematics
and denote them respectively as white and black blobs
as shown in Fig.\,\ref{fig:hkins},
analogously to the massless on-shell diagram notation
in \rrcite{hodges1980twistor,hodges1990feynman,hodges2005a,hodges2005b,arkani2010s-matrix,arkani2012positivegrassmannian,Atiyah:2017erd}.
Then momentum conservation
fully refines
their delta function supports
as
\begin{align}
    \label{eq:zkin}
\begin{split}
    \text{zig}:\quad
    \del{
        \lambda_1
    }{
        \lambda_2
    }
    \,\mem
    \dell{
        \rambda_1 
            {\mem-\mem} 
        \lambda\inv{1} k_3
    }{
        \rambda_2
    }
    \,,\\
    \text{zag}:\quad
    \dell{
        \lambda_1 
            {\mem-\mem}
        k_3\hem \rambda\inv{2}
    }{
        \lambda_2
    }
    \,\mem
    \del{
        \rambda_1
    }{
        \rambda_2
    }
    \,,
\end{split}
\end{align}
where $k_3{}_{\a\da}$ is the massless momentum of the third leg.
We have abbreviated 
contracted indices as
$(\lambda\inv{1} k_3)_{I\da} = (\lambda\inv{1})_I{}^\a\mem k_3{}_{\a\da}$.
These are \textit{complexified} kinematics
\footnote{
    Cf. Penrose's argument in \rrcite{penrose1976nonlinear,Penrose:1976js}
    that the notion of a ``single-graviton spacetime''
    is inherently complex.
},
so $\rambda_{1,2}$ here
would be
more appropriately
denoted as $\smash{\tilde{\lambda}_{1,2}}$.

Furthermore,
we stipulate a link between chirality and holomorphy:
the \textit{zig} and \textit{zag} kinematics
are realized for
\textit{positive} and \textit{negative} helicities of the massless leg,
respectively.
With this postulate,
the zig and zag kinematics embody
the scattering amplitudes statement of 
an analog of equivalence principle in half-flat backgrounds,
which we dub \textit{spinning equivalence principle}.

Suppose a spinning object traveling in a SD background spacetime.
What would be its simplest motion?
By definition, the left-handed spinor bundle is flat,
which one can locally trivialize.
If the gravitational forces are to
arise uniquely from this geometry,
the object's left-handed Lorentz frame would remain constant
in local laboratories.
This deduces the spinning equivalence principle:
the left-handed (right-handed) frame
of a spinning object
is covariantly constant
under dynamical evolution in SD (ASD) backgrounds.

Of course, in a modern understanding,
the equivalence principle is rather a statement about the minimality of a test object,
which can be broken by non-minimal couplings.
For example,
a scalar particle endowed with nonzero non-minimal Wilson coefficients
will not follow the geodesic trajectory
as desired by the equivalence principle.
Similarly, the spinning equivalence principle
rather proposes
the most ideal behavior which a spinning object may exhibit.

Remarkably, we find that such an ideal behavior is exhibited by \textit{black holes}.
To see this,
consider the massive half-Fourier transform of the zig kinematics to the twistor \textit{coherent state} basis
\footnote{
    The punchline here parallels that of \rcite{arkani2010s-matrix}.
    Also, a hidden assumption will be that the in and out states for the black hole are given in the twistor coherent states.
}:
\begin{align}
\begin{split}
    \label{eq:half+1}
    &
    \exp\hnem\Big({
        -i\hem\bmu_2
        \lambda_1
    }\Big)
    \,
    \exp\hnem\Big(\mem{
        i\hem(\lambda\inv{1}k_3 \hhnem+\nem\rambda_2)
        \mu_1
    }\Big)
    \\
    &
    =\,
    \BraKet{\bZ_2}{Z_1}
    \,
    \exp\hnem\Big(\mem{
        i\mem k_3\mem\mu_1 \lambda\inv{1}
    }\mem\Big)
    \,.
\end{split}
\end{align}
By the massive incidence relation in \eqref{eq:incidence},
the exponential factor computes to
$e^{ik_3z_1} {\:=\:} e^{ik_3x_1} e^{-k_3y_1}$,
where
the factor $e^{-k_3y_1}$
\footnote{
    Here,
    the minus sign
    is due to our conventions $\ve_{0123} = +1$
    and $S^{\m\n} = \ve^{\m\n\r\s} y_\r\hhem p_\s$:
    $y^\m = -a^\m$.
}
for receiving positive-helicity force carriers
is known to implement
the NJ shift property
at the amplitudes level
\cite{aho2020,chkl2019,Guevara:2018wpp,Guevara:2019fsj}.
Here, it is crucial that
the ``imaginary impact parameter'' $y_1^\m$ describes spin
due to the very (re)interpretation of 
complex coordinates in twistor theory
as spinspacetime coordinates,
as is deduced in \eqref{eq:incidence}.
Similarly,
half-Fourier transforming the zag kinematics
yields the factor 
$e^{ik_3\bz_1} {\:=\:} e^{ik_3x_1} e^{k_3y_1}$,
encoding the NJ shift for receiving negative-helicity force carriers.

In sum,
spinning black holes
are characterized by
the extremal complexified on-shell kinematics in Fig.\,\ref{fig:hkins},
which encode the spinning equivalence principle.
Notably, we could have also assumed
any multiplicity of massless quanta,
building up arbitrary half-flat backgrounds.
This derives the same-helicity spin exponentiation \cite{Johansson:2019dnu,Lazopoulos:2021mna,Aoude:2020onz}
at all multiplicities
via the half-Fourier transform.
Eventually, this case would arise from a gluing of the same-helicity on-shell diagrams.
It will be interesting if a smart gluing is viable for mixed helicities as well,
given the spurious pole problem 
\cite{ahh2017,chkl2019,Aoude:2022trd}.

A few remarks are in order.
Firstly,
in the above derivation of NJ shift,
the off-diagonal commutator structure
of the massive twistor space
(as the structure of half-Fourier transforms),
namely the K\"ahler geometry,
has played an important role.
To make this point more evident,
consider the following identity:
\begin{align}
\begin{split}
    \label{eq:expl}
    \Bra{\lambda_2\rambda_2}
        \exp\hnem\Big(\mem{
            i\mem 
            k_3 \mu \lambda^{-1}
        }\mem\Big)
    =
    \Bra{\lambda_2,\rambda_2 {\,+\mem} \lambda\inv{1}k_3}
    \,.
\end{split}
\end{align}
Overlapping with $|\lambda_1\rambda_1\rangle$
reproduces \eqref{eq:half+1}
as the matrix element of $e^{ik_3 z}$ in the twistor basis.
Crucially, the \textit{holomorphic} operator $e^{ik_3 z}$
shifts the \textit{anti-holomorphic} variable $\rambda$,
precisely due to the off-diagonal structure.

Another essential role is played by
the intrinsic association between chirality and holomorphy.
To see this, suppose we knew that the object is a spinning black hole from the beginning.
While sitting at the asymptotic infinity
with an experimentalist mindset,
we ``measure'' the S-matrix to deduce what background this black hole had interacted with.
If the S-matrix is \textit{holomorphic},
then, by the off-diagonal structure,
the \textit{anti-holomorphic} variable
$\rambda$
solely exhibits precession.
But crucially, the \textit{anti-holomorphic} spinor-helicity variable is \textit{right-handed}.
Due to this very endemic link between holomorphy and handedness,
the background must have been
filled with \textit{right-handed}, i.e., SD fields.

On a related note,
the exponent in \eqref{eq:expl}
describes the SD angular momentum
$J^+ = (z\mwedge p)^+$
times a factor,
providing a twistorial realization of the equations employed in \rcite{Guevara:2018wpp}.

For generic massive spinning objects,
the kinematics is not extremal.
For example,
$\lambda$ and $\rambda$
are shifted symmetrically (cf.~\rcite{aho2020})
when
the object's spin-induced multipole moments \cite{Levi:2015msa} vanish from the quadrupole order
\cite{ambikerr1}.
The complexified kinematics
we have studied
is sensitive to all multipole moments,
while Newman's explorations 
\cite{newman1974curiosity,newman1974collection,Newman:1973afx,Newman:2004ba,Newman:1973yu,Newman:2002mk,Newman:1976gc,ko1981theory}
focused only on the dipole coupling (gyromagnetic ratio).

Lastly,
we were able to deduce the essential physics
just by focusing on the delta function support,
similarly as in \rcite{arkani2012positivegrassmannian}.
A full-blown theory should not only incorporate the $mx^h$ prefactors \cite{ahh2017}
but also properly formulate
the independent massive little group orbits.
The first-quantized massive twistor particle \cite{ambikerr0,ambikerr1}
could provide a reference,
which concretely reproduces the delta function supports
considered above
\cite{ambikerr1}.

\section{
Conclusions}%
In this work, we established
the notion of spinspacetime as a universal consequence of Poincar\'e symmetry.
Its characteristic geometric structures,
which trace back to those of twistor space,
revealed that spinning black holes 
obey a spinning analog of equivalence principle.
The Newman-Janis shift factors 
are effortlessly obtained by
massive half-Fourier transforming
the extremal on-shell kinematics in \fref{fig:hkins}.

This revival of a curious concept
should trigger various future directions.
Firstly,
the theory of bulk spinspacetime will be developed,
paralleling the historical development of curved twistor theory \cite{penrose1973twistor}.
To this end, one expects
a curved-spacetime generalization or reincarnation
of 
Newman and Winicour \cite{newman1974curiosity}'s
Hodge duality between orbital and spin angular momenta in \eqref{eq:prop.J}.\footnote{
    Note that 
    we have realized
    this general-relativistic ``orbit-spin duality''
    in \Chaps{K3:PROBENJ}{K3:OSD1}:
    the local orbital angular momentum due to geodesic deviation
    turns into a spin angular momentum,
    giving rise to the general-relativistic holomorphic spinspacetime coordinates in the SD sector.
    In fact, we had developed and established the complete mathematical and physical theory of curved spinspacetime
    from this perspective of orbit-spin duality,
    which, unfortunately, had to be omitted from this thesis
    due to lack of space.
}
Secondly,
applications to PM gravity
will be pursued.
Lastly,
the massive on-shell diagrams
and their gluing
should be investigated,
reminiscently of the massless story
\cite{hodges1980twistor,hodges1990feynman,hodges2005a,hodges2005b,arkani2010s-matrix,arkani2012positivegrassmannian,Atiyah:2017erd,Witten:2003nn}.
It will be exciting if
a complete on-shell framework,
incorporating both massless and massive legs,
can serve as a powerful tool
for both theoretical and practical directions.

\chapter{Exact Black Hole Equations of Motion}\label{K3:BHEOM}

    We demonstrate how
    the framework of spinspacetime
    facilitates
    a systematic derivation of
    all-orders-in-spin EoM
    of spinning black holes
    in their point-particle effective theory.
    A model-independent derivation of BMT equations
    and their all-orders-in-spin extensions
    is demonstrated.
    A model-independent derivation of {\kerr} EoM
    is provided
    in general, non-SD external electromagnetic fields.
    Applications of the spinspacetime framework to
    dynamical symmetries
    are also discussed.
    A dynamical NJ shift
    derives
    exact, all-orders conserved charges of the spinning black hole probe
    from
    the conserved charges of the non-spinning probe.
    A maximally superintegrable
    subsector is identified for
    the {\kerr}-{\kerr} binary problem.

\begin{fullnote}
    Contents of this section are adapted from 
    \rcite{univ}, \fullcite{univ}.
\end{fullnote}

\section{
The Universal Poisson Bracket Relation of Spinning Particles}
\label{KB:FREE}

This section reviews a set of crucial facts shown by \rcite{sst-asym},
which enables direct identifications of
the universal geometrical structures
of massive spinning-particle phase spaces.
An additional analysis on symplectic leaves is provided.

\subsection{Derivation from Poincar\'e Symmetry}
\label{KB:FREE>FULL}

Suppose any Hamiltonian formulation of
a massive physical system
that enjoys global Poincar\'e symmetry.
\rcite{sst-asym} shows that
one can always construct
functions 
$\hx^\m$, $\hy^\m$, $p_\m$
on its phase space $\ps$
such that 
$p \mdot \hx = 0 = p \mdot \hy$
and
\begin{align}
\begin{split}
\label{M3*}
    \pb{\hx^\m}{\hx^\n}
    \,\,&=\,\,
        \frac{1}{-p^2}\, 
        \BB{\mem
            \hx^\m p^\n {-\,} p^\m \hx^\n
            + \ve^{\m\n\r\s} \hy_\r\mem p_\s
        }
    \,,\\
    \pb{\hx^\m}{\hy^\n}
    \,\,&=\,\,
        \frac{1}{-p^2}\,
            \hy^\m p^\n
    \,,\quad
    \pb{\hy^\m}{\hy^\n}
    \,=\,
        \frac{1}{-p^2}\, 
            \ve^{\m\n\r\s} \hy_\r\mem p_\s
    \,,\\
    \pb{\hx^\m}{p_\n}
    \,\,&=\,\,
        \delta^\m{}_\n - \frac{p^\m\hem p_\n}{p^2}
    \,,\quad
    \pb{\hy^\m}{p_\n}
    \,=\,0
    \,,\quad
    \pb{p_\m}{p_\n}
    \,=\, 0
    \,.
\end{split}
\end{align}
Here, $\pb{f}{g}$ denotes the Poisson bracket
between functions $f$ and $g$ on the phase space $\ps$.
This result applies to not only discrete systems but also continuous systems:
a compact mass distribution on flat spacetime
with a conserved stress-energy $T^{\m\n}$, for instance.
Yet, the interest of this work will be restricted specifically to one-particle systems.

Firstly,
Poincar\'e symmetry
implies that
one can construct
the Poincar\'e charges
as functions on the phase space $\ps$:
momentum $p_\m$ and total angular momentum $J^{\m\n}$.
They must satisfy the following Poisson brackets
as per the definition of the Poincar\'e algebra:
\begin{align}
\begin{split}
\label{poincare}
    \pb{J^{\m\n}}{J^{\r\s}}
    \,\,&=\,\,
          J^{\m\r}\mem \eta^{\n\s}
        - J^{\m\s}\mem \eta^{\n\r}
        + J^{\n\s}\mem \eta^{\m\r}
        - J^{\n\r}\mem \eta^{\m\s}
    \,,\\[0.15\baselineskip]
    \pb{J^{\m\n}}{p_\r}
    \,\,&=\,\,  
          \delta^\m{}_\r\mem p^\n 
        - \delta^\n{}_\r\mem p^\m 
    \,,\\[0.15\baselineskip]
    \pb{p_\m}{p_\n} \,\,&=\,\, 0
    \,.
\end{split}
\end{align}
Here, $\eta^{\m\n}$ is the flat inverse metric
of signature $(-,+,+,+)$.
The system being massive means that its classical states are not allowed to approach the zero locus of $p^2$
in $\ps$.

Secondly,
\textit{define} $\hx^\m$ and $\hy^\n$ as
\begin{align}
    \label{hxy-def}
    \hx^\m
    \,:=\,
        \frac{1}{p^2}\, J^{\m\n}\mem p_\n
    \,,\quad
    \hy^\m
    \,:=\,
        \frac{1}{-p^2}\, {*}J^{\m\n}\mem p_\n
    \,,
\end{align}
where ${*}J^{\m\n} = \frac{1}{2}\mem \ve^{\m\n\r\s} J_{\r\s}$
describes the Hodge star
in the convention $\ve_{0123} = +1$.
By direct computation,
one shows that
the Poincar\'e algebra in \eqref{poincare}
implies
the Poisson brackets in \eqref{M3*}
between $\hx^\m$, $\hy^\m$, and $p_\m$.
Also,
$p \mdot \hx = 0 = p \mdot \hy$
by the antisymmetry of $J^{\m\n}$ and ${*}J^{\m\n}$.
This concludes the proof.

On account of the well-known definition of the Pauli-Lubanski pseudovector,
the variable $\hy^\m$
in \eqref{hxy-def}
describes the spin pseudovector
normalized in units of length.
In this work,
we refer to this $\hy^\m$
as \textit{spin length pseudovector}.

In fact,
$\hy^\m$ can evaluate to zero in some systems;
take a free relativistic scalar particle
formulated in the cotangent bundle of Minkowski space,
for instance.
In this work,
we concern systems with nontrivial $\hy^\m$ only.

\subsection{Partly Reduced Formulation}
\label{KB:FREE>PART}

Physically,
the notion of mass dimension must exist in the phase space $\ps$.
Especially, the momentum $p_\m$
carries mass dimension $+1$
while the angular momentum $J^{\m\n}$
carries mass dimension $0$.
This implies that the dilatation charge $D$
would also exist
as a function on the phase space $\ps$,
such that the Poisson bracket with $D$
measures the mass dimension:
\begin{align}
    \label{dilatation}
    \pb{p_\m}{D}
    \,=\, 
        +p_\m
    \,,\quad
    \pb{J^{\m\n}}{D}
    \,=\,
        0
    \,.
\end{align}

\rcite{sst-asym} shows that
\eqrefs{poincare}{dilatation}
together imply the existence of 
functions $x^\m$, $\hy^\m$, and $p_\m$
on $\ps$
such that
$p \mdot \hy = 0$
and
\begin{align}
\begin{split}
\label{M3}
    \pb{x^\m}{x^\n}
    \,\,&=\,\,
        \frac{1}{-p^2}\, 
            \ve^{\m\n\r\s} \hy_\r\mem p_\s
    \,=\,
    \pb{\hy^\m}{\hy^\n}
    \,,\\
    \pb{x^\m}{\hy^\n}
    \,\,&=\,\,
        \frac{1}{-p^2}\,\BB{
            \hy^\m p^\n + p^\m \hy^\n 
        }
    \,,\\[0.15\baselineskip]
    \pb{x^\m}{p_\n}
    \,\,&=\,\,
        \delta^\m{}_\n
    \,,\quad
    \pb{\hy^\m}{p_\n}
    \,=\,0
    \,,\quad
    \pb{p_\m}{p_\n}
    \,=\, 0
    \,.
\end{split}
\end{align}

The proof is straightforward.
Define
\begin{align}
    \label{x-def}
    x^\m
    \,:=\,
        \frac{1}{p^2}\, 
        \BB{
            J^{\m\n}\mem p_\n
            - D\mem p^\m
        }
    \,,
\end{align}
which adds an additional term to $\hx^\m$.
Then direct computation verifies \eqref{M3}.

In terms of $x^\m$, $\hy^\m$, and $p_\m$,
the total angular momentum $J^{\m\n}$
and the dilatation charge $D$
are expressed as
\begin{align}
    \label{JD}
    J^{\m\n}
    \,=\,
        x^\m p^\n {-\,} p^\m x^\n 
        + \ve^{\m\n\r\s} \hy_\r\mem p_\s
    \,,\quad
    D
    \,=\,
        - p \mdot x
    \,.
\end{align}
From \eqref{JD}
and the Poisson bracket $\pb{x^\m}{p_\n} = \delta^\m{}_\n$ in 
\eqref{M3},
it is evident that
the interpretation of $x^\m$
is a reconstructed notion of
flat spacetime coordinates.

Again, the above result applies to any massive system.
In the context of
effective point-particle description of
compact bodies,
$x^\m$ in \eqref{x-def}
describes the Poincar\'e-covariant center coordinates.
Accordingly,
\eqref{JD}
describes the covariant split
of the total angular momentum $J^{\m\n}$
to orbital
($x^\m p^\n {-\,} p^\m x^\n$)
and
spin
($\ve^{\m\n\r\s} \hy_\r\mem p_\s$)
parts.

\subsection{Symplectic Leaves}
\label{KB:FREE>LEAVES}

Lastly, let us comprehend the above results
from a mathematical perspective.

A Poisson manifold is a manifold equipped with a Poisson bracket
$\pb{f}{g} = \Pi(df,dg)$
between functions $f$ and $g$,
which arises from a bivector $\Pi$.
A symplectic manifold is a manifold equipped with a symplectic form $\omega$,
which is a closed nondegenerate two-form.
Symplectic manifolds are Poisson manifolds, but the converse is not true.
This is because the Poisson bivector $\Pi$ can be degenerate,
unlike the symplectic form $\omega$.

Yet,
any Poisson manifold is foliated by 
\textit{symplectic leaves},
which are
(connected)
symplectic submanifolds of maximal dimension.
The dimension of symplectic leaves
is determined by the rank of the Poisson bivector $\Pi$.

Provided that $\hy^\m$ is nontrivial,
$\hx^\m$, $\hy^\m$, and $p_\m$
in \eqref{M3*}
together describe
a $10$-dimensional Poisson submanifold embedded in the phase space $\ps$.
Its symplectic leaves are $8$-dimensional.
To see this,
regard 
the Poisson brackets in \eqref{M3*}
as components of an antisymmetric matrix
and compute its rank: $8$.
Furthermore, examining its kernel shows that
each of these symplectic leaves
is the locus of 
\begin{align}
    -p^2 \,=\, m^2
    \,,\quad
    -p^2\mem \hy^2 \,=\, w^2
    \,,
\end{align}
where $m$ and $w$ are constants.
Physically,
the first equation computes the \textit{rest mass} $m$
whereas
the second equation computes the \textit{spin magnitude} $w$.

In this way, it follows that
one can always identify an $8$-dimensional symplectic submanifold $\ups_{8}(m,w)$
inside any phase space $\ps$
of a massive system
with nontrivial $\hy^\m$.
This symplectic submanifold
will be referred to as the \textit{8-dimensional universal phase space}
of the massive spinning system,
on which the rest mass and the spin magnitude
are fixed to constant values $m$ and $w$:
\begin{align}
    \label{ups8}
    \ps
    \,\,\supset\,\,
    \ups_{8}(m,w)
    \,.
\end{align}

In turn, $8$ sets the minimal bound
for the dimension of a massive spinning phase space.
It achieves three translational degrees of freedom and one spinning degree of freedom.
It can be seen that 
$\ups_8(m,w)$ is homeomorphic to $\R^6 {\,\times\,} S^2$,
which
describes the coadjoint orbit
\cite{kirillov1975elements,kostant1970orbits,souriau1970structure}
of the Poincar\'e group
\cite{souriau1970structure,bacry1967space,arens1971classical,Carinena:1989uw}.
It follows that 
$\ups_8(m,w)$ is equivalent to
Souriau's ``$8$-dimensional space of motion''
reviewed in \rcite{Damour:2024mzo}.

Similarly,
$x^\m$, $\hy^\m$, and $p_\m$ 
in \eqref{M3}
together describe an $11$-dimensional
Poisson submanifold in $\ps$
whose symplectic leaves are $10$-dimensional.
Computation shows that
each symplectic leaf is the locus of 
$-p^2\mem \hy^2 = w^2$.
This identifies the \textit{10-dimensional universal phase space} $\ups_{10}(w)$,
on which the spin magnitude
is fixed to a constant value $w$:
\begin{align}
    \label{ups10}
    \ps
    \,\,\supset\,\,
    \ups_{10}(w)
    \,.
\end{align}
It can be seen that $\ups_{10}(w)$
is homeomorphic to $\R^8 {\,\times\,} S^2$.

It is possible to reproduce
the $8$-dimensional universal phase space in \eqref{ups8}
from the $10$-dimensional universal phase space in \eqref{ups10}:
\begin{align}
    \label{quotient}
    \ups_{8}(m,w)
    \,\,=\,\,
        \Big\{\mem{
            (x,\hy,p)
            \in
                \ups_{10}(w)
        \,\Big|\,\hem
            p^2 + m^2 = 0
        }\,\Big\}
        \hem\Big/\nem\hnem
            \sim
    \,.
\end{align}
This describes a quotient along
one-parameter orbits
in $\ups_{10}(w)$,
\begin{align}
    \label{reparam}
    \bigbig{
        x^\m
        ,\mem
        \hy^\m 
        ,\mem
        p_\m
    }
    \,\,\sim\,\,
    \bigbig{
        x^\m + k\mem p^\m
        ,\mem
        \hy^\m 
        ,\mem
        p_\m
    }
    \,,
\end{align}
which is the Hamiltonian flow of $p^2 + m^2$.
Physically, \eqref{reparam} implements a translation of the reconstructed spacetime coordinates in \eqref{x-def}
along the momentum direction.

Mathematically,
quotients of the form \eqref{quotient}
could be called
\textit{symplectic quotients}:
impose a constraint \textit{and} quotient by its Hamiltonian flow
\cite{marsden1974reduction,meyer1973symmetries}.
The process of reducing a symplectic manifold to a smaller one
via symplectic quotient
will be referred to as \textit{symplectic reduction}.
Clearly,
symplectic reduction always reduces the dimension
by an even number.

The physicists' toolkit for symplectic reduction
is known as 
\textit{Dirac bracket} 
\cite{Dirac:1950pj,dirac1964lectures,Henneaux:1992ig}.
In this work,
we use Dirac brackets to 
explicitly realize symplectic quotients
as symplectic submanifolds.

It is a simple exercise to show that
the universal phase space 
$\ups_8(m,w)$ in \eqref{ups8}
is a symplectic submanifold of
$\ups_{10}(w)$ in \eqref{ups10},
by constructing the Dirac bracket
that fixes $p^2 + m^2$ and $p \mdot x$ to zero
within $\ups_{10}(w)$:
\begin{align}
    \label{upss}
    \ps
    \,\,\supset\,\,
    \ups_{10}(w)
    \,\,\supset\,\,
    \ups_8(m,w)
    \,.
\end{align}
Surely, this reduction recipe
applies universally in every massive spinning system.

\section{
Universality in Interacting Theory: Electromagnetism}
\label{INT1}

\Sec{KB:FREE} has explicitly reviewed that
any model of a massive spinning particle
describes
physical variables
$(x^\m, \hy^\m, p_\m)$
that exhibit the following
universal Poisson bracket relation
in the \textit{free theory}:
\begin{align}
\begin{split}
\label{xyp}
    \pb{x^\m}{x^\n}^\circ
    \,\,&=\,\,
        \frac{1}{-p^2}\, 
            \ve^{\m\n\r\s} \hy_\r\mem p_\s
    \,=\,
    \pb{\hy^\m}{\hy^\n}^\circ
    \,,\\
    \pb{x^\m}{\hy^\n}^\circ
    \,\,&=\,\,
        \frac{1}{-p^2}\,\BB{
            \hy^\m p^\n + p^\m \hy^\n 
        }
    \,,\\[0.15\baselineskip]
    \pb{x^\m}{p_\n}^\circ
    \,\,&=\,\,
        \delta^\m{}_\n
    \,.
\end{split}
\end{align}
Our objective now
is to
generalize such universal structures
in the \textit{interacting theory},
e.g.,
with electromagnetic interactions.
The output is model-agnostic derivations of
the physical EoM of massive spinning particles
coupled to electromagnetism.

\subsection{Couplings as Symplectic Perturbations}
\label{INT>SPT}

Our point of departure is
the celebrated fact in symplectic geometry
that
a manifestly gauge-invariant coupling
of particles to electromagnetic fields
is viable by
modifying the symplectic structure
in terms of the field strength.
This construction is due to Souriau \cite{souriau1970structure}
and also traces back to Feynman \cite{dyson1990feynman}.
Historically, this manifestly gauge-invariant method has served as an important milestone in the study of particle mechanics in the symplectic framework
\cite{woodhouse1997geometric,torrence1973gauge,guillemin1978equations,sternberg1978classical,guillemin1990symplectic}.
Unfortunately,
it appears to be less recognized in recent literature,
so we shall provide a brief review
(see also \rcite{csg}).

For the sake of concreteness,
let us suppose a simple example:
a massive scalar particle in Minkowski space.
This is a Hamiltonian system defined on
the cotangent bundle $T^*\mflat$.
In the free theory,
this phase space is
equipped with the symplectic form and Hamiltonian
\begin{align}
    \label{scalar.omega0}
    \omega^\circ \,=\,
        dp_\m \swedge dx^\m
    \,,\quad
    \phi_0 \,=\,
        \frac{1}{2m}\mem
        \BB{
            p^2 + m^2
        }
    \,.
\end{align}
We have started to put the accent $^\circ$
to explicitly indicate free-theory objects.
The canonical Poisson bracket reads
\begin{align}
    \label{scalar.pb0}
    \pb{x^\m}{x^\n}^\circ
    \,=\,
        0
    \,,\quad
    \pb{x^\m}{p_\n}^\circ
    \,=\,
        \delta^\m{}_\n
    \,,\quad
    \pb{p_\m}{p_\n}^\circ
    \,=\,
        0
    \,.
\end{align}

The Souriau-Feynman method of electromagnetic coupling
modifies the geometric data in \eqref{scalar.omega0} as
\begin{align}
    \label{scalar.omega}
    \omega \,=\,
        dp_\m \swedge dx^\m 
        \,+\,
        qF
    \,,\quad
    \phi_0 \,=\,
        \frac{1}{2m}\mem
        \BB{
            p^2 + m^2
        }
    \,,
\end{align}
where $F = \frac{1}{2}\mem F_{\m\n}(x)\mem dx^\m \swedge dx^\n$ is the field strength two-form,
and $q$ is the electric charge.
Crucially, the Hamiltonian is left unchanged
while the symplectic form is perturbed as
\begin{align}
    \label{omega-split}
    \omega \,=\,
        \omega^\circ \mem+\, \omega'
    \,,
\end{align}
with $\omega' = qF$.
As a result, the Poisson bracket in \eqref{scalar.pb0} is altered as
\begin{align}
    \label{scalar.pb}
    \pb{x^\m}{x^\n}
    \,=\,
        0
    \,,\quad
    \pb{x^\m}{p_\n}
    \,=\,
        \delta^\m{}_\n
    \,,\quad
    \pb{p_\m}{p_\n}
    \,=\,
        qF_{\m\n}(x)
    \,.
\end{align}

To see why \eqref{scalar.omega} achieves the electromagnetic coupling,
one can derive the EoM:
\begin{align}
\begin{split}
    \label{lorforce}
    \dot{x}^\m
    \mem=\mem
        \pb{x^\m}{\phi_0}
    \mem=\mem
        \frac{p^\m}{m}
    \,,\quad
    \dot{p}_\m
    \mem=\mem
        \pb{p_\m}{\phi_0}
    \mem=\mem
        \frac{q}{m}\,
        F_{\m\n}(x)\mem p^\n
    \,.
\end{split}
\end{align}
\eqref{lorforce} exactly describes 
the EoM of a charged scalar particle in electromagnetic background.
The modified $\pb{p_\m}{p_\n}$ bracket in \eqref{scalar.pb}
implements the Lorentz force law
in \eqref{lorforce}.
As a result, the intuition for noncanonical bracket $\pb{p_\m}{p_\n} \neq 0$
is the cyclotron precession exhibited by the momentum $p_\m$
in the external electromagnetic field.

Crucially,
this derivation of the Lorentz force law
does not invoke the electromagnetic gauge potential at all,
as is emphasized by both 
Souriau \cite{souriau1970structure}
and Feynman \cite{dyson1990feynman}.
As a result, $p_\m$
is the gauge-invariant, physical momentum
(also known as kinetic momentum).

The only consistency requirement in this framework
is the closure of the modified symplectic form in \eqref{scalar.omega}
\cite{souriau1970structure},
which holds via the Bianchi-type field equations $dF = 0$ on the background \cite{dyson1990feynman}.
Physically, this closure
amounts to the \textit{Liouville property} of classical time evolution:
conservation of classical probability
(as phase space measure) \cite{Kim:2025sey,csg}.

If one insists on keeping the canonical Poisson brackets,
one can perform
a gauge-dependent coordinate transformation 
in the phase space
(a change of worldline field basis)
to the canonical momenta.
In this case
the Hamiltonian is perturbed and solely encodes the electromagnetic interactions,
while the symplectic structure is brought back to the free-theory form.

Albeit canonical,
this formulation exhibits evident disadvantages
in both conceptual and practical aspects.
First, gauge invariance is grossly obscured.
The canonical momentum is not a gauge-invariant observable.
Second, the coupling is nonlinear.
In perturbation theory,
there arises a spurious quartic vertex
whose sole role is to fix gauge invariance.
Third, the derivation of the Lorentz force law
from the Hamiltonian EoM
becomes not so simple.

In this work,
we always fix the Hamiltonian in the free-theory form
such that the external field couplings are 
implemented solely by modifications of the symplectic structure
(declaration of worldline field basis).
When the symplectic form is modified in the form of \eqref{omega-split},
we say that the two-form $\omega'$
is a \textit{symplectic perturbation}.

We end with a useful formula.
The pointwise inverses of $\omega^\circ$ and $\omega$
define
the Poisson bivectors 
$\Pi^\circ$ and $\Pi$
of the free and interacting theories,
respectively.
Their relation is
\begin{align}
    \label{ppert}
    \omega \,=\,
        \omega^\circ \mem+\, \omega'
    \qiq
    \Pi
    \,=\,
        \Pi^\circ
        - \Pi^\circ\mem \omega'\, \Pi
    \,.
\end{align}
In writing down \eqref{ppert}, we have treated
$\omega$, $\omega^\circ$, $\omega'$, $\Pi$, $\Pi^\circ$
like antisymmetric matrices.

\subsection{Model-Agnostic Derivation of BMT Equations}
\label{INT>BMT}

By combining the ideas in \Secs{KB:FREE}{INT>SPT},
we establish
universal recipes
of coupling massive spinning particles to electromagnetism.

\paragraph{Universal Recipe for Minimal Coupling}

Suppose a free massive spinning particle
formulated in a symplectic manifold $\ps$.
Let $\omega^\circ$ be the symplectic form given to $\ps$.
Time evolution describes a Hamiltonian vector field on $\ps$,
which we denote as $T^\circ$.

The result in \Sec{KB:FREE>PART}
implies that
there exist functions $(x^\m, \hy^\m, p_\m)$
on $\ps$
that satisfy the universal Poisson brackets in \eqref{xyp}.
Without loss of generality, we assume that
these functions evolve under the free-theory time evolution as
\begin{align}
    \label{T0}
    T^\circ\act{
        x^\m
    }
    \,=\, \frac{p^\m}{m}
    \,,\quad
    T^\circ\act{
        \hy^\m
    }
    \,=\, 0
    \,,\quad
    T^\circ\act{
        p_\m
    }
    \,=\, 0
    \,.
\end{align}
Physically, this means that 
the free particle follows a straight-line trajectory
while carrying a constant spin pseudovector.

As an application of the Souriau-Feynman framework in \Sec{INT>SPT},
we suppose the following symplectic perturbation on $\ps$:
\begin{align}
    \label{EM.sp0}
    \omega'
    \,\,=\,\,
        qF
    \,\,=\,\,
        \frac{1}{2}\,
            qF_{\r\s}(x)\,
            dx^\r \swedge dx^\s
    \,.
\end{align}
The formula in \eqref{ppert} then implies that
the Poisson bivector is modified as
\begin{align}
\begin{split}
    \label{ppert.EM}
    \Pi(df,dg)
    \,=\,
        \Pi^\circ(df,dg)
        \mem-\mem
        \Pi^\circ(df,dx^\r)\,
            qF_{\r\s}(x)\,
        \Pi(dx^\s,dg)
    \,.
\end{split}
\end{align}
Iterating \eqref{ppert.EM} will provide an infinite series expansion for $\pb{f}{g} = \Pi(df,dg)$,
which governs how the Poisson bracket is modified under the symplectic perturbation:
$
    \pb{f}{g}
    =
        \pb{f}{g}^\circ
        \mem-\mem
        \pb{f}{x^\r}^\circ\,
            qF_{\r\s}(x)\,
        \pb{x^\s}{g}^\circ
        \mem+\mem
        \pb{f}{x^\r}^\circ\,
    $ $
            qF_{\r\s}(x)\,
        \pb{x^\s}{x^\k}^\circ\,
            qF_{\k\l}(x)\,
        \pb{x^\l}{g}^\circ
        \mem-\mem
        \cdots
$.

Let $T$ be the time-evolution Hamiltonian vector field 
in the interacting theory
such that the Hamiltonian EoM reads $\dot{f} = T\act{f}$.
\eqrefs{T0}{ppert.EM} together imply that
\begin{align}
    \label{ppert.EM0}
    \dot{f}
    \,=\,
        T^\circ\act{
            f
        }
        \mem-\mem
        \pb{f}{x^\r}^\circ\,
            qF_{\r\s}(x)\,
        \dot{x}^\s
    \,.
\end{align}
Plugging in the physical variables $(x^\m,\hy^\m,p_\m)$ to $f$ in \eqref{ppert.EM0},
we find
\begin{align}
\begin{split}
    \label{uBMT}
    \dot{x}^\m
    \,&=\,
            \frac{p^\m}{m} 
        \mem-\mem
        \frac{1}{m^2}\mem
        \BB{
            \ve^{\m\r\k\l} \hy_\k\mem p_\l
        }\mem
        \BB{
            qF_{\r\s}(x)\mem \dot{x}^\s
        }
    \,,\\
    \dot{\hy}^\m
    \,&=\,
        \frac{1}{m^2}\mem
        \BB{
            \hy^\m\mem p^\r
            + p^\m\mem \hy^\r
        }\mem
        \BB{
            qF_{\r\s}(x)\mem 
            \dot{x}^\s
        }
    \,,\\[0.12\baselineskip]
    \dot{p}^\m
    \,&=\,
        qF^\m{}_\n(x)\mem \dot{x}^\n
    \,,
\end{split}
\end{align}
where we have used \eqref{xyp}
to evaluate 
$\pb{x^\m}{x^\r}^\circ$,
$\pb{\hy^\m}{x^\r}^\circ$,
and
$\pb{p_\m}{x^\r}^\circ$.

\eqref{uBMT} defines the classical EoM of a consistent Hamiltonian system,
as the modified symplectic form due to \eqref{EM.sp0} is closed.
What is its physical interpretation, then?

\paragraph{BMT Equations}

The BMT \cite{Bargmann:1959gz} equations 
dictate the dynamics of charged massive spinning particles in external electromagnetic fields
at the linear order in electric charge $q$,
in the regime where derivative effects can be completely ignored:
$F(x) \gg \hy\mem \partial F(x), \hy^2\mem \partial^2 F(x), \cdots$.
In our conventions, they are
\begin{subequations}
\label{BMT}
\begin{align}
\label{BMT.x}
    \dot{x}^\m
    \,&=\,
        \frac{p^\m}{m}
        \mem-\mem
        \frac{q}{m}\,
        \bbsq{    
            \frac{g}{2}\,
                {*}\hnem F^\m{}_\n(x)
            + \bb{1 - \frac{g}{2}}\,
                \hdelta^\m{}_\k\,
                {*}\hnem F^\k{}_\n(x)
        }\, \hy^\n
    + \cdots
    \,,\\
\label{BMT.y}
    \dot{\hy}^\m
    \,&=\,
        \frac{q}{m}\,
        \bbsq{    
            F^\m{}_\n(x)
            - \bb{1 - \frac{g}{2}}\,
                \hdelta^\m{}_\k\mem
                F^\k{}_\n(x)
        }\, \hy^\n
    + \cdots
    \,,\\[0.12\baselineskip]
\label{BMT.p}
    \dot{p}^\m
    \,&=\,
        \frac{q}{m}\,
            F^\m{}_\n(x)\mem p^\n
    + \cdots
    \,,
\end{align}
\end{subequations}
where the ellipses signify that
terms of $\O(q^{1+n})$ or $\O(\partial^n\nem F)$
are discarded from the right-hand sides
for $n \geq 1$.

Many references present just \eqrefs{BMT.y}{BMT.p}
without explicitly stating the relation between $p^\m$ and $\dot{x}^\m$.
However, it should be emphasized that
\eqref{BMT.x}
must be specified 
together with \eqrefs{BMT.y}{BMT.p}
to fully define the time evolution
and ensure its consistency,
especially when working in the Hamiltonian formulation.

In \eqref{BMT}, the constant parameter $g$
is the gyromagnetic ratio,
also known as the $g$-factor.
The gyromagnetic ratio controls the spin-induced magnetic dipole moment.
The specific value $g{\:=\:}0$
has been identified as the minimal coupling
in the traditional sense,
in which case
the BMT equations in \eqref{BMT} read
\begin{subequations}
\label{BMT0}
\begin{align}
\label{BMT0.x}
    \dot{x}^\m
    \,&=\,
        \frac{p^\m}{m}
        \mem-\mem
        \frac{q}{m}\,
            \hdelta^\m{}_\k\mem
            {*}\hnem F^\k{}_\n(x)\mem \hy^\n
    + \cdots
    \,,\\
\label{BMT0.y}
    \dot{\hy}^\m
    \,&=\,
        - \frac{q}{m}\,
            \hp^\m\mem \hp_\k\hhem
            F^\k{}_\n(x)\mem  \hy^\n
    + \cdots
    \,,\\[0.12\baselineskip]
\label{BMT0.p}
    \dot{p}^\m
    \,&=\,
        \frac{q}{m}\,
            F^\m{}_\n(x)\mem p^\n
    + \cdots
    \,.
\end{align}
\end{subequations}

Notably,
the EoM derived in \eqref{uBMT} 
yield the $g {\:=\:} 0$ BMT equations in \eqref{BMT0}
through
plugging in $\dot{x}^\m = p^\m\nem/m + \O(q^1)$
and using an identity involving two epsilon tensors.
Therefore, 
the $g{\:=\:}0$ BMT equations
arise in any massive spinning particle model
by prescribing
the symplectic perturbation
in \eqref{EM.sp0}
realized at the physical center $x^\m$.

\paragraph{Universal Recipe for Dipolar Coupling}

To implement the dipolar coupling in our universal framework,
consider the following symplectic perturbation:
\begin{align}
    \label{EM.sp1}
    \omega'
    \,\,=\,\,
        \frac{1}{2}\,
            qF_{\m\n}(x)\,
            dx^\m \swedge dx^\n
        \,+\,
        q\mem 
        d\,\bb{
            c_1\,
            \hy^\r\,
            {*}F_{\r\s}(x)\,
            dx^\s
        }
    \,.
\end{align}
Here, $c_1$ is a constant.
\eqref{EM.sp1} is the most general symplectic perturbation
that can contribute to the $\O(q^1\hy^1)$ part of the EoM.
It is \textit{bootstrapped} by 
the physical principles of
\begin{align}
\label{bootstrap.EM}
\begin{split}
    \text{(a)}\,\,&\text{
        Liouville property%
    }
    \,,\\
    \text{(b)}\,\,&\text{
        Poincar\'e invariance%
    }
    \,,\\
    \text{(c)}\,\,&\text{
        Gauge invariance%
    }
    \,,\\
    \text{(d)}\,\,&\text{
        Parity invariance%
    }
    \,.
\end{split}
\end{align}

By applying the method demonstrated in \eqref{ppert.EM0},
one finds that 
the $\O(q^1y^1)$ contribution to the resulting EoM
arises from the combination
\begin{align}
    \label{EM.q1y1}
    - \frac{q}{m}\,
    \bbsq{    
        \pb{f}{x^\r}^\circ\,
        F_{\r\s}(x)\mem p^\s
        +
        c_1\,
        \pb{f}{\hy^\r}^\circ\,
        {*}F_{\r\s}(x)\mem p^\s
    }
    \,.
\end{align}
By examining \eqref{EM.q1y1} for $f = (x^\m,\hy^\m,p_\m)$,
one exactly reproduces the BMT equations in \eqref{BMT}
provided the identification
\begin{align}
    c_1
    \,=\,
        g/2
    \,.
\end{align}

Therefore, the BMT equations with the generic gyromagnetic ratio $g$
arise in any massive spinning particle model
by prescribing the symplectic perturbation in \eqref{EM.sp1}
in terms of
the physical center $x^\m$ and spin length pseudovector $\hy^\m$.

\subsection{Higher Multipoles and Beyond}
\label{APPL>POLES}

\paragraph{Linear Couplings}

As is glimpsed in \eqref{EM.sp1},
our universal framework in \Sec{INT1}
readily generalizes to all multipole orders.
For electromagnetism,
an ansatz that conforms to the physical principles in \eqref{bootstrap.EM} is
\begin{align}
\begin{split}
    \label{EM.spell}
    \omega
    \,=\,
        \omega^\circ
        \,&+\,
        \frac{1}{2}\,
            qF_{\m\n}(x)\,
            dx^\m \swedge dx^\n
    \\
        \,&+\,
        q\, d\mem\bbsq{\,
        \sum_{\ell=1}^\infty\mem
        \frac{c_\ell}{\ell!}\,\mem
            {*^\ell}\hnem F_{\r_1\s,\r_2\cdots\r_\ell}(x)\,
            \hy^{\r_1}\hy^{\r_2}\cdots\hy^{\r_\ell}\,
            dx^\s
        \mem}
        \vphantom{\bigg|}
    \,,
\end{split}
\end{align}
where $*^\ell$ means to apply the Hodge star $\ell$ times.
\eqref{EM.spell} achieves
the most general linear electromagnetic coupling,
where
the Wilson coefficient $c_\ell$ parameterizes
the spin-induced $2^\ell$-pole moment.
Note that the Hodge star implements parity invariance.
Thus
$c_{2k}$ are electric moments,
while
$c_{2k+1}$ are magnetic moments.

\paragraph{Nonlinear Couplings}

The above multipole coefficients $c_\ell$
dictate only the linear couplings
to external fields.
At the nonlinear level,
there are broad possibilities for generalization.
We may want to illuminate two categories of terms,
in particular:
\begin{enumerate}
    \item 
        Ordinary terms
        ($dx$),
        such as
        $(q^2/m^2)\mem 
            (\hem{
                \hy^\m F_{\m\r}(x)\hem F^\r{}_\n(x)\mem \hy^\n
            }\hem)
            \mem 
            (\hem{
                p_\k\mem dx^\k
            }\hhem)
        $.
    \item 
        Spin-derivative terms
        ($d\hy$),
        such as
        $   
            qF_{\m\n}(x)\mem \hy^\m\mem d\hy^\n
        $ or
        $
            \hy^\r\mem qF_{\m\n,\r}(x)\mem \hy^\m\mem d\hy^\n
        $.
\end{enumerate}
Here, 
we have enumerated
perturbations on the symplectic potential.

The systematic enumeration of all possible terms
at a given order
is an important problem,
although it goes beyond the present work's scope.
Generally speaking, this counting will depend on
the field basis choice
(symplectic versus Hamiltonian perturbations),
while computing scattering amplitudes can provide some guidance.
As a reminder, our symplectic perturbation approach
specifies a particular worldline field basis
by fixing the mass-shell constraint in the free-theory form.

The above two classes of terms
play different roles in worldline perturbation theory.
Consider the diagrammatic computation of two-quanta Compton amplitudes,
for instance.
The example term
        $(q^2/m^2)\mem 
            (\hem{
                \hy^\m F_{\m\r}(x)\hem F^\r{}_\n(x)\mem \hy^\n
            }\hem)
            \mem 
            (\hem{
                p_\k\mem dx^\k
            }\hhem)
        $
contributes
to the amplitude
as a \textit{contact vertex}.
In contrast,
the example terms
in the spin-derivative category
contribute essentially
via \textit{exchange channels}
due to propagation of worldline fluctuations,
as the background worldline
will exhibit $\dot{\hy} = 0$.

This point should also justify
the identification of the above spin-derivative terms
as ``nonlinear couplings.''
Although
they can be nominally linear in the curvature tensors
at face value,
they are still nonlinear in the sense that 
they do not contribute to the three-point amplitudes with one massless quantum.
Their effects on the classical EoM,
however,
could be just as significant as
the ordinary terms
at the same curvature order.

As shown by \rrcite{probe-nj,njmagic.1,njmagic.11},
spin-derivative terms are \textit{necessary}
for achieving black hole couplings
(see \eqref{rkerr}).
Also, they may not be so fundamentally different than the ordinary terms
in the spinspacetime perspective,
as spin is (the imaginary part of) spacetime.

\paragraph{Definition of Black Hole Coupling}

Physically, the nonlinear couplings will encode various tidal deformability effects, taking different values for each astrophysical object. 

A particularly intriguing case is black holes,
which have attracted considerable attention in the current literature.
The determination of the exact black hole coupling to all orders
has been a long-standing problem.
The central question reads
\begin{center}
    \vphantom{.}\llap{``\:}\textit{What defines black holes}
    among all massive \\ spinning objects
    in the point-particle effective theory?\rlap{\:''}
\end{center}
The problem stands unambiguously solved
at least 
in the following subsectors.
\begin{enumerate}
    \item 
        In \textsc{Linearized Gravity}:
        The Kerr black hole is defined by
        the \textit{unity of multipole coefficients},
        $C_\ell = 1$ for all $\ell = 2,3,4,\cdots$
        \cite{%
            Hansen:1974zz,Newman:1965tw-janis,Hernandez:1967zza,Thorne:1980ru%
        }
        (cf. \rrcite{ahh2017,Guevara:2018wpp,Guevara:2019fsj,chkl2019,aho2020}).
    \item
        In \textsc{Self-Dual Gravity}:
        The Kerr black hole is defined by
        a \textit{superintegrability}
        in type-D backgrounds
        with Killing-Yano tensors
        \cite{probe-nj}
        (cf. \rrcite{Johansson:2019dnu,Aoude:2020onz,Lazopoulos:2021mna}).
\end{enumerate}
We have reviewed this development in \Sec{NJB}.
If a worldline model passes the above tests (a) and (b),
then it qualifies as an effective description of the Kerr black hole
according to our best theoretical understanding 
at the current moment.

Criterion (a) is necessary for reproducing the NJ \cite{Newman:1965tw-janis} property
in linearized gravity.
The NJ shift
describes that 
the Kerr black hole is
in some sense
a Schwarzschild black hole
displaced into ``complex spacetime''
along an imaginary direction
set by its spin length pseudovector.
Criterion (b) arises via a nonlinear generalization of this property
in curved backgrounds \cite{probe-nj}.

The problem of constructing worldline models for black holes
may have been approached in model-specific manners
so far:
spherical top model \cite{gmoov},
massive twistor model \cite{njmagic.1,njmagic.11},
fermionic model \cite{bonocore2025higher},
etc.
Meanwhile, \rcite{probe-nj} has explicated
the model-independent nature of the problem 
by introducing a probe counterpart of the NJA \cite{Newman:1965tw-janis}
in terms of the universal variables $(x,\hy,p)$
and providing
the definition of black holes in terms of global symmetries
(criterion (b) above).

In this work, we hope to
illuminate this model-independent aspect
again.
For a quick demonstration,
we will concern the electromagnetic analog of the Kerr black hole:
The zero-gravitation limit 
$G_\mathrm{N} \too 0$
of the Kerr-Newman solution,
dubbed the {\kerr} solution
\cite{aho2020,Lynden-Bell:2002dvr}.
The {\kerr} solution
exhibits
the unity of multipole coefficients 
\cite{Newman:1965tw-janis,Newman:1973yu,Lynden-Bell:2002dvr}:
$c_\ell \eqq 1$ for all $\ell \eqq 1,2,3,\cdots$.
Its dynamics as a probe
enjoys an analogous superintegrability.

\section{
Simplicity of Black Hole Dynamics}
\label{APPL>BH}

\subsection{Hidden Complex Geometry of the Universal Phase Space}

Recall \Sec{KB:FREE>PART},
where we reviewed the fact \cite{sst-asym} that
the universal Poisson bracket relation in \eqref{M3}
holds as a direct implication of
Poincar\`e symmetry.

In fact,
we can now disclose to the reader that
the universal Poisson bracket relation
admits a remarkably simpler presentation \cite{sst-asym}.
Consider the complex combination,
\begin{align}
    \label{zcoords}
    z^\m
    \,=\,
        x^\m \mem+\hem i\hy^\m
    \,.
\end{align}
Straightforward algebra shows that \eqref{M3}
is equivalent to
\begin{subequations}
\label{M3-in-zp}
\begin{align}
    \label{zp-canonical}
    \pb{z^\m}{z^\n}^\circ
    \,\,&=\,\,
        0
    \,,\quad
    \pb{z^\m}{p_\n}^\circ
    \,=\,
        \delta^\m{}_\n
    \,,\quad
    \pb{p_\m}{p_\n}^\circ
    \,=\,
        0
    \,,\\
    \label{zzpb-vec}
    \pb{z^\m}{\bz^\n}^\circ
    \,\,&=\,\,
        \frac{1}{p^2}\,
        \BB{
            \delta^\m{}_\r\mem \delta^\n{}_\s
            + \delta^\m{}_\s\mem \delta^\n{}_\r
            + i\mem \ve^{\m\n}{}_{\r\s}
        \nem}\,
            (z^\r {\mem-\,} \bz^\r)\, p^\s
    \,,
\end{align}
\end{subequations}
which fits on two lines.
Many components are set to zero.
This rewriting
not only provides a succinct summary
of \eqref{M3}
but also reveals that
the components of $z^\m$ in \eqref{zcoords}
are Poisson-commutative,
despite the peculiar noncommutativity of 
the physical center
coordinates $x^\m$
in \eqref{M3}.

Following \cite{sst-asym},
we refer to $z^\m$ as \textit{spinspacetime coordinates},
as they unify spacetime $x^\m$ and spin $\hy^\m$
as real and imaginary parts.
Mathematically,
they are the holomorphic coordinates
for the complexified Minkowski space $\mflat^\C = (\C^4,\eta^\C)$
whose real section is the flat spacetime $\mflat = (\R^4,\eta)$.
The property that
\begin{align}
    \label{zigzag}
    \pb{z^\m}{z^\n}^\circ
    \,=\,0
    \,,\quad
    \pb{z^\m}{\bz^\n}^\circ
    \,\neq\,0
    \,,\quad
    \pb{\bz^\m}{\bz^\n}^\circ
    \,=\, 0
\end{align}
is referred to as the zig-zag structure,
as the nicknames ``zig'' and ``zag''
are used for referring to
``holomorphic''\:($z$) and ``anti-holomorphic''\:($\bz$),
respectively
\cite{sst-asym}.

It should be evident that
spinspacetime and its zig-zag structure exist \textit{universally} in
every special-relativistic massive spinning system.

\subsection{Inherent Link Between Holomorphy and Self-Duality}

The idea of spinspacetime 
traces back to
Newman and Winicour \cite{newman1974curiosity}.\footnote{
    See also  
    \rrcite{newman1974collection,newman1988remarkable,Newman:1973afx,Newman:2004ba,Newman:1973yu,Newman:2002mk,ko1981theory,grg207flaherty},
    which portray
    Newman's serious take on spinspacetime.
}
The observation was that
the self-dual part of the total angular momentum in \eqref{JD} is
\begin{align}
    \label{OSD}
    J
    \,=\,
        (x \wedge p)
        + {*}(\hy \wedge p)
    \qiq
    J^+
    \,=\,
        (z \wedge p)^+
    \,.
\end{align}
Here, the $+$ superscript
signifies self-dual projection by
$\tfrac{1}{2}\mem ( 1 \mminus\mem i\mem{*} )$.
The Hodge duality ${*}$ in \eqref{OSD} turns into the imaginary unit $+i$ in the self-dual sector,
so holomorphic coordinates $z^\m = x^\m + i\hy^\m$
arise from the self-dual angular momentum $J^+$.
In the same way, anti-holomorphic coordinates $\bz^\m = x^\m - i\hy^\m$
arise from the anti-self-dual angular momentum $J^-$.
Importantly, an inherent association arises between 
\textit{holomorphy and self-duality}
in spinspacetime,
consistently with the fact that $\hy^\m$ is a \textit{pseudo}vector.

Newman and Winicour \cite{newman1974curiosity}'s discussion on spinspacetime,
however,
does not examine
its Poisson structure
in \eqref{M3-in-zp}.
The modern reboot of the spinspacetime program \cite{sst-asym}
proposes that one should pay attention to the Poisson structure of spinspacetime
and its zig-zag property,
as they are physical, universal structures.

For instance,
it is instructive to employ the spinor notation,
in which case \eqref{M3-in-zp} is boiled down to a further succinct form:
\begin{align}
    \label{zzpb}
    \pb{z^{\da\a}}{p_\wrap{\b\db}}^\circ
    \,=\,
        \delta^\da{}_\wrap{\db}\mem \delta_\wrap{\b}{}^\a
    \,,\quad
    \pb{z^{\da\a}}{\bz^{\db\b}}^\circ
    \,=\,
        \frac{1}{-p^2}\,
            (z \mminus \bz)^{\da\b}\hem p^{\db\a}
    \,.
\end{align}
\eqref{zzpb} is a one-line summary of the entirety of universal Poisson brackets in \eqref{M3}.

The chiral nature of the zig-zag bracket
$\pb{z^{\da\a}}{\bz^{\db\b}}^\circ$ in \eqref{zzpb}
reflects the aforementioned association between 
holomorphy and self-duality,
which
dictates the precession behavior of spinning particles under influence of external fields
and derives the probe-level NJ shift
from an ideal spin precession behavior
\cite{ambikerr1,sst-asym}.

\begin{figure}[t]
    \centering
    \includegraphics[scale=1.5, valign=c]{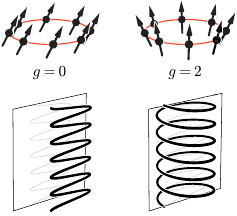}
    \caption{%
        Spinspacetime offers a new angle on spin precession.
        The Dirac gyromagnetic ratio $g \eqq 2$
        can be geometrically characterized
        as the synchronization of
        orbital and spin precessions in external fields.
        When seen from spinspacetime,
        such a motion appears as an isotropic spiral trajectory,
        pointing to the equal blending of all $dx, d\hy$ components in $dz \swedge dz$.
        In contrast, the traditional minimal coupling $g \eqq 0$
        translates to a flattened sinusoidal curve,
        which points to 
        the reality of $dx \swedge dx$.
        Spinspacetime identifies $g \eqq 2$
        as the simplest coupling.
    }
    \label{g=0-g=2}
\end{figure}

\subsection{The Simplest Symplectic Perturbation}

Spinspacetime opens a new chapter in relativity, offering a unique perspective on spinning-particle mechanics in four dimensions. It is now not only \textit{space} and \textit{time} but also \textit{spin} that are unified. 
This invites a different way of thinking:
spin is the imaginary part of spacetime.
Spacetime and spin may not be fundamentally distinct.
\fref{g=0-g=2} invites this shift.

An electromagnetic field on spacetime
takes the form $\tfrac{1}{2}\, F_{\m\n}(x)\mem dx^\m \swedge dx^\n$.
Spacetime $\mflat$ is embedded in spinspacetime $\mflat^\C$ as the real section.
Spacetime fields would somehow permeate into spinspacetime.
This will describe
$dz \swedge dz$,
$dz \swedge d\bz$,
or
$d\bz \swedge d\bz$
components.

For 
charged scalar particles,
electromagnetic fields
are symplectic perturbations
(Sec. \ref{INT>SPT}).
For 
charged spinning particles,
\textit{spinspacetime fields}
will be
symplectic perturbations.

For scalar particles, 
$\omega' = \frac{1}{2}\, qF_{\m\n}(x)\mem dx^\m \swedge dx^\n$
is said to be the minimal coupling
because it is the simplest symplectic perturbation
that one can possibly write down.
For spinning particles,
the simplest symplectic perturbation on \textit{spinspacetime}
will declare the true minimal coupling.

The spacetime field
splits into self-dual and anti-self-dual parts as
\begin{align}
    \label{Fsplit}
    \frac{1}{2}\, F_{\m\n}(x)\mem dx^\m \swedge dx^\n
    \,=\,
        \frac{1}{2}\, F^+_{\m\n}(x)\mem dx^\m \swedge dx^\n
    +
        \frac{1}{2}\, F^-_{\m\n}(x)\mem dx^\m \swedge dx^\n
    \,.
\end{align}
When studying how spacetime fields permeate into spinspacetime,
it can be helpful to examine their self-dual and anti-self-dual parts separately.

Recalling the earlier association,
suppose the self-dual field
is \textit{holomorphic}
on spinspacetime:\footnote{
    Note that the argument of $F^+_{\m\n}(z)$ here had to be also holomorphic,
    due to the principle of symplecticity (closure $d\omega' = 0$).
}
\begin{align}
    \label{sp.heaven}
    \omega'
    \,=\,
        \frac{1}{2}\, 
            qF^+_{\m\n}(z)\mem dz^\m \swedge dz^\n
    \,.
\end{align}
Holomorphic symplectic perturbations are
the simplest symplectic perturbations
on spinspacetime,
in the sense that it minimally deforms the universal Poisson brackets in \eqref{zigzag}.
Holomorphic symplectic perturbations
preserve the Poisson-commutativity of holomorphic coordinates,
due to the very zig-zag structure in \eqref{zigzag}
(recall \eqref{ppert}):
\begin{align}
    \label{zz=0}
    \pb{z^\m}{z^\n}
    \,\,=\,\,
        \pb{z^\m}{z^\n}^\circ
        \mem-\mem
        \pb{z^\m}{z^\r}^\circ\,
            \hem qF^+_{\r\s}(z)
        \,\pb{z^\s}{z^\n}
    \,\,=\,\,
        0
    \,.
\end{align}
In this manner, only the zag-zag bracket
is deformed by the zig symplectic perturbation:
\begin{align}
    \label{zigzag'}
    \pb{z^\m}{z^\n}
    \,=\,0
    \,,\quad
    \pb{z^\m}{\bz^\n}
    \,\neq\,0
    \,,\quad
    \pb{\bz^\m}{\bz^\n}
    \,\neq\, 0
    \,.
\end{align}

We should understand the physics due to 
the simplest symplectic perturbation in \eqref{sp.heaven}.
By straightforward application of
the methodology in \Sec{INT>BMT},
the EoM are derived from the universal Poisson brackets in \eqref{M3-in-zp}---%
equivalently \eqref{zzpb}, if one likes to work with spinors---%
that
\begin{align}
    \label{eom.heavenly.bz}
    \dot{z}^\m
    \,=\,
        \hp^\m
    \,,\quad
    \dot{\bz}^\m
    \,=\,
        \hp^\m
        - 
        \frac{2iq}{m}\, F^+{}^\m{}_\n(z)\mem \hy^\n
    \,,\quad
    \dot{p}^\m
    \,=\,
        \frac{q}{m}\, F^+{}^\m{}_\n(z)\mem p^\n
    \,.
\end{align}
We see that
the EoM for the zig coordinates are
completely ignorant of the external field,
due to the very Poisson-commutativity shown in \eqref{zz=0}:
$\dot{z}^\m = \hp^\m$.

\eqref{eom.heavenly.bz} translates to
\begin{align}
    \label{eom.heavenly}
    \dot{z}^\m
    \,=\,
        \hp^\m
    \,,\quad
    \dot{\hy}^\m
    \,=\,
        \frac{q}{m}\, F^+{}^\m{}_\n(z)\mem \hy^\n
    \,,\quad
    \dot{p}^\m
    \,=\,
        \frac{q}{m}\, F^+{}^\m{}_\n(z)\mem p^\n
    \,,
\end{align}
which describes that
the orbital ($\hp^\m$) and spin ($\hy^\m$) precessions are perfectly synchronized
by the cyclotron angular frequency 
$(q/m)\mem F^+{}^\m{}_\n(z)$.
Recalling the preliminary exploration in \fref{g=0-g=2},
this synchronization implies the gyromagnetic ratio $g \eqq 2$, which is $c_1 \eqq 1$.\footnote{
    If self-dual fields were anti-holomorphic in spinspacetime,
    then the particle exhibits $g = -2$,
    which deviates from the black hole behavior.
    See also \rcite{sst-asym} for 
    a physical interpretation of
    the synchronization of orbital and spin precessions
    for gravitational interactions
    (``spinning equivalence principle'').
}
In fact,
we will see shortly that
all multipole coefficients are unity:
$c_\ell \eqq 1$ for all $\ell \eqq 1,2,3,\cdots$.

\subsection{The Simplest Real Symplectic Perturbation}

The symplectic perturbation in \eqref{sp.heaven}
is formal,
giving rise to a complexified Poisson structure as in \eqref{zigzag'}.
For a real 
dynamics,
we add the missing complex conjugate
so that the anti-self-dual field is restored:
\begin{align}
    \label{sp.rkerr}
    \omega'
    \,&=\,
        \frac{1}{2}\:
            qF^+_{\m\n}(z)\mem dz^\m \swedge dz^\n
        +
        \frac{1}{2}\:
            qF^-_{\m\n}(\bz)\mem d\bz^\m \swedge d\bz^\n
    \,.
\end{align}
By methods in differential geometry \cite{gde},
it follows that \eqref{sp.rkerr}
describes
\cite{njmagic.1,njmagic.11}\footnote{
    To elaborate,
    the potential one-form for 
    \eqref{sp.rkerr} can be represented as
    $\theta' = \cos(\pounds_N)\mem (A_\m(x)\mem dx^\m) + \sinc(\pounds_N)\mem ({*}F_{\m\n}(x)\mem \hy^\m\mem dx^\n)$
    via $F^\pm = (1 \mp i\mem {*})F /2$,
    where $N$ is a vector field such that
    $\i_N dx^\m = \hy^\m$,
    $\i_N d\hy^\m = 0$,
    and
    $\i_N dp_\m = 0$.
}
\begin{align}
\begin{split}
    \label{rkerr}
    \omega'
    \,=\,
    {}&{}
        \frac{1}{2}\,
            qF_{\m\n}(x)\,
            dx^\m \swedge dx^\n
        \,+\,
        q\, d\mem\bbsq{\,
        \sum_{\ell=1}^\infty\mem
        \frac{1}{\ell!}\,\mem
            {*^\ell}\hnem F_{\r_1\s,\r_2\cdots\r_\ell}(x)\,
            \hy^{\r_1}\hy^{\r_2}\cdots\hy^{\r_\ell}\,
            dx^\s
        \mem}
    \,,\\[-0.1\baselineskip]
    {}&{}
        \,+\,
        q\, d\mem\bbsq{\,
        \sum_{\ell=1}^\infty\mem
        \frac{(\ell{\,-\mem}1)}{\ell!}\,\mem
            {*^\ell}\hnem F_{\r_1\s,\r_2\cdots\r_{\ell-1}}(x)\,
            \hy^{\r_1}\hy^{\r_2}\cdots\hy^{\r_{\ell-1}}\,
            d\hy^\s
        \mem}
    \,.
\end{split}
\end{align}

Compare \eqref{rkerr} with \eqref{EM.spell}.
For the linear couplings,
we identify the unity $c_\ell \eqq 1$ of multipole moments.
Thus \eqref{rkerr} passes the first test of spinning black holes,
criterion (a).
Furthermore,
\eqref{rkerr} also describes
nonlinear couplings
via
\textit{spin-derivative terms},
which are necessary for passing the second test of spinning black holes.\footnote{
    For instance,
    it can be diagrammatically shown \cite{njmagic.11}
    that the resulting same-helicity Compton amplitudes
    exhibit spin exponentiation for all multiplicities.
    Thus \eqref{sp.rkerr}
    is the symplectic perturbation
    of the {\kerr} spinning particle
    up to mixed-chirality terms
    (earthly deformations \cite{probe-nj})
    that vanish in the self-dual limit,
    such as $(q^2/m^2)\mem (\hy\mem F^+(x)\hem F^-(x)\mem\hy)\mem (p\mem dx)$.
}

The message here is that the simplest symplectic perturbation from spinspacetime
directly yields
the simplest (black hole) coupling.
This fact is obscured in typical spacetime-based approaches;
it is a unique insight that spinspacetime can provide.
Also, it should be clear that
this derivation of a black hole coupling
is model independent.

The traditional minimal coupling, $c_\ell \eqq 0$,
arises by
the simplest symplectic perturbation from the standards of spacetime:
$\tfrac{1}{2}\, qF_{\m\n}(x)\mem dx^\m \swedge dx^\n$.
The spinspacetime Poisson brackets in \eqref{M3-in-zp} shows that
this grossly complicates the dynamics of the particle,
as $dx \swedge dx$
contains all the
$dz \swedge dz$,
$dz \swedge d\bz$,
and
$d\bz \swedge d\bz$
components.

\subsection{Universal Derivation of Root-Kerr Equations of Motion}

Via \eqrefs{ppert}{M3-in-zp},
it is easy to 
derive the complete {\kerr} EoM 
due to the real symplectic perturbation in \eqref{sp.rkerr}:
\begin{align}
\begin{split}
    \label{eom.earthly}
    \dot{p}_\m
    \,&=\,
        qF^+_{\m\n}(z)\mem \dot{z}^\n
        + qF^-_{\m\n}(\bz)\mem \dot{\bz}^\n
    \,,\\
    \dot{z}^\m
    \,&=\,
        \hp^\m
        - \frac{2iq}{m^2}\,
        \BB{
            \hy^\m p^\n + p^\m \hy^\n
            -i\mem \ve^{\m\n\r\s} \hy_\r p_\s
        }\mem
        F^-_{\n\k}(\bz)\mem \dot{\bz}^\k
    \,.
\end{split}
\end{align}
The EoM for $\bz^\m$ follow by complex conjugation
since the dynamics is real.
The above equations have been known from the massive twistor model \cite{Kim:2024grz}.
Here, we have reproduced them in the model-agnostic fashion.
Note that \eqref{eom.earthly} describes \textit{all-orders-in-spin} EoM.

It is left as an exercise to
deduce the EoM for $x^\m$ and $y^\m$
by taking real and imaginary parts.
One may use an identity
$
\ve^{\m\n\r\s} \phi^\pm_{\n\k}\mem u^\k
= \mp\mem 3\hhem i\mem \phi^\pm{}^{[\r\s} u^{\m]}
$
to remove the epsilon tensors
if wanted,
where $\phi^+_{\m\n}$\,($\phi^-_{\m\n}$) is self-dual\:(anti-self-dual).

Overall, spinspacetime is a powerful tool
for making predictions on the all-orders-in-spin dynamics of spinning objects.

As demonstrated in \rcite{ambikerr1},
spinspacetime is also useful for
deriving the all-orders-in-spin EoM of particles with generic multipole moments.
In this case,
one perturbs away from the black hole coupling by
redefining the Wilson coefficients on the complex worldline.

\section{
Dynamical Symmetries and Conserved Charges}
\label{APPL>Q}

This section demonstrates how
model-ignorant conserved charges
are directly uncovered from
the universal Poisson and covariant Poisson bracket relations.
For simplicity, we limit our attention to spacetime isometries.

\subsection{Symmetry in Hamiltonian Mechanics}

Let us first review the definition of symmetry in Hamiltonian mechanics.
Suppose a Hamiltonian system defined by
a symplectic form $\omega$ and a Hamiltonian $H$
on a phase space.
A vector field $V$ implements
an infinitesimal symmetry transformation on
this system iff
both of the following conditions are met:
\begin{subequations}
\label{symdef}
\begin{align}
\label{symdef.symp}
    \text{Symplectic Condition}:
    &\quad
    \pounds_V\hem \omega
    \,=\,
        0
    \,,
    \,\,
    \\
\label{symdef.ham}
    \text{Hamiltonian Condition}:
    &\quad
    \pounds_V H
    \,=\,
        V\act{ H }
    \,=\,
        0
    \,.
    \,\,
\end{align}
\end{subequations}

Due to the Cartan formula
and the closure of $\omega$,
the symplectic condition
translates to $d(\i_V\hem\omega) = 0$,
which implies local existence of a function $Q$
such that
\begin{align}
    \label{symQ}
    \i_V\hem\omega \,=\, -dQ
    \qfq
    V \,=\, \pb{\blank}{Q}
    \,.
\end{align}
Here, $\pb{\blank}{f} = \omega^{-1}(\blank,df)$ denotes the Hamiltonian vector field of $f$.
This shows that symmetries in Hamiltonian mechanics
are realized as Hamiltonian actions on the phase space.
In other words, we have approached the Noether theorem
in the symplectic geometry language:
the function $Q$
is the Noether charge.

On the other hand, the Hamiltonian condition
encodes compatibility with the system's 
time-evolution generator
(or constraints if any).
Plugging in \eqref{symQ},
one finds that it encodes
Poisson commutativity of $Q$ and $H$:
\begin{align}
    \label{symH}
    \pb{H}{Q}
    \,=\,
    V\act{ H }
    \,=\,
        0
    \,.
\end{align}

We shall also note that the above definition
does not invoke the symplectic potential $\theta$.
Famously, the symplectic potential 
involves redundancies $\theta \sim \theta + \Del\theta$
for any closed one-form $\Del\theta$ on the phase space.
Thus it cannot provide an invariant definition of symmetry.

Instead, the role of the symplectic potential 
in this discussion
would be
a customary gadget
that provides a convenient formula
for the Noether charge.
By using the freedom $\theta \sim \theta + \Del\theta$,
one may find a particular $\theta$ such that
$\pounds_V\hhem \theta = 0$.
In this case, the Cartan formula asserts
$\i_V\hem\omega = -d\i_V\hhem \theta$,
leading to a concrete identification of the Noether charge as
\begin{align}
    \label{charge-from-theta}
    \pounds_V\hhem \theta \,=\, 0
    \qiq
        Q \,=\, \i_V\hhem \theta
    \,.
\end{align}
In fact, such a symmetry-preserving symplectic potential
$\theta$ will give rise to a natural, symmetry-adapted phase space action.

\subsection{Symmetry-Preserving Symplectic Perturbations}

Next, we develop a general theory of symmetries
in the framework of symplectic perturbations.
Suppose $(\ps,\omega^\circ,H)$ defines the free theory of a Hamiltonian system.
Suppose a vector field $V$ on $\ps$
implements an infinitesimal symmetry transformation on $(\ps,\omega^\circ,H)$,
whose Noether charge is $Q^\circ$:
\begin{align}  
\label{sym0}
    \i_V \omega^\circ
    \,=\,
        -dQ^\circ
    \,,\quad
    \pb{H}{Q^\circ}^\circ
    \,=\,
        V\act{ H }
    \,=\,
        0
    \,.
\end{align}

A closed two-form $\omega'$ on $\ps$
is a \textit{symmetry-preserving symplectic perturbation}
iff
\begin{align}
    \label{sym.SPT}
    \pounds_V\hem \omega'
    \,=\,
        0
    \,.
\end{align}
Again, 
through the Cartan formula
and $d\omega' \eqq 0$,
\eqref{sym.SPT}
translates to $d(\i_V\hem\omega') = 0$,
which implies local existence of a function $Q'$
such that
\begin{align}
    \label{sym'Q}
    \i_V\hem\omega' \,=\, -dQ'
    \,.
\end{align} 
Via \eqref{sym0}, \eqref{sym'Q} implies that
\begin{align}
    \label{sym1}
    \i_V\hem\omega \,=\, -dQ
    \,,\quad
    \pb{H}{Q}
    \,=\,
        V\act{
            H
        }
    \,=\,
        0
\end{align}
where $Q = Q^\circ + Q'$ and $\omega = \omega^\circ + \omega'$.

The implication of \eqrefs{sym0}{sym1}
is that
$V$,
the \textit{same} vector field from the free theory $(\omega^\circ,H)$,
implements an infinitesimal symmetry transformation
on the interacting theory $(\omega,H)$ as well,
yet via the modified Noether charge
$Q = Q^\circ + Q'$ in \eqref{sym1}:
\begin{align}
    \label{Vsame0}
    V
    \,=\,
        \pb{\blank}{Q^\circ}^\circ
    \,=\,
        \pb{\blank}{Q}
    \,.
\end{align}

Further, suppose the free theory admits a number of symmetry charges forming an algebra.
The above analysis shows that the symmetry actions on the phase space 
(as the vector fields)
remain the same
under symmetry-preserving symplectic perturbations.
As a result, the symmetry algebra remains the same
in the interacting theory.

\subsection{Scalar Particle in Electromagnetism}

For a concrete demonstration of the above abstract formalization,
we could revisit the familiar example of the charged scalar particle
described in \Sec{INT>SPT}.

The free theory is defined in \eqref{scalar.omega0},
which supposes
the phase space $T^*\mflat$
with coordinates $x^\m$ and $p_\m$.
The symplectic form is $\omega^\circ = dp_\m \swedge dx^\m$,
while the mass-shell constraint $\phi_0 = \frac{1}{2}\mem (\eta^{\m\n} p_\m p_\n \mplus m^2)$ serves as the Hamiltonian.
A well-known fact is that
\begin{align}
    \label{V0}
    V
    \,=\,
        K^\m(x)\, \frac{\partial}{\partial x^\m}
        - p_\n\mem K^\n{}_{,\m}(x)\, \frac{\partial}{\partial p_\m}
\end{align}
generates a symmetry transformation
on $(T^*\mflat,\omega^\circ,\phi_0)$
iff $K^\m(x)\mem \partial/\partial x^\m$
is a Killing vector of flat spacetime $\mflat \eqq (\R^4,\eta)$.
To elaborate,
the symplectic condition
is trivially satisfied by the ansatz in \eqref{V0},
identifying the Noether charge as
\begin{align}
    Q^\circ
    \,=\,
        p_\m\mem K^\m(x)
    \,.
\end{align}
The Hamiltonian condition
then yields the
Killing equation in flat spacetime,
\begin{align}
    \label{killing0}
    \pounds_K \eta
    \,=\, 0
    \qfq
    K_{\m,\n}(x) \,=\, -K_{\n,\m}(x)
    \,.
\end{align}

Now consider an electromagnetic field strength $F$
on flat spacetime
such that
\begin{align}
    \label{isometry-F}
    \pounds_K F \,=\, 0
    \,.
\end{align}
Via the Cartan formula
and the closure $dF = 0$,
\eqref{isometry-F} implies the local existence of a scalar field $\a$ such that \cite{Hughston:1972qf}
\begin{align}
    \label{symQ.EM}
    \i_K F \,=\, -d\a
    \qfq
    K^\m(x)\mem F_{\m\n}(x)
    \,=\,   
        -\a_{,\n}(x)
    \,.
\end{align}

Crucially, the condition in \eqref{isometry-F}
implies that
$\omega' \eqq qF$
(now viewed as a two-form on $\ps$
via the pullback of the bundle projection)
is a symmetry-preserving symplectic perturbation.
Namely, it is easy to show that $\pounds_V\hhem \omega' = 0$ by $\pounds_K F = 0$.
According to \eqref{sym1},
the correction to the Noether charge  
can be taken as
$Q' = q\hem \a$,
as
$-dQ' = \i_V\hem \omega' = -q\mem d\a$.
Therefore, the exact Noether charge
in the interacting theory is
\begin{align}
    \label{scalar.QEM}
    Q
    \,=\,
        p_\m\mem K^\m(x)
        + \a(x)
    \,.
\end{align}
\eqref{scalar.QEM} is precisely
the conserved charge of the charged scalar probe
\cite{Hughston:1972qf}.

To derive the perhaps more widely-recognized formula
$Q = (\hem{p_\m + qA_\m(x)}) $ $ K^\m(x)$,
note that the scalar field $\a$ in \eqref{symQ.EM} can be explicitly given as follows
if one is willing to make an explicit gauge choice:
\begin{align}
    \pounds_K A \,=\, 0
    \qiq
    \a(x)
    \,=\,
        q A_\m(x)\mem K^\m(x)
    \,.
\end{align}

\subsection{Spinning Particles in Electromagnetism}

Now we are ready to derive the \textit{universal} (model-ignorant) conserved charges
of massive spinning particles
in electromagnetism.

Suppose any Hamiltonian formulation of a free massive spinning particle
based on a phase space $(\ps,\omega^\circ)$.
Suppose infinitesimal symmetry $V$ exists for this particle.
Our ansatz for the symmetry action on the physical variables is
\begin{align}
\begin{split}
    \label{V0.act}
    V\act{ x^\m }
    \,&=\,
    \pb{x^\m}{Q^\circ}^\circ
    \,=\,
        K^\m(x)
    \,,\\
    V\act{ \hy^\m }
    \,&=\,
    \pb{\hy^\m}{Q^\circ}^\circ
    \,=\,
        K^\m{}_{,\n}(x)\mem \hy^\n
    \,,\\
    V\act{ p_\m }
    \,&=\,
    \pb{p_\m}{Q^\circ}^\circ
    \,=\,
        - p_\n\mem K^\n{}_{,\m}(x)
    \,.
\end{split}
\end{align}
To validate this ansatz,
we apply the two conditions in \eqref{symdef}.
First, the symplectic condition is solved by
taking the Noether charge as
\begin{align}
    \label{Q0.univ}
    Q^\circ 
    \,=\,
        p_\m\mem K^\m(x)
        - \frac{1}{2}\:
            \ve^{\m\n\r\s} \hy_\r\mem p_\s
            \mem K_{\m,\n}(x)
    \,,
\end{align}
which reproduces \eqref{V0.act} via the
universal Poisson bracket relation in \eqref{xyp}
provided that $K^\m(x)$ exhibits linearity:
$K^\m{}_{,\r\s}(x) = 0$.
Second, the Hamiltonian condition is 
satisfied if $K_{\m,\n}(x) = -K_{\n,\m}(x)$,
as \eqref{V0.act} implies that
the symmetry action on the mass-shell constraint
$\phi_0 \eqq \frac{1}{2}\,(p^2\mplus m^2)$
is $V\act{\phi_0} = -p_\n\mem K^\n{}_{,\m}(x)\mem p^\m$.
Crucially, since $Q^\circ$ in \eqref{Q0.univ} is composed solely of the physical variables $(x^\m,\hy^\m,p_\m)$,
it must Poisson-commute with any gauge constraints
of the model.
Consequently, the Hamiltonian action of \eqref{Q0.univ}
passes both the symplectic and Hamiltonian tests of symmetry in \eqref{symdef}.

Note that this fact could be shown
without writing down
an explicit formula for $V$
(which will depend on the microscopic implementations, i.e., the details of the model).

It follows that
the symplectic perturbation $\omega' \eqq qF$
due to the universal minimal coupling recipe 
in \eqref{EM.sp0}
is symmetry-preserving
as long as the electromagnetic background $F$
exhibits symmetry \`a la \eqref{isometry-F}.
Since \eqref{V0.act} implies
$\i_V dx^\m = V\act{x^\m} = K^\m(x)$,
we have
(with abuse of notation regarding the pullbacks due to bundle projection)
\begin{align}
\begin{split}
    \pounds_V \hem\omega'
    \,=\,
    d\mem \i_V \omega'
    \,&=\,
        q\mem d\BB{
            K^\m(x)\mem
                F_{\m\n}(x)\mem dx^\n
        }    
    \,,\\   
    \,&=\,
        q\mem d\mem \i_K F
    \,=\,
        q\mem \pounds_K F
    \,=\,
        0
    \,.
\end{split}
\end{align}
Therefore, the same vector field $V$
still defines a symmetry in the interacting theory.

It remains to find the Noether charge.
Since
$-dQ' = \i_V\hem \omega' = q\mem \i_KF = -q\mem d\a$,
so we can take $Q' = q\mem \a(x)$.
Thus,
we conclude that
minimally coupled 
massive spinning particles
in electromagnetism
universally exhibit the conserved charge
\begin{align}
    \label{Q.EM0}
    Q
    \,=\,
        p_\m\mem K^\m(x)
        - \frac{1}{2}\:
            \ve^{\m\n\r\s} \hy_\r\mem p_\s
            \mem K_{\m,\n}(x)
        + q\mem \a(x)
    \,,
\end{align}
where the gauge choice $\pounds_K A = 0$
will take $\a(x) = A_\m(x)\mem K^\m(x)$.

To reiterate,
the Poisson commutativity of
\eqref{Q.EM0}
with the various constraints of the particle
is automatic
since the infinitesimal symmetry transformation $V$ is left unchanged.
Still, it is a nice and interesting exercise to
reconfirm $\pb{H}{Q} = 0$
by direct computation,
by using the exact universal
EoM derived in \eqref{uBMT},
for instance.

Incorporating the multipolar couplings is not difficult.
For instance,
the non-minimal coupling
in \eqref{EM.sp1}
adds
$\theta' = (g/2)\mem q\, {*}F_{\m\n}(x) $ $\hy^\m\mem dx^\n$
on the symplectic potential.
Computation using \eqref{V0.act} shows that
$\pounds_V\hhem \theta' = 0$
by $\pounds_K {*}F = 0$.
Here, $\pounds_K$ preserves the Hodge star
because it preserves the flat metric.
This shows that $\theta'$ is a symmetry-preserving perturbation on the symplectic potential.
Based on our earlier discussion around \eqref{charge-from-theta}, 
this implies that the universal conserved charge
with the generic gyromagnetic ratio $g$ is
\begin{align}
\begin{split}
    Q
    \,=\,
    {}&{}
        p_\m\mem K^\m(x)
        - \frac{1}{2}\:
            \ve^{\m\n\r\s} \hy_\r\mem p_\s
            \mem K_{\m,\n}(x)
    \\
    {}&{}
        + q\mem \a(x)
        + \frac{g}{2}\:
            q\, {*}F_{\m\n}(x)\mem \hy^\m\mem K^\n(x)
    \,.
\end{split}
\end{align}

\section{
Dynamical Newman-Janis Shift of Conserved Charges}

As a glimpse toward the applications of the above EoM,
we provide a reproduction of
the fact \cite{njmagic.11} that 
an idealized version of the spinning black hole binary problem
exhibits superintegrability
via exact hidden symmetries.
We follow the model-independent presentation due to \rcite{probe-nj}.
This discussion confirms the criterion (b),
confirming that \eqref{rkerr} qualifies as an effective point-particle description of {\kerr}.

The dream of a theoretical relativist
is to understand the exact dynamics of the
Kerr-Kerr black hole binary system.
However, the problem is too hard.
As a first simplification,
we suppose that one of the black holes is significantly heavier than the other:
the zeroth self-force limit.
As a second simplification,
we suppose the electromagnetic analog:
the {\kerr}-{\kerr} binary problem.

As a result, we study the motion of a {\kerr} probe
in the background of the {\kerr} solution.
Since Newman and Janis \cite{Newman:1965tw-janis},
it has been available that the field strength of the {\kerr} solution takes the form
\begin{align}
    \label{bkg.F}
    F_{\m\n}(x)
    \,=\,
        \frac{Q}{8\pi|\vex \mminus i\vea|^3}\:
            {*}Y^+_{\m\n}(x)
    \mem+\mem
        \frac{Q}{8\pi|\vex \mplus i\vea|^3}\:
            {*}Y^-_{\m\n}(x)
    \,,
\end{align}
where $Q$ is the charge parameter,
$a^\m$ is a spacelike pseudovector encoding the ring radius,
and $u^\m$ is a timelike unit vector defining the stationary direction.
We have denoted
\begin{align}
    {*}Y^\pm_{\m\n}(x)
    \,=\,
        \bigbig{
            u \wedge (x\mp ia)
        }\hnem{}^\pm_{\m\n}
    \,,
\end{align}
which
encodes the position three-vector
from imaginary centers $\pm ia$.
As disclosed by \rcite{nja},
these centers are precisely
self-dual and anti-self-dual dyons.

The problem is still too hard.
With both self-dual and anti-self-dual fields,
the geometric series expansion of the Poisson bivector 
does not truncate at a finite order,
so \eqref{eom.earthly} describes time derivatives on its right-hand sides
in contrast to \eqref{eom.heavenly.bz}.

Hence we simplify the problem again by dropping
the anti-self-dual part of the field configuration in \eqref{bkg.F}:
\begin{align}
    \label{bkg.Fp}
    F_{\m\n}(x)
    \,=\,
        \frac{Q}{8\pi|\vex \mminus i\vea|^3}\:
            {*}Y^+_{\m\n}(x)
    \qiq
    {*}F_{\m\n}(x)
    \,=\,
        +i\mem F_{\m\n}(x)
    \,.
\end{align}
This extracts the self-dual part of the {\kerr} background in the precise sense;
see \rcite{note-sdtn} for the gauge potential.
It is not difficult to see that \eqref{bkg.Fp} describes
\begin{align}
    \vec{E}(\vex)
    \,=\,
        i\hem \vec{B}(\vex)
    \,=\,
        \frac{Q}{8\pi}\mem
        \frac{\vex \mminus i\vea}{|\vex \mminus i\vea|^3}
    \,,
\end{align}
which is the electromagnetic field of a dyon
with electric charge $Q/2$ and magnetic charge $-i\hem Q/2$:
a self-dual dyon.
The magnetic dipole moment of {\kerr} arises precisely like
a \textit{static} magnet made of monopoles:
$Q\vea = (-iQ/2)(2i\vea)$.
See \rcite{nja} for more details.

In short, we study
the motion of a {\kerr} probe
in the background of a self-dual dyon.
The relevance to the original problem
is that the self-dual dyon
is a part of {\kerr}.

In the self-dual dyon background,
we concern the combination
\begin{align}
    Y_{\m\n}(x)
    \,=\,
        Y^+_{\m\n}(x)
        \mem+\mem
        Y^-_{\m\n}(x)
    \,=\,
        \ve_{\m\n\r\s}\mem x^\r\hem u^\s
        - (u \wedge a)_{\m\n}
    \,.
\end{align}
It turns out that this defines a Killing-Yano tensor \cite{yano1952some} such that
\begin{align}
\begin{split}
    &
    Y_{\m\n,\r}(x)
    \,=\,
        \ve_{\m\n\r\s}\mem u^\s
    \,=\,
        Y_{[\m\n,\r]}(x)
    \,,\\
    &
    F^\m{}_\r(x)\mem Y^\r{}_\n(x)
    \,=\,
        Y^\m{}_\r(x)\mem F^\r{}_\n(x)
    \,.
\end{split}
\end{align}
The first equation is the definition statement,
while the second equation describes a commuting property with the field strength.

In this background,
the {\kerr} probe
obeys
the complexified EoM in \eqref{eom.heavenly}:
\begin{align}
    \label{eom.heavenly.sdd}
    \dot{z}^\m
    \,=\,
        \hp^\m
    \,,\quad
    \dot{p}^\m
    \,=\,
        \frac{q}{m}\, F^\m{}_\n(z)\mem p^\n
    \,,\quad
    \dot{\hy}^\m
    \,=\,
        \frac{q}{m}\, F^\m{}_\n(z)\mem \hy^\n
    \,,
\end{align}
Notably, the first two equations here
are identical to those of the charged scalar particle
up to the complexification $x \to z$,
which originated from the Poisson commutativity of holomorphic spinspacetime in \eqref{zz=0}.
Thus the well-known discourse on the scalar particle \cite{Hughston:1972qf} 
is straightforwardly recycled as
\begin{align}
\begin{split}
    \frac{d}{d\t}\mem\BB{
        Y^\m{}_\n(z)\mem p^\n
    }
    \,&=\,
        Y^\m{}_\n(z)\mem F^\n{}_\r(z)\mem p^\r
        + Y^\m{}_{\n,\r}(z)\mem p^\n\mem \hp^\r
    \,,\\
    \,&=\,
        F^\m{}_\n(z)\mem Y^\n{}_\r(z)\mem p^\r
    \,,
\end{split}
\end{align}
showing that the precessions of
$p$ and $Y(z)\mem p$ are synchronized.

Crucially,
the precession of the spin length pseudovector $\hy$ is also synchronized with $p$.
Now we find three vectors singing in unison,
$p$, $Y(z)\mem p$, and $\hy$.
It is then immediate that
\begin{align}
    \label{rkerr.RC}
    R \,=\, p_\m\mem Y^\m{}_\n(z)\mem \hy^\n
    \,,\quad
    C \,=\, - p_\m\mem Y^\m{}_\r(z)\mem Y^\r{}_\n(z)\mem p^\n
\end{align}
are conserved quantities:
the all-orders-in-spin completions of
the R\"udiger \cite{Rudiger:1981uu,Rudiger:1984er} and Carter \cite{Carter:1968ks} constants.

Moreover, the self-dual dyon background exhibits isometries $\R \times \SO(3)$.
Again, we immediately deduce the {\kerr} probe's conserved charge 
by direct dynamical NJ shift of the Coulomb probe's conserved charge in \eqref{scalar.QEM}:
\begin{align}
    \label{rkerr.Q}
    Q
    \,=\,
        p_\m\mem K^\m(z) + \a(z)
    \,.
\end{align}
The symmetry algebra between these NJ shifted Killing charges
are the same as in the non-spinning case,
crucially because of the Poisson commutativity
$\pb{z^\m}{z^\n} = 0$
that we have had to emphasize.

Using
\eqrefs{rkerr.RC}{rkerr.Q},
one can show that
the dynamics is exactly solvable (integrable).
See \rrcite{probe-nj,njmagic.11} to grasp more details for the superintegrability.

In summary, we have learned that
the self-dual sector of the {\kerr}-{\kerr} binary problem is exactly solvable,
based on
direct dynamical NJ shifts of the non-spinning probe's conserved charges
as well as
the synchronization of orbital and spin precessions.

This is yet an idealized discussion.
How should one approach the original problem of the {\kerr}-{\kerr} binary system, then?

A reasonable pathway may be perturbing away from the self-dual sector
where exact all-orders expressions for the conserved charges are known:
the googly agenda \cite{probe-nj}.

\section{
Conclusion and Outlook}

In this chapter,
we have demonstrated 
the utility of spinspacetime
in deriving exact spinning-particle EoM.
Spinspacetime facilitates systematic all-orders extensions of the BMT equations.
Moreover, spinspacetime 
proves to be a powerful tool for
deriving and studying 
black hole EoM.
The inherent association between chirality and holomorphy
that it entails
discovers hidden simplicities of spinning black hole mechanics.
The spinspacetime framework
also provides a model-independent 
approach to
conserved charges.
We established
dynamical NJ shift of conserved charges
for spinning black hole probes
and concretely (re)demonstrated
the maximal superintegrability
in a SD subsector of the black hole binary problem,
as a response to \Chap{K3:HYDROGEN}
where the simplicity of the SD sector is framed with the keyword ``hydrogen.''

Further promising developments
have been
underway.
The theory of curved spinspacetime is viable.
The derivation of spinning black hole EoM in YM and gravity backgrounds
can utilize the ``googly'' idea,
which is the chiral approach of understanding the full nonlinear dynamics
demonstrated in \Sec{Jcx}
applied to spinning particles.
In the spinspacetime framework,
this means to perturb away from the holomorphic worldline
where SD dynamics resums onto
(``perturbatively add zag on zig'').
Compton amplitudes are computed
via worldline methods,
again by implementing 
a chiral perturbation theory
in the perturbative googly spirit;
see \Chap{K3:COMPTON}.

\begin{subappendices}

\section[
    Definition of Black Hole Coupling
]{
Definition of Black Hole Coupling}
\label{NJB}

Eventually,
we are now in a position to
review the developments
surrounding the question,
\begin{center}
    \vphantom{.}\llap{``\:}\textit{What defines black holes}
    among all massive \\ spinning objects
    in the point-particle effective theory?\rlap{\:''}
\end{center}
We will only cover the IR approaches within the point-particle effective theory,
although a large volume of the current literature
pursues
the determination of black hole coupling
from the UV (i.e., matching with GR).

\smallskip
\begin{center}
\begin{minipage}{0.7\linewidth}
\vphantom{.}\llap{1.\: }
    In \textit{linearized gravity},
    the Kerr black hole is defined by
    the \textit{unity of multipole coefficients}
    (i.e., \textit{three-point spin exponentiation}).
\end{minipage}
\end{center}

This result is due to the classic works
\cite{%
    Hansen:1974zz,Geroch:1970cd,%
    Newman:1965tw-janis,janis1965structure,%
    Hernandez:1967zza,%
    Thorne:1980ru%
}
in the '60s and '70s:
the Kerr solution
is shown to exhibit
$C_\ell \eqq 1$ for all $\ell \eqq 1,2,3,\cdots$.

According to the traditional definition,
this seems like a grossly non-minimal coupling.
However, a modern analysis
\cite{ahh2017,Guevara:2018wpp,Guevara:2019fsj,chkl2019,aho2020}
shows that it is the simplest coupling
in terms of 
a large simplification of
the three-point graviton scattering amplitudes
(cf. \rcite{Holstein:2006wi}).
This simplicity is represented as
a complexification of the impact parameter
that adds spin length as an imaginary component
and is dubbed the \textit{three-point spin exponentiation}.

To elaborate,
these three-point amplitudes
describe the process of the black hole (as a point-particle source) sourcing a single massless quantum.
Thus,
they are synonymous to the linearized part of the Kerr metric
\cite{Duff:1973zz,Neill:2013wsa,vines2018scattering}.
In this manner,
it is shown that
the three-point spin exponentiation
is the NJ shift 
in linearized gravity \cite{aho2020}.

The NJ shift
refers to the property 
discovered by Newman and Janis \cite{Newman:1965tw-janis} in 1965
that the Kerr black hole is secretly
a Schwarzschild black hole
displaced into ``complex spacetime''
in some rough sense,
in which case the spin length (as ring radius) describes the imaginary direction.
This trick was the very method
that enabled the historical discovery of the Kerr-Newman solution \cite{Newman:1965my-kerrmetric}.
The aspects of NJ shift
in linearized gravity
have been well-studied from the modern scattering amplitudes perspective by \rrcite{ahh2017,Guevara:2018wpp,Guevara:2019fsj,chkl2019,aho2020},
while
its nonlinear aspect has been recently elucidated by \rcite{nja}.

\begin{center}
\begin{minipage}{0.7\linewidth}
\vphantom{.}\llap{2.\: }
    In \textit{SD gravity},
    the Kerr black hole is defined by
    the \textit{$n$-point spin exponentiation}.
\end{minipage}
\end{center}

At the nonlinear orders,
it has been unclear
what principle in the point-particle effective theory 
defines spinning black holes.
Meanwhile,
explorations via scattering amplitudes
investigated the $n$-point Compton scattering process
in which the Kerr black hole
receives $(n \mminus 2)$ gravitons,
where $n \geq 3$.
This research showed that
the $n$-point gravitational Compton amplitudes
can exhibit the spin exponentiation property
as a theoretically allowed possibility,
iff all the helicities of the gravitons are the same
\cite{Johansson:2019dnu,Aoude:2020onz,Lazopoulos:2021mna}.\footnote{
    The mixed-helicity Compton amplitudes are plagued by a contact term issue,
    first noticed in \rcite{ahh2017}.
}

Therefore, by extending the definition of the simplicity of scattering amplitudes,
it has been believed that
this ``$n$-point spin exponentiation''
for all $n = 3,4,5,6,\cdots$
may define the black hole coupling at all nonlinear orders
within the \textit{same-helicity} sector
as a beautiful possibility,
yet without a necessary physical reason.

This scattering amplitudes statement
is equivalent to the proposition that
the point-particle effective action of the Kerr black hole
is unique
(up to worldline field redefinitions)
in background SD spacetimes.
A SD spacetime 
describes a formal complexified limit of spacetime
in which the Riemann tensor becomes SD:
${*}R_{\m\n\r\s} = {+i\hem R_{\m\n\r\s}}$.
A well-established fact is that
the same-helicity gravitational Compton amplitudes
due to incoming positive-helicity gravitons
describe
scattering processes in 
SD background spacetimes;
see Figure 1 in \rcite{GravityMHVTwistors}
and also 
\rrcite{bialynicki1981note,ashtekar1986note}.

This unique effective action has been rigorously established in \rrcite{probe-nj,njmagic.1}.
The lesson is that
the $n$-point spin exponentiation
for positive-helicity gravitons
amounts to
a nonlinear generalization of
NJ shift
for black hole probes
in SD spacetimes.

In sum,
three-point spin exponentiation
is NJ shift 
in linearized gravity,
while
$n$-point spin exponentiation
is nonlinear NJ shift
in SD gravity.

It has been recently shown \cite{probe-nj} that
hidden symmetries (superintegrability) in the SD sector
implies the spin exponentiation property of same-helicity Compton amplitudes to all multiplicities
\cite{probe-nj}.\footnote{
    \rcite{probe-nj}
    also provided a physical interpretation of
    the $n$-point spin exponentiation:
    the helicity selection rule \cite{Adamo:2023fbj} exhibited by
    the SD Taub-Newman-Unti-Tamburino solution,
    which constitutes part of the Kerr black hole
    \cite{nja}.
}
This result has promoted
the $n$-point spin exponentiation
from a mere beauty statement
to a physical assertion based on global symmetries.

\begin{center}
\begin{minipage}{0.7\linewidth}
\vphantom{.}\llap{3.\: }
    In \textit{full gravity},
    the Kerr black hole is defined by
    \textit{(?)}.
\end{minipage}
\end{center}

The unique characterization in the SD sector
does not pinpoint the complete black hole couplings
in full nonlinear gravity
due to mixed-helicity interactions.
The extent of this ambiguity
is explicitly examined in
the analysis of \rcite{probe-nj}.

So far, the construction of the exact worldline effective theory of the Kerr black hole 
may have been 
based on model-specific perspectives.
An early sketch was given in \rcite{gmoov}
by extending
the Levi-Steinhoff action
in the spherical top model,
which identified the NJ shift 
as an important clue.
This work provided valuable intuitions and insights,
although
the subtleties of spin gauge redundancy,
geodesic deviation, and integrability of complex structures
are not 
clearly addressed
while
explicit expressions are limited to the linear-in-Riemann order.
In the meantime, \rcite{ambikerr1} provided an approach based on
the K\"ahler geometry of massive twistor space,
unearthing a unique geometrical perspective on the NJ shift
that was also demonstrated in \rcite{sst-asym}.

Eventually,
\rrcite{njmagic.1,njmagic.11}
have established
rigorous constructions of
the exact worldline actions for
Kerr and Kerr-Newman black holes
in the massive twistor model,
providing 
their explicit formulae
to all orders
via the systematic formalism of \rcite{gde}.
As clearly portrayed in 
\rcite{njmagic.1},
this construction
envisions
a curved generalization of
the magical feature of twistor particle theory
that spin is literally an imaginary
deviation in terms of the complexified incidence relation \cite{Shirafuji:1983zd,penrose:maccallum,newman1974curiosity}.
The dynamical NJ shift
is implemented by
a solid mathematical framework known as
adapted complex structure
\cite{guillemin1992grauert,halverscheid2002complexifications,hall2011adapted},
refining \rcite{gmoov}'s earlier sketch.

While 
\rrcite{njmagic.1,njmagic.11}
paved a top-down pathway
to the problem
from massive twistor theory,
\rcite{probe-nj} provided a complementary, bottom-up view
that emphasizes model independence.
\rcite{probe-nj}
defines
an EoM-level implementation of
the NJA
in terms of the universal variables $(x,\hy,p)$
of the spinning probes.
A corollary is
the aforementioned hidden symmetry in the SD sector.
Notably,
the conserved charges of the spinning probe
are directly obtained
by dynamically NJ shifting
the conserved charges of the scalar probe.

In this universal manner,
\rcite{probe-nj} also suggests
a natural proposal for uplifting 
the SD sector couplings
to the full nonlinear sector
in a rather model-independent fashion.
In this case, the principle that defines black holes
is ``orbit-spin duality,''
a generalization of an old idea due to Newman and Winicour \cite{newman1974curiosity}
in curved backgrounds
\cite{sst-asym,probe-nj,njmagic.1}.
A dedicated article will appear soon \cite{sodual}
for an elaboration on this point.

\end{subappendices}

\chapter{Black Hole Compton Amplitudes}
\label{K3:COMPTON}

    We compute the classical
    Compton scattering amplitudes
    of the nonabelian {\kerr} and Kerr solutions
    receiving two gluons or gravitons
    in YM theory or GR,
    to all orders in spin.
    Our amplitudes
    are completely free of any spurious pole
    and exhibit correct factorizations
    for all helicity configurations.
    They are systematically structured in terms of
    truncated exponential functions.
    In fact, our nonabelian {\kerr} amplitude
    seems to be unique
    when restricting to
    a certain class of ansatz
    due to Newton's interpolation formula.

    \smallskip
    \noindent
    Crucially, we provide a Lagrangian derivation of these amplitudes 
    based on explicit local-in-worldline-time point-particle effective actions
    that exactly resum onto holomorphic or anti-holomorphic worldlines
    in SD or ASD backgrounds,
    respectively.
    These resummations realize the probe NJ shift,
    boiling down to the same-helicity spin exponentiation to all orders in perturbation theory.
    At the technical level,
    we implement the worldline formalism in massive twistor space
    in the backgrounds of nonlinearly superposed plane waves in YM theory and GR
    while making a fruitful use of
    axial gauge and
    covariant CK duality.

    \smallskip
    \noindent
    Physically,
    the holomorphic worldline is naturally interpreted as
    the infrared counterpart of the ASD dyon or Taub-NUT instanton
    constituting {\kerr} and Kerr,
    based on a helicity selection rule.
    Accordingly,
    we implement a chiral perturbation theory
    in which the worldline actions are represented by
    perturbing away from
    the holomorphic worldline.
    In this formalism,
    spin exponentiation
    is manifestly guaranteed
    in the positive-helicity sector
    to all orders in perturbation theory.
    The negative-helicity modes are then perturbatively added in a systematic fashion.
    Notably,
    our formalism
    provides a direct realization of the twistor theorists' aspiration of
    achieving {\kerr} or Kerr
    by perturbing away from ASD dyon or ASD Taub-NUT:
    ``MHV-ing spinning black holes,''
    to say.

    \smallskip
    \noindent
    Our framework does not preclude
    adding mixed-chirality contact deformations
    for Kerr.
    In any case,
    our explicit worldline actions and their chiral presentations
    establish
    a unique and powerful formalism
    that facilitates
    a systematic addition of 
    mixed-chirality contact deformations
    without spoiling the same-helicity 
    spin exponentiation
    at any perturbative order.

\section{
    Introduction
}
\label{COMPTON>INT}

In \Sec{Ib3},
we addressed the startling fact in contemporary relativity
that no physicist is aware of the exact EoM of
spinning black holes
in their point-particle effective theory,
even in the probe limit.
This is
a fairly simple question,
at least conceptually,
and describes a one-body problem
in background, nondynamical geometry.
In particular,
any hope toward tackling the spinning black hole two-body problem
is obstructed by this fundamental issue already.

\begin{figure}[t]
   \centering
   \includegraphics[scale=1.8]{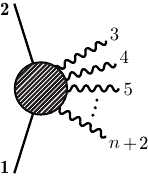}
   \caption{%
       The Compton scattering amplitude for massless multiplicity 
       $n$,
       which we refer to as 
       ``$n$-Compton amplitude.''
       The wavy lines represent massless quanta;
       the straight lines are massive legs.
       Note that we always presume equal masses
       for the massive legs
       in this thesis,
       since our scope is strictly restricted to conservative physics.
       This chapter
       investigates the cases
       where
       the straight lines
       represent spinning black holes
       in their point-particle effective theory.
   }
   \label{fig:int.Compton}
\end{figure}

As also mentioned
in \Sec{Ib3},
this problem was first formulated
in an invariant fashion
by AHH in 2017 \cite{ahh2017}.
Presuming sufficiently weak external fields,
the classical gravitational dynamics of an object
as a test particle is
completely
characterized by
specifying its gravitational $n$-Compton scattering amplitude
for all graviton multiplicities $n = 1,2,3,\cdots$
in the classical limit
(see \fref{fig:int.Compton}).
Evidently,
this characterization is not only gauge invariant but also field basis invariant.
Said in another way,
an object as a test particle
is uniquely defined by
a consistent collection of $n$-Compton amplitudes.
This consistency
must include conditions such as
correct residues on physical poles
and
the absence of unphysical poles,
in particular.

When one holds only a partial knowledge on the Compton data,
the object will be only unambiguously identified
up to a certain perturbative order.
For example,
it could suffice to restrict the multiplicity $n$ to $1$ and $2$
if one's goal is
to study gravitational physics up to the 2PM order
and not beyond.

AHH \cite{ahh2017}
provided a complete classification of
$1$-Compton amplitudes
for equal masses.
It was then established
that the ``simplest'' gravitational $1$-Compton amplitude
defines the Kerr black hole
in linearized gravity
\cite{Guevara:2018wpp,Guevara:2019fsj,chkl2019,aho2020}.
In fact, this rephrases
the unity $C_\ell = 1$ of Kerr multipole coefficients
known from classic works
\cite{%
    Hansen:1974zz,Geroch:1970cd,%
    Newman:1965tw-janis,janis1965structure,%
    Hernandez:1967zza,%
    Thorne:1980ru%
}.
After taking the classical limit for spin,
this simplicity describes
the celebrated ``spin exponentiation''
\cite{Guevara:2018wpp,Guevara:2019fsj}
property.
This means that the $1$-Compton amplitude
of the Kerr black hole,
say $\M_\text{spinning}(3^\pm)$
receiving a graviton 
$3^\pm$,
is related to
the $1$-Compton amplitude
$\M_\text{scalar}(3^\pm)$
of the minimally coupled scalar particle
(``Schwarzschild black hole'')
as
\begin{align}
\label{CompShift1}
\begin{split}
    \M_\text{spinning}(3^\pm)
    \,&=\,
        \M_\text{scalar}(3^\pm)
        \,
        \mathe^{\pm k_3\cdot a}
    \,,
\end{split}
\end{align}
where $a^\m = -y^\m$ 
encodes the spin length
of the Kerr black hole.

The natural aspiration of the amplitudes theorists
was then to
completely bootstrap the Compton data
for the Kerr black hole,
in which case the probe gravitational dynamics of 
the Kerr black hole
(or more precisely, the point-particle representing it in the effective theory)
is uniquely determined
to all orders in perturbation theory.
It was found that
the $n$-Compton amplitude
can be uniquely constructed
by gluing the $1$-Compton amplitude
via recursion
\cite{Johansson:2019dnu,Aoude:2020onz,Lazopoulos:2021mna},
iff all the incoming graviton helicities are the same.%
\footnote{
    In the relativist terminology,
    this means that
    the probe gravitational dynamics of Kerr
    could possibly be completely fixed
    in purely SD backgrounds.
    In \Chap{K3:PROBENJ} \cite{probe-nj},
    we have added a relativist axis on this discourse
    and formulated the probe NJA that uniquely determines the dynamics of Kerr in SD backgrounds.
    This probe NJA was then shown to be equivalent to 
    (i.e., is the relativist statement of)
    the same-helicity spin exponentiation.
}
Specifically,
the spin exponentiation in \eqref{CompShift1}
is generalized to
\begin{align}
\begin{split}
    \label{CompShiftSame}
    \M_\text{spinning}\nem\bigbig{
        3^+{\hem\cdots\mem}n^+
    \hnem}
    \,&=\,
        \M_\text{scalar}\nem\bigbig{
            3^+{\hem\cdots\mem}n^+
        \hnem}\,
        \mathe^{+(k_3+\cdots+k_n)\cdot a}
    \,,\\
    \M_\text{spinning}\nem\bigbig{
        3^-{\mem\hdots\mem}n^-
    \hnem}
    \,&=\,
        \M_\text{scalar}\nem\bigbig{
            3^-{\hem\cdots\mem}n^-
        \hnem}\,
        \mathe^{-(k_3+\cdots+k_n)\cdot a}
    \,.
\end{split}
\end{align}

Crucially, however,
AHH \cite{ahh2017} had shown
that
the gluing of the positive-helicity and negative-helicity $1$-Compton amplitudes of Kerr
by recursion
gives
a nonsensical $2$-Compton amplitude,
which is explicitly
\begin{align}
    \label{CompAHH!}
    \M\mathrlap{%
        \adjustbox{raise=2.1pt}{$^{\text{(AHH)}}$}%
    }_\text{spinning}(3^+4^-)
    \,=\,
        \M_\text{scalar}(3^+4^-)
        \,
            \mathe^{+k_3\cdot a-k_4\cdot a}
        \,
            \mathe^{\gamma}
    \,.
\end{align}
The first exponential factor,
$\mathe^{+k_3\cdot a-k_4\cdot a}$,
describes the expected spin exponentiations.
On the contrary,
the second exponential factor,
$\mathe^{\gamma}$,
arises in terms of a linear-in-spin parameter $\gamma$
($a^\m$ dotted with a certain null vector)
that diverges in the backward limit $\theta \too \pi$
as $\tan \tfrac{\theta}{2}$;
see, e.g., \rcite{fabian2}.
Hence the amplitude diverges without any exchange of a physical intermediate state.
It follows that
the $n$-Compton amplitude
will suffer from the same sort of problem
when any of the incoming helicity is different from another.

For general helicity $h$,
the minimally coupled 
$\M_\text{scalar}(3^+4^-)$
contains the suppressing factor $\cos^{2h} \tfrac{\theta}{2}$.
Hence,
to become physical,
the AHH 2-Compton amplitude in \eqref{CompAHH!}
must gain contact deformations
such that
the power of the problematic parameter $\gamma$
gets topped off at $2h$
or becomes even smaller.
Said in another way,
the AHH result implies that
the probe dynamics of
spinning black holes
could possibly be defined without ambiguity
up to $\O(a^{2h})$ at most,
which is $\O(a^4)$ in the case of gravity.\footnote{
    Note that this does not preclude the possibility of contact deformations starting at an earlier order.
    The same assessment had applied to the same-helicity spin exponentiation in \eqref{CompShiftSame} as well,
    since, strictly speaking,
    the amplitudes-oriented derivation in \rrcite{Johansson:2019dnu,Aoude:2020onz,Lazopoulos:2021mna} 
    describes a proposal for a possible ideal behavior.
    The same-helicity spin exponentiation 
    had previously been only
    a beauty argument
    before \rcite{probe-nj},
    which
    showed its necessity
    for a hidden dynamical symmetry in the SD sector:
    a physical principle, not mere aesthetics.
}

Since 2017,
a substantial body of work
has been devoted to
the problem of determining the $2$-Compton amplitude for Kerr,
demonstrating various perspectives such as
a pattern suggested at low orders
\cite{Bern:2022kto,Aoude:2022trd},
a gauge redundancy
\cite{Ochirov:2022nqz,Cangemi:2022bew,Cangemi:2023bpe},
black hole perturbation theory
\cite{fabian1,fabian2,zihan23,Saketh:2023bul},
or
string amplitudes
\cite{Cangemi:2022abk,Alessio:2025nzd}.

In this chapter,
we provide a Lagrangian derivation of the $2$-Compton amplitudes for helicities $h\eqq1$ and $2$,
based on
the explicit constructions of
{\kerr} and Kerr
worldline effective actions
in \Chaps{K3:PROBENJ}{K3:OSD1}
\cite{probe-nj,njmagic.1,njmagic.11}.
Since
our actions are
local in worldline time
and
exhibit correct linearized coupling,
we expect that the results will show
correct factorization
without spurious poles
by construction.
Our actions are also constructed 
for manifest resummations onto a holomorphic (anti-holomorphic) worldline
in SD (ASD) backgrounds,
so the same-helicity spin exponentiation is guaranteed to all orders.
Our mixed-helicity $2$-Compton amplitudes
show a unified structure
in terms of
Hermite interpolation of
the problematic factor
$\mathe^\gamma$
of AHH.

In principle,
the nonlinearity of YM theory and GR
hinders simultaneously manifesting the spin exponentiation
in the SD and ASD sectors.
To this end, we implement a ``googly'' solution:
perturbing away from the SD sector.
We will see how the use of an axial gauge
turns out to be
fruitful regarding this point.

Our derivations might also appear as top-down,
based on the dynamical generalization
of the factorization of {\kerr} or Kerr
into a
chiral dyon or Taub-NUT
pair
in \Chap{K3:NJA} \cite{nja}.
The holomorphic worldline is the infrared avatar of the ASD dyon or Taub-NUT constituting {\kerr} or Kerr.

\section{
Summary of Results}
\label{COMPTON>summary}

This section documents
our predictions on 
the all-orders-in-spin classical Compton scattering amplitudes
of {\kerr} and Kerr
receiving
two photons, gluons, or gravitons
in Maxwell theory, YM theory, or GR.
Our amplitudes
are free of any spurious pole
and
exhibit correct factorizations.

For same-helicity configurations,
we find guaranteed spin exponentiation.
For the opposite-helicity configuration,
we record the ``relative amplitude,''
\begin{align}
	\frac{
		\M_\text{spinning}(3^+4^-)
        \vphantom{\big|}
	}{
		\M_\text{scalar}(3^+4^-)\mem
		\mathe^{-\a/2}\mem \mathe^{\b/2}
        \vphantom{\big|}
	}
    \,.
\end{align}
The relevant kinematic invariants are
$\a,\b,\c$,
which
are dimensionless linear-in-spin parameters such that
the trivial spin exponentiation is
$
	\M_\text{spinning}(3^+4^-) = \M_\text{scalar}(3^+4^-)\mem
	\mathe^{-\a/2}\mem \mathe^{\b/2}
$
while
the AHH \cite{ahh2017} ansatz with the spurious pole
is 
$
	\M_\text{spinning}(3^+4^-) = \M_\text{scalar}(3^+4^-)\mem
	\mathe^{-\a/2}\mem \mathe^{\b/2}\mem
	\mathe^{\gamma}
$.
Explicitly,
\begin{align}
    \label{abcfirst}
    \a
    \,:=\,
        -2\hem k_3 \mdot a
    \,,\quad
    \b
    \,:=\,
        -2\hem k_4 \mdot a
    \,,\quad
    \c
    \,:=\,
        (-2k_3\mdot p\hhem)\,
        \frac{w\mdot a}{w\mdot p}
    \,,
\end{align}
where $k_3,k_4$ are the massless wavenumbers,
$p$ is the massive momentum,
and
$w_{\a\da} \propto 4_\a\hhem \bar{3}_\da$.
We also define the ``truncated exponential'' functions as
\begin{align}
    E_h(\xi)
    \,:=\,\mem
        \sum_{\ell=h}^\infty\,
            \frac{
                \xi^{\ell-h}
            }{\ell!}
    \,=\,
        \frac{1}{\xi^h}\mem
        \bb{
            \mathe^\xi
            - 
            \sum_{\ell=0}^{h-1}\,
                \frac{
                    \xi^{\ell}
                }{\ell!}
        \hem}
    \,.
\end{align}
\hrule\vspace{0.5pt}
\hrule\vspace{-5pt}
\paragraph{Maxwell Theory}
\begin{subequations}
\label{RelCompton}
\begin{align}
    \label{RelCompton:EM}
    1 + \gamma
\end{align}
\paragraph{YM Theory}\nem\nem%
(nonabelian contribution, color-stripped)
\begin{align}
    \label{RelCompton:YM}
    1 + \gamma\,
    \bb{
        \frac{\gamma+\beta}{\alpha+\beta}\mem
        E_1(\a)
        -
        \frac{\gamma-\alpha}{\alpha+\beta}\mem
        E_1(-\b)
    \nem}
\end{align}
\paragraph{GR}
\begin{align}
    \label{RelCompton:GR}
    1 + \gamma + \gamma^2\mem
    \bb{
        \frac{\gamma+\beta}{\alpha+\beta}\mem
        E_2(\a)
        -
        \frac{\gamma-\alpha}{\alpha+\beta}\mem
        E_2(-\b)
    \nem}
\end{align}
\end{subequations}
\hrule\vspace{0.5pt}
\hrule\vspace{-5pt}

\pagebreak

The YM and GR results in \eqrefs{RelCompton:YM}{RelCompton:GR} were obtained by the present author in the year 2024. 
The YM result was computed during academic travels in Europe in August and September.
The GR result was computed in November and finalized by early December.
These results were then immediately shared with several colleagues
in December,
including
Clifford Cheung, 
Jung-Wook Kim, 
Sangmin Lee, 
Donal O'Connell, 
Julio Parra-Martinez, 
Justin Vines, 
and 
Zihan Zhou. 
They were publicly presented in a talk \cite{IHEStalk25} at
Institut des Hautes Études Scientifiques on January 15, 2025.

\section{
Properties}

\subsection{Absence of Spurious Poles}

The fact that our amplitudes are free of any spurious pole
can be comprehended in a systematic fashion as follows.
The point of departure is
an intriguing identity
that rewrites the AHH relative amplitude:\footnote{
    The second line in \eqref{ahh-exp}
    computes a second divided difference 
    $
        \tfrac{1}{\a+\b}
        \bigbig{
            \tfrac{\mathrm{e}^\a - \mathrm{e}^\c}{\a-\c}
            -
            \tfrac{\mathrm{e}^\c - \mathrm{e}^{-\b}}{\c+\b}
        }
    $
    for the exponential function.
}
\begin{align}
\begin{split}
    \label{ahh-exp}
    \mathe^\gamma
    \mem\,=\,\mem
    {}&{}
        \frac{\b\mem \mathe^\a + \a\mem \mathe^{-\b}}{\a+\b}
        + \c\mem 
        \frac{\mathe^\a - \mathe^{-\b}}{\a+\b}
    \\[-0.1\baselineskip]
    {}&{}
        + (\c-\a)(\c+\b)\,
        \sum_{j=0}^\infty\,
            \c^{j}\,
            \frac{
                E_{j+1}(\a) - E_{j+1}(-\b)
            }{\a+\b}
    \,.
\end{split}
\end{align}
Based on \eqref{ahh-exp},
we can consider a truncation of the AHH answer
up to a finite power
$(h+\hnem1) = 1,2,3,\cdots$
in the parameter $\gamma$:
\label{AHHt}
\begin{align}
\begin{split}
    \label{AHHt.Original}
    \AHH[h]
    \mem\,:=\,\mem
    {}&{}
        \frac{\b\mem \mathe^\a + \a\mem \mathe^{-\b}}{\a+\b}
        + \c\mem 
        \frac{\mathe^\a - \mathe^{-\b}}{\a+\b}
    \\[-0.1\baselineskip]
    {}&{}
        + (\c-\a)(\c+\b)\,
        \sum_{j=0}^{h-1}\,
            \c^{j}\,
            \frac{
                E_{j+1}(\a) - E_{j+1}(-\b)
            }{\a+\b}
    \,.
\end{split}
\end{align}
Said in another way, $\AHH[h]$
gives a contact deformation of the AHH Compton amplitude
in terms of the difference
$
    \AHH[h] - \AHH[\infty]
    = (\c\mminus\a)(\c\mplus\b)$ $ \sum_{j=h}^\infty
        \c^j\mem 
        \smash{\bigbig{
            E_{j+1}(\a) - E_{j+1}(-\b)
        }}/(\a+\b)
$,
where $\AHH[\infty] = \mathe^\c$.
Also, it should be clear that $\AHH[h]$ is completely regular at $\a+\b = 0$.

Remarkably,
we find that
our relative amplitudes in \eqrefs{RelCompton:YM}{RelCompton:GR}
are
$\AHH[1]$ and $\AHH[2]$, respectively.
This should be clear from the fact that
\eqref{AHHt.Original} admits an alternative representation,
\begin{align}
\begin{split}
    \label{AHHt.Alt}
    \AHH[h]
    \mem\,=\,\mem
    {}&{}
    \bb{\mem
        \sum_{\ell=0}^{h-1}\,
        \frac{\c^\ell}{\ell!}
    }
    \,+\,
        \c^h\hem\bb{
            \frac{\c+\b}{\a+\b}\mem
                E_h(\a)
            - \frac{\c-\a}{\a+\b}\mem
                E_h(-\b)
        \nem}
    \,.
\end{split}
\end{align}
Consequently,
our amplitudes
are free from the notorious spurious pole
at $\gamma = \infty$.
Let us elaborate on this point below.

\begin{figure}[t]
    \centering
    \includegraphics[scale=1.9]{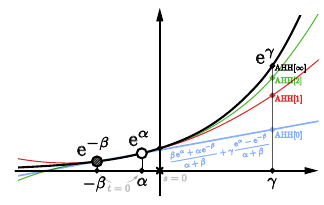}
    \caption{%
        The truncation in \eqref{AHHt.Original}
        arises in the context of Hermite interpolation.
        $\AHH[0]$ represents the unique line through
        $(-\b,\mathe^{-\b})$ and $(\a,\mathe^{\a})$.
        $\AHH[1]$ represents the unique parabola through
        $(-\b,\mathe^{-\b})$, $(\a,\mathe^{\a})$, and $(0,1)$.
        $\AHH[2]$ represents the unique cubic through
        $(-\b,\mathe^{-\b})$, $(\a,\mathe^{\a})$, and $(0,1)$
        that is tangent to $\mathe^\c$ at $(0,1)$.
    }
    \label{Hermite123}
\end{figure}

For helicity $h$,
the relative amplitude develops the notorious spurious pole
if it involves a power of $\gamma$
greater than 
$2h$,
which is the content of the famous AHH result \cite{ahh2017}.
Within the above particular truncated class,
the helicity-$1$ relative amplitude
can in principle maximally extend to $\AHH[1]$,
whereas the helicity-$2$ relative amplitude
can in principle maximally extend to $\AHH[3]$.
In particular, the expansion in spin length gives
\begin{align}
    \AHH[1]
    \,&=\,
        1 + \frac{\c^1}{1!}
        + \frac{\c^2}{2!}
        + \frac{1}{3!}\mem
            \BB{
                \c^2(\a\mminus\nem\b) + \c\a\b
            }
        + \O(a^4)
    \,,\\
    \nonumber
    \AHH[2]
    \,&=\,
        1 + \frac{\c^1}{1!}
        + \frac{\c^2}{2!}
        + \frac{\c^3}{3!}
        + \frac{1}{4!}\mem
            \BB{
                \c^3(\a\mminus\nem\b) + \c^2\a\b
            }
        + \O(a^5)
    \,,\\
    \AHH[3]
    \,&=\,
        1 + \frac{\c^1}{1!}
        + \frac{\c^2}{2!}
        + \frac{\c^3}{3!}
        + \frac{\c^4}{4!}
        + \frac{1}{5!}\mem
            \BB{
                \c^4(\a\mminus\nem\b) + \c^3\a\b
            }
        + \O(a^6)
    \,,
    \kern-0.45em
    \nonumber
\end{align}
so the contact deformations are introduced
from $\O(a^3)$, $\O(a^4)$, and $\O(a^5)$
for $\AHH[1]$, $\AHH[2]$, $\AHH[3]$, respectively.

Our helicity-$2$ relative amplitude in \eqref{RelCompton:GR}
is $\AHH[2]$,
which means that it introduces
an additional contact deformation
that is not strictly mandatory,
at $\c^4$.
This could appear as a peculiarity,
but we may
note that the Maxwell answer in \eqref{RelCompton:EM}
also has entailed such an extra deformation,
as is first noted in \rcite{Kim:2024grz}.
In fact,
it might be possible to argue
$\AHH[1]$ and $\AHH[2]$
from
the order of
nonlocality involved in the Dirac/Misner strings.\footnote{
    Note that
    $E_h(\xi) = 
        \int_0^1 d\eta_1\,
        \int_0^{\eta_1} d\eta_2\,
        \int_0^{\eta_2} d\eta_3\,
        \cdots
        \int_0^{\eta_{h-1}} d\eta_h\,
            \mathe^{\eta_h\xi}
    $
    describes the exponential factor
    integrated $h$ times.
    It will also be interesting to formulate orbit-spin duality at the amplitudes level.
}
On a related note,
it will be interesting if
a double copy relation could be established,
or a useful integral representation of the relative amplitudes
reflecting the open 
(Dirac/Misner, not fundamental)
string structure
is viable.

More generally,
one can ask
if the $\AHH[h]$ class of ansatz
arises uniquely as the result of a bootstrap.
Note that 
the identity in \eqref{ahh-exp}
admits a nice geometric interpretation:
see \frefs{Hermite123}{HermiteX}.

\begin{figure}[t]
    \centering
    \includegraphics[scale=1.9]{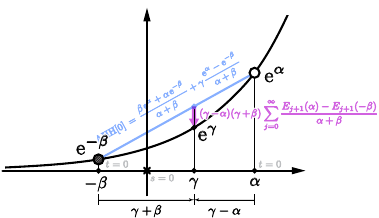}
    \caption{%
        A visualization of the identity in \eqref{ahh-exp}.
        This uses the same geometric interpretation
        as \fref{Hermite123}.
    }
    \label{HermiteX}
\end{figure}

\subsection{Correct Residues on Physical Poles}

The fact that our amplitudes exhibit 
correct residues on physical poles
can be shown as follows
(cf. Sec\,3.1 of \rcite{fabian2}).
To clarify,
we strictly work in the classical limit
where the massless wavenumbers $k_3,k_4$ take macroscopic values.
With the definitions
\begin{align}
    \label{cmandelstu}
    K_3
    \,:=\,
        2k_3\mdot p
    \,,\quad
    K_4
    \,:=\,
        2k_4\mdot p
    \,,\quad
    t
    \,:=\,
        -2k_3\mdot k_4
    \,,
\end{align}
the Mandelstam identity gives
$K_3 + K_4 = \hbar\hem t$
so that,
in the classical limit,
\begin{align}
    \label{K3=-K4}
    K_3 
    \,=\,
        -K_4
\end{align}
while $t$ can take any value.
\newpage

To begin with,
suppose the generic form of the spinning amplitude,
\begin{align}
    \label{Rans}
    \M_\text{spinning}(3^+4^-)
    \,=\,
        \M_\text{scalar}(3^+4^-)\mem \mathe^{-\a/2}\mathe^{\b/2}
        \,
        \mathcal{R}(\a,\b,\c)
    \,,
\end{align}
where the relative amplitude $\mathcal{R}(\a,\b,\c)$
is presumed to be regular at $\c=0$.

Firstly,
the $s$-channel factorization corresponds to the limit $K_3\to0$.
We import the factorization of the scalar amplitude
in the limit $K_3 \to 0$,
\begin{align}
    \label{schanfact0}
    \M_\text{scalar}(3^+4^-)
    \,\,\sim\,\,
        \M_\text{scalar}(3^+)
            \,\frac{1}{K_3}\,
        \M_\text{scalar}(4^-)
    \,.
\end{align}

Secondly,
the $t$-channel factorization
corresponds to the limit $t\to0$,
which
will be only discussed
for YM and GR amplitudes.
By complexification,
this boils down to either
$\lp{34} = 0$ or $\rp{34} = 0$.
In these cases,
the scalar amplitude factorizes as
\begin{subequations}
\label{tchanfact0}
\begin{align}
\label{tchanfact0.+}
    \lp{34} = 0
    &:\quad
    \M_\text{scalar}(3^+4^-)
    \,\,\sim\,\,
        \M_\text{scalar}(5^-)
            \,\frac{1}{-t}\,
        \M_\text{massless}(3^+4^-5^+)
    \,,\\
\label{tchanfact0.-}
    \rp{34} = 0
    &:\quad
    \M_\text{scalar}(3^+4^-)
    \,\,\sim\,\,
        \M_\text{scalar}(5^+)
            \,\frac{1}{-t}\,
        \M_\text{massless}(3^+4^-5^-)
    \,,
\end{align}
\end{subequations}
where
$\M_\text{massless}(3^+4^-5^\pm)$
are the three-point gluon or graviton amplitudes
(MHV and anti-MHV):
\begin{subequations}
\begin{align}
    \lket{3} \propto \lket{4} \propto \lket{5}
    &\quad\xleftrightarrow{\blank}\quad
    \M_\text{massless}(3^+4^-5^+)
    \,\propto\,
        \frac{\rp{35}^4}{\rp{34}\rp{45}\rp{35}}
    \,,\\
    \rket{3} \propto \rket{4} \propto \rket{5}
    &\quad\xleftrightarrow{\blank}\quad
    \M_\text{massless}(3^+4^-5^-)
    \,\propto\,
        \frac{\lp{45}^4}{\lp{34}\lp{45}\lp{35}}
    \,.
\end{align}
\end{subequations}

\begin{figure}[t]
    \centering
    \includegraphics[scale=1.7]{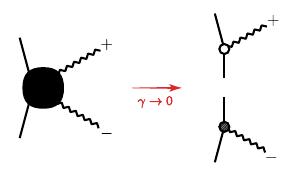}
    \caption{%
        $s$-channel factorization of black hole Compton amplitudes.
    }
    \label{fig:ComptonFactorization:s}
\end{figure}

We are now ready to discuss
the factorizations of 
the generic spinning amplitude 
$\M_\text{spinning}(3^+4^-)
=
    \M_\text{scalar}(3^+4^-)\mem \mathe^{-\a/2}\mathe^{\b/2}
    \,
    \mathcal{R}(\a,\b,\c)
$
in \eqref{Rans}.

Firstly,
the $s$-channel factorization
entails 
\begin{align}
    K_3 \hem\to\mem 0
    \qiq
    \c 
    \,=\,
        -K_3\mem \frac{w\mdot a}{w\mdot p}
    \,\to\,
        0
    \,.
\end{align}
With this understanding,
\eqref{schanfact0} implies that
the leading asymptotic of
$\M_\text{spinning}(3^+4^-)$
in the limit $K_3 \to 0$
is given by
\begin{align}
\label{schanfacta}
    \BB{\M_\text{scalar}(3^+)\mem \mathe^{-\a/2}}
        \,\frac{1}{K_3}\,
    \BB{\M_\text{scalar}(4^-)\mem \mathe^{\b/2}}
    \,
    \mathcal{R}(\a,\b,0)
    \,.
\end{align}
Crucially,
we see that all of the three relative amplitudes in
\eqref{RelCompton}
satisfy
\begin{align}
    \label{RelCond:s}
    \mathcal{R}(\a,\b,0)
    \,=\,
        1
    \,,
\end{align}
in which case \eqref{schanfacta}
precisely gives
the desired residue:
the product of
the spin-exponentiated three-point spinning amplitudes.

\begin{figure}[t]
    \centering
    \includegraphics[scale=1.7]{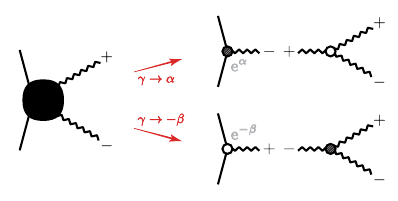}
    \caption{%
        $t$-channel factorizations of black hole Compton amplitudes.
    }
    \label{fig:ComptonFactorization:t}
\end{figure}

Secondly,
the $t$-channel factorization entails\footnote{
    Note also that the 
    Britto-Cachazo-Feng-Witten (BCFW)
    \cite{BCF,BCFW}
    \nomenclature{BCFW}{Britto-Cachazo-Feng-Witten}
    shift of the massless momenta
    for the $s$-channel factorization
    induces
    $\a \mapsto \a - \c$
    and
    $\b \mapsto \b + \c$.
}
\begin{subequations}
\begin{align}
    \lp{34} = 0
    &\,\,\implies\,\,
        w_{\a\da}
        \mem\propto\mem
        3_\a \bar{3}_\da
    \,\,\implies\,\,
        \c
        \,=\,
            (-2k_3\mdot p\hhem)\,
            \frac{k_3\mdot a}{k_3\mdot p}
        \,=\,
            \a
    \,,\\
    \rp{34} = 0
    &\,\,\implies\,\,
        w_{\a\da}
        \mem\propto\mem
        4_\a \bar{4}_\da
    \,\,\implies\,\,
        \c
        \,=\,
            (+2k_4\mdot p\hhem)\,
            \frac{k_4\mdot a}{k_4\mdot p}
        \,=\,
            -\b
    \,.
\end{align}
\end{subequations}
With this understanding,
\eqref{tchanfact0} implies that
the leading asymptotics of
$\M_\text{spinning}(3^+4^-)$
in the limit $t \to 0$
are given by
\begin{subequations}
\label{tchanfacta}
\begin{align}
\label{tchanfacta.+}
    \BB{
        \M_\text{scalar}(5^-)
        \mem \mathe^{(\a+\b)/2}
    }
    \BB{
        \mathe^{-\a}\mem
        \mathcal{R}(\a,\b,\a)
    }
            \,\frac{1}{-t}\,
        \M_\text{massless}(3^+4^-5^+)
    \,,\\
\label{tchanfacta.-}
    \BB{
        \M_\text{scalar}(5^+)
        \mem \mathe^{-(\a+\b)/2}
    }
    \BB{
        \mathe^{+\b}\mem
        \mathcal{R}(\a,\b,-\b)
    }
            \,\frac{1}{-t}\,
        \M_\text{massless}(3^+4^-5^-)
    \,,
\end{align}
\end{subequations}
for $\lp{34} = 0$ and $\rp{34} = 0$,
respectively.
Crucially,
both of the relative amplitudes in
\eqrefs{RelCompton:YM}{RelCompton:GR}
satisfy
the conditions
\begin{align}
    \label{RelCond:t}
    \mathcal{R}(\a,\b,\a)
    \,=\,
        \mathe^\a
    \,,\quad
    \mathcal{R}(\a,\b,-\b)
    \,=\,
        \mathe^{-\b}
    \,,
\end{align}
in which case \eqref{tchanfacta}
precisely establishes
the desired factorization into
the spin-exponentiated three-point spinning amplitude
times the massless MHV or anti-MHV amplitude.
To elaborate,
it is easy to see that
plugging in 
$\c = \a$ or $\c = -\b$
to $\AHH[h]$ in \eqref{AHHt.Original}
respectively
gives 
$\mathe^\a$ and $\mathe^{-\b}$,
which can be also directly seen
from the geometric interpretation given in
\fref{HermiteX}.

The above discussion
endows us with
a clear breakdown of
the $\AHH[h]$ ansatz
in \eqref{AHHt.Original}
and explicates how restrictive it is.
In general,
the ansatz for the relative amplitude
can be
\begin{align}
\begin{split}
    \label{RelQ}
    \mathcal{R}(\a,\b,\c)
    \,=\,
\begin{aligned}[t]
    {}&{}
        \bb{\nem{\hnem
            \frac{\b\mem \mathe^\a + \a\mem \mathe^{-\b}}{\a+\b}
            + \c\mem 
            \frac{\mathe^\a - \mathe^{-\b}}{\a+\b}
        }\nem}
    \\
    {}&{}
        + (\c\mminus\a)(\c\mplus\b)\mem
            \mathcal{Q}(\a,\b,\c)
    \,,
\end{aligned}
\end{split}
\end{align}
where $\mathcal{Q}(\a,\b,\c)$ is
polynomial in $\c$.
The first group of terms on the right-hand side
gives a ``linear interpolation''
between $\mathe^\a$ and $\mathe^{-\b}$,
while $(\c\mminus\a)(\c\mplus\b) \to 0$
when $\c \to \a$ or $-\b$
so that
the $t$-channel factorization condition in \eqref{RelCond:t}
is guaranteed.
Imposing the $s$-channel condition in \eqref{RelCond:s} 
then yields
\begin{align}
    \mathcal{R}(\a,\b,0)
    \,=\,
        1
    \qiq
    \mathcal{Q}(\a,\b,0)
    \,=\,
        \frac{E_1(\a) - E_1(-\b)}{\a+\b}
    \,.
\end{align}
Hence $\mathcal{Q}(\a,\b,\c)$ in \eqref{RelQ}
can be taken as
\begin{align}
    \mathcal{Q}(\a,\b,\c)
    \,=\,
        \frac{E_1(\a) - E_1(-\b)}{\a+\b}
    \hem+\mem
        \sum_{j=1}^\infty\,
            \c^j\mem
            q_j(\a,\b)
    \,.
\end{align}
Consequently,
the residual freedom for helicity $h$ boils down to
a set of functions $q_j(\a,\b)$
for $j = 1,2,\cdots$.
Eventually,
by demanding the absence of spurious poles as well,
we find that
the case $h=1$ allows for
\textit{no residual freedom}
whereas
the case $h=2$ allows for
choosing two functions,
$q_1(\a,\b)$ and $q_2(\a,\b)$.

\subsection{Comparison with Literature}

Arguably,
our amplitudes in \eqrefs{RelCompton:YM}{RelCompton:GR}
exhibit
a degree of elegance and simplicity
while pointing to
the suggestive truncation structure
as $\AHH[h]$.

Our amplitudes,
however,
do not agree with the amplitudes
obtained or advocated in
a group of works
\cite{%
    Ochirov:2022nqz,Cangemi:2022bew,Cangemi:2023bpe,%
    fabian1,fabian2,zihan23,Saketh:2023bul%
}.
We can highlight three points.
\begin{itemize}
    \item 
        The amplitudes in \rrcite{%
            Ochirov:2022nqz,Cangemi:2022bew,Cangemi:2023bpe,%
            fabian1,fabian2,zihan23,Saketh:2023bul%
        }
        seem to
        appear more complicated than our prediction
        when brought to the same kinematic basis.
    \item
        The amplitudes in \rrcite{%
            Ochirov:2022nqz,Cangemi:2022bew,Cangemi:2023bpe,%
            fabian1,fabian2,zihan23,Saketh:2023bul%
        },
        to our understanding,
        seem to
        attach a special meaning
        to the center of the black hole on the real spacetime.
        In our opinion,
        this dependence seems rather spurious or even unphysical
        in view of
        the factorization of black holes \cite{nja}
        shown in \Chap{K3:NJA}:
        there is nothing on the midpoint of the Dirac/Misner strings,
        which are gauge artifacts after all.
        Recall that the same-helicity spin exponentiations are 
        none other than
        localizations on the SD/ASD Taub-NUT worldlines.
    \item
        The amplitudes in \rrcite{%
            Ochirov:2022nqz,Cangemi:2022bew,Cangemi:2023bpe,%
            fabian1,fabian2,zihan23,Saketh:2023bul%
        }
        are dependent on 
        yet another kinematic invariant:
        the optical parameter,
        \begin{align}
            \label{opticalparam}
            \xi^{-1}
            \,:=\,
                \frac{
                    m^2\mem t
                }{{K_3}^2}
            \,.
        \end{align}
        As a result,
        the absence of the spurious pole
        is achieved 
        through a more involved mechanism,
        counting on a coordinated cancellation between
        the poles in
        $\c$ and the optical parameter,
        according to
        Sec.\,3.3 of \rcite{fabian2}.
\end{itemize}

\section{
Methodology}
\label{COMPTON>methodology}

This section outlines our methodology
for obtaining the classical Compton amplitudes
in \eqref{RelCompton}.
Technical details 
are left to 
\App{APP:WLFS}.

In short,
we carry out worldline perturbation theories of
the explicit first-order {\kerr} and Kerr actions in \Chap{K3:OSD1},
implemented in the massive twistor model.
The massive twistor space
is essentially a phase space
for a massive spinning particle;
recall \Chap{K3:SST}.

For Compton scattering
with gluons or gravitons,
the universal Feynman rules of symplectic sigma models
\cite{wlf-ps}
(reviewed in \App{CA:WLF>FEYN})
imply that
it suffices to compute
the interaction symplectic potential
$\theta'$ in the background of nonlinearly superposed plane waves---%
those obtained in \Sec{Jcx>NLIN}---%
and
the symplectic perturbation
$\omega'$ in the linearized background of a single plane wave.
The former gives a vertex
of fluctuation degree 
    (meaning the valence of worldline fluctuations involved)
zero
and interaction degree 
    (meaning the number of massless legs involved)
two.
The latter gives a vertex of
fluctuation degree one
and interaction degree one.
Vertices of the latter kind
describe a linear coupling,
so they can be simply obtained by directly
complexifying
(NJ shifting, spin exponentiating)
the corresponding vertices in the spinless case.

\subsection{Vertices and Master Formula for Compton Amplitude}

Explicitly,
given a sigma model with a symplectic target
equipped with coordinates $\zeta^i$
that are Darboux with respect to the free theory,
the vertices of fluctuation degree zero and one are respectively given by
\begin{align}
    \label{V-universal}
    \mathcal{V}^{(0)}
    \,=
        \int d\s\,\,
            \theta'_i\hhnem(\bm{\zeta}(\s)\hnem)\mem \dot{\bm{\zeta}}^i
    \,,\quad
    \mathcal{V}^{(1)}\hnem(\s)
    \,=\,
            \d\zeta^i(\s)\,
            \omega'_{ij}\hhnem(\bm{\zeta}(\s)\hnem)\mem \dot{\bm{\zeta}}^j
    \,,
\end{align}
assuming expansion around 
a free theory saddle
$\s \mapsto \bm{\zeta}^i(\s)$
whose time derivative is constant.
Plugging in the nonlinearly superposed plane wave backgrounds
generically yields the specific time dependence,
\begin{align}
\begin{split}
    \label{pwdevelop-theta,omega}
    \theta'_i\hhnem(\bm{\zeta}(\s)\hnem)
    \,&=\,
        \e_3\mem
            \theta'^3_i\, \mathe^{iK_3\s}
        +
        \e_4\mem
            \theta'^4_i\, \mathe^{iK_4\s}
        +
        \e_3\e_4\mem
            \theta'^{34}_i\, \mathe^{i(K_3+K_4)\s}
    \,,\\
    \omega'_{ij}\hhnem(\bm{\zeta}(\s)\hnem)
    \,&=\,
        \e_3\mem
            \omega'^3_{ij}\, \mathe^{iK_3\s}
        +
        \e_4\mem
            \omega'^4_{ij}\, \mathe^{iK_4\s}
        +
        \e_3\e_4\mem
            \omega'^{34}_{ij}\, \mathe^{i(K_3+K_4)\s}
    \,.
\end{split}
\end{align}
As a result, a universal master formula for the Compton amplitude arises that
\begin{align}
    \label{M-universal-proto}
    \M
    \,=\,
        \theta'^{34}_i\mem \dot{\bm{\zeta}}^i
        \mem+\mem
            \omega'^3_{ik}\mem
            \dot{\bm{\zeta}}^k
            \hem
            \omega'^4_{jl}\mem
            \dot{\bm{\zeta}}^l
        \,
        \expval{
            \d\zeta^i(K_3)\mem
            \d\zeta^j(-\hnem K_3)\hnem
        }
    \,.
\end{align}
Here, 
$
    \expval{
        \d\zeta^i(K_3)\mem
        \d\zeta^j(-\hnem K_3)\hnem
    }
$
describes
the worldline propagator
in the frequency domain,
incorporating all quadratic vertices
in the free theory.
Its causality prescription
does not matter much
at this order.\footnote{
    There exist concrete prescriptions in which 
    \eqref{M-universal-proto}
    can be legitimately
    called a scattering amplitude
    (in connection with \Chap{K3:SST} \cite{sst-asym})
    or holds a direct classical significance.
}

\subsection{Propagators}

The worldline propagators in the time domain
can be derived from frameworks described in
\App{APP:WLFS}
(see also \rcite{Kim:2024grz}),
using a free-theory saddle such that
\begin{align}
    \label{zzpsaddle}
    \bm{z}^\m(\s)
    \mem=\mem
        z^\m
        + 2p^\m\hem \s
    \,,\quad
    \bm{\bz}^\m(\s)
    \mem=\mem
        \bz^\m
        + 2p^\m\hem \s
    \,,\quad
    \bm{p}_\m(\s)
    \mem=\mem
        p_\m
    \,.
\end{align}
Here, $z^\m$, $\bz^\m$, and $p_\m$
are \textit{constants}
specifying the background worldline;
$x^\m = (z^\m \mplus \bz^\m)/2$ is the impact parameter
($b^\m$ in usual notation);
$y^\m = (z^\m \mminus \bz^\m)/2i$ is the spin length pseudovector
($-a^\m$ in usual convention).
With the definition
\begin{align}
    \expval{
        \d\zeta^i(K_3)\mem
        \d\zeta^j(K_4)\hnem
    }
    \,=
        \int
            d\s_3\mem d\s_4\,\,
        \mathe^{iK_3\s_3}\mem
        \mathe^{iK_4\s_4}\mem
        \expval{
            \d\zeta^i(\s_3)\mem
            \d\zeta^j(\s_4)\hnem
        }
    \,,
\end{align}
it follows that
\begin{subequations}
\label{Ints}
\begin{align}
    \label{Int1}
    \expval{
        \d z^\m\hhnem(K_3)\mem
        \d p_\n\hhnem(K_4)\hnem
    }
    \,&=\,
        \frac{
            -i\hem\delta^\m{}_\n
        }{K_3}
    \,,\\
    \label{Int2}
    \expval{
        \d z^\m\hhnem(K_3)\mem
        \d z^\n\hhnem(K_4)\hnem
    }
    \,&=\,
    \expval{
        \d \bz^\m\hhnem(K_3)\mem
        \d \bz^\n\hhnem(K_4)\hnem
    }
    \,=\,
        -\frac{
            2\hhem\eta^{\m\n}
        }{{K_3}^2}
    \,,\\
    \label{IntZZ}
    \expval{
        \d z^\m\hhnem(K_3)\mem
        \d \bz^\n\hhnem(K_4)\hnem
    }
    \,&=\,
        \frac{2}{m^2}\hem
        \BB{
            -i\hhem S^{\m\n}
            \mminus 2\hhhem y^{(\m} p^{\n)}
        \hnem}\hem
        \frac{1}{K_3}
        \hem-\hem
        \frac{
            2\hhem\eta^{\m\n} \mplus 4\hem
            \hp^\m \hp^\n
        }{{K_3}^2}
    \,,
\end{align}
on the support of $K_3 \mplus K_4 = 0$.
Here, $S^{\m\n} = \ve^{\m\n\r\s} y_\r p_\s$
and $\hp^\m = p^\m\hnem/m$.\footnote{
    In fact, 
    it turns out that
    the ``longitudinal'' components of \eqref{IntZZ}
    with respect to $p^\m$
    could be freely discarded,
    reminiscently of a Ward identity.
}
When working with the color extension,
the adjoint color variable describes
\begin{align}
    \label{Intqq}
    \expval{
        \d q_a\hhnem(K_3)\mem
        \d q_b\hhnem(K_4)\hnem
    }
    \,&=\,
        \frac{
            -i\hem
                q_c f^c{}_{ab}
        }{K_3}
    \,,
\end{align}
\end{subequations}
where $q_c$
refers to the background value:
$\bm{q}_a(\s) \eqq q_a$.
To clarify,
all of
$\d z^\m$, $\d \bz^\m$,
$\d p_\m$, and $\d q_a$
in \eqref{Ints}
are fluctuations of composite variables.

\section{
Interpretation}

The propagators in \eqrefs{Int2}{IntZZ}
admit a physical interpretation
in the spinspacetime framework.
The \textit{second-order} poles
represent \textit{inertia}
as the proportionality between
force and acceleration,
as is supported by its diagrammatic origin as well.
The \textit{first-order} pole in \eqref{IntZZ}
is evidently an incarnation
of the ``zig-zag'' spinspacetime bracket
in \eqrefsor{eq:zzpb}{zzpb-vec},
which will represent
an \textit{angular momentum} effect.

In sum,
intuitively speaking,
the above propagators
characterize the precise behavior
in which the
zig and zag points
``wobble''
when each of them is
``hit by a hammer''
by external perturbations.
Note that the zig and zag points are
where the SD and ASD couplings resum onto.
For instance,
recall that
the abelian {\kerr} action
cleanly splits into the interaction actions
coupling SD gauge potential to the zig worldline
and ASD gauge potential to the zag worldline,
the observation of which traces back to \rcite{gmoov}.
Overall, 
the perturbation theory 
encourages and prompts
us to imagine the zig and zag points as actual objects.

Remarkably,
a vivid picture emerges
by the revelation in
\Chap{K3:NJA} \cite{nja}:
the factorization of Kerr
into Taub-NUT pairs.
The zig point $z$
is precisely where the ASD Taub-NUT center is located at
in the complexified spacetime,
while
the zag point $\bz$
is precisely where the SD Taub-NUT center is located at.
In fact,
we call ASD Taub-NUT a ``zig''
and SD Taub-NUT a ``zag.''
In \textit{zig-zag theory},
zigs and zags 
were
particles that exhibit a helicity selection rule,
whose existence was first speculated out of
the zig-zag twistor particle program \cite{ambikerr1}
on account of the holomorphy of SD symplectic perturbation.
Zigs can only absorb positive-helicity massless quanta
and thus can only source negative-helicity massless quanta.
It was represented as a white blob,
on account of
the standard on-shell diagrams notation:
recall \fref{fig:hkins}.
Amusingly,
\fref{fig:factorization} discloses that
this white blob is \textit{real}
and actually resides inside the Kerr black hole.
It turns out that the zig is none other than the ASD Taub-NUT instanton.
Indeed,
the ASD Taub-NUT instanton
enjoys the proposed helicity selection rule
for zig
\cite{Adamo:2023fbj,Adamo:2024xpc,Adamo:2025fqt}.

\eqref{Int2} states 
the zig and zag have (their own) inertia.
\eqref{IntZZ} tells us that,
when one hits zag by inputting an ASD wave,
the zig first moves in a ``\textit{skew}'' direction
just as in the case of gyroscope motion---%
that of the spinning \textit{bicycle wheel}---%
as the first-order effect
(which is precisely the content of the zig-zag spinspacetime bracket),
then starts to drift/accelerate gradually by the second-order effect.
Again, this description is all about
how the white and black blobs in \fref{fig:factorization}
move and react to external forces
as actual physical entities,
or, more precisely,
their avatars in the infrared theory.

Note that the zig and zag locations are also 
Newman's ``complex centers of mass'' 
\cite{%
    newman1988remarkable,Newman:1973afx,Newman:2002mk,Newman:1973yu,newman1974curiosity,newman1974collection,Newman:2004ba,
    Newman:1976gc,ko1981theory,grg207flaherty%
},
in which case we learn that
Newman's provisional ideas
have been completed into a concrete picture:
a fully dynamical theory of the zig-zag system,
defined beyond SD backgrounds.

It is instructive to consider the case of Maxwell theory.
The {\kerr} solution is secretly a pair of a zig (ASD dyon) and a zag (SD dyon) \cite{nja,Newman:1965tw-janis}.
No complicated calculation is needed for
showing that 
the ASD dyon exhibits the precise helicity selection rule of a zig.
As a test particle,
a zig sees only the SD part of external fields
via its Lorentz force law
$
\smash{
    d\vec{p}/dt
}
= 
\smash{
    q (1 {\,-\,} i\vec{v}\mt) (\vec{E} {\,+\,} i\vec{B})
}
$.
As a background,
a zig always sources an ASD field
regardless of motion.
For a static zig, for instance, the fields are
$\vec{E} {\,=\,} q \vex/4\pi |\vex|^3$,
\smash{$\vec{B} {\,=\,} iq \vex/4\pi |\vex|^3$}.

\section{
Kinematic Bases for Relative Amplitudes}

As another preliminary,
let us also discuss
the kinematic invariants
and their relations
for each helicity configurations.
To begin with,
suppose 
two massless momenta $k_3,k_4$
and
their associated
bivector polarizations
$\vf_3,\vf_4$,
which can carry any helicity at this moment.
Suppose a massive momentum $p^2 = -m^2$
and a spin length pseudovector $y$
subject to the condition $p\mdot y = 0$.
The goal is 
to construct invariants out of these tensors
and identify a minimal basis.
In particular,
let us
seek the minimal basis for
dimensionless and little group neutral variables,
on which the relative amplitude can depend.\footnote{
    Regarding such invariants at higher multiplicities,
    we had explored 
    multi-photon scattering off
    abelian {\kerr}
    in Maxwell theory
    by taking twistorial variables as fundamental.
    It seems that
    the integrands can be represented
    in the
    Hodges diagram notation
    with infinity twistors and
    a slight massive extension,
    for which
    the massless dual twistors and twistors
    are
    respectively
    incident on the holomorphic or anti-holomorphic
    spinspacetime coordinates
    to systematically construct kinematic invariants.
    Amusingly, the $\gamma$ factor arises via
    the contraction $(\bW_3)^A (W_4)_\rmA \propto [\bar{3}|\mem{ z(\s_3) \mminus \bz(\s_4) }\mem|4\rangle$
    between massless twistor and dual twistor.
}

First of all,
the invariants can be organized in terms of
the number of polarizations involved
and the mass dimension,
for instance.
Schematically, they are
$kk$,
$kp$, $ky$,
$k\vf p$,
$y\vf k$,
$y\vf p$,
$\vf\vf$,
$p\vf\vf p$,
$y\vf\vf p$,
$y\vf\vf y$,
etc.,
together with 
the ``spin magnitude'' $y^2$.
Of course,
not all of these are independent.
By utilizing
the on-shell conditions on the bivector polarizations,
it can be shown that
\begin{subequations}
\label{kids}
\begin{align}
\label{kid.f3kp}
    k_4\vf_3p
    \,&=\,
        -\frac{
            \pffp
        }{2A_4}\mem
        (\c_{43}\mplus\b)
    \,,\\
\label{kid.f4kp}
    k_3\vf_4p
    \,&=\,
        +\frac{
            \pffp
        }{2A_3}\mem
        (\c_{34}\mminus\a)
    \,,\\
\label{kid.f3ky}
    +\frac{
        y\hhem\varphi_3\hnem k_4
    }{
        k_3y
    }
    \,&=\,
        \frac{\pffp}{K_3A_3\b}\mem
        \BB{
            \d - (\c_{34}\mminus\a)\mem\b
        }
    \,,\\
\label{kid.f4ky}
    -\frac{
        y\hhem\varphi_4\hnem k_3
    }{
        k_4y
    }
    \,&=\,
        \frac{\pffp}{K_3A_4\a}\mem
        \BB{
            \d + (\c_{43}\mplus\b)\mem\a
        }
    \,,\\
\label{kid.ff}
    \vf_3 \mdot \vf_4
    \,&=\,
        \frac{
            {\pffp}
        }{m^2\x}\mem
        \bb{\nem
            \frac{(\c_{34}\mminus\a)(\c_{43}\mplus\b)}{\d}
            -1
        \hnem}
    \,,\\
\label{kid.yffy}
    \yffy
    \,&=\,
        \frac{\pffp}{{K_3}^2}\mem
        \BB{
            \d + \c_{43}\a - \c_{34}\b + \a\b
        }
    \,,\\
\label{kid.ysq}
    t\hem y^2
    \,&=\,
        \frac{\c^2}{\xi}
        + \d
    \,.
\end{align}
\end{subequations}
Here, we have denoted
\begin{align}
    \label{A3,A4}
    A_3
    \mem:=\mem
        y\vf_3p
    \,,\quad
    A_4
    \mem:=\mem
        y\vf_4p
    \,,
\end{align}
while
$\pffp$
abbreviates
$p^\r \vf_3{}_\r{}^\m \vf_4{}_{\m\s} p^\s$
and so on.
Also,
$
\vf_3 \mdot \vf_4
\mem:=\mem
\tfrac{1}{2}\,
    \vf_3{}_{\m\n} \vf_4^{\m\n}
$.
We have identified
dimensionless 
and little group neutral
variables,
\begin{align}
\begin{split}
    \label{abcdbasis}
    \a
    \mem:=\mem
        2k_3\mdot y
    \,,\quad
    \b
    \mem:=\mem
        2k_4\mdot y
    &\,,\quad
    \c_{34}
    \mem:=\mem
        K_3\mem \frac{\yffp}{\pffp}
    \,,\\
    \d
    \mem:=\mem
        -2\,
            \frac{
                A_3A_4\hem t
            }{
                \pffp
            }
    &\,,\quad
    \c_{43}
    \mem:=\mem
        K_3\mem \frac{\ryffp}{\pffp}
    \,,
\end{split}
\end{align}
while using 
\begin{align}
    \label{cmandel2}
    K_3
    \mem:=\mem
        2k_3\mdot p
    \,,\quad
    \xi
    \mem:=\mem
        \frac{{K_3}^2}{m^2t}
\end{align}
as a substitute for
the Mandelstams in the classical limit
as introduced earlier in
\eqrefss{cmandelstu}{K3=-K4}{opticalparam}.
The relative amplitude will be a function of 
the dimensionless little group neutrals:
$\a,\b,\c_{34},\c_{43},\d$, $\x$
and $t\hem y^2$.

\subsection{Same-Helicity Configurations}
\label{HSAME}

The same-helicity configurations
are characterized by
the identity
\begin{align}
    \label{hsame:cond}
    \tfrac{1}{2}\mem
    \bigbig{
        \vf_3\vf_4 \mplus \vf_4\vf_3
    }\hnem{}_{\m\n}
    \,=\,
        -\eta_{\m\n}\mem
            (\vf_3\mdot\vf_4)
    \,,
\end{align}
which is easy to see via the spinor-helicity formalism.

The implications of \eqref{hsame:cond} are the following.
First,
$\c_{34} + \c_{43} = 0$
since $p\mdot y = 0$.
Hence we can set
\begin{align}
    \label{hsame:c}
    \c_{34}
    \mem=\mem
    -\c_{43}
    \mem=:\mem
        \c
    \,.
\end{align}
Second,
$\pffp = m^2\mem (\vf_3\mdot\vf_4)$,
so \eqref{kid.ff} gives
\begin{align}
    \label{hsame:d}
    \d
    \,=\,
        -
        \frac{
            (\c\mminus\a)
            (\c\mminus\b)
        }{
            1 + \xi
        }
    \,.
\end{align}
Third,
$y^2 = - (\yffy)/(\vf_3\mdot\vf_4)$,
which reproduces \eqref{kid.ysq}.
To conclude,
the minimal basis for the relative same-helicity Compton amplitude
can be given as
$(\a,\b,\c,\xi)$
with $\c_{34}=-\c_{43}=\c$,
or equivalently
$(\a,\b,\c,\d)$.

\subsection{Mixed-Helicity Configuration}
\label{HOPPO}

The mixed-helicity configuration,
$(+,-)$,
is characterized by
\begin{align}
    \label{hoppo:cond}
    \comm{\vf_3}{\vf_4}
    \,=\,
        0
    \,,\quad
    \vf_3\mdot\vf_4
    \,=\,
        0
    \,,
\end{align}
which are again trivial in the spinor-helicity formalism.

First of all, $\comm{\vf_3}{\vf_4} = 0$ implies that
\begin{align}
    \label{hoppo:c}
    \c_{34}
    \mem=\mem
    \c_{43}
    \mem=:\mem
        \c
    \,.
\end{align}
Second of all, $\vf_3\mdot\vf_4 = 0$ implies,
via \eqref{kid.ff},
that 
\begin{align}
    \label{hoppo:d}
    \d
    \,=\,
        (\c\mminus\a)(\c\mplus\b)
    \,.
\end{align}
Thirdly, \eqref{kid.yffy} boils down to
\begin{align}
    \yffy
    \,=\,
        \frac{\pffp}{{K_3}^2}\,
        \c^2
    \,.
\end{align}
To conclude,
the minimal basis for the relative mixed-helicity Compton amplitude
can be given as
$(\a,\b,\c,\xi)$,
with $\c_{34}=\c_{43}=\c$.
It is easily seen that this $\c$ reproduces the earlier definition in \eqref{abcfirst},
via $(\vf_3\vf_4)_{\m\n} \propto w_\m w_\n$:
\begin{align}
    \label{cdef-via-y}
    w_{\a\da}
    \mem\propto\mem
        4_\a \bar{3}_\da
    \qiq
    \c
    \mem=\mem
        K_3\,
        \frac{w\mdot y}{w\mdot p}
    \,.
\end{align}

Note that the spin-magnitude $y^2$
is not an independent invariant
in any helicity configurations,
which should be clear from \eqref{kid.ysq}
(cf. \rcite{fabian2}).
On a related note,
our scope here is strictly restricted to conservative dynamics.

\section{
Nonabelian Root-Kerr Compton Amplitudes}
\label{CYM}

We scatter the nonabelian {\kerr} particle,
implemented in the phase space $\mtflat \times \C^N$
where $\mtflat$ is the massive twistor space
and $\C^N$ is the fundamental representation space of $\SU(N)$,
off the nonlinearly superposed background of two plane waves
in $\G = \SU(N)$ YM theory.

\subsection{The Two-Gluon Background}

Earlier in \eqref{F34},
we had constructed
the on-shell two-gluon geometry
in YM theory
while using an axial gauge condition.
We adopted covariant CK duality \cite{CCK},
which crucially allows for
a structured/principled organization of perturbation theory
in terms of
SD and ASD sectors.
The field strength
was given by
(evaluating at the holomorphic position with hindsight)
\begin{align}
\begin{split}
    \label{CYM:Foriginal}
	F^{a\sprime}
	\,=\,
    {}&{}
		\frac{ig\hem\e_3}{\sqrt{2}}\,
			c_3^{a\sprime}\, \varphi_3\,
				\mathe^{ik_3z}
		+ \frac{ig\hem\e_4}{\sqrt{2}}\,
			c_4^{a\sprime}\, \varphi_4\,
				\mathe^{ik_4z}
    \\
    {}&{}
		+ g^2\mem \e_3\e_4\,
			c_{34}^{a\sprime}\mem
			\bb{
				\varphi_{34}
				- \varphi_3\mem
					\frac{
						\eta\hhem\varphi_4\hnem k_3
					}{
						k_4\eta
					}
				+ \varphi_4\mem
					\frac{
						\eta\hhem\varphi_3\hnem k_4
					}{
						k_3\eta
					}
			\nem}\mem
				\frac{1}{t}\,
				\mathe^{ik_{34}z}
        \,,
\end{split}
\end{align}
where
$c_{34} := \comm{c_3}{c_4}$ by the $\SU(N)$ Lie bracket
while
$\varphi_{34} := \comm{\varphi_3}{\varphi_4}$ by the $\so(1,3)$ Lie bracket.

In general,
one could orient the reference vector $\eta^\m$
along a generic direction
and see if its dependence drops out from the final results.
This will provide an explicit check of the gauge invariance of the Compton amplitudes.
To simplify calculations,
however,
one can also consider making a clever choice of $\eta^\m$.

Let us postpone the verification of gauge invariance
to \Sec{CYM>GI}.
With hindsight,
we find that
the spin length vector $y^\m$ of the \textit{background} worldline
qualifies as a perfect candidate
for the reference vector,
as it is nondynamical.
In fact,
the nonlocal operator $1/\eta\mdot\partial$
in axial gauge
is naturally associated with the Dirac string,
while 
we recall from \Chap{K3:NJA} \cite{nja} that
the {\kerr} solution represents
a pair of chiral dyons joined by a Dirac string oriented in the direction of $iy^\m$.

We refer to this particular gauge,
$\eta^\m \propto y^\m$,
as the ``aligned axial gauge.''
With this choice made, 
\eqref{CYM:Foriginal} boils down to
\begin{align}
\begin{split}
    \label{CYM:F}
	F^{a\sprime}
	\,=\,
    {}&{}
		\frac{ig\hem\e_3}{\sqrt{2}}\,
			c_3^{a\sprime}\, \varphi_3\,
				\mathe^{ik_3z}
		+ \frac{ig\hem\e_4}{\sqrt{2}}\,
			c_4^{a\sprime}\, \varphi_4\,
				\mathe^{ik_4z}
    \\
    {}&{}
		+ g^2\mem \e_3\e_4\,
			c_{34}^{a\sprime}\mem
			\bb{
				\varphi_{34}
				- \varphi_3\mem
					\frac{
						y\hhem\varphi_4\hnem k_3
					}{
						k_4y
					}
				+ \varphi_4\mem
					\frac{
						y\hhem\varphi_3\hnem k_4
					}{
						k_3y
					}
			\nem}\mem
				\frac{1}{t}\,
				\mathe^{ik_{34}z}
        \,,
\end{split}
\end{align}
where \eqrefs{kid.f3ky}{kid.f4ky}
can be applied.

\subsection{The Two-Gluon Symplectic Structure}
\label{CYM>SYMP}

The {\kerr} symplectic potential is given in \eqref{rkerr.googly},
in the googly form:
\begin{align}
    \label{CYM|theta}
    \theta_{\nem\kerr}
	\mem&=\,
        i\mem \bZ_I{}^\rmA\mem dZ_\rmA{}^I
        +
		i\mem \bpsi_{i'} D\psi^{i\sprime}
		- 2i\mem E_1\hnem(-2i\pounds_N)
		\act{
            q_{a'}\hem \i_NF^-{}^{a\sprime}
        }
    \,,
\end{align}
where we see that the truncated exponential function $E_1$ is in play,
in connection with the Dirac string.
With the identification of the free theory as
\begin{align}
    \label{CYM|theta0}
    \theta^\circ
    \mem=\,
        i\mem \bZ_I{}^\rmA\mem dZ_\rmA{}^I
        +
		i\mem \bpsi_{i'}\mem d\psi^{i\sprime}
    \,,
\end{align}
the interaction symplectic potential for {\kerr} is
\begin{align}
    \label{CYM|theta'}
    \kern-0.05em
    \theta'_{\nem\kerr}
    \mem=\mem
		q_{a'} A^{a\sprime}
		+
        \sum_{\ell=1}^\infty
            \frac{(-2i)^\ell}{\ell!}\,
                q_{a'}
                \bb{\nem
                    \bigbig{\hnem
                        P^-_\ell
                    \hnem}
                    {}^{a\sprime}{}_\s\, dz^\s
                    +
                    (\ell\mminus1)\mem
                    \bigbig{\hnem
                        P^-_{\ell-1}
                    \hnem}
                    {}^{a\sprime}{}_\s\, dy^\s
                \nem}
    \,,
    \kern-0.3em
\end{align}
so $\theta_{\nem\kerr} = \theta^\circ + \theta'_{\nem\kerr}$.
Crucially, \eqref{CYM|theta0} employs
the primed color variables.
In this perturbation scheme,
the color phase space $\C^N$
takes $\psi^{i\sprime}$ as coordinates.

To study the physics of the {\kerr} particle
in the two-gluon background,
one plugs in \eqref{CYM:Foriginal} to \eqref{CYM|theta'}.
To this end,
let 
$\mathbf{F} = \d\zeta^i\mem {\partial}/{\partial\zeta^i}$
be a section of $T(\mtflat \mtimes \C^N)$.
Let $\mathbf{T}$
be a Hamiltonian vector field on $\mtflat \times \C^N$
such that
\begin{align}
    \label{CYM|tev}
    \i_\mathbf{T}\hem dz^\m
    \mem=\mem
        2p^\m
    \,,\quad
    \i_\mathbf{T}\hem dy^\m
    \mem=\mem
        0
    \,,\quad
    \i_\mathbf{T}\hem dp_\m
    \mem=\mem
        0
    \,,\quad
    \i_\mathbf{T}\hem dq_{a\sprime}
    \mem=\mem
        0
    \,.
\end{align}
The astute reader will recognize that
$\mathbf{F}$ gives the notion of fluctuation,
whereas
$\mathbf{T}$ refers to
the time-evolution vector field of the free theory
described in \eqref{zzpsaddle}.
With this formulation,
the universal formulae in \eqref{V-universal}
amount to computing the following geometric expressions
on the saddle trajectory:
\begin{subequations}
\label{unv}
\begin{align}
\label{unv0}
    \mathcal{V}^{(0)}
    &\quad\rightsquigarrow\quad
    \i_\mathbf{T}\hem \theta'
    \,=\,
        \theta'\hhnem(\mathbf{T})
    \,,\\
\label{unv1}
    \mathcal{V}^{(1)}
    &\quad\rightsquigarrow\quad
    \i_\mathbf{T}\hem
    \i_\mathbf{F}\hem
        \omega'
    \,=\,
        \omega'\hhnem(\mathbf{F},\mathbf{T})
    \,.
\end{align}
\end{subequations}

\subsection{Vertices of Fluctuation Valence One}
\label{CYM>V1}

We begin with \eqref{unv1}.
In this case, we simply need to work in the linearized theory.
Consequently,
we can simply focus on deriving the positive-helicity vertices.
This is
because
the negative-helicity vertices are
trivially related to the positive-helicity ones
at the linearized order.

In any SD background,
the infinite sum in \eqref{CYM|theta'}
drops
so that the interaction symplectic form is simply
\begin{align}
    \label{CYM|omega'}
    \omega'_{\nem\kerr}
    \,=\,
            dq_{a\sprime} \wedge A^{a\sprime}
    		+ q_{a\sprime} \wedge dA^{a\sprime}
    \,.
\end{align}
For instance,
plugging in the background of a single SD plane wave $3^+$
gives
\begin{align}
    \label{CYM|ocalc}
    \kern-0.1em
    \omega'_{\nem\kerr}(\mathbf{F},\mathbf{T})
    \mem\Big|_{\e_3}
    =
    \sqrt{2}g\mem 
    \bb{\nem
        (\d q_{a\sprime}\, c^{a\sprime}\hem)
        \mem
            \frac{2A_3}{\a}
        +
        i\hem (q\mdot c)
        \,
            \d z^\m\hem
            (\vf_3p)_\m
    \hnem\nem}\,
    \mathe^{ik_3z}\,
    \mathe^{iK_3\s}
    \,.
    \kern-0.25em
\end{align}
From this calculation, we
deduce the following vertices:
\begin{subequations}
\label{CYMV1}
\begin{align}
\label{CYMV1:z}
    \adjustbox{valign=c}{\begin{tikzpicture}[]
        \node[empty] (o) at (0,0) {};
        \node[empty] (R) at (1,0) {};
        \draw[lin] (o)--(R);
        \node[wdot=$3^+$] (v) at (o) {};
        \node[f=$\d z^\m$] (f) at (R) {};
    \end{tikzpicture}}
    \,\,&=\,\,\,
        \sqrt{2}\hem
        i\hem
        (q\mdot c)\mem
            (\vf_3^+\hnem p)_\m
        \,\mathe^{ik_3z}
    \,,\\[.5\baselineskip]
\label{CYMV1:bz}
    \adjustbox{valign=c}{\begin{tikzpicture}[]
        \node[empty] (o) at (0,0) {};
        \node[empty] (R) at (1,0) {};
        \draw[lin] (o)--(R);
        \node[bdot=$4^-$] (v) at (o) {};
        \node[f=$\d \bz^\m$] (f) at (R) {};
    \end{tikzpicture}}
    \,\,&=\,\,\,
        \sqrt{2}\hem
        i\hem
        (q\mdot c)\mem
            (\vf_4^-\hnem p)_\m
        \,\mathe^{ik_4\bz}
    \,,\\[.5\baselineskip]
\label{CYMV1:q+}
    \adjustbox{valign=c}{\begin{tikzpicture}[]
        \node[empty] (o) at (0,0) {};
        \node[empty] (R) at (1,0) {};
        \draw[dlin] (o)--(R);
        \node[wdot=$3^+$] (v) at (o) {};
        \node[f=$\d q_{a\sprime}$] (f) at (R) {};
    \end{tikzpicture}}
    \,\,&=\,\,\,
        2\hnem\sqrt{2}\hem
        c^{a\sprime}\,
            \frac{A_3^+}{\a}
        \,\mathe^{ik_3z}
    \,,\\[-.1\baselineskip]
\label{CYMV1:q-}
    \adjustbox{valign=c}{\begin{tikzpicture}[]
        \node[empty] (o) at (0,0) {};
        \node[empty] (R) at (1,0) {};
        \draw[dlin] (o)--(R);
        \node[bdot=$4^-$] (v) at (o) {};
        \node[f=$\d q_{a\sprime}$] (f) at (R) {};
    \end{tikzpicture}}
    \,\,&=\,\,\,
        2\hnem\sqrt{2}\hem
        c^{a\sprime}\,
            \frac{A_4^-}{\b}
        \,\mathe^{ik_4\bz}
    \,.
\end{align}
\end{subequations}
Here, we have peeled off 
the YM coupling $g$
and
explicitly written down the fluctuation species,
while omitting the factors
$\mathe^{iK_3\s}$, $\mathe^{iK_4\s}$.
It is left as a straightforward exercise
to derive the negative-helicity vertices in \eqrefs{CYMV1:bz}{CYMV1:q-}
directly
from the exterior derivative of \eqref{CYM|theta'},
in which case incorporating the $dy$ terms 
is crucial.
By construction,
the vertices in \eqref{CYMV1} are identical to the scalar vertices obtained in \rcite{wlf-ps}
up to attaching the $\mathe^{ikz}$, $\mathe^{ik\bz}$ factors.

\subsection{Vertices of Fluctuation Valence Zero}
\label{CYM>V0}

The trickier part is \eqref{unv0},
which contracts \eqref{CYM|tev}
with \eqref{CYM|theta'}.
In this case,
we need to extract the part of $\theta'\hhnem(\mathbf{T})$
multilinear in both $\e_3$ and $\e_4$.
Luckily, the axial gauge with $\eta^\m \propto y^\m$
radically simplifies the extraction of this multilinear part of the $P$-tensors:
$y\mdot D = y\mdot \partial$.
As a result, we find
\begin{align}
    \label{CYMV0|ex0}
    \theta'_\kerr(\mathbf{T})\mem\Big|_{\e_3\e_4}
    \mem=\mem
		q_{a'}\mem
        \bb{
            A^{a\sprime}{}_\m(z)
            -2i\hem E_1(\a\mplus\b)\,
                y^\r F^-{}^{a\sprime}{}_{\r\m}(z)
        }\bigg|_{\e_3\e_4}
        \mem 2p^\m
    \,.
\end{align}
In fact, 
this can be directly deduced from
\eqref{CYM|theta}:
note how the $E_1$ structure
is directly inherited down.
By axial gauge,
\eqref{CYMV0|ex0}
boils down to
\begin{align}
\begin{split}
    \label{CYMV0|ex1}
    {}&{}
		-4i\hem q_{a'}\hem
        \bb{
            \frac{1}{\a\mplus\b}\,
                F^{a\sprime}{}_{\r\s}(z)
            + E_1(\a\mplus\b)\,
                F^-{}^{a\sprime}{}_{\r\s}(z)
        }\bigg|_{\e_3\e_4}
        \mem y^\r p^\s
    \,,\\
    {}&{}
    \,=\,
        \frac{
            -4i\hem q_{a'}
        }{\a\mplus\b}\mem
        \bb{
                F^+{}^{a\sprime}{}_{\r\s}(z)
            + F^-{}^{a\sprime}{}_{\r\s}(\bz)
        }\bigg|_{\e_3\e_4}
        \mem y^\r p^\s
    \,,
\end{split}
\end{align}
where the replacement of
$\mathe^{\a+\b}\mem F^-{}^{a\sprime}{}_{\r\s}(z)$
with
$F^-{}^{a\sprime}{}_{\r\s}(\bz)$
is valid
by virtue of the dependence
$F^-{}^{a\sprime}{}_{\r\s}(z) \mem\big|_{\e_3\e_4} \sim \mathe^{i(k_3+k_4)z}$.

\paragraph{Same-Helicity Configurations}

We had discussed in \eqref{collision34.YM++}
that the collision of two SD plane waves in YM theory
produces a purely SD mode.
In the same way,
the collision of two ASD plane waves in YM theory
produces a purely ASD mode.
Provided this fact,
the particular simplified form of \eqref{CYMV0|ex1}
in the aligned axial gauge
implies that
the $(+,+)$ and $(-,-)$ vertices
will 
be identical
up to swapping $z \leftrightarrow \bz$.
In this sense
it happens to be that
our perturbation theory
is treating the plus and minus helicities
symmetrically.
To emphasize, however,
this accidental simplification is crucially by virtue of our clever choice $\eta^\m \propto y^\m$
of the gauge condition:
see \Sec{CYM>GI}.

For the $(+,+)$ case,
plugging in \eqref{CYM:F} to
\eqref{CYMV0|ex1}
gives
\begin{align}
    \kern0.2em
    \frac{
        -4i g^2\hem
        (q\mdot c_{34})
    }{(\a\mplus\b)\hem t}
    \mem
        \bb{
            \yffpPP
            - \ryffpPP
            - A^+_3\mem
                \frac{
                    y\hhem\varphi^+_4\nem k_3
                }{
                    k_4y
                }
            + A^+_4\mem
                \frac{
                    y\hhem\varphi^+_3\nem k_4
                }{
                    k_3y
                }
        \nem}
        \mem\mathe^{ik_{34}z}
    \,.
    \kern-0.4em
\end{align}
Using 
the kinematic identities in 
\eqrefs{kid.f3kp}{kid.f4kp}
with the findings in \Sec{HSAME},
this boils down to
\begin{align}
    (q\mdot c_{34})
    \bbsq{
        - \frac{4ig^2\hem(\pffpPP)}{{K_3}\hem t}
    }
    \bb{\hnem
        1 \hnem+\hnem \frac{\d}{\a\b}
    \hnem}
        \mem\mathe^{ik_{34}z}
    \,.
\end{align}
As a result,
the same-helicity contact vertices are found as
\begin{subequations}
\label{CYMV0}
\begin{align}
\label{CYMV0:++}
    \adjustbox{valign=c}{\begin{tikzpicture}[]
        \node[empty] (o) at (0,0) {};
        \node[wdot=$3^+4^+$] (v) at (o) {};
    \end{tikzpicture}}
    \,\,\,
    \,\,&=\,\,\,
        (q\mdot c_{34})
        \bbsq{
            - \frac{4i\hem(\pffpPP)}{{K_3}\hem t}
        }
        \bb{\hnem
            1 \hnem+\hnem \frac{\d}{\a\b}
        \hnem}
            \mem\mathe^{ik_{34}z}
    \,,\\[.5\baselineskip]
\label{CYMV0:--}
    \adjustbox{valign=c}{\begin{tikzpicture}[]
        \node[empty] (o) at (0,0) {};
        \node[bdot=$3^-4^-$] (v) at (o) {};
    \end{tikzpicture}}
    \,\,\,
    \,\,&=\,\,\,
        (q\mdot c_{34})
        \bbsq{
            - \frac{4i\hem(\pffpMM)}{{K_3}\hem t}
        }
        \bb{\hnem
            1 \hnem+\hnem \frac{\d}{\a\b}
        \hnem}
            \mem\mathe^{ik_{34}\bz}
    \,.
\end{align}
\end{subequations}

\paragraph{Mixed-Helicity Configuration}

We had discussed in \eqref{collision34.YM+-}
that the collision of SD and ASD plane waves in YM theory
produces both SD and ASD modes.
Hence in the mixed-helicity configuration,
both terms in \eqref{CYMV0|ex1} are activated.

Noting that
$\comm{\vf^+_3}{\vf_4^-} = 0$,
plugging in \eqref{CYM:F} to
\eqref{CYMV0|ex1}
gives
\begin{align}
\begin{split}
    \label{CYMV0|calcPM1}
    {}&{}
    \frac{
        -4i g^2\hem
        (q\mdot c_{34})
    }{(\a\mplus\b)\hem t}
    \mem
        \bb{
            - A^+_3\mem
                \frac{
                    y\hhem\varphi^-_4\nem k_3
                }{
                    k_4y
                }
            \mem\mathe^{ik_{34}z}
            + A^-_4\mem
                \frac{
                    y\hhem\varphi^+_3\nem k_4
                }{
                    k_3y
                }
            \mem\mathe^{ik_{34}\bz}
        \nem}
    \\
    {}&{}
    \,=\,
        \frac{
            -4i g^2\hem
            (q\mdot c_{34})
        }{(\a\mplus\b)\hem t}
        \mem
            \bb{
                - A^+_3\mem
                    \frac{
                        y\hhem\varphi^-_4\nem k_3
                    }{
                        k_4y
                    }
                \mem\mathe^{-\b}
                + A^-_4\mem
                    \frac{
                        y\hhem\varphi^+_3\nem k_4
                    }{
                        k_3y
                    }
                \mem\mathe^{\a}
            \nem}
        \mem\mathe^{ik_3z} \mathe^{ik_4\bz}
    \,.
\end{split}
\end{align}
Again,
we use
the kinematic identities in 
\eqrefs{kid.f3kp}{kid.f4kp}
with the findings in \Sec{HOPPO}.
The resulting mixed-helicity contact vertex is
\begin{align}
\label{CYMV0:+-}
    \adjustbox{valign=c}{\begin{tikzpicture}[]
        \node[empty] (o) at (0,0) {};
        \node[ndot'=$3^+4^-$] (v) at (o) {};
    \end{tikzpicture}}
    \,\,\,
    \,\,&=\,\,\,
        (q\mdot c_{34})
        \bbsq{
            - \frac{4i\hem(\pffpPM)}{{K_3}\hem t}
        }
        \bb{\nem\nem
            \bb{\hnem
                1 \hnem+\hnem \frac{\d}{\a\b}
            \hnem}\mem
            \frac{\b\mem \mathe^\a + \a\mem \mathe^{-\b}}{\a+\b}
            + \c\mem 
            \frac{\mathe^\a - \mathe^{-\b}}{\a+\b}
        }
            \mem\mathe^{ik_3z}\hem \mathe^{ik_4\bz}
    \,.
\end{align}

\subsection{Derivation of Same-Helicity Compton Amplitudes}
\label{CYM>DER++}

Provided 
the Feynman rules in \Secss{COMPTON>methodology}{CYM>V0}{CYM>V1},
we are finally ready to
compute the {\kerr} Compton amplitudes.
We begin with the same-helicity cases,
in which case
we may simply work out the $(+,+)$ configuration
since the Feynman rules are symmetric
in the aligned axial gauge.

\paragraph{Abelian Contribution}

By using
the vertex in \eqref{CYMV1:z}
and the propagator in \eqref{Int2},
we immediately find
\begin{align}
    \label{CYMA:a++}
    \adjustbox{valign=c}{\begin{tikzpicture}[]
        \node[empty] (o) at (0,0) {};
        \node[empty] (R) at (1,0) {};
        \draw[lin] (o)--(R);
        \node[wdot=$3^+$] (v) at (o) {};
        \node[wdot=$4^+$] (v) at (R) {};
    \end{tikzpicture}}
    \,\,\,
    \,\,&=\,\,\,
        (q\mdot c_3)
        (q\mdot c_4)
        \hem
        \bbsq{
            - \frac{4\hem(\pffpPP)}{{K_3}^2}
        }
        \mem\mathe^{ik_{34}z}
    \,.
\end{align}
Upon stripping off the trivial product 
$
    (q\mdot c_3)
    (q\mdot c_4)
$
of color factors,
\eqref{CYMA:a++}
is precisely the scalar Compton amplitude
of abelian gauge theory
shown in \eqref{Compton0},
spin-exponentiated
to the zig position $z$.

\paragraph{Nonabelian Contribution}

By using
the vertex in \eqref{CYMV1:q+}
and the propagator in \eqref{Intqq},
we find
\begin{align}
    \label{CYMF:q++}
    \adjustbox{valign=c}{\begin{tikzpicture}[]
        \node[empty] (o) at (0,0) {};
        \node[empty] (R) at (1,0) {};
        \draw[dlin] (o)--(R);
        \node[wdot=$3^+$] (v) at (o) {};
        \node[wdot=$4^+$] (v) at (R) {};
    \end{tikzpicture}}
    \,\,\,
    \,\,&=\,\,\,
        (q\cdot c_{34})\mem
        \bbsq{
            - \frac{4i\hem(\pffpPP)}{{K_3}\hem t}
        }
        \bb{\hnem
            -\frac{\d}{\a\b}
        \hnem}
        \mem\mathe^{ik_{34}z}
    \,,
\end{align}
where
the definition of $\d$ in \eqref{abcdbasis}
is used.
Via the core calculation
\begin{align}
    \bb{\hnem
        1 \hnem+\hnem \frac{\d}{\a\b}
    \hnem}
    +
    \bb{\hnem
        -\frac{\d}{\a\b}
    \hnem}
    \,=\,
        1
    \,,
\end{align}
the sum of \eqrefs{CYMF:q++}{CYMV0:++}
is found as
\begin{align}
    \label{CYMA:na++}
    \adjustbox{valign=c}{\begin{tikzpicture}[]
        \node[empty] (o) at (0,0) {};
        \node[empty] (R) at (1,0) {};
        \draw[dlin] (o)--(R);
        \node[wdot=$3^+$] (v) at (o) {};
        \node[wdot=$4^+$] (v) at (R) {};
    \end{tikzpicture}}
    \,\,\,
    \,\,\,+\,\,\,
    \,\,\,
    \adjustbox{valign=c}{\begin{tikzpicture}[]
        \node[empty] (o) at (0,0) {};
        \node[wdot=$3^+4^+$] (v) at (o) {};
    \end{tikzpicture}}
    \,\,\,
    \,\,&=\,\,\,
        (q\mdot c_{34})\mem
        \bbsq{
            - \frac{4i\hem(\pffpPP)}{{K_3}\hem t}
        }
        \mem\mathe^{ik_{34}z}
    \,.
\end{align}
Upon dropping the color factor
$
    (q\mdot c_{34})
$,
\eqref{CYMA:na++} is precisely 
the color-stripped scalar Compton amplitude
of YM theory
shown in \eqref{Compton0},
spin-exponentiated
to the zig position $z$.
Note that the calculations here are isomorphic to
the calculations for 
the minimally coupled scalar particle
in \rcite{wlf-ps}.

\paragraph{Conclusion}

To conclude,
the same-helicity {\kerr} Compton amplitudes in YM theory are
\begin{align}
    {}&{}
    \M_{\kerr}(3^\pm4^\pm)
    \\[-0.05\baselineskip]
    \nonumber
    {}&{}
    \,=\,
    \lrp{\nem
        (q\mdot c_3)
        (q\mdot c_4)
        \hem
        \bbsq{
            - \frac{4\hem(\pffpPPMM)}{{K_3}^2}
        }
    +\mem
        (q\mdot c_{34})\mem
        \bbsq{
            - \frac{4i\hem(\pffpPPMM)}{{K_3}\hem t}
        }
    }
        \mathe^{ik_{34}z}
    \,,
\end{align}
where it is of course customary to
suppress the impact parameter: $x^\m\to0$.

\subsection{Derivation of Mixed-Helicity Compton Amplitude}
\label{CYM>DER+-}

\paragraph{Abelian Contribution}

By using
the vertices in \eqrefs{CYMV1:z}{CYMV1:bz}
and the propagator in \eqref{IntZZ},
we find
\begin{align}
    \label{CYMA:a+-.first}
    \adjustbox{valign=c}{\begin{tikzpicture}[]
        \node[empty] (o) at (0,0) {};
        \node[empty] (R) at (1,0) {};
        \draw[lin] (o)--(R);
        \node[wdot=$3^+$] (v) at (o) {};
        \node[bdot=$4^-$] (v) at (R) {};
    \end{tikzpicture}}
    \,\,\,
    \,\,&=\,\,\,
        (q\mdot c_3)
        (q\mdot c_4)
        \hem
        \bbsq{
            \frac{-4i\hem
                (p\hem\vf_3^+\nem S\vf_4^-\nem p)
            }{m^2K_3}
            +
            \frac{-4\hem
                (\pffpPM)
            }{{K_3}^2}
        }
        \mem\mathe^{ik_3z}
        \hem\mathe^{ik_4\bz}
    \,.
\end{align}
The expression in \eqref{CYMA:a+-.first}
was first found in \rcite{Kim:2024grz}.
Regarding the product
$-4i\hem (p\hem\vf_3^+\nem S\vf_4^-\nem p)$,
we use the self-duality condition
$i\hem\vf_3 = {*}\vf_3$
to compute instead
$-4\hem \bigbig{
    p\mem({*}\vf_3^+) 
    ({*}(y\swedge\hnem p)\hhnem)
    \vf_4^-\nem p
}$.
By eliminating the epsilon tensors,
it boils down to
$-4m^2\mem (\yffpPM)$.
In this way,
\eqref{CYMA:a+-.first}
is brought to
\begin{align}
    \label{CYMA:a+-}
    \adjustbox{valign=c}{\begin{tikzpicture}[]
        \node[empty] (o) at (0,0) {};
        \node[empty] (R) at (1,0) {};
        \draw[lin] (o)--(R);
        \node[wdot=$3^+$] (v) at (o) {};
        \node[bdot=$4^-$] (v) at (R) {};
    \end{tikzpicture}}
    \,\,\,
    \,\,&=\,\,\,
        (q\mdot c_3)
        (q\mdot c_4)
        \hem
        \bbsq{
            - \frac{4\hem(\pffpPM)}{{K_3}^2}
        }
        \bigbig{
            1 + \c
        }
        \mem\mathe^{ik_3z}
        \hem\mathe^{ik_4\bz}
    \,.
\end{align}
\eqref{CYMA:a+-} derives the relative amplitude $1+\c$
stated in \eqref{RelCompton:EM}.
We see that the zig-zag structure of spinspacetime
is crucial in generating the $\c$ term.
As remarked earlier in a footnote,
this physically represents a gyroscope effect:
a first-order communication between the zig point $z$ and the zag point $\bz$.

\paragraph{Nonabelian Contribution}

By using
the vertices in \eqrefs{CYMV1:q+}{CYMV1:q-}
and the propagator in \eqref{Intqq},
we find
\begin{align}
    \label{CYMF:q+-}
    \adjustbox{valign=c}{\begin{tikzpicture}[]
        \node[empty] (o) at (0,0) {};
        \node[empty] (R) at (1,0) {};
        \draw[dlin] (o)--(R);
        \node[wdot=$3^+$] (v) at (o) {};
        \node[bdot=$4^-$] (v) at (R) {};
    \end{tikzpicture}}
    \,\,\,
    \,\,&=\,\,\,
        (q\cdot c_{34})\mem
        \bbsq{
            - \frac{4i\hem(\pffpPM)}{{K_3}\hem t}
        }
        \bb{\hnem
            -\frac{\d}{\a\b}
        \hnem}
        \mem\mathe^{ik_{34}z}
    \,,
\end{align}
again using \eqref{abcdbasis}.
With $\d = (\c\mminus\a)(\c\mplus\b)$,
the core calculation
\begin{align}
    \bb{\nem\nem
        \bb{\hnem
            1 \hnem+\hnem \frac{\d}{\a\b}
        \hnem}\mem
        \frac{\b\mem \mathe^\a + \a\mem \mathe^{-\b}}{\a+\b}
        + \c\mem 
        \frac{\mathe^\a - \mathe^{-\b}}{\a+\b}
    \hnem}
    +
    \bb{\hnem
        -\frac{\d}{\a\b}
    \hnem}
    \,=\,
        \AHH[1]
\end{align}
establishes that
the sum of \eqrefs{CYMF:q+-}{CYMV0:+-}
is given by
\begin{align}
    \label{CYMA:na+-}
    \adjustbox{valign=c}{\begin{tikzpicture}[]
        \node[empty] (o) at (0,0) {};
        \node[empty] (R) at (1,0) {};
        \draw[dlin] (o)--(R);
        \node[wdot=$3^+$] (v) at (o) {};
        \node[bdot=$4^-$] (v) at (R) {};
    \end{tikzpicture}}
    \,\,\,
    \,\,\,+\,\,\,
    \,\,\,
    \adjustbox{valign=c}{\begin{tikzpicture}[]
        \node[empty] (o) at (0,0) {};
        \node[ndot'=$3^+4^-$] (v) at (o) {};
    \end{tikzpicture}}
    \,\,\,
    \,\,&=\,\,\,
        (q\mdot c_{34})\mem
        \bbsq{
            - \frac{4i\hem(\pffpPM)}{{K_3}\hem t}
        }
        \BB{
            \AHH[1]
        }
        \mem\mathe^{ik_3z}
        \hem\mathe^{ik_4\bz}
    \,.
\end{align}

\paragraph{Conclusion}

To conclude,
we have successfully derived
the mixed-helicity {\kerr} Compton amplitude
as
\begin{align}
    {}&{}
    \M_{\kerr}(3^+4^-)
    \\[-0.05\baselineskip]
    \nonumber
    {}&{}
    \,=\,
    \lrp{
    \begin{aligned}[c]
        {}&{}
            (q\mdot c_3)
            (q\mdot c_4)
            \hem
            \bbsq{
                - \frac{4\hem(\pffpPM)}{{K_3}^2}
            }
            \bigbig{
                1+\c
            }
        \\[-.05\baselineskip]
        {}&{}
        +\mem
            (q\mdot c_{34})\mem
            \bbsq{
                - \frac{4i\hem(\pffpPM)}{{K_3}\hem t}
            }\nem
            \lrp{\nem
            \begin{aligned}[c]
                {}&{}
                    \frac{\b\mem \mathe^\a + \a\mem \mathe^{-\b}}{\a+\b}
                    + \c\mem 
                    \frac{\mathe^\a - \mathe^{-\b}}{\a+\b}
                \\
                {}&{}
                    + (\c\mminus\a)(\c\mplus\b)\,
                    \frac{E_1(\a) \hnem-\nem E_1(-\b)}{\a+\b}
            \end{aligned}
            \nem\nem}
    \end{aligned}
    \nem\nem}
        \mathe^{ik_3z}
        \hem\mathe^{ik_4\bz}
    \,.
\end{align}
%
%
%
%
%
%
%
%
%
%
%
%
%
%
%
%
%
%
%
%
%
%
%
%
%
%
%
%
%
%
%
%
%
%
%
%
%
%
%
%
%
%


\subsection{Gauge Invariance}
\label{CYM>GI}

In \Secs{CYM>DER++}{CYM>DER+-},
we have shown
the derivation of
the {\kerr} Compton amplitudes
from the googly {\kerr} action
while employing the aligned axial gauge:
$\eta^\m \propto y^\m$.
Remarkably,
this particular choice of the reference vector
rendered the perturbation theory
symmetric
on the helicities.

The role of this subsection
is to describe the case
when the reference vector $\eta^\m$
is oriented along a generic direction.
The objective is twofold.
First,
we establish the gauge invariance
and overall consistency
of our formalism
by explicitly showing that
the reference-dependent terms
in the Feynman rules
cancel each other
in the amplitudes.
Second,
we explicitly clarify that
the perturbation theories
due to googly actions
are supposed to treat the helicities asymmetrically,
i.e.,
are inherently chiral in general.
When the googly action is realized around the 
zig worldline,
the minus-helicity spin exponentiation
onto zag
is nontrivial
and rather provides a nice consistency check.

To this end,
one should find
the general formula for the $P$-tensors
evaluated on the background in \eqref{CYM:Foriginal}
with the generic reference $\eta^\m$.
This is viable by carrying out brute-force calculations at a few low orders
and then observing a pattern,
for instance.
By resummation, the vertices are found as follows.
The linear vertices are
\begin{align}
\begin{split}
\label{GCYMV1}
    \adjustbox{valign=c}{\begin{tikzpicture}[]
        \node[empty] (o) at (0,0) {};
        \node[empty] (R) at (1,0) {};
        \draw[lin] (o)--(R);
        \node[wdot=$3^+$] (v) at (o) {};
        \node[f=$\d z^\m$] (f) at (R) {};
    \end{tikzpicture}}
    \,\,&=\,\,\,
        \sqrt{2}\hem
        i\hem
        (q\mdot c)\mem
            (\vf_3^+\hnem p)_\m
        \,\mathe^{ik_3z}
    \,,\\[.5\baselineskip]
    \adjustbox{valign=c}{\begin{tikzpicture}[]
        \node[empty] (o) at (0,0) {};
        \node[empty] (R) at (1,0) {};
        \draw[lin] (o)--(R);
        \node[bdot=$4^-$] (v) at (o) {};
        \node[f=$\d \bz^\m$] (f) at (R) {};
    \end{tikzpicture}}
    \,\,&=\,\,\,
        \sqrt{2}\hem
        i\hem
        (q\mdot c)\mem
            (\vf_4^-\hnem p)_\m
        \,\mathe^{ik_4\bz}
    \,,\\[.25\baselineskip]
    \adjustbox{valign=c}{\begin{tikzpicture}[]
        \node[empty] (o) at (0,0) {};
        \node[empty] (R) at (1,0) {};
        \draw[dlin] (o)--(R);
        \node[wdot=$3^+$] (v) at (o) {};
        \node[f=$\d q_{a\sprime}$] (f) at (R) {};
    \end{tikzpicture}}
    \,\,&=\,\,\,
        2\hnem\sqrt{2}\hem
        c^{a\sprime}\,
            \frac{A_3^+}{\a}
            \mem\bigbig{1\mem\mplus\hla{X_3}}
        \,\mathe^{ik_3z}
    \,,\\[-.1\baselineskip]
    \adjustbox{valign=c}{\begin{tikzpicture}[]
        \node[empty] (o) at (0,0) {};
        \node[empty] (R) at (1,0) {};
        \draw[dlin] (o)--(R);
        \node[bdot'=$4^-$] (v) at (o) {};
        \node[f=$\d q_{a\sprime}$] (f) at (R) {};
    \end{tikzpicture}}
    \,\,&=\,\,\,
        2\hnem\sqrt{2}\hem
        c^{a\sprime}\,
            \frac{A_4^-}{\b}
            \mem\bigbig{1\mem\mplus\hla{X_4}\hem \mathe^{-\b}}
        \,\mathe^{ik_4\bz}
    \,,
\end{split}
\end{align}
The contact vertices, after stripping off
$(q\mdot c_{34})$
and normalizing by the scalar Compton amplitude,
give
\begin{align}
\begin{split}
\label{GCYMV0}
    \adjustbox{valign=c}{\begin{tikzpicture}[]
        \node[empty] (o) at (0,0) {};
        \node[wdot=$3^+4^+$] (v) at (o) {};
    \end{tikzpicture}}
    \,\,\,
    \,\,&\rightsquigarrow\,\,\,
        \bb{\hnem
            1 \hnem+\hnem \frac{\d}{\a\b}\mem
            \bigbig{1\mem\mplus\hla{X_3}}
            \bigbig{1\mem\mplus\hla{X_4}}
        \nem}
            \mem\mathe^{ik_3z}\hem \mathe^{ik_4z}
    \,,\\[.1\baselineskip]
    \adjustbox{valign=c}{\begin{tikzpicture}[]
        \node[empty] (o) at (0,0) {};
        \node[bdot'=$3^-4^-$] (v) at (o) {};
    \end{tikzpicture}}
    \,\,\,
    \,\,&\rightsquigarrow\,\,\,
        \bb{\hnem
            1 \hnem+\hnem \frac{\d}{\a\b}\mem
            \bigbig{1\mem\mplus\hla{X_3}\hem \mathe^{-\a}}
            \bigbig{1\mem\mplus\hla{X_4}\hem \mathe^{-\b}}
        \nem}
            \mem\mathe^{ik_3\bz}\hem \mathe^{ik_4\bz}
    \,,\\[.1\baselineskip]
    \adjustbox{valign=c}{\begin{tikzpicture}[]
        \node[empty] (o) at (0,0) {};
        \node[ndot'=$3^+4^-$] (v) at (o) {};
    \end{tikzpicture}}
    \,\,\,
    \,\,&\rightsquigarrow\,\,\,
        \bb{\hnem
            \AHH[1]
            + 
                \frac{\d}{\a\b}\mem
                \bigbig{1\mem\mplus\hla{X_3}}
                \bigbig{1\mem\mplus\hla{X_4}\hem \mathe^{-\b}}
        \nem}
            \mem\mathe^{ik_3z}\hem \mathe^{ik_4\bz}
    \,.
\end{split}
\end{align}
We color-code
reference-dependent terms
in red.
It is easily seen that
\eqrefs{GCYMV1}{GCYMV0}
predict the same amplitudes as in
\Secs{CYM>DER++}{CYM>DER+-}.
This establishes the gauge invariance of our framework.
We have defined
\begin{align}
\begin{split}
    \hla{\a'}
    \,:=\,
        2k_3\eta
    \,,\quad
    \hla{\b'}
    \,:=\,
        2k_4\eta
    \,,\quad
    \hla{X_3}
    \,:=\,
        \frac{
            \eta\vf_3y
        }{
            \a'
        }\mem
        \frac{K_3}{A_3}
    \,,\quad
    \hla{X_4}
    \,:=\,
        \frac{
            \eta\vf_4y
        }{
            \b'
        }\mem
        \frac{K_4}{A_4}
    \,,
\end{split}
\end{align}
so
$X_3,X_4 \to 0$
as $\eta^\m$ gets aligned with $y^\m$.

Clearly, the vertices in \eqrefs{GCYMV1}{GCYMV0}
treat the plus and minus helicities on an unequal footing.
In particular,
the square-shaped vertices
in \eqrefs{GCYMV1}{GCYMV0}
fail to be localized at a single point,
unlike the round-shaped ones.
For instance,
$
\bigbig{1\mem\mplus\hla{X_4}\hem \mathe^{-\b}}
\mem\mathe^{ik_4\bz}
= \mathe^{ik_4\bz} + \hla{X_4}\hem \mathe^{ik_4z}
$
could be seen as describing
a mixture of
a vertex localizing at the zag position $\bz$
and 
a vertex localizing at the zig position $z$.
Similarly,
the $(-,-)$ vertex in \eqref{GCYMV0}
could be seen as
describing a mixture of interactions happening at
$(z,z)$, $(z,\bz)$, $(\bz,z)$, and $(\bz,\bz)$
for the third and fourth waves,
i.e., at all possible combination of locations.
Overall,
we see that 
the nonabelian {\kerr} particle
always interacts with incoming SD waves
via the zig position $z$
but receives incoming ASD waves 
either at the zig position $z$
or the zag position $\bz$.
This describes a mild form of nonlocality,
over a discrete set of points.\footnote{
    In any case, our {\kerr} particle seems to
    interact with
    the external environment
    always through either $z$ or $\bz$,
    and never $x$.
    The amplitudes advocated in \rrcite{%
        Ochirov:2022nqz,Cangemi:2022bew,Cangemi:2023bpe,%
        fabian1,fabian2,zihan23,Saketh:2023bul%
    },
    in contrast,
    seem to have a tendency to add
    $x = (z+\bz)/2$ as an additional interaction point.
}

For the 
square vertex in \eqref{GCYMV1}
and the $(-,-)$ vertex in \eqref{GCYMV0},
this nonlocality
will be considered 
a \textit{gauge artifact},
since it completely disappears in the aligned axial gauge.
For the $(-,+)$ contact vertex in \eqref{GCYMV0},
however,
even the aligned axial gauge yields
a superposition of multipoint localizations
due to the nature of $\AHH[1]$.
It seems unclear whether there can exist a perturbation theory
that represents the $(+,-)$ amplitude
as a sum 
in which each term is local in spinspacetime.

Physically,
the {\kerr} interaction action
is rather
a faithful implementation of
the chiral dyon pair model.
Amusingly,
the nonabelian Dirac string term
in \eqref{stheta1},
\begin{align}
    {}&{}
    \int 
        \sinc(\pounds_N)\bigbig{
            q_a\mem \i_N {*}F^a
        }
    \,=\,
        \int_{-1}^{+1} \frac{d\eta}{2}\,
        \int 
            \mathe^{i\eta\pounds_N}
            \bigbig{
                q_a\mem \i_N {*}F^a
            }
    \,,\\
    {}&{}
    \,=\,
        \frac{1}{2i}
        \int
            q_b(\t)\mem
            W^b{}_a\hnem\bigbig{\nem
                X(\t,0),
                X(\t,\eta)
            \nem}\mem
            {*}F^a{}_{\m\n}\hnem\bigbig{\nem
                X(\t,\eta)
            \nem}\hem
            \frac{\partial X^\m\hnem(\t,\eta)}{\partial \eta}
            \hem
            \frac{\partial X^\n\hnem(\t,\eta)}{\partial \tau}
            \,
                d\eta\hem d\tau
    \,,\nonumber
\end{align}
could be compared to
the topological surface operator
in \Chap{K1}, \eqref{1form.U},
\begin{align}
    \label{topsrfre}
    \int_\S\mem {*}F^a \mem \vn_a
    \,,
\end{align}
with $\vn_a$ made dynamical by
parallel transportation of a dynamical worldline profile $q_a(\t)$ across the worldsheet:
\begin{align}
    \vn_a(\t,\eta)
    \,=\,
        \frac{1}{2i}\mem
            q_b(\t)\mem
            W^b{}_a\hnem\bigbig{\nem
                X(\t,0),
                X(\t,\eta)
            \nem}
    \,.
\end{align}
Recalling the topological nature of 
the surface operator in \eqref{topsrfre},
the same-helicity amplitudes would localize on
a single worldline
(see \rcite{gmoov} for a prototype discussion in Maxwell theory).

The discrete nonlocality
of the $(+,-)$ amplitude
and its physical content
stand out as interesting in this respect.
Rather than attributing it
directly to the Dirac worldsheet,
we might wish to understand 
this nonlocality in complexified spacetime
as a consequence of a dynamical, physical interaction 
between the nonabelian SD and ASD dyons,
via inertia and the angular momentum effect.

\section{
Kerr Compton Amplitudes}
\label{CGR}

We scatter the Kerr particle,
in its massive twistor implementation,
off the nonlinearly superposed background of two gravitational waves in GR.

We follow the same methodology as in \Sec{CYM}.
For reasons of space,
we will simply perform the calculation in the aligned axial gauge,
despite the fact that 
the gauge invariance of our formalism
can be verified
in generic axial gauges.

\subsection{The Two-Graviton Background}

Earlier in \eqref{vf34.GR},
we had constructed
the on-shell two-graviton geometry
in generic axial gauge.
Here, we implement
the aligned axial gauge.
The spin length vector $y^\m$ at $x$
and
its parallel propagation to the zig point $z$, $y^{\m\sprime}$,
must agree 
at the level of their background value
by the very scattering setup.
Our aligned axial gauge will be
$\eta^{\m\sprime} \propto y^{\m\sprime}$
via \textit{spacetime} (not tetrad) components.

For instance,
let us reproduce \eqref{vf34.GR.oppo}
in the aligned axial gauge
for the mixed-helicity Compton scattering:
\begin{align}
\begin{split}
    \label{CGR|F}
    \vf_{34}^{\ta\sprime\mem\tb\sprime}
    \,\,=\,\mem
    {}&{}
        \vf_3^+{}^{\ta\sprime} 
        \vf_3^+{}^{\tb\sprime}
        \,
        \bb{\nem
            \frac{y\hhem\varphi_4\hnem k_3}{k_4y}
        \nem}^{\nem\nem\hnem2}
    \mem+\,
        \vf_4^-{}^{\ta\sprime} 
        \vf_4^-{}^{\tb\sprime}
        \,
        \bb{\nem
            \frac{y\hhem\varphi_3\hnem k_4}{k_3y}
        \nem}^{\nem\nem\hnem2}
    \,.
\end{split}
\end{align}
Again, the kinematic identities in 
\eqrefs{kid.f3ky}{kid.f4ky}
will be applied.

\subsection{The Two-Graviton Symplectic Structure}

The Kerr symplectic potential
coupled to full GR
is shown in \eqref{theta-earth}.
By a worldline field redefinition,
it can be equivalently presented as
\eqref{kerr.googly},
which manifests the resummation of SD graviton couplings onto the holomorphic (zig) worldline in spinspacetime.
This worldline field basis will be referred to as the googly basis.

Via discarding a total derivative
(cf. the gymnastics in \Sec{OSD1>CHIRAL}),
\eqref{kerr.googly} can be brought to
\begin{align}
\begin{split}
    \label{CGR|thetafull}
	\theta_\Kerr
	\mem=\,
    {}&{}
		p_{m'}\hhem e^{m\sprime}
        - 2i\mem
            \bar{\lambda}_{I\da'}\mem
            y^{\da\sprime\a\sprime}\hem
            D\lambda_{\a\sprime\mem}{}^I
    \\
    {}&{}
        + (-2i)^2\mem
            E_2(-2i\pounds_N)\act{
                p_{m'}
                (\i_N\mem R^-{}^{m\sprime}{}_{n'})\hem y^{n\sprime}
            }
    \,,
\end{split}
\end{align}
where $p_{\a\sprime\hem\da\sprime} = -\lambda_{\a\sprime\mem}{}^I\hem \trambda_{I\da'}$
is a composite.
To reiterate, \eqref{CGR|thetafull} represents the exact Kerr symplectic potential in generic, non-SD backgrounds.
The first term 
$p_{m'}\hhem e^{m\sprime}$
in the right-hand side
implements the mass monopole
while the second term
$
    - 2i\mem
        \trambda_{I\da'}\mem
        y^{\da\sprime\a\sprime}\hem
        D\lambda_{\a\sprime\mem}{}^I
$
is associated with dipoles
as discussed below \eqref{NJMAGIC|g2a}.
The last term,
crucially
involving the $E_2$ function,
implements higher multipoles
in terms of the ASD $Q$-tensors.

Notably,
in this googly presentation,
we see that SD gravitational modes
couple to the Kerr particle
only through the coframe $e^{m'}$:
the SD spin connection completely drops out 
from the dipole part,
while the higher multipole terms
insert an explicit ASD projector
on the curvature two-form
$R^{m\sprime}{}_{n'}$.
Eventually, 
it follows that
the perturbation theory 
in the SD sector
is isomorphic to that of the minimally coupled scalar particle
(symplectic potential $p_m\hem e^m$)
via the zig complexification.

Concretely,
to \textit{define} perturbation theory,
let $h^{m\sprime} = e^{m\sprime} - \delta^{m\sprime}{}_{\m'}\hem dz^{\m\sprime}$
be the tetrad perturbation
around the Kronecker vacuum.
With this split, we identify the free theory part of \eqref{CGR|thetafull} as
\begin{align}
\begin{split}
    \label{CGR|theta0}
	\theta^\circ
    \mem=\,
        - \trambda_{I\da'}\mem dz^{\da\sprime\hem\a\sprime}\mem
        \lambda_{\a\sprime\mem}{}^I\hem 
        - 2i\mem
            \trambda_{I\da'}\mem
            y^{\da\sprime\a\sprime}\hem
            d\lambda_{\a\sprime\mem}{}^I
    \,.
\end{split}
\end{align}
Based on this particular physical motivation,
we now make the crucial identification
of the ``nominal'' zag coordinates as
\begin{align}
    \label{nominalzag}
    w^{\da\sprime\hem\a\sprime}
    \,:=\,
        z^{\da\sprime\hem\a\sprime}
        - 2i\hem
            y^{\da\sprime\a\sprime}
    \,.
\end{align}
In fact, we had already encountered this idea 
in \eqref{NJMAGIC|muvariables}.
Via the algebraic worldline field redefinition
(construction of the mu variables)
\begin{align}\begin{split}
    \label{nominalmu}
	\mu^{\da\sprime\mem I}
	\,=\,
		z^{\da\sprime\a}\hem\hhem \lambda_\a{}^I
	\,,\quad
	\tmu_I{}^{\a\sprime}
	\,=\,
		\trambda_{I\da'}\mem w^{\da\sprime\hem\a\sprime}
	\,=\,
		\trambda_{I\da'}
		\bigbig{
			z^{\da\sprime\a} {\hem-\mem} 2iy^{\da\sprime\a}
		}
	\,,
\end{split}\end{align}
\eqref{CGR|theta0} is brought to
\begin{align}
    \label{CGR|freetwistorbasis}
	\theta^\circ
	\,&=\,
    	- \trambda_{I\da'}\hem d\mu^{\da\sprime\mem I}
    	+ \tmu_I{}^{\a\sprime}\hem d\lambda_{\a\sprime\mem}{}^I
	\,,
\end{align}
which exactly describes the complexified free massive twistor model
via the unpriming worldline field redefinition
$\trambda_{I\da'} \mapsto \trambda_{I\da}$,
$\mu^{\da\sprime\mem I} \mapsto \mu^{\da I}$,
etc.
Crucially,
\eqref{CGR|freetwistorbasis}
identifies Darboux coordinates
so that the free theory is quadratic.

Finally, the interaction symplectic potential
arises by the difference between
\eqrefs{CGR|thetafull}{CGR|theta0},
which computes to
\begin{align}
    \label{CGR|theta'}
    \theta'_\Kerr
	\mem&=\,
\begin{aligned}[t]
		&
		p_{m'}\hhem h^{m\sprime}
        - 2i\mem p_{m'}\hem \gamma^-{}^{m\sprime}{}_{n'}\hhem y^{n\sprime}
        \\
        &
        +\mem\hhem \sum_{\ell=2}^\infty
\begin{aligned}[t]
{}&{}
    	 		\frac{(-2i)^\ell}{\ell!}
    			\sum_{p=1}^{\lfloor{\ell/2}\rfloor}\kern-0.2em
    			\sum_{\a \in \Omega_p(\ell)}\kern-0.2em
    	 			\bbsq{
    	 				\prod_{i=1}^p
    	 				\binom{
    	 					\bigbig{
    	 						\sum_{j=i}^p \a_j
    	 					} \mminus 2
    	 					\hem
    	 				}{
    	 					\a_i \mminus 2
    	 				}
    	 				\nem\nem
    	 			}\mem
\\
{}&{}
    	 		\lrp{
    	 		\begin{aligned}[c]
    	 			&
				\bigbig{\hnem
					(Q^-_{\a_1}\hnem)\mem Q_{\a_2}\, {\nem\cdots\mem} Q_{\a_p}
				\hnem}
				{}^{m\sprime}{}_{s'}\mem e^{s\sprime}
				\\
				&
				+ (\a_p\mminus2)\mem
				\bigbig{\hnem
					(Q^-_{\a_1}\hnem)\mem Q_{\a_2}\, {\nem\cdots\mem} Q_{\a_{p-1}} Q_{\a_p - 1}
				\hnem}
				{}^{m\sprime}{}_{s'}\mem Dy^{s\sprime}
    	 		\end{aligned}}
 	 \,.
\end{aligned}
\end{aligned}
\end{align}
For the purposes of this chapter,
it suffices to truncate the $Q$-tensor tower
up to the second order.
As a result,
the relevant terms inside \eqref{CGR|theta'}
fall into two groups.
The first group originates from a covariant tower of multipoles,
\begin{align}
\begin{split}
    \label{CGR|theta'Q}
    \theta'^{(Q)}_\Kerr
    =\mem
    {}&{}
		p_{m'}\hhem h^{m\sprime}
        - 2i\mem p_{m'}\hem \gamma^-{}^{m\sprime}{}_{n'}\hhem y^{n\sprime}
    \\
    {}&{}
    + 
    \sum_{\ell=2}^\infty\,
        \frac{(-2i)^\ell}{\ell!}\,\mem
        p_{m'}\hhnem
        \bb{\nem
            (Q^-_{\ell})^{m\sprime}{}_{s'} \mem{e^{s\sprime}}
            + (\ell{\mem-\mem}2)\mem (Q^-_{\ell-1})^{m\sprime}{}_{s'} \mem{Dy^{s'}}
        \nem}
    \,,
\end{split}
\end{align}
whereas the second group
inserts curvature-squared operators:
\begin{align}
    \label{CGR|theta'QQ}
    \theta'^{(QQ)}_\Kerr
    =\mem
    \sum_{\ell=4}^\infty
        \frac{(-2i)^\ell}{\ell!}
    \mem
    \sum_{j=2}^{\ell-2}
        \binom{\ell{\mem-\mem}2}{j}\,
        p_{m\sprime}
        \lrp{\nem\nem\hnem
        \begin{aligned}[c]
        {}&{}
            (Q^-_{\ell-j} Q^{\vphantom{+}}_{j})^{m\sprime}{}_{s'} \mem{e^{s\sprime}}
        \\[-.1\baselineskip]
        {}&{}
            + (j{\mem-\mem}2)\mem (Q^-_{\ell-j} Q^{\vphantom{+}}_{j-1})^{m\sprime}{}_{s'} \hem{Dy^{s'}}
        \end{aligned}
        \nem\nem\nem}
    \,.
\end{align}
Let us refer to
the terms in
\eqrefs{CGR|theta'Q}{CGR|theta'QQ}
as ``level-one'' and ``level-two'' terms,
respectively.
While 
\eqref{CGR|theta'Q} could be compared to \eqref{CYM|theta'},
higher-level terms such as those in
\eqref{CGR|theta'QQ} 
seem characteristic to gravity.

Lastly,
we identify the fluctuation and time evolution vector fields,
$\mathbf{F}$ and $\mathbf{T}$.
We define perturbation theory 
in the field basis in \eqref{CGR|freetwistorbasis},
so
\begin{align}
    \mathbf{F}
    \,\,=\,\,
        \delvec{\lambda_{\a\sprime\mem}{}^I}
        + \delvec{\mu^{\da\sprime\mem I}}
        + \delvec{\trambda_{I\da'}}
        + \delvec{\tmu_I{}^{\a\sprime}}
\end{align}
with 
$
\d{\lambda_{\a\sprime\mem}{}^I},
\d{\mu^{\da\sprime\mem I}},
\d{\trambda_{I\da'}},
\d{\tmu_I{}^{\a\sprime}}
$
declared as fundamental.
The time-evolution vector field is,
in the conventions of this thesis,
\begin{align}
    \label{CGR|T}
    \mathbf{T}
    \,\,=\,\,
        \det(\lambda)\hnem \det(\rambda)\mem 
        \mem
        \bb{
            (\rambda^{-1})^{\da\sprime\mem I}
            \frac{\partial}{\partial
                \mu^{\da\sprime\mem I}
            }
            +
            (\lambda^{-1})_I{}^{\a\sprime}
            \frac{\partial}{\partial
                \tmu_I{}^{\a\sprime}
            }
        }
    \,,
\end{align}
which implies
\begin{align}
    \label{CGR|tev}
    \i_\mathbf{T}\hem dz^{m\sprime}
    \mem=\mem
        2p^{m\sprime}
    \,,\quad
    \i_\mathbf{T}\hem dy^{m\sprime}
    \mem=\mem
        0
    \,,\quad
    \i_\mathbf{T}\hem dp_{m\sprime}
    \mem=\mem
        0
    \,.
\end{align}
Note that 
$z^{m\sprime} := \delta^{m\sprime}{}_{\m'} z^{\m\sprime}$
by the flat background tetrad.

\subsection{Vertices of Fluctuation Valence One}
\label{CGR>V1}

In any SD background,
the Kerr interaction symplectic potential in \eqref{CGR|theta'}
collapses to $p_{m'} h^{m\sprime}$,
which is isomorphic to the minimally coupled scalar particle's $p_m\hem h^m$
via the zig complexification.
In this case, the interaction symplectic form is simply
\begin{align}
    \label{CGR|omega'}
    \omega'_\Kerr
    \,=\,
        dp_{m\sprime} \wedge h^{m\sprime}
        + p_{m'} dh^{m\sprime}
    \,.
\end{align}
Evaluating \eqref{CGR|omega'}
in the background of a single SD graviton
readily derives the positive-helicity linearized vertices
as simple NJ shifts of the scalar vertices produced in \rcite{wlf-ps}.
By linearity,
the negative-helicity linearized vertices
are then also found to be isomorphic.

In this way, one deduces all linearized vertices of
fluctuation degree one 
for the Kerr particle
as
\begin{subequations}
\label{CGRV1}
\begin{align}
\label{CGRV1:z}
    \adjustbox{valign=c}{\begin{tikzpicture}[]
        \node[empty] (o) at (0,0) {};
        \node[empty] (R) at (1,0) {};
        \draw[lin] (o)--(R);
        \node[wdot=$3^+$] (v) at (o) {};
        \node[f=$\d z^{m\sprime}$] (f) at (R) {};
    \end{tikzpicture}}
    \,\,&=\,\,\,
        2i\,
        \frac{A_3^+}{\a}\mem
            (\vf_3^+\hnem p)_{m'}
        \,\mathe^{ik_3z}
    \,,\\
\label{CGRV1:bz}
    \adjustbox{valign=c}{\begin{tikzpicture}[]
        \node[empty] (o) at (0,0) {};
        \node[empty] (R) at (1,0) {};
        \draw[lin] (o)--(R);
        \node[bdot=$4^-$] (v) at (o) {};
        \node[f=$\d \mathrlap{w^{m\sprime}}\phantom{z^{m\sprime}}$] (f) at (R) {};
    \end{tikzpicture}}
    \,\,&=\,\,\,
        2i\,
        \frac{A_4^-}{\b}\mem
            (\vf_4^-\hnem p)_{m'}
        \,\mathe^{ik_4\bz}
    \,,\\[.1\baselineskip]
\label{CGRV1:p+}
    \adjustbox{valign=c}{\begin{tikzpicture}[]
        \node[empty] (o) at (0,0) {};
        \node[empty] (R) at (1,0) {};
        \draw[lin] (o)--(R);
        \node[swdot=$3^+$] (v) at (o) {};
        \node[f=$\d p_{m\sprime}$] (f) at (R) {};
    \end{tikzpicture}}
    \,\,&=\,\,\,
        4\,
        \frac{A_3^+}{\a^2}\mem
            (y\vf_3^+)^{m\sprime}
        \,\mathe^{ik_3z}
    \,,\\[-.05\baselineskip]
\label{CGRV1:p-}
    \adjustbox{valign=c}{\begin{tikzpicture}[]
        \node[empty] (o) at (0,0) {};
        \node[empty] (R) at (1,0) {};
        \draw[lin] (o)--(R);
        \node[sbdot=$4^-$] (v) at (o) {};
        \node[f=$\d p_{m\sprime}$] (f) at (R) {};
    \end{tikzpicture}}
    \,\,&=\,\,\,
        4\,
        \frac{A_4^-}{\b^2}\mem
            (y\vf_4^-)^{m\sprime}
        \,\mathe^{ik_4\bz}
    \,.
\end{align}
\end{subequations}
Again, here we have peeled off 
the gravitational coupling $\k$
and
explicitly wrote down the fluctuation species,
while omitting the factors
$\mathe^{iK_3\s}$, $\mathe^{iK_4\s}$.
To clarify,
$\bz^{m\sprime}$ in $\mathe^{ik\bz}$
is defined as $z^{m\sprime} \mminus 2iy^{m\sprime}$
with the background worldline's 
holomorphic impact parameter
$z^{m\sprime}$
and
spin length vector
$y^{m\sprime}$.

\subsection{Vertices of Fluctuation Valence Zero}
\label{CGR>V0}

To derive vertices of fluctuation valence zero,
one contracts
the interaction symplectic potential in 
\eqref{CGR|theta'}
with the free-theory time evolution vector field in \eqref{CGR|T}.

\paragraph{Level Two}

For Compton scattering,
the curvature-squared operators
in \eqref{CGR|theta'QQ}
directly boil down to
\textit{gauge-invariant} contact terms
at each spin order.
Hence they can be computed and treated separately.
Simple calculation shows that
\eqref{CGR|theta'QQ} 
gives the vertices
\begin{subequations}
\label{CGRV0:QQ}
\begin{align}
\label{CGRV0:+-QQ}
    \adjustbox{valign=c}{\begin{tikzpicture}[]
        \node[empty] (o) at (0,0) {};
        \node[ndot=$3^+4^-$] (v) at (o) {};
        \node[ta=$(2)$] at (o) {};
    \end{tikzpicture}}
    \,\,\,
    \,\,&\rightsquigarrow\,\,\,
        \bb{\nem{
            -\d\hem\c^2\mem
            \mathcal{S}(\b,\a)
        }\hnem}
        \mem\mathe^{ik_{34}z}
    \,,\\[.2\baselineskip]
\label{CGRV0:--QQ}
    \adjustbox{valign=c}{\begin{tikzpicture}[]
        \node[empty] (o) at (0,0) {};
        \node[ndot=$3^-4^-$] (v) at (o) {};
        \node[ta=$(2)$] at (o) {};
    \end{tikzpicture}}
    \,\,\,
    \,\,&\rightsquigarrow\,\,\,
        \bb{\nem{
            - \d\mem
            \BB{
                \d + \a\b
                -\hnem \c\hem(\a\mplus\hnem\b)
            \hhnem\hnem}
            \bigbig{
                \mathcal{S}(\b,\a)
                + \mathcal{S}(\a,\b)
            \hnem}
        }\hnem}
        \mem\mathe^{ik_{34}z}
    \,,
\end{align}
\end{subequations}
after normalizing by the scalar Compton amplitude.
Here, we have defined
\begin{align}   
\begin{split}
\label{Sfunction}
    {}&{}
    \mathcal{S}(\b,\a)
    \,:=\,
        \sum_{\ell=4}^\infty
        \frac{1}{\ell!}\,
        \sum_{j=2}^{\ell-2}\mem
        \binom{\ell{\mem-\mem}2}{j}\mem
            \b^{\ell-j-2} \a^{j-2}
    \,,\\[.1\baselineskip]
    {}&{}
    \,=\,
        \frac{
            E_2(\a\mplus\hnem\b)
            - E_2(\b)
            - \a\hem E'_2(\b)
        }{\a^2}
    \,=
        \int_0^1 d\eta\,\,
            \eta^2(1\nem\mminus\eta)\,
            \mathe^{\eta\b}
            E_2(\eta\a)
    \,.
%
\end{split}
\end{align}
    Clearly, \eqref{Sfunction} directly transcribes the structure of the level-two $Q$-tensor tower in \eqref{CGR|theta'QQ}:
    the combinatorics of Jacobi propagators in \Chap{K3:GDE} \cite{gde}.
    Note also the identity
\begin{align}
    \label{Ehprime}
    E'_h(\xi)
    \,=\,
        \frac{1}{\xi}\mem\BB{
            E_{h-1}(\xi) - h\mem E_h(\xi)
        }
    \,,
\end{align}
    which holds for $h=1,2,3,\cdots$
    with $E_0(\xi) := \mathe^\xi$.

\paragraph{Level One}

The level-one terms in \eqref{CGR|theta'Q}
require relatively more complicated manipulations,
but
brute-force calculation shows that
they give\footnote{
    In the first line of \eqref{CGRV0:+-Q},
    $(1 + \d/\a\b - \d\c/\a\b^2 + \d\c/\a^2\b)$
    arises from the monopole part.
    The contribution from the dipole part is
    $\c^2(\c\mplus\hnem\b)/\a^2$,
    which happens to not involve a $1/(\a+\b)$ factor.
    Importantly, 
    the formula in \eqref{axialgravitysol.A}
    involves a coframe $e^B{}_\n$
    that does not disappear by the axial gauge.
    If one mistakenly ignores this contribution,
    one obtains a wrong result
    $\c^2(\c\mplus\hnem\b)^2\nem/\a^2(\a\mplus\hnem\b)$.
}
\newpage

\phantom{.}
\vspace{-.8\baselineskip}
\begin{subequations}
\label{CGRV0:lev1}
\begin{align}
\label{CGRV0:++Q}
    \kern0.5em
    \adjustbox{valign=c}{\begin{tikzpicture}[]
        \node[empty] (o) at (0,0) {};
        \node[wdot=$3^+4^+$] (v) at (o) {};
    \end{tikzpicture}}
    \,\,\,
    \,\,\rightsquigarrow\,\,\,
        {}&{}
        \bb{
            1 + \frac{\d}{\a\b}
            + \frac{\d\c}{\a\b^2}
            + \frac{\d\c}{\a^2\b}
        }
        \mem\mathe^{ik_{34}z}
    \,,\\[.2\baselineskip]
\begin{split}
\label{CGRV0:+-Q}
    \adjustbox{valign=c}{\begin{tikzpicture}[]
        \node[empty] (o) at (0,0) {};
        \node[ndot'=$3^+4^-$] (v) at (o) {};
        \node[ta={$(1)$}] at (o) {};
    \end{tikzpicture}}
    \,\,\,
    \,\,\rightsquigarrow\,\,\,
    {}&{}\kern-.36em
        \lrp{
            \begin{aligned}[c]
            {}&{}
                1 + \frac{\d}{\a\b}
                - \frac{\d\c}{\a\b^2}
                + \frac{\d\c}{\a^2\b}
                + \frac{\c^2\hem(\c\mplus\hnem\b)}{\a^2}\mem
                    E_1(\a\mplus\hnem\b)
            \\
            {}&{}
                + \d\hem\c^2\mem
                    \mathcal{S}(\b,\a)
            \end{aligned}
        \hnem}
        \mathe^{ik_{34}z}
    \,,
\end{split}
    \\[.2\baselineskip]
\begin{split}
\label{CGRV0:--Q}
    \adjustbox{valign=c}{\begin{tikzpicture}[]
        \node[empty] (o) at (0,0) {};
        \node[ndot'=$3^-4^-$] (v) at (o) {};
        \node[ta={$(1)$}] at (o) {};
    \end{tikzpicture}}
    \,\,\,
    \,\,\rightsquigarrow\,\,\,
        {}&{}
        \bb{
            1 + \frac{\d}{\a\b}
            + \frac{\d\c}{\a\b^2}
            + \frac{\d\c}{\a^2\b}
        }
        \mem\mathe^{ik_{34}\bz}
    \\[-.05\baselineskip]
        {}&{}
        +
        \bb{
            \d\mem
            \BB{
                \d + \a\b
                -\hnem \c\hem(\a\mplus\hnem\b)
            \hhnem\hnem}
            \bigbig{
                \mathcal{S}(\b,\a)
                + \mathcal{S}(\a,\b)
            \hnem}
        \nem}
        \mem\mathe^{ik_{34}z}
    \,.
\end{split}
\end{align}
\end{subequations}
Again,
we have normalized by the scalar Compton amplitude.
We see that,
even with the aligned axial gauge,
the plus and minus helicities are treated asymmetrically in this perturbation theory,
unlike in the earlier YM case.
While the $(+,+)$ vertex in \eqref{CGRV0:++Q}
is local in spinspacetime,
the $(-,-)$ vertex in \eqref{CGRV0:--Q}
is not localized at a single complexified point.

\subsection{Derivation of Compton Amplitudes for All Helicities}
\label{CGR>DER++}

Given
the Feynman rules in \Secss{COMPTON>methodology}{CGR>V0}{CGR>V1},
we are finally ready to
compute the Kerr Compton amplitudes
for any helicity configuration.

\paragraph[Helicities $(+,+)$]{Helicities $\bm{(+,+)}$}

From the vertices in \eqrefss{CGRV1:z}{CGRV1:p+}{CGRV0:++Q},
the $(+,+)$ amplitude is computed as
\begin{align}
\begin{split}
    \label{CGRA:++}
    {}&{}
    \adjustbox{valign=c}{\begin{tikzpicture}[]
        \node[empty] (o) at (0,0) {};
        \node[empty] (R) at (1,0) {};
        \draw[lin] (o)--(R);
        \node[wdot=$3^+$] (v) at (o) {};
        \node[wdot=$4^+$] (v) at (R) {};
    \end{tikzpicture}}
    \,\,\,
    \,\,\,+\,\,\,
    \,\,\,
    \adjustbox{valign=c}{\begin{tikzpicture}[]
        \node[empty] (o) at (0,0) {};
        \node[empty] (R) at (1,0) {};
        \draw[lin] (o)--(R);
        \node[wdot=$3^+$] (v) at (o) {};
        \node[swdot=$4^+$] (v) at (R) {};
    \end{tikzpicture}}
    \,\,\,
    \,\,\,+\,\,\,
    \,\,\,
    \adjustbox{valign=c}{\begin{tikzpicture}[]
        \node[empty] (o) at (0,0) {};
        \node[empty] (R) at (1,0) {};
        \draw[lin] (o)--(R);
        \node[swdot=$3^+$] (v) at (o) {};
        \node[wdot=$4^+$] (v) at (R) {};
    \end{tikzpicture}}
    \,\,\,
    \,\,\,+\,\,\,
    \,\,\,
    \adjustbox{valign=c}{\begin{tikzpicture}[]
        \node[empty] (o) at (0,0) {};
        \node[wdot=$3^+4^+$] (v) at (o) {};
    \end{tikzpicture}}
    \,\,\,
\,\,\,
    \,\,\rightsquigarrow\,\,\,
\,\,\,
        \mem\mathe^{ik_{34}z}
    \,.
    \vphantom{\bigg|_0}
\end{split}
\end{align}
The core calculation here is
\begin{align}
    \label{++cancellation}
    \bb{\nem\nem- \frac{\d}{\a\b}\nem\nem}
    +
    \bb{\nem\nem- \frac{\d\c}{\a\b^2}\nem\nem}
    +
    \bb{\nem\nem- \frac{\d\c}{\a^2\b}\nem\nem}
    +
    \bb{\nem
        1 + \frac{\d}{\a\b}
        + \frac{\d\c}{\a\b^2}
        + \frac{\d\c}{\a^2\b}
    \nem}
    \,=\,
        1
    \,.
\end{align}
Note how the second-order and first-order propagators contribute in this calculation.
We conclude that
\begin{align}
    \label{CGRM:++}
    \M_\Kerr(3^+4^+)
    \,=\,
        \bbsq{
            - \frac{4\hem(\pffpPP)^2}{{K_3}^2\hem t}
        }
        \mem\mathe^{ik_{34}z}
    \,.
\end{align}

\paragraph[Helicities $(+,-)$]{Helicities $\bm{(+,-)}$}

Next, we consider the case of prime interest: mixed helicities.
From the vertices in \eqrefss{CGRV1}{CGRV0:+-QQ}{CGRV0:+-Q},
the $(+,-)$ amplitude is computed as the following.
\begin{align}
\begin{split}
    \label{CGRA:+-}
    {}&{}
    \adjustbox{valign=c}{\begin{tikzpicture}[]
        \node[empty] (o) at (0,0) {};
        \node[empty] (R) at (1,0) {};
        \draw[lin] (o)--(R);
        \node[wdot=$3^+$] (v) at (o) {};
        \node[bdot=$4^-$] (v) at (R) {};
    \end{tikzpicture}}
    \,\,\,
    \,\,\,+\,\,\,
    \,\,\,
    \adjustbox{valign=c}{\begin{tikzpicture}[]
        \node[empty] (o) at (0,0) {};
        \node[empty] (R) at (1,0) {};
        \draw[lin] (o)--(R);
        \node[wdot=$3^+$] (v) at (o) {};
        \node[sbdot=$4^-$] (v) at (R) {};
    \end{tikzpicture}}
    \,\,\,
    \,\,\,+\,\,\,
    \,\,\,
    \adjustbox{valign=c}{\begin{tikzpicture}[]
        \node[empty] (o) at (0,0) {};
        \node[empty] (R) at (1,0) {};
        \draw[lin] (o)--(R);
        \node[swdot=$3^+$] (v) at (o) {};
        \node[bdot=$4^-$] (v) at (R) {};
    \end{tikzpicture}}
    \,\,\,
    \,\,\,+\,\,\,
    \,\,\,
    \adjustbox{valign=c}{\begin{tikzpicture}[]
        \node[empty] (o) at (0,0) {};
        \node[ndot'=$3^+4^-$] (v) at (o) {};
        \node[ta=$(1)$] at (o) {};
    \end{tikzpicture}}
    \,\,\,
    \,\,\,
    \,\,\,+\,\,\,
    \,\,\,
    \adjustbox{valign=c}{\begin{tikzpicture}[]
        \node[empty] (o) at (0,0) {};
        \node[ndot=$3^+4^-$] (v) at (o) {};
        \node[ta=$(2)$] at (o) {};
    \end{tikzpicture}}
    \vphantom{\bigg|_0}
\end{split}
\end{align}
\begin{subequations}
The first three diagrams give
\begin{align}
    \label{CGR|mpc1}
    \bb{
        - \frac{\d}{\a\b}\mem
            \bigbig{1+\c}
        + \frac{\d\c}{\a\b^2}
        - \frac{\d\c}{\a^2\b}
    }
    \mem\mathe^{ik_3z}
    \hem\mathe^{ik_4\bz}
    \,,
\end{align}
while the last two diagrams,
via a nice cancellation of $\mathcal{S}(\a,\b)$,
give
\begin{align}
    \label{CGR|mpc2}
    \bb{
        1 + \frac{\d}{\a\b}
        - \frac{\d\c}{\a\b^2}
        + \frac{\d\c}{\a^2\b}
        + \frac{\c^2\hem(\c\mplus\hnem\b)}{\a^2}\mem
            E_1(\a\mplus\hnem\b)
    }\mem \mathe^{-\b}
    \mem\mathe^{ik_3z}
    \hem\mathe^{ik_4\bz}
    \,.
\end{align}
\end{subequations}
The sum of \eqrefs{CGR|mpc1}{CGR|mpc2}
readily computes to
$
    \AHH[2]\hem
    \mem\mathe^{ik_3z}
    \hem\mathe^{ik_4\bz}
$.
Hence we conclude that
\begin{align}
    \label{CGRM:+-}
    {}&{}
    \M_\Kerr(3^+4^-)
    \\[-.1\baselineskip]
    {}&{}
    =\mem
        \bbsq{
            - \frac{4\hem(\pffpPM)^2}{{K_3}^2\hem t}
        }\nem\nem
        \lrp{\nem\nem
        \begin{aligned}[c]
            {}&{}
                \frac{\b\mem \mathe^\a + \a\mem \mathe^{-\b}}{\a+\b}
                + \c\mem 
                \frac{\mathe^\a - \mathe^{-\b}}{\a+\hnem\b}
            \\
            {}&{}
                + (\c\mminus\hnem\a)(\c\mplus\hnem\b)\,
                \frac{E_1(\a) \hnem-\nem E_1(-\b)}{\a+\hnem\b}
            \\
            {}&{}
                + (\c\mminus\hnem\a)(\c\mplus\hnem\b)\,
                \c\mem
                \frac{E_2(\a) \hnem-\nem E_2(-\b)}{\a+\hnem\b}
        \end{aligned}
        \nem\nem}
        \nem
        \mathe^{ik_3z}
        \hem\mathe^{ik_4\bz}
    \,.
    \nonumber
\end{align}

\paragraph[Helicities $(-,-)$]{Helicities $\bm{(-,-)}$}

Finally, we compute the $(-,-)$ amplitude
as a rather nontrivial consistency check of our chiral perturbation theory.
From the vertices in \eqrefsss{CGRV1:bz}{CGRV1:p-}{CGRV0:--QQ}{CGRV0:--Q},
the $(-,-)$ amplitude is computed as the following.
\begin{align}
\begin{split}
    \label{CGRA:--}
    {}&{}
    \adjustbox{valign=c}{\begin{tikzpicture}[]
        \node[empty] (o) at (0,0) {};
        \node[empty] (R) at (1,0) {};
        \draw[lin] (o)--(R);
        \node[bdot=$3^-$] (v) at (o) {};
        \node[bdot=$4^-$] (v) at (R) {};
    \end{tikzpicture}}
    \,\,\,
    \,\,\,+\,\,\,
    \,\,\,
    \adjustbox{valign=c}{\begin{tikzpicture}[]
        \node[empty] (o) at (0,0) {};
        \node[empty] (R) at (1,0) {};
        \draw[lin] (o)--(R);
        \node[bdot=$3^-$] (v) at (o) {};
        \node[sbdot=$4^-$] (v) at (R) {};
    \end{tikzpicture}}
    \,\,\,
    \,\,\,+\,\,\,
    \,\,\,
    \adjustbox{valign=c}{\begin{tikzpicture}[]
        \node[empty] (o) at (0,0) {};
        \node[empty] (R) at (1,0) {};
        \draw[lin] (o)--(R);
        \node[sbdot=$3^-$] (v) at (o) {};
        \node[bdot=$4^-$] (v) at (R) {};
    \end{tikzpicture}}
    \,\,\,
    \,\,\,+\,\,\,
    \,\,\,
    \adjustbox{valign=c}{\begin{tikzpicture}[]
        \node[empty] (o) at (0,0) {};
        \node[ndot'=$3^-4^-$] (v) at (o) {};
        \node[ta=$(1)$] at (o) {};
    \end{tikzpicture}}
    \,\,\,
    \,\,\,
    \,\,\,+\,\,\,
    \,\,\,
    \adjustbox{valign=c}{\begin{tikzpicture}[]
        \node[empty] (o) at (0,0) {};
        \node[ndot=$3^-4^-$] (v) at (o) {};
        \node[ta=$(2)$] at (o) {};
    \end{tikzpicture}}
    \vphantom{\bigg|_0}
\end{split}
\end{align}
While
the first four diagrams
achieve
the same cancellation mechanism as in \eqref{++cancellation},
it is clear that the $\mathcal{S}$-function terms cancel nicely.
Thus, we find
\begin{align}
    \label{CGRM:--}
    \M_\Kerr(3^-4^-)
    \,=\,
        \bbsq{
            - \frac{4\hem(\pffpMM)^2}{{K_3}^2\hem t}
        }
        \mem\mathe^{ik_{34}\bz}
    \,.
\end{align}
Despite the fact that
there is no symmetry between the $(+,+)$ and $(-,-)$ configurations
at the level of individual Feynman rules,
the total amplitude in \eqref{CGRM:--}
is exactly conjugate to
the $(+,+)$ amplitude in \eqref{CGRM:++}.

Note also how
the gauge-invariant contact vertex from level two,
\eqref{CGRV0:--QQ},
has canceled a part of 
the level-one vertex in \eqref{CGRV0:--Q}
that will be generically gauge dependent.
In any case,
gauge dependencies must cancel
without invoking the level-two vertices
at this order.

\subsection{Manifest Reality; Earthly Deformations}

Finally,
suppose we carried out the perturbation theory with the earthly Kerr action in \eqref{theta-earth}.
This perturbation scheme
treats the SD and ASD modes on an equal footing
and reproduces the same amplitudes.
Unfortunately,
the same-helicity
spin exponentiation is never manifest at the Lagrangian level
and is portrayed as a nontrivial property
relying on mystical cancellations.

Another evident point regarding this approach
is that it utilizes the real section
of complexified spacetime
as a spurious ``reference'' structure.
It declares
the (geodetic) midpoint between the chiral dyons or Taub-NUT centers
as a particularly distinguished point
of a certain importance,
the physical rationale of which may appear thin
given the topological nature of the Dirac/Misner string.
It should be understood that there is nothing in between the SD and ASD Taub-NUT instantons,
so any wave factor in the final answer
shall presumably describe either $\mathe^{ikz}$ or $\mathe^{ik\bz}$
(and not $\mathe^{ikx}$).
Of course, 
a balanced view will arise because
one could also argue that
the Kerr solution is supposed to represent 
a real spacetime
(although by considering, e.g., all-plus Compton amplitudes one inevitably is embracing complexification).

In sum, there lies a tension between 
manifest spin exponentiation
and
manifest reality.
The manifestly real Kerr Lagrangian
is given in \eqref{theta-earth}.
The manifestly spin-exponentiating Kerr Lagrangian
is given in \eqref{kerr.googly}.

In connection with this point,
let us summarize and compare
how the level-two terms contribute
to the relative amplitude
in each perturbation scheme.
The contact vertex
$\theta'^{(QQ)}_\Kerr\hnem(\mathbf{T})$
evaluates to
\begin{subequations}
\begin{align}
    \label{CGR|lev2:googly}
    \text{Googly}&:\quad
    2\mem
    \sum_{\ell=4}^\infty
        \frac{(-2i)^\ell}{\ell!}
    \,
    \sum_{j=2}^{\ell-2}
        \binom{\ell{\mem-\mem}2}{j}\,
        \hem
        p_{m'}
        (Q^-_{\ell-j} Q^{\vphantom{+}}_{j})^{m\sprime}{}_{s'} 
        {p^{s\sprime}}
    \,,\\
    \label{CGR|lev2:earthly}
    \text{Symmetric}&:\quad
    2\mem
    \sum_{\ell=4}^\infty
        \frac{1}{\ell!}
    \,
    \sum_{j=2}^{\ell-2}
        \binom{\ell{\mem-\mem}2}{j}\,
        \hem
        p_m\hem
        (\hhnem({\star}^\ell Q^{\vphantom{+}}_{\ell-j} )\hhem
        Q^{\vphantom{+}}_{j})^m{}_s
        \hem
        p^s
    \,.
\end{align}
\end{subequations}
Consequently, with
\begin{align}
\begin{split}
    C_0
    \,:=\,
    {}&{}
        8\hem A_3 A_4\,
            (\yffy)
    \,,\\
    \,=\,
    {}&{}
        \Bigg\{{\renewcommand{\arraystretch}{1.1}
        \begin{array}{ll}
            \d\hem\c^2
            &
            (\text{mixed})
            \,,
        \\
            \d\mem
            \bigbig{
                \d + \a\b
                -\hnem \c\hem(\a\mplus\hnem\b)
            \hhnem\hnem}
            &
            (\text{same})
            \,,
        \end{array}}
\end{split}
\end{align}
the contributions to the relative amplitude are given by
\begin{subequations}
\begin{align}
\begin{split}
    \text{Googly, $3^+4^+$}:
    &\quad
        0
    \,,\\
    \text{Googly, $3^+4^-$}:
    &\quad
        {-C_0}\,
            \mathcal{S}(\b,\a)
            \mem\mathe^{-\b}
    \,,\\
    \text{Googly, $3^-4^-$}:
    &\quad
        {-C_0}\mem\hhem
            \bigbig{
                \mathcal{S}(\b,\a)
                + \mathcal{S}(\a,\b)
            \hnem}
            \mem\mathe^{-\a-\b}
    \,,
\end{split}\\[.02\baselineskip]
\begin{split}
    \text{Symmetric, $3^+4^+$}:
    &\quad
        {-\tfrac{1}{16}\mem C_0}\mem
            \bigbig{
                \mathcal{S}(-\tfrac{\b}{2},-\tfrac{\a}{2})
                + \mathcal{S}(-\tfrac{\a}{2},-\tfrac{\b}{2})
            \hnem}
            \mem\mathe^{\a/2+\b/2}
    \,,\\
    \text{Symmetric, $3^+4^-$}:
    &\quad
        {-\tfrac{1}{16}\mem C_0}\mem
            \bigbig{
                \mathcal{S}(\tfrac{\b}{2},\tfrac{\a}{2})
                + \mathcal{S}(-\tfrac{\a}{2},-\tfrac{\b}{2})
            \hnem}
            \mem\mathe^{\a/2-\b/2}
    \,,\\
    \text{Symmetric, $3^-4^-$}:
    &\quad
        {-\tfrac{1}{16}\mem C_0}\mem
            \bigbig{
                \mathcal{S}(\tfrac{\b}{2},\tfrac{\a}{2})
                + \mathcal{S}(\tfrac{\a}{2},\tfrac{\b}{2})
            \hnem}
            \mem\mathe^{-\a/2-\b/2}
    \,.
\end{split}
\end{align}
\end{subequations}
Note how the symmetric scheme
induces the half-scaled factors $\a/2$, $\b/2$ in both signs,
instating a spurious expansion point in the middle of $z$ and $\bz$.

Said in another way,
if the level-two operators
are dropped
at the Lagrangian level,
the Compton amplitude gains a contact deformation 
by 
$0$, $+C_0\, \mathcal{S}(\b,\a)$, 
\smash{$+C_0\mem 
    \bigbig{
        \mathcal{S}(\b,\a)
        + \mathcal{S}(\a,\b)
    \hnem}
$}, 
etc.
Obviously,
this must not happen in the googly scheme,
as it violates reality.
In the symmetric scheme,
on the other hand,
one might be allured to drop the $QQ$ tower of operators
by associating them with tidal deformability,
although it is a grossly worldline field basis dependent operation.
Crucially, this is also forbidden
if the same-helicity spin exponentiation is a nonnegotiable feature of spinning black hole amplitudes.
As discussed in \Chap{K3:PROBENJ} \cite{probe-nj},
the best possibility is to add
curvature-squared operators on the worldline
that explicitly implements a mixed-chirality projection,
such as 
$
    p_m\hem
    (Q^+_{\ell-j}Q^-_{j})^m{}_s
    \hem
    p^s
$.
Since this deformation starts at $\O(a^4)$,
it could be asked if 
there exists a natural deformation prescription that achieves $\AHH[3]$,
or the amplitudes advocated in 
\rrcite{%
    Ochirov:2022nqz,Cangemi:2022bew,Cangemi:2023bpe,%
    fabian1,fabian2,zihan23,Saketh:2023bul%
}.

\section{
Conclusions}

In this chapter,
we provided a Lagrangian derivation of 
the nonabelian {\kerr} and Kerr Compton amplitudes
in YM theory and GR,
based on explicit effective worldline actions
that guarantee the same-helicity spin exponentiation.
Our results are all orders in spin,
exhibit correct residues on physical factorization channels,
and are completely free of unphysical poles,
for all helicity configurations.
They also obey a systematic organization
in terms of truncated exponential functions
and Hermite interpolation.
In other words, we obtained a new concrete proposal regarding the longstanding puzzle of AHH \cite{ahh2017}.

The physical significance of this calculation
is that 
it gives the
gauge invariant and field basis invariant
characterization
of spinning black hole dynamics 
up to the second order in coupling,
in terms of scattering amplitudes.
This finally addresses
the persistent curiosity toward
the exact spinning-black-hole EoM 
in a fully invariant fashion.

Conceptually,
our construction implements a dynamical Taub-NUT or chiral dyon pair.
At the technical level,
this physical interpretation stipulates particular towers of curvature operators
with certain fixed coefficients
that have origins in the combinatorics behind geodesic or gauge-covariant translations.
Arguably,
our Lagrangians
carry out a unique minimal off-shell extrapolation
from the three-point/same-helicity couplings,
in the sense that
Dirac/Misner strings carry
no physical degrees of freedom.

In the most critical assessment,
our construction is no more than an idealized model of spinning black holes,\footnote{
    the ``zig-zag model,'' so to speak
}
which is yet plausible in the sense of guaranteed 
same-helicity spin exponentiation
to all orders in perturbation theory.
To our knowledge,
the explicit all-orders construction of worldline effective actions for spinning black holes
that exhibit this property
is new,
and our mixed-helicity Compton amplitudes have not been proposed in the previous literature.

Even with this critical evaluation,
we emphasize that
our effective actions
will serve as a \textit{useful} starting point of
future (higher-order, higher-multiplicity, all-orders-in-spin)
worldline effective theory calculations
for spinning black holes,
if the same-helicity spin exponentiation is 
a nonnegotiable feature.
The addition of mixed-chirality deformations
can be carried out systematically.

A further critical investigation,
in connection with
black hole perturbation theory results
\cite{fabian1,fabian2,zihan23,Saketh:2023bul},
will be expected.
It is interesting to
imagine carrying out Kerr perturbation theory
with the factorized Taub-NUT picture
such that the boundary conditions at the ASD and SD Taub-NUT centers
correspond to the behavior of the zig and zag points in the effective theory,
although the role of the horizon may become less emphasized.
In any case,
it will be fruitful to promote
cross-pollination between the UV picture of \Chap{K3:NJA} \cite{nja}
and the IR picture of \Chaps{K3:PROBENJ}{K3:OSD1} \cite{probe-nj,njmagic.1},
as has already happened a number of times in this thesis.

Formalism-wise,
we demonstrated a chiral way of comprehending Kerr
by expanding around a holomorphic worldline onto which the SDGR couplings become resummed,
whose physical interpretation is the effective theory avatar of ASD Taub-NUT.
As mentioned several times,
the idea itself,
of achieving Kerr by perturbing away from SD or ASD Taub-NUT,
is an old one
and has been advocated by twistor theorists
(see, e.g., \rrcite{Adamo:2023fbj,Guevara:2023wlr}).
To our knowledge, however,
its full-blown implementation
in the worldline perturbation theory
is new.
On a related note,
it should be clear that
our explorations
draw
conceptual precursors and insights from
the discourse on
MHV amplitudes \cite{arkani2012positivegrassmannian},
first-quantized massive twistor models
\cite{Albonico:2022pmd,Albonico:2023kpc},
twistor particle program
\cite{Penrose:1974di,Perjes:1974ra,penrose1977twistor,Perjes:1976sy,perjes1976evidence,hughston1976twistor,hughston1979twistors,perjes1979unitary,hughston1980programme,perjes1982internal,perjes1982introduction,tod1975dissertation,tod1976two,tod1977some},
and the various provisional ideas of Newman
\cite{%
    newman1988remarkable,Newman:1973afx,Newman:2002mk,Newman:1973yu,newman1974curiosity,newman1974collection,Newman:2004ba,
    Newman:1976gc,ko1981theory,grg207flaherty%
}.

\begin{subappendices}

\section{
Scalar Compton Amplitudes}

For reference,
we record the classical Compton amplitudes
for the minimally coupled scalar particle
below.

\phantom{.}
\hrule\vspace{0.5pt}
\hrule\vspace{-5pt}
\paragraph{Maxwell Theory}
\begin{subequations}
\label{Compton0}
\begin{align}
    \label{Compton0:EM}
    \M_\text{scalar}
    \,\,\,=\,\,\,
    - \frac{4\hem(\pffp)}{{K_3}^2}
\end{align}
\paragraph{YM Theory}\nem\nem%
(nonabelian contribution, color-stripped)
\begin{align}
    \label{Compton0:YM}
    \M_\text{scalar}
    \,\,\,=\,\,\,
    - \frac{4i\hem(\pffp)}{{K_3}\hem t}
\end{align}
\paragraph{GR}
\begin{align}
    \label{Compton0:GR}
    \M_\text{scalar}
    \,\,\,=\,\,\,
    - \frac{4\hem(\pffp)^2}{{K_3}^2\hhem t}
\end{align}
\end{subequations}
\hrule\vspace{0.5pt}
\hrule\vspace{-5pt}

\end{subappendices}

\pagebreak
\bibliographystyle{utphys-modified}
\bibliography{references}

\appendix
\chapter{More on Self-Dual Spacetime}
\label{APP:PLEB}

For interested readers,
we provide a brief review of
Pleba\'nski's second heavenly formalism
and
the basic scaffolding of curved twistor theory,
in connection with \Sec{Jcx}.

\section{
Pleba\'nski's Second Heavenly Equation}
\label{APP:PLEBHE}

This appendix reviews
the derivation of 
Pleba\'nski's second heavenly equation
governing SD spacetimes
\cite{plebanski1975some,dunajski2000hyper,Adamo:2021bej,boyer1983geometry,hull1991geometry}.

When working in the complexified, holomorphic setup,
spacetime is an oriented complex four-manifold $\M$
endowed with a holomorphic metric $g$.
Based on the decomposition of its tangent bundle into SD and ASD spinor bundles,
\begin{align}
    T\M 
    \,\cong\,
    {S^+\nem\M}\hem {\,\otimes\,} {S^-\nem\M}
    \,,
\end{align}
the Levi-Civita connection $\nabla$ of $g$
splits into SD and ASD parts,
defining connections on $S^\pm\nem\M$.

In this thesis, we defined that
a spacetime $\M$ is SD 
if $S^-\nem\M$ is flat.
This means $\M$ is hyperk\"ahler
(structure group $\SL(2,\mathbb{C}) \cong \Sp(2,\mathbb{C})$)
and is automatically Ricci-flat.

Let $e^{\da\a}$ be the coframe 
due to a local trivialization of $T\M$.
In a SD spacetime $\M$,
the ASD spin connection coefficients
can be locally set to zero.
This statement translates to
the closure of
the {\Pleb} \cite{Plebanski:1977zz,Capovilla:1991qb}
area two-forms,
$
\Sigma^{\a\b}
:= \te_\wrap{\da\db}\mem e^{\da\a} \swedge e^{\db\b}
$:
\begin{align}
    \label{closure-asd}
    d\Sigma^{\a\b} = 0
    \,.
\end{align}
Let 
$\o^\a {\:\doteq\:} \delta_0{}^\a$ and $\i^\a {\:\doteq\:} \delta_1{}^\a$
be constant spinors forming an ASD dyad.
Since $\Sigma^{\a\b}$ is closed,
there locally exists
functions $q^\da$ and $p_\da$
such that
\begin{align}
    \label{darboux}
    \Sigma^{01} = dq^\da \swedge dp_\da
    \,,\quad
    \Sigma^{11} = \te_\wrap{\da\db}\mem dq^\da \swedge dq^\db
    \,,
\end{align}
on account of the Darboux theorem \cite{darboux1882probleme}.
These functions are the very second heavenly coordinates of {\Pleb} \cite{plebanski1975some}:
\begin{align}
    \label{xcoords}
    x^{\da0} \,:=\, p^\da
    \,,\quad
    x^{\da1} \,:=\, q^\da
    \,.
\end{align}

Certainly, 
this coordinate choice is a sort of a gauge-fixing that trivializes a part of the self-duality conditions in \eqref{closure-asd},
so
it remains to examine the condition
$d\Sigma^{00} {\:=\:} 0$.

To this end, note that
the coframe that yields \eqref{darboux}
takes the form
\begin{align}
    \label{tet-e}
    e^{\da0} 
    \,=\, \L^\da{}_\db\mem 
        (dp^\db {\mem+\,} \k\mem \Phi^\db{}_\dc\mem dq^\dc)
    \,,\quad
    e^{\da1} 
    \,=\, \L^\da{}_\db\mem
        dq^\db
    \,,
\end{align}
where 
$\L^\da{}_\db$ is $\SL(2,\mathbb{C})$-valued,
$\Phi^\da{}_\db$ is traceless,
and $\k$ is a customary constant.
Indices are raised and lowered as 
\smash{$q^\da = \te^{\da\db} q_\db$},
\smash{$q_\da = \te_\wrap{\da\db}\mem q^\db$}.
With \eqref{tet-e} and $d\Sigma^{00} = 0$,
the following can be shown.
First of all,
$\partial_{0\da} \Phi^\da{}_\wrap{\db} {\:=\:} 0$,
which guarantees the local existence of a scalar function $\Phi$ such that
$\Phi_\wrap{\da\db} {\:=\:} \partial_\wrap{0\da} \partial_\wrap{0\db} \Phi$.
Second of all,
the very equation
\begin{align}
    \label{heavenly}
    \square\hem \Phi
    \,=\,
        \k\,\mem
            \Phi_\wrap{\da\db}\mem \Phi^{\da\db}
    \,.
\end{align}
This derives the result that
the sole dynamical geometry of SD spacetime
is governed by
Pleba\'nski's second heavenly equation \cite{plebanski1975some}.

In \eqref{heavenly},
$\Box$
is the Laplacian associated with a flat metric 
$\eta = -2\hem dp_\da \modot dq^\da$
on $\M$.
To elaborate, this flat limit
arises as the $\k\to0$ limit
of the curved metric
which the coframe in \eqref{tet-e} defines:
\begin{align}
    \label{metg}
    g
    \,&=\,
        -2\hem
        \e_\wrap{\a\b}\mem \te_\wrap{\da\db}\mem
            e^{\da\a} {\:\odot\:} e^{\db\b}
    \,=\,
        - 2\mem dp_\da {\:\odot\:} dq^\da
        + 2\hem\k\mem \Phi_\wrap{\da\db}\mem dq^\da {\:\odot\:} dq^\db
    \,.
\end{align}
Explicitly,
the Laplacian operators associated with
the flat metric $\eta$
and the curved metric $g$
are
\begin{align}
    \label{boxes}
    \Box
    \,:=\,
        -2\mem \frac{\partial^2}{\partial p_\da \partial q^\da}
    \,,\quad
    \Box_\Phi
    \,:=\,
        \Box
        \mem-\mem 2\hem\k\mem \Phi_\wrap{\da\db}\mem
            \frac{\partial^2}{\partial p_\da \partial p_\db}
    \,,
\end{align}
respectively.
It follows that
linearized perturbations around 
a solution $\Phi$ of \eqref{heavenly}
are harmonic with respect to the curved Laplacian $\Box_\Phi$
\cite{dunajski2000hyper}.

Note that the coframe in \eqref{tet-e} is unimodular,
and the frame vector fields dual to \eqref{tet-e} are
\begin{align}
    \label{tet-E}
    E_\wrap{0\da}
    \,=\, (\L^{-1})^\db{}_\da\mem
        \frac{\partial}{\partial p^\db}
    \,,\quad
    E_\wrap{1\da}
    \,=\, (\L^{-1})^\db{}_\da\mem
        \bigg(\mem{
            \frac{\partial}{\partial q^\db}
            - \k\mem \Phi^\dc{}_\db\mem \frac{\partial}{\partial p^\dc}
        }\mem\bigg)
    \,.
\end{align}
Note also that the SD Weyl tensor components are given by
the fourth-order derivatives of the {\Pleb} scalar $\Phi$
with respect to the $p$-coordinates.

\section{
Self-Dual Spacetime as a Cotangent Bundle}
\label{COTANGENT}

The coordinate choice in \eqref{xcoords}
is characterized by the Darboux formatting of the ASD two-forms 
$\Sigma^{01}$ and $\Sigma^{11}$
shown
in \eqref{darboux}.
Hence diffeomorphisms of $\M$
shall be restricted
such that
$\Sigma^{01}$ and $\Sigma^{11}$
are preserved.

Consequently,
the SD spacetime $\M$
is understood
as the cotangent bundle $T^*\N$
of a two-dimensional symplectic manifold $\N$
\cite{dunajski2000hyper,boyer1983geometry,hull1991geometry}:
\begin{align}
    \label{splicing}
    \M \,\cong\, T^*\N
    \,.
\end{align}
Here, the base $\N$
is coordinatized by $q^\da$
and endowed with
$\Sigma^{11} = \te_\wrap{\da\db}\mem dq^\da \swedge dq^\db$
as the symplectic form.
The cotangent fibers are coordinatized by $p_\da$
and soldered by $\Sigma^{01} = dq^\da \swedge dp_\da$.

Specifically,
it can be seen that
any vector field in $\M$
that preserves $\Sigma^{01}$ and $\Sigma^{11}$
is given by a linear combination of
\begin{align}
    \label{Xuv}
    X_u
    \,=\,
            - u^{\da}(q)\mem
            \frac{\partial}{\partial q^\da}
            + p_\db\mem 
            u^\db{}_\da(q)\mem
            \frac{\partial}{\partial p_\da}
    \,,\quad
    Y_v
    \,=\,
        v_\da(q)\mem
        \frac{\partial}{\partial p_\da}
    \,,
\end{align}
for $u, v \in \Cinfty(\N)$.
Here, we have denoted
\begin{align}
    \label{qnotation}
    f_\wrap{\da}(q)
    \,:=\,
        \frac{\partial}{\partial q^\da}\mem
            f(q)
    \,,\quad
    f_\wrap{\da\db\cdots}(q)
    \,:=\,
        \frac{\partial}{\partial q^\da}
        \frac{\partial}{\partial q^\db}
    \mem f(q)
    \,,\quad
    \cdots
    \,,
\end{align}
if $f(q)$ is a sole function of $q^\da$.
Clearly,
$X_u$ describes
the uplift of a symplectomorphism on $\mathcal{N}$
while
$Y_v$ describes
a certain translation of the zero section
of $T^*\N$.

As a result,
it follows that
the residual diffeomorphisms
form a Lie group
$\mathrm{SDiff}(\mathcal{N}) {\mem\ltimes\mem} (\Gamma(T^*\hnem\mathcal{N}),+)$,
which describes a subgroup of $\SDiff(\M)$:
\begin{align}
\begin{split}
    [X_{u}, X_{v}] = X_{
        u_\da v^\da
    }
    \,,\quad
    [X_{u}, Y_{v}] = Y_{
        u_\da v^\da
    }
    \,,\quad
    [Y_{u}, Y_{v}] = 0
    \,.
\end{split}
\end{align}
Here, $\SDiff(\M)$ denotes the volume-preserving diffeomorphism group of $\M$,
while $\SDiff(\N)$ describes the symplectomorphism group of $\N$.

The $2+2$ splicing
of spacetime
in \eqref{splicing}
underlies
the connection between
SD gravity and
two-dimensional models
based on area-preserving diffeomorphisms
\cite{castro2004moyal,park19922d,hull1991geometry}.

In this context,
consider how the infinitesimal coordinate transformations in \eqref{Xuv}
affect the Pleba\'nski scalar $\Phi$.
As the heavenly equation in \eqref{heavenly}
should be preserved,
the change in the Pleba\'nski scalar $\Phi$
should be
harmonic with respect to the curved Laplacian $\Box_\Phi$.

For instance, take 
$\Lambda^\da{}_\db = \delta^\da{}_\db$
and compute
$\pounds_{X_{\kappa u}} e^{\db\a} + \k\mem u^\da{}_\db\mem e^{\db\a}$
and 
$\pounds_{Y_{\kappa v}} e^{\da\a}$.
The former
is reproduced by $\Phi \mapsto \Phi + \delta_u \Phi$
while the latter
is reproduced by $\Phi \mapsto \Phi + \delta_v \Phi$,
where
\begin{subequations}
\label{Phivar.uv}
\begin{align}
    \delta_u \Phi
    \,&=\,
        \frac{1}{6}\,
            p_\da\mem p_\db\mem p_\dc\mem u^{\da\db\dc}(q)
        + \k\mem X_u\act{ \Phi }
    \,,\\
    \delta_v \Phi
    \,&=\,
        \frac{1}{2}\,
            p_\da\mem p_\db\mem v^{\da\db}(q)
        + \k\mem Y_v\act{ \Phi }
    \,.
\end{align}
\end{subequations}
Note that $e^{\da\a} \mapsto (\delta^\da{}_\db \mplus \k\mem u^\da{}_\db)\mem e^{\db\a}$
is undone
by adjusting
the SD frame $\Lambda^\da{}_\db$.

Straightforward computation verifies that
the variations in \eqref{Phivar.uv}
are elements of $\ker(\Box_\Phi)$
but not $\ker(\Box)$.
Nonetheless,
they leave
the heavenly equation in \eqref{heavenly}
invariant.

When deriving \eqref{Phivar.uv},
one integrates a second derivative
$\partial^2\nem/\partial p_\da \partial p_\db$.
The kernel for this operation describes
\begin{align}
    \label{Phivar.ab}
    \delta_a \Phi
    \,=\,
        a(q)
    \,,\quad
    \delta_b \Phi
    \,=\,
        p_\da\mem b^\da(q)
    \,,
\end{align}
where $a,b \in \Cinfty(\N)$.
The variations in \eqref{Phivar.ab}
are elements of
not only $\ker(\Box)$ 
but also $\ker(\Box_\Phi)$.
Thus,
they leave
each side of the heavenly equation in \eqref{heavenly}
invariant.
Parenthetically,
we might remark that \eqref{Phivar.ab}
is analogous to the symmetry of Galileons \cite{hinterbichler2015hidden}.

Two solutions 
to the heavenly equation in \eqref{heavenly}
that can be mapped to each other
via \eqrefs{Phivar.uv}{Phivar.ab}
should be identified as the same solution,
as they will describe the same spacetime geometry.
In other words,
\eqrefs{Phivar.uv}{Phivar.ab}
describe the gauge redundancies of the heavenly equation in \eqref{heavenly}.
See \rcite{dunajski2000hyper}
for a further discussion.

\section{
Curved Twistor Theory}
\label{REVIEW>T}

Finally, let us sketch the basics of curved twistor theory
\cite{Penrose:1976js,GravityMHVTwistors,ward1991twistor,huggett1994introduction}.
In particular,
in \Sec{Jcx}
(around \eqref{LaxLL}),
we mentioned
the Lax pair formulation of SDGR
and its connection to curved twistor space.
We should explicate on this point.

Recall that
the {\Pleb} approach to SD spacetime in \App{APP:PLEBHE}
utilizes a fixed ASD direction $\o_\a$,
which essentially is a gauge fixing for SDGR:
null axial gauge, i.e., lightcone gauge.
On the contrary,
the idea here is to treat
all ASD directions on an equal footing
by means of an auxiliary ASD spinor variable $\l_\a \inn \C^2$,
which can be thought of as projectively coordinatizing a Riemann sphere $\CP^1$.

\subsection{Lax Pair Formulation of Self-Dual Gravity}

There are two versions of this construction.
The version of
Pleba\'nski and Gindikin \cite{plebanski1975some,gindikin1986construction}
works in terms of differential forms.
The version due to
Mason and Newman \cite{MasonNewman:1989,chakravarty1991canonical,chacon2019self}
works in terms of vector fields.

The first version starts with
a four-manifold $\M$
equipped with a nondegenerate volume form $\nu$.
If $\M$ admits a set of two-forms $\Sigma^{\a\b}$
such that
\begin{align}
    \label{S(l)}
    d\Sigma(\lambda) \,=\, 0
    \:\:\:\:\text{and}\:\:\:\:
    \Sigma(\lambda) \wedge \Sigma(\lambda)
    \,=\, 0
    \,\quad
    \forall\mem \lambda_\a \in \C^2
    \,,
\end{align}
where $\Sigma(\lambda) := \lambda_\a\lambda_\b \hem\Sigma^{\a\b}$,
it can be shown that
a coframe $e^{\da\a}$ exists such that
$\Sigma^{\a\b} = \te_\wrap{\da\db}\mem e^{\da\a} \swedge e^{\db\b}$,
which defines a SD metric\footnote{
    Note that this metric can also be directly obtained from $\Sigma^{\a\b}$
    via the Urbantke formula
    \cite{Urbantke:1984eb,Capovilla:1991qb}.
} 
$
    g
    =
        \e_\wrap{\a\b}\mem \te_\wrap{\da\db}\mem
            e^{\da\a} {\:\odot\:} e^{\db\b}
$ on $\M$.

The second version 
also starts with a four-manifold $\M$
equipped with a nondegenerate four-form $\nu$,
but assumes the existence of four linearly independent vector fields $V_{\a\da}$ on it.
For $\l_\a \in \C^2$,
one then constructs
the vector fields $L_\da := \lambda^\a\hem V_{\a\da}$.
If $L_\da$ form a Lax pair
such that
\begin{align}
    \label{MNLax}
    [L_\da,L_\db] \,=\, 0
    \:\:\:\:\text{and}\:\:\:\:
    \pounds_{L_\da} \nu \,=\, 0
    \qquad
    \forall\mem \lambda_\a \in \C^2
    \,,
\end{align}
then the vector fields 
$E_{\a\da} = f^{-1}\hem V_{\a\da}$
give an orthonormal frame
of a SD Riemannian structure on $\M$
with $f^2 = i\mem \nu(V_\wrap{0\doz},V_\wrap{0\diz},V_\wrap{1\doz},V_\wrap{1\diz})$.
Conversely,
an SD spacetime $(\M,g)$
locally admits an orthonormal frame $E_{\a\da}$
and a nonvanishing function $f$
such that
$L_\da = f\l^\a\hem E_{\a\da}$ form
a volume-preserving Lax pair as in \eqref{MNLax}.
for some volume form $\nu$.
This is the Mason-Newman result
that we had reviewed in \Sec{Jc.MN}, in fact.

The above two approaches
can be related as
$\te^{\da\db} (\i_{L_\da} \i_{L_\db} \hem\nu) = i\mem \Sigma(\lambda)$
and 
$\i_{L_\da} \Sigma(\lambda) = 0$.
In {\Pleb}'s second heavenly coordinates,
the scaling factor is trivial as $f = 1$,
so $L_\da = \l^\a\hem E_{\a\da}$.

Crucially,
the vanishing Lie bracket
$[L_\da,L_\db] = 0$
in \eqref{MNLax}
implies that 
translations along SD null directions
in $\M$
are integrable
for each fixed $\l_\a$,
forming 
SD null surfaces
known as $\a$-surfaces.
The space of $\a$-surfaces
is the \textit{twistor space} $\PT$.
As will become self-evident,
the twistor space has complex dimension three.

\subsection{Deformed Incidence Relation}
\label{REVIEW>DFIB}

To establish the precise relation between
the twistor space $\PT$ and the spacetime $\M$,
it is helpful to employ
what is known as
the correspondence space,
the concept of which arises by promoting 
the auxiliary spinor variable $\l_\a$ to coordinates of an extended space.

The correspondence space
$\F \eqq {\mathbb{P} S^-\nem\M}$
is the projectivization 
of
the ASD spinor bundle,
which arises by removing the zero section and stipulating the identification
$\lambda_\a \sim r\mem \lambda_\a$
for $r \in \C^*$.
A function on $\F$
means
a function $f(x,\lambda)$ on $S^-\nem\M$
homogeneous in $\lambda_\a$.
It descends to the twistor space
$\PT$ iff 
it is constant along the $\a$-surfaces, meaning
$L_\da\act{
    f(x,\lambda)
} = 0$
with $L_\da = \l^\a\hem E_{\a\da} \in \Gamma(T\F)$.

Moreover, 
one sees that
$\Sigma(\l)$ in \eqref{S(l)} defines a two-form on $\F$
that descends to the twistor space.
In particular, it can be brought to a Darboux form
$\Sigma(\lambda) = \te_{\smash{\da\db}\vphantom{\b}}\mem d_x F^\da\hnem(x,\lambda) \swedge d_x F^\db\hnem(x,\lambda)$,
where $d_x$ denotes the 
``partial'' exterior derivative 
that discards $d\lambda_\a$.
Then $\i_{L_\dc} \Sigma(\lambda) = 0$ gives
\begin{align}
    \label{LF}
    L_\dc\act{
        F^\da\hnem(x,\lambda)
    } \,=\, 0
    \,,
\end{align}
so $F^\da(x,\lambda)$ 
gives a function on $\F$,
with degree one in $\lambda_\a$,
that descends to $\PT$.

The incidence relation refers to the equation
\begin{align}
    \label{REVIEW:incidence}
    \mu^\da
    \,=\,
        F^{\da}(x,\lambda)
    \,.
\end{align}
For each fixed $\lambda_\a$ and $\mu^\da$,
\eqref{REVIEW:incidence} defines an $\a$-surface in $\M$.
It follows that
the pair $Z_\rmA = (\lambda_\a,\mu^\da)$
provides
projective coordinates for the twistor space $\PT$.
Each point of $\PT$ represents the equivalence class $[Z]$
of $Z_\rmA = (\lambda_\a,\mu^\da)$
under identifications by rescalings.

It is instructive to consider
the twistor space of
flat spacetime $\mflat$.
Let $x^{\da\a}$ be lightcone coordinates of $\mflat$.
$\a$-surfaces in $\mflat$ describe
$x^{\da\a} \sim x^{\da\a} + y^\da\hem \lambda^\a$
for fixed $\lambda^\a$
and variable $y^\da$,
along which
\begin{align}
    \label{REVIEW:incidence.flat}
    \mu^\da \,=\, x^{\da\a}\hem \lambda_\a
\end{align}
are constant.
\eqref{REVIEW:incidence.flat} is the incidence relation for flat spacetime.
The flat twistor space is
\begin{align}
    \label{pt}
    \pt
    \,=\,
        \Big\{\mem{
            [(\lambda_\a , \mu^\da)] \in \CP^3
        \,\Big|\,
            \lambda_\a \neq 0
        }\mem\Big\}
    \,.
\end{align}

Upon introduction of spacetime curvature,
the incidence relation in \eqref{REVIEW:incidence.flat}
develops a ``general-relativistic correction''
and gets deformed into the generic form in \eqref{REVIEW:incidence}.
In this sense, \eqref{REVIEW:incidence} is sometimes referred to as
``deformed incidence relation''
for an explicit disambiguation \cite{GravityMHVTwistors}.

The celebrated nonlinear graviton construction of Penrose \cite{Penrose:1976js} states that,
the curved twistor space $\PT$ of a SD spacetime $\M$ 
locally arises by
deforming the complex structure of 
flat twistor space $\pt$.
This deformation is known to utilize
the holomorphic Poisson structure,
\begin{align}
    \label{holPoisson}
    I
    \,=\,
        \te^{\da\db}\mem
        \frac{\partial}{\partial \m^\da}
        \mwedge
        \frac{\partial}{\partial \m^\db}
    \,.
\end{align}
The origin of \eqref{holPoisson}
can be traced back to the fact that
the Gindikin $2$-form
$\Sigma(\lambda) = \te_{\smash{\da\db}\vphantom{\b}}\mem d_x F^\da\hnem(x,\lambda) \swedge d_x F^\db\hnem(x,\lambda)$
corresponds to
$\te_\wrap{\da\db}\mem d\m^\da \swedge d\m^\db$.

\subsection{Double Fibration}

More precisely,
the geometry of the incidence relation is discussed in the picture known as double fibration,
shown below.
\begin{align}
\label{fibration}
\adjustbox{valign=c}{\begin{tikzpicture}
    \node[b] (o) at (0,0) {};
    \node[b] (a) at (-1.5, -1.5) {};
    \node[b] (b) at ( 1.5, -1.5) {};
    \node[b] (bundle) at ($(o)$) {\vphantom{|}};
    \node[b] (Bundle) at ($(bundle)$) {$\mathclap{
        \:\:
        \F
    }\phantom{|}$};
    \node[b] (base a) at ($(Bundle)+(a)$) {\vphantom{|}};
    \node[b] (Base a) at ($(base a)$) {$
    \mathclap{
        \PT
    }\phantom{|}$};
    \node[b] (base b) at ($(Bundle)+(b)$) {\vphantom{|}};
    \node[b] (Base b) at ($(base b)$) {$\mathclap{
        \M
    }\phantom{|}$};
    \draw[->] (bundle)--(base a) node[midway,above,pos=0.62] 
    {\scriptsize $\pi_1\:\:$};
    \draw[->] (bundle)--(base b) node[midway,above,pos=0.62] {\scriptsize $\,\,\pi_2$};
\end{tikzpicture}}
\end{align}
The map $\pi_1 : \F \to \PT : (x^\m,[\lambda_\a]) \mapsto [(\lambda_\a,F^\da(x,\lambda))]$
formalizes the incidence relation in \eqref{REVIEW:incidence}.
The map
$\pi_2: \F \to \M : (x^\m,[\lambda_\a]) \mapsto x^\m$
is the bundle projection.
Via this double fibration,
geometric data in twistor space is transcribed to spacetime
and vice versa.

A well-known duality between twistor space and spacetime
is discussed as follows.
Each twistor $[Z] \in \PT$
defines the $\a$-surface in spacetime as
\begin{align}
    \mathcal{S}_{[Z]}
    \,=\,
    \pi_2(\pi_1^{-1}([Z]))
    \,\subset\,
        \M
    \,.
\end{align}
Meanwhile, each spacetime point $x \inn \M$
defines a holomorphically embedded $\CP^1$ in the twistor space $\PT$,
known as the twistor line $\LL_x$:
\begin{align}
    \label{twistorline}
    \LL_x
    \,:=\,
        \pi_1(\pi_2^{-1}(x))
    \,\subset\,
    \PT
    \,.
\end{align}
In this context,
the incidence relation
provides an explicit parameterization of the twistor line:
$[Z_\rmA] = [(\l_\a,F^\da(x,\l))]$
for $\l_\a \in \C^2$.
Finally,
it follows that
\begin{align}
    [Z] \,\in\, \LL_x \,\subset\, \PT
    \qfq
        x \,\in\, \mathcal{S}_{[Z]} \,\subset\, \M
    \,,
\end{align}
meaning that
an $\a$-surface $\mathcal{S}_{[Z]}$ is incident at 
the spacetime point $x$
iff the twistor $[Z]$ lies on the twistor line $\LL_x$.
The vivid picture behind this statement
is that
the twistor line $\LL_x$
essentially represents
the lightcone, i.e., \textit{sky} at $x$
(as the space $\CP^1 \cong \mathrm{S}^2$ of null directions).
Evidently, a twistor $[Z]$ appears on the sky $\LL_x$
for an observer at $x$
iff $x$ lies on the pathway of the light ray $[Z]$.

To conclude,
the twistor space $\PT$
is the space of SD null surfaces in a SD spacetime $\M$.
The precise relationship between
$\PT$ and $\M$
arises via the correspondence space
and the double fibration.
This framework essentially traces back to the Lax pair formulation of SDGR.
The twistor-space counterpart
of spacetime point $x \in \M$
is the twistor line $\LL_x$,
which gets deformed into ``non-straight curves''
upon introduction of gravity.
Accordingly,
in twistor space,
SDGR is formulated as deformations of the complex structure.

Twistor space finds various applications in physics.
For instance, the Penrose transform \cite{Penrose:1969ae}
famously constructs massless on-shell fields
(zero-rest-mass fields)
from 
meromorphic functions 
on twistor space
(more precisely, representatives of twistor cohomology classes)
via contour integral
\cite{penrose:maccallum,Atiyah:2017erd,penrose:1986spinors2}.

\section{
Canonical Transformations on Fibers}

The redundancies of the heavenly framework,
discussed in \App{COTANGENT},
corresponds to Hamiltonian vector fields in twistor space
with respect to the holomorphic Poisson structure in \eqref{holPoisson}.
This Poisson structure
is equipped to
each fiber of 
$\pi_0 : \PT \to \CP^1 : [(\lambda_\a,\mu^\da)] \mapsto [\lambda_\a]$.

To approach this fact concretely,
one may use
an explicit formula for the deformed incidence relation
in {\Pleb}'s second heavenly coordinates
\cite{dunajski2000hyper,Adamo:2021bej}:
\begin{align}
    \label{deformed-incidence}
    F^{\da}\hnem(x,\lambda)
    \,=\,
        x^{\da\a}\hem \lambda_\a
        \mem+\mem
        \k\,
            \frac{\partial \Phi}{\partial p_\da}
            \mem\frac{{\l_0}^2}{{\l_1}}
        \mem+\mem
        \k\,
            \frac{\partial \Phi}{\partial q_\da}
            \mem\frac{{\lambda_0}^3}{{\lambda_1}^2}
        \hem
        \mem+\mem
        \cdots
    \,.
\end{align}
This explicit formula 
describes a series in $\l_0/\l_1$
and is valid 
in the affine patch $\lambda_1 \neq 0$ of $\CP^1$.
It arises in a framework
known as Dunajski-Mason recursion
\cite{dunajski2000hyper,dunajski1996heavenly,dunajski1997recursion}.

Based on \eqref{deformed-incidence},
one can directly investigate
the transformation behavior of $F^\da(x,\l)$ under
the redundancies identified in \App{COTANGENT}.
Computation verifies that it amounts to
a Poisson diffeomorphism
with respect to \eqref{holPoisson}
that transforms
the $\m$-fibers of $\pi_0 : \PT \to \CP^1$ as
\begin{align}
    \te_\wrap{\da\db}\,
        \delta\hnem\m^\da\mem
        d\m^\db
    \,=\,
        dh(\m)
    \,,
\end{align}
where the Hamiltonian,
after pullback by $\pi_1 : \F \too \PT$,
exhibits the expansion
$
    h(F(x,\lambda)\hhnem) =
    \k\hem
    {\lambda_1}^2 $ $ \smash{\bigbig{
        h^{(0)}(x)
        - h^{(1)}(x)\mem \zeta
        + h^{(2)}(x)\mem \zeta^2
        - h^{(3)}(x)\mem \zeta^3
        + \cdots
    }}
$:
\begin{align}
\begin{split}
    \label{TwistorHams}
    h^{(0)}
    \,&=\,
        u(q)
    \,,\quad
    h^{(1)}
    \,=\,
        p_\da\mem u^\da(q) + v(q)
    \,,\\
    h^{(2)}
    \,&=\,
        \tfrac{1}{2}\, p_\da p_\db\hem u^{\da\db}(q)
        + \k\mem u_\da(q)\mem \Phi^\da
        + p_\da\mem v^\da(q)
        - b(q)
    \,,\\
    h^{(3)}
    \,&=\,
    \begin{aligned}[t]
    {}&{}
        \tfrac{1}{6}\, p_\da p_\db p_\dc\mem u^{\da\db\dc}(q)
        + \k\hem X_u\act{ \Phi }
        + \tfrac{1}{2}\, p_\da p_\db\mem v^{\da\db}(q)
        + \k\hem X_v\act{ \Phi }
    \\
    {}&{}
        - a(q)
        - p_\da\mem b^\da(q)
    \,.
    \end{aligned}
\end{split}
\end{align}
In fact,
a quick derivation of \eqref{TwistorHams}
is viable by series-expanding
$u(F(x,\l))$, $v(F(x,\l))$, etc. 

\chapter{Worldline Formalisms}
\label{APP:WLFS}

This appendix reviews
the computation or definition of
fundamental objects in scattering theory
in first-quantized frameworks.

\section{
S-Matrix Element; Scattering Amplitudes}
\label{WLF}

\subsection{The Classic Worldline Formalism}

The celebrated 
worldline formalism 
\cite{Feynman:1948ur,Feynman:1950ir,Schwinger:1951nm,vH,polchinski1985worldlineformalism,Bern:1990cu,Bern:1991aq,Strassler:1992zr,Schubert:2001he,schubert2012lectures,schwartz2014qft,witten2015every}
provides a direct realization of the idea that
Feynman diagrams visualize spacetime trajectories of quantum particles.
In particular,
it establishes that
the transition amplitude
(as partition function)
for a first-quantized relativistic particle,
moving from one spacetime point to another,
derives an off-shell propagator in field theory.
The formalism is also known for
providing powerful pathways toward
loop amplitudes or effective actions
and has also been identified as
the infinite-tension counterpart of string-theoretic computations
\cite{Bern:1990cu,Bern:1991aq,Strassler:1992zr,Schubert:2001he}.

To elaborate,
the paradigmatic calculation in 
this formalism
gives rise to the Schwinger parameter representation of the free off-shell propagator
(see, e.g., \rcite{Schwinger:1951nm,vH,polchinski1985worldlineformalism}):
\begin{align}
    \label{Schwinger}
    \frac{-i}{p^2 + m^2 - i\epsilon}
    \,=\,
    \int_0^\infty dT\,\,
        \mathe^{-iT(p^2 + m^2)}
    \,.
\end{align}
Here, the right-hand side arises by
a faithful
sum over histories
for a worldline
traveling from one spacetime point to another,
which involves
a sum over the moduli space of einbeins $\ein$
on a finite interval $[0,1]$.
This refers to the path integral,
\begin{align}
	\label{recipe.xx}
	\int
	\frac{\D{\ein}}{\vol(\Gauge)}
	\quad\longrightarrow\quad
    	\int_0^\infty\nem dT\,
    	\int
    	\D{\ein}\,\,
    	\delta[\ein {\mem-\,} 2T]
	\,,
\end{align}
where the path integral measure is crucially quotiented
on account of the one-dimensional diffeomorphism redundancy on the worldline.

To extract tree-level scattering amplitudes
in the worldline formalism,
the conventional practice
simply
imports the LSZ reduction formula
as an \textit{external input} from field theory.
For instance,
the multi-photon Compton amplitudes
are obtained by
first computing the off-shell propagators
in the backgrounds of superposed electromagnetic plane waves
and then amputating two external legs
via LSZ reduction.

Computationally,
however,
this post-processing might not be 
entirely trivial
in practice.
Besides the technicalities of
Gaussian and Fourier integrals,
a central tension lies in the fact that
the primary output of the worldline formalism
is an off-shell propagator in position space,
whereas scattering amplitudes are on-shell objects examined in momentum space.
Thus, if one's goal is to obtain tree-level scattering amplitudes,
the original workflow
might have offered a slightly roundabout path,
although conceptually faithful and direct.

\subsection{Reduction by Topology}

Regarding this point,
it had been observed
\cite{Fabbrichesi:1993kz,fradkin1964electrodynamics,milekhin1964doubly,barbashov1965functional,pervushin1970method,Laenen:2008gt}
that employing noncompact worldline topologies,
such as half-line or full line,
can automate the LSZ reductions.
This observation traces back to Fradkin \cite{fradkin1964electrodynamics},
according to \rcite{Fabbrichesi:1993kz},
and has been revived in a modern literature
as well \cite{Bonocore:2020xuj,Mogull:2020sak}.
\rlap{\phantom{\cite{Mogull:2020sak,Jakobsen:2021smu,Jakobsen:2022psy}}}

\begin{figure}[t]
	\centering
	\begin{align*}
        {\renewcommand{\arraystretch}{1.5}
        \begin{array}{ccc}
            \includegraphics[valign=c,scale=0.5]{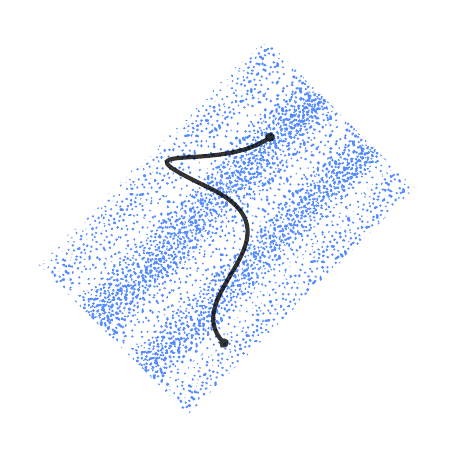}
            &
            \includegraphics[valign=c,scale=0.55]{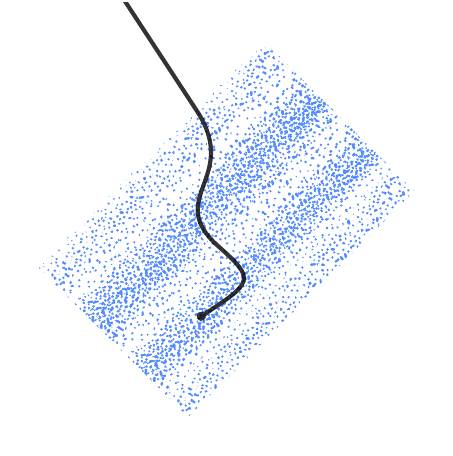}
            &
            \includegraphics[valign=c,scale=0.55]{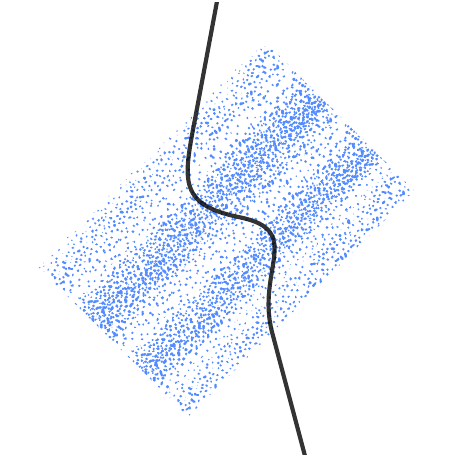}
            \\
            \text{\footnotesize
                    (a)
                    Interval
                }
            &\,
            \text{\footnotesize
                    (b)
                    Half-Line
                }
            &\,
            \text{\footnotesize
                    (c)
                    Full Line
                }
        \end{array}}
    \end{align*}
	\caption[
        Various worldline topologies for
        particle propagation in sandwich geometry.
    ]{
        Various worldline topologies for
        particle propagation in sandwich geometry.
        (Figure adapted from \rcite{wlf-ps}).
    }
	\label{fig:sand}
\end{figure}

Notably,
\rcite{Bonezzi:2025iza}
provided a top-down derivation of this fact
from the moduli space of einbein for each noncompact topology,
while \rcite{wlf-ps},
released in the same year as \rcite{Bonezzi:2025iza},
had
also independently 
discovered the same conclusion
through a physical motivation
(an exploration of boundary conditions in the first-order formulation, in connection with manifest gauge invariance).
Crucially,
based on the geometric semantics of LSZ reduction as
bulk-to-boundary propagation \cite{Cheung:2022pdk},
it was realized that
the first-quantized probability amplitude
for the interval, half-line, and full line topologies
can respectively be interpreted as
a
bulk-to-bulk,
bulk-to-boundary,
and
boundary-to-boundary propagator,
the last of which is
identified with scattering amplitude
\cite{wlf-ps}.
It follows that
states escaping to the boundaries must be on-shell
for consistency and finiteness of results.

The mantra goes as follows.
What have escaped to the boundary are on-shell states.
Bulk-to-bulk processes are virtual and are off-shell.
LSZ reduction 
sends
the bulk endpoints of an off-shell propagator
to the boundaries,
making the particle register the infinities at far past and future.

\fref{fig:sand} visualizes
the three worldline topologies,
interval $[0,1]$,
half-line $[0,\infty)$, 
and 
full line $(-\infty,+\infty)$.
The recipe for the einbein path integral
is given 
for each topology
as follows.\footnote{
    Here, we ignore the Faddeev-Popov determinant
    for simplicity.
}

For the interval topology,
the recipe is well-known 
from classic literature 
(see, e.g., \rcite{polchinski1985worldlineformalism})
and is shown in \eqref{recipe.xx}.
The moduli space of einbeins is one-dimensional
and is coordinatized by
the diffeomorphism-invariant total length
$\int_0^1 \ein(\s)\, d\s = 2T$,
and
this $T$ is dubbed the Schwinger proper time.
In \eqref{recipe.xx},
$\delta[\ein {\mem-\,} 2T]$ is 
the delta functional
fixing the einbein to a constant profile, $\ein(\s) = 2T$.
Consequently,
one performs
an ordinary integral over the Schwinger proper time $T$.

For the half-line topology, 
the recipe is 
\cite{Bonezzi:2025iza,wlf-ps}
\begin{align}
	\label{recipe.px}
	\int
	\frac{\D{\ein}}{\vol(\Gauge)}
	\quad\longrightarrow\quad
	\int
	\D{\ein}\,\,
		\delta[\ein {\mem-\,} \ein_\infty]
	\,,
\end{align}
which gauge-fixes the einbein to a nonzero constant $\ein_\infty$.
This is because
the moduli space collapses to a single point:
all half-lines are diffeomorphic to each other.\footnote{
    This assumes sum over complete metrics
    such that the total length of the half-line is infinite,
    in which case the einbein has to asymptote to the nonzero constant $\ein_\infty$.
}
In fact,
by the very reparametrization invariance,
the value of $\ein_\infty$ is irrelevant:
one can rescale it away to any other value,
of which the partition function is completely ignorant.
Consequently,
one simply drops the einbein path integral entirely,
in practice.

For the full line topology, the recipe reads
\cite{Bonezzi:2025iza,wlf-ps}
\begin{align}
	\label{recipe.pp}
	\int
	\frac{\D{\ein}}{\vol(\Gauge)}
	\quad\longrightarrow\quad
	\frac{1}{\deltabar(0)}\,
	\int
		\D{\ein}\,\,
		\delta[\ein {\mem-\,} \ein_\infty]
	\,,
\end{align}
where $\deltabar(0) = \int_{\mathbb{R}} d\s = \vol(\mathbb{R})$
is the volume of the space of constant translations.
In this case,
the gauge quotient is bigger than the integral itself.
While all full lines are diffeomorphic to each other,
there is a residual gauge redundancy
formed by
diffeomorphisms that asymptote to a common constant value at both ends,
the space of which is $\mathbb{R}$.\footnote{
    Here, we again presume sum over complete metrics
    and take the two ends of the full line as distinguishable.
}
Again, the value of the constant $\ein_\infty$
in \eqref{recipe.pp}
can be arbitrary.
Consequently,
one drops the einbein path integral
while also discarding
an infinite volume factor $\deltabar(0)$.

\rcite{wlf-ps} gives a friendly discussion on 
how the iconic calculation in \eqref{Schwinger}
gets revisited
with the noncompact worldlines.
Concretely,
suppose the computation of 
the massive off-shell propagator of a charged particle
dressed with one photon leg.
Let the momenta for the massive legs be
$p_1$ and $p_2$
while $k$ is the wavenumber for the massless leg.

In the classic version employing the interval topology,
the core part of the computation describes
\begin{align}
    \label{SchwingerDemo.xx}
    \int_0^\infty dT\,\,
        \mathe^{-iTX_2}
    \int_0^1 d\s\,\,
        T\mem \mathe^{iTK\s}
    \,=\,
        \frac{-i}{X_2 - i\epsilon}\mem
        \frac{-i}{X_2 \hem\mminus K - i\epsilon}
    \,,
\end{align}
where we have denoted
$X_2 = {p_2}^2 + m^2$
and $K = 2k\mdot p_2$.
This produces two propagators
for the two off-shell massive legs.
Hence two LSZ reductions are performed by hand.

In comparison, the core part of the half-line approach describes
\begin{align}
    \label{SchwingerDemo.xp}
    \int_0^\infty d\s\,\,
        \mathe^{iK\s}
    \,=\,
        \frac{-i}{- K - i\epsilon}
    \,,
\end{align}
provided that one imposes $X_2 = 0$ proactively.
This produces one off-shell massive pole,
namely $1/({p_1}^2 + m^2)$.
In this case, only one LSZ reduction
needs to be performed manually.

Lastly,
in the case of the full line topology,
the core calculation describes
\begin{align}
    \label{SchwingerDemo.pp}
    \frac{1}{\deltabar(0)}
    \int_{-\infty}^\infty d\s\,\,
        \mathe^{iK\s}
    \,=\,
        \frac{\deltabar(K)}{\deltabar(0)}
    \,=\,
        1
    \,,
\end{align}
in which case one imposes both
${p_1}^2 + m^2 = 0$
and
${p_2}^2 + m^2 = 0$
from the beginning,
so $K = 0$.
Again,
what have escaped to the boundary are on-shell states.
Here, all the LSZ reductions are
completely automated.\footnote{
    Yet, one comment is that,
    in the case of the conventional scalar particle,
    there is no free-theory saddle
    that changes the value of the momentum,
    so the calculation using full line topology
    could be a bit tricky or subtle to carry out.
    Notably, the massive twistor worldline
    avoids this issue
    in an amusing way.
}

Note that,
when utilizing noncompact topologies,
it is desirable to have
definite-momentum states
be realized at the asymptotic infinities.
In fact, 
based on this connection between topologies and boundary conditions,
the key motivation and idea of \rcite{wlf-ps}
was to utilize phase space actions
to have momentum as a basic variable on the worldline,
automating both LSZ reductions and Fourier transforms.

To sum up,
the above geometrical discourse \cite{Bonezzi:2025iza,wlf-ps,Bastianelli:2025xne}
achieves 
a direct formulation of LSZ reduction
\textit{within} first quantization.

When equipped with this automation of LSZ reduction,
the worldline formalism
becomes perfectly suited for
computing Compton scattering amplitudes efficiently.
In particular,
\rcite{wlf-ps} concretely showed that computing the boundary-to-boundary propagator in the backgrounds of nonlinearly superposed plane waves in YM theory and GR
derives the gluon and graviton Compton scattering amplitudes
of a massive probe,
while also providing 
an explicit higher-order exploration
in Maxwell theory
up to six points.

\subsection{Universal Feynman Rules for Phase Space Actions}
\label{CA:WLF>FEYN}

The classic worldline formalism 
\cite{Feynman:1948ur,Feynman:1950ir,Schwinger:1951nm,vH,polchinski1985worldlineformalism,Bern:1990cu,Bern:1991aq,Strassler:1992zr,Schubert:2001he,schubert2012lectures,schwartz2014qft,witten2015every,townsend2004string}
had conventionally preferred
working with
configuration space (second-order) actions.
On the other hand,
\rcite{wlf-ps} demonstrated the use of phase space (first-order) actions.

Specifically,
\rcite{wlf-ps} worked out the systematics of
the perturbation theory of
generic one-dimensional sigma models onto a symplectic target.
The Feynman rules are organized in terms of elements of symplectic geometry.
In particular,
some key features
that are helpful to remember
are reported as follows.
\begin{enumerate}
	\item 
		The bare propagator
		encodes the free theory's Poisson bracket.
	\item
		Vertices arise from Hamiltonian and symplectic perturbations
        (dubbed Hamiltonian and symplectic vertices, respectively).
	\item 
		Vertices sourcing no worldline fluctuation
		compute the interaction action about the background trajectory.
	\item
		Vertices sourcing one worldline fluctuation
		arise from corrections to the Hamiltonian EoM.
		Especially, the linearized symplectic vertex of valence one
		is quickly derived by
		a formula analogous to ``Lorentz force.''
	\item
		The symplectic vertices
        are classified into two types:
        regular and pinched.
\end{enumerate}

To see these facts,
suppose 
$(\theta^\circ,\omega^\circ = d\theta^\circ, H^\circ)$
and
$(\theta,\omega = d\theta, H)$
define
the free theory and the interacting theory
on an even-dimensional phase space $\P$
in terms of first-order Lagrangians,
equipped with coordinates $\zeta^i$
that are Darboux with respect to $\omega^\circ$:
\begin{align}
		L^\circ[\zeta] &=
		\theta^\circ_i(\z)\mem \dot{\z}^i
		- H^\circ(\z)
		\,,\quad
		\label{psL}
		L[\zeta] =
		\theta_i(\z)\mem \dot{\z}^i
		- H(\z)
		\,.
\end{align}
\rcite{wlf-ps} finds that
the all-order variation of $L[\z]$
can be given as
\begin{align}
	\begin{split}
		\label{Lparts}
		L[\z {\,+\,} \dz]
		\,=\,
		L^\circ[\z]
		&
		+ L_\text{bd}[\z,\dz]
		+ \dz^i\mem \Big(\,{
			\omega^\circ_{ij}(\z)\mem \dot{\z}^j
			- H^\circ_{,i}\hnem(\z)
		}\,\Big)
		\\
		&
		+ L_\text{prop}[\z,\dz]
		+ L_\text{sv}[\z,\dz]
		+ L_\text{hv}[\z,\dz]
		\,,
	\end{split}
\end{align}
where terms inside the bracket vanish
upon plugging in 
a saddle 
of the free theory
as a background,
say $\s \mapsto \bm{\z}^i\hnem(\s)$.
The boundary term is chosen as
\begin{align}
	\label{Lbd}
	L_\text{bd}[\z,\dz]
	\,=\,
	\frac{d}{d\s}\hem \bigg[\,\mem{
		\sum_{n=0}^\infty\,
		\frac{1}{(n\hem{\mem+\mem}1)!}\,
		\theta_{j,i_1\cdots i_n}\hnem(\z)
		\mem \dz^j
		\mem \dz^{i_1}{\cdots\mem}\dz^{i_n}
	}\mem\bigg]
	\,.
\end{align}

\paragraph{Propagator}

The worldline propagator is determined from the term
\begin{align}
	\label{Lprop}
	L_\text{prop}[\z,\dz]
	\,=\, \frac{1}{2}\, \omega^\circ_{ij}\,
	\dz^i\mem
	\delta\dot{\z}^j
	\,,
\end{align}
which is quadratic in the worldline fluctuation $\dz^i$.
Via Dirac's correspondence,
it follows that the worldline propagator must satisfy
\begin{align}
	\lim_{\e\to0^+}
	\Big(\,{
		G^{ij}\hnem(\e,0)
		-
		G^{ji}\hnem(\e,0)
	}\,\Big)
    \,=\,
        -(\omega^\circ{}^{-1})^{ij}
    \,,
\end{align}
in any causality/boundary prescriptions.
For example,
\begin{align}
	\label{generic.prop}
	G^{ij}\hnem(\s_1,\s_2)
	\,=\,
		(\omega^\circ{}^{-1})^{ij}\,
		\frac{1}{-\partial_{\s_1}}\mem
        \delta(\s_1{\mem-\,}\s_2)
\end{align}
qualifies as a Green's function,
where the anti-derivative
$(-\partial_{\s_1})^{-1}\mem \delta(\s_1{\mem-\,}\s_2)$
is resolved into either of the following:
\begin{subequations}
\begin{align}
	\label{greens.sym}
	\Theta(\s_1,\s_2)
	\,&=\,
	\bigg\{
	\begin{array}{ll}
		-1/2
		&
		\quad\text{if}\quad
		\s_1 > \s_2
		\,,\\
		+1/2
		&
		\quad\text{if}\quad
		\s_1 < \s_2
		\,,
	\end{array}\\
	\label{greens.ret}
	\Theta_>(\s_1,\s_2)
	\,&=\,
	\bigg\{
	\begin{array}{ll}
		-1
		&
		\quad\text{if}\quad
		\s_1 > \s_2
		\,,\\
		0
		&
		\quad\text{if}\quad
		\s_1 < \s_2
		\,,
	\end{array}
	\\
	\label{greens.adv}
	\Theta_<(\s_1,\s_2)
	\,&=\,
	\bigg\{
	\begin{array}{ll}
		0
		&
		\quad\text{if}\quad
		\s_1 > \s_2
		\,,\\
		+1
		&
		\quad\text{if}\quad
		\s_1 < \s_2
		\,.
	\end{array}
\end{align}
\end{subequations}
In general, however,
it is not always the case that
$G^{ij}(\s_1,\s_2)$ 
can be written
in the form 
in \eqref{generic.prop}.

\paragraph{Vertices}

Suppose the free-interaction splits,
\begin{align}
    \theta
    \,=\,
        \theta^\circ
        \mem+\mem
        \theta'
    \,,\quad
    H
    \,=\,
        H^\circ
        \mem+\mem
        H'
    \,.
\end{align}
The Hamiltonian vertices arise from
\begin{align}
\begin{split}
    {}&{}
	L_\text{hv}[\z,\dz]
    \\
    {}&{}
	\,=\,
	- H'(\z)
	- H'_{,i}(\z)\mem \dz^i
	-
	\sum_{n=2}^\infty\,
	\frac{1}{n!}\mem
	H_{,i_1i_2\cdots i_n\nem}\hnem(\z)\mem \dz^{i_1}\hem \dz^{i_2}\hem \cdots \dz^{i_n}
	\,,
\end{split}
\end{align} 
which should give familiar result.
The symplectic vertices,
on the other hand,
arise from
\begin{subequations}
\begin{align}
	L_\text{sv}[\z,\dz]
	\,&=\,
		L_\text{sv}^\text{regular}[\z,\dz]
		+
		L_\text{sv}^\text{pinched}[\z,\dz]
	\,,\\[0.25\baselineskip]
	\label{Lsv.regular}
	L_\text{sv}^\text{regular}[\z,\dz]
	\,&=\,
	\theta'_i\hnem(\z)\mem \dot{\z}^i
	-
	\sum_{n=1}^\infty\,
	\frac{1}{n!}\,
		\dot{\z}^j\mem
		\omega'_{ji_n,i_1\cdots i_{n-1}\hnem}\hnem(\z) 
		\, \dz^{i_1}{\hhnem\cdots\mem}\dz^{i_n}
	\,,\\
	\label{Lsv.pinched}
	L_\text{sv}^\text{pinched}[\z,\dz]
	\,&=\,
	-
	\sum_{n=2}^\infty\,
	\frac{n\hem{\mem-\mem}1}{n!}\,
	\omega'_{ji_1,i_2\cdots i_{n-1}\hnem}\hnem(\z)
	\, 
	\delta\dot{\z}^j\mem
	\dz^{i_1}{\hhnem\cdots\mem}\dz^{i_{n-1}}
	\,.
\end{align}
\end{subequations}

In particular,
the symplectic vertex with
no worldline fluctuation
is simply the interaction action
due to the symplectic perturbation,
\begin{align}
	\label{generic.int0}
	\adjustbox{valign=c}{\begin{tikzpicture}[]
        \node[ndot,minimum size=3.1pt] (O) at (0,0) {};
	\end{tikzpicture}}\,\,\hem
	\,\,\,&=\,\,\,
	\int d\s\,\,
	\theta'_i\hnem(\bm{\z}(\s)\hnem)\mem \dot{\bm{\z}}^i\hnem(\s)
	 \,,
\end{align}
whereas
the symplectic vertex with $n=1$ worldline fluctuation
is the ``Lorentz force due to symplectic perturbation'':
\begin{align}
	\label{generic.int1}
	\adjustbox{valign=c}{\begin{tikzpicture}[]
        \node[ndot,minimum size=3.1pt] (O) at (0,0) {};
        \node[empty] (F) at (-1,0) {};
        \draw[lin] (O)--(F);
        \node[empty] (tF) at (F) {\llap{\small$\color{feyntext}
                \delta\z^i\hnem(\s)
            $}\,\,\,\mem
        };
	\end{tikzpicture}}\,\,
	\,\,\,&=\,\,\,
    	\omega'_{ij}\hhnem(\bm{\z}(\s)\hnem)\,
        \dot{\bm{\z}}^j\hnem(\s)
	 \,.
\end{align}

Note that the derivation of \eqrefs{generic.int0}{generic.int1}
itself
does not require complicated calculation.
For instance,
\eqref{generic.int1} can be directly seen from the Hamiltonian EoM.
Otherwise,
consider the ``$\mathbf{T}$-and-$\mathbf{F}$ formalism''
in \Sec{CYM>SYMP}:
$\mathbf{T}$ gives the free theory's time evolution vector field
while
$\mathbf{F}$ formalizes the fluctuation.
The first-order variation of the interaction symplectic potential gives
\begin{align}
    \pounds_{\mathbf{F}}\hem \theta'
    \,=\,
        \i_{\mathbf{F}}\hem \omega'
        + d\hem (\i_{\mathbf{F}}\hem \theta')
    \qiq
    \cont{\pounds_{\mathbf{F}}\hem \theta'}{\mathbf{T}}
    \,=\,
        \i_{\mathbf{T}} \i_{\mathbf{F}}\hem \omega'
        + \mathbf{T}\act{
            (\i_{\mathbf{F}}\hem \theta')
        }
    \,.
\end{align}
Since 
$\mathbf{T}\act{
    (\i_{\mathbf{F}}\hem \theta')
}$ gives a boundary term when 
integrated after pullback onto the background worldline,
the vertex of fluctuation valence one is
\begin{align}
    \mathcal{V}^{(1)}
    \,=\,
        \i_{\mathbf{T}} \i_{\mathbf{F}}\hem \omega'
    \,=\,
        \omega'(\mathbf{F},\mathbf{T})
    \,,
\end{align}
which precisely gives \eqref{generic.int1}
after the pullback;
recall \eqref{unv1}.

\section{
S-Symplectomorphism; Scattering Generators}
\label{SSYMP}

The worldline formalism covered in \App{WLF}
computes scattering \textit{amplitudes}
as transition amplitudes of first-quantized particles.
It implements a path integral calculation with
definite 
initial and final conditions,
and thus computes S-matrix elements
with kets and bras:
\begin{align}
    \label{1Amp}
    \M(p_2,p_1)
    \,\,=\,\mem
        \Bra{p_2}\hem
        \hat{T}
        \mem\Ket{p_1}
    \,.
\end{align}
By ``classical amplitude,''
it means the $\hbar\to0$ limit of \eqref{1Amp}
if well-defined.

In this appendix, we review an alternative formalism
that computes a different object:
\begin{align}
    \label{1Omega}
    \Omega
    \,\,=\,\,
        \lim_{\hbar\to0}\mem
        \mathcal{Q}^{-1}\hnem\bigbig{\mem
            i\hbar
            \log \hat{S}
        \mem}
    \,.
\end{align}
This framework has been recently established through works
\cite{Kim:2024grz,Gonzo:2024zxo,KKKL,Kim:2025hpn,Kim:2025olv,Kim:2025gis,Kim:2025sey,PSFOR,Damgaard:2021ipf,Damgaard:2023ttc},
with an early identification in \rcite{ambikerr1}\footnote{
    ``a dequantization of the S-matrix
    will be a time-ordered action of the deformed Hamiltonian vector fields as infinitesimal
    diffeomorphisms on the physical phase space''
}
and roots in the formalism of, e.g., \rrcite{Mogull:2020sak,Jakobsen:2021smu,Jakobsen:2022psy}.
In fact, it traces back to works
\cite{hunziker1968s,simon1971wave,herbst1974classical,sokolov1979classical,osborn1980levinson,narnhofer1981canonical,thirring1981classical} 
around the '70-80s
as well.
The object $\Omega$ in \eqref{1Omega}
has been dubbed
classical scattering \textit{generator}.
It provides a nice in-in formalism
well-suited for classical problems.

Via Hamiltonian action,
the classical scattering generator
is shown to generate
the map from
the initial phase space to the final phase space
in classical scattering,
which we dub \textit{S-symplectomorphism} \cite{Kim:2025sey,PSFOR}
or classical S-map.
Said in another way,
the scattering generator
is an ``effective Hamiltonian''
that reproduces the entire interaction-picture time evolution from far past to far future
within dimensionless unit time.
The impulses of
classical observables are computed via the pullback map of the S-symplectomorphism
\cite{PSFOR},
which unfolds into
the so-called nested bracket formula \cite{Kim:2024grz,Gonzo:2024zxo,KKKL}:
\begin{align}
    \label{impulseformula}
    \mathit{\Delta}f
    \,=\,
        \pb{f}{\Omega}
        + \frac{1}{2!}\mem
            \pb{\pb{f}{\Omega}}{\Omega}
        + \frac{1}{3!}\mem
            \pb{\pb{\pb{f}{\Omega}}{\Omega}}{\Omega}
        + \cdots
    \,.
\end{align}

One of the landmark successes of
the modern amplitudes program
is indisputably
its fruitful applications to classical physics.
In particular,
the celebrated formalism of Kosower, Maybee, and O'Connell \cite{KMOC}
established the systematics of extracting classical observables from the S-matrix
via the $\hbar\to0$ limit.
The S-symplectomorphism program
seeks to provide a purely and manifestly classical framework,
in which the scattering generator
emerges as a new currency,
while also greatly clarifying 
the relation with the quantum S-matrix.

Historically,
it seems that
the idea of 
S-symplectomorphism\footnote{
    The terminology ``symplectomorphism'' is being preferred over ``canonical transformation'' here
    due to a standardized contextual meaning of
    ``generator of canonical transformation''
    (as on-shell action, generating function).
}
had been proposed 
and investigated in works
\cite{hunziker1968s,simon1971wave,herbst1974classical,sokolov1979classical,osborn1980levinson,narnhofer1981canonical,thirring1981classical}
over the '70s and '80s,
under the name ``(canonical) S-transformation.''
Interesting applications can be found in,
e.g.,
\rrcite{osborn1980levinson,thirring1981classical}.

To be precise,
there are various physical situations in which the phase space is a degenerate Poisson manifold.
In such cases,
the terminologies
``classical S-map'' or ``Poisson S-diffeomorphism''
will be adequate,
although 
one could always work in the symplectic setup
in principle
since
every Poisson manifold admits a symplectic realization.

\subsection{Purely Classical Definition}

Concretely,
suppose a classical system formulated on a phase space $\P$,
endowed with a symplectic structure $\omega$ for simplicity.
A principle of Hamiltonian mechanics 
reads that
time evolution is a symplectomorphism.
Suppose $U^\circ(t_1,t_2)$ and $U(t_1,t_2)$
are the time-evolution symplectomorphisms
from the phase space at time $t_2$ to the phase space at time $t_1$,
due to the free and interacting theory of the system,
respectively.
We restrict our attention to perturbative interactions
with a coupling parameter $\e$,
so $\lim_{\e\to0} U(t_1,t_2) = U^\circ(t_1,t_2)$.

The S-symplectomorphism,
realized at a fixed time $t_0$ of one's choice (usually $t_0 = 0$),
is defined as
\begin{align}
    \label{Ssymp-def0}
    S
    \,\,\mem\,=\mem
        \lim_{t_\pm \to \pm\infty}
        \BB{
            U^\circ\hnem(t_0,t_+)
            \circ
            U(t_+,t_-)
            \circ
            U^\circ\hnem(t_-,t_0)
        }
    \,,
\end{align}
provided a well-defined limit.

To see the physical significance of the formula in \eqref{Ssymp-def0},
we may borrow the framework of the Liouville equation
in classical statistical mechanics,
in which case
the time-evolution symplectomorphisms
act on the time-dependent probability distribution $\rho(t)$,
representing the state of a classical ensemble,
as
\begin{align}
    \label{rhoevolU}
    \rho(t_1)
    \,=\,
        \rho(t_2)
        \circ
            U(t_2,t_1)
    \,.
\end{align}
With \eqref{rhoevolU},
it can be seen that
\eqref{Ssymp-def0} implies
\begin{align}
    \label{trhoevolS}
    \trho(+\infty)
    \,=\,
        \trho(-\infty) \circ S^{-1}
    \,=\,
        S^*\act{
            \trho(-\infty)
        }
    \,.
\end{align}
Firstly, we have denoted
\begin{align}
    \label{CIPdef}
    \tilde{f}(t)
    \,:=\,
        f(t)
        \circ
        U^\circ(t,t_0)
    \,.
\end{align}
\newpage\noindent
This is the definition of
the interaction picture in classical mechanics
tracing back to
Campbell \cite{Campbell:1975nn}.
Secondly,
we have adopted
the differential operator representation of diffeomorphisms
\cite{vershikgelfand1975diffreps}
by the induced left action,\footnote{
    We use this notation within this appendix;
    this is the pullback by $S^{-1}$.
}
\begin{align}
    \label{pullbackdef}
    S^*\act{ f }
    \,=\,
        f \circ S^{-1}
    \,.
\end{align}

With these understandings,
\eqref{trhoevolS} establishes that
the induced left action $S^*$ by 
the S-symplectomorphism $S$
maps the initial classical state $\trho(-\infty)$
to the final state $\trho(+\infty)$
in the interaction picture.
Recalling that
the S-matrix $\hat{S}$
maps the initial quantum state $\Ket{\psi(-\infty)}$
to the final quantum state $\Ket{\psi(+\infty)}$,
one sees that the S-symplectomorphism
provides a classical counterpart of the S-matrix.
As will be reviewed in \App{SSYMP>DEQUANT},
the precise relation between the S-symplectomorphism and the S-matrix
can be established \cite{PSFOR}
by means of 
the phase space formulation of quantum mechanics
\cite{Moyal:1949sk,Groenewold:1946kp,weyl1927quantenmechanik,Wigner:1932eb,Curtright:2011vw,zachos2005quantum,
leonhardt1997measuring}.

It is not difficult to see that the definition in \eqref{CIPdef}
conforms to the expected concept of the interaction picture in classical mechanics.
In fact, it simply means to plug in the free theory time evolution,
such as 
the straight-line motion 
in nonrelativistic mechanics:
$\tilde{f}(x,p;t) = f(x\mplus pt/m,p;t)$.

In \eqref{trhoevolS},
$S^*$ gives the differential operator
representation of $S$ \cite{vershikgelfand1975diffreps},
acting on $\trho(-\infty)$.
The definition of the classical scattering generator $\Omega$ is
\begin{align}
    \label{1Odef}
    X_\Omega
    \,=\,
        \log S^*
    \qiq
    S^*
    \,=\,
        \exp\hnem\BB{
            X_\Omega\nem
        }
    \,,
\end{align}
with the identification of vector fields with first-order differential operators.
Namely, $\Omega$ is the generator of the S-symplectomorphism
in the senses of Hamiltonian action and exponential map.
The technical ambiguity in $\Omega$
due to addition of constant functions
will be fixed shortly.
Crucially, the exponential representation in \eqref{1Odef}
manifests symplecticity of classical scattering.

\subsection{Concrete Definition and Computation by Magnus Series}

To be more concrete,
suppose the time-evolution symplectomorphisms
$U^\circ(t_1,t_2)$
and
$U(t_1,t_2)$
arise by Hamiltonian evolutions 
due to 
$H^\circ$ and $H(t)$,
respectively.
Suppose the free-interaction split
\begin{align}
    \label{1hamsplit}
    H(t)
    \,=\,
        H^\circ
        \mem+\mem
        \e\mem V(t)
    \,.
\end{align}
As is well-known,
the Liouville equation is given by
\begin{align}
    \label{liouvilleiff}
    \rho(t_1)
    \,=\,
        U(t_1,t_2)^*\act{
            \rho(t_2)
        }   
    \qfq
    \dot{\rho}(t)
    \,=\,
        X_{H(t)}\act{
            \rho(t)
        }
    \,.
\end{align}
Here, $X_f = \pb{f}{\blank}$ denotes the Hamiltonian vector field of $f$.

In \eqref{liouvilleiff},
the time-evolution diffeomorphism $U(t_1,t_2)$ is concretely represented
as a \textit{differential operator} $U(t_1,t_2)^*$,
which can be
explicitly computed
by Dyson series \cite{Dyson:1949ha}.

After implementing the interaction picture,
it can be shown that
\begin{align}
    \label{liouville.trho}
    \dot{\trho}(t)
    \,=\,
        \e\mem
        X_{\tV(t)}\act{
            \trho(t)
        }
    \,,
\end{align}
where $\tV(t) := V(t) \circ U^\circ(t,t_0)$.
\eqrefs{trhoevolS}{liouville.trho}
lead to the Dyson series representation of
the S-symplectomorphism,
\begin{align}
    \label{1Ssymp:Dyson}
    S^*
    \,=\,
        \mathrm{T}
		\exp\hnem\bb{\mem
            \e\nem
			\int dt\,\,
				X_{\tV(t)}
		\nem}
    \,.
\end{align}
Computing \eqref{1Ssymp:Dyson} could yet be
not practically ideal,
since it generically produces a giant collection of differential operators of all degree
and will violate symplecticity perturbatively 
upon truncation at each finite order.

Since \eqref{1Ssymp:Dyson} represents a symplectomorphism,
its actual content simply boils down to a single Hamiltonian function,
namely
the classical scattering generator
$\Omega$ in \eqref{1Odef}:
\begin{align}
    X_\Omega
    \,=\,
        \log 
        \mathrm{T}
		\exp\hnem\bb{\mem
            \e\nem
			\int dt\,\,
				X_{\tV(t)}
		\nem}
    \,.
\end{align}
Notably, 
the Magnus series \cite{magnus1954exponential}
facilitates the explicit evaluation of $\Omega$ as
\begin{align}
\begin{split}
	\label{1O:Magnus}
	\Omega
	\mem\,=\,\mem
    {}&{}
        \e\nem
		\int dt_1\,\,
			\tV(t_1)
	\mem+\mem
		\frac{\e^2}{2}
		\int_{t_1>t_2}\nem d^2t\,\,
			\pb{\tV(t_1)}{\tV(t_2)}
    \\
    {}&{}
		+
		\frac{\e^3}{6}
		\int_{t_1>t_2>t_3}\nem d^3t\,\,
			\lrp{
            \begin{aligned}[c]
            {}&{}
				\pb{ \tV(t_1) }{ \pb{ \tV(t_2) }{ \tV(t_3) } }
            \\[-.02\baselineskip]
            {}&{}
				+
				\pb{ \tV(t_3) }{ \pb{ \tV(t_2) }{ \tV(t_1) } }
            \end{aligned}}
		+ \cdots
	\,,
\end{split}
\end{align}
where the default integral domains are
$(-\infty,+\infty)$.
To obtain \eqref{1O:Magnus},
one uses the Lie algebra homomorphism
$\comm{X_f}{X_g} = X_{\pb{f}{g}}$
and also 
selects the natural default value
$\Omega \to 0$ in the $\e\to0$ limit
given the perturbative context.

In sum, 
the above exposition
establishes a purely classical definition of the S-symplecto\-morphism $S$
and the classical scattering generator $\Omega$,
concretely and explicitly 
by means of Dyson and Magnus series.
Notably, the classical scattering generator
serves as an efficient data type
that is easy to handle
while
encoding the full information of classical scattering.

Diagrammatically, \eqref{1O:Magnus}
amounts to using
the retarded Green's function
arising from the Poisson bracket between
free-theory trajectories \cite{KKKL}.

Although we have assumed a Hamiltonian deformation scenario in \eqref{1hamsplit}
for concreteness,
the definitions in \eqrefs{Ssymp-def0}{1Odef}
are coordinate-free,
implying worldline field basis invariance.
The symplectic deformation scenario stands much subtler,
but it is being observed 
by examples
that the same diagrammatic rules can be used
(see, e.g., \rcite{Kim:2025olv}).

\subsection{Via Dequantization}
\label{SSYMP>DEQUANT}

The phase space formulation of quantum mechanics
\cite{Moyal:1949sk,Groenewold:1946kp,weyl1927quantenmechanik,Wigner:1932eb,Curtright:2011vw,zachos2005quantum,
leonhardt1997measuring}
is based on the idea of ``quantization map,''
which associates each function on a phase space $\P$
with an operator acting on a Hilbert space $\Hilb$:
\begin{align}
    \label{QuantMap}
	\Q
	\,\,\,:\,\,\,
	\Cinfty(\P) 
	\,\,\to\,\,
	\Ops(\Hilb)
	\,.
\end{align}
For simplicity and concreteness,
we also restrict our interest
to the cases in which
this quantization map is invertible:
\begin{align}
    \label{DeQuantMap}
	\Q^{-1}
	\,\,\,:\,\,\,
	\Ops(\Hilb)
	\,\,\to\,\,
	\Cinfty(\P) 
	\,.
\end{align}
As a result, each operator $\hat{f}$ is associated with a unique phase space function $f = \Q^{-1}\hnem(\hat{f})$,
known as the \textit{symbol} of $\hat{f}$.
The inverse map $\Q^{-1}$ is colloquially referred to as the dequantization map.

Well-known examples of the quantization/dequantization maps include
the Weyl and Wigner transforms \cite{weyl1927quantenmechanik,Wigner:1932eb}
as well as
Glauber-Sudarshan $P$-map
\cite{glauber1963quantum,sudarshan1963equivalence}
and
Husimi $Q$-representation \cite{husimi1940some}.
For spinning systems,
elegant formalisms exist via compact phase spaces
(see, e.g., \rcite{Varilly:1989sv}),
while we underscore the fact that
the massive twistor formalism
arguably provides the best framework for quantizing/dequantizing relativistic spin in four dimensions
via its coherent states and Darboux nature.

The phase space formulation of quantum mechanics
describes the category of noncommutative manifolds,
for which the star product is given by
\begin{align}
    f
    \,=\,
        \Q^{-1}\hnem(\hat{f})
    \,,\quad
    g
    \,=\,
        \Q^{-1}\hnem(\hat{g})
    \qiq
    f \star g
    \,=\,
        \Q^{-1}\hnem(\hat{f}\hat{g})
    \,.
\end{align}
Automorphisms of the noncommutative ring of smooth functions
(i.e., maps preserving the pointwise addition and the star product of smooth functions)
will be identified as fuzzy diffeomorphisms.

The fuzzy Poisson bracket is defined as
\begin{align}
    \label{fuzPB}
    \pb{f}{g}^\star
    \,=\,
        \frac{1}{i\hbar}\,
        \Q^{-1}\hnem(\comm{\hat{f}}{\hat{g}})
    \,=\,
        \frac{1}{i\hbar}\mem
        \BB{
            f \star g - g \star f
        }
    \,.
\end{align}
The fuzzy Hamiltonian vector field is defined as
\begin{align}
    X^\star_f
    \,=\,
        \pb{f}{\blank}^\star
    \qiq
    \comm{X^\star_f}{X^\star_g}
    \,=\,
        X^\star_{\pb{f}{g}^\star}
    \,,
\end{align}
which is a derivation on the noncommutative ring.
In general,
a map on $\Cinfty(\P)$
will be called a fuzzy vector field on $\P$
if it is a derivation on the noncommutative ring.
Concretely, this could describe a giant sum of differential operators of all degree.

The quantum scattering generator
refers to
the logarithm of the S-matrix:
\begin{align}
    \hat{\Omega}
    \,=\,
        i\hbar\mem \log\hat{S}
    \mem\,\,\,\in\,\,\,
        \Ops(\Hilb)
    \qiq
    \hat{S}
    \,=\,
        \exp\hnem\bb{
            \frac{1}{i\hbar}\mem
            \hat{\Omega}
        }
    \,.
\end{align}
In the phase space formulation,
its symbol is given by
\begin{align}
    \label{1Omega*}
    \Omega^\star
    \,\,:=\,\,
        \mathcal{Q}^{-1}\hnem\bigbig{\mem
            i\hbar
            \log \hat{S}
        \mem}
    \,.
\end{align}
Consider the adjoint action of the S-matrix
on quantum observables:
\begin{align}
    \Ad{\hat{S}}\mem \hat{\O}
    \,=\,
        \exp\hnem\bb{\nem{
            \frac{1}{i\hbar}\mem
            \ad{\hat{\Omega}}
        }\nem}\mem
        \hat{\O}
    \,.
\end{align}
Upon dequantization,
it follows that
\begin{align}
    \label{dequantize-Adaction}
    \Q^{-1}\hnem\BB{
        \Ad{\hat{S}}\mem \hat{\O}
    }
    \,=\,
        \exp\hnem\BB{\nem{
            X^\star_{\Omega^\star}
            \nem
        }\nem}\act{
            \O
        }
    \,.
\end{align}
Here, $\O = \Q^{-1}\hnem(\hat{\O})$.
Since $X^\star_{\Omega^\star}$ is a derivation on the noncommutative ring,
i.e., is a fuzzy vector field,
its exponentiation in \eqref{dequantize-Adaction}
defines an automorphism,
i.e., a fuzzy diffeomorphism:
\begin{align}
    \label{1Ssymp*}
    S^\star
    \,:=\,
        \exp\hnem\BB{\nem{
            X^\star_{\Omega^\star}
            \nem
        }\nem}
    \,\,\,\in\,\,\,
        \Diff^\star(\P)
    \,.
\end{align}
This means that
\begin{align}
    S^\star\act{
        f + g
    }
    \,=\,
        S^\star\act{f}
        + S^\star\act{g}
    \,,\quad
    S^\star\act{
        f \star g
    }
    \,=\,
        S^\star\act{f} \star S^\star\act{g}
    \,.
\end{align}
We refer to $S^\star$ in \eqref{1Ssymp*}
as the fuzzy S-diffeomorphism.
This is essentially the phase space formulation of the (adjoint) S-matrix.

In typical situations,
the symbol of the quantum scattering generator
admits a well-defined classical limit,
dubbed classical scattering generator
(reproducing \eqref{1Omega}):
\begin{align}
    \label{1Omega(re)}
    \Omega
    \,\,:=\,\,
        \lim_{\hbar\to0}\mem
            \Omega^\star\nem
    \,\,=\,\,
        \lim_{\hbar\to0}\mem
        \mathcal{Q}^{-1}\hnem\bigbig{\mem
            i\hbar
            \log \hat{S}
        \mem}
    \,.
\end{align}
Accordingly,
the automorphism in \eqref{1Ssymp*}
also admits a well-defined classical limit,
say $S^*$:
\begin{align}
    \label{1Ssymp}
    S^*
    \,\,:=\,\,
        \lim_{\hbar\to0}\mem
            S^\star
    \,\,\,\in\,\,\,
        \Diff(\P,\pb{\blank}{\nem\blank})
    \,.
\end{align}
Since the classical limit of the fuzzy Poisson bracket in \eqref{fuzPB}
defines the Poisson bracket of the classical theory,
$S^*$ in \eqref{1Ssymp}
is an automorphism of the Poisson algebra,
i.e., represents a Poisson diffeomorphism $S$.
Explicitly, 
\eqref{1Ssymp*} shows that
it is the exponentiation of the Hamiltonian vector field of the classical scattering generator:
\begin{align}
    \label{def-Ssymp}
    S^*
    \,\,=\,\,
        \exp\hnem\BB{\nem{
            X_\Omega
        }\nem}
    \,\,\,\in\,\,\,
        \Diff(\P,\pb{\blank}{\nem\blank})
    \,.
\end{align}
In sum,
the S-symplectomorphism $S$
is the dequantization
of the adjoint action of the S-matrix,
up to the differential operator representation.

Finally,
the impulse of quantum observables
in quantum scattering
is defined from
the adjoint action of the inverse S-matrix:\footnote{
    To be more precise, 
    the quantum impulse should be defined
    as the expectation value of this $\Del\hat{\O}$
    via a density matrix.
    The symbol of the density matrix
    evolves by the dequantization of the adjoint S-matrix,
    i.e., the fuzzy S-diffeomorphism.
}
\begin{align}
    \label{1DelOhat}
    \Del \hat{\O}
    \,=\,
        \hat{S}^{-1} \hat{O}\mem \hat{S}
        \mem-\mem \hat{O}
    \,=\,
        \Ad{\hat{S}^{-1}} \hat{O}
        \mem-\mem \hat{O}
    \,.
\end{align}
It should be clear that the dequantization of
\eqref{1DelOhat}
is
\begin{align}
    \label{impulseformula-deriv}
    \Del \O
    \,=\,
        (S^{-1})^*\act{
            \O
        }
        \mem-\mem \O
    \,=\,
        \mathe^{-X_\Omega}\hhnem
        \act{\O}
        \mem-\mem \O
    \,=\,
        \mathe^{\pb{\blank}{\Omega}}\hhnem
        \act{\O}
        \mem-\mem \O
    \,,
\end{align}
which reproduces \eqref{impulseformula}
when expanded.
That is,
the impulse of classical observables in
a classical scattering process
is computed via
the pullback of the S-symplectomorphism.
This fact should be clear by recalling \eqref{pullbackdef},
in which case \eqref{impulseformula-deriv} is brought to
\begin{align}
    \Del\O
    \,=\,
        \O \circ S
        \mem-\mem \O
    \,.
\end{align}

\subsection{In Deformation Quantization}

The classical scattering generator
is well-defined on Poisson manifolds \cite{Kim:2025sey}.
Flipping the perspective,
one could consider starting from the classical scattering map $S$
and $\hbar$-deforming it into a quantum S-diffeomorphism $S^\star$,
without presuming 
the existence of 
the operators $\hat{S}$ and $\hat{\Omega}$
on a concrete Hilbert space.

In formal (not necessarily strict) deformation quantization
\cite{bayen1977quantum,bayen1978deformationI,bayen1978deformationII,berezin1975general,fedosov1994simple,kontsevich,weinstein1995deformation},
$S^\star$ and $\Omega^\star$
will be defined as formal power series in $\hbar$.
This essentially means to ignore the convergence properties,
and for physicists,
work perturbatively in the loop expansion
yielding asymptotic series.

\printindex

\end{document}